\documentclass[a4paper,
12pt,
openany]{memoir} 

\newcommand\authorname {Gonzague Paul-Constantin Hugues-Antoine \textsc{Radureau}}

\newcommand\FrenchTitle {Apprentissage machine pour l'hydrodynamique radiative en astrophysique}
\newcommand\EnglishTitle {Machine learning for radiative hydrodynamics in astrophysics}

\newcommand\FrKeywords {Hydrodynamique radiative, Intelligence Artificielle, Choc radiatifs, Simulation numérique}
\newcommand\EngKeywords {Radiative hydrodynamics, Artificial Intelligence, Radiative shocks, Numerical simulation}

\newcommand\field {Sciences de la Planète et de l'Univers }
\newcommand\university {Université Côte d'Azur }
\newcommand\lab {Laboratoire J.-L. Lagrange}

\newcommand\supervisor {Claire \textsc{Michaut}}

\usepackage{stmaryrd}        
\SetSymbolFont{stmry}{bold}{U}{stmry}{m}{n}
\usepackage{verbatim}        

\usepackage[english]{datetime2}

\usepackage{pifont}   
\usepackage{calc,soul} 
\usepackage{geometry}
\usepackage{calligra}     

\usepackage{tikz}          
\usetikzlibrary{shapes.geometric, arrows.meta, positioning, calc}

\usepackage{ragged2e}
\justifying

\usepackage[colorlinks=true,allcolors=black]{hyperref} 
\usepackage{footnote} 
\usepackage{tcolorbox} 

\usepackage{multirow}                   
\usepackage{array}                      
\usepackage{subcaption}                 
\usepackage{arydshln}

\usepackage{enumitem}
\usepackage{enumerate}

\usepackage[intlimits]{amsmath} 
\usepackage{empheq} 

\let\iiint\relax

\usepackage{esint}  

\usepackage{bbm}
\usepackage{amssymb}            
\usepackage{tensor}            
\usepackage{bm}                 

\settrimmedsize{297mm}{210mm}{*}
\settypeblocksize{634pt}{448.13pt}{*} 
\setulmargins{4cm}{*}{*} 
\setlrmargins{*}{*}{1.5} 
\setmarginnotes{17pt}{51pt}{\onelineskip} 
\setheadfoot{2\onelineskip}{2\onelineskip} 
\setheaderspaces{*}{2\onelineskip}{*} 
\checkandfixthelayout

\makepagestyle{myvf} 
\makeoddfoot{myvf}{}{\thepage}{} 
\makeevenfoot{myvf}{}{\thepage}{} 
\makeheadrule{myvf}{\textwidth}{\normalrulethickness} 
\makeevenhead{myvf}{\small\textsc{\leftmark}}{}{} 
\makeoddhead{myvf}{}{}{\small\textsc{\rightmark}}
\OnehalfSpacing 
\makeatletter 

\newsavebox{\feline@chapter} 
\newcommand\feline@chapter@marker[1][4cm]{%
	\sbox\feline@chapter{%
		\resizebox{!}{#1}{\fboxsep=1pt%
			\colorbox{gray}{\color{white}\thechapter}%
		}}%
		\rotatebox{90}{
			\resizebox{%
				\heightof{\usebox{\feline@chapter}}+\depthof{\usebox{\feline@chapter}}}%
			{!}{\scshape\so\@chapapp}}\quad%
		\raisebox{\depthof{\usebox{\feline@chapter}}}{\usebox{\feline@chapter}} 
} 

\newcommand\feline@chm[1][4cm]{%
	\sbox\feline@chapter{\feline@chapter@marker[#1]}%
	\makebox[0pt][c]{
		\makebox[1cm][r]{\usebox\feline@chapter}%
	}}

\makechapterstyle{daleifmodif}{

	\renewcommand\printchapternum{\null\hfill\feline@chm[2.5cm]\par} 

} 
\makeatother 

\chapterstyle{daleifmodif}

\setsecnumdepth{subsection}

\usepackage[acronyms,toc,nonumberlist]{glossaries-extra} 
\setabbreviationstyle[acronym]{long-short}

\newacronym{mlp}{MLP}{Multi-Layer Perceptron}
\newacronym{vae}{VAE}{Variational Autoencoders}
\newacronym{pinn}{PINN}{Physics-Informed Neural Networks}
\newacronym{ai}{AI}{Artificial Intelligence}
\newacronym{lte}{LTE}{Local Thermodynamic Equilibrium}
\newacronym{neuhpc}{NeuHPC}{Neural High-Performance Computing}
\newacronym{lbfgs}{L-BFGS}{Limited-memory Broyden-Fletcher-Goldfarb-Shano}
\newacronym{bfgs}{BFGS}{Broyden-Fletcher-Goldfarb-Shano}
\newacronym{adam}{Adam}{Adaptative Moment Estimation}
\newacronym{sgd}{SGD}{Stochastic Gradient Descent}
\newacronym{pde}{PDE}{Partial Differential Equations}
\newacronym{cnn}{CNN}{Convolutional Neural Networks}
\newacronym{rnn}{RNN}{Recurent Neural Networks}
\newacronym{lstm}{LSTM}{Long Short-Term Memory}
\newacronym{gan}{GAN}{Generative Adversarial Networks}
\newacronym{bnn}{BNN}{Bayesian Neural Networks}
\newacronym{hades}{HADES}{Hydrodynamique Adaptée à la Description des Ecoulements Supersoniques}
\newacronym{cfl}{CFL}{Courant–Friedrichs–Lewy}
\newacronym{icf}{ICF}{Inertial Confinement Fusion}
\newacronym{fld}{FLD}{Flux Limited Diffusion}
\newacronym{vtef}{VTEF}{Variable Tensor Eddington Factor}
\newacronym{ode}{ODE}{Ordinary Differential Equation}
\newacronym{hlle}{HLLE}{Harten-Lax-van Leer-Einfeld}
\newacronym{hll}{HLL}{Harten-Lax-van Leer}
\newacronym{cmb}{CMB}{Cosmological Background}
\newacronym{ska}{SKA}{Square Kilometer Array}

\makenoidxglossaries

\usepackage[T2A,T1]{fontenc}
\usepackage[russian,english]{babel}	
\usepackage{newtxtext} 
\usepackage{tempora}   
\usepackage{newtxmath} 

\DeclareFontFamilySubstitution{T2A}{ntxtlf}{Tempora-TLF}

\usepackage{csquotes}
\usepackage[
  backend=biber,
  sorting=none,         
  maxnames=3,           
  minnames=1,
  giveninits=true       
]{biblatex} 

\newcommand{\clearemptydoublepage}{\newpage{\thispagestyle{empty}\cleardoublepage}}

\usepackage{lettrine}
\newcommand{\initialletter}[1]{%
    \lettrine[lines=3,lhang=0.33,nindent=0em]{
        \color{gray}
            {\textsc{#1}}}{}}

\newcommand{\HRule}{\rule{\linewidth}{0.5mm}}

\newcommand{\quotes}[1]{``#1''}

\newcommand{\refmark}[1]{\overset{\circ}{\mathrm{#1}}}

\newcommand{\sgn}[1]{\mathrm{sgn}(#1)}

\newcommand{\starsect}[1]{
\vspace{\baselineskip}

\noindent \textbf{\ding{118} #1}

\vspace{\baselineskip}
}

\newcommand{\substarsect}[1]{
\vspace{\baselineskip}

\noindent \textbf{\underline{#1}}

\vspace{\baselineskip}
}

\newcommand{\closeinterv}[2]{\left[ #1\,;\,#2 \right]}
\newcommand{\openinterv}[2]{\left] #1\,;\,#2 \right[}
\newcommand{\openrinterv}[2]{\left[ #1\,;\,#2 \right[}
\newcommand{\openlinterv}[2]{\left] #1\,;\,#2 \right]}

\allowdisplaybreaks

\newcommand{\vectorr}[1]{\overrightarrow{#1}}                                                        
\newcommand{\arctanh}[1]{\textnormal{arctanh} #1}                                         

\newcommand*\dif{\mathop{}\!\mathrm{d}}

\newcommand*\nablav{\vectorr{\nabla}}

\newcommand*\tensorr[1]{\mathbbm{#1}}

\newcommand*\identity{\mathbbm{I}_\mathrm{d}}

\newcommand*\indicatrix{\mathbbm{1}}

\newcommand*\vectorIA[1]{\boldsymbol{#1}}

\newcommand{\bcdot}{\raisebox{0.25ex}{\tikz\filldraw[black,x=2pt,y=2pt] (0,0) circle (0.6);}}

\newcommand*\Li[1]{\mathrm{Li}_{#1}}

\newcommand*\tending[2]{\underset{#1\to#2}{\longrightarrow}}

\newcommand*\entiere[1]{\left \lfloor #1 \right \rfloor}

\newcommand{\asinh}{\mathrm{asinh}}

\newcommand{\secref}[1]{\ref{#1}.~\textit{\nameref{#1}}}

\newcolumntype{L}[1]{>{\raggedright\arraybackslash}p{#1}}
\newcolumntype{C}[1]{>{\centering\arraybackslash}p{#1}}
\newcolumntype{R}[1]{>{\RaggedLeft\arraybackslash}p{#1}}  

\newcommand\Tstrut{\rule{0pt}{2.5ex}}        
\newcommand\Bstrut{\rule[-1.1ex]{0pt}{0pt}}  
\newcommand{\TBstrut}{\Tstrut\Bstrut}        

\makeindex

\begin{document}
\frontmatter
\pagenumbering{roman}

\begin{adjustwidth}{-2 cm}{-3 cm}
    \begin{tikzpicture}[remember picture, overlay]
        \node[anchor=north west] at ([xshift=-1.48mm, yshift=5mm] current page.north west)
            {\includegraphics[width=\paperwidth]{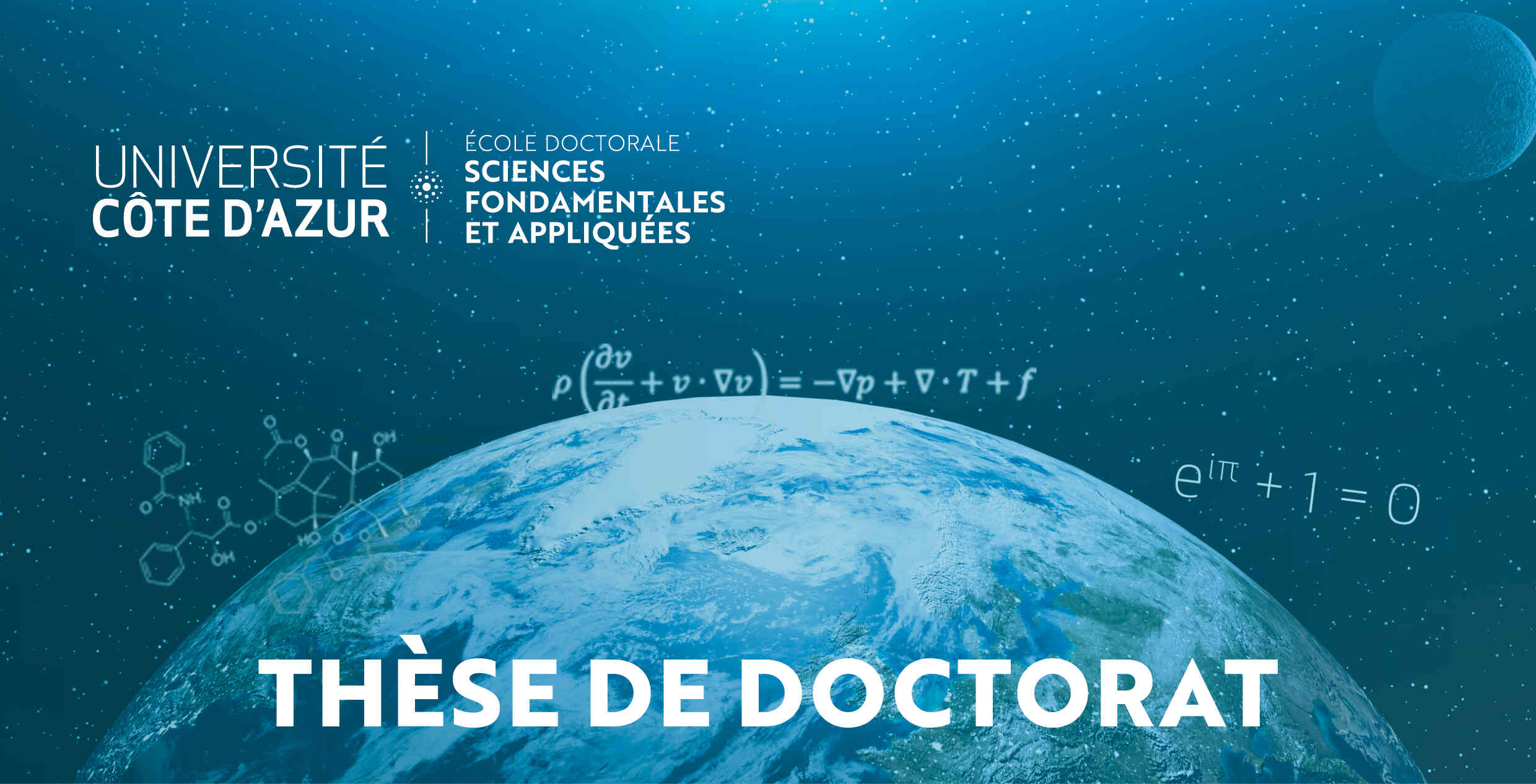}};
    \end{tikzpicture}
    
    \vspace*{6cm} 
    
    \begin{center}    
        \parbox{1.3\textwidth}{\centering \LARGE \bfseries \EnglishTitle\par}\vspace{0.6cm} 

        \parbox{1.2\textwidth}{\centering \large \authorname \par}\vspace{0.4cm} 

        \parbox{1.2\textwidth}{\centering \normalsize \lab \par}\vspace{0.4cm} 

        \parbox{1.2\textwidth}{\centering \normalsize \textit{Translated by the author, based on the official version archived on theses.fr (ID~2025COAZ5041).}}\vspace{0.6cm}

        \parbox{1.2\textwidth}{
        \begin{minipage}[t]{0.37\textwidth}
            \begin{flushleft} 
                \small
                \textbf{Presented with a view to obtaining the degree of doctor in}\\
                \field d'\university

                \textbf{Supervised by:} \supervisor

                \textbf{defended the:} 30 Septembre 2025
            \end{flushleft}
        \end{minipage}
        \hspace{1.5cm}
        \begin{minipage}[t]{0.63\textwidth}
            \begin{flushleft} 
                \small
                \textbf{Before the jury composed of:}\\
                Jérôme Breil,\\
                \textit{Research Director CEA, CEA-CESTA, Le Barp, France}\\
                Emmanuel Franck,\\
                \textit{Research Scientist, INRIA, Strasbourg, France}\\
                Stéphane Mazevet,\\
                \textit{Director of OCA, Nice, France}\\
                Florence Hubert,\\
                \textit{Professor, I2M, Marseille, France}\\
                Serge Bouquet,\\
                \textit{Emeritus Research Director CEA, Paris, France}\\
                Claire Michaut,\\
                \textit{Research Director, OCA, Nice, France}\\
                Andrew Comport,\\
                \textit{Research Scientist, I3S, Sophia-Antipolis, France}
            \end{flushleft}
        \end{minipage}
        }

        \vspace{1cm}

        \parbox{1.2\textwidth}{
            \centering
            \includegraphics[height=1.5cm]{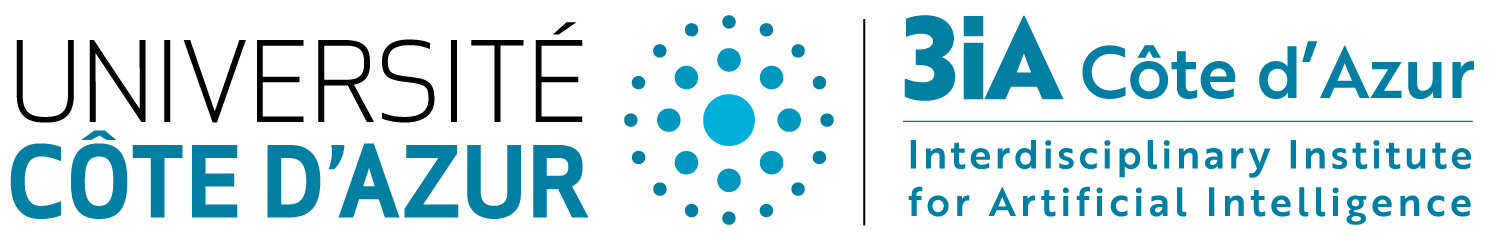}
        }
    \end{center}
\end{adjustwidth}

\thispagestyle{empty}
\clearemptydoublepage

\begin{adjustwidth}{-2 cm}{-3 cm}
    \begin{center}
        \parbox{\textwidth}{\centering \huge \textsc{\EnglishTitle}\par}
    \end{center}
\end{adjustwidth}

\vspace{1.5cm}

\Large \noindent \textbf{\textsc{Jury:}}

\vspace{0.2cm}

\Large \noindent \textbf{\textsc{Chair of the Jury}}

\vspace{3mm}

\normalsize
\noindent Stéphane Mazevet,\\
\noindent \textit{Director of OCA, Nice, France}

\vspace{0.4cm}

\Large \noindent \textbf{\textsc{Reviewers}}

\vspace{3mm}

\normalsize
\noindent Jérôme Breil,\\
\noindent \textit{Research Director, CEA, Le Barp, France}

\vspace{5mm}

\noindent Emmanuel Franck,\\
\noindent \textit{Research Scientist, INRIA, Strasbourg, France}

\vspace{0.4cm}

\noindent \Large \textbf{\textsc{Examiners}}

\vspace{3mm}

\normalsize
\noindent Florence Hubert,\\
\noindent \textit{Professor, I2M, Marseille, France}

\vspace{5mm}

\noindent Serge Bouquet,\\
\noindent \textit{Emeritus Research Director, CEA, Paris, France}

\vspace{0.4cm}

\Large \noindent \textbf{\textsc{Supervisor}}

\vspace{3mm}

\normalsize
\noindent Claire Michaut,\\
\noindent \textit{Research Director, OCA, Nice, France}

\vspace{0.4cm}

\Large \noindent \textbf{\textsc{Co-supervisor}}

\vspace{3mm}

\normalsize
\noindent Andrew Comport,\\
\noindent \textit{Research Scientist, I3S, Sophia-Antipolis, France}

\newpage
\addcontentsline{toc}{chapter}{Abstract} 
\pdfbookmark[0]{Abstract}{Abstract}

\thispagestyle{empty}

{\vspace*{-3cm}
\centering 

\HRule   

\vspace{-0.1cm}

\Large \textbf{\EnglishTitle}

\vspace{-0.1cm}

\HRule}

\raggedright
\normalsize \textbf{Abstract}\\

\justifying
\small 
Radiation hydrodynamics studies the interaction between the motion of a high-temperature hypersonic plasma and the radiation it emits or absorbs. This coupling is central to many astrophysical phenomena, particularly those linked to the processes of accretion and ejection. The HADES code has been developed to model such systems. It provides a realistic model of radiative transport, which is particularly crucial for optically intermediate media. The code HADES couples the hydrodynamic equations with M1-gray or M1-multigroup radiative transfer models.

However, radiative hydrodynamics simulations remain extremely costly in terms of computing time, due to two main limitations. Firstly, the M1-multigroup model is based on a closure relation that has no analytic form, requiring costly numerical estimates. Secondly, the Courant-Friedrichs-Lewy condition strongly constrains the explicit schemes used in HADES, imposing very small time steps. In order to overcome these limitations, two complementary strategies based on Artificial Intelligence were developed during the course of this thesis.

The first involves training a neural network of the Multi-Layer Perceptron type to approximate the closure relation of the M1-multigroup model. This innovative approach achieves excellent accuracy while reducing the computational cost by a factor $3\,000$, making it to date the most efficient method known for this type of calculation. This substantial increase in performance has enabled us to carry out simulations of radiative shocks - shock waves in which the radiation directly influences the structure and dynamics of the front - with an unprecedented level of accuracy. In particular, we have been able to quantify the influence of a detailed spectral description of the radiation: the more detailed the description, the more the shock is slowed down, and the larger the size of the radiative precursor.

The second approach is based on the use of Physics-Informed Neural Networks to solve the radiative hydrodynamics equations directly and extrapolate the simulations beyond their initial time domain. These networks, which are also based on Multi-Layer Perceptrons, explicitly integrate the physical equations into their cost function, thereby guiding learning towards solutions that are consistent with the fundamental laws of physics. Initial tests, carried out on purely hydrodynamic shock configurations, show that this method can deal effectively with discontinuities and faithfully reconstruct the evolution of the shock at later times. On the other hand, the application to radiative shocks is proving more difficult~: the networks are struggling to extrapolate the profiles correctly, due to the high residuals associated with the radiative hydrodynamics equations, which indicates that the physical constraints are not being properly respected. This last phase still requires in-depth investigation in order to understand the origin of these difficulties and to propose appropriate improvement strategies.

\vspace{0.1cm}

\noindent \raggedright \small
\textbf{Keywords~:} \EngKeywords

\newpage
\thispagestyle{empty}

{\vspace*{-3cm}

\centering 

\HRule   

\vspace{-0.1cm}

\Large \textbf{\FrenchTitle}

\vspace{-0.1cm}

\HRule}  

\raggedright
\normalsize \textbf{Résumé}\\

\vspace{0.4cm}

\justifying
\small 
L'hydrodynamique radiative étudie l'interaction entre le mouvement d'un plasma hypersonique souvent à haute température et le rayonnement qu'il émet ou absorbe. Ce couplage est central dans de nombreux phénomènes astrophysiques, notamment ceux liés aux processus d'accrétion et d'éjection. Pour modéliser de tels systèmes, le code HADES a été développé. Il offre une modélisation réaliste du transport radiatif, particulièrement cruciale pour les milieux optiquement intermédiaires. Le code HADES couple les équations de l'hydrodynamique avec des modèles de transfert radiatif de type M1-gris ou M1-multigroupe. 

Cependant, les simulations en hydrodynamique radiative demeurent extrêmement coûteuses en temps de calcul, en raison de deux limitations principales. D'une part, le modèle M1-multigroupe repose sur une relation de fermeture qui n'admet pas de forme analytique, ce qui impose des estimations numériques coûteuses. D'autre part, la condition de Courant–Friedrichs–Lewy contraint fortement les schémas explicites utilisés dans HADES, imposant des pas de temps très réduits. Afin de surmonter ces limitations, deux stratégies complémentaires basées sur l'Intelligence Artificielle ont été développées au cours de cette thèse. 

La première consiste à entraîner un réseau de neurones de type Multi-Layer Perceptron pour approcher la relation de fermeture du modèle M1-multigroupe. Cette approche novatrice atteint une excellente précision tout en réduisant le coût de calcul d'un facteur $3\,000$, faisant d'elle à ce jour la méthode la plus efficace connue pour ce type de calcul. Ce gain substantiel de performance nous a permis de réaliser des simulations de chocs radiatifs, des ondes de choc où le rayonnement influence directement la structure et la dynamique du front, avec un niveau de précision inédit. En particulier, nous avons pu quantifier l'influence d'une description spectrale fine du rayonnement~: plus cette description est détaillée, plus le choc est ralenti, et plus la taille du précurseur radiatif augmente. 

La seconde approche repose sur l'utilisation des Physics-Informed Neural Networks pour résoudre directement les équations de l'hydrodynamique radiative et extrapoler les simulations au-delà de leur domaine temporel initial. Ces réseaux, également basés sur des Multi-Layer Perceptrons, intègrent explicitement les équations physiques dans leur fonction de coût, orientant ainsi l'apprentissage vers des solutions cohérentes avec les lois fondamentales de la physique. Les premiers tests, menés sur des configurations de chocs purement hydrodynamiques, montrent que cette méthode permet de traiter efficacement les discontinuités et de reconstituer fidèlement l'évolution du choc à des temps ultérieurs. En revanche, l'application aux chocs radiatifs s'avère plus délicate~: les réseaux peinent à extrapoler correctement les profils, en raison de résidus élevés associés aux équations de l'hydrodynamique radiative, ce qui indique un mauvais respect des contraintes physiques. Cette dernière phase nécessite encore des investigations approfondies afin de comprendre l'origine de ces difficultés et de proposer des stratégies d'amélioration adaptées.

\vspace{0.1cm}

\noindent \raggedright \small
\textbf{Mots clés~:} \FrKeywords

\chapter*{Acknowledgements}
\phantomsection
\addcontentsline{toc}{chapter}{Acknowledgements} 
\pdfbookmark[1]{Acknowledgements}{Acknowledgements}

\normalsize 
\setlength{\parindent}{20pt} 
\setlength{\parskip}{4pt} 
\justifying

\begin{SingleSpace}
    Je souhaite tout d'abord exprimer ma profonde gratitude à Claire Michaut, pour m'avoir proposé ce sujet de thèse et pour m'avoir initié à l'univers de l'hydrodynamique radiative. Merci pour ton encadrement, ta disponibilité et tous tes précieux conseils tout au long de ces années. J'adresse également mes remerciements à Andrew Comport, qui, malgré des difficultés personnelles, m'a grandement aidé à développer les méthodes présentées dans ce manuscrit, en partageant avec moi de nombreuses idées et en m'orientant vers des pistes existantes à explorer.

    Je remercie chaleureusement mes rapporteurs, Jérôme Breil et Emmanuel Franck, pour avoir accepté d'évaluer ce travail et pris le temps d'en rédiger un rapport. J'adresse également mes remerciements à l'ensemble du jury, Stéphane Mazevet, Florence Hubert et Serge Bouquet, pour leur présence et leur participation à ma soutenance.
    
    Je suis reconnaissant aux membres de mon comité de suivi de thèse, André Ferrari et Frédéric Paletou, pour le temps qu'ils ont consacré chaque année à suivre l'avancement de mes travaux et pour leur écoute bienveillante. Je souhaite remercier tout particulièrement Frédéric Paletou~: sans lui, je n'aurais probablement jamais entrepris une thèse en hydrodynamique radiative. Ses cours de M2 sur le transfert radiatif et ses nombreuses digressions sur leurs applications m'ont profondément inspiré et donné l'envie d'approfondir ce domaine. Mes remerciements vont également à Malik Chami, avec qui j'ai pu avoir des échanges enrichissants sur l'avenir de l'intelligence artificielle dans nos disciplines, et qui m'a plusieurs fois débloqué sur des questions techniques, malgré la distance entre nos champs de recherche.
    
    Je tiens à remercier le professeur Vladimir Mikhailovitch Lipunov, qui m'a donné mes tout premiers cours de transfert radiatif à l'Université de Moscou. Malgré l'exigence de son enseignement, ces cours ont suscité en moi une véritable passion pour la discipline. Je remercie également le professeur Alexeï Nikolaevitch Rubtsov, qui m'a fait découvrir les méthodes d'intelligence artificielle appliquées à l'astrophysique et éveillé un vif intérêt pour ce domaine en plein essor.
    \begin{otherlanguage}{russian}
    Без вашего преподавания эта диссертация была бы невозможна. Я хотел бы выразить вам самую искреннюю благодарность.
    \end{otherlanguage}
    
    Je souhaite exprimer ma reconnaissance à Andrea Chiavassa, pour son aide précieuse dans la recherche d'opportunités après la thèse, et pour nos nombreuses discussions sur l'hydrodynamique radiative et les codes de simulation. J'espère vivement que nos chemins se recroiseront à l'avenir et que nous aurons l'occasion de collaborer.

    Je remercie l'Institut 3IA Côte d'Azur et l'Université Côte d'Azur pour avoir financé cette thèse et rendu possible tout le travail présenté dans ce manuscrit.
    
    Mes remerciements s'adressent également au frère Manuel-Marie, pour ses conseils éclairés sur le sens du travail de recherche et la gestion du temps. Je tiens à exprimer ma gratitude à Sophie et Loïc, qui animent avec bienveillance l'aumônerie de Nice et ont été des hôtes exceptionnels au foyer Saint Jean-Paul II et bienheureux Christian Chessel.

    Je souhaite remercier de tout cœur ma famille, et plus particulièrement ma mère, pour son soutien indéfectible tout au long de ces années. J'adresse également un chaleureux merci à Alessandra Marelli, pour sa présence et son soutien constant.
    
    Je suis profondément reconnaissant envers Khaled, qui régale chaque jour les membres de l'Observatoire avec ses plats délicieux à la cantine, et qui crée des moments de détente inoubliables grâce à ses blagues et à la bonne humeur qu'il nous transmet quotidiennement. Sans ton enthousiasme et ton talent, la vie à l'Observatoire de la Côte d'Azur ne serait pas la même~!

    Je remercie également Corentin Pecontal, qui m'a grandement aidé lors des tests de PINNs durant les derniers mois de ma thèse, alors que je rédigeais ce manuscrit. J'ai réellement apprécié ton indépendance, ta curiosité et ton enthousiasme : j'espère que nous aurons l'occasion de collaborer de nouveau~!
    
    Enfin, j'adresse mes remerciements les plus chaleureux à tous les collègues et amis rencontrés à l'Observatoire~: Paul, Louise, Manon, Elisa, Chloé, Gabriel, Marie, Katherine, Kate, Adrien, Fabiola et Carlo. Votre présence et votre camaraderie ont rythmé et enrichi ces trois années de thèse. Un grand merci également à tous mes amis niçois pour les soirées dansantes, les balades et tous les souvenirs partagés. Je pense en particulier à Carlotta, Campe, Mariam, Ségolène, Benoît, Guillaume, Alban, Claire, Timothée et Pierre~: merci pour tous ces moments de joie. Je souhaite enfin remercier chaleureusement Anna Roussel, dont le soutien et les nombreuses idées m'ont accompagné une grande partie de cette thèse.
\end{SingleSpace}
\clearpage

\newpage
\vspace*{0.2\textheight}

\normalsize

\centering{\enquote{\textit{Ce que vous faites dans la vie résonne dans l'éternité.}}}\bigbreak

\hfill Film Gladiator

\vspace{5cm}

\begin{center}
    \enquote{\textit{Oui, sans cesse un monde se noie\\
    Dans les feux d'un nouveau soleil,\\
    Les cieux sont toujours dans la joie;\\
    Toujours un astre a son réveil,\\
    Partout où s'abaisse ta vue,\\
    Un soleil levant te salue,\\
    Les cieux sont un hymne sans fin! }}
\end{center}\bigbreak

\RaggedRight Hymne du matin, Harmonies poétiques et religieuses, Lamartine

\justifying
\clearemptydoublepage

\addcontentsline{toc}{chapter}{Table of Content}
\renewcommand{\contentsname}{Table of Content}
\maxtocdepth{paragraph}
\tableofcontents*

\newpage
\listoffigures
\addtocontents{lof}
{\par\nobreak\textbf{{\scshape Figure} \hfill Page}\par\nobreak}

\newpage
\listoftables
\addtocontents{lot}
{\par\nobreak\textbf{{\scshape Table} \hfill Page}\par\nobreak}

\newpage
\glsaddall[types={\acronymtype}]
\printnoidxglossary[type=\acronymtype, title=Acronyms, toctitle=List of Acronyms]
\clearemptydoublepage

\mainmatter


\chapter{Introduction} \label{ch:chapitre1} 

\initialletter{A}strophysics constitutes a true natural laboratory in which physical theories can be tested under extreme conditions, often inaccessible on Earth. It thus enables us to refine our models, verify the validity of our understanding of the fundamental laws of physics in such regimes, and deepen our knowledge of how the Universe operates. This field has, for instance, revealed that time is not an absolute quantity but a relative one: it depends in particular on the velocity of the observer and on the gravitational field in which they evolve. By its very nature, astrophysics is interdisciplinary, drawing on all branches of physics, since astrophysical phenomena involve processes arising from quantum mechanics, electromagnetism, thermodynamics, nuclear physics, and relativity. However, observations of astrophysical objects alone are insufficient to capture the full complexity of the physical mechanisms at play. A rigorous theoretical modelling is required to interpret these observations and formulate general laws. This constant dialogue between theory and observation is a key driver of scientific progress in the field. Thus, the increased precision of Mercury's orbital measurements revealed an anomaly inexplicable by Newtonian mechanics, which led Einstein to develop the theory of general relativity~\cite{einstein_1915}. Among other things, this theory predicts the existence of gravitational waves, minute distortions of space-time, whose existence was experimentally confirmed in 2016~\cite{abbott_2016}.

In particular, the extreme densities and temperatures reached during violent phenomena such as supernovae or accretion around young stars produce intense radiation that interacts with the surrounding matter and alters the structure of the observed flows. These complex interactions require precise modelling, falling within the domain of radiative hydrodynamics. Understanding such phenomena demands a combination of complementary approaches. Astrophysical observation plays a crucial role because, as mentioned earlier, it gives access to phenomena that cannot be reproduced on Earth. When feasible, laboratory reproduction of these processes also constitutes a fundamental tool, offering a controlled environment in which to isolate the mechanisms at work and verify that the physical laws invoked indeed account for what is observed in the cosmos. To establish a meaningful correspondence between terrestrial experiments and astrophysical situations, it is necessary to define appropriate scaling laws~\cite{ryutov_2000,bouquet_2010,falize_2011,tranchant_2025}. Finally, and this is the central focus of this thesis, the development of theoretical models constitutes an indispensable pillar for addressing the full richness of the physical processes involved. Such modelling may be analytical, allowing one to isolate simple and well-understood mechanisms, even if this means neglecting nonlinear effects or complex interactions. It may also take the form of numerical simulations, which preserve a more complete formulation of the physical equations and allow one to explore more realistic configurations while maintaining a framework that remains interpretable from a theoretical standpoint.

\section*{Scientific context}

In this thesis, we focus on the modelling of violent phenomena in which radiative processes play a predominant role in the observed dynamics. But under what physical conditions do such radiative effects become significant? There is no sharp boundary separating a regime dominated by radiation from one in which radiation can be neglected. However, it is possible to introduce two dimensionless numbers that allow to assess the relative importance of radiative effects in a given configuration. The first is the Mihalas number~$\mathrm{R}$, which measures the ratio between the internal energy $u$ of a gas and the radiative energy $\mathrm{E}_R$. In the case of a monoatomic gas in radiative equilibrium\footnote{See definition in Chapter~\secref{ch:chapitre2}.}, this ratio can be expressed as:
\begin{equation*}
    \mathrm{R} = \frac{u}{\mathrm{E}_R} \approx \frac{1}{36} \frac{N}{\mathrm{T}^3} \;\;\mathpunct{.}
\end{equation*}

This expression shows that the higher the temperature $\mathrm{T}$ of the medium, the more dominant the radiative energy becomes in the energy budget. Conversely, the larger the gas density $N$, the less significant the radiative effects are. The gas is said to be radiation-dominated when \mbox{$\mathrm{R} \ll 1$}. The second dimensionless number is the Boltzmann number $\mathrm{Bo}$, which compares the energy flux transported by advection $\mathrm{F}$ with the energy flux transported by radiation $\mathrm{F}_R$. It is given by:
\begin{equation*}
    \mathrm{Bo} = \frac{\mathrm{F}}{\mathrm{F}_R} \approx \frac{1}{9} \frac{v}{c} \frac{N}{\mathrm{T}^3} = 4 \frac{v}{c} \mathrm{R} \;\;\mathpunct{.}
\end{equation*}

In the applications considered here, the fluids are non-relativistic, meaning that their velocity $v$ is much smaller than the speed of light $c$. As a result, radiative transport mechanisms become important well before radiative energy becomes dominant in the overall energy balance. To sum up, the phenomena studied in this thesis correspond to hot and sufficiently low-density media in which the transport of radiative flux plays a crucial role, often even before the regime in which radiation dominates energetically is reached.

\begin{figure}
    \begin{subfigure}[t]{0.49\textwidth}
        \centering
        \includegraphics[height=6cm]{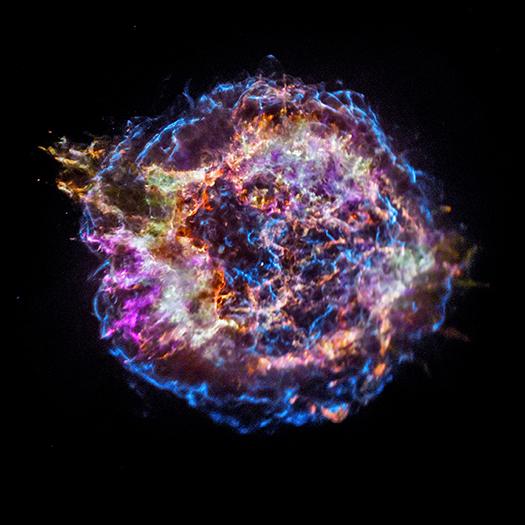}
        \caption{Cassiopeia A supernova remnant.}
        \label{fig:SNR}
    \end{subfigure}
    \hfill
    \begin{subfigure}[t]{0.49\textwidth}
        \centering
        \includegraphics[height=6cm]{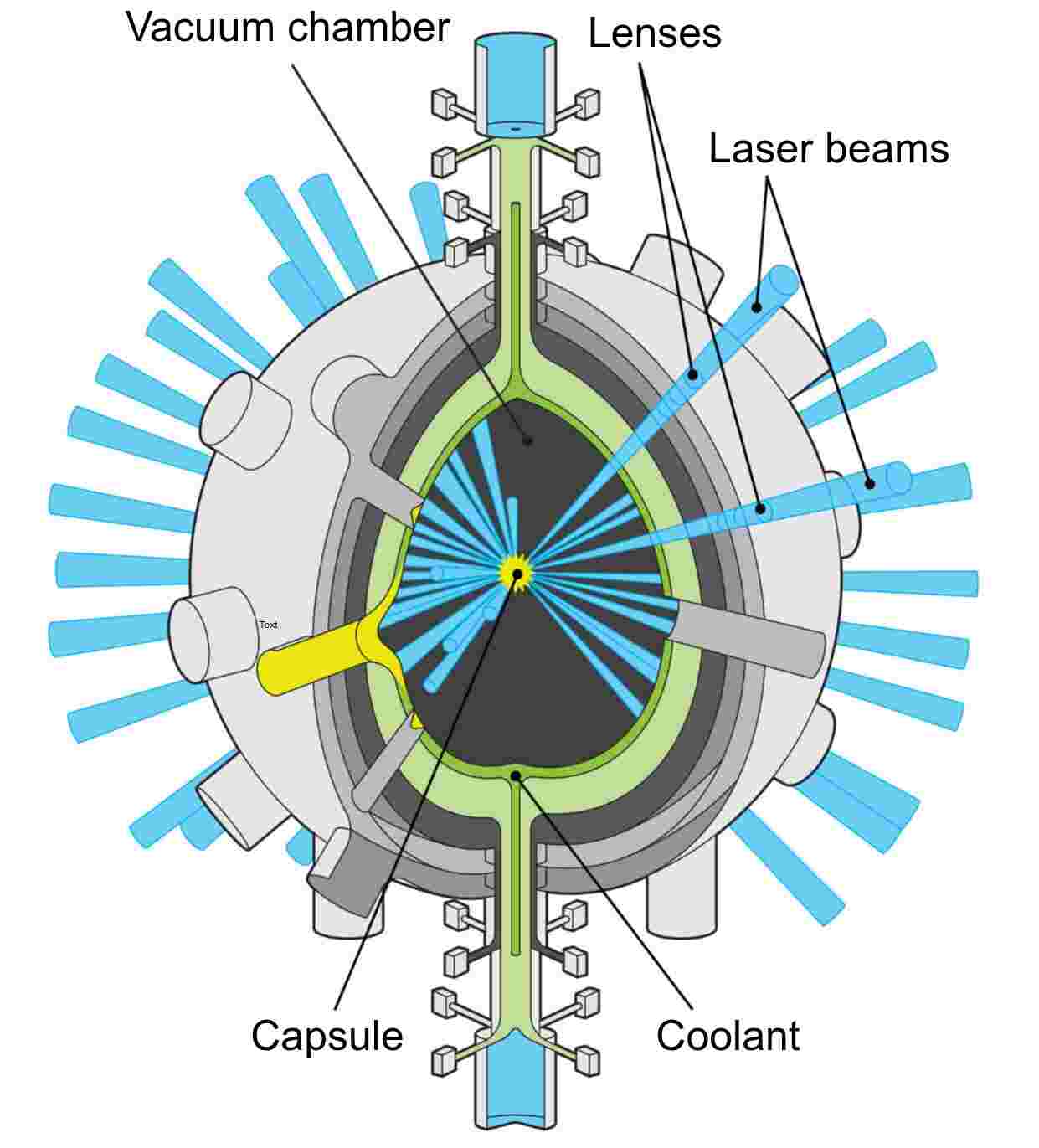}
        \caption{Descriptive diagram of Inertial Confinement Fusion.}
        \label{fig:FCI}
    \end{subfigure}
    \caption{Examples of applications in the field of radiative hydrodynamics.}
    \label{fig:applications}
\end{figure}

Such physical conditions arise naturally in astrophysics, a field in which radiative hydrodynamics plays a key role in understanding the dynamical and energetic phenomena affecting stars and other stellar objects. In stars, for example, nuclear fusion produces intense radiation that continually interacts with the surrounding plasma. This interaction profoundly influences the internal structure, stability, and evolution of stars. Radiative transfer modelling is especially crucial for studying mass loss and the chemical composition of red supergiants, whose extended outer envelopes are subject to complex radiative mechanisms~\cite{chiavassa_2011, kravchenko_2018}. Supernova remnants, formed from the cataclysmic explosions of dying stars, are also strongly shaped by the coupled effects of radiation and hydrodynamics. The interaction between shock waves and radiation, both during the initial explosion and throughout the subsequent evolution of the ejecta, modulates their dynamics, can induce instabilities, and largely determines their observable emission~\cite{wang_2001, miniere_2018}. Many questions remain open regarding the observed structures, such as the filaments present in some supernova remnants (see figure~\ref{fig:SNR}); more accurate modelling that includes radiative effects could provide new insights into these unresolved issues.

The relevance of these phenomena is not purely astrophysical: they also have major implications in the field of energy production, in particular for the development of clean and sustainable sources. A prominent example is nuclear fusion, envisioned as an energy alternative that produces little waste and emits no carbon dioxide. The most advanced method to date for reaching fusion conditions is the \gls{icf}, which achieved a historic milestone in 2022 with the first net‐energy ignition at the National Ignition Facility, where the energy produced by fusion exceeded the energy delivered to the system~\cite{abu-shawareb_2022, zylstra_2022}. In this approach, a capsule containing a mixture of deuterium and tritium is placed in a vacuum chamber and compressed and heated to extremely high temperatures by a radiative shock generated by X‐ray radiation produced from powerful laser beams interacting with a hohlraum target (see figure~\ref{fig:FCI}). This process is extremely sensitive to hydrodynamic instabilities, particularly Rayleigh–Taylor type instabilities, similar to those encountered in supernovae, which can compromise the efficiency of the implosion by causing substantial energy losses~\cite{remington_2013, casner_2019}. When poorly controlled, these instabilities can prevent the conditions required for fusion from being reached~\cite{smalyuk_2020}. A detailed understanding of radiation–matter interactions, both in astrophysical settings and in laboratory environments, therefore constitutes a major scientific and technological challenge.

\section*{Existing numerical codes}

Once the physical situations requiring radiative hydrodynamics modelling have been identified, it becomes necessary to determine how to describe these phenomena realistically while maintaining a reasonable computational cost. This involves choosing appropriate approximations that account for the characteristics of the media under consideration, as well as numerical limitations. Two classical simplifications are often used, especially in astrophysics: the diffusion approximation, suitable for optically thick media, and the free‐streaming approximation, valid for optically thin ones. However, these models are not valid in intermediate regimes, where more sophisticated approaches are required to faithfully capture the physics of radiative transfer.

Among these advanced approaches, the first category comprises statistical Monte Carlo methods. These consist of simulating radiative transport through the stochastic emission of photon packets, whose direction and frequency are randomly sampled. The opacity of the medium then determines the probability of interactions (absorption or scattering) between the photons and the matter. These methods have the advantage of being able to describe all opacity regimes, without restrictive assumptions about the radiation–matter coupling. However, they are extremely costly in optically thick media, where photons undergo a lot of interactions before being absorbed. Examples of codes using this approach include MC3D~\cite{wolf_2003}, RADMC~\cite{pascucci_2004}, CRASH~\cite{maselli_2003}, and AREPO-MCRT~\cite{smith_2020}.

A second class of methods relies on the spatial–angular discretisation of the specific intensity\footnote{Definition given in Chapter~\secref{ch:chapitre2}.}, i.e. the fundamental quantity describing radiation. These methods, referred to as angular discretisation schemes, approximate integrals over the solid angle by sums over a finite set of discrete directions. They offer good accuracy at a moderate numerical cost. However, they are subject to two well-known artefacts: ray effects, caused by angular discretisation, may generate artificial structures in free-streaming regions; and numerical diffusion (\textit{false scattering}), induced by spatial discretisation and interpolation, may produce an unphysical smoothing of the solution~\cite{coelho_2002}. These artefacts can be mitigated through finer spatial and angular meshes. A representative example of this approach is the Athena++ code~\cite{jiang_2014, jiang_2021, jiang_2022}.

To overcome some of these difficulties, the long- and short-characteristics methods have been developed. These techniques integrate the radiative transfer equation along the trajectories of light rays. The long-characteristics method performs this integration from each grid point to the domain boundaries, accounting for boundary conditions. Although highly accurate, it is costly because it requires to integrate over long distances across the mesh. Conversely, the short-characteristics method restricts the integration to the neighbouring cell, drastically reducing computational cost. However, the necessary interpolations introduce increased numerical diffusion. The FLASH code illustrates this approach~\cite{rijkhorst_2006}.

Another family of methods relies on solving the moment equations of the specific intensity. These methods solve the first angular moments of the radiative transfer equation and then close the system by introducing a relation between higher- and lower-order moments. The simplest model in this class is the \gls{fld} model, which relies on the two lowest moments (radiative energy and flux) and assumes a closure valid only in optically thick media. Within this framework, the radiative flux is always aligned with the gradient of radiative energy, an assumption that becomes incorrect as soon as the medium is even moderately transparent. Codes such as ZEUS2D-FLD~\cite{turner_2001}, COSMOS~\cite{anninos_2003}, $\text{CO}_5\text{BOLD}$~\cite{freytag_2012, freytag_2013, freytag_2017}, and a module of FLASH~\cite{chatzopoulos_2019} are based on this model. To improve accuracy in intermediate opacity regimes, the \gls{vtef} formalism was developed. This method also relies on the first three moments (radiative energy, flux, and pressure) but determines the closure relation linking pressure to energy via the Eddington tensor, computed by solving for the specific intensity from the full radiative transfer equation. This approach provides a precise estimate of the closure relation, though it is not strictly local with respect to radiative quantities. It captures radiation–matter coupling more effectively, but is numerically costly and complex to implement. Examples of codes implementing this method include ZEUS2D~\cite{stone_1992}, ZEUSMP~\cite{hayes_2006}, and TITAN~\cite{gehmeyr_1994}.

Finally, the model studied in this thesis is the M1 model, which also relies on the first three moments of the specific intensity. It proposes a local closure derived from an analytical formalism, in which the radiation field is assumed to be symmetric about the direction of the radiative flux. This assumption is equivalent to considering \glslink{lte}{local thermodynamic equilibrium} for the radiation, which allows expressing the Eddington tensor as a purely local function of the radiative quantities. The M1 model constitutes a good compromise between physical accuracy and computational cost: it is more general than the \gls{fld} model, notably by providing a better description of intermediate transport regimes, while being far less expensive than Monte Carlo, explicit angular discretisation, or \gls{vtef} approaches. It is used in several reference codes, such as HERACLES~\cite{gonzalez_2007, vaytet_2011}, AREPO-RT~\cite{kanan_2019}, and, most importantly for this thesis, the \glslink{hades}{\textsc{\glsentryshort{hades}} code (for \glsentrylong{hades})}\glsunset{hades}~\cite{nguyen_2011_these, michaut_2011, michaut_2017}.

The \gls{hades} code is a massively parallelised Fortran simulation tool based on finite-volume numerical schemes. It enables modelling radiative transfer using two approaches: the M1-gray model, used for global radiation–matter interaction studies, and the M1-multigroup model, which offers a more refined treatment of spectral effects. This code has been widely used to study various stellar physics phenomena, such as supernova remnants~\cite{cavet_2010_these, miniere_2014_these, saincir_2019_these, gintrand_2019_these}, accretion shocks in binary systems~\cite{busschaert_2013_these}, jets from young stars~\cite{falize_2008_these}, and, more fundamentally, radiative shocks~\cite{boireau_2005_these}. Although the M1 model makes radiative hydrodynamics simulations more accessible than more expensive methods (Monte Carlo, angular discretisation, \gls{vtef}), these simulations remain extremely demanding in computational resources, particularly when spectral effects are included through the M1-multigroup model. This is due to two main reasons:

\begin{itemize}
    \item \textbf{Computation of the Eddington factor\footnote{Details provided in Section~\secref{sec:dependence_chig}.} in the M1-multigroup model}\\
    In this model, the Eddington factor has no explicit analytical expression. The code therefore relies on numerical search algorithms to estimate it accurately. Although reliable, these methods are particularly costly in computational time.
    \item \textbf{The \gls{cfl} condition}\\
    To ensure numerical stability when solving the radiative hydrodynamics equations, the time step $\Delta t$ and spatial step $\Delta x$ must satisfy a \gls{cfl} type condition: $c \Delta t/\Delta x \le 1$, where $c$ is the speed of light. This constraint forces the use of very small time steps, significantly increasing the computational cost of simulations.
\end{itemize}

The objective of this thesis is therefore to explore the possibility of reducing the cost of such simulations through approaches derived from \gls{ai}, and in particular through the use of neural networks.

\section*{Artificial Intelligence}

Neural networks are machine-learning algorithms whose architecture is inspired by the functioning of the human brain. Their origin dates back to 1943, when Walter Pitts and Warren McCulloch proposed a simplified mathematical model to represent thought processes and decision-making in human cognition~\cite{mcculloch_1943}. This model associated elementary processing units, artificial neurons, with logical functions, constituting an initial attempt to formalise human reasoning. These pioneering works quickly attracted the attention of prominent researchers. Among them, Alan Turing drew inspiration from this approach to develop the famous \quotes{Turing test}~\cite{turing_1950}, designed to assess machine intelligence. This test, which consists in determining whether a human interlocutor can distinguish a machine from a human being through a conversational exchange, marked the beginning of the exploration of the capabilities of \gls{ai}.

Over the following decades, research in \gls{ai} and machine learning gave rise to a wide variety of neural network architectures, each tailored to specific types of problems. Among the most emblematic are:

\begin{itemize}
    \item \textbf{\glslink{mlp}{\glsentryshort{mlp} (\glsentrylong{mlp})} networks:} this is the simplest and most widespread architecture. An \glsxtrshort{mlp} consists of several layers of interconnected artificial neurons arranged hierarchically. It can model complex nonlinear relationships between inputs and outputs and remains widely used for classification and regression tasks;

    \item \textbf{Generative Adversarial Networks (\glslink{gan}{\glsentryshort{gan}, \glsentrylong{gan}}):} introduced in the 2010s, \glsxtrshort{gan}s rely on a dual architecture composed of a generator and a discriminator trained in competition. The generator attempts to produce data realistic enough to fool the discriminator, while the latter tries to distinguish real data from synthetic data. This competitive dynamic leads the generator to progressively improve the realism of the generated samples, enabling innovative applications in image synthesis, data augmentation, and simulation;

    \item \textbf{Variational Autoencoders (\glslink{vae}{\glsentryshort{vae}, \glsentrylong{vae}}):} these probabilistic generative models learn a compact and continuous representation of the input data. They are particularly effective for anomaly detection, as they allow the identification of samples that deviate significantly from the learned distribution, thereby revealing atypical behaviours or errors.
\end{itemize}

These architectures illustrate the diversity and richness of neural network approaches. Each model offers specific advantages depending on the nature of the data and the intended application. For a more comprehensive presentation of these concepts, the reader is referred to standard references such as Bishop (1995)~\cite{bishop_1995} and Wilamowski (2009)~\cite{wilamowski_2009}. Today, \gls{ai} techniques are experiencing rapid growth in astrophysics, where their applications can generally be grouped into three main categories:

\begin{enumerate}
    \item \textbf{Instrumentation control and robotics:} in space missions, \gls{ai} enhances the planning of observations onboard Martian rovers or observational satellites~\cite{alizadeh_2024, thangavel_2024, youssef_2024, ildirimzade_2025}. In ground-based observatories, it eases the automation of optical-instrument calibration, fault diagnosis, and optimisation of image quality~\cite{huang_2024};

    \item \textbf{Data processing:} \gls{ai} plays a central role in analysing observational data. It enables, for instance, the extraction of the \gls{cmb} signal from astrophysical and instrumental foregrounds~\cite{norgaard-nielsen_2008, baccigalupi_2000, farsian_2020, aylor_2020, casas_2022}, the detection of gravitational waves in complex and non-stationary noise backgrounds~\cite{prajapati_2024}, or the optimisation of massive data-flow management for the \gls{ska} project by reducing storage needs through intelligent detection of objects and anomalies during acquisition~\cite{mazumder_2023, sortino_2023};

    \item \textbf{Simulation and modelling:} more directly related to this thesis, \gls{ai} is used to improve the accuracy and efficiency of numerical simulations. Applications include super-resolution of physical fields~\cite{li_2021b} or the generation of adaptive meshes, computed more rapidly than with traditional techniques~\cite{foucart_2023}.
\end{enumerate}

A detailed review of \gls{ai} applications in astrophysics is provided in the article by Smith (2023)~\cite{smith_2023}. Despite its growing adoption, \gls{ai} remains a relatively young tool that must be carefully adapted and integrated into the specificities of each domain, particularly in the context of numerical simulations. In this regard, a new research field has recently emerged: \gls{neuhpc}~\cite{karimabadi_2023}, which aims to integrate \gls{ai} models directly into high-performance computing codes. The present thesis is fully aligned with this approach.

Among the neural network strategies applied to numerical simulation, a particularly notable class is the \gls{pinn}s. Introduced by Raissi et al. (2019)~\cite{raissi_2019}, building upon the pioneering work of Lagaris et al. (1998)~\cite{lagaris_1998}, \gls{pinn}s constitute an innovative approach for solving problems governed by \gls{pde}s. The core idea is to approximate the solution of such a problem using a neural network trained by minimising a loss function comprising several components: the \gls{pde} residual, the initial conditions, and the boundary conditions when available. The spatial and temporal derivatives required to evaluate the residual are computed through automatic differentiation~\cite{baydin_2017}, a technique based on decomposing any computer program into a sequence of elementary operations and standard functions. This class of networks is distinctive in that it embeds physical laws directly into the learning process, significantly reducing the amount of required data while ensuring that the predicted solutions remain consistent with the fundamental equations governing the studied problem.

\section*{Contributions}

\noindent This thesis is organised around three main scientific contributions, each corresponding to a clearly identified research direction:

\begin{enumerate}
    \item The development of an efficient method for computing the closure relation in the M1-multigroup model, enabling the incorporation of the spectral character of radiation into radiative hydrodynamics simulations (see Chapters~\secref{ch:chapitre2} and~\secref{ch:chapitre3});
    
    \item The analysis of the impact of a fine spectral modelling of radiation on the structure of radiative shocks (see Chapter~\secref{ch:chapitre4});
    
    \item A first exploration of the potential of \glslink{pinn}{\glsentrylong{pinn} (\glsentryshort{pinn})} to extrapolate radiative hydrodynamics simulation data to later instants (see Chapter~\secref{ch:chapitre5}).
\end{enumerate}

\noindent The first two parts of this work have led to the publication of the following articles:
\vspace{-0.5em}

\begin{itemize}
    \item Radureau, G., Michaut, C. \& Comport, A. I., AI-based computation method for the Eddington factor in the M1-multigroup model, \textit{Phys. Rev. E} \textbf{111}, 035301 (2025);
    \item Radureau, G. \& Michaut, C., Impact of frequency-dependent radiation on the dynamics and structure of radiative shocks, \textit{Phys. Rev. E} \textbf{111}, 045213 (2025).
\end{itemize}

Beyond its scientific contributions, this manuscript also aims to serve as a useful reference for those wishing to deepen or expand the use of \gls{ai} in the context of radiative hydrodynamics simulations. With this in mind, I have sought to adopt an approach as pedagogical as possible, both in the presentation of the foundations of radiative hydrodynamics and in the introduction of the \gls{ai} methods employed. The manuscript is structured around four main chapters, in addition to the present introduction, as well as four technical appendices. These various components introduce the key concepts used in this work, present the results obtained, and detail the methodological aspects necessary for the reproducibility of the proposed approaches.

Chapter~\ref{ch:chapitre2} is devoted to the physical modelling of radiative hydrodynamics used in this study. The M1-multigroup model, as implemented in the \gls{hades} code, is described with an emphasis on the assumptions and simplifications adopted.

Chapter~\ref{ch:chapitre3} introduces the main \gls{ai} techniques applied to numerical simulation, and then presents an original method developed during this thesis to efficiently estimate the closure relation of the M1-multigroup model using neural networks.

Chapter~\ref{ch:chapitre4} analyses the influence of a rigorous spectral modelling of radiation on the internal structure of radiative shocks, relying on the Eddington factor computation method developed in the preceding chapter.

Finally, Chapter~\ref{ch:chapitre5} presents preliminary attempts to use \glsxtrlong{pinn}s for the extrapolation of radiative shock simulations, thereby opening the way to new perspectives for applying neural networks in radiative hydrodynamics.
\clearemptydoublepage

\chapter{Radiative hydrodynamics} \label{ch:chapitre2}

\initialletter{A}strophysical fluids, such as those found in stars or supernovae, interact strongly with the surrounding radiation. Their flow must therefore be described by the equations of fluid mechanics, such as the Euler or Navier–Stokes equations. However, to model these interactions accurately, it is also necessary to account for the propagation of radiation and its exchanges of energy and momentum with matter. The combined study of these phenomena falls within the field of radiative hydrodynamics. In this context, the modelling of matter–radiation interactions depends on the degree of equilibrium reached by the system. Three levels of equilibrium may be considered:

\begin{enumerate}
    \item \textbf{Chemical equilibrium:} this is achieved when chemical kinetics have evolved toward a steady state, meaning that all chemical reactions are balanced and that the concentration of each species remains constant in time;
    
    \item \textbf{Thermal equilibrium:} this is achieved when all species constituting the fluid share a single temperature, indicating that internal energy exchanges are sufficiently rapid to homogenise the thermal distribution;
    
    \item \textbf{Radiative equilibrium:} this is achieved when matter is in equilibrium with radiation, implying that the radiation temperature is identical to that of the matter.
\end{enumerate}

When a fluid is simultaneously in chemical and thermal equilibrium, it is said to be in \gls{lte}. This concept is fundamental, as it determines how matter–radiation interactions must be modelled. Indeed, depending on whether \gls{lte} and radiative equilibrium are assumed, the complexity of the model can vary significantly.

In this chapter, we will assume that the matter is in \gls{lte}, while developing a theoretical framework that does not rely on the assumption of radiative equilibrium, in order to better capture the out-of-equilibrium effects frequently observed in many astrophysical environments. To clearly introduce the modelling assumptions adopted and to lay the groundwork for the approach used in the \gls{hades} code, I will begin by presenting the equations of classical hydrodynamics in the absence of interaction with radiation. I will then describe the radiative transport model in a fluid at rest, before detailing how interactions between radiation and a moving fluid are accounted for in the model implemented in \gls{hades}.

\section{Hydrodynamics} \label{sec:hydrodynamique}

Let us begin by establishing the general framework for describing a moving fluid, assuming that it does not interact with any external element. In such a context, the fluid evolves autonomously, without exchanging mass, energy, or momentum with its environment. It therefore satisfies the fundamental conservation principles:

\begin{enumerate}
    \item \textbf{Mass conservation:}
    The fluid neither gains nor loses matter during its evolution. This principle is expressed by the continuity equation, which states that the total mass contained in a moving fluid element remains constant;

    \item \textbf{Momentum conservation:}  
    In the absence of external forces, the momentum of the fluid is preserved, in accordance with the fundamental principle of dynamics. This is described by the Euler equations (for an inviscid fluid) or by the Navier–Stokes equations (when viscosity is taken into account);

    \item \textbf{Energy conservation:}  
    The total energy of the fluid (including kinetic, internal, and potential energy) remains constant in the absence of any external interaction. This conservation is expressed by the energy equation, which links variations of internal energy to heat transfer and the work of forces.
\end{enumerate}

These laws form the foundation of fluid dynamics and allow the behaviour of a fluid to be described within a purely hydrodynamic framework. We now examine the modelling of such fluids before introducing the influence of interactions with radiative phenomena.

\subsection{Mass conservation} \label{sec:mass_conservation}

Let $V$ be a control volume in space and $m$ the total mass of the fluid contained within it. This mass can be expressed as follows:
\begin{equation}
    \label{eq:mass}
    m(t) = \iiint \nolimits_V \rho(t, \vectorr{x}) \dif V \;\;\mathpunct{,}
\end{equation}

\noindent where \mbox{$\rho(t, \vectorr{x})$} denotes the mass density of the fluid at a point $\vectorr{x}$ within the control volume $V$ at time $t$. Since the mass of the fluid is conserved over time, we can write:
\begin{equation}
    \label{eq:mass_conservation_base}
    \frac{\dif m(t)}{\dif t} = \frac{\dif}{\dif t} \iiint \nolimits_V \rho(t, \vectorr{x}) \dif V = 0 \;\;\mathpunct{.}
\end{equation}

However, according to the Reynolds transport theorem, the time derivative of the integral can be written as:
\begin{equation*}
    \frac{\dif}{\dif t} \iiint \nolimits_V \rho \dif V = \iiint \nolimits_V \partial_t\rho \dif V + \oiint \nolimits_S \rho \vectorr{v} \cdot \vectorr{n} \dif S \;\;\mathpunct{,}
\end{equation*}

\noindent with $\vectorr{v}$ the velocity field of the fluid, $S$ the boundary surface of the control volume $V$, and $\vectorr{n}$ the unit normal vector to the surface element $\dif S$. Using the Green–Ostrogradsky theorem, the surface integral can be rewritten as a volume integral. Thus, we obtain the expression:
\begin{equation*}
    \oiint \nolimits_S \rho \vectorr{v} \cdot \vectorr{n} \dif S = \iiint \nolimits_V \nablav \cdot ( \rho \vectorr{v}) \dif V \;\;\mathpunct{.}
\end{equation*}

\noindent Thus, the mass conservation equation~\eqref{eq:mass_conservation_base} can be rewritten in the form:
\begin{equation*}
     \iiint \nolimits_V \left ( \partial_t\rho  + \nablav \cdot ( \rho \vectorr{v}) \right ) \dif V = 0 \;\;\mathpunct{.}
\end{equation*}

Considering that this equation is valid for any control volume $V$, we can deduce the \textit{continuity equation}, also called the \textit{local conservation equation of mass}:
\begin{equation}
    \label{eq:mass_conservation}
     \partial_t\rho  + \nablav \cdot \left ( \rho \vectorr{v} \right ) = 0 \;\;\mathpunct{.}
\end{equation}

It is important to emphasise that the transition from the integrated equation to the local equation is valid \quotes{almost everywhere}, in the sense of Lebesgue integrals. This means that at the location of a discontinuity, such as a shock, the local equation is no longer valid. This subtlety will be crucial when discussing of \glslink{pinn}{\textit{Physics-Informed Neural Networks}} (see Section~\secref{sec:intro_PINNs}).

\subsection{Momentum conservation}  \label{sec:momentum_conservation}

To establish the conservation law for momentum, we rely on Newton's second law of motion. This principle states that the time derivative of the momentum of the particles, $\vectorr{p_m}$, is equal to the sum of the external forces $\vectorr{F}_{ext}$ acting on them.
\begin{equation}
    \label{eq:PFD}
     \frac{\dif \vectorr{p_m}}{\dif t} = \sum \vectorr{F}_{ext} \;\;\mathpunct{.}
\end{equation}

In the present case, the momentum of a fluid contained in a control volume $V$ corresponds to the sum of the momenta of the particles that compose it. It can therefore be expressed as:
\begin{equation*}
    \vectorr{p_m} = \iiint \nolimits_V \rho \vectorr{v} \dif V \;\;\mathpunct{.}
\end{equation*}

For the same reasons as in Section~\secref{sec:mass_conservation}, using the Reynolds transport theorem as well as the Green–Ostrogradsky theorem, and writing $\vectorr{v} = (v_1, v_2, v_3)$, the time derivative of the momentum can be written as:
\begin{equation*}
     \frac{\dif \vectorr{p_m}}{\dif t} = \frac{\dif}{\dif t} \iiint \nolimits_V \rho \vectorr{v} \dif V = 
     \begin{bmatrix}
        \iiint \nolimits_V \left (  \partial_t (\rho v_1) + \nablav \cdot (\rho v_1 \vectorr{v}) \right )\\
        \iiint \nolimits_V \left (  \partial_t (\rho v_2) + \nablav \cdot (\rho v_2 \vectorr{v}) \right )\\
        \iiint \nolimits_V \left (  \partial_t (\rho v_3) + \nablav \cdot (\rho v_3 \vectorr{v}) \right )\\
    \end{bmatrix}
    \;\;\mathpunct{.}
\end{equation*}

Considering the coordinates $\vectorr{x} = (x_1, x_2, x_3)$, the divergence term in the previous equation can be expanded as follows:
\begin{align*}
     \begin{bmatrix}
        \nablav \cdot (\rho v_1 \vectorr{v})\\
        \nablav \cdot (\rho v_2 \vectorr{v})\\
        \nablav \cdot (\rho v_3 \vectorr{v})\\
    \end{bmatrix} &=
    \begin{bmatrix}
        \partial_{x_1} (\rho v_1^2) + \partial_{x_2} (\rho v_1 v_2) + \partial_{x_3} (\rho v_1 v_3)\\
        \partial_{x_1} (\rho v_2 v_1) + \partial_{x_2} (\rho v_2^2) +  \partial_{x_3} (\rho v_2 v_3)\\
        \partial_{x_1} (\rho v_3 v_1) + \partial_{x_2} (\rho v_3 v_2) + \partial_{x_3} (\rho v_3^2)\\
    \end{bmatrix} = 
    \nablav \cdot \left (\rho 
    \begin{bmatrix}
        v_1\\
        v_2\\
        v_3
    \end{bmatrix}
    \begin{bmatrix}
        v_1~~v_2~~v_3
    \end{bmatrix} \right )\\ 
    &= \nablav \cdot (\rho \vectorr{v} \otimes \vectorr{v})
    \;\;\mathpunct{,}
\end{align*}

where $\otimes$ denotes the \textit{dyadic product}. Thus, the time derivative of the momentum can be written more concisely as follows:
\begin{equation*}
     \frac{\dif \vectorr{p_m}}{\dif t} = \iiint \nolimits_V \left (  \partial_t(\rho \vectorr{v}) + \nablav \cdot (\rho \vectorr{v} \otimes \vectorr{v}) \right ) \dif V \;\;\mathpunct{.}
\end{equation*}

We still need to determine the external forces acting on the fluid. We distinguish between \textit{surface forces} $\vectorr{F}_S$, which act on the boundary of the control volume $V$, and \textit{body forces} $\vectorr{F}_V$, which act on every particle of the fluid.

For the surface forces, we may distinguish between \textit{pressure forces} and \textit{viscous stress forces}, which arise from the interaction of the particles at the boundary of the control volume $V$ with their environment. In the present case, we consider only the \textit{pressure forces} $p$, which take the following form:
\begin{equation*}
     \vectorr{F}_S = \oiint \nolimits_S -p \vectorr{n} \dif S \;\;\mathpunct{.}
\end{equation*}

\noindent  According to the Green-Ostrogradski theorem, this force can be rewritten in the form:
\begin{equation*}
     \vectorr{F}_S = -\iiint \nolimits_V \nablav p \dif V \;\;\mathpunct{.}
\end{equation*}

Regarding the \textit{body forces}, various contributions may be taken into account, such as gravitational forces, electromagnetic forces, and, in particular, the radiative force. For now, we will neglect body forces, and later detail in Section~\secref{sec:Hydrodad_equations} how to include the radiative force. Finally, the fundamental principle of dynamics can be rewritten in the form:
\begin{equation*}
     \iiint \nolimits_V \left (  \partial_t(\rho \vectorr{v}) + \nablav \cdot (\rho \vectorr{v} \otimes \vectorr{v} + p \identity) \right ) \dif V = 0 \;\;\mathpunct{,}
\end{equation*}

\noindent where $\identity$ denotes the identity matrix. By considering that this relation holds for any control volume $V$, we obtain the \textit{momentum conservation equation}:
\begin{equation}
    \label{eq:momentum_conservation}
    \partial_t(\rho \vectorr{v}) + \nablav \cdot \left (\rho \vectorr{v} \otimes \vectorr{v} + p \identity \right ) = 0 \;\;\mathpunct{.}
\end{equation}

Once again, the transition from the integrated equation to the local equation is valid \quotes{almost everywhere}, in the sense of Lebesgue integrals.

\subsection{Total energy conservation} \label{sec:energy_conservation}

The conservation law for the total energy $\mathrm{E}_{tot}$ of a system is based on the first law of thermodynamics, which states that the variation of the total energy of the fluid within a control volume $V$ is equal to the sum of the power of the non-conservative external forces $\mathcal{P}_e$ and the heating power exchanged with the fluid $\mathcal{P}_c$. This relation can therefore be written in the following form:
\begin{equation}
    \label{eq:1er_principe_thermo}
   \frac{\dif \mathrm{E}_{tot}}{\dif t} = \mathcal{P}_e + \mathcal{P}_c \;\;\mathpunct{.}
\end{equation}

First, the total energy is the sum of the energy contained within a control volume $V$ and can be expressed as:
\begin{equation*}
   \frac{\dif \mathrm{E}_{tot}}{\dif t} = \frac{\dif}{\dif t} \iiint \nolimits_V \mathrm{E} \dif V \;\;\mathpunct{,}
\end{equation*}

\noindent where $\mathrm{E}$ denotes the total energy density of the fluid. Analogously to what was presented in Section~\secref{sec:mass_conservation}, the time derivative of the total energy can be written as follows:
\begin{equation*}
   \frac{\dif \mathrm{E}_{tot}}{\dif t} = \iiint \nolimits_V  \left ( \partial_t \mathrm{E} + \nablav \cdot (\mathrm{E} \vectorr{v}) \right ) \dif V \;\;\mathpunct{.}
\end{equation*}

The power term associated with non-conservative forces may include the power of the gravitational force, the power of the electric field, or the power of pressure forces. In the present case, we consider only the power of the pressure forces, which can be written as follows:
\begin{equation*}
   \mathcal{P}_e = \oiint \nolimits_S  -p \vectorr{v} \cdot \vectorr{n} \dif S \;\;\mathpunct{,}
\end{equation*}

\noindent which, according to the Green–Ostrogradsky theorem, can be rewritten in the following form:
\begin{equation*}
   \mathcal{P}_e = -\iiint \nolimits_V  \nablav \cdot (p \vectorr{v}) \dif V \;\;\mathpunct{.}
\end{equation*}

Concerning the heating power transferred to the fluid, one may account for thermal conduction as well as radiative energy losses or gains. For the moment, we neglect both phenomena ($\mathcal{P}_c = 0$), but we will later detail in Section~\secref{sec:Hydrodad_equations} how to include energy exchanges between the fluid and the radiation. Thus, the first law of thermodynamics can be rewritten as:
\begin{equation*}
   \iiint \nolimits_V  \left ( \partial_t \mathrm{E} + \nablav \cdot ((\mathrm{E}+p) \vectorr{v}) \right ) \dif V = 0 \;\;\mathpunct{.}
\end{equation*}

Since this integral is valid for any control volume $V$, we obtain the \textit{total energy conservation equation}:
\begin{equation}
    \label{eq:energy_conservation}
    \partial_t \mathrm{E} + \nablav \cdot \left ((\mathrm{E}+p) \vectorr{v} \right ) = 0 \;\;\mathpunct{.}
\end{equation}

Once again, the transition from the integrated equation to the local equation is valid \quotes{almost everywhere}, in the sense of Lebesgue integrals.

\subsection{Euler's equations} \label{sec:Euler_equations}

Finally, we have obtained the conservation equations for mass~\eqref{eq:mass_conservation}, momentum~\eqref{eq:momentum_conservation}, and total energy~\eqref{eq:energy_conservation}, which govern the evolution of a perfect fluid. These equations neglect the effects of gravity, electromagnetic forces, radiation, and thermal exchanges with the environment. They form a system of coupled equations known as the \textit{Euler equations}, which play a central role in fluid mechanics. These equations depend on four fundamental quantities: the mass density $\rho$, the velocity field $\vectorr{v}$, the pressure $p$, and the total energy density $\mathrm{E}$, and can be summarised as follows:
\begin{equation}
    \label{eq:euler}
    \left\{
    \begin{array}{lll}
        \partial_t\rho  + \nablav \cdot ( \rho \vectorr{v}) &= 0 \;\;\mathpunct{,} \\
        \partial_t(\rho \vectorr{v}) + \nablav \cdot (\rho \vectorr{v} \otimes \vectorr{v} + p \identity ) &= 0 \;\;\mathpunct{,} \\
        \partial_t \mathrm{E} + \nablav \cdot ((\mathrm{E}+p) \vectorr{v}) &= 0 \;\;\mathpunct{.}
    \end{array}
    \right.
\end{equation}

However, we have only three equations to describe four quantities, which makes the system insufficiently constrained. To complete the system and make the model solvable, it is necessary to introduce a fourth equation that relates the missing variables and allows the system to be closed. To do so, we will express the total energy density of the fluid in terms of the other hydrodynamic quantities. Moreover, since temperature plays a central role in describing the interactions of the fluid with radiation, we will also detail how to compute the equation of state of the fluid, which relates the pressure to the fluid's temperature.

\starsect{Equation of state}

Let us begin by establishing the expression of the equation of state, which relates the pressure to the temperature of the fluid. By neglecting interactions between the particles composing the gas, the partial pressure $p_i$ of each particle $i$ can be described using the ideal gas equation of state.
\begin{equation}
    p_i = n_i k_B \mathrm{T}_i \;\;\mathpunct{,}
\end{equation}

\noindent where $k_B$ is the Boltzmann constant, $n_i$ is the number density of particle $i$, and $\mathrm{T}_i$ is its temperature. The total pressure and total density can then be expressed respectively as follows:
\begin{align*}
    \begin{array}{lll}
        p &= \sum_i p_i \;\;\mathpunct{,}\\
        \rho &= \sum_i m_i n_i \;\;\mathpunct{,}
    \end{array}
\end{align*}

\noindent where $m_i$ is the mass of particle $i$. Assuming \gls{lte}, which implies a common temperature for all particles (\mbox{$\forall i,~\mathrm{T}_i = \mathrm{T}$}), the global equation of state of the fluid becomes:
\begin{equation}
    \label{eq:eq_etat}
    p = \frac{\rho k_B \mathrm{T}}{\mu m_H} \;\;\mathpunct{,}
\end{equation}

\noindent where $m_H$ is the mass of the hydrogen atom and $\mu m_H$ is the mean atomic weight of the medium. The quantity $\mu$ is defined by the relation:
\begin{equation}
    \label{eq:mu_expression}
    \mu = \frac{\sum_i n_i m_i}{\sum_i n_i m_H} \;\;\mathpunct{.}
\end{equation}

\noindent We may distinguish four different cases for computing $\mu$:

\begin{enumerate}
    \item \textbf{Pure, non-ionised gas:} when a gas consists of particles of mass $m_a$, the mean atomic weight is given by:
    \begin{equation}
        \mu = \frac{m_a}{m_H} \;\;\mathpunct{,}
    \end{equation}

    \item \textbf{Pure, ionised gas:} if the gas consists of neutral particles $a$, ions $i$, and electrons $e$, the mean atomic weight is:
    \begin{equation}
        \mu = \frac{m_e n_e + m_i n_i + m_a n_a}{(n_e+n_i+n_a) m_H} \;\;\mathpunct{,}
    \end{equation}

    where $n_e$, $n_i$, and $n_a$ are the number densities of electrons, ions, and neutral atoms, respectively. Since the electron mass is very small compared to that of ions and neutral atoms, and since neutral atoms have approximately the same mass as ions (\mbox{$m_e \ll m_i, m_a$ and $m_a \approx m_i$}), one may neglect terms involving the electron mass and consider the mass of ions and neutral atoms to be equal. This simplifies the expression for $\mu$:
    \begin{equation}
        \mu = \frac{m_i}{(1+Z) m_H} = \frac{\mu^*}{(1+Z)}  \;\;\mathpunct{,}
    \end{equation}

    where \mbox{$Z = n_e/(n_i+n_a)$} is the \textit{ionisation degree} and $\mu^* = m_a/m_H$. Under \gls{lte}, $Z$ depends only on the density and temperature of the medium. Consequently, the equation of state for an ionised gas is classically written as:
    \begin{equation}
        \label{eq:eq_etat_ionisee}
        p = (1+Z) \frac{\rho k_B \mathrm{T}}{\mu^* m_H} \;\;\mathpunct{,}
    \end{equation}

    \item \textbf{Non-reactive, non-ionised mixture:} no particular simplification exists for $\mu$, and $\mu m_H$ represents the average mass of the mixture's particles. The mean atomic weight $\mu$ remains constant over time and is computed using expression~\eqref{eq:mu_expression};

    \item \textbf{Non-reactive, ionised mixture:} the equation of state becomes more complex due to the evolution of the species' ionisation degree, which depends on time and on the fluid state;

    \item \textbf{Reactive, potentially ionised mixture:} the equation of state becomes even more complex, as it depends not only on time and on the fluid state, potentially accounting for ionisation, but also on ongoing chemical reactions. No simplification is possible for computing the mean atomic weight $\mu$. This case deviates from the \gls{lte} approximation and provides a more accurate description of non-equilibrium phenomena.
\end{enumerate}

In this work, although the regimes studied in radiative hydrodynamics generally involve ionised gases, we restrict ourselves to the case of pure, non-ionised gases (case 1), for which the mean atomic weight $\mu$ remains constant throughout the flow. This choice is justified by the main objective of the thesis, namely to improve the computation of radiation transport and to study its specific influence on the fluid dynamics.

\starsect{Total energy of the fluid}

It is necessary to express the total energy of the fluid in order to relate it to the hydrodynamic quantities and to close the Euler equations~\eqref{eq:euler}. It consists of three main contributions: the macroscopic kinetic energy $\mathrm{E}_{k,macro}$, the macroscopic potential energy $\mathrm{E}_{p,macro}$, and the internal energy $u$:
\begin{equation}
    \mathrm{E} = \mathrm{E}_{k,macro} + \mathrm{E}_{p,macro} + u \;\;\mathpunct{,}
\end{equation}

\noindent Let us detail the physics involved in each of these terms:

\begin{itemize}
    \item \textbf{Macroscopic kinetic energy:} It is associated with the bulk motion of the fluid and can be expressed as follows:
    \begin{equation*}
        \mathrm{E}_{k,macro} = \frac{1}{2} \rho ||\vectorr{v}||^2 \;\;\mathpunct{,}
    \end{equation*}

    where $||\vectorr{v}||$ denotes the norm of the velocity field.
    \item \textbf{Macroscopic potential energy:} It arises from external forces acting on the fluid, such as gravitational or electromagnetic fields. In the present case, none of these effects are considered, and one has \mbox{$\mathrm{E}_{p,macro}=0$}.
    \item \textbf{Internal energy:} It represents the total energy within the gas and is composed of two contributions:
    \begin{enumerate}
        \item \textbf{Internal kinetic energy:} It can be expressed as a function of the gas temperature as follows:
        \begin{equation*}
            \mathrm{E}_{k,int} = \sum_i \mathrm{E}_{k,i}  \;\;\mathpunct{,}
        \end{equation*}

        where $\mathrm{E}_{k,i}$ is the average kinetic energy associated with particles $i$, and for a perfect gas it is given by:
        \begin{equation*}
            \mathrm{E}_{k,i} = \frac{\ell_i  n_i k_B \mathrm{T}_i}{2} \;\;\mathpunct{,}
        \end{equation*}

        where $\ell_i$ is the number of degrees of freedom of particle $i$ (3 for a monoatomic particle, 5 for a diatomic particle). In the same way as in the equation of state, assuming that all particles have the same temperature, the total internal energy can be written as:
        \begin{equation*}
            \mathrm{E}_{k,int} = \rho \frac{\ell  k_B}{2 \mu m_H} \mathrm{T} = \rho c_v \mathrm{T} \;\;\mathpunct{,}
        \end{equation*}

        where \mbox{$\ell = (\sum_i \ell_i n_i)/(\sum_i n_i)$} is the average number of degrees of freedom of the particles, $c_v$ is the specific heat capacity at constant volume, and $\mu$ is the same quantity as detailed in the section \textit{Equation of state}.
        
        \item \textbf{Internal potential energy:} It accounts for chemical binding energies, particle state transitions, or nuclear transitions. In the case of a pure ionized gas, this energy corresponds to the ionization energy of the particles. In the present case, particle ionization is not taken into account, so this energy is zero.
    \end{enumerate}

    Thus, in the case where particle ionization, chemical transformations, or particle state transitions are not taken into account, the internal energy is given by:
    \begin{equation*}
        u = \rho \frac{\ell  k_B}{2 \mu m_H} \mathrm{T} = \rho c_v \mathrm{T} \;\;\mathpunct{.}
    \end{equation*}

    Lets note that the enthalpy of the fluid can thus be expressed as:
    \begin{equation*}
        h = u + p = \rho \frac{(\ell + 2)  k_B}{2 \mu m_H} \mathrm{T} = \rho c_p \mathrm{T} \;\;\mathpunct{,}
    \end{equation*}

    where $c_p$ is the specific heat capacity at constant pressure. The adiabatic index $\gamma$ is defined as the ratio \mbox{$c_p/c_v$}, which can be expressed as a function of the number of degrees of freedom of the particles.
    \begin{equation*}
        \gamma = \left. c_p \middle/ c_v \right. = \frac{\ell+2}{\ell} \;\;\mathpunct{.}
    \end{equation*}

    The specific heat capacity at constant volume can then be written as:
    \begin{equation*}
        c_v = \frac{k_B}{(\gamma-1) \mu m_H} \;\;\mathpunct{.}
    \end{equation*}

    Based on considerations of the internal degrees of freedom of the particles, the adiabatic index takes the value \mbox{$\gamma = 5/3$} for a monoatomic gas, and \mbox{$\gamma = 7/5 = 1.4$} for a diatomic gas. By contrast, for fluids composed of more complex particles, mixtures of different species, real or non-ideal gases, or partially or fully ionized plasmas, the value of $\gamma$ may deviate from these two standard values. It may even evolve in time, depending on the local hydrodynamic conditions of the fluid, such as temperature or density, which directly influence the degrees of freedom that are effectively excited.
\end{itemize}

\noindent Finally, the total energy considered here can be written as:
\begin{equation}
    \label{eq:energie_totale}
    \mathrm{E} = \frac{1}{2} \rho ||\vectorr{v}||^2 + \rho c_v \mathrm{T} \;\;\mathpunct{,}
\end{equation}

\noindent or, equivalently, using the equation of state developed in the previous section and the expression of $c_v$:
\begin{equation}
    \label{eq:energie_totale2}
    \mathrm{E} = \frac{1}{2} \rho ||\vectorr{v}||^2 + \frac{p}{\gamma-1}\;\;\mathpunct{.}
\end{equation}

\section{Radiative transfer}  \label{sec:Radiation}

In astrophysics, flows interact with radiation in different ways depending on the properties of the medium. In many cases, radiation propagates freely through space and acts only as an energy loss mechanism for the fluid. Such media are referred to as optically thin. In this situation, cooling functions can be introduced to model these energy losses. This framework is particularly relevant for supernova remnants~\cite{falle_1981,smith_1989,blondin_1989,miniere_2018} as well as for accretion processes~\cite{falize_2009,busschaert_2013}.

Conversely, some media exhibit very rapid interactions between radiation and matter, making the radiative equilibrium approximation relevant. This regime is typical of star formation~\cite{commercon_2011}, stellar pulsations~\cite{baker_1962,christy_1966,shibahashi_2005}, and protoplanetary disks~\cite{flock_2013}.

However, these two extreme cases are not sufficient to describe the full complexity of radiation–matter interactions. There exist intermediate situations in which radiation interacts significantly with matter, but too slowly to reach local thermal equilibrium with its surroundings. This phenomenon is encountered in particular at stellar surfaces, where radiation undergoes a transition between an optically thick medium (the stellar atmosphere) and an optically thin medium (the interstellar medium), as well as in protoplanetary disks, which are optically thick in their interior but allow radiation to escape in certain regions~\cite{fuksman_2021}. Moreover, these simplified descriptions neglect the dependence of radiative behavior on photon frequency, which limits the accuracy of predictions of observable radiative fluxes. To overcome these limitations, we develop in this section a model that enables a more realistic description of radiation propagation and its interactions with matter, while remaining within the framework of \gls{lte}.

Finally, to simplify the mathematical expressions, we explicitly omit the spatio-temporal dependencies on $\vectorr{x}$ and $t$, although all the quantities considered generally vary in time and space.

\subsection{Radiative transfer equation}  \label{sec:radiative_equation}

\starsect{Specific intensity}

\begin{figure}
    \begin{center}
        \begin{minipage}[t]{0.45\linewidth}
            \centering
            \includegraphics[width=\textwidth]{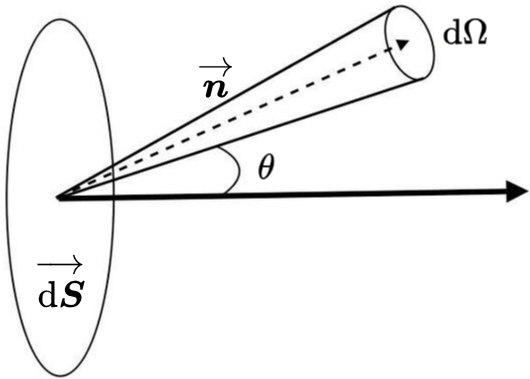}
        \end{minipage}
        \caption{Schematic representation of a radiation beam crossing the surface $\vectorr{\dif S}$ and propagating within the solid angle $\dif \Omega$.}
        \label{fig:intensite_specifique}
    \end{center}
\end{figure}

In order to describe radiation, it is essential to define quantities that characterize the radiation field. The most important one is the specific intensity $I_\nu(\vectorr{n})$, which is evaluated at a time $t$, at a position $\vectorr{x}$, and describes radiation propagating in a direction $\vectorr{n}$ (see figure~\ref{fig:intensite_specifique}). The amount of radiative energy with frequencies between $\nu$ and \mbox{$\nu + \dif \nu$}, crossing the surface $\vectorr{\dif S}$, and propagating within a solid angle $\dif \Omega$, at position $\vectorr{x}$ and time $t$, depends on the specific intensity as follows:
\begin{equation}
    \label{eq:intensite_specifique}
    \dif \mathrm{E}_\nu = I_\nu(\vectorr{n})~\vectorr{n} \cdot  \vectorr{\dif S} \dif \Omega \dif \nu \dif t \;\;\mathpunct{.}
\end{equation}

Here, the vector $\vectorr{\dif S}$ is oriented along the normal to the considered surface. It is important to note that, in the absence of interactions with matter, the specific intensity remains constant along the direction of propagation of the radiation. By contrast, when interactions with matter occur, radiation may lose or gain energy, leading to a variation of the specific intensity. It is therefore necessary to model these interactions in order to correctly describe the evolution of the radiation intensity within the medium.

\starsect{Absorption, emission and scattering}

When radiation propagates through a material medium, it can interact with matter in three different manners, thereby modifying its behavior: through photon absorption, emission, or scattering.

\begin{figure}
    \begin{subfigure}[t]{0.32\textwidth}
        \centering
        \includegraphics[height=4cm]{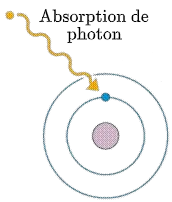}
        \caption{Photon absorption}
        \label{fig:absorption}
    \end{subfigure}
    \hfill
    \begin{subfigure}[t]{0.32\textwidth}
        \centering
        \includegraphics[height=4cm]{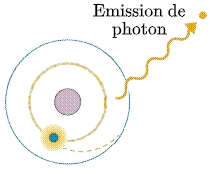}
        \caption{Photon emission}
        \label{fig:emission}
    \end{subfigure}
    \hfill
    \vspace{1cm}
    \begin{subfigure}[t]{0.32\textwidth}
        \centering
        \includegraphics[height=4cm]{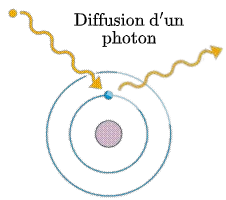}
        \caption{Photon scattering}
        \label{fig:diffusion}
    \end{subfigure}
    \caption{Types of interactions between light and matter.}
    \label{fig:interactions}
\end{figure}

\begin{enumerate}
    \item \textbf{Photon absorption (figure~\ref{fig:absorption}):} photons from the radiation field can be absorbed by matter, transferring their energy to the medium. This can increase the temperature of the material or induce changes in the state of matter, such as molecular dissociation or ionization;
    \item \textbf{Photon emission (figure~\ref{fig:emission}):} after absorbing energy, matter can re-emit photons. This may occur during energy transitions in the atoms or molecules of the medium, thereby releasing photons;
    \item \textbf{Photon scattering (figure~\ref{fig:diffusion}):} photons can also be deflected from their original trajectory as they interact with particles in the medium. Scattering can be elastic, when the photon energy remains unchanged, or inelastic, when it is modified.
\end{enumerate}

These interactions affect both the intensity and the direction of the radiation and must be taken into account to model radiative transport through a medium. Let us consider radiation propagating through a layer of matter of thickness $\dif \ell$, over which photons are absorbed. The amount of specific intensity lost is:
\begin{equation}
    \label{eq:absorption}
    \dif I_\nu = -\chi_\nu(\vectorr{n}) I_\nu \dif \ell \;\;\mathpunct{,}
\end{equation}

\noindent where $\chi_\nu(\vectorr{n})$ is called the absorption coefficient and is composed of two components: \textit{pure absorption} $\kappa_\nu (\vectorr{n})$ and \textit{pure scattering} $\sigma_\nu(\vectorr{n})$,
\begin{equation}
    \label{eq:coef_absorption}
    \chi_\nu(\vectorr{n}) = \kappa_\nu (\vectorr{n}) + \sigma_\nu(\vectorr{n})\;\;\mathpunct{.}
\end{equation}

Matter can also emit photons as the light beam traverses a layer of material of thickness $\dif \ell$. This gain in photons is modeled by the following equation:
\begin{equation}
    \label{eq:emission}
    \dif I_\nu = \eta_\nu(\vectorr{n}) \dif \ell \;\;\mathpunct{,}
\end{equation}

\noindent where $\eta_\nu(\vectorr{n})$ is called the \textit{emission coefficient} and is also composed of two parts: \textit{thermal emission} $\eta^t_\nu(\vectorr{n})$ and \textit{scattering emission} $\eta^s_\nu(\vectorr{n})$,
\begin{equation}
    \label{eq:coef_emission}
    \eta_\nu(\vectorr{n}) = \eta^t_\nu (\vectorr{n}) + \eta^s_\nu(\vectorr{n}) \;\;\mathpunct{.}
\end{equation}

All these coefficients provide a general description of the interactions between matter and radiation. Their values are determined by four fundamental physical processes~\cite{sobolev_1985, rybicki_lightman_1979}:
\begin{enumerate}
    \item \textbf{Free–free processes (\textit{bremsstrahlung}):} a free electron, accelerated or decelerated through interaction with an ion or another electron, emits a photon, producing continuous radiation. Conversely, a free electron may absorb a photon in the presence of an ion, thereby increasing its kinetic energy;
    \item \textbf{Bound–free processes (\textit{photoionization}):} an atom or ion can absorb a photon with sufficient energy to eject an electron, leading to ionization of the particle;
    \item \textbf{Free–bound processes (\textit{recombination}):} conversely, a free electron can be captured by an ion or an atom, resulting in the emission of a photon whose energy equals the difference between the initial and final energy states of the electron;
    \item \textbf{Bound–bound processes:} an atom or molecule may absorb or emit a photon when one of its electrons changes energy level. These processes are responsible for the characteristic emission and absorption spectra of the elements.
\end{enumerate}

In a medium at rest, the absorption coefficient $\kappa_\nu$ and the thermal emission coefficient $\eta^{t}_\nu$ are isotropic. When a photon is scattered, the energy absorbed through the pure scattering coefficient $\sigma_\nu$ is re-emitted via the scattering emission coefficient $\eta^{s}_\nu$. These two coefficients are therefore related. In general, determining $\eta^{s}_\nu$ is not trivial, as this coefficient strongly depends on the microscopic properties of the particles present. However, under certain simplifying assumptions, it is possible to derive an analytical expression. These assumptions are:

\begin{enumerate}
    \item The pure scattering coefficient $\sigma_\nu$ is isotropic (which is often the case);
    \item Photons are scattered isotropically;
    \item Scattering is elastic and does not modify the photon energy,
\end{enumerate}

\noindent The analytical expression of this coefficient is then:
\begin{equation*}
    \eta^s_\nu(\vectorr{n}) = \frac{\sigma_\nu}{4\pi} \int \nolimits_{4 \pi} I_\nu(\vectorr{n}') \dif \Omega' \;\;\mathpunct{.}
\end{equation*}

\noindent Furthermore, if radiation is assumed to be in \gls{lte}, Kirchhoff's law allows the thermal emission coefficient to be expressed as follows:
\begin{equation*}
    \eta^t_\nu(\vectorr{n}) = \kappa_\nu B_\nu(\mathrm{T}) \;\;\mathpunct{,}
\end{equation*}

\noindent where $B_\nu(\mathrm{T})$ is the Planck function, given by:
\begin{equation*}
    B_\nu(\mathrm{T}) = \frac{2 h \nu^3}{c^2} \left [ e^{-\frac{h \nu}{k_B \mathrm{T}}} - 1 \right ]^{-1} \;\;\mathpunct{,}
\end{equation*}

where $h$ is Planck's constant, $c$ is the speed of light, $\nu$ is the radiation frequency, and $\mathrm{T}$ is the temperature. Finally, the total emission coefficient can be written entirely in terms of the absorption coefficients as:

\begin{equation}
    \label{eq:emission_approx}
    \eta_\nu = \eta^t_\nu + \eta^s_\nu = \kappa_\nu B_\nu(\mathrm{T}) + \frac{\sigma_\nu}{4\pi} \int \nolimits_{4 \pi} I_\nu( \vectorr{n}') \dif \Omega' \;\;\mathpunct{.}
\end{equation}

\starsect{Expression of the radiative transfer equation}

The radiative transfer equation describes the evolution of the specific intensity of radiation as it propagates through a material medium. It expresses the balance between emission and absorption processes that modify this intensity. This balance can be written in the following form:
\begin{equation*}
    \dif I_\nu(\vectorr{n}) = \eta_\nu(\vectorr{n}) \dif \ell -\chi_\nu(\vectorr{n}) I_\nu(\vectorr{n}) \dif \ell \;\;\mathpunct{,}
\end{equation*}

\noindent or, equivalently, in Cartesian coordinates:
\begin{equation}
    \label{eq:transfer_rad}
    \left (\frac{1}{c} \partial_t + \vectorr{n} \cdot \nablav \right ) I_\nu(\vectorr{n}) = \eta_\nu(\vectorr{n}) -\chi_\nu(\vectorr{n}) I_\nu(\vectorr{n})   \;\;\mathpunct{.}
\end{equation}

Taking into account all the assumptions previously mentioned regarding radiation-matter interactions, the radiative transfer equation can also be written as:
\begin{equation}
    \begin{aligned}
        \label{eq:transfer_rad_approx}
        \left (\frac{1}{c} \partial_t + \vectorr{n} \cdot \nablav \right ) I_\nu(\vectorr{n}) =~&\kappa_\nu (B_\nu(\mathrm{T}) - I_\nu(\vectorr{n})) +\\
        &\sigma_\nu \left ( \frac{1}{4\pi}\int \nolimits_{4 \pi} I_\nu(\vectorr{n}') \dif \Omega' - I_\nu(\vectorr{n}) \right ) \;\;\mathpunct{.}
    \end{aligned}
\end{equation}

\subsection{The moment equations} \label{sec:M1}

In practice, the specific intensity of radiation is not directly observable. Instead, one has access to integrated quantities, such as the radiative flux, which corresponds to the amount of energy transported by radiation through a given surface, or the radiative pressure, which represents the mechanical force exerted by light on matter. This effect plays a key role in a wide range of phenomena, such as the dynamics of orbiting satellites or the operation of solar sails. In order to reformulate the radiative transfer equation~\eqref{eq:transfer_rad} in a more directly usable form, we therefore seek to express it in terms of three fundamental quantities: radiative energy, radiative flux, and radiative pressure.

The monochromatic radiative energy represents the energy density of photons of frequency $\nu$ at a given point in space. It corresponds to the zeroth-order moment of the specific intensity and is expressed as:
\begin{equation}
    \label{eq:Erad}
    \mathrm{E}_\nu = \frac{1}{c} \int \nolimits_{4 \pi} I_\nu(\vectorr{n}) \dif \Omega \;\;\mathpunct{.}
\end{equation}

The monochromatic radiative flux quantifies the energy transported by photons of frequency $\nu$ through a given surface. It corresponds to the first-order moment of the specific intensity and is written as:
\begin{equation}
    \label{eq:Frad}
    \vectorr{F_\nu} = \int \nolimits_{4 \pi} I_\nu(\vectorr{n}) \vectorr{n} \dif \Omega \;\;\mathpunct{.}
\end{equation}

Finally, the monochromatic radiative pressure represents the stress exerted by radiation on matter when it interacts with photons of frequency $\nu$. It corresponds to the second-order moment of the specific intensity and is expressed as:
\begin{equation}
    \label{eq:Prad}
    \tensorr{P}_\nu = \int \nolimits_{4 \pi} I_\nu(\vectorr{n}) \vectorr{n} \otimes \vectorr{n}  \dif \Omega \;\;\mathpunct{.}
\end{equation}

To obtain the monochromatic zeroth-order moment equation (referred to as the \textit{radiation energy equation}), one integrates the radiative transfer equation~\eqref{eq:transfer_rad} over all solid-angle directions. Using definitions~\eqref{eq:Erad} and~\eqref{eq:Frad}, this equation can be written as:
\begin{equation}
    \label{eq:ordre_0}
    \partial_t \mathrm{E}_\nu + \nablav \cdot \vectorr{F_\nu} = -c S_\nu^0  \;\;\mathpunct{.}
\end{equation}

To obtain the monochromatic first-order moment equation (referred to as the \textit{radiation momentum equation}), the radiative transfer equation is multiplied by \mbox{$\vectorr{n}/c$} and integrated over all solid-angle directions. This equation reads:
\begin{equation}
    \label{eq:ordre_1}
    \partial_t (c^{-2} \vectorr{F_\nu}) + \nablav \cdot \tensorr{P}_\nu = -\vectorr{S}_\nu  \;\;\mathpunct{.}
\end{equation}

The terms $S_\nu^0$ and $\vectorr{S}_\nu$ correspond to source terms modeling the interactions between light and matter and are given by:
\begin{align}
    S_\nu^0 &= -c^{-1} \int \nolimits_{4 \pi} \left( \eta_\nu(\vectorr{n}) - \chi_\nu(\vectorr{n}) I_\nu(\vectorr{n}) \right) \dif \Omega \;\;\mathpunct{,} \label{eq:S0_general_spectral} \\
    \vectorr{S}_\nu &= -c^{-1} \int \nolimits_{4 \pi} \left( \eta_\nu(\vectorr{n}) - \chi_\nu(\vectorr{n}) I_\nu(\vectorr{n}) \right) \vectorr{n}  \dif \Omega \;\;\mathpunct{.} \label{eq:S_general_spectral}
\end{align}

Working in the comoving frame and adopting the assumptions stated previously regarding radiation–matter interactions, these source terms can be written as:
\begin{align}
    S_\nu^0 &= \kappa_\nu \left ( \mathrm{E}_\nu - \frac{4 \pi}{c} B_\nu(\mathrm{T}) \right ) \;\;\mathpunct{,}
    \label{eq:S0_simplified_spectral}\\
    \vectorr{S}_\nu &= \chi_\nu \vectorr{F_\nu}/c \;\;\mathpunct{.} \label{eq:S_simplified_spectral}
\end{align}

The system of equations~\eqref{eq:ordre_0} and~\eqref{eq:ordre_1} describes the full set of radiative phenomena and is equivalent to the radiative transfer equation~\eqref{eq:transfer_rad}. However, it is underdetermined, as it involves three quantities for only two equations. To close the system, it is therefore necessary to introduce an additional relation, referred to as a \textit{closure relation}. Several models propose different approaches to define this relation.

\begin{enumerate}
    \item \textbf{The isotropic Flux-Limited Diffusion model~\cite{levermore_1981}:}\\
    In this model, radiation is assumed to be close to radiative equilibrium. Only the radiative energy equation~\eqref{eq:ordre_0} is retained, while the evolution of the radiative flux is neglected, leading to the approximation \mbox{$\vectorr{F_\nu} = c/\chi_\nu \nablav \cdot \tensorr{P}_\nu$}. Assuming isotropic radiation, the radiative pressure is written as \mbox{$\tensorr{P}_\nu = 1/3 \mathrm{E}_\nu \identity$}. This model is the fastest to solve numerically, but it does not guarantee the fundamental physical constraint on the radiative flux, namely \mbox{$||\vectorr{F_\nu}|| \leq c \mathrm{E}_\nu$}. An improvement consists in introducing a flux limiter $\lambda$, which modifies the expression of the radiative flux such that \mbox{$\vectorr{F_\nu} = c \lambda/(3 \chi_\nu) \nablav \mathrm{E}_\nu$}, where the factor $\lambda$ is chosen to ensure that the physical constraint \mbox{$||\vectorr{F_\nu}|| \leq c \mathrm{E}_\nu$} is satisfied.
    
    \item \textbf{The P1 model~\cite{olson_2000}:} \\  
    This model still assumes that radiation is close to radiative equilibrium, while retaining both the radiative energy equation~\eqref{eq:ordre_0} and the radiative flux equation~\eqref{eq:ordre_1}. It imposes a closure based on the assumption of isotropic radiation, \mbox{$\tensorr{P}_\nu = 1/3 \mathrm{E}_\nu \identity$}. Unlike the previous model, it guarantees the inequality \mbox{$||\vectorr{F_\nu}|| \leq c \mathrm{E}_\nu$} and allows for more general situations to be treated, but it loses accuracy when the radiation exhibits strong anisotropy (radiative equilibrium not valid anymore).
    
    \item \textbf{The M1 model~\cite{levermore_1983, dubroca_1999}:}\\
    This model generalizes the description of radiative transfer by relaxing the radiative equilibrium assumption. It retains both the radiative energy and flux equations~\eqref{eq:ordre_0} and~\eqref{eq:ordre_1}, while allowing for a certain degree of radiation anisotropy. Assuming that the radiation has an axis of symmetry aligned with the direction of the radiative flux, the radiative pressure can be expressed as:
    \begin{equation}
        \label{eq:Pnu_closure}
        \tensorr{P}_\nu = \tensorr{D}_\nu \mathrm{E}_\nu  \;\;\mathpunct{,}
    \end{equation}
    
    where $\tensorr{D}_\nu$ is called the \textit{Eddington tensor} and is defined by the relation~\cite{levermore_1983}:
    \begin{equation}
        \label{eq:Dnu_closure}
        \tensorr{D}_\nu = \frac{1-\chi_{R,\nu}}{2} \identity + \frac{3 \chi_{R,\nu} - 1}{2} \frac{\vectorr{F_\nu} \otimes \vectorr{F_\nu}}{||\vectorr{F_\nu}||^2} \;\;\mathpunct{,}
    \end{equation}
    
    with $\chi_{R,\nu}$ the \textit{Eddington factor}, whose expression depends on the adopted assumptions. Several authors have proposed different formulations for $\chi_{R,\nu}$:
    \begin{itemize}
        \item[-] \textbf{Kershaw (1976)~\cite{kershaw_1976}} proposed a second-order polynomial expression, which is simple but does not account for the physical nature of the radiation field;
        \item[-] \textbf{Minerbo (1978)~\cite{minerbo_1978}} proposed an approach based on the maximization of radiative entropy, assuming that \mbox{$\mathrm{E}_\nu<<4\pi h \nu^3/c^3$};
        \item[-] \textbf{Levermore (1979, 1981)~\cite{levermore_1979,pomraining_1981}} relied on the isotropic diffusion model;
        \item[-] \textbf{Pomraning (1973)~\cite{pomraning_1973}} developed a formulation adapted to relativistic fluids and applicable to radiative quantities integrated over all frequencies.
    \end{itemize}
    This model satisfies the inequality \mbox{$||\vectorr{F_\nu}|| \leq c \mathrm{E}_\nu$} and efficiently describes situations in which radiation is strongly anisotropic, making it one of the most accurate approaches. However, in 2D or 3D, its assumption of symmetry around the flux axis limits its ability to correctly represent complex configurations, such as two intersecting light beams\footnote{In this scenario, the M1 model predicts that the two light beams \quotes{collide}, which is not physically realistic.}. Despite this limitation, this model will be used, and its features will be detailed in the following sections.
\end{enumerate}

From a numerical standpoint, it is impossible to represent radiation propagation at all frequencies $\nu$, as this would require solving an infinite number of equations. An approach that enables the numerical solution of these equations consists in discretizing the electromagnetic spectrum into frequency \quotes{groups}. This corresponds to the M1-multigroup model.

\begin{figure}
    \begin{center}
        \begin{minipage}[t]{\linewidth}
            \centering
            \includegraphics[width=\textwidth]{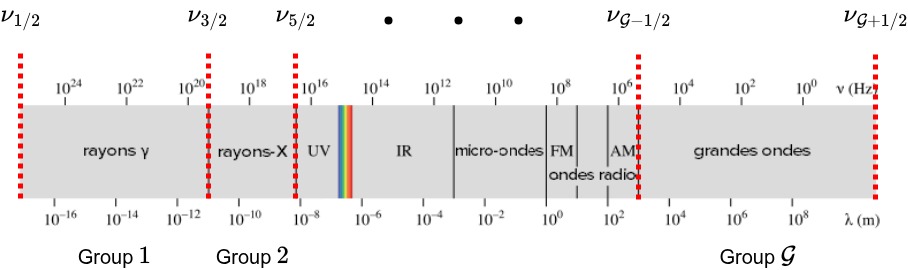}
            \caption{Electromagnetic spectrum. Example of discretization into $\mathcal{G}$ groups: the 1\textsuperscript{st} group includes gamma rays, the 2\textsuperscript{nd} includes X-rays, and the last includes long-wavelength radiation.}
        \label{fig:spectre_electromag}
        \end{minipage}
    \end{center}
\end{figure}

\starsect{The M1-multigroup model}

The M1-multigroup model relies on a spectral discretization of radiation into $\mathcal{G}$ frequency groups. This approach makes it possible to capture the spectral variability of radiative transfer by considering several frequency bands separately. Before addressing the detailed formulation of the model, it is necessary to define the structure of the frequency groups. One thus introduces a set of $\mathcal{G}$ frequency intervals \mbox{$\left \{ [\nu_{g-1/2}, \nu_{g+1/2}]\right \}_{g \in \llbracket 1, \mathcal{G} \rrbracket}$} covering the entire electromagnetic spectrum, with boundary conditions \mbox{$\nu{1/2} = 0$} and \mbox{$\nu_{\mathcal{G}+1/2} = +\infty$}. Each group \mbox{$g \in \llbracket 1, \mathcal{G} \rrbracket$} is defined by a pair of frequencies \mbox{$(\nu_{g-1/2}, \nu_{g+1/2})$}, which delimit the corresponding frequency range. Figure~\ref{fig:spectre_electromag} illustrates an example of a partition of the electromagnetic spectrum into several groups: the first group describes the behavior of photons in the gamma-ray domain, the second in the X-ray domain, and so on. The overall behavior of radiation within a given group is then described from the contributions of photons with frequencies lying between these two bounds. This model, developed by Turpault~\cite{turpault_2002,turpault_2005}, makes it possible to account for the spectral behavior of radiation within a numerical simulation code. It constitutes an improvement over the M1-gray model, which treats radiation as a whole without distinguishing between different spectral components. The latter is in fact a particular case of the M1-multigroup model, in which only a single group covering the entire spectrum is considered, with frequency bounds $\nu_{1/2} = 0$ and $\nu_{3/2} = +\infty$.

Before detailing the equations of the model, we introduce a few fundamental definitions that will be used in this section. Let us begin by defining the Planck and Rosseland mean opacities associated with each frequency group $g$, according to the following expressions:
\begin{align}
    \label{eq:opac_moyenne_Planckg}\
    \kappa_{P,g} &= \left. \left [ \int \nolimits_{\nu_{g-1/2}}^{\nu_{g+1/2}} \kappa_\nu B_\nu(\mathrm{T}) \dif \nu \right ] \middle/ \left [ \int \nolimits_{\nu_{g-1/2}}^{\nu_{g+1/2}} B_\nu(\mathrm{T}) \dif \nu \right ] \right. \;\;\mathpunct{,} \\
    \label{eq:opac_moyenne_Rosselandg}
    \kappa_{R,g} &= \left. \left [ \int \nolimits_{\nu_{g-1/2}}^{\nu_{g+1/2}} (\partial B_\nu(\mathrm{T}) / \partial \mathrm{T}) \dif \nu \right ] \middle/ \left [ \int \nolimits_{\nu_{g-1/2}}^{\nu_{g+1/2}} \chi_\nu^{-1} (\partial B_\nu(\mathrm{T}) / \partial \mathrm{T}) \dif \nu \right ] \right. \;\;\mathpunct{.}
\end{align}

\noindent Let us also define the operator $<\bcdot>_g$ as follows:
\begin{equation}
    \label{eq:op_<>g}
    <\bcdot>_g =  c^{-1} \int \nolimits_{\nu_{g-1/2}}^{\nu_{g+1/2}} \int \nolimits_{4 \pi} \bcdot~\dif \Omega \dif \nu \;\;\mathpunct{.}
\end{equation}

The M1-multigroup model aims to describe the global behavior of radiation within each spectral group $g$ (see figure~\ref{fig:spectre_electromag}). To this end, $\mathcal{G}$ sets of radiative quantities associated with group $g$ are introduced, including the energy $\mathrm{E}_g$, the flux $\vectorr{F_g}$, and the radiative pressure $\tensorr{P}_g$, defined as follows:
\begin{align}
    \mathrm{E}_g &= \int \nolimits_{\nu_{g-1/2}}^{\nu_{g+1/2}} \mathrm{E}_\nu \dif \nu \;\;\mathpunct{,} \label{eq:Erad_g} \\ 
    \vectorr{F_g} &= \int \nolimits_{\nu_{g-1/2}}^{\nu_{g+1/2}} \vectorr{F_\nu} \dif \nu \;\;\mathpunct{,} \label{eq:Frad_g} \\
    \tensorr{P}_g &= \int \nolimits_{\nu_{g-1/2}}^{\nu_{g+1/2}} \tensorr{P}_\nu \dif \nu \;\;\mathpunct{.} \label{eq:Prad_g}
\end{align}

Moreover, we define the total radiative energy $\mathrm{E}_R$, the total radiative flux $\vectorr{F_R}$, and the total radiative pressure $\tensorr{P}_R$ as:
\begin{align}
    \mathrm{E}_R &= \int \nolimits_{0}^{+\infty} \mathrm{E}_\nu \dif \nu = \sum_{g=1}^{\mathcal{G}}  \mathrm{E}_g \;\;\mathpunct{,} \label{eq:Erad_tot} \\ 
    \vectorr{F_R} &= \int \nolimits_{0}^{+\infty} \vectorr{F_\nu} \dif \nu = \sum_{g=1}^{\mathcal{G}}  \vectorr{F_g} \;\;\mathpunct{,} \label{eq:Frad_tot} \\
    \tensorr{P}_R &= \int \nolimits_{0}^{+\infty} \tensorr{P}_\nu \dif \nu = \sum_{g=1}^{\mathcal{G}}  \tensorr{P}_g \;\;\mathpunct{.} \label{eq:Prad_tot}
\end{align}

By integrating the monochromatic radiative transfer equations~\eqref{eq:ordre_0} and~\eqref{eq:ordre_1} over the frequency interval \mbox{$\closeinterv{\nu_{g-1/2}}{\nu_{g+1/2}}$}, one obtains the radiative transfer equations within group $g$:
\begin{equation}
    \label{eq:eqs_g}
    \forall g \in \llbracket 1, \mathcal{G} \rrbracket,~
    \left\{
    \begin{array}{lll}
        \partial_t \mathrm{E}_g + \nablav \cdot \vectorr{F_g} &= -c S^0_g \;\;\mathpunct{,}\\
        \partial_t (c^{-2} \vectorr{F_g}) + \nablav \cdot \tensorr{P}_g &= -\vectorr{S}_g \;\;\mathpunct{,}
    \end{array}
    \right.
\end{equation}

\noindent where $S^0_g$ and $\vectorr{S}_g$ are the source terms describing interactions between photons of each group $g$ and matter. Their general expressions are:
\begin{align}
    S^0_g &= -<\left (\eta_\nu(\vectorr{n}) -\chi_\nu(\vectorr{n}) I_\nu(\vectorr{n}) \right )>_g \;\;\mathpunct{,}
    \label{eq:S0_general_g}\\
    \vectorr{S}_g &= -< \left ( \eta_\nu(\vectorr{n}) -\chi_\nu(\vectorr{n}) I_\nu(\vectorr{n}) \right ) \vectorr{n}  >_g \;\;\mathpunct{.} \label{eq:S_general_g}
\end{align}

\noindent Adopting the assumptions previously introduced regarding radiation–matter interactions, these terms can be written as:
\begin{align}
    S^0_g &= \kappa_{E,g} \mathrm{E}_g - \kappa_{P,g} a_R \theta_g^4(\mathrm{T}) \;\;\mathpunct{,} \label{eq:S0_simplified_g} \\
    \vectorr{S}^i_g &= \chi_{F,g}^i \vectorr{F_g}^i/c \;\;\mathpunct{,} \label{eq:S_simplified_g}
\end{align}

\noindent where $a_R$ is the radiation constant and $\theta_g$ represents, in a sense, the radiative temperature of group $g$ at radiative equilibrium, and is given by:
\begin{equation}
    \theta_\mathrm{g}(\mathrm{T}) = \left ( \frac{<B_\nu(\mathrm{T})>_g}{a_R} \right)^{1/4} \;\;\mathpunct{.}
    \label{eq:temperature_multigroup}
\end{equation}

\noindent The opacities $\kappa_{E,g}$ and $\chi_{F,g}^i$ are energy- and flux-weighted opacities for group $g$ and are defined as:
\begin{align}
    \label{eq:opac_moyenne_Eg}
    \kappa_{E,g} &= \left. \left [ \int \nolimits_{\nu_{g-1/2}}^{\nu_{g+1/2}} \kappa_\nu \mathrm{E}_\nu \dif \nu \right ] \middle/ \left [ \int \nolimits_{\nu_{g-1/2}}^{\nu_{g+1/2}} \mathrm{E}_\nu \dif \nu \right ] \right. \;\;\mathpunct{,}\\
    \label{eq:opac_moyenne_Fg}
    \chi_{F,g}^i &= \left. \left [ \int \nolimits_{\nu_{g-1/2}}^{\nu_{g+1/2}} \chi_\nu \vectorr{F_\nu}^i \dif \nu \right ] \middle/ \left [ \int \nolimits_{\nu_{g-1/2}}^{\nu_{g+1/2}} \vectorr{F_\nu}^i \dif \nu \right ] \right. \;\;\mathpunct{.}
\end{align}

It is clear that directly computing the mean opacities $\kappa_{E,g}$ and $\chi_{F,g}^i$ is difficult, as their evaluation requires solving the spectral moment equations~\eqref{eq:ordre_0} and~\eqref{eq:ordre_1}, as well as precise knowledge of the spectral opacities of matter. This approach, in addition to being complex, would significantly increase the computational cost associated with solving the M1-multigroup model~\eqref{eq:eqs_g}. To simplify this issue, Mihalas~\cite{mihalas_1999} proposed replacing $\kappa_{E,g}$ and $\chi_{F,g}^i$ with the Planck and Rosseland mean opacities, $\kappa_{P,g}$ and $\kappa_{R,g}$, respectively. In the present framework, we assume that the fluid is in \gls{lte}, which implies that these opacities depend only on the temperature and density of the fluid.

This approximation is justified near the radiative equilibrium regime, where the radiative energy tends toward the Planck function (\mbox{$\mathrm{E}_\nu \propto B_\nu(\mathrm{T})$}), and where the radiative flux can be expressed as \mbox{$\vectorr{F_\nu} \approx \left. c \middle/ \chi_\nu \right. \nablav\cdot\tensorr{P}_\nu \propto \left. \partial B_\nu(\mathrm{T}) \middle/ (\chi_\nu \partial \mathrm{T}) \right.$}. It is particularly well suited to optically thick media, in which radiation efficiently thermalizes with matter.

However, this approximation becomes less valid in optically thin regimes, where radiation is more directional and departs from radiative equilibrium. Nevertheless, within the M1-multigroup framework, the finer the spectral groups, the more the mean opacities converge toward the spectral opacities: one has \mbox{$\kappa_{P,g},~\kappa_{E,g} \rightarrow \kappa_\nu$} and \mbox{$\kappa_{R,g},~\chi_{F,g}^i \rightarrow \chi_\nu$}, which reduces the impact of this approximation. In this context, the source terms then take the following form:
\begin{align}
    S^0_g &= \kappa_{P,g} (\mathrm{E}_g - a_R \theta_g^4(\mathrm{T})) \;\;\mathpunct{,} \label{eq:S0_simplified2_g}\\
    \vectorr{S}^i_g &= \kappa_{R,g} \vectorr{F_g}^i/c \;\;\mathpunct{,} \label{eq:S_simplified2_g}
\end{align}

It remains to define the closure relation for each group in order to solve the system. In the present approach, as before, radiation is assumed to be symmetric around the direction of the radiative flux $\vectorr{F_g}$ within each group. Under this assumption, the closure relation takes a form analogous to that given in~\eqref{eq:Pnu_closure}:
\begin{equation}
    \label{eq:Pg_closure}
    \tensorr{P}_g = \tensorr{D}_g \mathrm{E}_g  \;\;\mathpunct{,}
\end{equation}

\noindent where $\tensorr{D}_g$ is the \textit{Eddington tensor} associated with group $g$ and is given by~\cite{levermore_1983}:
\begin{equation}
    \label{eq:Dg_closure}
    \tensorr{D}_g = \frac{1-\chi_g}{2} \identity + \frac{3 \chi_g - 1}{2} \frac{\vectorr{F_g} \otimes \vectorr{F_g}}{||\vectorr{F_g}||^2} \;\;\mathpunct{,}
\end{equation}

\noindent where $\chi_g$ denotes the \textit{Eddington factor} associated to the group $g$. Its determination requires formulating an assumption on the shape of the specific intensity in order to render the problem mathematically tractable. In the following section, we present the developments that enable its numerical computation.

\subsection{The Eddington factor} \label{sec:dependence_chig}

Within the framework of the M1-multigroup model, the objective is to determine an expression for the specific intensity that minimizes the radiative entropy, in order to close the system of equations for each group and to derive an expression for the Eddington factor. The radiative entropy density is given by~\cite{fort_1997}:
\begin{equation}
    \label{eq:densite_entropie_R}
     h_R(I) = \frac{2 k_B \nu^2}{c^3} [n \ln{(n)} - (n+1) \ln{(n+1)}] \;\;\mathpunct{,}
\end{equation}

\noindent where $n$ is the occupation number, defined by \mbox{$n = \frac{c^2}{2 h \nu} I$}. The total radiative entropy $H_R(I)$ is then defined as
\begin{equation}
    \label{eq:entropie_R}
     H_R(I) = \sum_{g=1}^{\mathcal{G}}< h_R(I) >_g \;\;\mathpunct{,}
\end{equation}

As seen in equations~\eqref{eq:Erad_g} and~\eqref{eq:Frad_g}, the specific intensity is related to the radiative energy and flux in each group. We therefore seek to retain only the specific intensity $\mathcal{I}_\nu$ that minimizes the radiative entropy, i.e. the specific intensity that is the solution of the following minimization problem:
\begin{equation}
    \label{eq:minimisation_entropie_mg}
     H_R(\mathcal{I}_\nu) = \min_I \left \{ H_R(I)~|~\forall g \in \llbracket 1, \mathcal{G} \rrbracket,~<I>_g = \mathrm{E}_R,~<c \vectorr{n} I>_g=\vectorr{F_g} \right \} \;\;\mathpunct{.}
\end{equation}

\noindent It can be shown that the intensity satisfying this problem can be written as~\cite{turpault_2003_these}:
\begin{equation}
    \label{eq:I_specifique_mg}
    \mathcal{I}_\nu = \sum_{g=1}^{\mathcal{G}} \indicatrix_g(\nu) ~\mathcal{I}_{\nu,g} \;\;\mathpunct{,}
\end{equation}

\noindent where \mbox{$\indicatrix_g(\nu):\tensorr{R}^+ \rightarrow \{0, 1\}$} is the characteristic function, equal to $1$ for frequencies $\nu$ in \mbox{$\closeinterv{\nu_{g-1/2}}{\nu_{g+1/2}}$} and $0$ otherwise, and where $\mathcal{I}_{\nu,g}$ denotes the specific intensity in group $g$, given by:
\begin{equation}
    \label{eq:I_specifique2_mg}
    \mathcal{I}_{\nu,g} = \frac{2 h \nu^3}{c^2} \left [ \exp \left ( \frac{h \nu}{k_B} \vectorr{m}_g \cdot \vectorr{\alpha}_g \right ) - 1 \right ]^{-1} \;\;\mathpunct{,} 
\end{equation}

\noindent where \mbox{$\vectorr{\alpha}_g = (\alpha_{0,g}, \vectorr{\alpha_{1,g}})$} are the Lagrange multipliers of group $g$ associated with problem~\eqref{eq:minimisation_entropie_mg}, $\vectorr{m}_g$ is a four-vector defined as \mbox{$\vectorr{m}_g = (1, \vectorr{n}_g)$}, and $\vectorr{n}_g$ is the propagation direction of light in group $g$. By setting \mbox{$\vectorr{\alpha_{1,g}} \cdot \vectorr{n}_g = \alpha_{0,g} \beta_g \cos(\theta_g)$}, the specific intensity in group $g$ can also be written as:
\begin{equation}
    \label{eq:I_specifique_g2}
     \mathcal{I}_{\nu,g} = \frac{2 h \nu^3}{c^2} \left [ \exp\left ( \frac{h \nu}{k_B} \alpha_{0,g} (1 + \beta_g \cos(\theta_g)) \right ) -1 \right ]^{-1} \;\;\mathpunct{,}
\end{equation}

In this context, $\alpha_{0,g}$ represents the inverse of a temperature and takes values in $\mathbbm{R}^+$, while $\beta_g$ measures the degree of anisotropy of the radiation in group $g$, with values in the interval \mbox{$\openinterv{-1}{1}$}. It is important to note that if $\alpha_{0,g}$ and $\beta_g$ are identical for all groups (i.e. if \mbox{$\forall g \in \llbracket 1, \mathcal{G} \rrbracket$}, \mbox{$\alpha_{0,g} = \alpha_0$} and \mbox{$\beta_g = \beta$}), one recovers the specific intensity of the M1-gray model.

However, unlike the M1-gray model, no general analytical formula exists for the Eddington factor $\chi_g$ in this framework. Nevertheless, Turpault showed that, in the M1-multigroup model, for any physically admissible pair \mbox{($\mathrm{E}_g$, $||\vectorr{F}_g||$)}, there exists a unique pair of Lagrange multipliers \mbox{($\alpha_{0,g}$, $\beta_g$)}~\cite{turpault_2003_these}. Consequently, the Eddington factor can be computed numerically using a root-finding algorithm, such as the bisection–Newton method or a line-search method, described in Appendix~\secref{appendice:Radiation_numerique}.

In the next section, I will analyze the behavior of the Eddington factor in the M1-multigroup model, in accordance with the study presented in the corresponding article~\cite{radureau_2025a}. In addition, in Section~\secref{sec:Eddington_mg}, I will present a method based on \gls{ai}, developed during this thesis, that enables an efficient computation of $\chi_g$.

To begin with, let us examine the parameters that influence the Eddington factor in the M1-multigroup model. The radiative energy $\mathrm{E}_g$, the norm of the radiative flux $||\vectorr{F_g}||$, and the radiative pressure component \mbox{$\mathrm{P}_g = \chi_g \mathrm{E}_g$} in the direction of the radiative flux can be expressed as follows:
\begin{align}
    \mathrm{E}_g~~~ &= \frac{1}{c} \int \nolimits_{\nu_{g-1/2}}^{\nu_{g+1/2}} \int \nolimits_{4 \pi} \mathcal{I}_{\nu}~\mathrm{d}\Omega~\mathrm{d}\nu \;\;\mathpunct{,} \label{eq:Erad_int} \\
    \mathpunct||\vectorr{F_g}|| &= \int \nolimits_{\nu_{g-1/2}}^{\nu_{g+1/2}} \int \nolimits_{4 \pi} \mathcal{I}_{\nu}~\cos(\theta)~\mathrm{d}\Omega~\mathrm{d}\nu  \;\;\mathpunct{,} \label{eq:Frad_int}\\
    \mathrm{P}_g~~ &= \frac{1}{c} \int \nolimits_{\nu_{g-1/2}}^{\nu_{g+1/2}} \int \nolimits_{4 \pi} \mathcal{I}_{\nu}~\cos^2(\theta)~\mathrm{d}\Omega~\mathrm{d}\nu  \;\;\mathpunct{,} \label{eq:Prad_int}
\end{align}

or, equivalently, by noting that \mbox{$\mathrm{d}\Omega=\sin(\theta) \dif \theta \dif \phi$}, setting \mbox{$\mu = \cos(\theta)$}, and using the expression of the specific intensity in group $g$, one can write:
\begin{align}
    \mathrm{E}_g~~~ &= \frac{4 h \pi}{c^3} \int \nolimits_{\nu_{g-1/2}}^{\nu_{g+1/2}} \int \nolimits_{-1}^{1} \nu^3 \left [ e^{\frac{h \nu}{k_B} \alpha_{0,g} ( 1+ \beta_g \mu )} - 1 \right ]^{-1}~\mathrm{d}\mu~\mathrm{d}\nu \;\;\mathpunct{,} \label{eq:Er_g_calc}\\
    \mathpunct||\vectorr{F_g}|| &= \frac{4 h \pi}{c^2} \int \nolimits_{\nu_{g-1/2}}^{\nu_{g+1/2}} \int \nolimits_{-1}^{1} \nu^3 \mu \left [ e^{\frac{h \nu}{k_B}  \alpha_{0,g} ( 1+ \beta_g \mu )} - 1 \right ]^{-1}~\mathrm{d}\mu~\mathrm{d}\nu  \;\;\mathpunct{,} \label{eq:Fr_g_calc}\\
    \mathrm{P}_g~~ &= \frac{4 h \pi}{c^3} \int \nolimits_{\nu_{g-1/2}}^{\nu_{g+1/2}} \int \nolimits_{-1}^{1} \nu^3 \mu^2 \left [ e^{\frac{h \nu}{k_B}  \alpha_{0,g} ( 1+ \beta_g \mu )} - 1 \right ]^{-1}~\mathrm{d}\mu~\mathrm{d}\nu  \;\;\mathpunct{.} \label{eq:Pr_g_calc}
\end{align}

\noindent It is then possible to distinguish four particular cases depending on the spectral bounds of the group under consideration:

\begin{enumerate}
    \item \textbf{\mbox{$\boldsymbol{\nu_{g-1/2}=0}$} and \mbox{$\boldsymbol{\nu_{g+1/2}=+\infty}$}:} this corresponds to the M1-gray model, in which the Eddington factor $\chi_R$ admits an analytical expression~\cite{pomraning_1973}:
    \begin{equation}
        \label{eq:facteur_Eddington_gray}
         \chi_R = \frac{3+4~\mathrm{f}_R^{~2}}{5+2\sqrt{4 - 3~\mathrm{f}_R^{~2}}} \;\;\mathpunct{,}
    \end{equation}
    where $\mathrm{f}_R$ is the reduced flux, defined as \mbox{$\mathrm{f}_R = ||\vectorr{F_R}||/(c~\mathrm{E}_R)$}, with $\vectorr{F_R}$ and $\mathrm{E}_R$ denoting the radiative flux and radiative energy integrated over the entire electromagnetic spectrum, respectively;
    \item \textbf{\mbox{$\boldsymbol{\nu_{g-1/2}=0}$} and \mbox{$\boldsymbol{\nu_{g+1/2}<+\infty}$}:} Eddington factor for the first group;
    \item \textbf{\mbox{$\boldsymbol{\nu_{g-1/2}>0}$} and \mbox{$\boldsymbol{\nu_{g+1/2}=+\infty}$}:} Eddington factor for the last group;
    \item \textbf{\mbox{$\boldsymbol{\nu_{g-1/2}>0}$} and \mbox{$\boldsymbol{\nu_{g+1/2}<+\infty}$}:} general Eddington factor for all remaining groups.
\end{enumerate}

In the following sections, I will analyze the parameters that influence the Eddington factor in each of these cases.

\starsect{Case \mbox{$\boldsymbol{\nu_{g-1/2}>0}$} and \mbox{$\boldsymbol{\nu_{g+1/2}<\infty}$}}

In this case, it is possible to perform the change of variable \mbox{$\tilde{\nu}=\nu / \nu_{m,g}$}, where $\nu_{m,g}$ is the geometric mean of the frequency bounds of the group (\mbox{$\nu_{m,g}=\sqrt{\nu_{g-1/2}~\nu_{g+1/2}}$}), in integrals~\eqref{eq:Er_g_calc}, \eqref{eq:Fr_g_calc}, and \eqref{eq:Pr_g_calc}. The radiative quantities can then be expressed as follows:
\begin{align*}
    \frac{k_B^4~\mathrm{E}_g}{h^4~\nu_{m,g}^4} &= a_R \frac{15}{\pi^4} \int \nolimits_{\delta_g^{1/2}}^{\delta_g^{-1/2}} \int \nolimits_{-1}^{1} \tilde{\nu}^3 \left [ e^{\tilde{\nu}~\widetilde{\alpha_{0,g}}~(1 + \beta_g \mu )} - 1 \right ]^{-1}~\mathrm{d}\mu~\mathrm{d}\tilde{\nu} \;\;\mathpunct{,}\\
    \frac{k_B^4~||\vectorr{F_g}||}{h^4~\nu_{m,g}^4} &= a_R c \frac{15}{\pi^4} \int \nolimits_{\delta_g^{1/2}}^{\delta_g^{-1/2}} \int \nolimits_{-1}^{1} \tilde{\nu}^3 \mu \left [ e^{\tilde{\nu}~\widetilde{\alpha_{0,g}} (1 + \beta_g \mu )} - 1 \right ]^{-1}~\mathrm{d}\mu~\mathrm{d}\tilde{\nu} \;\;\mathpunct{,}\\
    \frac{k_B^4~\mathrm{P}_g}{h^4~\nu_{m,g}^4} &= a_R \frac{15}{\pi^4} \int \nolimits_{\delta_g^{1/2}}^{\delta_g^{-1/2}} \int \nolimits_{-1}^{1} \tilde{\nu}^3 \mu^2 \left [ e^{\tilde{\nu}~\widetilde{\alpha_{0,g}}~(1 + \beta_g \mu )} - 1 \right ]^{-1}~~\mathrm{d}\mu~\mathrm{d}\tilde{\nu} \;\;\mathpunct{,}
\end{align*}

\noindent where \mbox{$\delta_g=\nu_{g-1/2}/\nu_{g+1/2}$} is a parameter taking values in \mbox{$\openinterv{0}{1}$}, referred to as the \textit{group narrowness}, and \mbox{$\widetilde{\alpha_{0,g}}=h \nu_{m,g} \alpha_{0,g}/k_B$} is the Lagrange multiplier $\alpha_{0,g}$ made dimensionless using the frequency $\nu_{m,g}$. In other words, by introducing the reduced flux \mbox{$\mathrm{f}_g = ||\vectorr{F_g}||/(c \mathrm{E}_g)$} and the Eddington factor \mbox{$\chi_g = \mathrm{P}_g/\mathrm{E}_g$}, one can write:
\begin{align}
    \mathcal{T}_g &= \left ( \frac{15}{\pi^4} \int \nolimits_{\delta_g^{1/2}}^{\delta_g^{-1/2}} \int \nolimits_{-1}^{1} \tilde{\nu}^3 \left [ e^{\tilde{\nu}~\widetilde{\alpha_{0,g}}~(1 + \beta_g \mu )} - 1 \right ]^{-1}~\mathrm{d}\mu~\mathrm{d}\tilde{\nu} \right )^{1/4} \;\;\mathpunct{,} \label{eq:Tg_expr}\\
    \mathrm{f}_g &= \frac{\int \nolimits_{\delta_g^{1/2}}^{\delta_g^{-1/2}} \int \nolimits_{-1}^{1} \tilde{\nu}^3 \mu \left [ e^{\tilde{\nu}~\widetilde{\alpha_{0,g}}~(1 + \beta_g \mu )} - 1 \right ]^{-1}~\mathrm{d}\mu~\mathrm{d} \tilde{\nu}}{\int \nolimits_{\delta_g^{1/2}}^{\delta_g^{-1/2}} \int \nolimits_{-1}^{1} \tilde{\nu}^3 \left [ e^{\tilde{\nu}~\widetilde{\alpha_{0,g}}~(1 + \beta_g \mu )} - 1 \right ]^{-1}~\mathrm{d}\mu~\mathrm{d}\tilde{\nu}} \;\;\mathpunct{,} \label{eq:fg_expr}\\
    \chi_g &= \frac{\int \nolimits_{\delta_g^{1/2}}^{\delta_g^{-1/2}} \int \nolimits_{-1}^{1} \tilde{\nu}^3 \mu^2 \left [ e^{\tilde{\nu}~\widetilde{\alpha_{0,g}}~(1 + \beta_g \mu )} - 1 \right ]^{-1}~\mathrm{d}\mu~\mathrm{d}\tilde{\nu}}{\int \nolimits_{\delta_g^{1/2}}^{\delta_g^{-1/2}} \int \nolimits_{-1}^{1} \tilde{\nu}^3 \left [ e^{\tilde{\nu}~\widetilde{\alpha_{0,g}}~(1 + \beta_g \mu )} - 1 \right ]^{-1}~\mathrm{d}\mu~\mathrm{d}\tilde{\nu}} \;\;\mathpunct{,} \label{eq:chig_expr}
\end{align}

\noindent where \mbox{$\mathcal{T}_g=k_B \mathrm{T}_g /(h \nu_{m,g})$} is the dimensionless radiative temperature and \mbox{$\mathrm{T}_g=(\mathrm{E}_g / a_R)^{1/4}$} is the radiative temperature of group $g$. As for any physically admissible pair \mbox{($\mathrm{E}_g$, $||\vectorr{F_g}||$)}, there exists a unique pair of Lagrange multipliers \mbox{($\alpha_{0,g}$, $\beta_g$)}. The equations above show that, for any pair \mbox{($\mathcal{T}_g$, $\mathrm{f}_g$)}, there exists a unique pair of Lagrange multipliers \mbox{($\alpha_{0,g}$, $\beta_g$)}, which then allows the Eddington factor to be computed. One therefore concludes that, in the present case, the Eddington factor depends only on three parameters:

\begin{figure}
    \begin{center}
        \begin{minipage}[t]{0.6\linewidth}
            \centering
            \includegraphics[width=\textwidth]{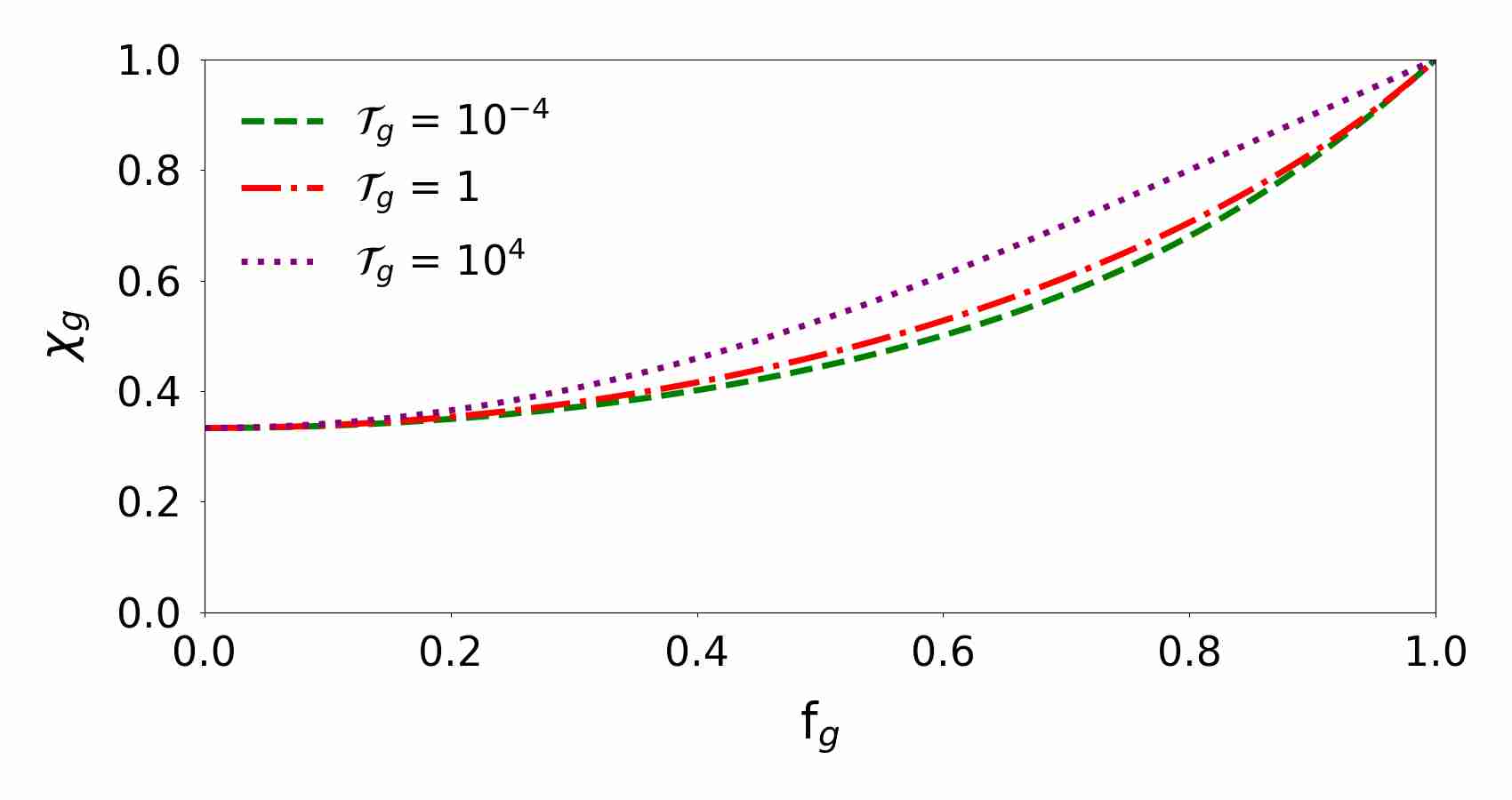}
            \caption{Evolution of the Eddington factor $\chi_g$ at different radiative temperatures $\mathcal{T}_g$, using a group narrowness $\delta_g = 10^{-4}$.}
             \label{fig:fg_dep}
        \end{minipage}
    \end{center}
\end{figure}

\begin{enumerate}
    \item \textbf{The dimensionless radiative temperature $\bm{\mathcal{T}_g}$:} comparison between the equivalent radiative energy of the group, $k_B \mathrm{T}_g$, and the geometric mean of the photon energy levels within the group, $h \nu_{m,g}$;
    \item \textbf{The reduced flux $\pmb{\mathrm{f}_{g}}$:} degree of anisotropy of the radiation within group $g$. Its values lie in $\openrinterv{0}{1}$ (\mbox{$\mathrm{f}_g=0$}: isotropic radiation, \mbox{$\mathrm{f}_g \rightarrow 1$}: highly directional radiation);
    \item \textbf{The group narrowness $\boldsymbol{\delta_g}$:} indicator of the separation between the frequency bounds of the group, $\nu_{g-1/2}$ and $\nu_{g+1/2}$, defined by the relation \mbox{$\delta_g = \nu_{g-1/2}/\nu_{g+1/2}$}, with values in the interval $\openinterv{0}{1}$ (\mbox{$\delta_g \rightarrow 0$}: widely separated frequency bounds, \mbox{$\delta_g \rightarrow 1$}: very close frequency bounds).
\end{enumerate}

In order to gain insight into the dependence of the Eddington factor on the reduced flux, I plotted its evolution for different values of the dimensionless radiative temperature in figure~\ref{fig:fg_dep}. Two common features can be observed across all curves:

\begin{enumerate}
    \item For any value of $\mathcal{T}_g$, $\chi_g$ always varies from $1/3$ at \mbox{$\mathrm{f}_g=0$} to $1$ at \mbox{$\mathrm{f}_g=1$}, and has a zero slope at \mbox{$\mathrm{f}_g=0$};
    \item $\mathcal{T}_g$ influences the evolution of $\chi_g$ only at intermediate values of $\mathrm{f}_g$.
\end{enumerate}

Overall, the Eddington factor depends more strongly on the reduced flux $\mathrm{f}_g$ than on the radiative temperature and the group narowness. However, this dependence does exist, and neglecting it leads to errors in the estimation of the Eddington factor of up to 16~\% for values of $\mathrm{f}_g = 0.65$, as shown in figure~\ref{fig:fg_dep}. Let us now analyze in more detail the dependence on $\mathcal{T}_g$ and $\delta_g$.

\begin{figure*}
    \begin{subfigure}[t]{0.48\textwidth}
        \centering
        \includegraphics[width=\textwidth]{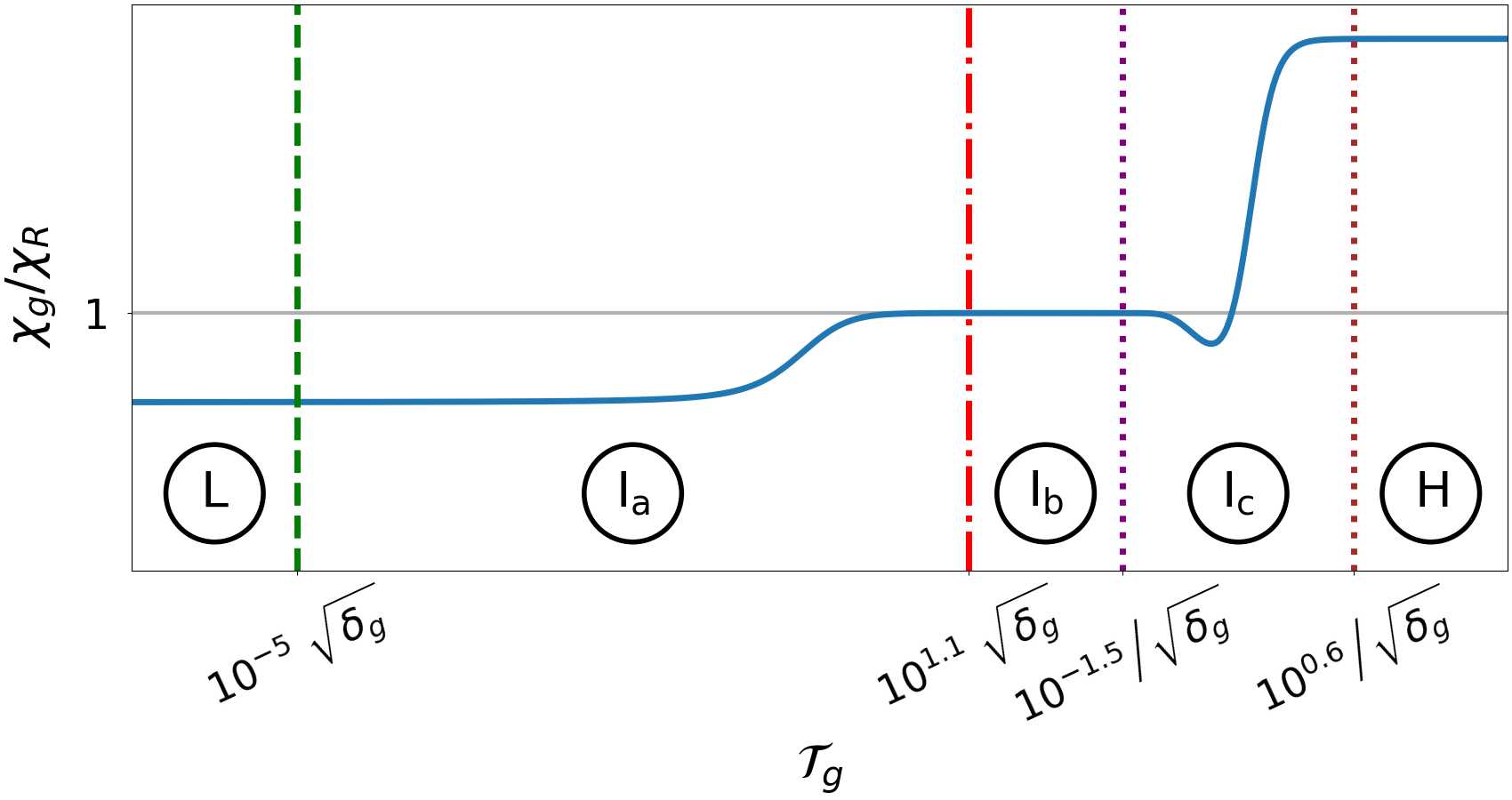}
        \caption{\mbox{$\delta_g=10^{-4}$}. L, H: asymptotic domains; $\mathrm{I_a}$: low-to-intermediate radiative temperatures; $\mathrm{I_b}$: domain in which $\chi_R$ is valid; $\mathrm{I_c}$: high-intermediate radiative temperatures.}
        \label{fig:dnu_1e-4}
    \end{subfigure}
    \hfill
    \begin{subfigure}[t]{0.48\textwidth}
        \centering
        \includegraphics[width=\textwidth]{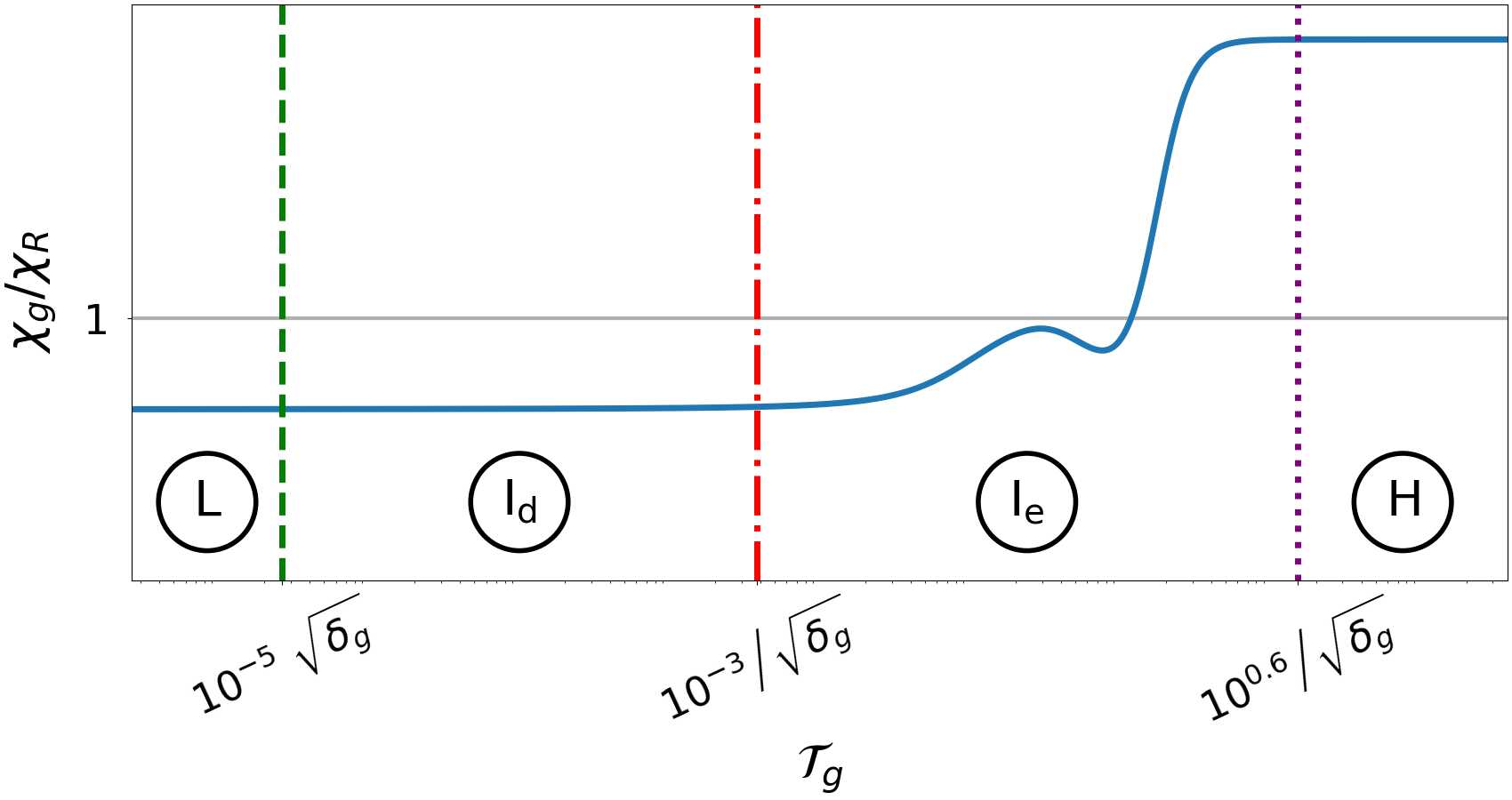}
        \caption{\mbox{$\delta_g=0.07$}. L, H: asymptotic domains; $\mathrm{I_d}$: low-to-intermediate radiative temperatures; $\mathrm{I_e}$: intermediate-to-high radiative temperatures.}
        \label{fig:dnu_7e-2}
    \end{subfigure}
    \hfill
    \begin{subfigure}[t]{0.48\textwidth}
        \centering
        \includegraphics[width=\textwidth]{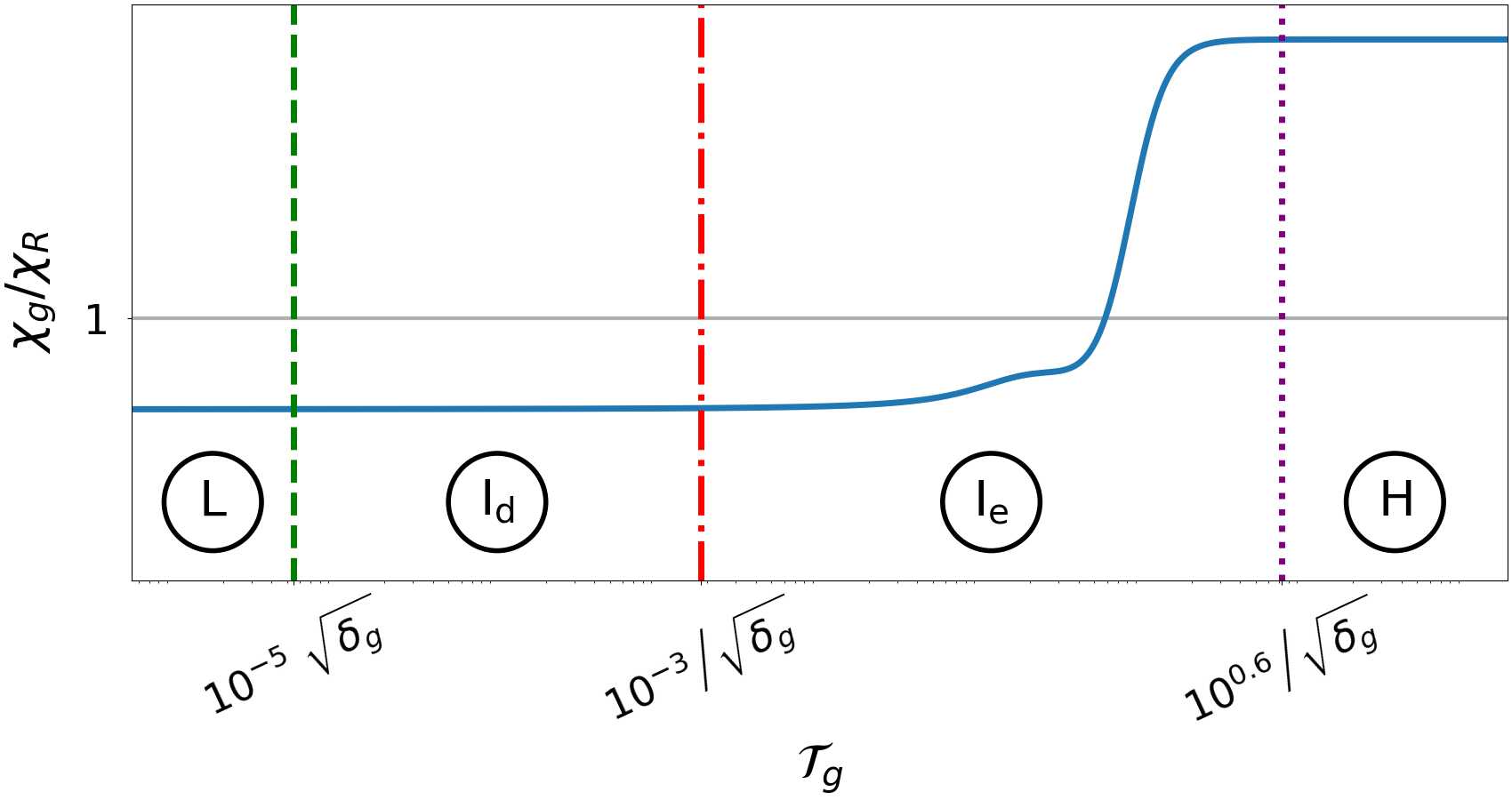}
        \caption{\mbox{$\delta_g=0.3$}. L, H: asymptotic domains; $\mathrm{I_d}$: low-to-intermediate radiative temperatures; $\mathrm{I_e}$: intermediate-to-high radiative temperatures.}
        \label{fig:dnu_3e-1}
    \end{subfigure}
    \hfill
    \begin{subfigure}[t]{0.48\textwidth}
        \centering
        \includegraphics[width=\textwidth]{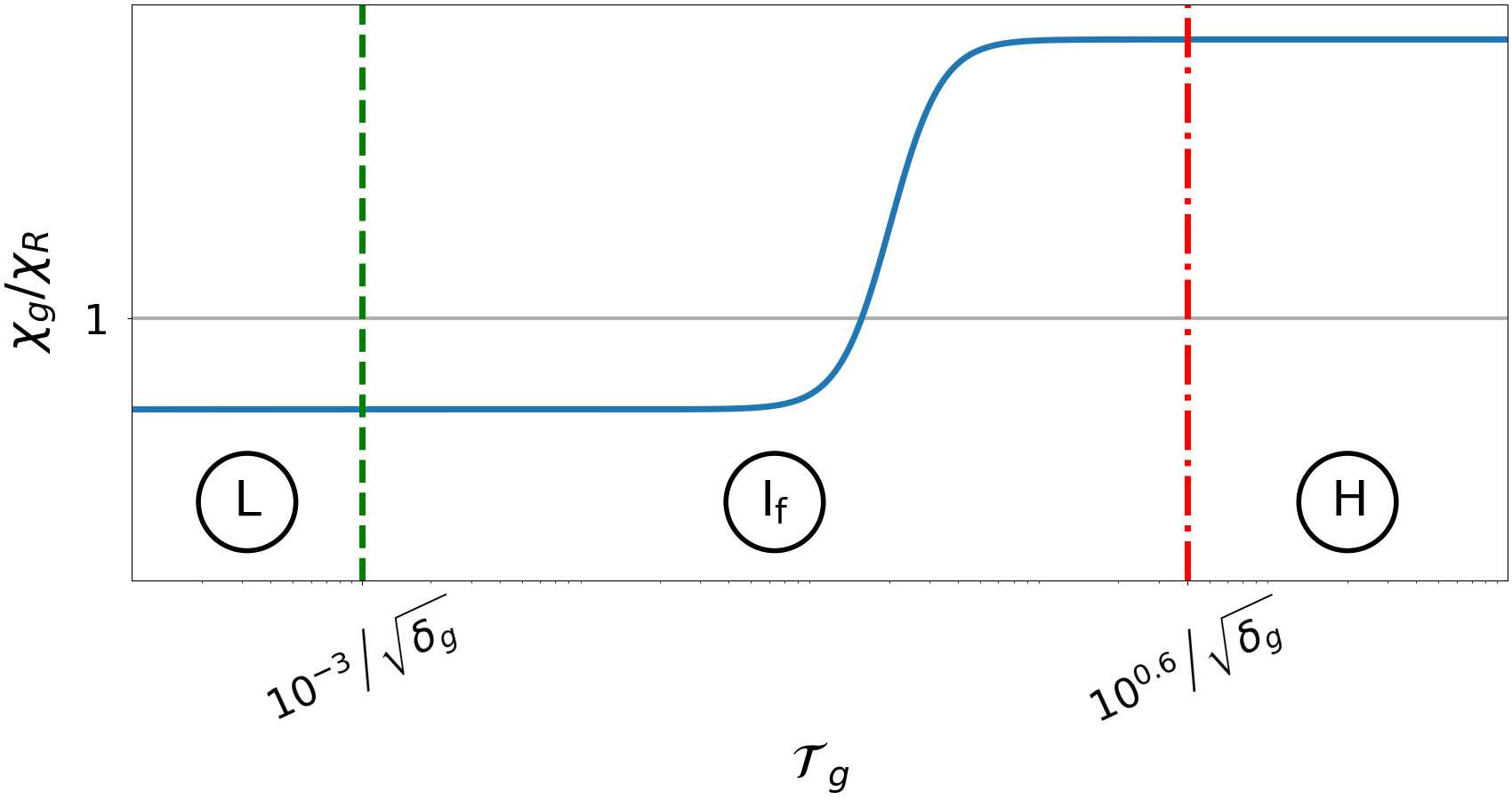}
        \caption{\mbox{$\delta_g=0.99$}. L, H: asymptotic domains; $\mathrm{I_f}$: intermediate radiative temperatures.}
        \label{fig:dnu_0.99}
    \end{subfigure}
    \caption{Ratio between the Eddington factor of the M1-multigroup model, $\chi_g$, and that of the M1-gray model, $\chi_R$, for a reduced flux \mbox{$\mathrm{f}_g = 0.65$}, at different values of the group finesse $\delta_g$.}
    \label{fig:Tg_dep}
\end{figure*}

The ratio of the Eddington factor $\chi_g$ in the M1-multigroup model to its gray expression is shown as a function of the radiative temperature $\mathcal{T}_g$ for different values of the group narowness $\delta_g$ (see figure~\ref{fig:Tg_dep}). Qualitatively, when the group is wide (\mbox{$\delta_g<<1$}), the presence of a central \quotes{plateau} can be observed within domain $\mathrm{I_b}$ (see figure~\ref{fig:dnu_1e-4}). This plateau lies approximately in the interval $\mathcal{T}g \in \closeinterv{\sqrt{\delta_g}}{1/\sqrt{\delta_g}}$, which corresponds to radiative temperatures such that $k_B \mathrm{T}_g \in \closeinterv{h \nu_{g-1/2}}{h \nu_{g+1/2}}$. As the group becomes narrower, i.e. as $\delta_g$ increases, the size of the \quotes{plateau} decreases until it disappears (see figures~\ref{fig:dnu_7e-2} and~\ref{fig:dnu_3e-1}). When the frequencies become nearly identical, two distinct regimes emerge: one at low temperature and one at high temperature. The transition between these regimes occurs at lower radiative temperatures as the frequencies approach one another (\mbox{$\delta_g \rightarrow 1$}, see figure~\ref{fig:dnu_0.99}). Overall, three main regimes can be identified in figure~\ref{fig:Tg_dep}:

\begin{enumerate}
    \item \textbf{Domain L:} low radiative temperature. In this domain, \mbox{$\mathcal{T}_g<<\sqrt{\delta_g}$}. $\chi_g$ depends only on $\mathrm{f}_g$. This domain corresponds to that described by Minerbo~\cite{minerbo_1978};
    \item \textbf{Domain I:} intermediate radiative temperatures. $\chi_g$ depends strongly on $\mathcal{T}_g$, $\mathrm{f}_g$, and $\delta_g$;
    \item \textbf{Domain H:} high radiative temperature. In this domain, \mbox{$\mathcal{T}_g>>1 / \sqrt{\delta_g}$}. The value of $\chi_g$ depends only on $\mathrm{f}_g$.
\end{enumerate}

The intermediate domain I, based on the group narrowness, can be further subdivided into three cases:

\begin{enumerate}
    \item \textbf{Case \mbox{$\boldsymbol{\delta_g \rightarrow 0}$}:} wide group. Three intermediate domains can be identified (see figure~\ref{fig:dnu_1e-4}):
    \begin{itemize}
        \item[-] \textbf{$\boldsymbol{\mathrm{I_a}}$:} $\chi_g$ depends strongly on $\mathcal{T}_g$ and $\mathrm{f}_g$;
        \item[-] \textbf{$\boldsymbol{\mathrm{I_b}}$:} corresponds to the domain in which the Eddington factor of the M1-gray model is valid. This domain is located approximately in the interval $\mathcal{T}_g \in \closeinterv{\sqrt{\delta_g}}{1/\sqrt{\delta_g}}$, and its width is therefore linked to $\delta_g$, i.e. it decreases as $\delta_g$ increases;
        \item[-] \textbf{$\boldsymbol{\mathrm{I_c}}$:} $\chi_g$ depends strongly on $\mathcal{T}_g$ and $\mathrm{f}_g$.
    \end{itemize}
    \item \textbf{Intermediate \mbox{$\boldsymbol{\delta_g}$}:} in domains $\mathrm{I_d}$ and $\mathrm{I_e}$, $\chi_g$ depends on $\mathcal{T}_g$, $\mathrm{f}_g$, and $\delta_g$ (see figures~\ref{fig:dnu_7e-2} and~\ref{fig:dnu_3e-1});
    \item \textbf{Case \mbox{$\boldsymbol{\delta_g \rightarrow 1}$}:} narrow group. In domain $\mathrm{I_f}$, $\chi_g$ is observed to depend only on $\mathcal{T}_g$ and $\mathrm{f}_g$.
\end{enumerate}

\begin{figure}
    \begin{center}
        \begin{minipage}[t]{0.6\linewidth}
            \centering
            \includegraphics[width=\textwidth]{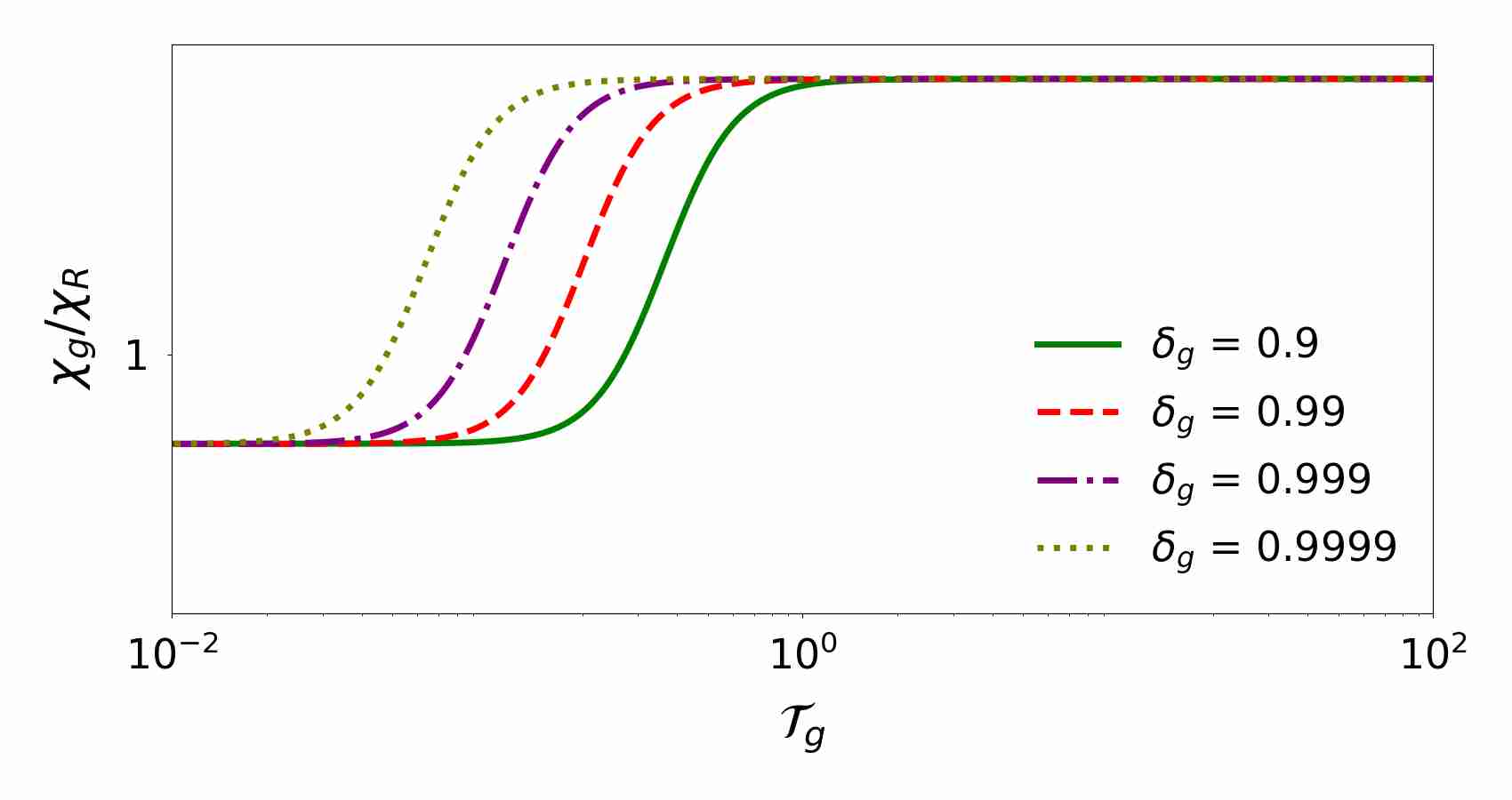}
            \caption{Evolution of the ratio between the Eddington factor of the M1-multigroup model, $\chi_g$, and that of the M1-gray model, $\chi_R$, as a function of $\delta_g$, illustrating the shift of the curve as $\delta_g$ approaches 1.}
             \label{fig:T_shift}
        \end{minipage}
    \end{center}
\end{figure}

In the case \mbox{$\delta_g \rightarrow 1$}, figure~\ref{fig:T_shift} shows the curve of the Eddington factor $\chi_g$ as a function of the dimensionless radiative temperature $\mathcal{T}_g$ for values of the group narrowness $\delta_g$ close to 1. The shape of the curve remains unchanged but is shifted toward lower radiative temperatures. In order to understand this behavior, one can seek an approximate expression for the Eddington factor at very high radiative temperatures $\mathcal{T}_g$. Let us assume that the dimensionless Lagrange multiplier $\widetilde{\alpha_{0,g}}$ satisfies \mbox{$\widetilde{\alpha_{0,g}}<<\sqrt{\delta_g}$} (i.e. \mbox{$\alpha_{0,g} << k_B/(h \nu_{g+1/2})$}). The specific intensity in group $g$ can then be written as:
\begin{equation}
    \mathcal{I}_{\nu,g} (\mu) \approx \frac{2 k_B}{c^2}\frac{\nu^2}{\alpha_{0,g} (1 + \beta_g \mu)} \;\;\mathpunct{,}
\end{equation}

\noindent which allows the following expressions for the radiative temperature, the reduced flux, and the Eddington factor to be derived:
\begin{align}
    \mathrm{T}_g &= \left ( \frac{5 h^3}{\pi^4 k_B^3} \frac{\nu_{g+1/2}^3 - \nu_{g-1/2}^3}{\alpha_{0,g}} \frac{\arctanh(\beta_g)}{\beta_g} \right )^{1/4} \;\;\mathpunct{,} \label{eq:Tg_H}\\
     \mathrm{f}_g &= - \frac{\arctanh(\beta_g) - \beta_g}{\beta_g \arctanh(\beta_g)} \;\;\mathpunct{,} \label{eq:fg_H}\\
     \chi_g &= \frac{\arctanh(\beta_g) - \beta_g}{\beta_g^2 \arctanh(\beta_g)} \;\;\mathpunct{.} \label{eq:chig_H}
\end{align}

Equations~\eqref{eq:fg_H} and~\eqref{eq:chig_H} show that, in this case, the Eddington factor depends only on the reduced flux $\mathrm{f}_g$ and makes it possible to recover the correct expression for domain H. Since \mbox{$\alpha_{0,g} << k_B/(h \nu_{g+1/2})$}, one can derive a condition on $\mathcal{T}_g$ that guarantees the validity of this approximation:
\begin{equation*}
    \mathrm{T}_g >> \frac{h \nu_{g+1/2}}{\pi k_B}\left ( 5 (1 - \delta_g^3) \frac{\arctanh(\beta_g)}{\beta_g} \right )^{1/4} \;\;\mathpunct{.}
\end{equation*}

Since \mbox{$\arctanh(\beta_g)/\beta_g\ge1$} for \mbox{$\beta_g \in \openinterv{-1}{1}$}, this inequality can be rewritten in terms of $\mathcal{T}_g$ as:
\begin{equation}
    \mathcal{T}_g >> \frac{(1 - \delta_g^3)^{1/4}}{\sqrt{\delta_g}} \;\;\mathpunct{.} \label{eq:H_condition}
\end{equation}

This inequality shows that the expression of the Eddington factor in domain H extends to progressively lower radiative temperatures. This explains the behavior observed in figure~\ref{fig:T_shift}.

These observations apply in the general case where \mbox{$\nu_{g-1/2}>0$} and \mbox{$\nu_{g+1/2}<+\infty$}. However, if \mbox{$\nu_{g-1/2}=0$} or \mbox{$\nu_{g+1/2}=+\infty$}, the group narrowness $\delta_g$ can no longer be defined, nor can the dimensionless radiative temperature $\mathcal{T}_g$. Let us now examine what happens in these two cases.

\starsect{Case \mbox{$\boldsymbol{\nu_{g-1/2}=0}$} and \mbox{$\boldsymbol{\nu_{g+1/2}<+\infty}$}}

\begin{figure}
    \begin{center}
        \begin{minipage}[t]{0.6\linewidth}
            \centering
            \includegraphics[width=\textwidth]{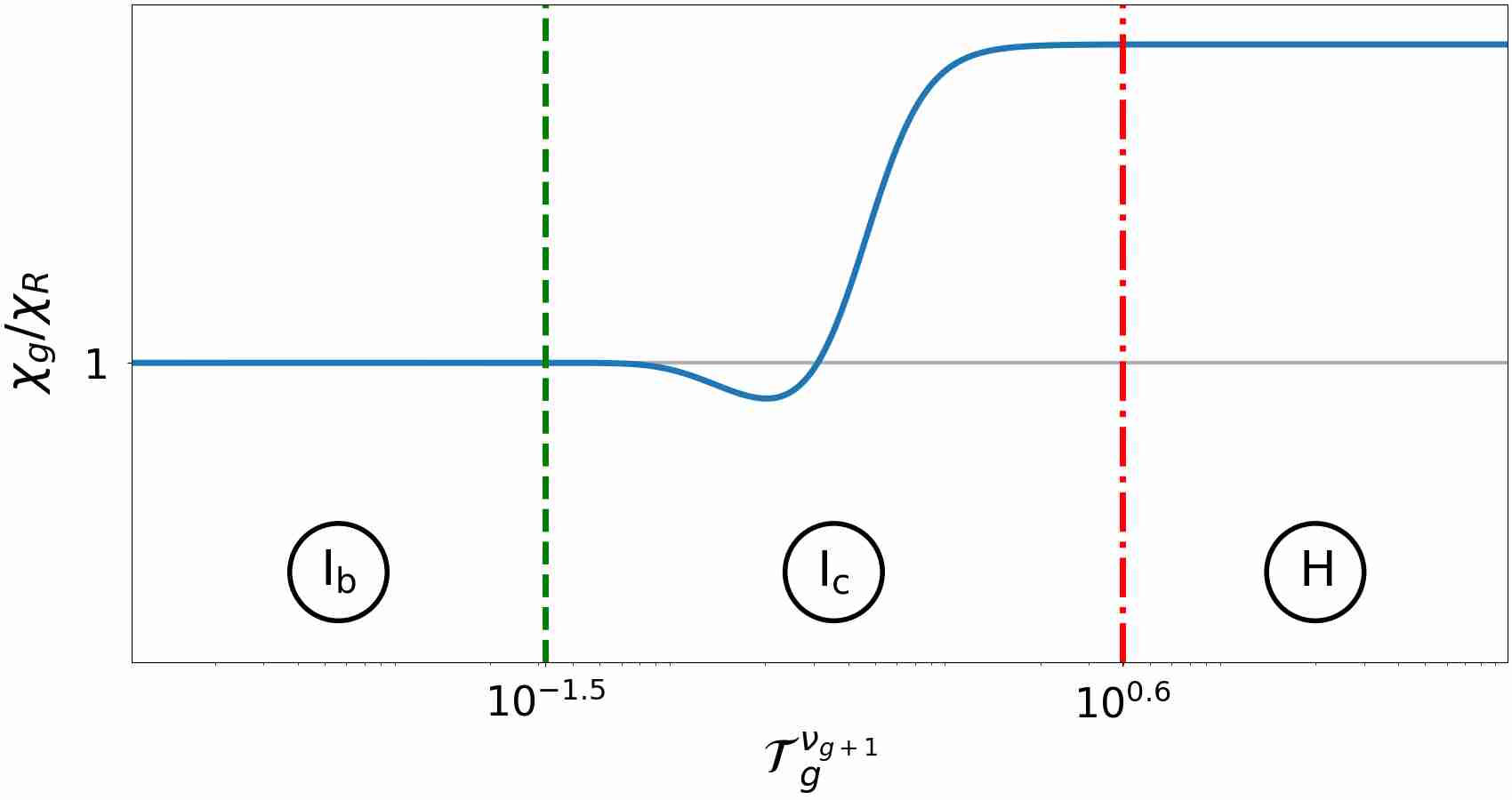}
            \caption{Ratio between the Eddington factor of the M1-multigroup model, $\chi_g$, and that of the M1-gray model, $\chi_R$, for \mbox{$\mathrm{f}_g=0.65$} and \mbox{$\nu_{g-1/2}=0$}. $\mathrm{I_b}$: domain in which $\chi_R$ is valid; $\mathrm{I_c}$: intermediate-to-high radiative temperatures; H: asymptotically high radiative temperatures.}
            \label{fig:numin0}
        \end{minipage}
    \end{center}
\end{figure}

In the case where \mbox{$\nu_{g-1/2}=0$}, one can perform the change of variable \mbox{$\tilde{\nu}=\nu / \nu_{g+1/2}$}, which allows the radiative temperature, the reduced flux, and the Eddington factor to be expressed as:
\begin{align}
    \frac{k_B \mathrm{T}_g}{h \nu_{g+1/2}} &= \left ( \frac{15}{\pi^4} \int \nolimits_{0}^{1} \int \nolimits_{-1}^{1} \tilde{\nu}^3 \left [ e^{\tilde{\nu}~\overline{\alpha_{0,g}}~(1 + \beta_g \mu )} - 1 \right ]^{-1}~\mathrm{d}\mu~\mathrm{d}\tilde{\nu} \right )^{1/4} \;\;\mathpunct{,} \label{eq:Tg_expr_nug0}\\
    \mathrm{f}_g &= \frac{\int \nolimits_{0}^{1} \int \nolimits_{-1}^{1} \tilde{\nu}^3 \mu \left [ e^{\tilde{\nu}~\overline{\alpha_{0,g}}~(1 + \beta_g \mu )} - 1 \right ]^{-1}~\mathrm{d}\mu~\mathrm{d} \tilde{\nu}}{\int \nolimits_{0}^{1} \int \nolimits_{-1}^{1} \tilde{\nu}^3 \left [ e^{\tilde{\nu}~\overline{\alpha_{0,g}}~(1 + \beta_g \mu )} - 1 \right ]^{-1}~\mathrm{d}\mu~\mathrm{d}\tilde{\nu}} \;\;\mathpunct{,} \label{eq:fg_expr_nug0}\\
    \chi_g &= \frac{\int \nolimits_{0}^{1} \int \nolimits_{-1}^{1} \tilde{\nu}^3 \mu^2 \left [ e^{\tilde{\nu}~\overline{\alpha_{0,g}}~(1 + \beta_g \mu )} - 1 \right ]^{-1}~\mathrm{d}\mu~\mathrm{d}\tilde{\nu}}{\int \nolimits_{0}^{1} \int \nolimits_{-1}^{1} \tilde{\nu}^3 \left [ e^{\tilde{\nu}~\overline{\alpha_{0,g}}~(1 + \beta_g \mu )} - 1 \right ]^{-1}~\mathrm{d}\mu~\mathrm{d}\tilde{\nu}} \;\;\mathpunct{,} \label{eq:chig_expr_nug0}
\end{align}

\noindent where \mbox{$\overline{\alpha_{0,g}} = h \nu_{g+1/2} \alpha_{0,g} / k_B$} is the dimensionless Lagrange multiplier, defined using the frequency $\nu_{g+1/2}$. Using these expressions, one can introduce the dimensionless radiative temperature \mbox{$\mathcal{T}_g^{\nu_{g+1/2}} = k_B \mathrm{T}_g / (h \nu_{g+1/2})$}, which must be used in this case. This temperature can also be defined in the general case where \mbox{$\nu_{g-1/2}>0$} and \mbox{$\nu_{g+1/2}<+\infty$}, and it is related to the radiative temperature $\mathcal{T}_g$ by the relation \mbox{$\mathcal{T}_g^{\nu_{g+1/2}} = \mathcal{T}_g \sqrt{\delta_g}$}.

In this context, one can deduce that the Eddington factor depends only on two variables: the reduced flux $\mathrm{f}_g$ and the dimensionless radiative temperature $\mathcal{T}_g^{\nu_{g+1/2}}$. The evolution of the Eddington factor as a function of the radiative temperature $\mathcal{T}_g^{\nu_{g+1/2}}$ is shown in figure~\ref{fig:numin0}. In this figure, one can identify domains $\mathrm{I_b}$, $\mathrm{I_c}$, and H, previously observed in the case of a wide group (\mbox{$\delta_g << 1$}). The only difference in the computation of the Eddington factor in this case lies in the choice of the dimensionless radiative temperature used.

\starsect{Case \mbox{$\boldsymbol{\nu_{g-1/2}>0}$} and \mbox{$\boldsymbol{\nu_{g+1/2}=+\infty}$}}

\begin{figure}
    \begin{center}
        \begin{minipage}[t]{0.6\linewidth}
            \centering
            \includegraphics[width=\textwidth]{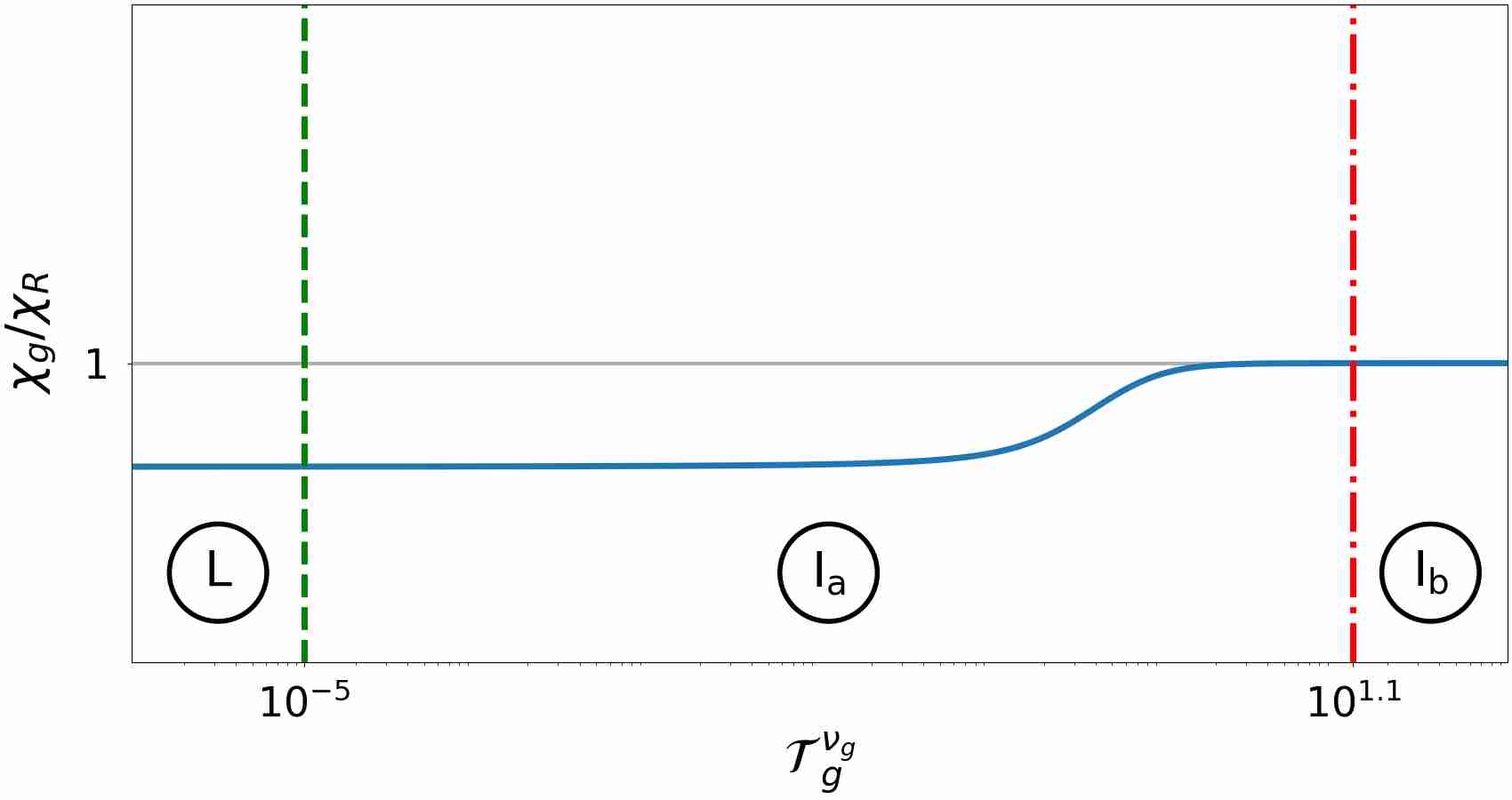}
            \caption{Ratio between the Eddington factor of the M1-multigroup model, $\chi_g$, and that of the M1-gray model, $\chi_R$, for \mbox{$\mathrm{f}_g=0.65$} and \mbox{$\nu_{g+1/2}\rightarrow\infty$}. L: asymptotically low radiative temperatures; $\mathrm{I_a}$: low-to-intermediate radiative temperatures; $\mathrm{I_b}$: domain in which $\chi_R$ is valid.}
            \label{fig:numaxinf}
        \end{minipage}
    \end{center}
\end{figure}

In the case where \mbox{$\nu_{g+1/2}=+\infty$}, one can instead perform the change of variable \mbox{$\tilde{\nu}=\nu / \nu_{g-1/2}$}, which allows the radiative temperature, the reduced flux, and the Eddington factor to be expressed as:
 \begin{align}
    \frac{k_B \mathrm{T}_g}{h \nu_{g-1/2}} &= \left ( \frac{15}{\pi^4} \int \nolimits_{1}^{+\infty} \int \nolimits_{-1}^{1} \tilde{\nu}^3 \left [ e^{\tilde{\nu}~\widehat{\alpha_{0,g}}~(1 + \beta_g \mu )} - 1 \right ]^{-1}~\mathrm{d}\mu~\mathrm{d}\tilde{\nu} \right )^{1/4} \;\;\mathpunct{,} \label{eq:Tg_expr_nugp1inf}\\
    \mathrm{f}_g &= \frac{\int \nolimits_{1}^{+\infty} \int \nolimits_{-1}^{1} \tilde{\nu}^3 \mu \left [ e^{\tilde{\nu}~\widehat{\alpha_{0,g}}~(1 + \beta_g \mu )} - 1 \right ]^{-1}~\mathrm{d}\mu~\mathrm{d} \tilde{\nu}}{\int \nolimits_{1}^{+\infty} \int \nolimits_{-1}^{1} \tilde{\nu}^3 \left [ e^{\tilde{\nu}~\widehat{\alpha_{0,g}}~(1 + \beta_g \mu )} - 1 \right ]^{-1}~\mathrm{d}\mu~\mathrm{d}\tilde{\nu}} \;\;\mathpunct{,} \label{eq:fg_expr_nugp1inf}\\
    \chi_g &= \frac{\int \nolimits_{1}^{+\infty} \int \nolimits_{-1}^{1} \tilde{\nu}^3 \mu^2 \left [ e^{\tilde{\nu}~\widehat{\alpha_{0,g}}~(1 + \beta_g \mu )} - 1 \right ]^{-1}~\mathrm{d}\mu~\mathrm{d}\tilde{\nu}}{\int \nolimits_{1}^{+\infty} \int \nolimits_{-1}^{1} \tilde{\nu}^3 \left [ e^{\tilde{\nu}~\widehat{\alpha_{0,g}}~(1 + \beta_g \mu )} - 1 \right ]^{-1}~\mathrm{d}\mu~\mathrm{d}\tilde{\nu}} \;\;\mathpunct{,} \label{eq:chig_expr_nugp1inf}
\end{align}

\noindent where \mbox{$\widehat{\alpha_{0,g}} = h \nu_{g-1/2} \alpha_{0,g} / k_B$} is the dimensionless Lagrange multiplier, defined using the frequency $\nu_{g-1/2}$. From these expressions, one can introduce the dimensionless radiative temperature \mbox{$\mathcal{T}_g^{\nu_{g-1/2}} = k_B \mathrm{T}_g / (h \nu_{g-1/2})$}, which must be used in this case. This temperature can also be defined in the general case where \mbox{$\nu_{g-1/2} > 0$} and \mbox{$\nu_{g+1/2} < +\infty$}, and it is related to the radiative temperature $\mathcal{T}_g$ by the relation \mbox{$\mathcal{T}_g^{\nu_{g-1/2}} = \mathcal{T}_g / \sqrt{\delta_g}$}.

In this case, one can deduce that the Eddington factor depends only on two variables: the reduced flux $\mathrm{f}_g$ and the radiative temperature $\mathcal{T}_g^{\nu_{g-1/2}}$. The evolution of the Eddington factor as a function of the radiative temperature $\mathcal{T}_g^{\nu_{g-1/2}}$ is shown in figure~\ref{fig:numaxinf}. In this figure, one finds domains L, $\mathrm{I_a}$, and $\mathrm{I_b}$, previously observed in the case of a wide group ($\delta_g << 1$). The only difference lies in the choice of the radiative temperature used.

\section{Coupling} \label{sec:couplage}

Up to this point, the radiative model has been formulated within the framework of a fluid at rest. However, in radiation hydrodynamics models, light propagates through a moving fluid, which induces Doppler effects that modify the energy and momentum exchanged between the fluid and the radiation. Two choices of reference frame are then possible to describe the radiative quantities:

\begin{enumerate}
    \item \textbf{The comoving frame}, in which the fluid is locally at rest. In this framework, matter–radiation interactions (emission and absorption) are isotropic, so that the source terms of the radiative transfer equations~\eqref{eq:eqs_g} remain unchanged. However, transforming the radiative quantities from the comoving frame to the laboratory frame introduces additional terms into the equations. In particular, advection effects appear in the form of terms such as \mbox{$\nablav \cdot (\vectorr{v} \mathrm{E}_g)$} for the radiative energy and \mbox{$\nablav \cdot (\vectorr{v} \otimes \vectorr{F_g})$} for the radiative flux~\cite{mihalas_1999, buchler_1979, lowrie_2001}. Moreover, the Doppler effect induces a change in photon frequency between the two frames. Within the M1-multigroup model, this implies transfers of energy and momentum between spectral groups, represented by the terms \mbox{$\nablav \vectorr{v}: \int \nolimits_{\nu_{g-1/2}}^{\nu_{g+1/2}} \partial_\nu (\nu \tensorr{P}\nu) \dif \nu$} and \mbox{$\nablav \vectorr{v}: \int \nolimits_{\nu_{g-1/2}}^{\nu_{g+1/2}} \partial_\nu (\nu \tensorr{Q}_\nu) \dif \nu$}. These terms make the solution of the radiative transfer equations in the M1-multigroup model particularly complex~\cite{vaytet_2011};
    \item \textbf{The laboratory frame}, in which the fluid is in motion. In this frame, the radiative transfer equations~\eqref{eq:eqs_g} retain their usual form, without the introduction of advection terms or energy transfers between spectral groups in the M1-multigroup model. By contrast, the source terms become more complex, as they must include Doppler effects and geometric aberrations~\cite{mihalas_2001}.
\end{enumerate}

\noindent In this work, we adopt this second approach by expressing the radiative quantities in the laboratory frame.

\subsection{Lorentz transformation} \label{sec:Lorentz}

Before proceeding, let us recall the expression of the Lorentz transformation. Consider two reference frames, $\mathcal{R}$ and $\mathcal{R}'$, where $\mathcal{R}'$ moves with velocity $\vectorr{v}$ relative to $\mathcal{R}$. In special relativity, the transformation of a four-vector $\tensor{X}{^\alpha}$ from the reference frame $\mathcal{R}$ to the reference frame $\mathcal{R}'$ is expressed as follows:
\begin{equation}
    X'^\beta = \tensor{\Lambda}{^\beta_\alpha} X^\alpha \;\;\mathpunct{,}
\end{equation}

\noindent where $X'^\beta$ is the four-vector in the reference frame $\mathcal{R}'$, and $\tensor{\Lambda}{^\beta_\alpha}$ is the Lorentz tensor, given by:
\begin{equation}
    \tensor{\Lambda}{^\beta_\alpha} = 
    \begin{bmatrix}
        \begin{array}{c|ccc}
            \gamma_u & -\gamma_u \vectorr{\beta_u}^T\\
            \hline
            & \\
            -\gamma_u \vectorr{\beta_u} & \identity + \frac{\gamma_u^2}{\gamma_u+1} \vectorr{\beta_u} \otimes \vectorr{\beta_u} \\
            & \\
        \end{array}
    \end{bmatrix} \;\;\mathpunct{,}
\end{equation}

\noindent with:
\begin{align*}
    &&\gamma_u = \frac{1}{\sqrt{1 - \beta_u^2}} \;\;\mathpunct{,} &&\vectorr{\beta_u} = \vectorr{v}/c  \;\;\mathpunct{,} &&\beta_u=||\vectorr{\beta_u}|| \;\;\mathpunct{.} 
\end{align*}

The quantity $\gamma_u$ is called the \textit{Lorentz factor}. Let us first consider the case of the M1-multigroup model. We then define the second-order radiation energy–momentum tensor of group $g$, denoted $\tensorr{R}_g = (R_g^{\alpha \beta})_{\alpha, \beta}$, as:
\begin{equation}
    R_g^{\alpha \beta} = c^{-1} \int \nolimits_{\nu_{g-1/2}}^{\nu_{g+1/2}} \int \nolimits_{4 \pi} I_\nu (\vectorr{n}) n^\alpha n^\beta \dif \Omega \dif \nu  \;\;\mathpunct{,}
\end{equation}

\noindent where \mbox{$n^\alpha = (n^0, \vectorr{n}) = (1, \vectorr{n})$}. This tensor can also be written in the form:
\begin{equation}
    R_g^{\alpha \beta} = 
    \begin{bmatrix}
        \begin{array}{c|ccc}
            \mathrm{E}_g&  \vectorr{F_g}^T/c \\
            \hline
            & \\
            \vectorr{F_g}/c & \tensorr{P}_g \\
            & \\
        \end{array}
    \end{bmatrix} \;\;\mathpunct{.}
\end{equation}

To transform this tensor from the comoving frame to the laboratory frame, it is necessary to apply the Lorentz transformation to each contravariant index. Knowing that the laboratory frame moves with velocity $\vectorr{v}$ relative to the comoving frame, the radiative quantities in the latter, namely $\mathrm{E}_g$, $\vectorr{F_g}$, and $\tensorr{P}_g$, can be expressed as functions of the radiative quantities in the laboratory frame according to the following relations:
\begin{align}
    \mathrm{E}_{g,0} =& \gamma_u^2 \left [ \mathrm{E}_g - \frac{2 \vectorr{v} \cdot \vectorr{F_g}}{c^2} + \vectorr{\beta_u}^T \tensorr{P}_g \vectorr{\beta_u} \right ] \;\;\mathpunct{,} \label{eq:Er_comobile} \\
    \vectorr{F_{g,0}} =& \gamma_u \left [ \vectorr{F_g} - \left \{ \tensorr{P}_g + \gamma_u \identity  \left ( \mathrm{E}_g - \frac{2 \gamma_u + 1}{\gamma_u+1} \frac{\vectorr{v} \cdot \vectorr{F_g}}{c^2} + \frac{\gamma_u \vectorr{\beta_u}^T \tensorr{P}_g \vectorr{\beta_u}}{\gamma_u+1} \right ) \right \} \vectorr{v} \right ] \;\;\mathpunct{,} \label{eq:Fr_comobile} \\
    \tensorr{P}_{g,0} = &\tensorr{P}_g - \gamma_u  \frac{\vectorr{v} \otimes \vectorr{F_g} + \vectorr{F_g}\otimes \vectorr{v}}{c^2} - \gamma_u^2 \left [ \frac{\tensorr{P}_g \vectorr{\beta_u} \otimes \vectorr{\beta_u} + \vectorr{\beta_u} \otimes \vectorr{\beta_u} \tensorr{P}_g}{\gamma_u+1} + \right. \nonumber \\
    &\left.  \vectorr{\beta_u} \otimes \vectorr{\beta_u} \left \{ \mathrm{E}_g - \frac{2 \gamma_u}{\gamma_u+1} \left ( \frac{\vectorr{v} \cdot \vectorr{F_g}}{c^2} - \frac{\gamma_u}{2} \vectorr{\beta_u}^T \tensorr{P}_g \vectorr{\beta_u}  \right )\right \} \right ] \;\;\mathpunct{.} \label{eq:Pr_comobile}
\end{align}

Let us now define the four-vector of source terms of the radiative transfer equation as \mbox{$S_g^\alpha = (S^0_g, \vectorr{S}_g)$}. By applying the Lorentz transformation, the source terms in the laboratory frame can be expressed in terms of those in the comoving frame through the following relations:
\begin{align}
    S^0_g & =\gamma_u \left ( S^0_{g,0} + \vectorr{\beta_u} \cdot \vectorr{S_{g,0}} \right ) \;\;\mathpunct{,} \label{eq:S0_lab} \\
    \vectorr{S_g} &= \vectorr{S_{g,0}} + \gamma_u \vectorr{\beta_u} \left (S^0_{g,0} +  \frac{\gamma_u \vectorr{\beta_u} \cdot \vectorr{S_{g,0}}}{\gamma_u+1} \right ) \;\;\mathpunct{.} \label{eq:S_lab}
\end{align}

As we saw in Section~\secref{sec:Radiation}, in the comoving frame the source term can be written as:
\begin{align*}
    S^0_{g,0} &= \kappa_{P,g} (\mathrm{E}_{g,0} - a_R \theta_g^4(\mathrm{T})) \;\;\mathpunct{,}\\
    \vectorr{S_{g,0}} &= \kappa_{R,g} \vectorr{F_{g,0}}/c \;\;\mathpunct{,}
\end{align*}

\noindent where $\kappa_{P,g}$ and $\kappa_{R,g}$ are the Planck and Rosseland opacities of group $g$ in the comoving frame. Using transformations~\eqref{eq:S0_lab} and~\eqref{eq:S_lab}, the source terms can therefore be rewritten in the laboratory frame as:
\begin{align*}
    S^0_g & =\gamma_u \left ( \kappa_{P,g} (\mathrm{E}_{g,0} - a_R \theta_g^4(\mathrm{T})) + \kappa_{R,g} \frac{\vectorr{v} \cdot \vectorr{F}_{g,0}}{c^2} \right ) \;\;\mathpunct{,}  \\
    \vectorr{S_g} &= \kappa_R \vectorr{F}_{g,0}/c + \gamma_u \vectorr{\beta_u} \left (\kappa_{P,g} (\mathrm{E}_{g,0} - a_R \theta_g^4(\mathrm{T})) +  \frac{\gamma_u \kappa_{R,g} }{\gamma_u+1} \frac{\vectorr{v} \cdot \vectorr{F_{g,0}}}{c^2} \right ) \;\;\mathpunct{.}
\end{align*}

Finally, by rewriting this source term in terms of the radiative quantities expressed in the laboratory frame and using transformations~\eqref{eq:Er_comobile}, \eqref{eq:Fr_comobile}, and~\eqref{eq:Pr_comobile}, one obtains the following expression for the source terms in the laboratory frame~\cite{mihalas_2001}:
\begin{align}
    S^0_g =&\gamma_u \left [ \left ( \gamma_u^2 \left (\kappa_{P,g} - \beta_u^2 \kappa_{R,g} \right ) \mathrm{E}_g - \kappa_{P,g} a_R \theta_g^4(\mathrm{T}) \right ) + \right. \nonumber\\
    & \left. \gamma_u^2 \left \{ \left ( (1 + \beta_u^2) \kappa_{R,g} - 2 \kappa_{P,g} \right ) \vectorr{v} \cdot (c^{-2} \vectorr{F_g}) - \left (\kappa_{R,g} - \kappa_{P,g} \right )\vectorr{\beta_u}^T \tensorr{P}_g \vectorr{\beta_u}  \right \} \right ] \;\;\mathpunct{,}  \label{eq:S0_lab_tot}\\
    \vectorr{S_g} =& \gamma_u \left [ \kappa_{R,g} \vectorr{F_g}/c - \kappa_{R,g} \tensorr{P}_g \vectorr{\beta_u} - \vectorr{\beta_u} \left\{ \left ( \gamma_u^2 (\kappa_{R,g} - \kappa_{P,g}) \mathrm{E}_g  + \kappa_{P,g} a_R \theta_g^4(\mathrm{T})  \right ) - \right. \right. \nonumber\\
    &\left. \left.   \gamma_u^2 (\kappa_{R,g} - \kappa_{P,g}) \left ( 2 \vectorr{v} \cdot (c^{-2} \vectorr{F_g}) - \vectorr{\beta_u}^T \tensorr{P}_g \vectorr{\beta_u} \right ) \right \} \right ] \;\;\mathpunct{.} \label{eq:S_lab_tot}
\end{align}

In this work, we assume that the fluid is non-relativistic (\mbox{$\beta_u<<1$}). Therefore, these source terms can be expanded to first order in $\beta$, noting that \mbox{$\gamma_u =1+\mathcal{O}(\beta_u^2)$}:
\begin{align}
    S^0_g =&\kappa_{P,g} \left ( \mathrm{E}_g - a_R \theta_g^4(\mathrm{T}) \right ) + (\kappa_{R,g} - 2 \kappa_{P,g}) \vectorr{v} \cdot (c^{-2} \vectorr{F_g}) \;\;\mathpunct{,} \label{eq:S0_lab_tot_approx} \\
    \vectorr{S_g} =& \kappa_{R,g} \vectorr{F_g}/c - \kappa_{R,g} \tensorr{P}_g \vectorr{\beta_u} - \vectorr{\beta_u} \left \{ (\kappa_{R,g} - \kappa_{P,g}) \mathrm{E}_g  + \kappa_{P,g} a_R \theta_g^4(\mathrm{T})  \right \} \;\;\mathpunct{.} \label{eq:S_lab_tot_approx}
\end{align}

\subsection{Radiative hydrodynamics equations} \label{sec:Hydrodad_equations}

We can now combine the hydrodynamics equations~\eqref{eq:euler} and the radiative transfer equations into a coupled system, forming what is known as the \textit{radiation hydrodynamics equations}. These equations describe the evolution of a fluid interacting with radiation.

Let us first present these equations within the framework of the M1-multigroup model for the description of radiation. In this context, the source terms $c S^0_g$ and $\vectorr{S_g}$ introduced previously (equations~\eqref{eq:S0_lab_tot_approx} and~\eqref{eq:S_lab_tot_approx}) represent, respectively, the exchanges of energy and momentum between the fluid and the radiation contained in each frequency group. These exchanges are taken into account by adding the total contribution of these interactions as source terms on the right-hand side of the hydrodynamics equations~\eqref{eq:euler} and by incorporating them into the radiative transfer equations~\eqref{eq:eqs_g}, formulated in the laboratory frame. One thus obtains the complete system of radiation hydrodynamics equations:
\begin{equation}
    \label{eq:hydro_rad_mg}
    \left\{
    \begin{array}{llll}
        \partial_t\rho  + \nablav \cdot ( \rho \vectorr{v}) &= 0\;\;\mathpunct{,} \\
        \partial_t(\rho \vectorr{v}) + \nablav \cdot (\rho \vectorr{v} \otimes \vectorr{v} + p \identity ) &= \sum_{g=1}^\mathcal{G} \vectorr{S_g}\;\;\mathpunct{,} \\
        \partial_t \mathrm{E} + \nablav \cdot ((\mathrm{E}+p) \vectorr{v}) &= \sum_{g=1}^\mathcal{G} c S^0_g\;\;\mathpunct{,}\\
        \partial_t \mathrm{E}_g + \nablav \cdot \vectorr{F_g} &= -c S^0_g, &\forall~g~\in~\llbracket1,~\mathcal{G}\rrbracket \;\;\mathpunct{,} \\
        \partial_t (c^{-2} \vectorr{F_g}) + \nablav \cdot \tensorr{P}_g &= -\vectorr{S_g}, &\forall~g~\in~\llbracket1,~\mathcal{G}\rrbracket \;\;\mathpunct{.}
    \end{array}
    \right.
\end{equation}

In the particular case of the M1-gray model, the radiation hydrodynamics equations take a simpler form:
\begin{equation}
    \label{eq:hydro_rad_gray}
    \left\{
    \begin{array}{lll}
        \partial_t\rho  + \nablav \cdot ( \rho \vectorr{v}) &= 0 \;\;\mathpunct{,} \\
        \partial_t(\rho \vectorr{v}) + \nablav \cdot (\rho \vectorr{v} \otimes \vectorr{v} + p \identity ) &= \vectorr{S} \;\;\mathpunct{,} \\
        \partial_t \mathrm{E} + \nablav \cdot ((\mathrm{E}+p) \vectorr{v}) &= c S^0 \;\;\mathpunct{,} \\
        \partial_t \mathrm{E}_R + \nablav \cdot \vectorr{F_R} &= -c S^0 \;\;\mathpunct{,} \\
        \partial_t (c^{-2} \vectorr{F_R}) + \nablav \cdot \tensorr{P}_R &= -\vectorr{S} \;\;\mathpunct{,}
    \end{array}
    \right.
\end{equation}

\noindent where $S^0$ and $\vectorr{S}$ are the source terms of the single group considered, integrated over the entire electromagnetic spectrum.

\subsection{Asymptotic cases} \label{sec:particular_cases}

Now that a general radiation hydrodynamics model has been established, it is essential to verify whether it is able to recover the well-known asymptotic limits: the optically thin regime and the optically thick regime. To this end, one can first identify several characteristic lengths defined from the source terms~\eqref{eq:S0_lab_tot_approx} and~\eqref{eq:S_lab_tot_approx}:
\begin{align*}
    \lambda_{th} &= 1/\max_\nu(\kappa_\nu)  \;\;\mathpunct{,} \\
    \lambda_{tr} &= 1/\max_\nu(\chi_\nu) \;\;\mathpunct{,} \\
    \lambda_{dyn,1} &= 1/\max_\nu(|\chi_\nu-2\kappa_\nu|) \;\;\mathpunct{,} \\
    \lambda_{dyn,2} &= 1/\max_\nu(\sigma_\nu) \;\;\mathpunct{.} \\
\end{align*}

\noindent These characteristic lengths, in turn, allow four associated characteristic times to be defined:
\begin{align}
    \tau_{th} &= \frac{\lambda_{th}}{c}  \;\;\mathpunct{,} \tag{temps de thermalisation}\\
    \tau_{tr} &= \frac{\lambda_{tr}}{c} \;\;\mathpunct{,} \tag{temps de transport}\\
    \tau_{dyn,1} &= \frac{\lambda_{dyn,1}}{\beta_u c} \;\;\mathpunct{,}  \tag{temps dynamique 1}\\
    \tau_{dyn,2} &= \frac{\lambda_{dyn,2}}{\beta_u c} \;\;\mathpunct{.}  \tag{temps dynamique 2}
\end{align}

\noindent These characteristic times play a key role in the dynamics of radiation–matter interactions:
\begin{itemize}
    \item $\tau_{th}$ corresponds to the time required for the gas to reach thermal equilibrium with the radiation,
    \item $\tau_{tr}$ represents the characteristic timescale of radiative transport,
    \item the times $\tau_{dyn,1}$ and $\tau_{dyn,2}$ quantify the dynamical interaction between radiation and matter in the presence of nonzero fluid velocities.
\end{itemize}

\noindent Three important observations can be made:
\begin{enumerate}
    \item \textbf{Hierarchy of timescales:}\\
    In the non-relativistic framework (\mbox{$\beta_u \ll 1$}), one observes that \mbox{$\tau_{th}, \tau_{tr} \ll \tau_{dyn,1}, \tau_{dyn,2}$}. This means that radiation has time to thermalize and to be transported before dynamical effects associated with fluid motion (Doppler effect, relativistic distortions) become significant. Thus, on short timescales, radiation behaves as if it were in a rest frame;

    \item \textbf{Contribution of higher-order terms in $\boldsymbol{\beta_u}$:}\\
    If the expansion of the source terms $S_0$ and $\vectorr{S}_g$ had not been truncated at first order in $\beta_u$, additional contributions would appear, associated with characteristic times of the form \mbox{$\tau_n = \lambda / (\beta_u^n c)$} with \mbox{$n \geq 2$}. However, as long as one considers situations in which the characteristic durations are sufficiently short, the impact of these terms remains negligible on the global dynamics of the fluid and the radiation;

    \item \textbf{Case where scattering is neglected (\mbox{$\boldsymbol{\sigma_\nu \approx 0}$}):}\\
    By neglecting photon scattering and assuming that the absorption coefficient is independent of photon frequency (\mbox{$\kappa_\nu \approx \chi_\nu = \kappa$}), the characteristic times simplify:
    \begin{align*} 
        &&\tau_{tr} \approx \tau_{th} = \tau = \frac{1}{\kappa c} \;\;\mathpunct{,} &&\tau_{dyn,1} \approx \frac{\tau}{\beta_u} \;\;\mathpunct{,}  &&\tau_{dyn,2} &\to +\infty. 
    \end{align*}

    In this case, the behavior of the system depends on the ratio between the characteristic time $t$ of the situation under consideration and the timescales defined above~\cite{krumholz_2007}:
    \begin{enumerate}
        \item \textbf{\mbox{$\boldsymbol{t<<\tau}$}: Optically thin regime (free streaming)}\\
        Radiation propagates freely with negligible interaction with matter. This regime corresponds to the optically thin case;
        \item \textbf{\mbox{$\boldsymbol{t>>\tau}$} and \mbox{$\boldsymbol{t<<\tau/\beta_u}$}: Static diffusion regime}\\
        Radiation is in thermal equilibrium with matter, but its dynamical interaction with the fluid remains negligible;
        \item \textbf{\mbox{$\boldsymbol{t>>\tau}$}, \mbox{$\boldsymbol{t>>\tau/\beta_u}$}, and \mbox{$\boldsymbol{t<<\tau/\beta_u^2}$}: Dynamic diffusion regime}\\
        The interaction between radiation and gas becomes dominant, and radiative transport is essentially governed by fluid advection. This regime remains within the validity limits of the model developed here.
    \end{enumerate}
\end{enumerate}

Thus, depending on the characteristic timescale considered, the physical situation differs, and different approximations may be employed. Let us now present the two main existing approximations used to describe radiation.

\starsect{Optically thin case}

In the interstellar medium, matter is very often optically thin because of the low densities and temperatures that prevail there. This implies weak interactions between radiation and matter, which translates into small opacities $\kappa_P$ and $\kappa_R$. Consequently, the characteristic thermalization and transport times, denoted $\tau_{th}$ and $\tau_{tr}$, respectively, are very long. In other words, radiation takes a long time to exchange energy with the surrounding matter and to propagate through the medium. It is therefore reasonable to assume that, in this regime, the influence of radiation on the fluid dynamics remains negligible.

Let us now consider a situation in which, initially, radiation is in radiative equilibrium with matter, which implies that the radiative flux is nearly zero (\mbox{$\vectorr{F_R} \approx 0$}) and that the radiative energy satisfies \mbox{$\mathrm{E}_R = a_R \mathrm{T}_i^4$}, where $\mathrm{T}_i$ is the initial temperature of the fluid. Suppose that, at some time, the fluid is heated, reaching a temperature $\mathrm{T_f}$ such that $\mathrm{T_f}^4 \gg \mathrm{T}_i^4$. In this case, the source terms describing matter–radiation interactions, given by expressions~\eqref{eq:S0_lab_tot_approx} and~\eqref{eq:S_lab_tot_approx}, can be approximated by:
\begin{align}
    S^0 \approx&-\kappa_P(\rho,\mathrm{T}) a_R \mathrm{T}^4 = -\Lambda(\rho,\mathrm{T}) \;\;\mathpunct{,} \label{eq:S0_mince} \\
    \vectorr{S}~ \approx& 0 \;\;\mathpunct{,} \label{eq:S_mince}
\end{align}

\noindent where $\kappa_P$ is the Planck mean opacity, defined by equation~\eqref{eq:opac_moyenne_Planckg}, taking \mbox{$\nu_{g-1/2}=0$} and \mbox{$\nu_{g+1/2}=+\infty$}. One recognizes here the typical form of the cooling function \mbox{$\Lambda(\rho,\mathrm{T})$}, which depends only on the gas density $\rho$ and its temperature $\mathrm{T}$. Assuming that the Planck opacity can be described by a power law in $\rho$ and $\mathrm{T}$, this cooling function can be written in the general form:
\begin{equation}
    \Lambda(\rho,\mathrm{T}) \approx \Lambda_0 \rho^\alpha \mathrm{T}^\beta \;\;\mathpunct{,} \label{eq:fct_refroidissement} 
\end{equation}

\noindent where $\Lambda_0$ is a constant that depends in particular on the radiation constant $a_R$, while $\alpha$ and $\beta$ are exponents characterizing the physical processes involved. For instance, to model bremsstrahlung cooling in a hot interstellar medium (\mbox{$\mathrm{T} > 10^7$ K}), one would choose \mbox{$\alpha = 2$} and \mbox{$\beta = 0.5$}. By contrast, for cooling via inverse Compton scattering of electrons, the commonly used values are \mbox{$\alpha = 1$} and \mbox{$\beta = 1$}.

Since we assume that the gas follows the ideal-gas equation of state~\eqref{eq:eq_etat}, it is also possible to express this cooling function in terms of the thermal pressure $p$ and the density $\rho$ as follows:
\begin{equation}
    \Lambda(\rho,p) \approx \Lambda_0' \rho^\epsilon p^\zeta \;\;\mathpunct{,} \label{eq:fct_refroidissement2} 
\end{equation}

\noindent where $\Lambda_0'$ is a constant related to $\Lambda_0$ through \mbox{$\Lambda_0' = \Lambda_0 \mu m_H/k_B$}. The exponents $\epsilon$ and $\zeta$ are related to $\alpha$ and $\beta$ by \mbox{$\epsilon = \alpha - \beta$} and \mbox{$\zeta = \beta$}. In this framework, the simplified radiation hydrodynamics equations take the form:
\begin{equation}
    \label{eq:hydrorad_cooling}
    \left\{
    \begin{array}{llll}
        \partial_t\rho  + \nablav \cdot ( \rho \vectorr{v}) &= 0\;\;\mathpunct{,} \\
        \partial_t(\rho \vectorr{v}) + \nablav \cdot (\rho \vectorr{v} \otimes \vectorr{v} + p) &= 0\;\;\mathpunct{,} \\
        \partial_t \mathrm{E} + \nablav \cdot ((\mathrm{E}+p) \vectorr{v}) &= -\Lambda(\rho,p)\;\;\mathpunct{.}
    \end{array}
    \right.
\end{equation}

This formulation is widely used because of its numerical simplicity: by avoiding the explicit resolution of the radiative transfer equations, it significantly reduces both the complexity of the calculations and the computational cost.

\starsect{Optically thick case}

Conversely, in dense regions such as stellar interiors, the medium is highly opaque and interacts strongly with radiation. In this regime, radiation is in near-equilibrium with matter, since the characteristic thermalization and transport times are very short compared with the dynamical timescales under consideration (\mbox{$t \gg \tau_{th},\tau_{tr}$}). One can then assume that radiative equilibrium is reached instantaneously and express the radiative pressure as well as the radiative flux as follows:
\begin{align*}
    &&\tensorr{P}_{R,0} =\frac{1}{3} \identity \mathrm{E}_{R,0} \;\;\mathpunct{,}&& \vectorr{F}_{R,0}=& -\frac{c}{3 \kappa_R} \nablav \mathrm{E}_{R,0} \;\;\mathpunct{,}    
\end{align*} 

\noindent where $\kappa_R$ is the Rosseland opacity defined by~\eqref{eq:opac_moyenne_Rosselandg}, with \mbox{$\nu_{g-1/2}=0$} and \mbox{$\nu_{g+1/2}=+\infty$}. These expressions reflect the fact that, in an optically thick medium, radiation interacts strongly with matter and propagates by diffusion. The radiative pressure is isotropic, while the radiative flux follows a diffusion law, proportional to the gradient of the radiative energy, with a diffusion coefficient given by \mbox{$c/(3 \kappa_R)$}.

Using these approximations, it becomes possible to retain only the radiative energy equation to describe radiation. Keeping only terms of order \mbox{$\mathcal{O}(v/c)$} as well as certain terms of order \mbox{$\mathcal{O}((v/c)^2)$}, and expressing the radiative quantities in the comoving frame, the radiation hydrodynamics equations can be rewritten in the form~\cite{saincir_2019_these}:
\begin{equation}
    \label{eq:hydro_rad_opaque}
    \left\{
    \begin{array}{llll}
        \partial_t\rho  + \nablav \cdot ( \rho \vectorr{v}) &= 0\;\;\mathpunct{,} \\
        \partial_t(\rho \vectorr{v}) + \nablav \cdot (\rho \vectorr{v} \otimes \vectorr{v} + p) &= -\frac{1}{3} \nablav \mathrm{E}_{R,0}\;\;\mathpunct{,} \\
        \partial_t \mathrm{E} + \nablav \cdot ((\mathrm{E}+p) \vectorr{v}) &= c \kappa_P (\mathrm{E}_{R,0} - a_R \mathrm{T}^4) + \frac{1}{3} \vectorr{v} \cdot \nablav \mathrm{E}_{R,0}\;\;\mathpunct{,}\\
        \partial_t \mathrm{E}_{R,0} + \nablav \cdot \left ( \frac{4}{3} \vectorr{v} \mathrm{E}_{R,0} \right ) &= \nablav \cdot \left ( \frac{c}{3 \kappa_R} \nablav \mathrm{E}_{R,0} \right ) - c \kappa_P (\mathrm{E}_{R,0} - a_R \mathrm{T}^4) - \frac{1}{3} \vectorr{v} \cdot \nablav \mathrm{E}_{R,0}\;\;\mathpunct{.}
    \end{array}
    \right.
\end{equation}

Compared with the complete system~\eqref{eq:hydro_rad_gray}, this simplified model contains only four equations instead of five, which greatly facilitates its numerical solution. It thus avoids the explicit treatment of detailed radiative transfer, while still capturing the essential interactions between matter and radiation in an optically thick medium. This formalism describes both the static diffusion regime and the dynamic diffusion regime, as introduced by Krumholz~\cite{krumholz_2007}. However, it does not allow one to treat regions where radiative equilibrium is locally broken, as is the case in radiative shocks (see Section~\secref{sec:intro_choc_rad}).

\section{Synthesis}

In this chapter, I have presented the model used to simulate radiation hydrodynamics. This model relies on several simplifying assumptions:

\begin{itemize}
    \item the gas is ideal, inviscid, and without thermal diffusion,
    \item it is assumed to be non-ionized,
    \item gravitational effects are neglected,
    \item the gas and the radiation are in \gls{lte},
    \item photon scattering is assumed to be elastic and isotropic, with an isotropic scattering coefficient,
    \item the fluid is treated within the non-relativistic framework.
\end{itemize}

These assumptions make it possible to derive the radiation hydrodynamics equations based on the M1-multigroup model for the description of radiative transfer. Since the Eddington factor associated with this model does not admit an analytical expression but can be determined numerically, I have studied its dependencies and variations in order to identify its main characteristics. The next chapter introduces \gls{ai} and describes the work I have carried out to leverage it with the aim of improving the efficiency of radiation hydrodynamics simulations.
\clearemptydoublepage

\chapter{Calculation of the Eddington factor with Artificial Intelligence} \label{ch:chapitre3}

\initialletter{T}he \gls{hades} code~\cite{nguyen_2011_these, michaut_2011, michaut_2017}, developed to solve the radiation hydrodynamics equations, models the interaction between radiation and matter outside radiative equilibrium. It was designed to limit approximations in the treatment of the radiation field as much as possible. Two radiative transfer models are implemented in the code: the M1-gray model and the M1-multigroup model, which make it possible to solve the equations introduced in the previous chapter using classical spatial and temporal discretization methods. We first present the numerical methods used in this code and rewrite the radiation hydrodynamics equations in the following form:
\begin{empheq}[left=\empheqlbrace]{align*}
    \partial_t \mathbf{q}_{h} + \nabla \cdot \mathbf{F}(\mathbf{q}_{h}) &= \mathbf{s}_{h} 
    \quad \text{(hydrodynamics)} \\
    \partial_t \mathbf{q}_{r} + \nabla \cdot \mathbf{F}(\mathbf{q}_{r}) &= \mathbf{s}_{r} 
    \quad \text{(radiative transfer)}
\end{empheq}

The algorithm used in \gls{hades} to solve the radiation hydrodynamics equations consists of two main steps~\cite{nguyen_2011_these}:
\begin{enumerate}
    \item Update of the radiative quantities $\mathbf{q}_{r}$ over two time steps $\Delta t$,
    \item Update of the hydrodynamic quantities $\mathbf{q}_{h}$ over two time steps $\Delta t$.
\end{enumerate}
    
\noindent Each of these steps is further decomposed into three substeps:

\begin{enumerate}
    \item Resolution of the \gls{ode} $\partial_t \mathbf{q} = \mathbf{s}$ over a time step $\Delta t/2$ (explicit scheme for hydrodynamics, implicit scheme for radiation),
    \item Resolution of the \glsreset{pde}\gls{pde} $\partial_t \mathbf{q} + \nabla \cdot \mathbf{F} (\mathbf{q}) = 0$ over a time step $\Delta t$ (for instance, an \gls{hlle} Riemann solver for hydrodynamics and an \gls{hll} Riemann solver for radiation),
    \item Resolution of the \gls{ode} $\partial_t \mathbf{q} = \mathbf{s}$ over a time step $\Delta t/2$ (explicit scheme for hydrodynamics, implicit scheme for radiation).
\end{enumerate}

\noindent This algorithm guarantees second-order accuracy in the solution of the radiation hydrodynamics equations. However, two major limitations remain:

\begin{enumerate}
    \item \textbf{Computation of the Eddington factor in the M1-multigroup model}\\
    In this model, the Eddington factor does not admit an explicit analytical expression. The code therefore relies on numerical root-finding algorithms to estimate it accurately. Although these methods provide precise values of the Eddington factor, they are particularly expensive in terms of computational cost;
    \item \textbf{\glslink{cfl}{\glsentrylong{cfl} (\glsentryshort{cfl})} condition}\\
    To ensure numerical stability when solving the radiation hydrodynamics equations, the time step $\Delta t$ and the spatial step $\Delta x$ must satisfy a \gls{cfl}-type condition: \mbox{$c \Delta t / \Delta x \leq 1$}, where $c$ denotes the speed of light. This constraint forces the use of very small time steps, which significantly increases the computational cost of the simulations.
\end{enumerate}

In order to overcome these limitations and accelerate the simulations, I explored the potential contribution of \gls{ai}. This technology has attracted growing interest and is experiencing rapid development in fields as diverse as medicine, robotics, and space exploration. More recently, it has been integrated into numerical simulation workflows, giving rise to an emerging research field known as \gls{neuhpc}~\cite{karimabadi_2023}. This field develops solutions along two main directions:

\begin{enumerate}
    \item \textbf{Inverse modeling:} The goal is to discover the analytical or differential equations underlying the data. This approach includes symbolic regression\footnote{Python library: \url{https://github.com/MilesCranmer/PySR}, which aims at identifying explicit analytical expressions; Julia library: \url{https://github.com/MilesCranmer/SymbolicRegression.jl}}~\cite{angelis_2023}, as well as the use of \gls{pinn}s to infer the differential equations governing the observed phenomena;
    \item \textbf{Simulation:} \gls{ai} is used to solve known differential equations, either by fully integrating them using machine-learning techniques or by combining traditional numerical methods with \gls{ai}-based approaches.
\end{enumerate}

In this work, I focus more specifically on the second direction, with the aim of developing more efficient methods to solve the radiation hydrodynamics equations. Three broad categories of neural-network applications in simulation can be identified:

\begin{enumerate}
    \item \textbf{\gls{ai} solver:} These methods consist in using \gls{ai} to solve differential equations end to end. For instance, \gls{pinn}s approximate the solution by treating the neural network as a function that approximates the solution of the underlying differential equations~\cite{raissi_2019,jagtap_2020a,cuomo_2022}. Another strategy relies on neural operators, which learn to map the solution at time $t$ to that at time \mbox{$t + \Delta t$}, using architectures such as Graph Neural Networks (GNNs)~\cite{scarselli_2009,alet_2019,brandstetter_2022,horie_2022} or Fourier Neural Operators (FNOs)~\cite{li_2021,li_2024};
    \item \textbf{Closure relations:} In this case, neural networks are used to determine or improve the closure relations of the equations. This approach can be applied within simplified models, for example for turbulence modeled by the RANS or LES equations~\cite{kurz_2022,taghizadeh_2020,chen_2024}, or to establish an unknown closure relation~\cite{harada_2022};
    \item \textbf{Error correction:} This approach consists in training a neural network on data generated by very high-resolution simulations and then using it to correct the errors made in low-resolution simulations. This procedure makes it possible to drastically reduce computational cost while maintaining high accuracy in the results~\cite{kochkov_2021}.
\end{enumerate}

In this work, I explored two complementary approaches to reduce the computational cost of radiation hydrodynamics simulations. The first consists in developing an \gls{ai}-based method to efficiently estimate the Eddington factor involved in the closure relation of the M1-multigroup model (see Section~\secref{sec:Eddington_mg}). The second aims at investigating the feasibility of a neural solver, based on \gls{pinn}s, capable of extrapolating a simulation from the initial data of an existing simulation, while relying directly on the radiation hydrodynamics equations (see Chapter~\secref{ch:chapitre5}).

In the present chapter, I will first present the main \gls{ai} techniques used for training neural networks, before detailing the proposed method for accelerating the evaluation of the Eddington factor.

\section{Steps to generate a functional neural network} \label{sec:intro_IA}

When addressing a problem in \gls{ai}, several fundamental questions must be posed and resolved during the development of the method.

\begin{itemize}
    \item \textbf{Which data should be used?}\\
    This involves determining the nature, quality, format, and quantity of the data required to train the network.
    \item \textbf{How should the model error be measured (cost function)?}\\
    The choice of the loss function (evaluation of the neural network's performance) is essential to guide the learning process.
    \item \textbf{Which optimization algorithm should be used?}\\
    The optimizer adjusts the parameters of the neural network in order to minimize the loss function.
    \item \textbf{Which neural network architecture should be adopted?}\\
    A neural network can be viewed as a more or less complex function, characterized by a large number of adjustable parameters. The choice of its architecture—that is, the way in which the computational units are organized and connected—determines the form of this function. Among the common architectures in \gls{ai}, one finds in particular \gls{mlp}s, \gls{gan}s, or \gls{lstm}s. This decision depends on the type of relationship to be modeled, the complexity of the problem, as well as constraints in terms of computational resources and available data.
    \item \textbf{Should a particular strategy be adopted to promote generalization?}\\
    This involves helping the neural network produce accurate predictions on configurations it has not seen during training.
\end{itemize}

\begin{figure}
    \begin{center}
        \begin{minipage}[t]{0.95\linewidth}
            \centering
            \includegraphics[width=\textwidth]{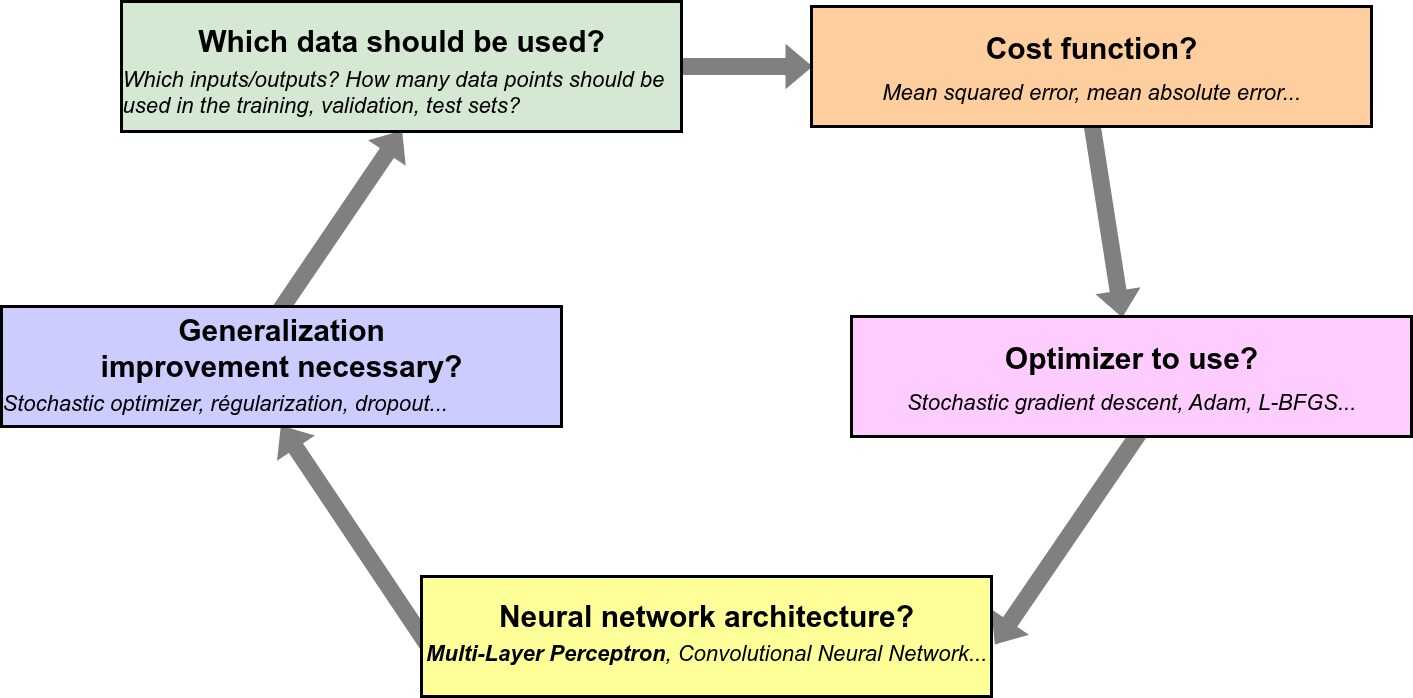}
            \caption{Diagram illustrating the reasoning process involved in developing an \glslink{ai}{artificial intelligence} strategy.}
            \label{fig:questions_IA}
        \end{minipage}
    \end{center}
\end{figure}

These choices are not independent and must be adjusted iteratively. An initial combination of assumptions is tested and then modified in light of the results obtained, until convergence toward a solution deemed satisfactory. This process is illustrated in figure~\ref{fig:questions_IA} in the form of a decision loop. Indeed, answering one question (for example, \textit{which data should be used?}) makes it possible to address the next one (for example, \textit{which architecture should be adopted?}), and so on, up to the evaluation of the effectiveness of the choices made. This cycle is repeated several times with different combinations before retaining the most successful trial.

In this case, since the objective is to reduce the computational cost of numerical simulations, I imposed a strong constraint on the architecture: I opted for \gls{mlp}s, one of the simplest architectures, yet sufficiently expressive to approximate complex nonlinear relationships while remaining computationally inexpensive.

In this section, I will first introduce \gls{mlp}s, before detailing the possible choices regarding the loss function, the optimization strategy, as well as techniques aimed at improving generalization capability.

\subsection{Multi-Layer Perceptrons (MLP)}\label{sec:MLP}

\begin{figure}
    \begin{center}
        \begin{minipage}[t]{0.38\linewidth}
            \centering
            \includegraphics[width=\textwidth]{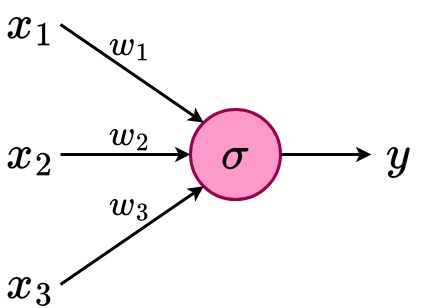}
            \caption{Diagram of a perceptron}
            \label{fig:perceptron}
        \end{minipage}
    \end{center}
\end{figure}

An \gls{mlp} can be viewed as a complex mathematical function, having a large number of adjustable parameters depending on the problem being addressed. It is composed of elementary units called \emph{perceptrons} (see figure~\ref{fig:perceptron}), whose formulation was first introduced by Rosenblatt~(1958)~\cite{rosenblatt_1958}. Each perceptron receives $n$ inputs $x_1$, $x_2$, …, $x_n$, associated with \emph{weights} $w_1$, $w_2$, …, $w_n$, as well as a \emph{bias} $b$. It produces an output~$y$, defined by the following expression:
\begin{equation*}
    y=\sigma\left ( \sum_{i=1}^{n} w_i x_i + b \right ) \;\;\mathpunct{,}
\end{equation*}

\begin{figure}
    \begin{subfigure}[t]{0.32\textwidth}
        \centering
        \includegraphics[width=\textwidth]{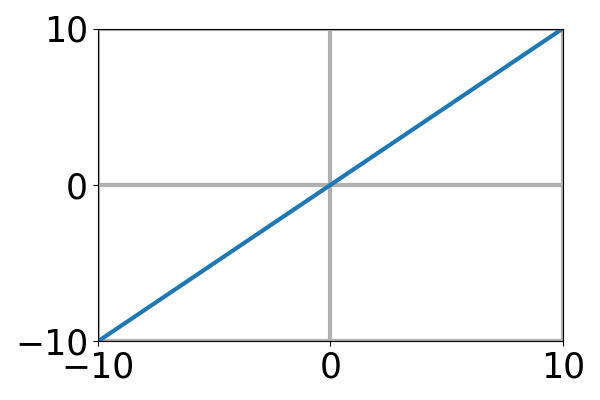}
        \caption{Identity function}
        \label{fig:linear}
    \end{subfigure}
    \hfill
    \begin{subfigure}[t]{0.32\textwidth}
        \centering
        \includegraphics[width=\textwidth]{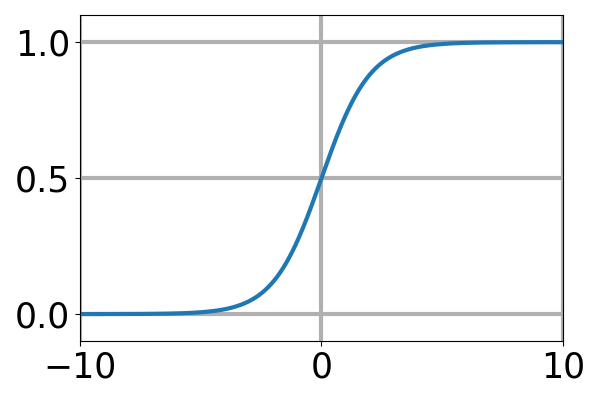}
        \caption{Sigmoid function}
        \label{fig:sigmoid}
    \end{subfigure}
    \hfill
    \begin{subfigure}[t]{0.32\textwidth}
        \centering
        \includegraphics[width=\textwidth]{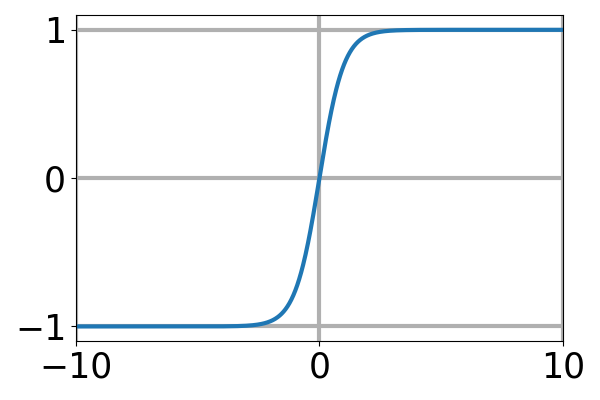}
        \caption{Hyperbolic tangent function}
        \label{fig:tanh}
    \end{subfigure}
    \hfill
    \begin{subfigure}[t]{0.32\textwidth}
        \centering
        \includegraphics[width=\textwidth]{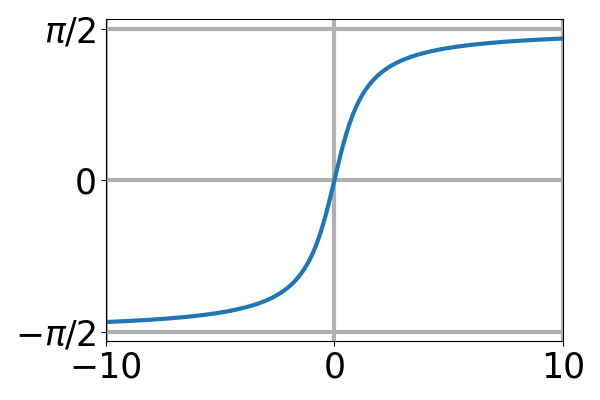}
        \caption{Arctangent function}
        \label{fig:arctan}
    \end{subfigure}
    \hfill
    \begin{subfigure}[t]{0.32\textwidth}
        \centering
        \includegraphics[width=\textwidth]{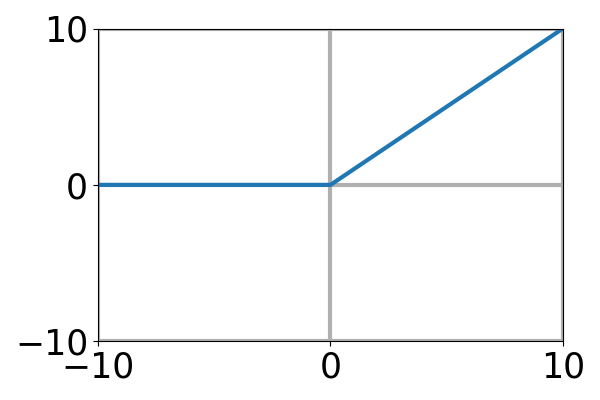}
        \caption{ReLU function}
        \label{fig:relu}
    \end{subfigure}
    \hfill
    \begin{subfigure}[t]{0.32\textwidth}
        \centering
        \includegraphics[width=\textwidth]{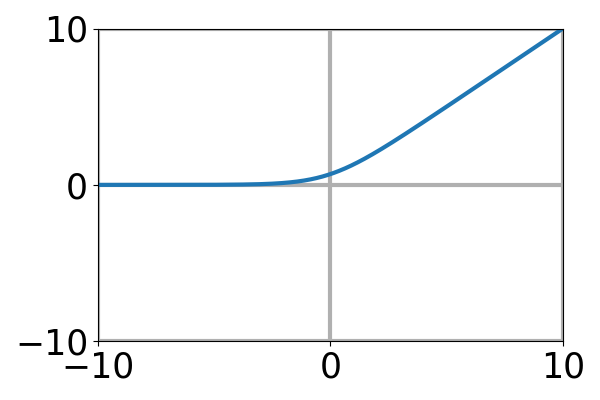}
        \caption{SoftPlus function}
        \label{fig:softplus}
    \end{subfigure}
    \caption{Examples of activation functions that can be used in neural networks.}
    \label{fig:fcts_activation}
\end{figure}

\noindent where $\sigma$ denotes an \emph{activation function}, which may be linear or nonlinear. The introduction of nonlinearity allows the perceptron to capture complex behaviors and to model nonlinear phenomena. Several activation functions are commonly used:

\begin{itemize}
    \item the identity function (see figure~\ref{fig:linear})
    \begin{equation*}
        \sigma(x) = x \in \mathbb{R} \;\;\mathpunct{,}
    \end{equation*}
    \item the sigmoid function (see figure~\ref{fig:sigmoid})
    \begin{equation*}
        \sigma(x) = \frac{1}{1+e^{-x}} \in \closeinterv{0}{1} \;\;\mathpunct{,}
    \end{equation*}
    \item the hyperbolic tangent function (see figure~\ref{fig:tanh})
    \begin{equation*}
        \sigma(x) = \tanh(x) = \frac{1-e^{-x}}{1+e^{-x}} \in \closeinterv{-1}{1} \;\;\mathpunct{,}
    \end{equation*}
    \item the arctangent function (see figure~\ref{fig:arctan})
    \begin{equation*}
        \sigma(x) = \mathrm{Arctan}(x) \in \closeinterv{-\pi/2}{\pi/2} \;\;\mathpunct{,}
    \end{equation*}
    \item the Rectified Linear Unit (ReLU) (see figure~\ref{fig:relu})
    \begin{equation*}
        \sigma(x) =
        \begin{cases}
            0~~\mathrm{if}~~x <   0\\
            x~~\mathrm{if}~~x \ge 0
        \end{cases}
        \in \mathbb{R^+}\;\;\mathpunct{,}
    \end{equation*}
    \item the soft rectified linear unit (SoftPlus) (see figure~\ref{fig:softplus})
    \begin{equation*}
        \sigma(x) = \ln(1+e^{-x}) \in \mathbb{R^+} \;\;\mathpunct{,}
    \end{equation*}
\end{itemize}

Although most of these activation functions are nonlinear, they generally do not allow the representation of discontinuities. Some alternative variants have nevertheless been proposed, notably by DellaSanta (2023)~\cite{dellasanta_2023}, which explicitly introduce discontinuities into the activation function in order to endow the network with specific properties. The adoption of such functions, however, requires particular care, both from a theoretical and a numerical standpoint, due to the consequences they may have on training stability and the local differentiability of the model.

However, a single perceptron is a relatively simple function with limited capacity to represent complex functions. To overcome this limitation, multiple perceptrons are grouped together to form a \emph{layer}. By subsequently stacking several layers of perceptrons, one obtains what is known as a neural network of type \gls{mlp}. Three types of layers are generally distinguished:

\begin{enumerate}
    \item \textbf{The input layer}, which receives the input data and transmits it to the network;
    \item \textbf{The hidden layers}, located between the input layer and the output layer, which extract increasingly abstract and complex features from the data. More than one hidden layer may be used;
    \item \textbf{The output layer}, which produces the final results of the network, depending on the task (classification, regression, etc.).
\end{enumerate}

\begin{figure}
    \begin{center}
        \begin{minipage}[t]{0.8\linewidth}
            \centering
            \includegraphics[width=\textwidth]{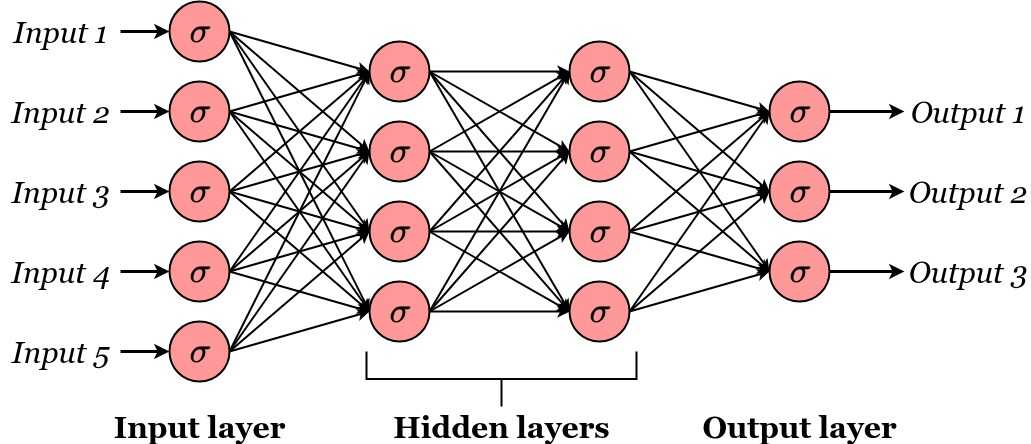}
        \end{minipage}
        \caption{Example of an \glsxtrshort{mlp} with five inputs and three outputs. The network consists of an input layer, two hidden layers, and an output layer. The red circles represent the neurons (or perceptrons), and $\sigma$ denotes the activation function applied to each of them. The same activation function may be used for all neurons, or different functions may be employed depending on the specific needs of the model.}
        \label{fig:nn_example}
    \end{center}
\end{figure}

Figure~\ref{fig:nn_example} illustrates an example of an \gls{mlp} neural network, consisting of an input layer with five neurons, an output layer with three neurons, and two hidden layers each containing four neurons. An \gls{mlp} neural network has a large number of parameters (weights and biases of the perceptrons) that must be adjusted. Hornik (1989)~\cite{hornik_1989} showed that this type of network is a universal function approximator, meaning that it can theoretically approximate any function, provided that a sufficiently large architecture and an appropriate parameter configuration are available. The output produced by this network, referred to as the \textit{prediction}, depends directly on these parameters. The process of optimizing the parameters to achieve the best possible performance on a given task constitutes the essence of neural network \textit{training}.

\subsection{Training} \label{sec:entrainement}

Training consists in optimizing the weights and biases of the perceptrons in the neural network in order to minimize the error made by the network. First, one must select the data used to train the neural network. In general, these data are divided into three datasets:
\begin{enumerate}
    \item \textbf{The training set}, which contains labeled examples used to train the neural network;
    \item \textbf{The validation set}, which is used to evaluate the generalization capability of the model and to tune the hyperparameters in order to avoid overfitting\footnote{See Section~\secref{sec:sous_sur_apprentissage} for the definition of overfitting.};
    \item \textbf{The test set}, which is completely independent of the previous datasets and is used to assess the final performance of the model on unseen data.
\end{enumerate}

To evaluate the overall quality of the predictions produced by the neural network, a \textit{cost function} is introduced. This function measures the global error made by the model. Let us assume that the dataset contains $N$ samples, that $y^{(i)}$ denotes the expected output of the neural network for the input $x^{(i)}$, and that $h_{\vectorIA{\theta}}(x^{(i)})$ represents the prediction produced by the network for the same input.

Several cost functions are commonly used in artificial intelligence; some of them are presented below.

\begin{itemize}
    \item \textbf{Mean Squared Error (MSE)}\\
    MSE is the most commonly used option. It strongly penalizes large errors, which makes it sensitive to outliers.
    \begin{equation*}
        \mathcal{L}(Y, h_{\vectorIA{\theta}}(X)) = \frac{1}{N} \sum_{i=1}^{N} \left [ h_{\vectorIA{\theta}}(x^{(i)}) - y^{(i)}\right ]^2 \;\;\mathpunct{.}
    \end{equation*}
    \item \textbf{Mean Absolute Error (MAE)}\\
    Unlike MSE, MAE penalizes errors linearly, which makes it more robust to outliers. It is frequently used in regression tasks.
    \begin{equation*}
        \mathcal{L}(Y, h_{\vectorIA{\theta}}(X)) = \frac{1}{N} \sum_{i=1}^{N} \left | h_{\vectorIA{\theta}}(x^{(i)}) - y^{(i)}\right | \;\;\mathpunct{.}
    \end{equation*}
    \item \textbf{Logarithmic Mean Squared Error (LMSE)}\\
    This cost function is particularly well suited when the data are strictly positive and when minimizing relative error is more important than minimizing absolute error. It is often used in problems where target values span several orders of magnitude.
    \begin{equation*}
        \mathcal{L}(Y, h_{\vectorIA{\theta}}(X)) = \frac{1}{N} \sum_{i=1}^{N} \left [ \ln \left (h_{\vectorIA{\theta}}(x^{(i)}) \right ) - \ln \left (y^{(i)} \right )\right ]^2 \;\;\mathpunct{.}
    \end{equation*}
    \item \textbf{Binary Cross-Entropy}\\
    Used in binary classification tasks, this function measures the divergence between the predicted probabilities and the true labels $y^{(i)}$, which are assumed to belong to \mbox{$\{0,1\}$}. It requires a network whose outputs lie in the interval $\closeinterv{0}{1}$.
    \begin{equation*}
        \mathcal{L}(Y, h_{\vectorIA{\theta}}(X)) = -\frac{1}{N} \sum_{i=1}^{N} \left [ y^{(i)} \ln \left (h_{\vectorIA{\theta}}(x^{(i)}) \right ) + (1-y^{(i)}) \ln \left (1-h_{\vectorIA{\theta}}(x^{(i)}) \right )\right ] \;\;\mathpunct{.}
    \end{equation*}
\end{itemize}

Next, in order to train the neural network, it is essential to choose an appropriate optimization algorithm capable of determining the weights and biases that minimize the value of the cost function defined above. These algorithms adjust the network parameters to achieve good performance on the training data while preserving generalization capability on the validation and test sets. Some commonly used optimizers are presented below.

\begin{itemize}
    \item \textbf{Gradient Descent}\\
    This classical method updates the parameters using the gradient of the cost function computed over the entire training dataset. The parameter update is written as:
    \begin{equation*}
        \vectorIA{\theta}_{t+1} = \vectorIA{\theta}_t - \eta \nablav_{\vectorIA{\theta}} \mathcal{L} \left (Y, h_{\vectorIA{\theta}_t}(X) \right ) \;\;\mathpunct{,}
    \end{equation*}
    where $\vectorIA{\theta}_t$ denotes the neural network parameters at iteration $t$, $\eta$ is the \textit{learning rate}, which must be tuned, $\nablav_{\vectorIA{\theta}}$ is the gradient with respect to the network parameters $\theta$, and $\mathcal{L}$ is the cost function evaluated over the entire training dataset. Each update corresponds to one \textit{epoch}. At each epoch, the value of the cost function $\mathcal{L}$ decreases. However, this method is sensitive to noisy data, which can degrade the quality of the final neural network.
    
    \item \textbf{\gls{sgd}}\\
    Rather than computing the gradient over the full dataset, \gls{sgd} updates the neural network parameters using each individual training example. For a training set containing $N$ samples, the update corresponding to the $i$-th example is:
    \begin{equation*}
        \vectorIA{\theta}_{t+1} = \vectorIA{\theta}_t - \eta \nablav_{\vectorIA{\theta}} \mathcal{L} \left (y^{(i)},h_{\vectorIA{\theta}_t}(x^{(i)}) \right ) \;\;\mathpunct{,}
    \end{equation*}
    One epoch therefore corresponds to $N$ updates. This method is, however, more computationally expensive and difficult to parallelize.

    \item \textbf{Mini-batch Gradient Descent}\\
    This approach lies between classical gradient descent and \gls{sgd}. It divides the training dataset into small batches (\textit{mini-batches}) of size $K$ and performs one update per mini-batch. If $\mathcal{B}_K$ denotes the set of all mini-batches, the updates are written as:
    \begin{equation*}
        \forall (X_K,Y_K) \in \mathcal{B}_K,~\vectorIA{\theta}_{t+1} = \vectorIA{\theta}_t - \eta \nablav_{\vectorIA{\theta}} \mathcal{L} \left (Y_K,h_{\vectorIA{\theta}_t}(X_K) \right ) \;\;\mathpunct{,}
    \end{equation*}
    Thus, $N/K$ updates are performed per epoch. This method reduces computation time compared to \gls{sgd}, is well suited to parallelization, and retains a stochastic character, making training less sensitive to noisy data than classical gradient descent.
    
    \item \textbf{\gls{adam}}\\
    Proposed by Kingma in 2014~\cite{kingma_2014}, \gls{adam} adapts the learning rate for each parameter by combining mini-batch gradient descent with adjustments based on the moments of the gradient. Two variables are defined:
    \begin{align*}
        \vectorIA{v}_t = \beta_1 \vectorIA{v}_{t-1} + (1-\beta_1) \vectorIA{g}_t \;\;\mathpunct{,} &&\vectorIA{s}_t = \beta_2 \vectorIA{s}_{t-1} + (1-\beta_2) \vectorIA{g}_t^2 \;\;\mathpunct{,}
    \end{align*}
    where $\vectorIA{g}_t$ is the gradient of the cost function evaluated on the mini-batch data, and $\beta_1$ and $\beta_2$ are hyperparameters (usually set to \mbox{$\beta_1=0.9$} and \mbox{$\beta_2=0.999$}). In the \gls{adam} algorithm, normalized values of $\vectorIA{v}_t$ and $\vectorIA{s}_t$ are used to avoid bias:
    \begin{align*}
        \vectorIA{\hat{v}}_t = \frac{\vectorIA{v}_{t}}{1-\beta_1^t}\;\;\mathpunct{,} &&\vectorIA{\hat{s}}_t = \frac{\vectorIA{s}_{t}}{1-\beta_2^t}\;\;\mathpunct{.}
    \end{align*}
    The update of the neural network parameters is then given by:
    \begin{equation*}
        \vectorIA{\theta}_{t+1} = \vectorIA{\theta}_t - \eta_t \vectorIA{\hat{v}}_t \;\;\mathpunct{,}
    \end{equation*}
    with
    \begin{equation*}
        \eta_t = \frac{\eta}{\sqrt{\vectorIA{\hat{s}}_t} + \epsilon} \;\;\mathpunct{,}
    \end{equation*}
    where $\epsilon$ is a small constant (e.g. $10^{-6}$) introduced to avoid division by zero. \gls{adam} is widely used due to its fast convergence.
    
    \item \textbf{\gls{lbfgs}}\\
    This quasi-Newton method uses an approximation of the Hessian matrix to accelerate convergence while limiting memory usage. In Newton's method, the update would be computed as:
    \begin{equation*}
        \vectorIA{\theta}_{t+1} = \vectorIA{\theta}_t - \left [ \nablav_{\vectorIA{\theta}}^{2} \mathcal{L}\left (Y, h_{\vectorIA{\theta}_t}(X) \right ) \right ]^{-1} \nablav_{\vectorIA{\theta}} \mathcal{L}\left (Y, h_{\vectorIA{\theta}_t}(X) \right ) \;\;\mathpunct{,}
    \end{equation*}
    where \mbox{$\left [ \nablav_{\vectorIA{\theta}}^{2} \mathcal{L}\left (Y, h_{\vectorIA{\theta}_t}(X) \right ) \right ]^{-1}$} and \mbox{$\nablav_{\vectorIA{\theta}}\mathcal{L}\left (Y, h_{\vectorIA{\theta}_t}(X) \right )$} denote respectively the inverse Hessian and the gradient of the cost function with respect to the neural network parameters, evaluated on the training dataset. \gls{lbfgs} replaces the Hessian with a matrix $\mathbb{B}_t$ (generally initialized as the identity matrix), which approximates the Hessian. The algorithm then proceeds through the following steps:
    \begin{enumerate}
        \item Computation of the descent direction:
        \begin{equation*}
            \vectorIA{p}_t = - \mathbb{B}_t^{-1} \nablav_{\vectorIA{\theta}} \mathcal{L}\left (Y, h_{\vectorIA{\theta}_t}(X) \right ) \;\;\mathpunct{,}
        \end{equation*}
        \item Line search to determine the step size $\alpha_t$ such that \mbox{$\alpha_t=\mathrm{argmin}(\mathcal{L}(\vectorIA{\theta}_t+\alpha_t\vectorIA{p}_t))$},
        \item Update of the neural network parameters:
        \begin{equation*}
            \vectorIA{\theta}_{t+1} = \vectorIA{\theta}_t + \alpha_t \vectorIA{p}_t \;\;\mathpunct{,}
        \end{equation*}
        \item Computation of the vectors $\vectorIA{s}_t$ and $\vectorIA{y}_t$:
        \begin{align*}
            \vectorIA{s}_t = \vectorIA{\theta}_{t+1} - \vectorIA{\theta}_t\;\;\mathpunct{,} &&\vectorIA{y}_t = \nablav_{\vectorIA{\theta}} \mathcal{L}\left (Y, h_{\vectorIA{\theta}_{t+1}}(X) \right ) - \nablav_{\vectorIA{\theta}} \mathcal{L}\left (Y, h_{\vectorIA{\theta}_t}(X) \right ) \;\;\mathpunct{,}
        \end{align*}
        \item Update of the inverse Hessian approximation:
        \begin{equation*}
            \mathbb{B}_{t+1}^{-1} = \mathbb{B}_t^{-1} + \frac{(\vectorIA{s}_t^T \vectorIA{y}_t + \vectorIA{y}_t^T \mathbb{B}_t^{-1} \vectorIA{y}_t) \vectorIA{s}_t \vectorIA{s}_t^T}{(\vectorIA{s}_t^T \vectorIA{y}_t)^2} - \frac{\mathbb{B}_t^{-1} \vectorIA{y}_t \vectorIA{s}_t^T  + \vectorIA{s}_t \vectorIA{y}_t^T \mathbb{B}_t^{-1} }{\vectorIA{s}_t^T \vectorIA{y}_t} \;\;\mathpunct{.}
        \end{equation*}
    \end{enumerate}
    \gls{lbfgs} converges much faster than the previous algorithms, but may suffer from the same sensitivity to noisy data as gradient descent. It is often used to refine the parameters after an initial optimization phase using the \gls{adam} optimizer.
\end{itemize}

\subsection{Underfitting and overfitting} \label{sec:sous_sur_apprentissage}

At the end of the training of a neural network, three different situations may arise:

\begin{itemize}
    \item \textbf{Underfitting:} The neural network fails to correctly solve the problem, including on the training set itself. This phenomenon may result from an overly simple architecture, which is unable to capture the complexity of the problem, or from an early termination of the training process;
    \item \textbf{Overfitting:} The neural network fits the training data perfectly but fails to provide reliable predictions on unseen cases. This may occur if the network architecture is too complex or if the training is continued for too long. In this case, the network is said to fail to \textit{generalize};
    \item \textbf{Satisfactory learning:} The neural network produces predictions with a similar level of error on the training set and on new data. The network is then said to generalize correctly.
\end{itemize}

\begin{figure}
    \begin{center}
        \begin{minipage}[t]{0.7\linewidth}
            \centering
            \includegraphics[width=\textwidth]{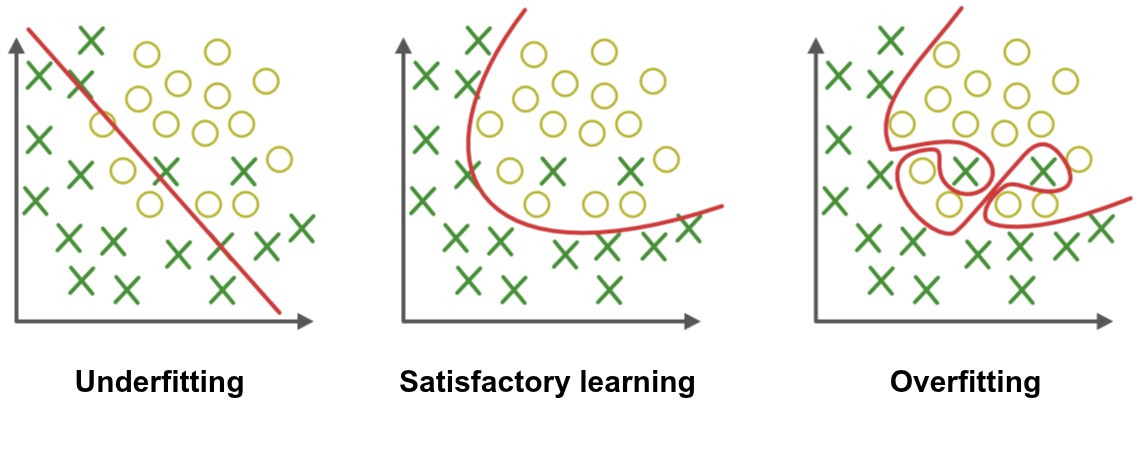}
        \end{minipage}
        \caption{Different possible outcomes after the training of a neural network.}
        \label{fig:overfit}
    \end{center}
\end{figure}

Figure~\ref{fig:overfit} illustrates three typical configurations encountered in the classification of points into two sets: underfitting, satisfactory learning, and overfitting. In the case of underfitting, the function used is too simple, which prevents the model from properly separating the two groups of points. Conversely, in the case of overfitting, the model adapts excessively to the training data, perfectly classifying all points, but at the expense of poor generalization: it also classifies as belonging to the sets certain points that are clearly outliers. Finally, the best compromise is achieved by a model capable of correctly classifying the majority of points while discerning those that do not belong to the main sets.

To limit underfitting, the network should be trained for a sufficient number of epochs, until the solution converges toward a stable error during training. It is also important to use an architecture that is sufficiently expressive to capture the dynamics of the problem, while avoiding excessive complexity.

In the case of overfitting, several strategies can be implemented. The first consists in using a stochastic optimization algorithm, such as \gls{adam}, which is known to improve generalization capabilities. Two additional techniques are also commonly employed:

\begin{enumerate}
    \item A common approach consists in adding a regularization term $\mathcal{L}^(\vectorIA{\theta})$ to the cost function, which explicitly depends on the parameters of the neural network. This term aims to encourage the weights and biases to remain close to zero, thereby improving the generalization capability. The modified cost function is then expressed:
    \begin{equation*}
        \mathcal{L}^R(Y, h_{\vectorIA{\theta}}(X)) = \mathcal{L}(Y, h_{\vectorIA{\theta}}(X)) + \mathcal{L}^*(\vectorIA{\theta}) \;\;\mathpunct{.}
    \end{equation*}
    Two common forms of regularization are used:
    \begin{enumerate}
        \item \textbf{L1 regularization:} penalizes the sum of the absolute values of the parameters:
        \begin{equation*} 
            \mathcal{L}^*(\vectorIA{\theta}) = \lambda \sum_i |\theta_i| \;\;\mathpunct{.} 
        \end{equation*} 
        It promotes a more parsimonious model by inducing sparsity through the cancellation of some parameters, which may improve the interpretability of the neural network;
        \item \textbf{L2 regularization:} penalizes the sum of the squares of the weights:
        \begin{equation*} 
            \mathcal{L}^*(\vectorIA{\theta}) = \lambda \sum_i \theta_i^2 \;\;\mathpunct{.} 
        \end{equation*} 
        It reduces the magnitude of the weights without setting them exactly to zero, making the network more stable with respect to variations in the input data.
    \end{enumerate}
    In both cases, $\lambda$ is a hyperparameter that must be tuned depending on the problem being addressed, and the $\theta_i$ denote the parameters of the neural network;
    \item \textbf{Dropout:} This technique consists in randomly deactivating a certain percentage of perceptrons at each training iteration. In practice, during the parameter update step, a subset of neurons is temporarily ignored, which prevents the network from relying too strongly on relationships that are specific to the training data. This approach helps limit overfitting and improves the robustness of the model. For more details on dropout, see Baldi (2014)~\cite{baldi_2014}.
\end{enumerate}

\section{Eddington factor in the M1-multigroup model} \label{sec:Eddington_mg}

Let us briefly recall the definition of the Eddington factor.

\begin{tcolorbox}[colback=gray!5!white,colframe=black!75!black,title=Definition of the Eddington factor $\chi_g$]
    Within the framework of the M1-multigroup model for radiative transfer, the Eddington factor $\chi_g$ associated with the spectral group $g$, defined between the frequencies $\nu_1$ and $\nu_2$, is a scalar function that relates the radiative pressure tensor $\mathbb{P}_g$ to the radiative energy $\mathrm{E}_g$ of the group. The closure relation reads:
    \[
    \mathbb{P}_g= \left [\frac{1 - \chi_g}{2} ~ \identity + \frac{3 \chi_g - 1}{2} \frac{\vectorr{F_g} \otimes
    \vectorr{F_g}}{||\vectorr{F_g}||^2} \right ] \mathrm{E}_g \;\;\mathpunct{,}
    \]
    where $\vectorr{F_g}$ is the radiative flux vector of the group and $\identity$ denotes the identity tensor.\\

    This expression makes it possible to express the radiative pressure knowing lower-order moments and thereby to close the system of the moment equations. The Eddington factor $\chi_g$ depends on three parameters:\begin{itemize}
        \item the reduced flux \mbox{$\mathrm{f}_g=||\vectorr{F_g}||/(c~\mathrm{E}_g)$},
        \item the dimensionless radiative temperature $\mathcal{T}_g = k_B \mathrm{E}_g^{1/4} / (a_R^{1/4} h \widetilde{\nu})$,
        \item the group narrowness \mbox{$\delta_g = \nu_1/\nu_2$}, when $\nu_1>0$ and $\nu_2<+\infty$.
    \end{itemize}
    
    The reference frequency $\widetilde{\nu}$ may be chosen as $\nu_1$, $\nu_2$, or $\sqrt{\nu_1 \nu_2}$, depending on the values of $\nu_1$ and $\nu_2$.\\
        
    The limiting values of $\chi_g$ correspond to well-identified physical regimes:
    \[
    \chi_g \tending{\mathrm{f}_g}{0} \frac{1}{3} \quad \text{(isotropic regime)} \quad \text{et}
    \quad \chi_g \tending{\mathrm{f}_g}{1} 1 \quad \text{(free-streaming regime)}.
    \]
\end{tcolorbox}

As indicated in Section~\secref{sec:M1}, there is no general analytical formula for computing the Eddington factor within the framework of the M1-multigroup model. However, it can be computed numerically (see Appendix~\secref{appendice:algorithmes_de_recherche}). In this section, I present several methods used in the literature to perform this computation, followed by the innovative method based on \gls{ai} that I developed during my PhD. Finally, I compare the performance of this new approach with that of existing methods.

\subsection{\quotes{Historical} methods} \label{sec:methodes_Eddington}

Historically, several approaches have been developed to estimate the Eddington factor in the M1-multigroup model. These methods aim to circumvent the absence of a general analytical formula by exploiting different computational and approximation strategies. Three main families of techniques have thus emerged:

\begin{enumerate}
    \item search algorithms;
    \item interpolation of precomputed values;
    \item the use of the Eddington factor expression from the M1-gray model.
\end{enumerate}

Each of these methods present advantages and limitations, which affect the accuracy and efficiency of radiative simulations, as will be shown in the following sections.

\starsect{Search algorithms}

As discussed in Section~\secref{sec:dependence_chig}, in the general case the radiative energy, the reduced flux, and the Eddington factor all depend on the Lagrange multipliers $\alpha_{0,g}$ and $\beta_g$, as well as on the frequency bounds of the group $\nu_{g-1/2}$ and $\nu_{g+1/2}$:
\begin{align*}
    \mathrm{E}_g = e(\alpha_{0,g}, \beta_g; \nu_{g-1/2}, \nu_{g+1/2})\;\;\mathpunct{,}\\
    \mathrm{f}_g = f(\alpha_{0,g}, \beta_g; \nu_{g-1/2}, \nu_{g+1/2})\;\;\mathpunct{,}\\
    \chi_g = c(\alpha_{0,g}, \beta_g; \nu_{g-1/2}, \nu_{g+1/2}) \;\;\mathpunct{.}
\end{align*}

Turpault~\cite{turpault_2002} showed that, for each physically admissible pair of radiative energy and reduced flux \mbox{$(\mathrm{E}_g, \mathrm{f}_g)$}, there exists a unique pair of Lagrange multipliers \mbox{$(\alpha_{0,g}, \beta_g)$}. A classical approach therefore consists in using a search algorithm to determine these multipliers and then deducing the associated Eddington factor $\chi_g$. Two search algorithms have been developed to solve this problem:

\begin{enumerate}
    \item \textbf{The Dichotomy-Newton algorithm}, introduced in the PhD thesis of Hung Chinh Nguyen (2011)~\cite{nguyen_2011_these}, which combines a bisection search to estimate $\beta_g$ within the interval \mbox{$\openinterv{-1}{1}$}, followed by the Newton method to determine $\alpha_{0,g}$ in $\mathbb{R}^+$;
    \item \textbf{The line-search algorithm}, developed in the course of this thesis, which simultaneously searches $\beta_g$ and $\alpha_{0,g}$ in the space \mbox{$\openinterv{-1}{1} \times \mathbb{R}^+$} in order to accelerate convergence and reduce computational cost.
\end{enumerate}

Additional details on the implementation of these search algorithms are provided in Appendix~\secref{appendice:algorithmes_de_recherche}. Although this approach yields highly accurate estimates of the Eddington factor, with a relative error on the order of $10^{-5}~\%$, it remains computationally expensive, even when using the line-search algorithm, which converges faster than the Dichotomy-Newton algorithm. Moreover, in some cases the algorithm may fail to converge, leading to simulation failure.

\starsect{Interpolation}

Due to the prohibitive computational cost of search algorithms, an alternative approach based on the precomputation and interpolation of Eddington factors was proposed by Turpault in 2003, and later adopted and refined within the \gls{hades} code by Nguyen in 2011~\cite{turpault_2003_these, nguyen_2011_these}. This approach aims to reconcile accuracy and computational efficiency by replacing systematic numerical resolution with interpolation over a grid of precomputed values.

In the \gls{hades} code, interpolation is performed using cubic splines, based on a grid of precomputed Eddington factor values for a given group. These values are obtained using the search algorithms presented previously and are stored for regularly spaced values of the logarithm of the radiative energy and the reduced flux (that is, $\ln(\mathrm{E}_g)$ and $\mathrm{f}_g$). This method enables a rapid estimation of the Eddington factor while maintaining good accuracy, provided that the grid is sufficiently dense and properly calibrated.

However, this approach has a major limitation: it relies on a good a priori knowledge of the orders of magnitude of the radiative quantities encountered during a simulation. In the context of the M1-multigroup model, these quantities are difficult to estimate, since they depend on spectral interactions with matter, which vary from one group to another and evolve over time. Insufficient coverage of the interpolation grid can then lead to significant errors in the estimation of the Eddington factor, thereby compromising the accuracy of the simulations.

\starsect{Analytical expression of the M1-gray model}

In the literature, some authors, such as Vaytet in his work on the implementation of the M1-multigroup model in the HERACLES code~\cite{vaytet_2011, vaytet_2012}, adopt a simplified approach by neglecting the dependence of the Eddington factor on the dimensionless radiative temperature $\mathcal{T}_g$ and on the group narrowness $\delta_g$. Instead, they use the analytical expression derived from the M1-gray model (see equation~\eqref{eq:facteur_Eddington_gray}):
\begin{equation*}
     \chi_g \approx \frac{3+4 \mathrm{f}_g^2}{5+2\sqrt{4 - 3 \mathrm{f}_g^2}} \;,
\end{equation*}

This method has the advantage of being extremely fast, since it is fully analytical. However, the expression used is not generalizable within the M1-multigroup framework and can introduce significant errors, reaching up to 16~\% relative error in the computation of the Eddington factor. While this approach may remain reasonably accurate for broad frequency groups, its precision rapidly degrades as the spectral resolution increases, leading to increasingly inaccurate predictions for narrow groups.

\subsection{Description of the new method based on \texorpdfstring{\glslink{ai}{Artificial Intelligence}}{AI}}  \label{sec:IA_Eddington}

\begin{figure}
    \begin{center}
        \begin{minipage}[t]{\linewidth}
            \centering
            \includegraphics[width=\textwidth]{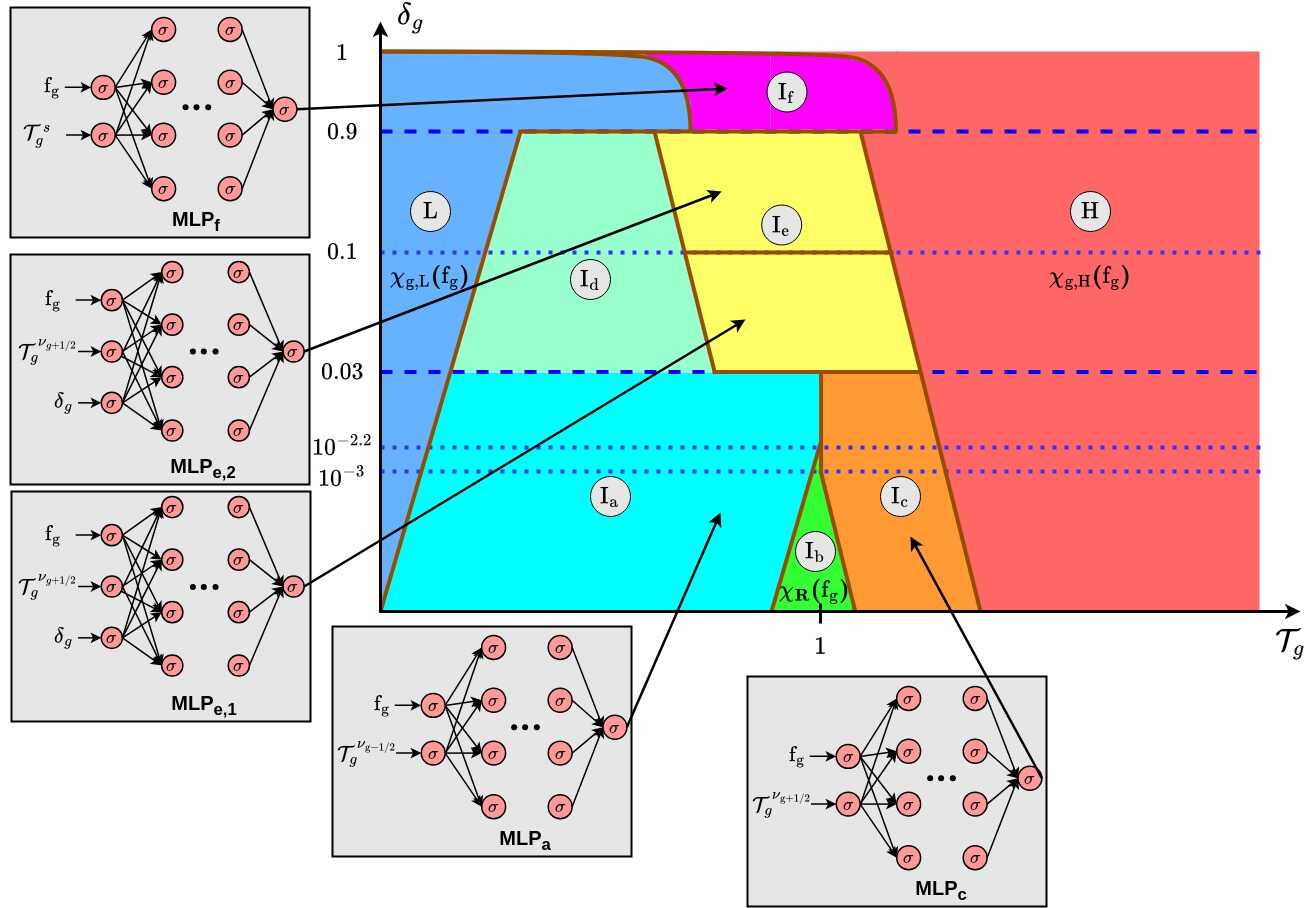}
        \end{minipage}
         \caption{Schematic overview of the developed method, for the case where \mbox{$\nu_{g-1/2}>0$} and \mbox{$\nu_{g+1/2}<+\infty$}. \mbox{$\chi_{g,\mathrm{L}}$} and \mbox{$\chi_{g,\mathrm{H}}$} correspond respectively to the polynomials defined by equations~\eqref{eq:polynome_lowT} and~\eqref{eq:polynome_highT}, used to compute $\chi_g$ in domains L and H, while $\chi_R$ denotes the Eddington factor of the M1-gray model (see equation~\eqref{eq:facteur_Eddington_gray}), used in domain $\mathrm{I_b}$. The \glsxtrshort{mlp}s are used to approximate the closure in the intermediate domains $\mathrm{I_a}$, $\mathrm{I_c}$, $\mathrm{I_d}$, $\mathrm{I_e}$, and $\mathrm{I_f}$. The boundaries of domains L, H, and $\mathrm{I_a}$ to $\mathrm{I_f}$ are defined in table~\ref{tab:domains}. The brown lines delineate the domains of applicability of the different approaches.}
        \label{fig:method_recap}
    \end{center}
\end{figure}

The objective of this study is to develop a new method based on \gls{ai} to accurately estimate the Eddington factor while maintaining a reasonable computational cost. Before implementing this approach, it is essential to select the \gls{ai} methodology best suited to the problem. Three types of models were considered:

\begin{enumerate}
    \item \textbf{Symbolic regression~\cite{smidt_2009}} aims at identifying mathematical expressions that fit the data. This approach has the advantage of producing interpretable models that are optimized for numerical evaluation. However, my experiments showed that it is difficult to simultaneously capture all the dependencies of the Eddington factor on the different parameters of the problem;
    
    \item \textbf{Gaussian processes~\cite{hida_1993}} are probabilistic models that provide predictions together with an uncertainty quantification. They are particularly effective trained on small datasets, but their computational cost becomes prohibitive using larger datasets. In the present case, the prediction time associated with Gaussian processes proved to be too high;
    
    \item \textbf{Neural networks (introduced previously)} are well suited to complex nonlinear relationships. They offer excellent accuracy and good generalization capabilities, at the cost of reduced interpretability.
\end{enumerate}

\begin{figure}
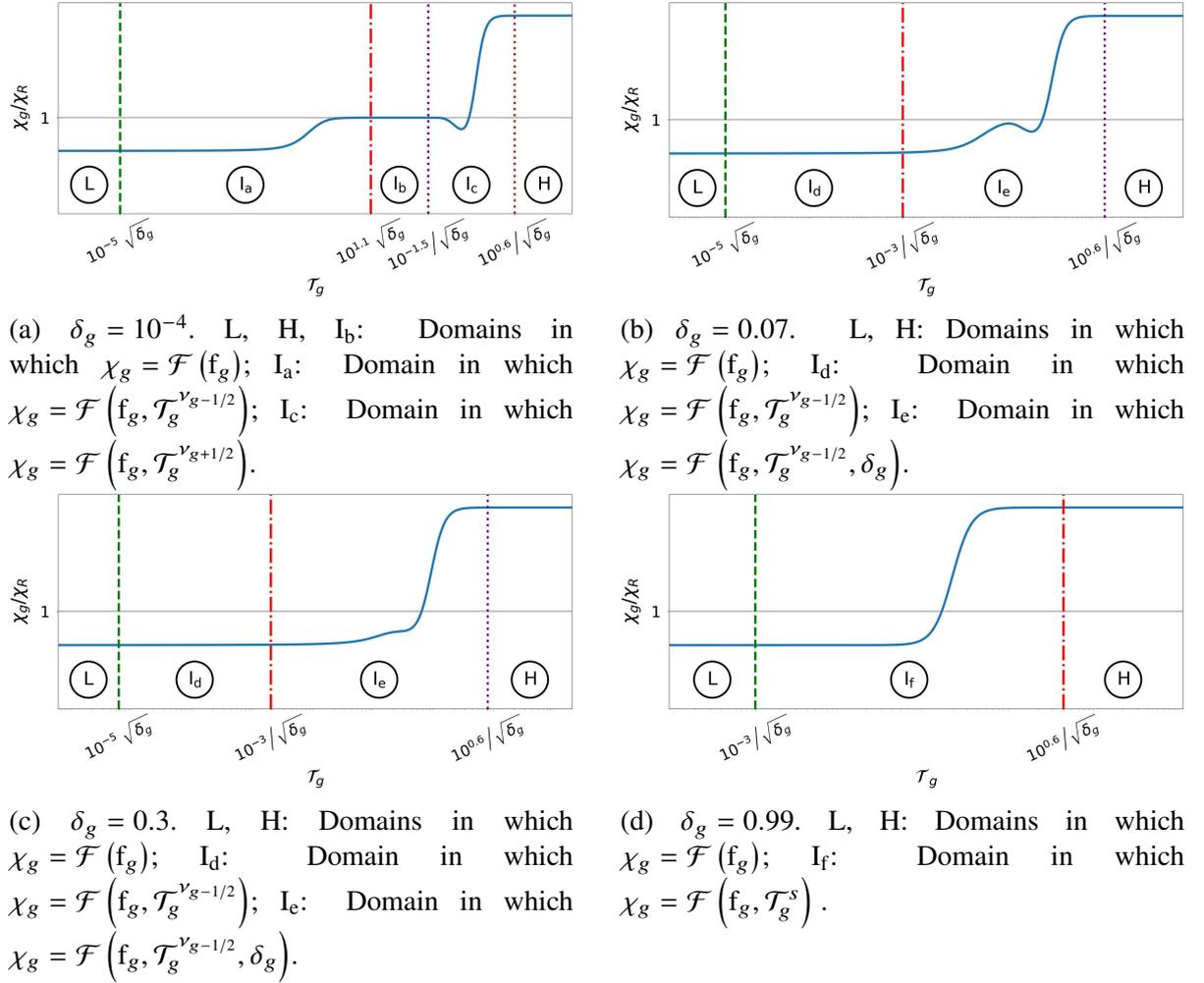

    \begin{subfigure}[t]{0.48\textwidth}
        \centering
        \includegraphics[width=\textwidth]{Images/chapitre_2/Tg_dep_d0.0001.jpg}
        \caption{\mbox{$\delta_g=10^{-4}$.} L, H, $\mathrm{I_b}$: Domains in which \mbox{$\chi_g = \mathcal{F} \left (\mathrm{f}_g \right )$;} $\mathrm{I_a}$: Domain in which \mbox{$\chi_g=\mathcal{F} \left (\mathrm{f}_g, \mathcal{T}_g^{\nu_{g-1/2}} \right )$;} $\mathrm{I_c}$: Domain in which \mbox{$\chi_g=\mathcal{F} \left (\mathrm{f}_g, \mathcal{T}_g^{\nu_{g+1/2}} \right )$}.}
        \label{fig:dnu_1e-4_deps}
    \end{subfigure}
    \hfill
    \begin{subfigure}[t]{0.48\textwidth}
        \centering
        \includegraphics[width=\textwidth]{Images/chapitre_2/Tg_dep_d0.07.jpg}
        \caption{\mbox{$\delta_g=0.07$}. L, H: Domains in which \mbox{$\chi_g=\mathcal{F} \left (\mathrm{f}_g \right )$;} $\mathrm{I_d}$: Domain in which \mbox{$\chi_g=\mathcal{F} \left (\mathrm{f}_g, \mathcal{T}_g^{\nu_{g-1/2}} \right )$;} $\mathrm{I_e}$: Domain in which \mbox{$\chi_g=\mathcal{F} \left (\mathrm{f}_g, \mathcal{T}_g^{\nu_{g-1/2}}, \delta_g \right )$.}}
        \label{fig:dnu_7e-2_deps}
    \end{subfigure}
    \hfill
    \begin{subfigure}[t]{0.48\textwidth}
        \centering
        \includegraphics[width=\textwidth]{Images/chapitre_2/Tg_dep_d0.3.jpg}
        \caption{\mbox{$\delta_g=0.3$.} L, H: Domains in which \mbox{$\chi_g=\mathcal{F} \left (\mathrm{f}_g \right )$;} $\mathrm{I_d}$: Domain in which \mbox{$\chi_g=\mathcal{F} \left (\mathrm{f}_g, \mathcal{T}_g^{\nu_{g-1/2}} \right )$;} $\mathrm{I_e}$: Domain in which \mbox{$\chi_g=\mathcal{F} \left (\mathrm{f}_g, \mathcal{T}_g^{\nu_{g-1/2}}, \delta_g \right )$}.}
        \label{fig:dnu_3e-1_deps}
    \end{subfigure}
    \hfill
    \begin{subfigure}[t]{0.48\textwidth}
        \centering
        \includegraphics[width=\textwidth]{Images/chapitre_2/Tg_dep_d0.99.jpg}
        \caption{\mbox{$\delta_g=0.99$.} L, H: Domains in which \mbox{$\chi_g=\mathcal{F} \left (\mathrm{f}_g \right )$;} $\mathrm{I_f}$: Domain in which \mbox{$\chi_g=\mathcal{F} \left (\mathrm{f}_g, \mathcal{T}_g^s \right )$} .}
        \label{fig:dnu_0.99_deps}
    \end{subfigure}
    \caption{Ratio between the Eddington factor of the M1-multigroup model, $\chi_g$, and that of the M1-gray model, $\chi_R$, for a reduced flux $\mathrm{f}_g = 0.65$, in the case where $\nu_{g-1/2}>0$ and $\nu_{g+1/2}<+\infty$, for different values of the group narrowness $\delta_g$.}
    \label{fig:Tg_dep_IA}
\end{figure}

In light of these considerations, the use of neural networks, and more specifically a \gls{mlp}-type architecture, was selected as the preferred approach (see Radureau (2025)~\cite{radureau_2025a} for further details). Two main strategies can be considered in this context. The first consists in training a single large network capable of predicting the Eddington factor over the entire domain, in the spirit of general-purpose approaches such as ChatGPT. However, this solution does not meet the objectives set here, as it involves large models whose predictions are computationally expensive. The second strategy relies on the use of more compact networks, specialized on specific subdomains, whose task is essentially limited to interpolation. With the goal of achieving numerical efficiency while maintaining a high level of accuracy, this second approach was therefore favored.

Figure~\ref{fig:method_recap} illustrates the organization of the method, distinguishing the different strategies adopted depending on the domain. In domains L and H, the Eddington factor is evaluated using polynomials. In domain $\text{I}_b$, the analytical expression from the M1-gray model is used. The remaining domains are handled using five distinct \gls{mlp}s, each with two or three inputs: one network specialized for domains $\mathrm{I_a}$ and $\mathrm{I_d}$; one for domain $\mathrm{I_c}$; one for $\mathrm{I_f}$; and two for domain $\mathrm{I_e}$: one for cases where $\delta_g \leq 0.03$, and the other for cases where $\delta_g > 0.03$. The remainder of this section is devoted to a detailed description of this approach.

\begin{figure}
    \begin{subfigure}[t]{0.48\textwidth}
        \centering
        \includegraphics[width=\textwidth]{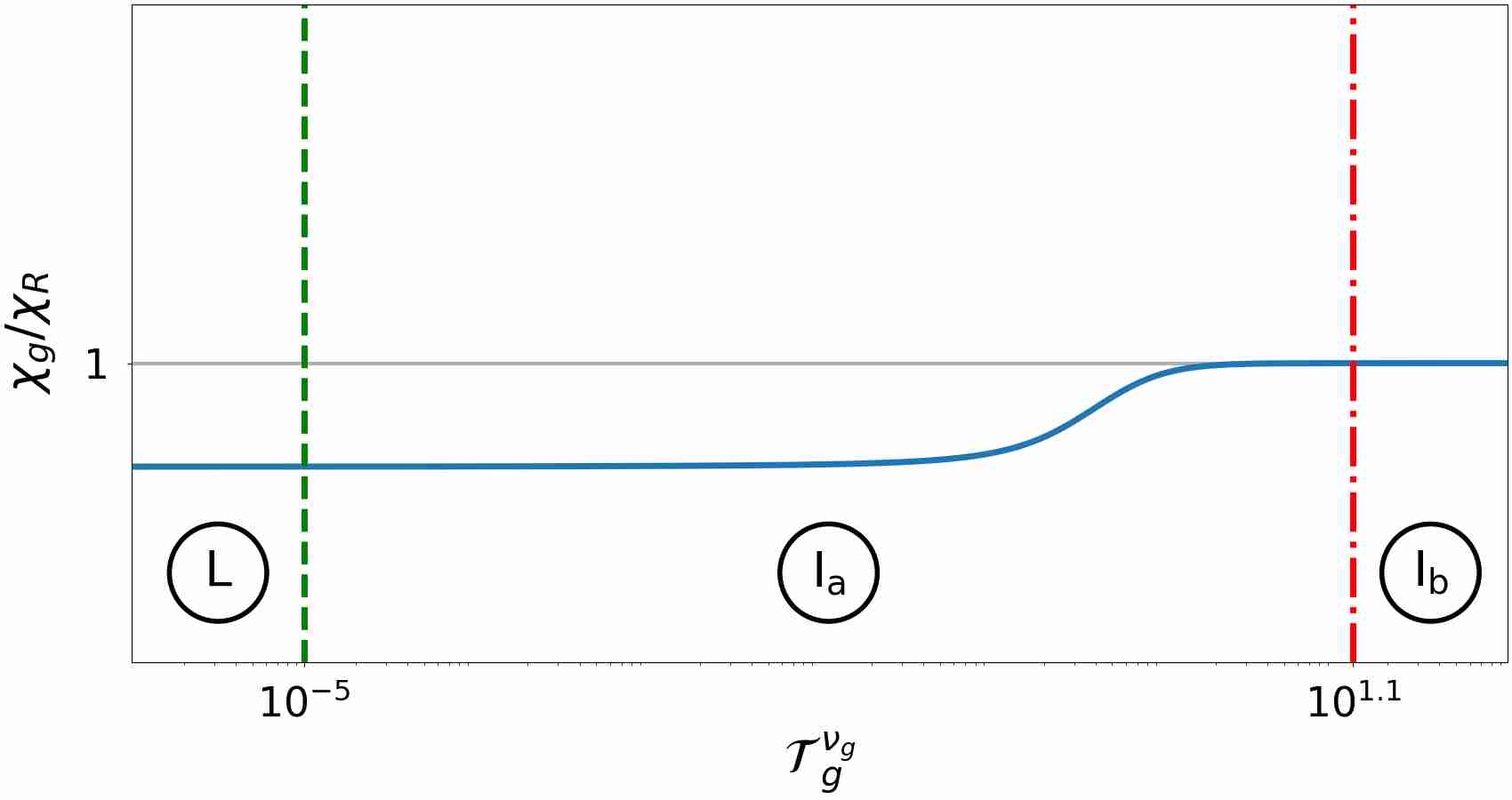}
        \caption{Case where $\nu_{g+1/2} = +\infty$. L, $\mathrm{I_b}$: Domains in which $\chi_g=\mathcal{F} \left (\mathrm{f}_g \right )$: $\mathrm{I_a}$: Domain in which $\chi_g=\mathcal{F} \left (\mathrm{f}_g, \mathcal{T}_g^{\nu_{g-1/2}} \right )$.}
        \label{fig:numaxinf_dep}
    \end{subfigure}
    \hfill
    \begin{subfigure}[t]{0.48\textwidth}
        \centering
        \includegraphics[width=\textwidth]{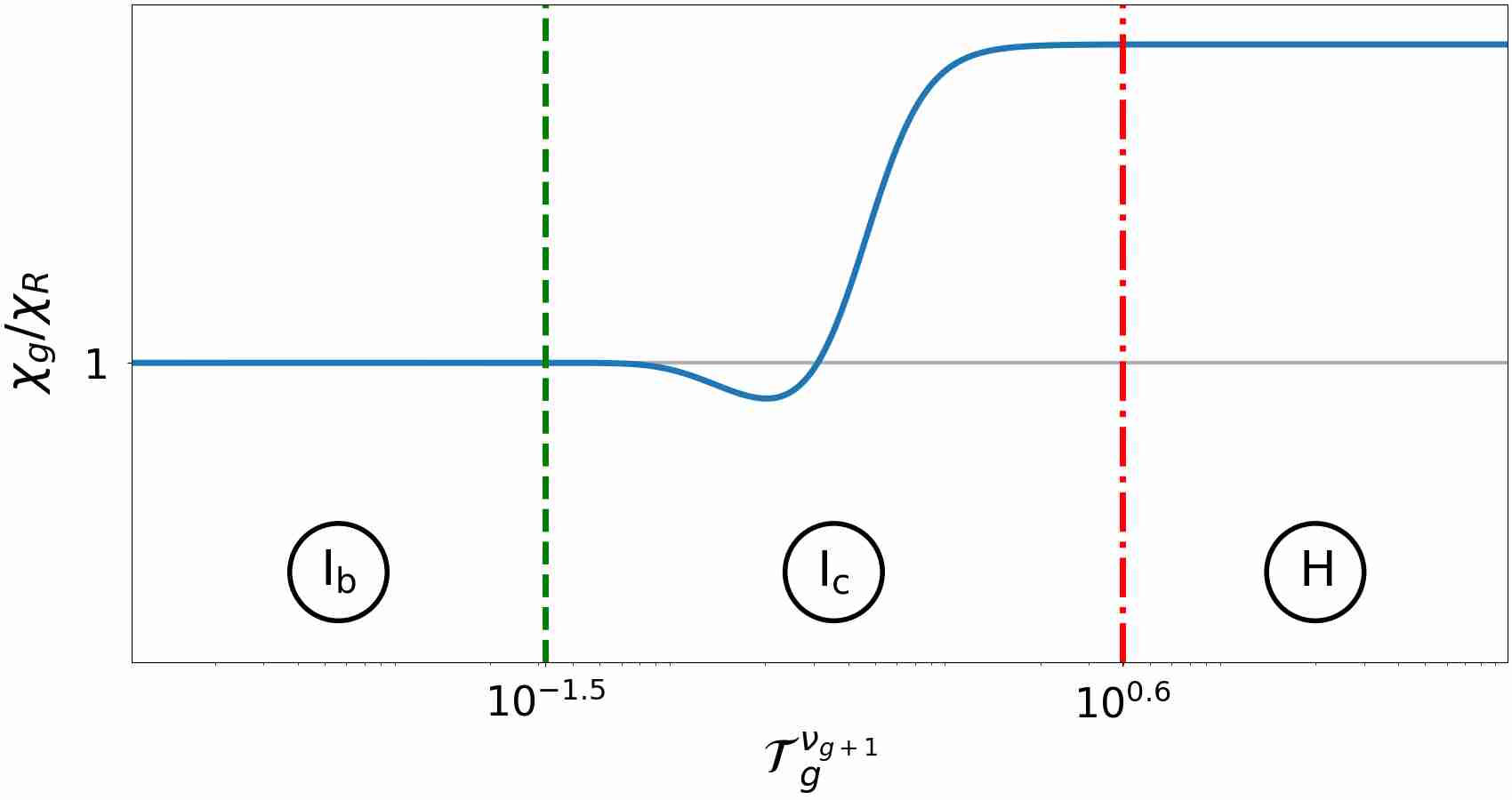}
        \caption{Case where $\nu_{g-1/2} = 0$. H, $\mathrm{I_b}$: Domains in which $\chi_g=\mathcal{F} \left (\mathrm{f}_g \right )$: $\mathrm{I_c}$: Domain in which $\chi_g=\mathcal{F} \left (\mathrm{f}_g, \mathcal{T}_g^{\nu_{g+1/2}} \right )$.}
        \label{fig:numin0_dep}
    \end{subfigure}
    \caption{Ratio between the Eddington factor of the M1-multigroup model, $\chi_g$, and that of the M1-gray model, $\chi_R$, for a reduced flux $\mathrm{f}_g = 0.65$, in the cases where $\nu_{g-1/2}=0$ and where $\nu_{g+1/2}=+\infty$.}
    \label{fig:Tg_dep_other_IA}
\end{figure}

To design this method, I first analyzed the dependence of the Eddington factor on the influential parameters, as detailed in Section~\secref{sec:dependence_chig}. The main results of this analysis are summarized below:

\begin{itemize}
    \item \textbf{When \mbox{$\mathbf{\nu_{g-1/2}>0}$} and \mbox{$\mathbf{\nu_{g+1/2}<+\infty}$} (figure~\ref{fig:Tg_dep_IA})}, the Eddington factor depends on the dimensionless radiative temperature $\mathcal{T}_g$ (or, equivalently, $\mathcal{T}_g^{\nu{g+1/2}}$ or $\mathcal{T}_g^{\nu{g-1/2}}$), on the reduced flux $\mathrm{f}_g$, and on the group narrowness $\delta_g$;
    \item \textbf{When \mbox{$\mathbf{\nu_{g-1/2}=0}$} (figure~\ref{fig:numin0_dep})}, the Eddington factor depends only on the dimensionless radiative temperature $\mathcal{T}_g^{\nu_{g+1/2}}$ and on the reduced flux $\mathrm{f}_g$;
    \item \textbf{When \mbox{$\mathbf{\nu_{g+1/2}=+\infty}$} (figure~\ref{fig:numaxinf_dep})}, it depends on the dimensionless radiative temperature $\mathcal{T}_g^{\nu_{g-1/2}}$ and on the reduced flux $\mathrm{f}_g$.
\end{itemize}

On this basis, I identified several domains in which the Eddington factor exhibits distinct dependencies:

\begin{itemize}
    \item \textbf{domains L and H (figures~\ref{fig:Tg_dep_IA} and~\ref{fig:Tg_dep_other_IA}):} asymptotic cases in which the Eddington factor depends only on the reduced flux $\mathrm{f}_g$,
    \item \textbf{domain $\mathbf{I_b}$ (figures~\ref{fig:dnu_1e-4_deps} and~\ref{fig:Tg_dep_other_IA}):} the Eddington factor depends only on the reduced flux $\mathrm{f}_g$, and the analytical expression of the Eddington factor from the M1-gray model is valid,
    \item \textbf{domain $\mathbf{I_a}$ (figures~\ref{fig:dnu_1e-4_deps} and~\ref{fig:numaxinf_dep}):} the Eddington factor depends only on the reduced flux $\mathrm{f}_g$ and on the dimensionless radiative temperature $\mathcal{T}_g^{\nu_{g-1/2}}$,
    \item \textbf{domain $\mathbf{I_c}$ (figures~\ref{fig:dnu_1e-4_deps} and~\ref{fig:numin0_dep}):} the Eddington factor depends only on the reduced flux $\mathrm{f}_g$ and on the dimensionless radiative temperature $\mathcal{T}_g^{\nu_{g+1/2}}$,
    \item \textbf{domain $\mathbf{I_d}$ (figures~\ref{fig:dnu_7e-2_deps} and \ref{fig:dnu_3e-1_deps}):} the Eddington factor depends only on the reduced flux $\mathrm{f}_g$, on the dimensionless radiative temperature $\mathcal{T}_g^{\nu_{g-1/2}}$, and very weakly on the group narrowness $\delta_g$. I neglected this last dependence and assumed that the Eddington factor values in this domain are the same as in domain $\mathrm{I_a}$, which is accurate within a relative error of $10^{-3}~\%$,
    \item \textbf{domain $\mathbf{I_e}$ (figures~\ref{fig:dnu_7e-2_deps} and \ref{fig:dnu_3e-1_deps}):} the Eddington factor depends strongly on the reduced flux $\mathrm{f}_g$, on the dimensionless radiative temperature $\mathcal{T}_g^{\nu_{g+1/2}}$, and on the group narrowness $\delta_g$,
    \item \textbf{domain $\mathbf{I_f}$ (figure~\ref{fig:dnu_0.99_deps}):} the Eddington factor depends on the reduced flux $\mathrm{f}_g$ and on a dimensionless radiative temperature $\mathcal{T}_g^s$, which will be referred to as the \textit{shifted temperature}, and which accounts for the shift described in equation~\eqref{eq:H_condition}, Section~\secref{sec:dependence_chig}.
\end{itemize}

\noindent First, let us define the temperature $\mathcal{T}_g^*$, related to the dimensionless temperature $\mathcal{T}_g$ by:
\begin{equation*}
    \mathcal{T}_g^* = \frac{\mathcal{T}_g \sqrt{\delta_g}}{(1-\delta_g^3)^{1/4}} = \frac{\mathcal{T}_g^{\nu_{g-1/2}}}{(1-\delta_g^3)^{1/4}} \;\;\mathpunct{.}
\end{equation*}

\noindent The shifted temperature can then be defined from $\mathcal{T}_g^*$ as follows\footnote{This definition differs slightly from the one presented in the article~\cite{radureau_2025a}, which relied on linear and quadratic adjustments. This difference does not affect the results.}:
\begin{equation}
    \label{eq:Tg_s_def}
    \mathcal{T}_g^s = \mathcal{T}_g^* \frac{(1-0.99^3)^{1/4}}{\sqrt{0.99}} = \frac{\mathcal{T}_g \sqrt{\delta_g/0.99}}{\left \{(1-\delta_g^3)\middle / (1-0.99^3)\right \}^{1/4}} \;\;\mathpunct{.}
\end{equation}
This expression ensures that \mbox{$\mathcal{T}_g^s = \mathcal{T}_g$} when \mbox{$\delta_g = 0.99$}.

The boundaries between the different domains H, L, etc. were determined empirically, so that the relative errors induced by the dependency simplifications remain below $10^{-3}~\%$. For domains L and H, polynomial expressions are used; for domain $\mathrm{I_b}$, the analytical expression from the M1-gray model; and for all other domains, neural networks are used. These choices are summarized in table~\ref{tab:domains}.

\begin{table}[ht]
    \centering
    \begin{tabular}{C{4mm}C{15mm}C{15mm}C{25mm}C{25mm}C{40mm}}
        \hline
         \TBstrut & \multicolumn{2}{c}{Range of $\delta_g$} & \multicolumn{2}{c}{Range of $\mathcal{T}_g$} & \multicolumn{1}{c}{Method$~^{**}$} \\
          & \multicolumn{1}{c}{Min} & \multicolumn{1}{c}{Max} & \multicolumn{1}{c}{Min} & \multicolumn{1}{c}{Max} & \TBstrut\\
        \hline
        \hline
        \Tstrut L & $0$               & $0.9$ & $0$                    & $10^{-5} \sqrt{\delta_g}$         & $\mathrm{P}$    \\
        \Bstrut  & $0.9$             & $1$         & $0~^{*}$            & $\left. 10^{-3} \middle /\sqrt{\delta_g} \right.$$~^{*}$ &  Input: $\mathrm{f}_g$ \\
        \hline
        \Tstrut H & $0$               & $0.9$ & $\left. 10^{0.6} \middle / \sqrt{\delta_g} \right.$       & $+\infty$         & $\mathrm{P}$        \\
        \Bstrut & $0.9$             & $1$         & $\left. 10^{0.6} \middle / \sqrt{\delta_g} \right.$$~^{*}$ & $+\infty$$~^{*}$        &  Input: $\mathrm{f}_g$ \\
        \hline
        \Tstrut $\mathrm{I_a}$ & $0$               & $10^{-2.2}$      & $10^{-5} \sqrt{\delta_g}$    & $10^{1.1} \sqrt{\delta_g}$       & $\mathrm{MLP}$ \\
        \Bstrut & $10^{-2.2}$ & $0.03$ & $10^{-5} \sqrt{\delta_g}$ & $1$ & Input: $\mathrm{f}_g$, $\mathcal{T}_g^{\nu_{g-1/2}}$ \\
        \hline
        \Tstrut $\mathrm{I_b}$ & $0$               & $10^{-3}$ & $10^{1.1} \sqrt{\delta_g}$   & $\left. 10^{-1.5} \middle / \sqrt{\delta_g} \right.$    & $\chi_R$ \\
         & $10^{-3}$ & $10^{-2.2}$ & $10^{1.1} \sqrt{\delta_g}$   & $1$    &  \\
        \hline
        \Tstrut $\mathrm{I_c}$ & $0$               & $10^{-3}$      & $\left. 10^{-1.5} \middle / \sqrt{\delta_g} \right.$  & $\left. 10^{0.6} \middle / \sqrt{\delta_g} \right.$     & $\mathrm{MLP}$ \\
        \Bstrut & $10^{-3}$ & $0.03$ & $1$ & $\left. 10^{0.6} \middle / \sqrt{\delta_g} \right.$ & Input: $\mathrm{f}_g$, $\mathcal{T}_g^{\nu_{g+1/2}}$ \\
        \hline
        \Tstrut $\mathrm{I_d}$ & $0.03$            & $0.9$       & $10^{-5} \sqrt{\delta_g}$  & $\left. 10^{-3} \middle / \sqrt{\delta_g} \right.$      & $\mathrm{MLP}$ \\
        \Bstrut & & & & & Input: $\mathrm{f}_g$, $\mathcal{T}_g^{\nu_{g+1/2}}$ \\
        \hline
        \Tstrut $\mathrm{I_e}$ & $0.03$            & $0.9$       & $\left. 10^{-3} \middle / \sqrt{\delta_g} \right.$  & $\left. 10^{0.6} \middle / \sqrt{\delta_g} \right.$     & $\mathrm{MLP}$ \\
        \Bstrut & & & & & Input: $\mathrm{f}_g$, $\delta_g$, $\mathcal{T}_g^{\nu_{g+1/2}}$ \\
        \hline
        \Tstrut $\mathrm{I_f}$ & $0.9$             & $1$         & $\left. 10^{-3} \middle / \sqrt{\delta_g} \right.$$~^{*}$  & $\left. 10^{0.6} \middle / \sqrt{\delta_g} \right.$$~^{*}$ & $\mathrm{MLP}$ \\
        \Bstrut & & & & & Input: $\mathrm{f}_g$, $\mathcal{T}_g^s$ \\
        \hline
        \multicolumn{6}{p{144mm}}{\hspace{-5pt}$^{*}$\footnotesize Min/max values of the shifted temperature $\mathcal{T}_g^s$ (see equation~\ref{eq:Tg_s_def}).}\\
        \multicolumn{6}{p{144mm}}{\hspace{-5pt}$^{**}$\footnotesize P: polynomial, \glsxtrshort{mlp}: neural network, $\chi_R$: Eddington factor of the M1-gray model. The variables listed after “Input:” correspond to the inputs used for each method.}
    \end{tabular}
    \caption{Limits for each domain and the method used to compute the Eddington factor.}
    \label{tab:domains}
\end{table}

Before detailing the method, it is crucial to examine the impact of an error in the estimation of the Eddington factor on the components of the radiative pressure tensor. In the context of 2D simulations performed with the \gls{hades} code, since the radiative pressure tensor is symmetric, it is sufficient to compute three of its components to fully determine the tensor: the components $\mathbb{P}_{g,xx}$, $\mathbb{P}_{g,yy}$, and $\mathbb{P}_{g,xy}$. Using equation~\eqref{eq:Pg_closure}, each component of the radiative pressure tensor can be expressed as:
\begin{align*}
    \mathbb{P}_{g,xx} &= \left \{ \frac{1- \chi_g}{2} + \frac{3 \chi_g - 1}{2} \frac{\mathrm{F}_{g,x}^2}{||\mathrm{\bm{F}}_g||^2} \right \} \mathrm{E}_g \;\;\mathpunct{,}\\
    \mathbb{P}_{g,yy} &= \left \{ \frac{1- \chi_g}{2} + \frac{3 \chi_g - 1}{2} \frac{\mathrm{F}_{g,y}^2}{||\mathrm{\bm{F}}_g||^2} \right \} \mathrm{E}_g \;\;\mathpunct{,}\\
    \mathbb{P}_{g,xy} &= \frac{3 \chi_g - 1}{2} \frac{\mathrm{F}_{g,x} \mathrm{F}_{g,y}}{||\mathrm{\bm{F}}_g||^2} \mathrm{E}_g \;\;\mathpunct{,}
\end{align*}

When considering an error $\delta\chi_g$ in the estimation of the Eddington factor, the relative error on the components of the radiative pressure can be expressed as:
\begin{align*}
    \left |\frac{\delta \mathbb{P}_{g,xx}}{\mathbb{P}_{g,xx}}\right | &= \left |\frac{3 x^2 - 1}{(3x^2 - 1) \chi_g + 1 - x^2}\right | \delta \chi_g \;\;\mathpunct{,}\\
    \left |\frac{\delta \mathbb{P}_{g,yy}}{\mathbb{P}_{g,yy}}\right | &= \left |\frac{3 y^2 - 1}{(3y^2 - 1) \chi_g + 1 - y^2}\right | \delta \chi_g \;\;\mathpunct{,}\\
    \left |\frac{\delta \mathbb{P}_{g,xy}}{\mathbb{P}_{g,xy}}\right | &= \frac{ \delta \chi_g}{ \chi_g - 1/3} \;\;\mathpunct{,}
\end{align*}

\noindent where \mbox{$x=\mathrm{F}_{g,x}/||\mathrm{\bm{F}}_g||$} and \mbox{$y=\mathrm{F}_{g,y}/||\mathrm{\bm{F}}_g||$}. The quantities \mbox{$|\delta \mathbb{P}_{g,xx}/\mathbb{P}_{g,xx}|$} and \mbox{$|\delta \mathbb{P}_{g,yy}/\mathbb{P}_{g,yy}|$} reach their maximum relative error when \mbox{$x=0$} or $1$ and \mbox{$y=0$} or $1$, respectively. In this case, the maximum relative errors can be written as:
\begin{equation}
    \label{eq:err_pressure}
    \left\{
    \begin{aligned}
        \left |\frac{\delta \mathbb{P}_{g,xx}}{\mathbb{P}_{g,xx}}\right |_{\max} &= \frac{\delta \chi_g}{\min\left \{ \chi_g, 1-\chi_g\right \}} \;\;\mathpunct{,} \\
        \left |\frac{\delta \mathbb{P}_{g,yy}}{\mathbb{P}_{g,yy}}\right |_{\max} &= \frac{\delta \chi_g}{\min\left \{ \chi_g, 1-\chi_g\right \}} \;\;\mathpunct{,} \\
        \left |\frac{\delta \mathbb{P}_{g,xy}}{\mathbb{P}_{g,xy}}\right |_{\max} &= \frac{\delta \chi_g}{ \chi_g - 1/3} \;\;\mathpunct{.}
    \end{aligned}
    \right.
\end{equation}

These expressions highlight the importance of an accurate estimation of the Eddington factor, particularly when the reduced flux $\mathrm{f}_g$ is close to $0$ or $1$, in order to minimize relative errors in the components of the radiative pressure tensor. When $\mathrm{f}_g$ is close to $0$, the component $\mathbb{P}_{g,xy}$ also tends toward zero. Moreover, if $\mathrm{f}_g$ is close to $0$ and the radiative flux in the $x$ or $y$ direction is weak, the components $\mathbb{P}_{g,xx}$ or $\mathbb{P}_{g,yy}$ also approach zero, which explains the large relative errors observed in such situations. The maximum error on the components of the radiative pressure tensor will therefore be evaluated using the expressions derived above. In the following, I will detail the methods I have developed to compute the Eddington factor in the different domains, taking this impact into account.

\starsect{Asymptotically high temperatures (domain H, $\boldsymbol{k_B \mathrm{T}_g \gg h \nu_{g+1/2}}$)}

In this case, equations~\eqref{eq:fg_H} and~\eqref{eq:chig_H}, presented in Section~\secref{sec:dependence_chig}, simplify in such a way that the reduced flux and the Eddington factor can be expressed as follows:
\begin{align}
     \mathrm{f}_g &= - \frac{\arctanh(\beta_g) - \beta_g}{\beta_g \arctanh(\beta_g)} \;\;\mathpunct{,} \label{eq:fg_highT} \\
     \chi_g &= \frac{\arctanh(\beta_g) - \beta_g}{\beta_g^2 \arctanh(\beta_g)} \;\;\mathpunct{.} \label{eq:chig_highT}
\end{align}

Here, it is possible to use a simple bisection algorithm to determine the Lagrange multiplier $\beta_g$ within the interval $\openinterv{-1}{1}$, and then to compute the associated Eddington factor. However, performing this search during a simulation is not optimal in terms of computational cost. Therefore, I chose to use a 12th-order polynomial, whose expression is:
\begin{align}
    \begin{split}
        \chi_{g,\mathrm{H}}~(\mathrm{f}_g) = &1/3 + \mathrm{f}_g^2 - \mathrm{f}_g^3/3~~+ \\
        &~~6.5813886320063~(\mathrm{f}_g^4-2 \mathrm{f}_g^3+\mathrm{f}_g^2)~- \\
        &~44.5593808930324~(\mathrm{f}_g^5-3 \mathrm{f}_g^3+2\mathrm{f}_g^2)~~+ \\
        &179.9251066153469~(\mathrm{f}_g^6-4 \mathrm{f}_g^3+3\mathrm{f}_g^2)~~- \\
        &463.2145547920471~(\mathrm{f}_g^7-5 \mathrm{f}_g^3+4\mathrm{f}_g^2)~~+ \\
        &776.0741150675088~(\mathrm{f}_g^8-6 \mathrm{f}_g^3+5\mathrm{f}_g^2)~~- \\
        &841.0048410765069~(\mathrm{f}_g^9-7 \mathrm{f}_g^3+6\mathrm{f}_g^2)~~+ \\
        &566.6515584347650~(\mathrm{f}_g^{10}-8 \mathrm{f}_g^3+7\mathrm{f}_g^2)~- \\
        &215.1853394583736~(\mathrm{f}_g^{11}-9 \mathrm{f}_g^3+8\mathrm{f}_g^2)~+ \\
        &~35.1130440991836~(\mathrm{f}_g^{12}-10 \mathrm{f}_g^3+9\mathrm{f}_g^2) \;\;\mathpunct{.}
    \end{split} \label{eq:polynome_highT}
\end{align}

\begin{figure}
    \begin{center}
        \begin{minipage}[t]{0.6\linewidth}
            \centering
            \includegraphics[width=\textwidth]{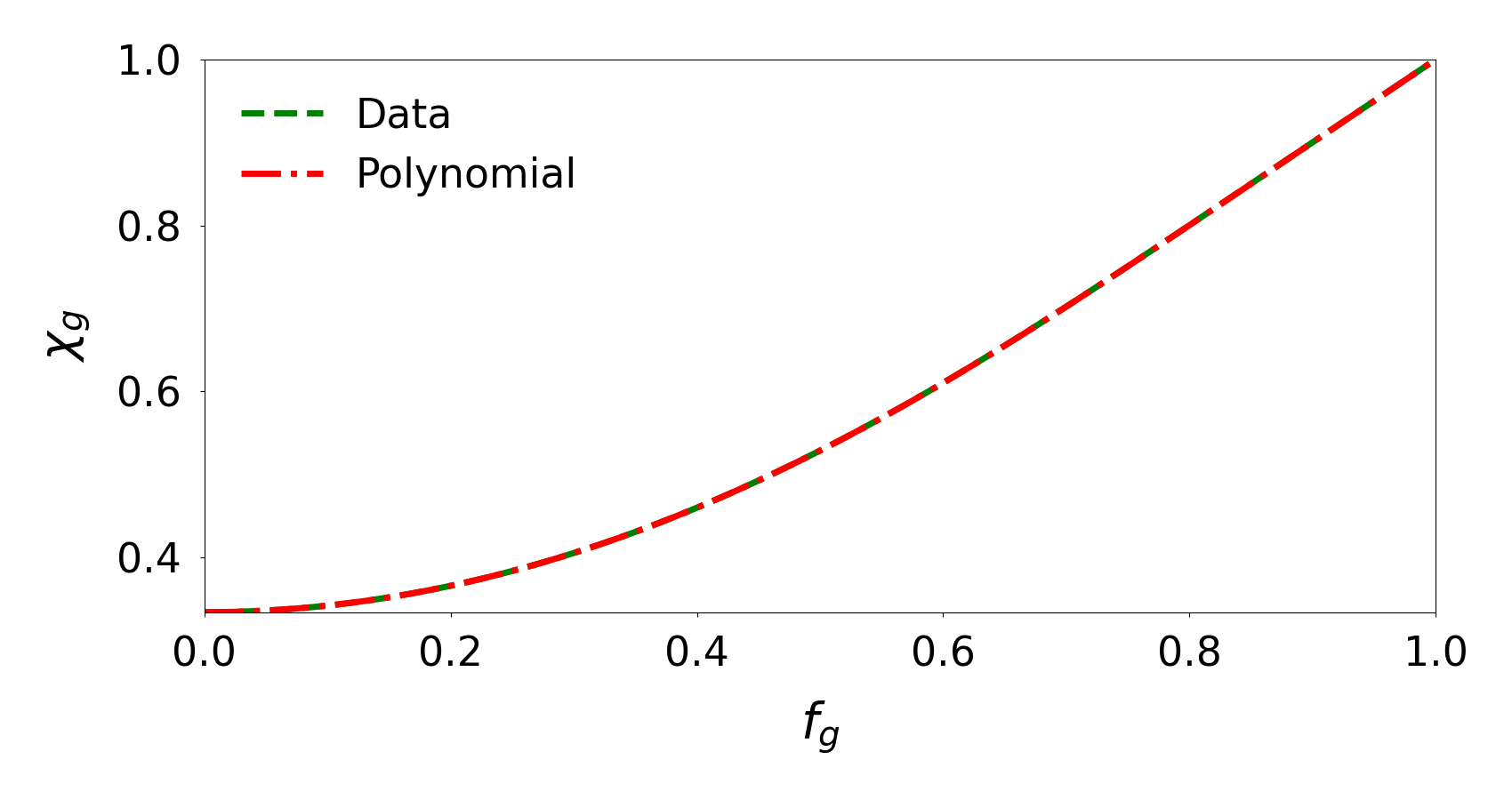}
        \end{minipage}
    \end{center}
    \caption{Comparison of the Eddington factor computed using the polynomial expression $\chi_{g,\mathrm{H}}$, defined by equation~\eqref{eq:polynome_highT}, with the reference data obtained by solving equations~\eqref{eq:fg_highT} and~\eqref{eq:chig_highT} using a bisection algorithm.}
    \label{fig:comp_highT_methods}
\end{figure}

In this formulation, the Eddington factor was constrained to take the value $1/3$ at \mbox{$\mathrm{f}_g = 0$} and $1$ at \mbox{$\mathrm{f}_g = 1$}. In addition, I imposed the conditions \mbox{$\chi_{g,\mathrm{H}}'(0) = 0$} and \mbox{$\chi_{g,\mathrm{H}}'(1) = 1$} in order to ensure appropriate regularity at the boundaries of the domain. This expression guarantees excellent accuracy, with a relative error below $10^{-3}~\%$ in the estimation of the Eddington factor. The maximum observed error is of the order of $10^{-2}~\%$ for the $\mathbb{P}_{g,xx}$ and $\mathbb{P}_{g,yy}$ components of the radiative pressure tensor, and of $10^{-1}~\%$ for the $\mathbb{P}_{g,xy}$ component. Figure~\ref{fig:comp_highT_methods} shows a comparison between the results obtained by numerically solving equations~\eqref{eq:fg_highT} (green curve) and~\eqref{eq:chig_highT} using a bisection algorithm, and those obtained from the polynomial approximation~\eqref{eq:polynome_highT} (red curve). Very good agreement between the two curves is observed.

\starsect{Asymptotically low temperatures (domain L, $\boldsymbol{k_B \mathrm{T}_g \ll h \nu_{g-1/2}}$)}

When the radiative temperature is very low, there is unfortunately no simplified analytical expression that allows the Eddington factor $\chi_g$ to be computed directly as a function of the reduced flux $\mathrm{f}_g$ alone, unlike in the previous case. This means that, in this regime, the radiative temperature still influences $\chi_g$. However, an analysis of the curves shown in figures~\ref{fig:Tg_dep_IA} and~\ref{fig:Tg_dep_other_IA} indicates that this dependence becomes very weak in this limit.

In order to ensure an efficient computation of the Eddington factor in this domain, I therefore chose to use an 11th-order polynomial, neglecting the residual dependence of $\chi_g$ on the radiative temperature. The retained expression is the following:
\begin{align}
    \begin{split}
        \chi_{g,\mathrm{L}}(\mathrm{f}_g) = &1/3~+~2\mathrm{f}_g^2/3~+ \\
        &~~0.3058945350580~(\mathrm{f}_g^3-\mathrm{f}_g^2)~~- \\
        &~~3.5960491961545~(\mathrm{f}_g^4-\mathrm{f}_g^2)~~+ \\
        &~24.1130825450229~(\mathrm{f}_g^5-\mathrm{f}_g^2)~~- \\
        &~92.2832025267787~(\mathrm{f}_g^6-\mathrm{f}_g^2)~~+ \\
        &220.1853600902954~(\mathrm{f}_g^7-\mathrm{f}_g^2)~~- \\
        &329.3727903123292~(\mathrm{f}_g^8-\mathrm{f}_g^2)~~+ \\
        &299.9957746394199~(\mathrm{f}_g^9-\mathrm{f}_g^2)~~- \\
        &151.3384873479037~(\mathrm{f}_g^{10}-\mathrm{f}_g^2)~+ \\
        &~32.2668624654447~(\mathrm{f}_g^{11}-\mathrm{f}_g^2) \;\;\mathpunct{.}
    \end{split}\label{eq:polynome_lowT}
\end{align}

\begin{figure}
    \begin{center}
        \begin{minipage}[t]{0.6\linewidth}
            \centering
            \includegraphics[width=\textwidth]{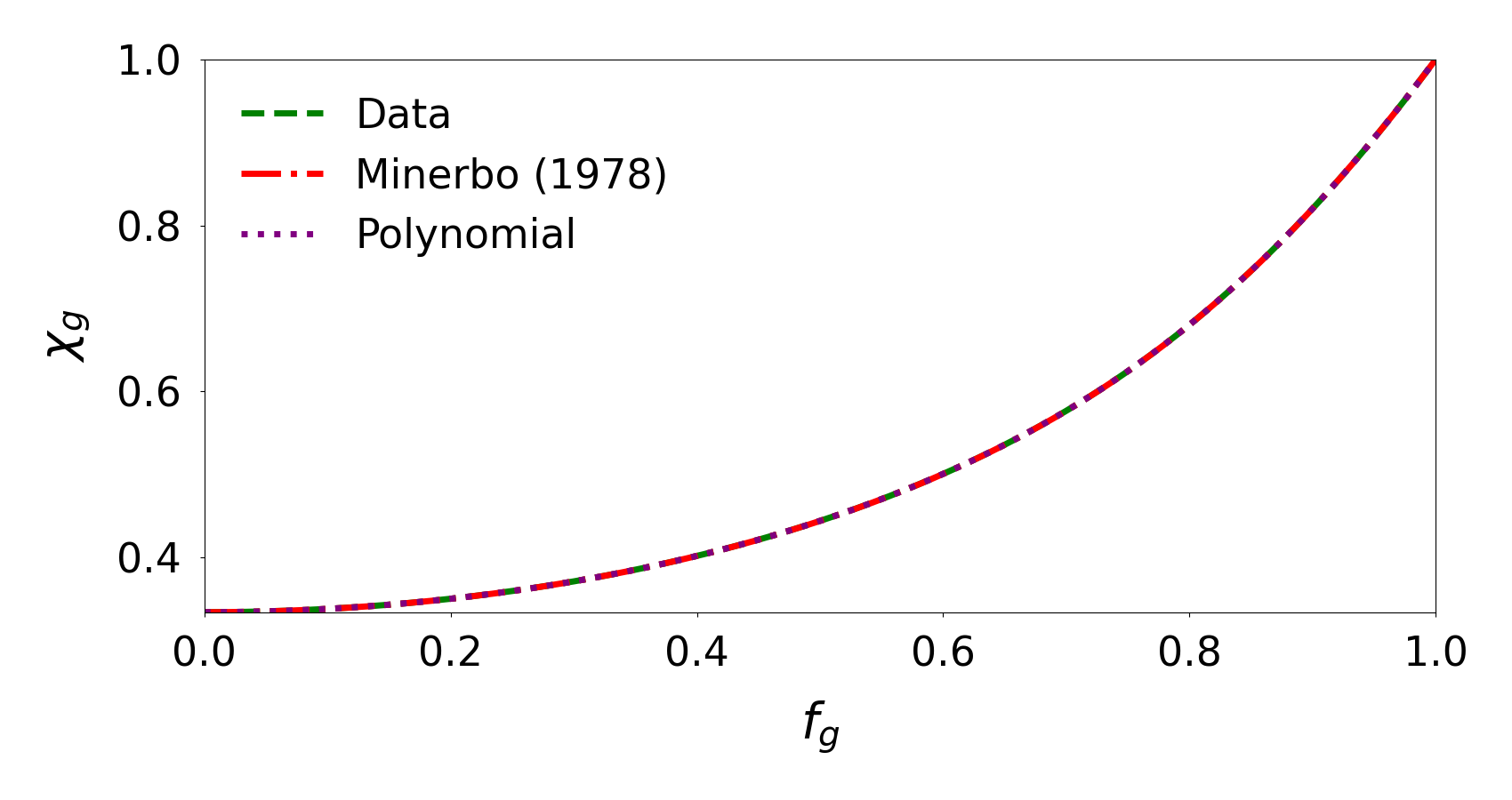}
        \end{minipage}
         \caption{Comparison of the computation of the Eddington factor between the polynomial expression $\chi_{g,\mathrm{L}}$, defined by equation~\eqref{eq:polynome_lowT}, and reference data obtained by solving equations~\eqref{eq:Tg_expr}, \eqref{eq:fg_expr}, and~\eqref{eq:chig_expr} using a line-search algorithm, for $\mathcal{T}_g = 10^{-8}$ and $\delta_g = 10^{-4}$. This comparison also includes results obtained by solving the equations proposed by Minerbo (1978) using a Newton algorithm.}
        \label{fig:comp_lowT_methods}
    \end{center}
\end{figure}

This polynomial expression was constructed so as to satisfy the following conditions: \mbox{$\chi_{g,\mathrm{L}}(0) = 1/3$}, \mbox{$\chi_{g,\mathrm{L}}(1) = 1$}, and \mbox{$\chi_{g,\mathrm{L}}'(0) = 0$}. The bounds of domain L were chosen in order to guarantee a relative error below $10^{-3}~\%$ in the estimation of $\chi_g$. The maximum error on the components $\mathbb{P}_{g,xx}$ and $\mathbb{P}_{g,yy}$ of the radiative pressure tensor is of the order of $10^{-1}~\%$, while that on the component $\mathbb{P}_{g,xy}$ can reach up to $1~\%$.

Figure~\ref{fig:comp_lowT_methods} compares the estimates obtained from this polynomial approximation (purple curve) with the values of $\chi_g$ computed using the line-search algorithm (green curve). An excellent agreement is observed between the two, confirming the validity of the approximation. It is also worth noting that the expression proposed by Minerbo (1978)~\cite{minerbo_1978} (red curve) provides values of the Eddington factor that are very close to those obtained numerically, with a maximum relative error of $10^{-3}~\%$ (see figure~\ref{fig:comp_lowT_methods}). This model therefore constitutes a relevant approximation in this specific case, although it cannot be exploited within the adopted approach, since it also requires a search algorithm.

\starsect{Intermediate domains I}

For all intermediate domains I, except for domain $\mathrm{I_b}$ (see figures~\ref{fig:Tg_dep_IA} and~\ref{fig:Tg_dep_other_IA}), I used different neural networks specialized for each domain:

\begin{itemize}
    \item \textbf{domain $\mathbf{I_a}$:} neural network $\mathrm{MLP_a}$ with two inputs: $\mathrm{f}_g$ and $\mathcal{T}_g^{\nu_g}$,
    \item \textbf{domain $\mathbf{I_c}$:} neural network $\mathrm{MLP_c}$ with two inputs: $\mathrm{f}_g$ and $\mathcal{T}_g^{\nu_{g+1}}$,
    \item \textbf{domain $\mathbf{I_d}$:} I used the neural network $\mathrm{MLP_a}$, which predicts accurate values of the Eddington factor in this domain,
    \item \textbf{domain $\mathbf{I_e}$:} I use two neural networks, $\mathrm{MLP_{e,1}}$ and $\mathrm{MLP_{e,2}}$, in order to reduce the required network size and, consequently, the prediction time. $\mathrm{MLP_{e,1}}$ is used for $\delta_g \in \openlinterv{0.03}{0.1}$ and $\mathrm{MLP_{e,2}}$ for $\delta_g \in \openlinterv{0.1}{0.9}$. These two neural networks have three inputs: $\mathrm{f}_g$, $\delta_g$, and $\mathcal{T}_g^{\nu_{g+1}}$,
    \item \textbf{domain $\mathbf{I_f}$:} neural network $\mathrm{MLP_f}$ with two inputs: $\mathrm{f}_g$ and $\mathcal{T}_g^s$ (defined by equation~\eqref{eq:Tg_s_def}).
\end{itemize}

\noindent The remainder of this section is devoted to detailing the answers I provided to the questions listed in Section~\secref{sec:intro_IA}, with the aim of obtaining reliable neural networks.

\substarsect{Which data should be used?}

\begin{table}[ht]
    \centering
    \begin{tabular}{C{1.2cm}L{2.1cm}C{3.5cm}C{3.5cm}C{3cm}}
        \hline
        \hline
        \multicolumn{2}{c}{} & Training dataset & Validation dataset & Range \TBstrut \\
        \hline
        \Tstrut \multirow{3}{*}{$\mathrm{MLP}_a$} & $\log_{10}(\mathcal{T}_g^{\nu_{g}})$ & $500$ & $1~000$ & $\closeinterv{-5}{1.1}$ \\
         & $\mathrm{f}_g$ & $100$ & $1~000$ & $\closeinterv{0}{1}$ \\
         & \# data & $49~695$ & $993~372$ & - \Bstrut \\
        \hdashline
        
        \TBstrut \multirow{3}{*}{$\mathrm{MLP}_c$} & $\log_{10}(\mathcal{T}_g^{\nu_{g+1}})$ & $300$ & $1~000$ & $\closeinterv{-1.5}{0.6}$ \\
        \TBstrut & $\mathrm{f}_g$ & $100$ & $1~000$ & $\closeinterv{0}{1}$ \\
        \TBstrut & \# data & $28~621$ & $951~821$ & - \Bstrut \\
        \hdashline
        
        \Tstrut \multirow{4}{*}{$\mathrm{MLP}_{e,1}$} & $\log_{10}(\mathcal{T}_g^{\nu_{g+1}})$ & $100$ & $150$ & $\closeinterv{-3}{0.6}$ \\
        \TBstrut & $\mathrm{f}_g$ & $100$ & $100$$~^{*}$ & $\closeinterv{0}{1}$ \\
        \TBstrut & $\delta_g$ & $60$ & $100$ & $\closeinterv{0.03}{0.1}$ \\
        \TBstrut & \# data & $590~160$ & $1~467~561$ & - \Bstrut \\
        \hdashline
        
        \Tstrut \multirow{4}{*}{$\mathrm{MLP}_{e,2}$} & $\log_{10}(\mathcal{T}_g^{\nu_{g+1}})$ & $100$ & $150$ & $\closeinterv{-3}{0.6}$ \\
        \TBstrut & $\mathrm{f}_g$ & $100$ & $100$$~^{*}$ & $\closeinterv{0}{1}$ \\
        \TBstrut & $\delta_g$ & $60$ & $100$ & $\closeinterv{0.1}{0.9}$ \\
        \TBstrut & \# data & $589~910$ & $1~465~462$ & - \Bstrut \\
        \hdashline
        
        \Tstrut \multirow{3}{*}{$\mathrm{MLP}_f$} & $\log_{10}(\mathcal{T}_g^s)$ & $300$ & $1~000$ & $\closeinterv{-3}{0.6}$ \\
        \TBstrut & $\mathrm{f}_g$ & $100$ & $1~000$ & $\closeinterv{0}{1}$ \\
        \TBstrut & \# data & $28~712$ & $955~143$ & - \Bstrut \\
        \hline
        \hline
        \multicolumn{5}{p{14cm}}{\hspace{-5pt}$^*$\footnotesize The interval $\mathrm{f}_g \in \closeinterv{10^{-5}}{0.99999}$ was used in the validation datasets of $\mathrm{MLP}_{e,1}$ and $\mathrm{MLP}_{e,2}$ to vary the values of $\mathrm{f}_g$.}\\
    \end{tabular}
    \caption{Description of the datasets used for training and validation of the neural networks.}
    \label{tab:datasets}
\end{table}

To train the neural networks, I generated training and validation datasets using the line-search algorithm that I developed (Appendix~\secref{appendice:algorithmes_de_recherche}). Table~\ref{tab:datasets} summarizes the parameters used to generate these datasets. A linear spacing was used for all quantities. I draw the reader's attention to the fact that the expected amount of data does not correspond to the amount of data actually obtained, since the line-search algorithm does not always converge (see Appendix~\secref{subappendice:comparaison_algorithmes_de_recherche}). Here, I used between 2 and 35 times more data in the validation set than in the training set in order to evaluate the performance of the neural networks at intermediate values and to assess their interpolation capabilities. Finally, to build a test set, a reference Marshak wave simulation using this method will be performed in Section~\secref{sec:performance_Eddington}, in order to evaluate both the accuracy of the method and the prediction error.

\substarsect{How should the model error be measured (cost function)?}

Given the variable impact of errors in the Eddington factor on the components of the radiative pressure tensor (see equation~\ref{eq:err_pressure}), I use a modified mean squared error loss function, giving priority to the regions where the Eddington factor has a stronger influence on the relative error of the radiative pressure components. To this end, I defined two loss functions:
\begin{align}
    \mathcal{L}_1(x_1, x_2, \chi_{g, data}, \chi_{g,pred}) &= \left ( \frac{\chi_{g,pred} - \chi_{g, data}}{\alpha_f~\alpha_T }\right )^2 \;\;\mathpunct{,}\\
    \mathcal{L}_2(x_1, x_2, x_3, \chi_{g, data}, \chi_{g,pred}) &= \left ( \frac{\chi_{g,pred} - \chi_{g, data}}{\alpha_f~\alpha_T~\alpha_\nu}\right )^2 \;\;\mathpunct{,}
\end{align}

\noindent where $\chi_{g,data}$ and $\chi_{g,pred}$ are, respectively, the Eddington factor known from the data and the one predicted by the neural network; $x_1$, $x_2$, and $x_3$ are the neural network inputs, whose expressions are given by equations~\eqref{eq:imput_MLPa}–\eqref{eq:imput_MLPf}; and $\alpha_f$, $\alpha_T$, and $\alpha_\nu$ are factors defined as:
\begin{align}
    \alpha_T &= \min(x_1, 1-x_1) + \epsilon_T \;\;\mathpunct{,} \\
    \alpha_f &= \min(1-x_2, \chi_{g, data}-1/3) + \epsilon_f \;\;\mathpunct{,} \\
    \alpha_\nu &= \min(x_3, 1-x_3) + \epsilon_\nu  \;\;\mathpunct{,}
\end{align}

\noindent where $\epsilon_T$, $\epsilon_f$, and $\epsilon_\nu$ are constants introduced to prevent these factors from vanishing. I set these constants to $0.1$, $10^{-3}$, and $0.1$, respectively. These factors serve two purposes:
\begin{itemize}
    \item $\alpha_T$ and $\alpha_\nu$ increase the weight of regions located at the interfaces between the domains defined in table~\ref{tab:domains} and illustrated in figure~\ref{fig:method_recap}. They therefore help mitigate the discontinuities induced by the use of multiple specialized neural networks;
    \item $\alpha_f$ increases the importance of regions where an error in the Eddington factor induces a significant relative error in the radiative pressure (see equation~\ref{eq:err_pressure}).
\end{itemize}

\noindent The loss function $\mathcal{L}_1$ is used to train the neural networks $\mathrm{MLP_a}$, $\mathrm{MLP_c}$, and $\mathrm{MLP_f}$, while $\mathcal{L}_2$ is used for $\mathrm{MLP_{e,1}}$ and $\mathrm{MLP_{e,2}}$.

\substarsect{Which optimization algorithm should be used?}

To train the neural networks, I used the Flux library in Julia, which is well known for its versatility in neural network modeling. The \gls{lbfgs} optimizer, provided by Julia's Optim library, is employed because of its fast convergence and its efficient minimization of the loss function. Although \gls{lbfgs} does not provide the stochastic properties that are often beneficial for promoting generalization, comparisons performed with the classical \gls{adam} optimizer show that networks trained with \gls{lbfgs} retain good generalization capabilities in the scenarios considered. I will discuss the generalization performance in more detail in the section \textit{Should a specific strategy be adopted to promote generalization?} The models are trained until the neural network parameters exhibit variations smaller than $10^{-32}$, that is, until the predicted values vary by less than $10^{-32}$, or until the gradient norm falls below $10^{-8}$.

\substarsect{Which neural network architecture should be used?}

\begin{figure}
    \begin{center}
        \begin{subfigure}[t]{0.7\textwidth}
            \centering
            \includegraphics[width=\textwidth]{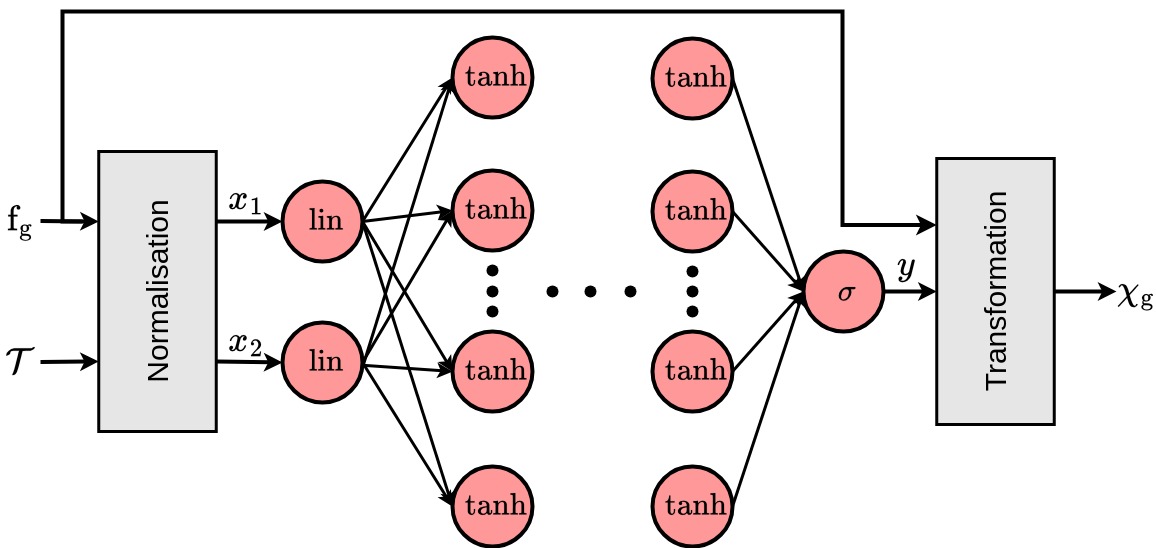}
            \caption{\glsxtrshort{mlp} architecture used for $\mathrm{MLP_a}$, $\mathrm{MLP_c}$, and $\mathrm{MLP_f}$.}
            \label{fig:MLP_2inputs}
        \end{subfigure}
        \hfill
        \begin{subfigure}[t]{0.7\textwidth}
            \centering
            \includegraphics[width=\textwidth]{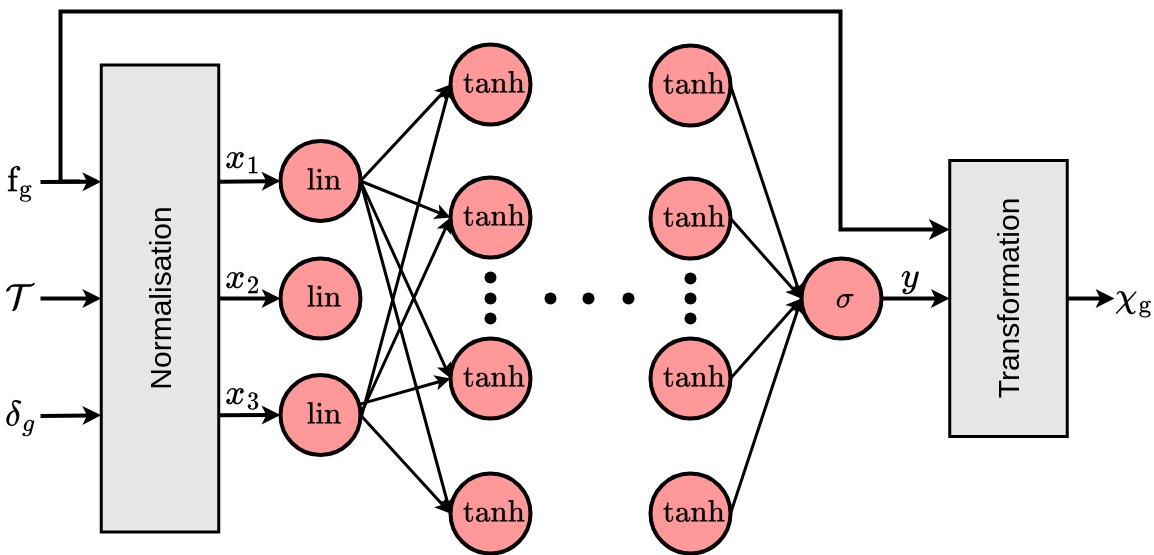}
            \caption{\glsxtrshort{mlp} architecture used for $\mathrm{MLP_{e,1}}$ and $\mathrm{MLP_{e,2}}$.}
            \label{fig:MLP_3inputs}
        \end{subfigure}
        \caption{\glsxtrshort{mlp} architectures used. $\mathrm{lin}$, $\mathrm{tanh}$, and $\sigma$ denote respectively the identity, hyperbolic tangent, and sigmoid activation functions. The \quotes{Normalization} and \quotes{Transformation} layers respectively apply equations~\eqref{eq:imput_MLPa} to~\eqref{eq:imput_MLPf}, and~\eqref{eq:sortie_MLP}. $\mathcal{T}$ is the dimensionless radiative temperature, defined differently depending on the MLP considered: $\mathcal{T}_g^{\nu_{g-1/2}}$ for $\mathrm{MLP_a}$, $\mathcal{T}_g^{\nu_{g+1/2}}$ for $\mathrm{MLP_c}$, $\mathrm{MLP_{e,1}}$, and $\mathrm{MLP_{e,2}}$, and $\mathcal{T}_g^s$ for $\mathrm{MLP_f}$.}
        \label{fig:MLP_archi}
    \end{center}
\end{figure}

All these neural networks do not take the quantities mentioned above directly as inputs, but rather normalized quantities, in order to help them avoid giving too much importance to some inputs relative to others.
\begin{itemize}
    \item Inputs for $\mathrm{MLP}_a$: 
    \begin{equation}
        \label{eq:imput_MLPa}
        \left\{
        \begin{array}{llll}
            x_1 &= \frac{\log_{10}\left (\mathcal{T}_g^{\nu_{g-1/2}}\right ) - \log_{10}\left (\mathcal{T}_{g,\min}^{\nu_g-1/2}\right )}{\log_{10}\left (\mathcal{T}_{g,\max}^{\nu_g-1/2}\right ) - \log_{10}\left (\mathcal{T}_{g,\min}^{\nu_g-1/2}\right )} \;\;\mathpunct{,}\\
            x_2 &= \mathrm{f}_g \;\;\mathpunct{.}
        \end{array}
        \right.
    \end{equation}

    \item Inputs for $\mathrm{MLP}_c$:
    \begin{equation}
        \label{eq:imput_MLPc}
        \left\{
        \begin{array}{llll}
            x_1 &= \frac{\log_{10}\left (\mathcal{T}_g^{\nu_{g+1/2}}\right ) - \log_{10}\left (\mathcal{T}_{g,\min}^{\nu_{g+1/2}}\right )}{\log_{10}\left (\mathcal{T}_{g,\max}^{\nu_{g+1/2}}\right ) - \log_{10}\left (\mathcal{T}_{g,\min}^{\nu_{g+1/2}}\right )} \;\;\mathpunct{,}\\
            x_2 &= \mathrm{f}_g \;\;\mathpunct{.}
        \end{array}
        \right.
    \end{equation}
    
    \item Inputs for $\mathrm{MLP}_{e,1}$ et $\mathrm{MLP}_{e,2}$:
    \begin{equation}
        \label{eq:imput_MLPe}
        \left\{
        \begin{array}{llll}
            x_1 &= \frac{\log_{10}\left (\mathcal{T}_g^{\nu_{g+1/2}}\right ) - \log_{10}\left (\mathcal{T}_{g,\min}^{\nu_{g+1/2}}\right )}{\log_{10}\left (\mathcal{T}_{g,\max}^{\nu_{g+1/2}}\right ) - \log_{10}\left (\mathcal{T}_{g,\min}^{\nu_{g+1/2}}\right )} \;\;\mathpunct{,}\\
            x_2 &= \mathrm{f}_g \;\;\mathpunct{,}\\
            x_3 &= \frac{\delta_g - \delta_{g,\min}}{\delta_{g,\max} - \delta_{g,\min}} \;\;\mathpunct{.}
        \end{array}
        \right.
    \end{equation}

    \item Inputs for $\mathrm{MLP}_f$:
    \begin{equation}
        \label{eq:imput_MLPf}
        \left\{
        \begin{array}{llll}
            x_1 &= \frac{\log_{10}\left (\mathcal{T}_g^{s}\right ) - \log_{10}\left (\mathcal{T}_{g,\min}^{s}\right )}{\log_{10}\left (\mathcal{T}_{g,\max}^{s}\right ) - \log_{10}\left (\mathcal{T}_{g,\min}^{s}\right )} \;\;\mathpunct{,}\\
            x_2 &= \mathrm{f}_g \;\;\mathpunct{.}
        \end{array}
        \right.
    \end{equation}
\end{itemize}

\noindent The quantities $X_{\min}$ and $X_{\max}$ represent the minimum and maximum values over the considered domain, and $x_i$ denotes an input of the neural networks.\\
\noindent Moreover, these neural networks do not directly predict the Eddington factor $\chi_g$, but rather a quantity related to it, according to the following expression:
\begin{equation}
    \chi_g = \frac{1}{3} + \frac{2}{3}~\mathrm{f}_g^2~\left ( 1 + \frac{(1 - \mathrm{f}_g) (3~y - 2)}{2} \right ) \;\;\mathpunct{.} \label{eq:sortie_MLP}
\end{equation}

\noindent Here, $y$ denotes the output of the neural network, which has been empirically verified to always lie within the interval $\closeinterv{0}{1}$ in the data. This construction ensures that the predicted Eddington factor satisfies the expected physical conditions: it tends toward $1/3$ when \mbox{$\mathrm{f}_g = 0$}, toward $1$ when \mbox{$\mathrm{f}_g = 1$}, and exhibits a zero slope at \mbox{$\mathrm{f}_g = 0$}, in accordance with its expected behavior (see figure~\ref{fig:fg_dep} and the definition recalled in Section~\secref{sec:Eddington_mg}). Consequently, the network output $y$ primarily affects the intermediate values of the Eddington factor, where its dependence on $\mathrm{f}_g$ is the strongest.

For all these neural networks, I considered various activation functions for the hidden and output layers and found that they did not have a significant impact. Therefore, I ultimately used hyperbolic tangent activation functions for the hidden layers, a linear activation function for the input layer, and a sigmoid activation function for the output layer (see the summary figure~\ref{fig:MLP_archi}). Only two hyperparameters remain to be tuned: the number of hidden layers and the number of neurons per hidden layer. These two parameters are selected via best-performing model selection (see table~\ref{tab:archi_MLP} for the final adopted meta-parameters).

\begin{figure}
    \begin{center}
        \begin{minipage}[t]{0.65\linewidth}
            \centering
            \includegraphics[width=\textwidth]{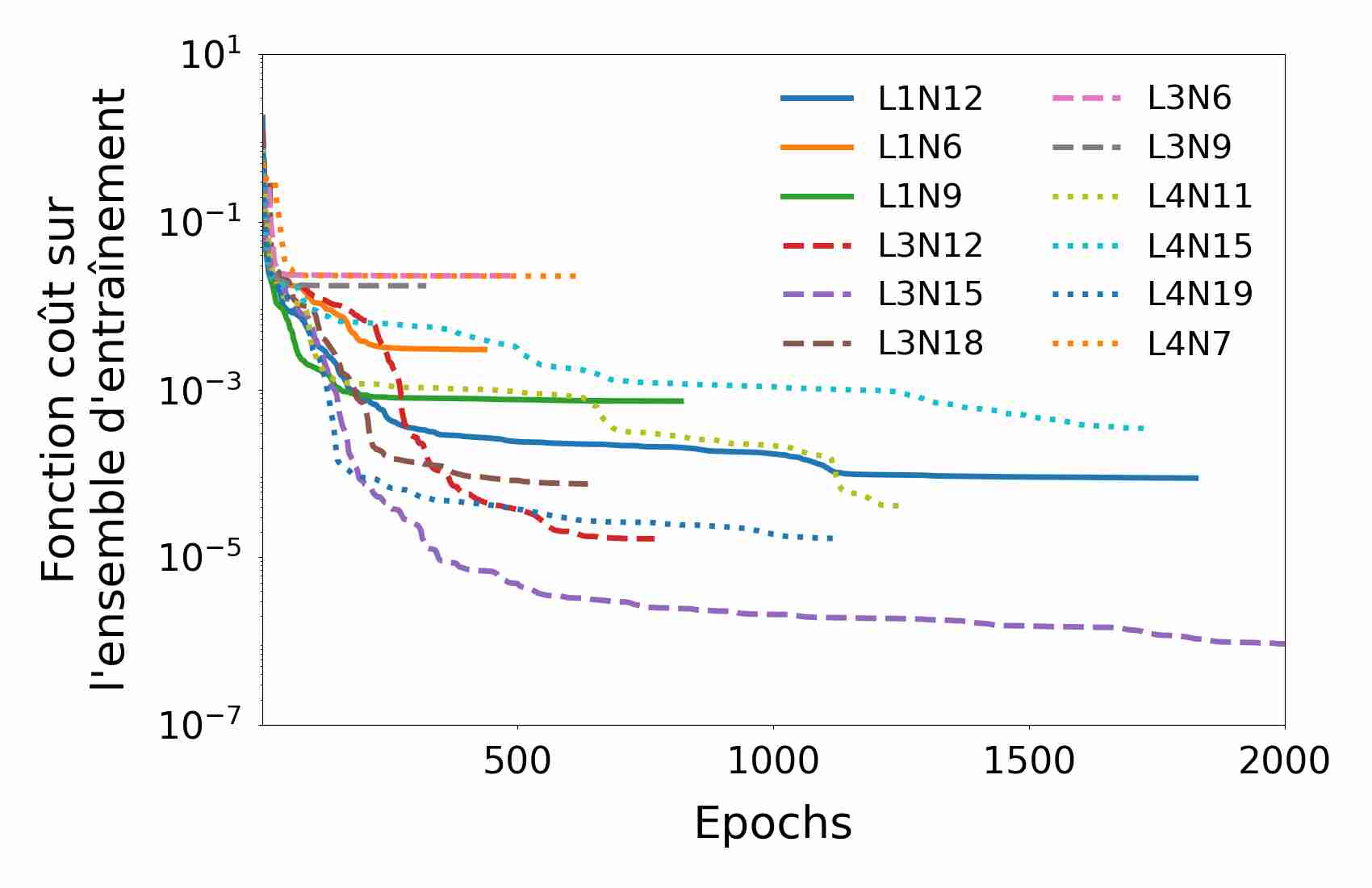}
            \caption{Learning curves of the different architectures tested for $\mathrm{MLP}_a$. $L$ denotes the number of hidden layers and $N$ the total number of perceptrons. For example, the architecture L1N6 consists of 1 hidden layer with 6 perceptrons.}
            \label{fig:LC}
        \end{minipage}
    \end{center}
\end{figure}

For each neural network $\mathrm{MLP}_a$, $\mathrm{MLP}_b$, etc., several architectures were generated by varying the number of hidden layers and the number of neurons per hidden layer, while keeping the models sufficiently simple in order to obtain fast predictions (less then 150 parameters to train). Training was performed on the training set specified in table~\ref{tab:datasets}. To evaluate the performance of each tested architecture, learning curves were plotted, showing the evolution of the cost function computed on the training set over the epochs. These curves provide two key pieces of information:
\begin{enumerate}
    \item They make it possible to verify that training has been fully completed (absence of underfitting), which is reflected by a plateau at the end of training;
    \item They allow identification of the architecture that reaches the minimum value of the cost function at the end of training.
\end{enumerate}

In the case of the neural network $\mathrm{MLP}_a$ (see figure~\ref{fig:LC}), all tested architectures exhibit complete training, and the one consisting of three hidden layers with 15 neurons each proves to be the most accurate, reaching the lowest value of the cost function at convergence. The same methodology was applied to select the architectures of the other neural networks. The final selected hyperparameters are summarized in table~\ref{tab:archi_MLP}.

\begin{table}[ht]
    \centering
    \begin{tabular}{L{3.5cm} C{3.5cm}C{3.5cm}C{3.5cm}}
        \hline
        \hline
        \TBstrut Neural network & Total \# of layers & Total \# of neurons & \# of parameters \\
        \hline
        \Tstrut $\mathrm{MLP}_a$ & $3$ & $15$ & $77$ \\
        $\mathrm{MLP}_c$          & $3$ & $15$ & $77$  \\
        $\mathrm{MLP}_{e,1}$      & $4$ & $24$ & $146$ \\
        $\mathrm{MLP}_{e,2}$      & $3$ & $19$ & $116$ \\
        $\mathrm{MLP}_f$  & $3$ & $15$ & $77$ \Bstrut \\
        \hline
        \hline
    \end{tabular}
    \caption{Architecture of the neural networks used.}
    \label{tab:archi_MLP}
\end{table}

\substarsect{Should a specific strategy be adopted to promote generalization?}

To evaluate the generalization capability of the selected neural networks, I performed a 10-fold cross-validation. This procedure consists in dividing the chosen dataset (here, the data from the training set, as described in table~\ref{tab:datasets}) into 10 subsets of equal size. At each iteration, one of these subsets is used as the validation set, while the remaining nine are used for training. This process is repeated 10 times, so that each subset is used once as the validation set. For each tested architecture, I computed the mean and the standard deviation of a representative performance score in order to assess the robustness of the generalization.

\begin{table}[ht]
    \centering
    \begin{tabular}{L{3.2cm}C{4.5cm}C{4.5cm}}
        \hline
        \hline
        Block number & Training set score & Validation set score \TBstrut \\
        \hline
        1  & $1.2 \times 10^{-4}$ & $2.7 \times 10^{-4}$ \\
        2  & $4.9 \times 10^{-5}$ & $4.8 \times 10^{-5}$ \\
        3  & $3.9 \times 10^{-3}$ & $1.1 \times 10^{-3}$ \\
        4  & $1.5 \times 10^{-4}$ & $1.9 \times 10^{-4}$ \\
        5  & $2.1 \times 10^{-4}$ & $4.7 \times 10^{-4}$ \\
        6  & $2.1 \times 10^{-4}$ & $5.6 \times 10^{-4}$ \\
        7  & $8.0 \times 10^{-5}$ & $5.3 \times 10^{-4}$ \\
        8  & $6.8 \times 10^{-4}$ & $1.0 \times 10^{-3}$ \\
        9  & $2.1 \times 10^{-2}$ & $1.1 \times 10^{-2}$ \\
        10 & $8.1 \times 10^{-5}$ & $2.4 \times 10^{-4}$ \\
        \hline
        Mean score & $2.7 \times 10^{-3}$ & $1.6 \times 10^{-3}$ \\
        Standard deviation & $6.7 \times 10^{-3}$ & $3.4 \times 10^{-3}$ \\
        \hline
        \hline
    \end{tabular}
    \caption{Results of the 10-fold cross-validation for $\mathrm{MLP}_a$.}
    \label{tab:kfolds}
\end{table}

The selected score is the mean absolute relative error on the estimation of the Eddington factor, defined as \mbox{$|\Delta \chi_g| / \chi_{g,\mathrm{true}}$}. The results of this cross-validation for the network $\mathrm{MLP}_a$ are presented in table~\ref{tab:kfolds}. On average, the network exhibits a good generalization capability, as evidenced by a mean score on the validation sets comparable to that observed on the training sets, and below $0.01~\%$, which is satisfactory. However, some folds (notably folds 7, 8, and 9) show significant discrepancies, indicating either localized overfitting or a difficulty of the model in adapting to certain specific sub-distributions of the data. In addition, the relatively large standard deviation observed in the validation scores reflects a non-negligible variability in performance across the different data splits, highlighting the sensitivity of the model to the distribution of the training set, but which generally leads to a neural network with good generalization properties. Similar observations were made for the other neural networks, with the exception of the network $\mathrm{MLP}_c$, for which the addition of L2 regularization to the loss function was necessary to ensure good generalization.

\begin{figure*}
    \begin{subfigure}[t]{0.49\textwidth}
        \centering
        \includegraphics[width=\textwidth]{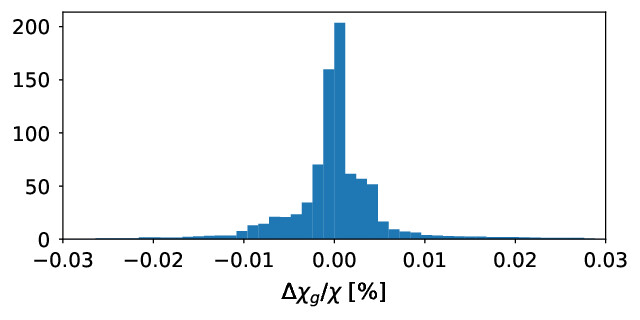}
        \caption{Training set.}
        \label{fig:training_err}
    \end{subfigure}
    \hfill
    \begin{subfigure}[t]{0.49\textwidth}
        \centering
        \includegraphics[width=\textwidth]{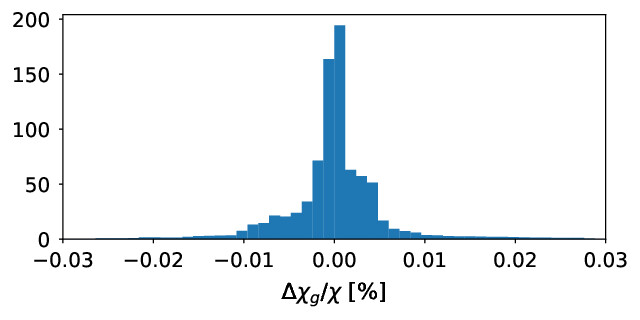}
        \caption{Validation set.}
        \label{fig:validation_err}
    \end{subfigure}
    \caption{Comparison of the distribution of the relative error of the $\mathrm{MLP}_a$ prediction for the training set and the validation set.}
    \label{fig:comp_distrib_err}
\end{figure*}

To more finely assess the interpolation performance of the final network, I analyzed its behavior on the training and validation sets described in table~\ref{tab:datasets}. The validation set, approximately twenty times larger than the training set, makes it possible to test the network on intermediate points that are absent from the training data; these points represent the majority of cases encountered in practice. I examined the distributions of the relative errors on both sets for each trained network in order to verify the absence of overfitting.

In the case of the network $\mathrm{MLP}_a$ (see figure~\ref{fig:comp_distrib_err}), the distributions are very similar and centered around 0, which indicates a good generalization capability of the final neural network. Similar results were observed for the other architectures. Finally, the maximum errors measured on the validation sets remain sufficiently small for the requirements of the simulations.

\subsection{Performance of the proposed method}  \label{sec:performance_Eddington} 

\begin{figure}
    \begin{center}
        \begin{minipage}[t]{0.7\linewidth}
            \centering
            \includegraphics[width=\textwidth]{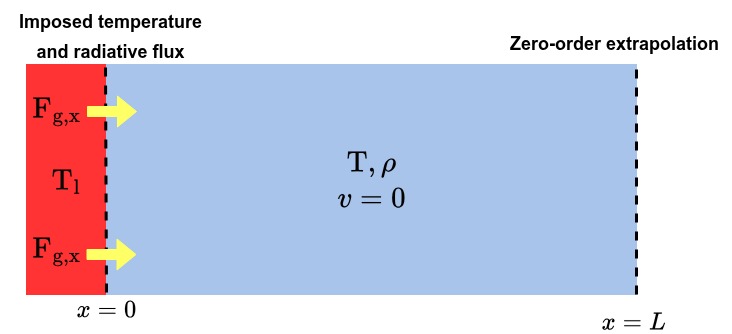}
            \caption{Initial and boundary conditions for the Marshak wave simulation.}
            \label{fig:marshak}
        \end{minipage}
    \end{center}
\end{figure}

In order to evaluate the performance of this new \gls{ai}-based method, I carried out a classical one-dimensional Marshak wave simulation using the \gls{hades} code, in which a radiative flux is emitted from a heated region (see figure~\ref{fig:marshak}). Since the code is strictly two-dimensional, the configuration was set up in plane-parallel geometry in order to reproduce an effectively one-dimensional case.

We then compare the simulation time as well as the error on the radiative pressure obtained using four different approaches for computing the Eddington factor:

\begin{table}[ht]
    \centering
    \begin{tabular}{L{2cm}L{2cm}C{2.5cm}C{2.5cm}C{2.5cm}}
        \hline
        \hline
        \multicolumn{2}{l}{Number of groups} & $\nu_{g-1/2}$ [eV] & $\nu_{g+1/2}$ [eV] & $\delta_g$ \TBstrut \\
        \hline
        \Tstrut 2 groups & - group 1 &  $0$                   & $1.0 \times 10^{-1}$ & - \\
                  & - group 2 &  $1.0 \times 10^{-1}$  & $+\infty$            & - \Bstrut \\
        \Tstrut 3 groups & - group 1 &  $0$                  & $1.0 \times 10^{-3}$ & - \\
                  & - group 2 &  $1.0 \times 10^{-3}$ & $1.0 \times 10^{-1}$ & $1.0 \times 10^{-2}$\\
                  & - group 3 &  $1.0 \times 10^{-1}$ & $+\infty$            & - \Bstrut \\
        \Tstrut 4 groups & - group 1 &  $0$                  & $1.0 \times 10^{-3}$ & - \\
                  & - group 2 &  $1.0 \times 10^{-3}$ & $2.0 \times 10^{-1}$ & $5.0 \times 10^{-2}$ \\
                  & - group 3 &  $2.0 \times 10^{-1}$ & $4.0 \times 10^{-1}$ & $5.0 \times 10^{-1}$ \\
                  & - group 4 &  $4.0 \times 10^{-1}$ & $+\infty$            & - \Bstrut \\
        \Tstrut 5 groups & - group 1 &  $0$                  & $1.0 \times 10^{-3}$ & - \\
                  & - group 2 &  $1.0 \times 10^{-3}$ & $1.0 \times 10^{0}$  & $1.0 \times 10^{-3}$\\
                  & - group 3 &  $1.0 \times 10^{0}$  & $1.1 \times 10^{0}$  & $9.1 \times 10^{-2}$\\
                  & - groupe 4 &  $1.1 \times 10^{0}$  & $1.0 \times 10^{1}$  & $1.1 \times 10^{-2}$\\
                  & - group 5 &  $1.0 \times 10^{1}$  & $+\infty$            & - \Bstrut \\
        \hline
        \hline
    \end{tabular}
    \caption{Definition of the groups}
    \label{tab:groupes}
\end{table}

\begin{enumerate}
    \item The line search algorithm described in Appendix~\secref{appendice:algorithmes_de_recherche};
    \item Interpolation of the Eddington factor from precomputed values~\cite{nguyen_2011_these};
    \item The analytical expression of the Eddington factor for the M1-gray model (equation~\ref{eq:facteur_Eddington_gray});
    \item The \gls{ai}-based method that I described in this section.
\end{enumerate}

In this simulation, we consider a domain of size \mbox{$L_x = 1$ m}, discretized into 1 000 cells in the x direction, and of size \mbox{$L_y = 3$ mm}, discretized into 3 cells in the y direction. The domain contains a fluid initially at rest, with a temperature of \mbox{$\mathrm{T}=300$ K} and a density of \mbox{$\rho=8.2 \times 10^{-9}$ kg/$\mathrm{m^3}$}. On its left boundary, the fluid is heated from an imposed temperature of \mbox{$\mathrm{T}_l=5~780$ K}, accompanied by an incoming radiative flux governed by the following expression:
\begin{equation}
    \forall g \in \llbracket 1, \mathcal{G} \rrbracket, ~\forall x \in \mathbb{R}_{-}^*,~\mathrm{F}_{g,x}(x)=c \left (\mathrm{E}_{g}(x)-\mathrm{E}_{g}(0) \right ) \;\;\mathpunct{,} \label{eq:flux_marshak}
\end{equation}

\noindent where $\mathcal{G}$ denotes the total number of groups. At the initial time, the radiation is in equilibrium with the gas. The latter is characterized by an atomic mass \mbox{$\mu m_H = 1$ u} and an adiabatic index \mbox{$\gamma = 5/3$}. The simulation covers a physical duration of \mbox{$13.3$ ns}, during which different group configurations, presented in table~\ref{tab:groupes}, are used. Figure~\ref{fig:marshak} summarizes the initial and boundary conditions.

\starsect{Computational time}

Table~\ref{tab:time} presents a comprehensive comparison of the simulation times for the different multigroup configurations described in Section~\secref{sec:methodes_Eddington}. The \gls{ai}-based method demonstrates remarkable efficiency, outperforming the line-search algorithm by a factor of 1 000 to 3 000 in terms of computational time. This approach is, however, slightly slower than the method using the analytical expression of the M1-gray model, which remains the fastest reference. Nevertheless, its computational cost remains comparable, exceeding that of the analytical M1-gray expression by only a factor of 1 to 3.

\begin{table}[ht]
    \centering
    \begin{tabular}{L{2cm}C{2.9cm}C{2.9cm}C{2.9cm}C{2.9cm}}
        \hline
        \hline
        \Tstrut Number of groups & M1-gray model expression & Interpolation & \gls{ai} method & Line-search algorithm \Bstrut \\
        \hline
        \TBstrut 2 groups & $3.11$~min & $3.70$~min & $4.18$~min  & $3.58$~jours \\
        \TBstrut 3 groups & $3.92$~min & $4.69$~min & $4.92$~min  & $7.05$~jours \\
        \TBstrut 4 groups & $4.55$~min & $5.65$~min & $10.37$~min & $22.6$~jours \\
        \TBstrut 5 groups & $5.53$~min & $6.88$~min & $7.72$~min  & $16.6$~jours \\
        \hline
        \hline
    \end{tabular}
    \caption{Comparison of simulation times (CPU time).}
    \label{tab:time}
\end{table}

It should be noted that the simulation times are higher for the 4-group configuration than for the others. This additional cost is mainly due to the use of the neural networks $\mathrm{MLP_{e,1}}$ and $\mathrm{MLP_{e,2}}$, whose more complex architecture (see table~\ref{tab:archi_MLP}) leads to a longer prediction time. As a result, the method I developed proves to be particularly efficient in terms of speed for a small number of groups, a situation in which the use of the neural networks $\mathrm{MLP_{e,1}}$ and $\mathrm{MLP_{e,2}}$ is not required.

\starsect{Precision of the method}

\begin{figure*}
    \begin{subfigure}[t]{0.49\textwidth}
        \centering
        \includegraphics[width=\textwidth]{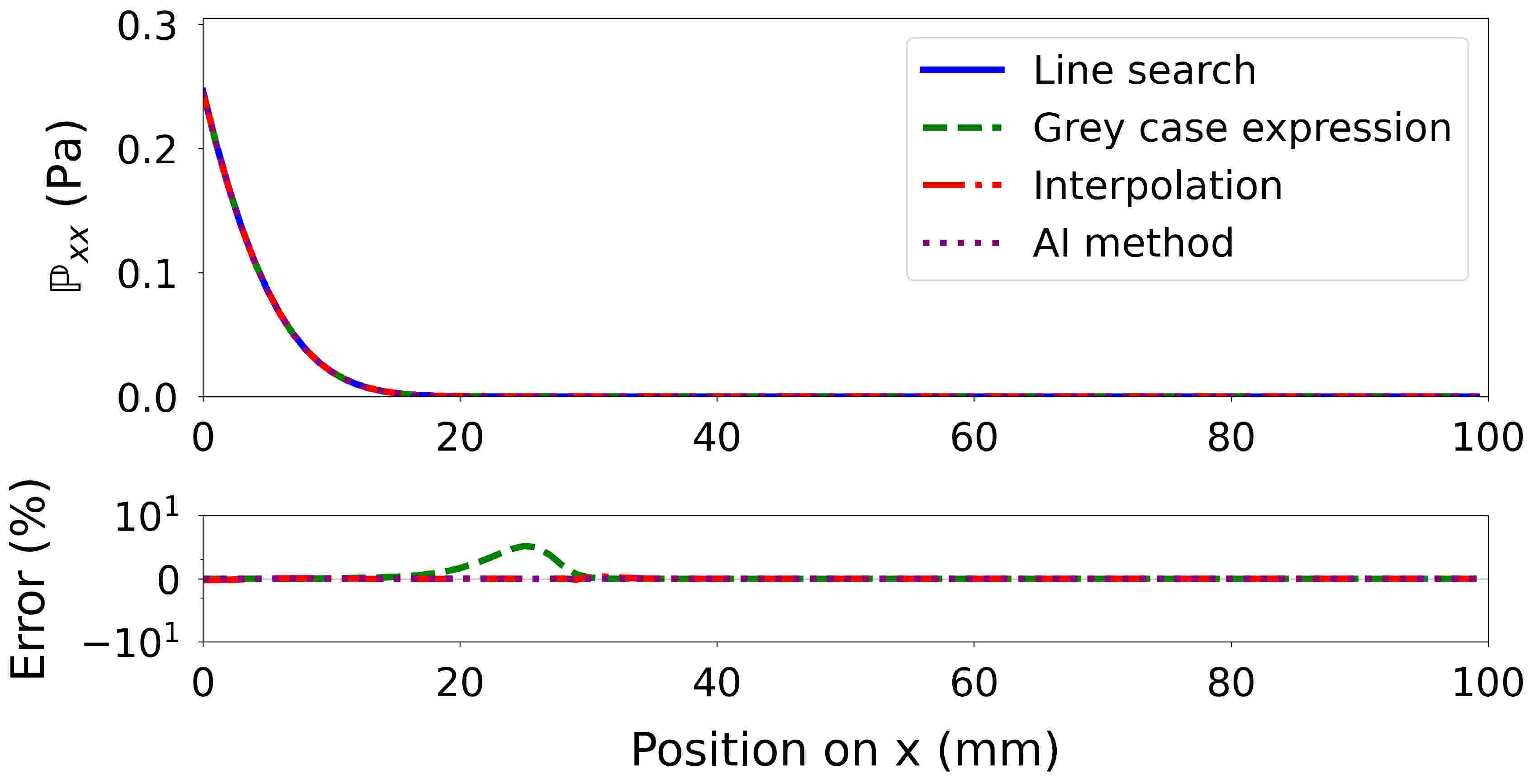}
        \caption{$2$ groups}
        \label{fig:2groups}
    \end{subfigure}
    \hfill
    \begin{subfigure}[t]{0.49\textwidth}
        \centering
        \includegraphics[width=\textwidth]{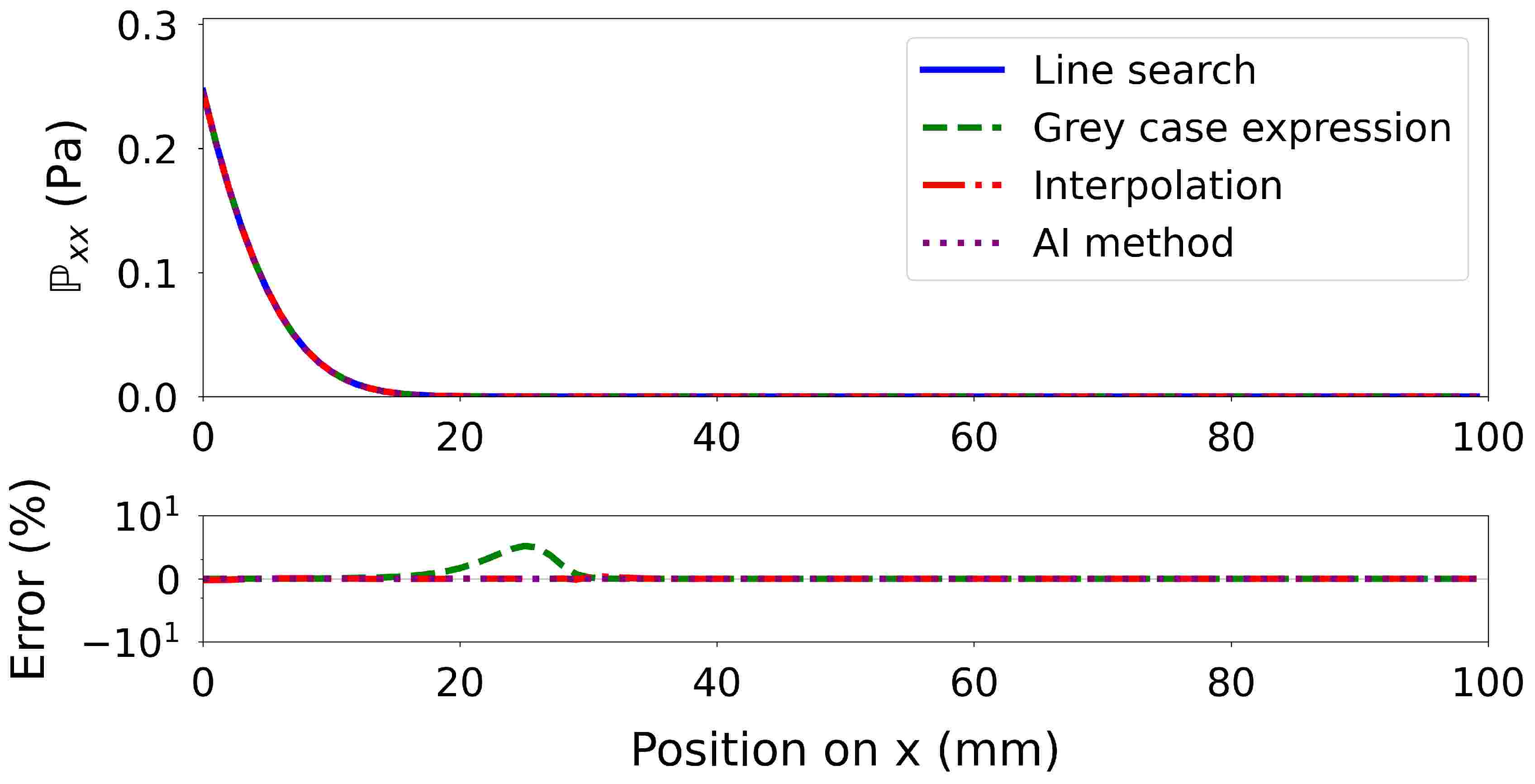}
        \caption{$3$ groups}
        \label{fig:3groups}
    \end{subfigure}
    \hfill
    \begin{subfigure}[t]{0.49\textwidth}
        \centering
        \includegraphics[width=\textwidth]{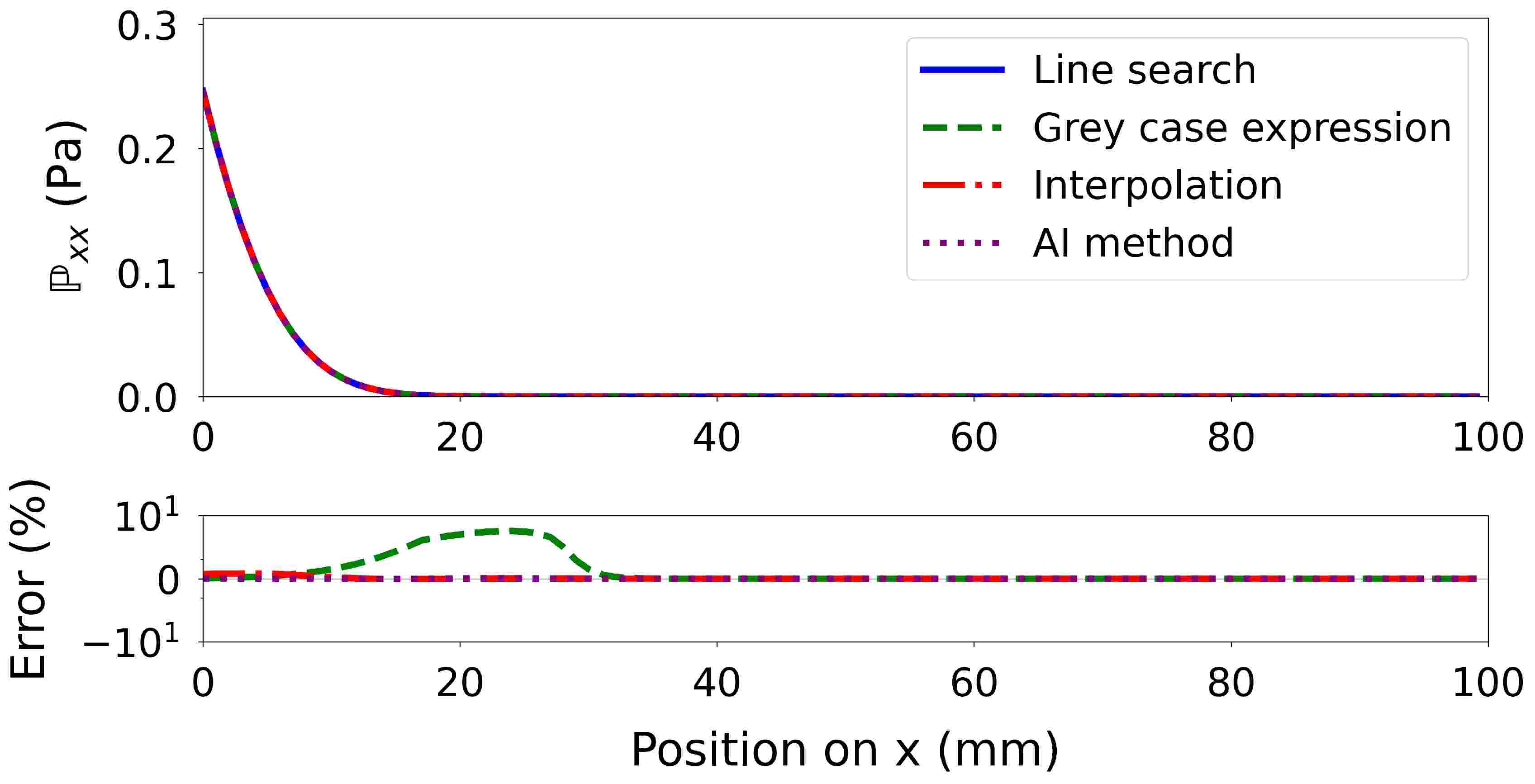}
        \caption{$4$ groups}
        \label{fig:4groups}
    \end{subfigure}
    \hfill
    \begin{subfigure}[t]{0.49\textwidth}
        \centering
        \includegraphics[width=\textwidth]{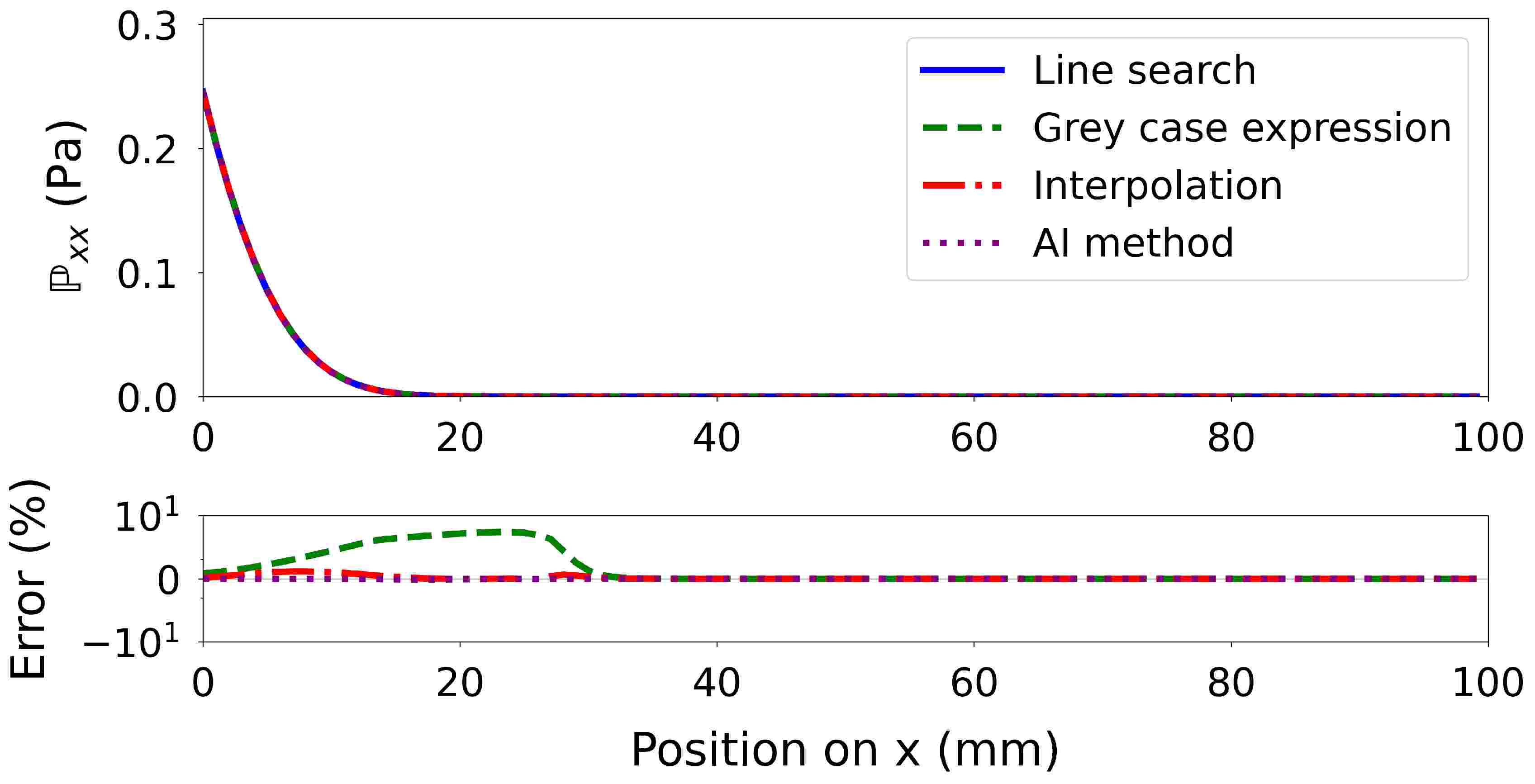}
        \caption{$5$ groups}
        \label{fig:5groups}
    \end{subfigure}
    \caption{Component $\mathbb{P}_{R,xx}$ of the total radiative pressure at time $t = 0.133~\text{ns}$. The relative error is given by the following formula: \mbox{($\mathbb{P}_{xx,\mathrm{method}} - \mathbb{P}_{xx,\mathrm{line~search}}$)/$\mathbb{P}_{xx,\mathrm{line~search}} \times 100$}.}
    \label{fig:valid_simu}
\end{figure*}

In figure~\ref{fig:valid_simu}, the profile of the total radiative pressure, defined as $\mathbb{P}_{R,xx} = \sum_g \mathbb{P}_{g,xx}$, is shown as a function of the spatial position $x$ at time \mbox{$t = 0.133~\text{ns}$}, as obtained using the four methods for computing the Eddington factor (top panels). The figure also displays the relative error on the radiative pressure, computed with respect to that obtained using the line-search algorithm, for simulations employing the analytical expression of the M1-gray model, interpolation, and the \gls{ai} method developed here (bottom panels).

The analysis of the curves shows that the \gls{ai}-based method I developed exhibits a relative error below $10^{-3}~\%$ on the Eddington factor. In comparison, the use of the M1-gray model expression can lead to errors of up to 9~\%, notably in the five-group simulation at the position $x=25~\text{mm}$, where the reduced flux reaches its maximum value. The error can reach 0.1~\% in the case of the Eddington factor interpolation method, in particular for the five-group configuration at the same location.

These results highlight the significant improvement in accuracy provided by this new approach. Moreover, one can observe that the error associated with the M1-gray model expression as well as with the interpolation method tends to increase with the number of groups, unlike the \gls{ai}-based method, whose accuracy remains stable.

\starsect{Conclusions on the method}

In conclusion, I have developed a neural-network-based approach capable of predicting the Eddington factor within the M1-multigroup model with a mean error between $10^{-3}$~\% and $10^{-2}$~\%. This method is approximately 3 000 times faster than the line search algorithm, while remaining only 1 to 3 times slower than using the Eddington factor derived from the M1-gray model. An important advantage of this approach is that it does not require any prior knowledge of the radiative energy or the reduced flux, unlike methods based on the interpolation of precomputed Eddington factors, making it one of the most efficient solutions currently available.

However, the simulations carried out so far, including those with a large number of groups or in two dimensions, have not revealed significant differences in the hydrodynamic quantities between the different closure methods. This can be explained by the fact that simulations in which radiative pressure dominates, outside radiative equilibrium, remain too computationally expensive with the \gls{hades} code. Nevertheless, this new method constitutes a reference for future work relying on the M1-multigroup model, as it enables an accurate and efficient estimation of the Eddington factor, thereby paving the way for simulations with an unprecedented level of accuracy. In the following chapter, I will apply this method to the study of radiative shocks in order to concretely assess its impact on the structure and dynamics of these phenomena.

\section{Synthesis}

In this chapter, I first briefly introduced neural networks, presenting their operating principles, the fundamentals of training, as well as the key steps involved in their design and selection. I then detailed the method I developed during my thesis, based on \gls{ai}, and aimed at estimating the Eddington factor of the M1-multigroup model in a general manner.

This approach achieves a mean prediction error between $10^{-3}$~\% and $10^{-2}$~\%, while being approximately 3 000 times faster than a line search algorithm, and only 1 to 3 times slower than using the Eddington factor from the M1-gray model. It also has the advantage of requiring no prior knowledge of the radiative energy or the reduced flux, unlike methods based on the interpolation of precomputed data, which makes it one of the most efficient solutions currently available. However, the simulations conducted so far have not revealed any notable differences in the hydrodynamic quantities, regardless of the number of groups or the dimensionality of the simulation (1D or 2D). This lack of discrepancy can be explained by the current inability to explore configurations out of radiative equilibrium dominated by radiative pressure, whose simulation remains particularly computationally expensive with the \gls{hades} code. Despite this, the proposed method constitutes a solid reference for future work relying on the M1-multigroup formalism, by providing an accurate and fast estimation of the Eddington factor.
\clearemptydoublepage

\chapter{Study of radiative shocks} \label{ch:chapitre4}

\initialletter{R}adiative shocks constitute a particular category of shock waves in a fluid, in which radiation plays a central role in the dynamics and structure of the shock. Unlike purely hydrodynamic shocks, where energy and momentum are essentially conserved within the fluid, radiative shocks convert a significant portion of them into radiation. Radiative energy transport profoundly influences their behavior: the energy deposited in the shocked region can excite the rotational–vibrational states of molecules~\cite{godard_2019}, and may even ionize atoms in extreme cases. This intense radiation emission in turn leads to additional heating and ionization of the surrounding medium, establishing a complex coupling between fluid dynamics, radiative transfer, and thermodynamic processes. These interactions significantly modify the structure of the shock, affecting the temperature, pressure, and density profiles~\cite{mihalas_1999, zeldovitch_2002, drake_2006}. Radiative shocks are ubiquitous in many astrophysical environments, making their study crucial for understanding a wide range of phenomena. They play a key role in accretion and ejection processes, particularly during gravitational collapse or supernovae, where high fluid velocities generate shocks accompanied by strong radiation. This radiation is often difficult to observe directly, which makes numerical studies essential for their interpretation. Such shocks occur in contexts as diverse as supernova remnants~\cite{michaut_2012, miniere_2018}, stellar atmospheres~\cite{fadeyev_2002, chiavassa_2011, kravchenko_2018, kervella_2009}, star formation through accretion processes~\cite{stahler_1980, commercon_2011, busschaert_2013}, astrophysical jets~\cite{raga_1999, hajime_2022, loupias_2009}, stellar mass-loss mechanisms~\cite{gillet_2014, gillet_2014b}, and symbiotic stars~\cite{imamura_1985, falize_2009}.

Their study relies simultaneously on analytical, experimental, and numerical approaches. Analytical models have been proposed to describe characteristic structures of the shock, such as the cooling layer or the radiative precursor~\cite{bouquet_2000, michaut_2004, drake_2007, lowrie_2008}. However, these models rest on simplifying assumptions intended to handle the complexity of the underlying equations, and therefore cannot capture the full impact and interplay of all the physics involved. Advances in high-energy laser facilities have made it possible to reproduce radiative shocks experimentally in the laboratory~\cite{bozier_1986, keiter_2002, bouquet_2004, koenig_2006}. Nevertheless, these experiments remain limited due to the presence of instabilities, observed experimentally~\cite{drake_1999} and interpreted theoretically as axial~\cite{chevalier_1982} or lateral instabilities~\cite{vishniac_1983}. In this context, numerical simulations play a central role by providing predictive tools that are invaluable for interpreting observations and exploring fundamental physical mechanisms.

Thanks to the improvements made in the \gls{hades} code concerning the calculation of the Eddington factor, I undertook a more detailed analysis of the influence of an accurate treatment of radiation on the structure and dynamics of radiative shocks. In this chapter, I will begin by presenting the current state of knowledge on radiative shocks, before discussing the simulations I have carried out in order to assess the impact of accurate radiation modeling on the evolution and internal structure of radiative shocks.

\section{Introduction to radiative shocks}  \label{sec:intro_choc_rad}

The behavior of radiative shocks depends strongly on the interaction between radiation and gas, an interaction that varies with the opacity of the medium. In optically thick media, this interaction is dominant: radiation efficiently exchanges energy and momentum with matter, thereby deeply modifying the shock dynamics. Conversely, in optically thin media, its influence becomes negligible and the shock behaves in an essentially hydrodynamic manner. Five parameters can be used to characterize one-dimensional radiative shocks~\cite{michaut_2009}:
\begin{enumerate} 
    \item \textbf{Mach number $\mathbf{M}$:}\\
    The Mach number characterizes the ratio between the velocity of a fluid and the speed of sound in that same medium. In the context of shock waves, it quantifies the intensity of the shock by comparing the relative velocity between the fluid and the shock front with the propagation speed of acoustic waves. It is written:
    \begin{equation}
        \label{eq:Mach_nb}
        \mathrm{M}=\frac{|v_{up} - v_{shock}|}{c_S} \;\;\mathpunct{,}
    \end{equation}
    where $v_{up}$ is the fluid velocity upstream of the shock, $v_{shock}$ is the shock-front velocity, and $c_s$ is the local upstream sound speed. In an ideal gas, one has \mbox{$c_S = \sqrt{\gamma~p_{up} / \rho_{up}}$}. The Mach number determines the strength of the shock: the larger $\mathrm{M}$ is, the stronger the shock. By definition, a shock can occur only if \mbox{$\mathrm{M}>1$}.
    \item \textbf{Cooling parameter $\boldsymbol{\chi}$:}\\
    This parameter compares the ability of the medium to lose its thermal energy through radiation to the dynamical evolution of the fluid. It is defined as the ratio between the radiative cooling time $t_{\text{cool}}$ and the characteristic dynamical time $t_{\text{dyn}}$ of the downstream fluid. In an optically thin medium:
    \begin{equation}
        \label{eq:cooling_param_mince}
        \chi = \frac{t_{cool}}{t_{\text{dyn}}} = \frac{1}{\gamma-1} \frac{k_B}{\mu m_H} \frac{\rho_{down}~v_{down}~\mathrm{T}_{down}}{\Lambda(\rho_{down},\mathrm{T}_{down})~L} \;\;\mathpunct{,}
    \end{equation}
    and in an optically thick medium:
    \begin{equation}
         \label{eq:cooling_param_epais}
        \chi = \frac{1}{\gamma-1} \frac{k_B}{\mu m_H} \frac{\rho_{down}~v_{down}~\mathrm{T}_{down}}{F_{R,down}} \;\;\mathpunct{,}
    \end{equation}
    where $\rho_{down}$, $v_{down}$, $\mathrm{T}_{down}$, and $F_{R,down}$ denote respectively the density, velocity, temperature, and radiative flux of the downstream fluid in the shock frame, and \mbox{$L=v_{down}~t_{dyn}$} is a characteristic length. The function \mbox{$\Lambda(\rho,\mathrm{T})$} represents the cooling function used in optically thin modeling (see equation~\ref{eq:hydrorad_cooling}). The constants $\gamma$, $k_B$, and $\mu m_H$ correspond respectively to the adiabatic index, Boltzmann's constant, and the mean molecular mass of the fluid. The smaller this parameter is, the more important the effects of radiative cooling are downstream.
    \item \textbf{Optical depth $\boldsymbol{\tau}$ (upstream/downstream):}\\
    This quantity measures how transparent a medium is to radiation. It is defined as the integral of a mean opacity over the entire spectrum, $\kappa_{ref}$, along a path $s$:
    \begin{equation}
        \label{eq:epaisseur optique}
        \tau = \int \nolimits_{s_0}^{s} \kappa_{ref}(s') \dif s' \;\;\mathpunct{.}
    \end{equation}
   One may consider either the Planck mean opacity $\kappa_P$ or the Rosseland mean opacity $\kappa_R$, depending on the radiative regime. For a homogeneous medium over a characteristic length $L$, this expression simplifies to:
    \begin{equation}
        \label{eq:epaisseur optique2}
        \tau = \kappa_{ref} L \;\;\mathpunct{.}
    \end{equation}
    A medium with \mbox{$\tau \ll 1$} is called optically thin, whereas a medium with \mbox{$\tau \gg 1$} is optically thick.

    \item \textbf{Mihalas number $\mathbf{R}$ (upstream/downstream):}\\
    The Mihalas number compares the internal energy of the gas $u$ to the radiation energy $\mathrm{E}_R$. It is expressed as:
    \begin{equation}
        \label{eq:mihalas_param}
        \mathrm{R} = \frac{u}{\mathrm{E}_R} = \frac{1}{\gamma-1} \frac{k_B}{a_R~\mu m_H} \frac{\rho }{\mathrm{T}^3~g(\tau)} \;\;\mathpunct{,}
    \end{equation}
    where $\rho$, $v$, and $\mathrm{T}$ are the hydrodynamic quantities upstream or downstream of the shock, $a_R$ is the radiation constant, and $g(\tau)$ is the ratio between the actual radiation energy $\mathrm{E}_R$ and that of a blackbody \mbox{$a_R T^4$}. One has \mbox{$g(\tau) \to 1$} when the medium is optically thick, and $g(\tau) \sim \tau$ in the optically thin regime. The smaller $\mathrm{R}$ is, the more significant the radiative effects are.

    \textbf{Boltzmann number $\mathbf{Bo}$ (upstream/downstream):}\\
    The Boltzmann number is the ratio between the internal energy flux transported by the fluid and the energy flux transported by radiation. It is given by:
    \begin{equation}
        \label{eq:bolzmann_param}
        \mathrm{Bo} = \frac{(u + p)v}{F_R} = \frac{\gamma}{\gamma-1} \frac{k_B}{\sigma~\mu m_H} \frac{\rho~v}{\mathrm{T}^3~f(\tau)} \;\;\mathpunct{,}
    \end{equation}
    where $\sigma$ is the Stefan–Boltzmann constant and $f(\tau)$ corresponds to the ratio between the actual radiative flux $F_R$ and the blackbody flux \mbox{$\sigma T^4$}. One has \mbox{$f(\tau) \to 1$} when the medium is optically thick, and \mbox{$f(\tau) \sim \tau$} in the optically thin regime. The smaller this parameter is, the more important radiative energy transport becomes.
\end{enumerate}

\begin{figure}
    \begin{subfigure}[t]{0.28\textheight}
        \centering
        \includegraphics[width=\textwidth]{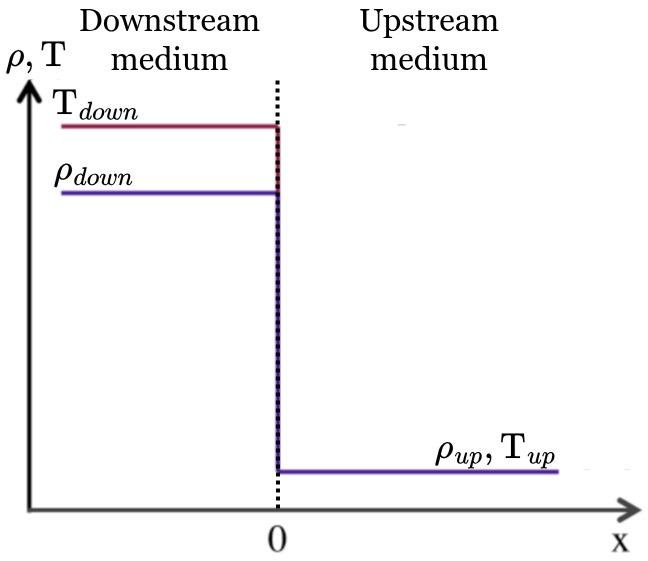}
        \caption{Hydrodynamic shock.}
        \label{fig:hydro_shock}
    \end{subfigure}
    \hfill
    \begin{subfigure}[t]{0.33\textheight}
        \centering
        \includegraphics[width=\textwidth]{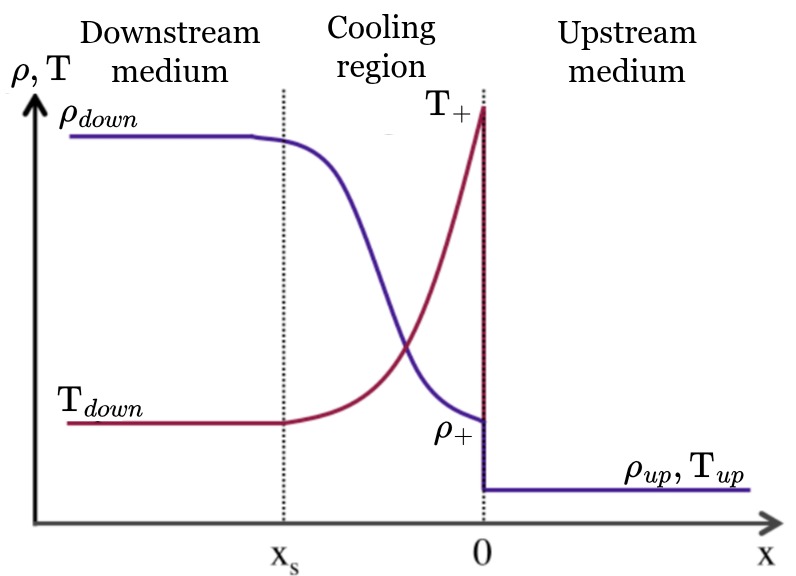}
        \caption{Optically thin shock.}
        \label{fig:cooling_shock}
    \end{subfigure}
    \hfill
    \begin{subfigure}[t]{0.33\textheight}
        \centering
        \includegraphics[width=\textwidth]{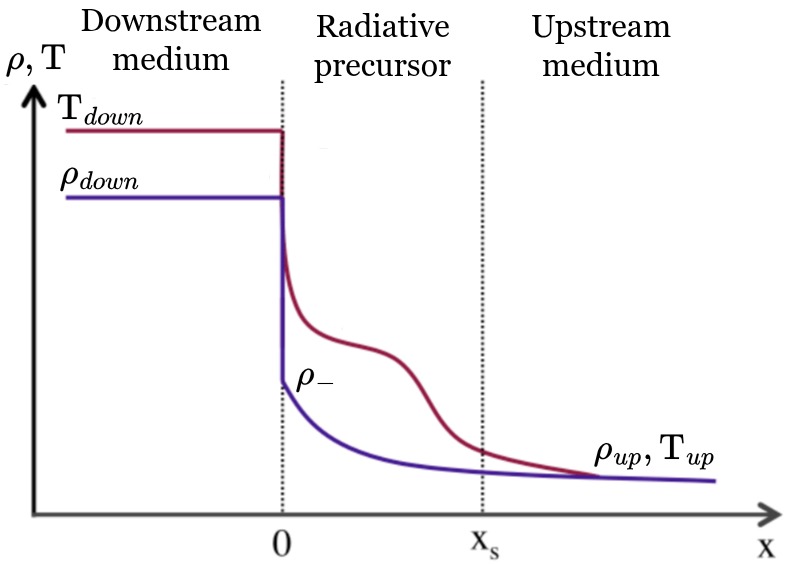}
        \caption{Optically thick shock, for Mach numbers \mbox{$\mathrm{M}<2.4 \, \mathrm{M}_{rad}$}.}
        \label{fig:thick_shock}
    \end{subfigure}
    \hfill
    \begin{subfigure}[t]{0.33\textheight}
        \centering
        \includegraphics[width=\textwidth]{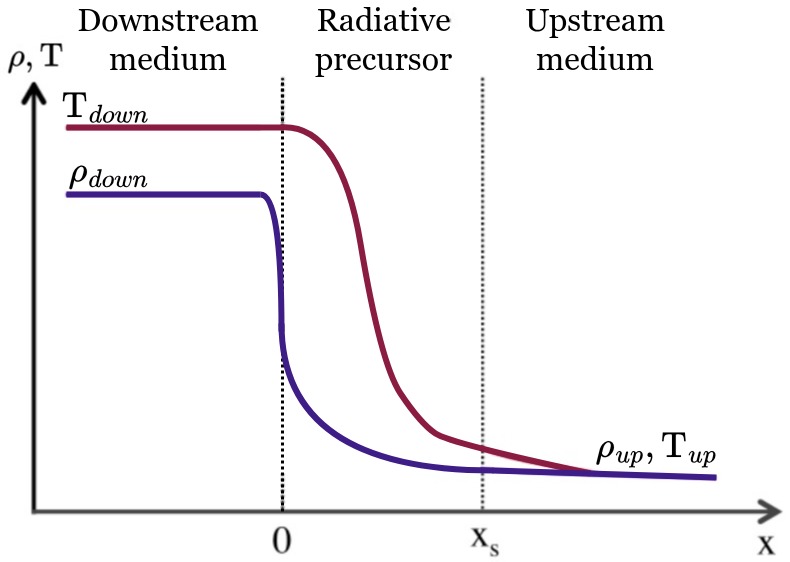}
        \caption{Optically thick shock, for Mach numbers \mbox{$\mathrm{M}>2.4 \, \mathrm{M}_{rad}$}.}
        \label{fig:thick_shock_Mrad}
    \end{subfigure}
    \caption{Different types of radiative shock structures. The red line represents the temperature, and the blue line represents the density. The position \mbox{$x=0$} corresponds to the location of the shock.}
    \label{fig:shock_struct}
\end{figure}

\noindent Depending on the values taken by these five parameters, the following classes of one-dimensional radiative shocks can be distinguished~\cite{bouquet_2000, drake_2007, michaut_2009}:
\begin{enumerate}
    \item \textbf{Purely hydrodynamic shock:} Radiation is negligible (\mbox{$\chi \to +\infty$}, \mbox{$\mathrm{Bo} \to +\infty$}, \mbox{$\mathrm{R} \to +\infty$}, \mbox{$\tau = 0$}). The shock structure reduces to a discontinuity separating the upstream and downstream regions, and the Euler equations are sufficient to describe the situation. The temperature and density profiles in this case therefore display a discontinuity and are shown in figure~\ref{fig:hydro_shock};

    \item \textbf{Optically thin shock:} The medium is optically thin, which allows photons emitted downstream of the shock front to escape freely. This radiative emission then acts as an efficient mechanism for dissipating energy. As a result, a cooling region forms immediately downstream of the shock. This regime is typically characterized by \mbox{$\chi \ll 1$}, \mbox{$\mathrm{Bo} \gg 1$}, and \mbox{$\mathrm{R} \gg 1$}. The associated density and temperature profiles are illustrated in figure~\ref{fig:cooling_shock}. In this case, the loss of energy through photon emission leads to cooling of the fluid, manifested by a temperature peak followed by a rapid drop in the cooling region.

    \item \textbf{Hybrid shock:} This type of shock occurs when the medium is optically thick on one side and optically thin on the other. In the optically thick region, radiation interacts strongly with the fluid (\mbox{$\chi \ll 1$}, \mbox{$\mathrm{Bo} \ll 1$}), whereas it escapes freely in the optically thin region, leading to significant losses of energy and momentum. The density and temperature profiles are then similar to those observed in the optically thin case (see figure~\ref{fig:cooling_shock}), but the downstream temperature is much higher than the upstream one (\mbox{$\mathrm{T}_{down} \gg \mathrm{T}_{up}$});

    \item \textbf{Optically thick radiative shock dominated by radiative flux:} The medium is optically thick on both sides of the shock, and energy transfer is mainly carried by the radiative flux (\mbox{$\mathrm{Bo} \ll 1$}, \mbox{$\mathrm{R} \gg 1$}). A \textit{radiative precursor} forms upstream, resulting from fluid–radiation interactions, without any significant contribution from radiation energy or pressure. This type of shock may occur at relatively low Mach numbers, typically when \mbox{$\mathrm{M} < \mathrm{M}_{rad} \approx 0.25~(\mu m_H)^{-1/6}~\gamma^{-1/2}~\mathrm{T}_{amont}^{-1/2}~\rho_{amont}^{1/6}$}. The corresponding temperature and density profiles are shown in figure~\ref{fig:thick_shock}. In this regime, photons interact strongly with matter and heat it upstream of the shock, creating a \quotes{bulge} in the temperature profile, visible in the figure, corresponding to the radiative precursor;

    \item \textbf{Optically thick radiative shock dominated by radiation:} This regime is similar to the previous one, but here it is the radiation energy and pressure that dominate the shock dynamics (\mbox{$\mathrm{Bo} \ll 1$}, \mbox{$\mathrm{R} \ll 1$}). This type of shock appears for Mach numbers greater than \mbox{$\mathrm{M}_{rad}$}. When \mbox{$\mathrm{M}_{rad} < \mathrm{M} < 2.4~\mathrm{M}_{rad}$}, the density profile exhibits a discontinuity, and the density and temperature profiles remain close to those observed in the flux-dominated case (see figure~\ref{fig:thick_shock}). Conversely, for \mbox{$\mathrm{M} > 2.4~\mathrm{M}_{rad}$}, the density profile becomes continuous. The density and temperature profiles are then modified: as illustrated in figure~\ref{fig:thick_shock_Mrad}, the density varies continuously, while the temperature still shows a \quotes{bulge} upstream of the shock, a signature of the radiative precursor.
\end{enumerate}

In the case of optically thick radiative shocks, a radiative precursor develops upstream of the shock front as a result of the intense radiation emitted by the compressed material. According to Paul Drake's book \textit{\quotes{High-Energy-Density Physics: Fundamentals, Inertial Fusion and Experimental Astrophysics}}~\cite{drake_2006}, this precursor exhibits a structure composed of two physically distinct regions: the diffusive precursor and the transmissive precursor. The diffusive precursor corresponds to the zone in which radiation interacts strongly with matter, leading to a state close to radiative equilibrium. Radiative energy transport there occurs mainly by diffusion, a mechanism favored by the high opacity of the medium, and the gas temperature increases progressively due to radiation absorption. Conversely, the transmissive precursor corresponds to a region in which the material is not in radiative equilibrium. Radiation propagates more rapidly than the ability of the matter to adjust thermally, which leads to thermal decoupling: the radiation temperature is already high, while the gas temperature remains close to the upstream value. This asymmetry results in a ratio \mbox{$\mathrm{T}_R/\mathrm{T}$} significantly different from unity, indicating a nonequilibrium regime. Depending on the structure of the precursor and the intensity of the shock, two types of radiative shocks are distinguished: \textit{subcritical} shocks and \textit{supercritical} shocks.

 \begin{figure}
        \begin{subfigure}[t]{0.54\textwidth}
            \centering
            \includegraphics[width=\textwidth]{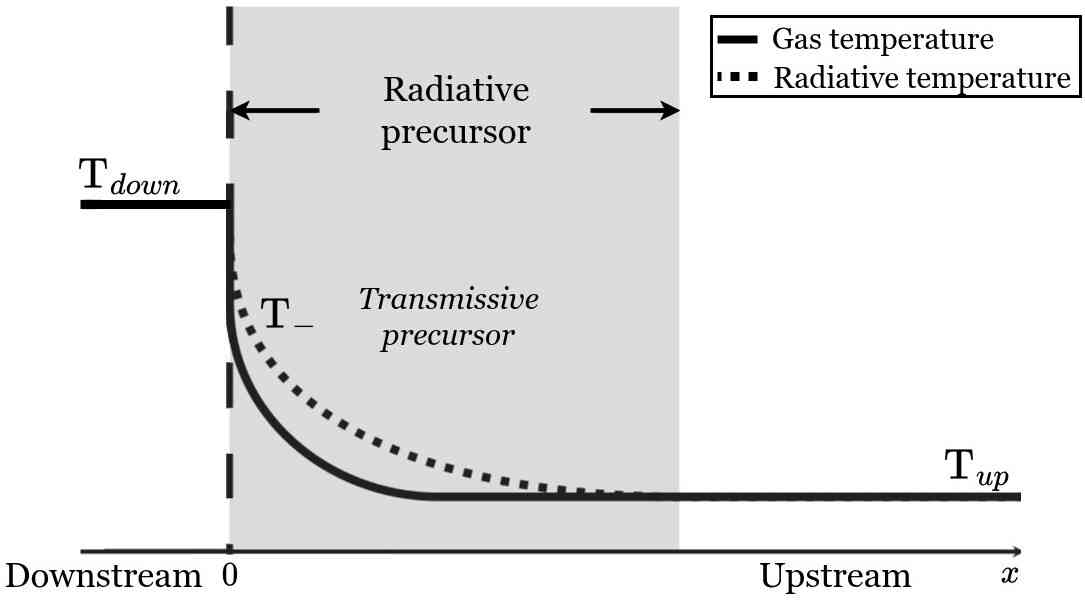}
            \caption{Precursor of a subcritical shock.}
            \label{fig:subcritical}
        \end{subfigure}
    \hfill
        \begin{subfigure}[t]{0.4\textheight}
            \centering
            \includegraphics[width=\textwidth]{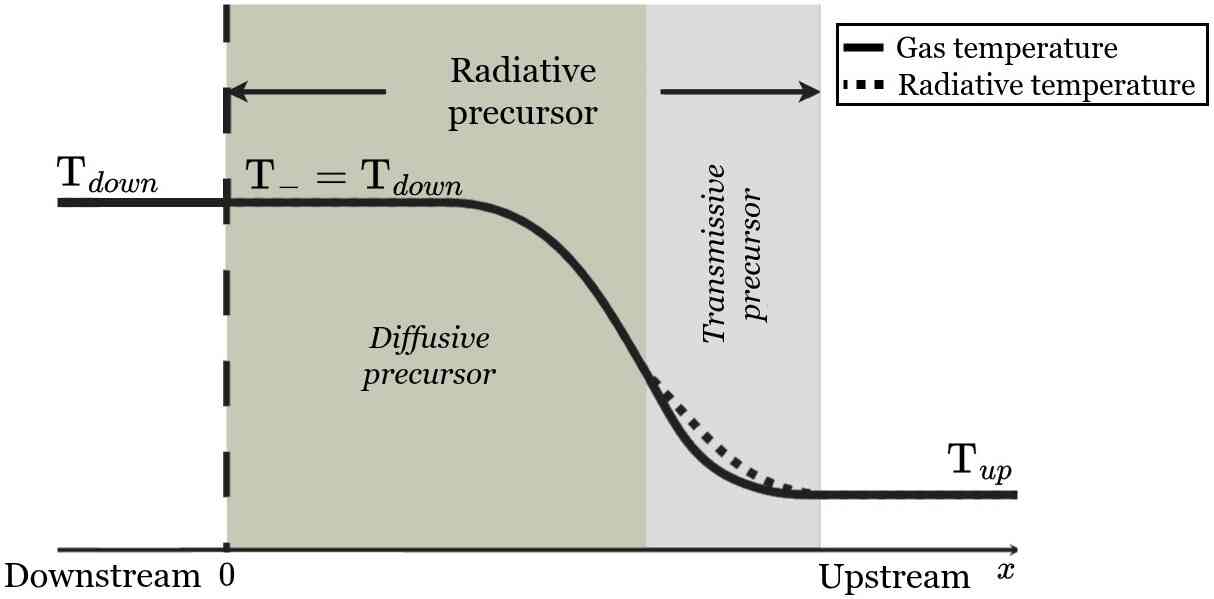}
            \caption{Precursor of a supercritical shock.}
            \label{fig:supercritical}
        \end{subfigure}
    \caption{Types of radiative precursor structures in the case of optically thick shocks. The position \mbox{$x=0$} corresponds to the location of the shock.}
    \label{fig:precursor_types}
\end{figure}

\begin{itemize}
    \item In a subcritical shock, the temperature $\mathrm{T}_-$ just upstream of the front is lower than the downstream temperature $\mathrm{T}_{down}$. The precursor is then essentially transmissive, and no region is in radiative equilibrium (see figure~\ref{fig:subcritical}). This type of shock occurs for Mach numbers below a critical value, denoted $\mathrm{M}_{crit}$. To date, no reliable analytical expression allowing one to compute this critical value $\mathrm{M}_{crit}$ has been established in the literature.
    
    \item In a supercritical shock, the temperature $\mathrm{T}_-$ is equal to $\mathrm{T}_{down}$, which indicates radiative heating strong enough to establish local radiative equilibrium. The radiative precursor then consists of a diffusive precursor close to the shock front, followed by a more distant transmissive precursor (see figure~\ref{fig:supercritical}). These shocks occur for Mach numbers greater than $\mathrm{M}_{crit}$.
\end{itemize}

\section{Influence of the spectral nature of light}  \label{sec:simus_spectral}

Despite existing classifications, radiative shocks remain imperfectly understood, largely because of the complexity of the physical interactions they involve, particularly when it comes to realistically modeling the couplings between radiation and matter. To date, no model is able to accurately predict the size and structure of the different regions of the radiative precursor, especially when two- or three-dimensional configurations are considered. Only approximations valid in one-dimensional cases exist for estimating the jump relations, and the evolution of the shock propagation speed is still poorly understood.

Furthermore, previous studies have highlighted the decisive influence of spectral effects on the shape of the precursor and on the cooling region~\cite{vaytet_2012}. However, such studies most often fall within the framework of a stationary shock, in which the simulation is initialized with a discontinuity whose upstream and downstream conditions are determined from approximate models. This approach therefore introduces, from the outset, prior assumptions about the jump relations, which may bias the analysis.

In addition, this type of configuration does not allow one to examine the actual dynamics of a propagating shock, particularly the effects induced by light–matter interactions, such as radiative braking. These approaches also frequently rely on numerical approximations intended to reduce computation time, which can impair the fidelity of opacity modeling and the accuracy of the closure relation within the M1-multigroup model.

In this context, I devoted part of my PhD work to studying in greater detail the impact of accurate spectral modeling of radiation on radiative shocks, by computing the closure relation of the M1-multigroup model with a level of precision never previously achieved. To do so, I used the method I developed (see Section~\ref{sec:Eddington_mg}), without introducing any simplifications on the opacities, and relying on a simulation capable of capturing both the dynamics and the structure of the shock.

This section summarizes the main results of this study, published in the article \quotes{\textit{Impact of frequency-dependent radiation on the dynamics and structure of radiative shocks}}~\cite{radureau_2025b}.

\subsection{Simulation description} \label{sec:description_simu}

\begin{figure}
    \centering
    \includegraphics[width=0.6\textwidth]{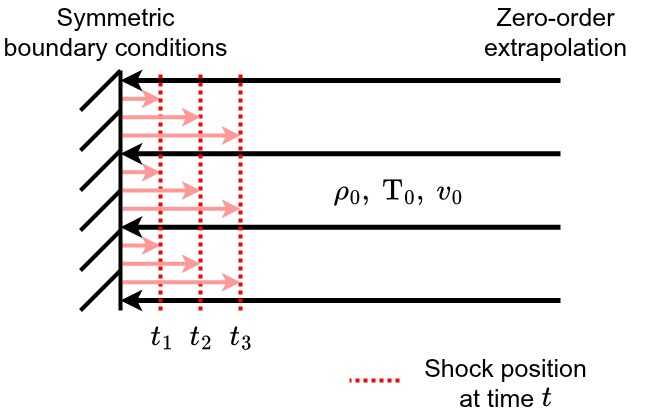}
    \caption{Schematic illustration of the simulation of shock formation at the interface with a rigid wall. The simulation starts with homogeneous hydrodynamic conditions: density $\rho_0$, temperature $\mathrm{T}_0$, and velocity $v_0$. At later times $t_1<t_2<t_3$, a shock front forms and gradually propagates away from the wall.}
    \label{fig:wall_shock}
\end{figure}

We performed simulations with the \gls{hades} code of the evolution of a one-dimensional radiative shock generated by a fluid moving toward a rigid wall. Figure~\ref{fig:wall_shock} illustrates this configuration. The initial conditions are homogeneous for the hydrodynamic quantities. A symmetric boundary condition is imposed on the left boundary in order to represent the presence of the wall (zero velocity and zero radiative flux at the wall), while a zero-order extrapolation condition is applied at the right boundary (no imposed condition). At later times, the interaction between the incoming fluid and the wall generates a shock front that progressively propagates away from the wall. This type of configuration has been widely used to study optically thin radiative shocks~\cite{chevalier_1982, sutherland_2010}, and constitutes a reference case for validating radiative-hydrodynamics codes~\cite{ensman_1994}.

Two shock configurations were considered: \quotes{Mach~4} and \quotes{Mach~8}. The initial hydrodynamic parameters are given in table~\ref{tab:init}. For the \quotes{Mach~4} simulation, a 10~cm domain was discretized into 2~000~cells, whereas the \quotes{Mach~8} simulation was carried out on a 20~cm domain with the same number of cells.

Because \gls{hades} is a two-dimensional code, a second direction must be defined. Its size was set to 0.15~mm for the \quotes{Mach~4} configuration and to 0.3~mm for the \quotes{Mach~8} configuration, and it was discretized on 3~cells in both cases. Periodic boundary conditions were imposed in this direction, thus allowing a plane-parallel simulation. The calculations were continued until the shock reached a steady state.

\begin{table}[ht]
    \centering
    \begin{tabular}{L{1.8cm} C{2.9cm} C{2.9cm} C{2.9cm} C{2.9cm}}
        \hline
        \hline
        \TBstrut Case & Density [kg/$\mathrm{m^3}$] & Velocity [m/s] & Temperature [K] & Temperature [eV] \\
        \hline
        \Tstrut \quotes{Mach~4} & 1 &  -5~622.82 & 11~604.52 & 1 \\
        \Bstrut \quotes{Mach~8} & 1 & -11~824.97 & 11~604.52 & 1 \\
        \hline
        \hline
    \end{tabular}
    \caption{Initial hydrodynamic quantities.}
    \label{tab:init}
\end{table}
 
The medium considered in this study consists of pure argon (\mbox{$\mu m_H$ = 39.948~u}). The effects of ionization on the mean atomic mass and on the adiabatic index were neglected. Although this simplification tends to overestimate the downstream temperature, it remains justified in the context of the objective of this study, which is to analyze the impact of the spectral behavior of radiation on the physics of radiative shocks.

In all simulations based on this model, a constant group narrowness $\delta_g$\footnote{The group narrowness $\delta_g$ is defined in Section~\secref{sec:dependence_chig}.} was adopted for each spectral group. This choice makes it possible to quantitatively evaluate the influence of using narrower groups on the structure of the radiative shock, which would not be possible with groups of different sizes. Only the first and last groups have different narrownesses: the first is smaller, because the TOPS opacity tables did not contain data below 1.47~eV, while the last is wider in order to capture residual radiation at high frequencies, even though this radiation does not play a predominant role in the phenomena studied. Additional details regarding the group partitioning are presented in table~\ref{tab:freqs}. The following radiation models were studied:
\begin{enumerate}
    \item M1-gray model;
    \item M1-multigroup model with a constant group narrowness of $\delta_g = 0.13$ (4 groups);
    \item M1-multigroup model with a constant group narrowness of $\delta_g = 0.50$ (9 groups);
    \item M1-multigroup model with a constant group narrowness of $\delta_g = 0.68$ (16 groups).
\end{enumerate}

\noindent For the \quotes{Mach~8} simulation, we tested an additional frequency spacing:
\begin{enumerate}\addtocounter{enumi}{4}
    \item M1-multigroup model with a constant group narrowness of $\delta_g = 0.78$ (\mbox{23 groups}).
\end{enumerate}

\begin{table}[ht]
    \centering
    \begin{tabular}{L{2cm} C{2cm} l c L{6cm}}
        \hline
        \hline
        \TBstrut Simulation & \# groups & \multicolumn{3}{c}{Group frequency bounds} \\
        \hline
        \Tstrut $\delta_g = 0.13$ & 4 & Group 1 &:& $\nu \in \closeinterv{1.47~\mathrm{eV}}{7.94~\mathrm{eV}}$ \\
        & & Groups 2-3 &:& Group narrowness $\delta_g=0.13$\\
        & & Group 4 &:& $\nu \in \closeinterv{500~\mathrm{eV}}{16~\mathrm{keV}}$ \Bstrut \\
        \Tstrut $\delta_g = 0.50$ & 9 & Group 1 &:& $\nu \in \closeinterv{1.47 \mathrm{eV}}{3.98 \mathrm{eV}}$\\
        & & Groups 2-8 &:& Group narrowness $\delta_g=0.50$\\
        & & Group 9 &:& $\nu \in \closeinterv{500~\mathrm{eV}}{16~\mathrm{keV}}$ \Bstrut \\
        \Tstrut $\delta_g = 0.68$& 16 & Group 1 &:& $\nu \in \closeinterv{1.47 \mathrm{eV}}{2.17 \mathrm{eV}}$\\
        & & Groups 2-15 &:& Group narrowness $\delta_g=0.68$\\
        & & Group 16 &:& $\nu \in \closeinterv{500~\mathrm{eV}}{16~\mathrm{keV}}$ \Bstrut \\
        \Tstrut $\delta_g = 0.78$ & 23 & Group 1 &:& $\nu \in \closeinterv{1.47~\mathrm{eV}}{7.94~\mathrm{eV}}$\\
        & & Groups 2-22 &:& Group narrowness $\delta_g=0.78$\\
        & & Group 23 &:& $\nu \in \closeinterv{300~\mathrm{eV}}{16~\mathrm{keV}}$ \Bstrut \\
        \hline
        \hline
    \end{tabular}
    \caption{Frequency bounds of each group used in the M1-multigroup simulations.}
    \label{tab:freqs}
\end{table}

This last simulation enabled us to further investigate the impact of narrower groups on the dynamics of the \quotes{Mach~8} shock. We used the TOPS opacity tables\footnote{\url{https://aphysics2.lanl.gov/apps/}} for argon, provided by Los Alamos National Laboratory. These tables supply the Rosseland and Planck mean opacities for the M1-gray and M1-multigroup models as functions of gas density and temperature, under the \gls{lte} assumption.

\subsection{Results}

\begin{figure}
    \begin{subfigure}[t]{0.45\textwidth}
        \centering
        \includegraphics[width=\textwidth]{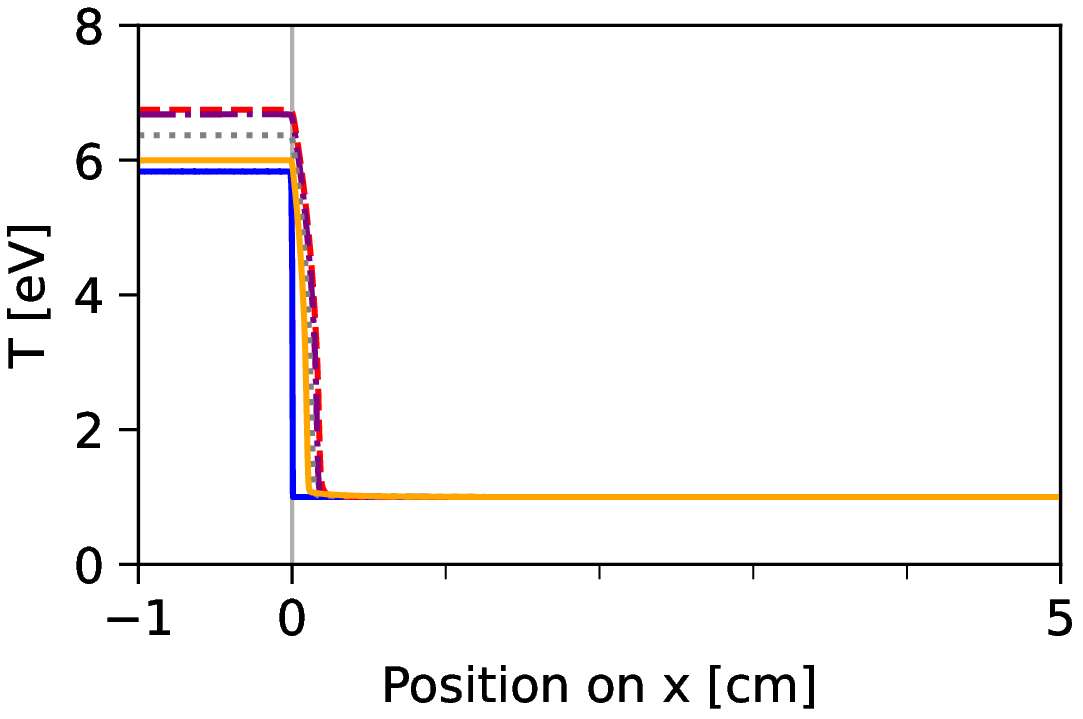}
        \caption{Gas temperature.}
        \label{fig:T_M4}
    \end{subfigure}
    \hfill
    \begin{subfigure}[t]{0.45\textwidth}
        \centering
        \includegraphics[width=\textwidth]{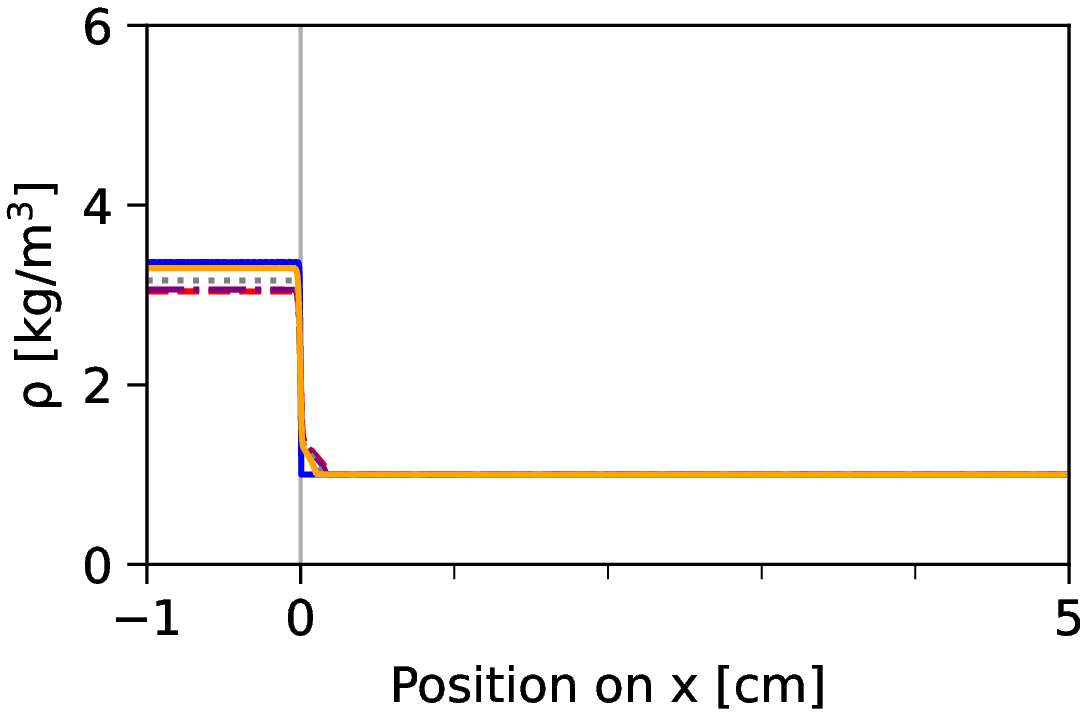}
        \caption{Density.}
        \label{fig:rho_M4}
    \end{subfigure}
    \hfill
    \begin{subfigure}[t]{0.45\textwidth}
        \centering
        \includegraphics[width=\textwidth]{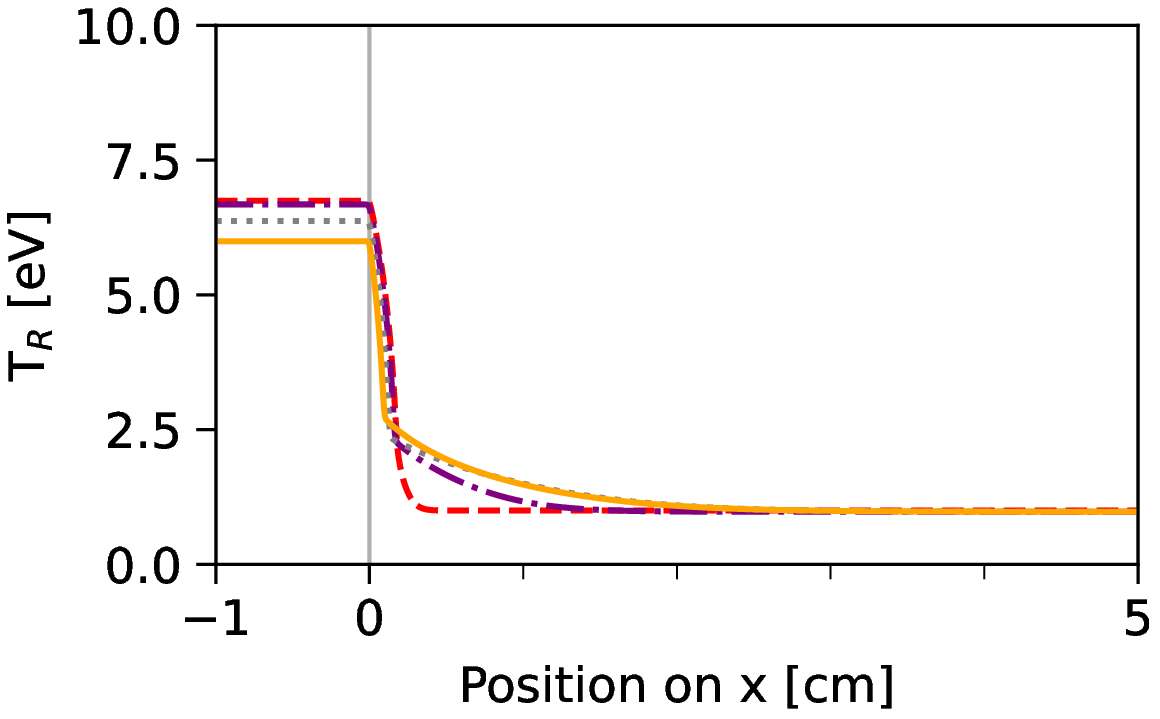}
        \caption{Radiative temperature.}
        \label{fig:Tr_M4}
    \end{subfigure}
    \hfill
    \begin{subfigure}[t]{0.45\textwidth}
        \centering
        \includegraphics[width=\textwidth]{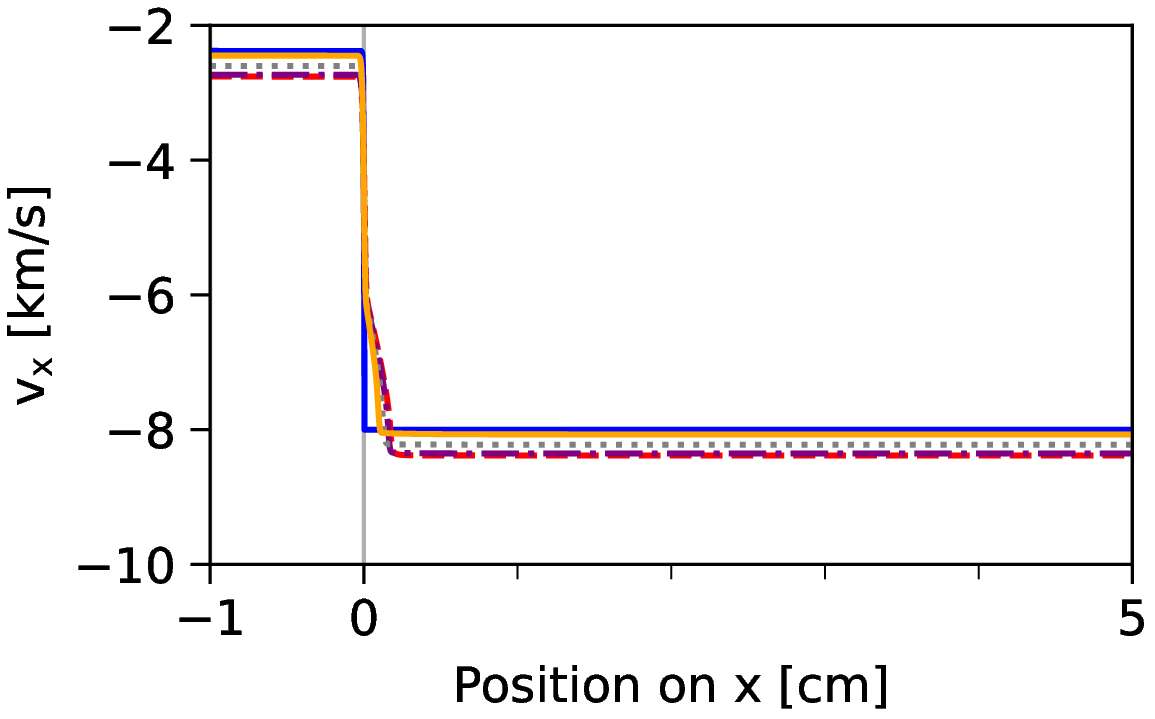}
        \caption{Velocity.}
        \label{fig:vx_M4}
    \end{subfigure}
    \hfill
    \begin{subfigure}[t]{0.45\textwidth}
        \centering
        \includegraphics[width=\textwidth]{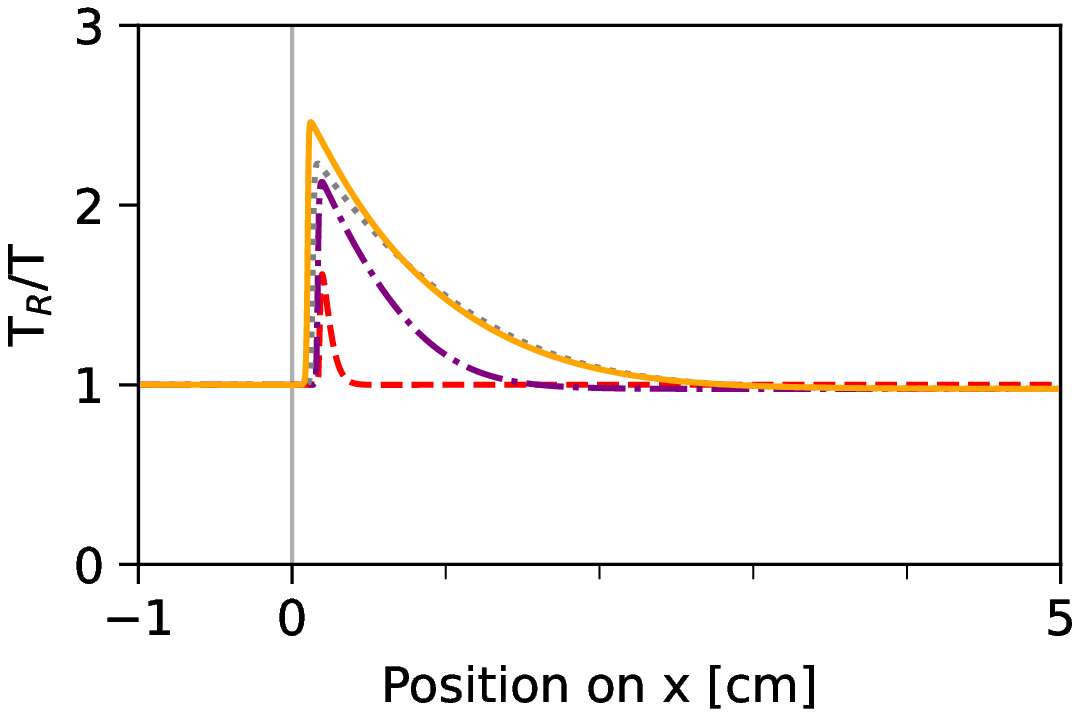}
        \caption{Ratio of radiative temperature to gas temperature.}
        \label{fig:TronT_M4}
    \end{subfigure}
    \hfill
    \begin{subfigure}[t]{0.45\textwidth}
        \centering
        \includegraphics[width=\textwidth]{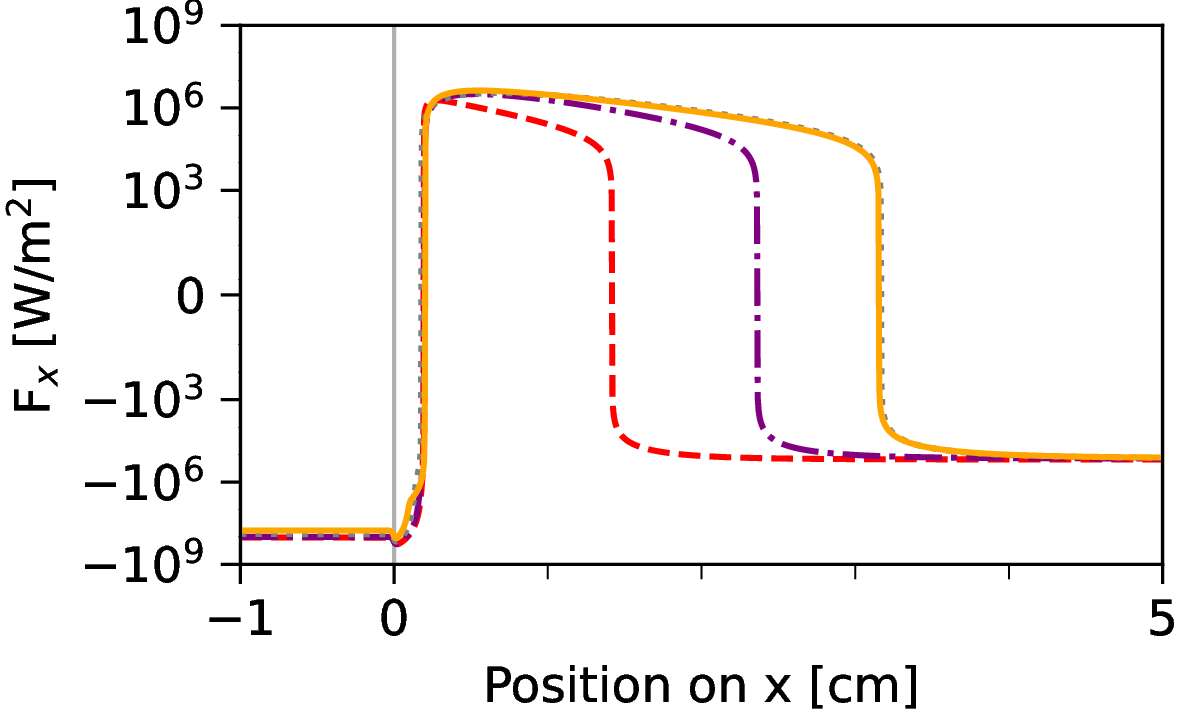}
        \caption{Radiative flux.}
        \label{fig:Fx_M4}
    \end{subfigure}
    \begin{center}
        \begin{subfigure}[t]{0.55\textwidth}
            \includegraphics[width=\textwidth]{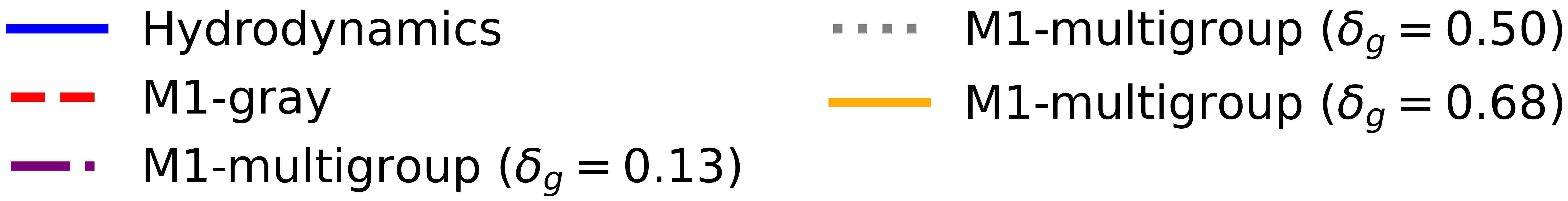}
        \end{subfigure}
    \end{center}
    \caption{Profiles of the physical quantities for the \quotes{Mach~4} simulation. The position \mbox{$x=0$ cm} corresponds to the location of the shock. All these data are expressed in the shock frame.}
    \label{fig:physics_choc_M4}
\end{figure}

\begin{figure}
    \begin{subfigure}[t]{0.45\textwidth}
        \centering
        \includegraphics[width=\textwidth]{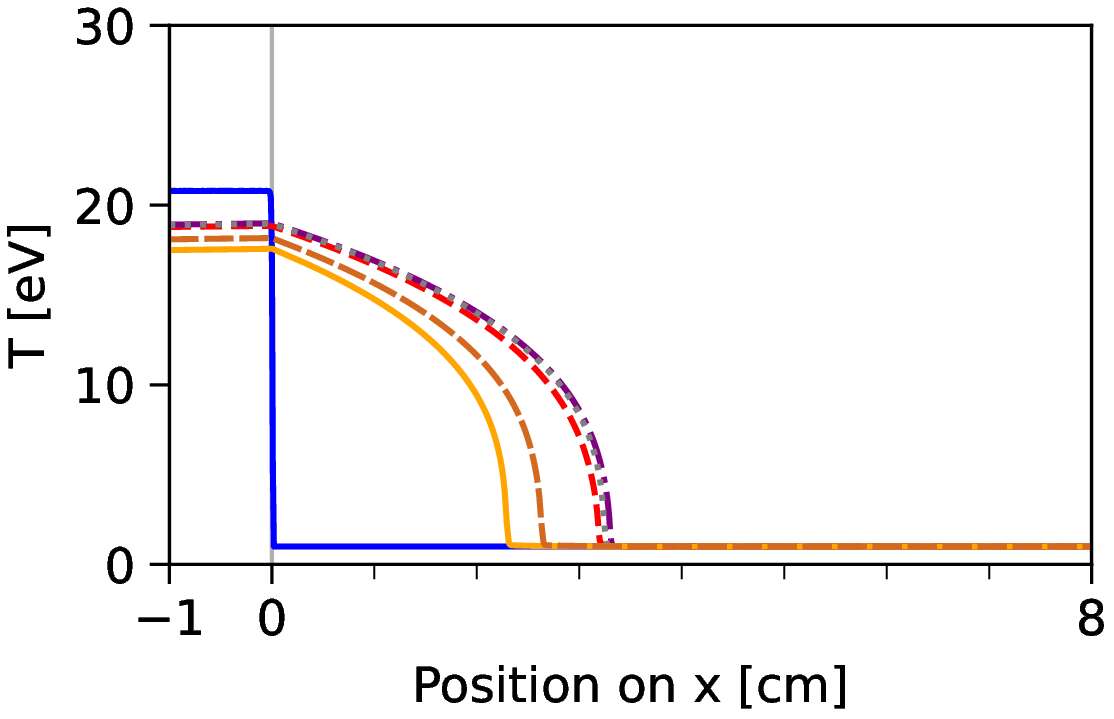}
        \caption{Gas temperature.}
        \label{fig:T_M8}
    \end{subfigure}
    \hfill
    \begin{subfigure}[t]{0.45\textwidth}
        \centering
        \includegraphics[width=\textwidth]{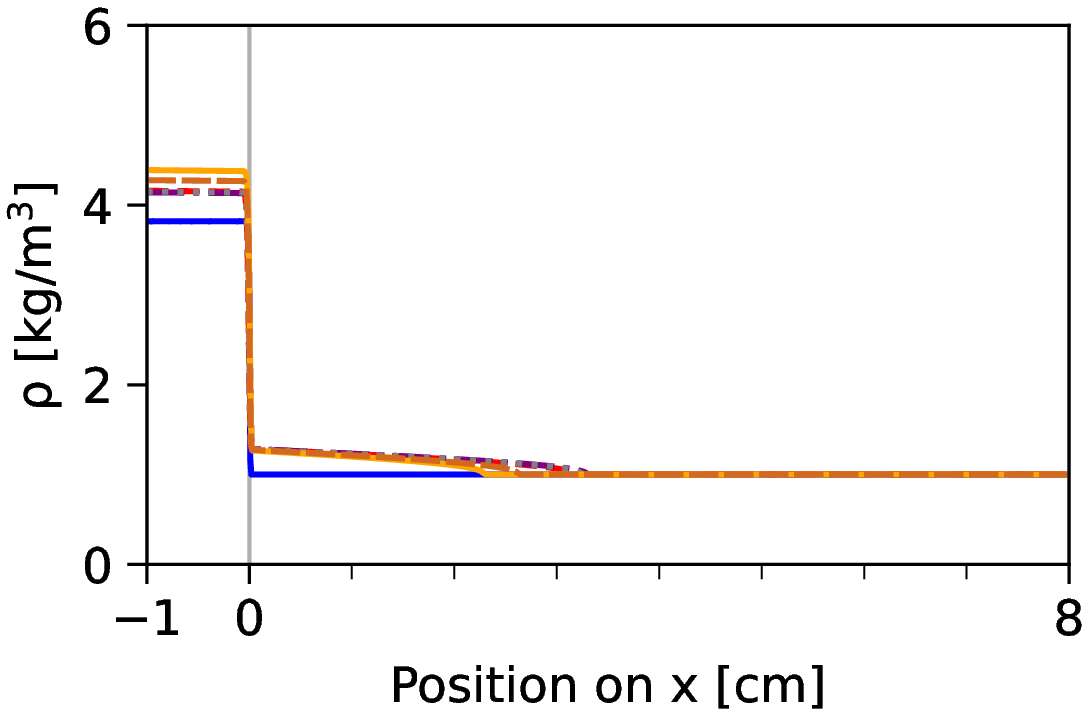}
        \caption{Density.}
        \label{fig:rho_M8}
    \end{subfigure}
    \hfill
    \begin{subfigure}[t]{0.45\textwidth}
        \centering
        \includegraphics[width=\textwidth]{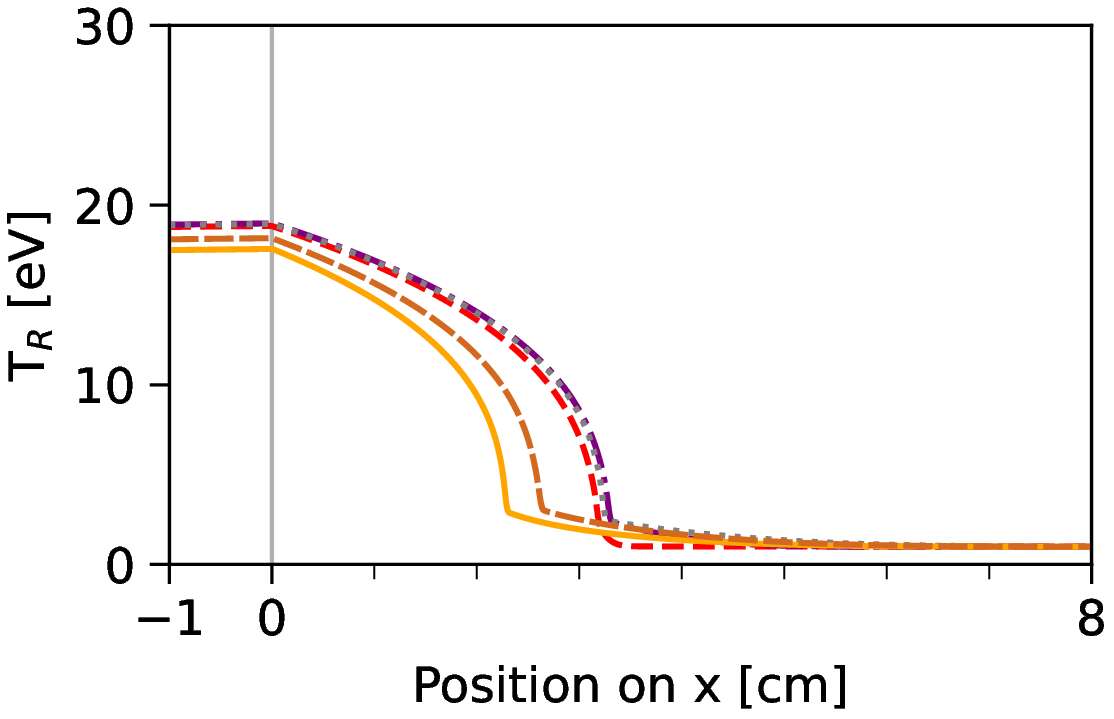}
        \caption{Radiative temperature.}
        \label{fig:Tr_M8}
    \end{subfigure}
    \hfill
    \begin{subfigure}[t]{0.45\textwidth}
        \centering
        \includegraphics[width=\textwidth]{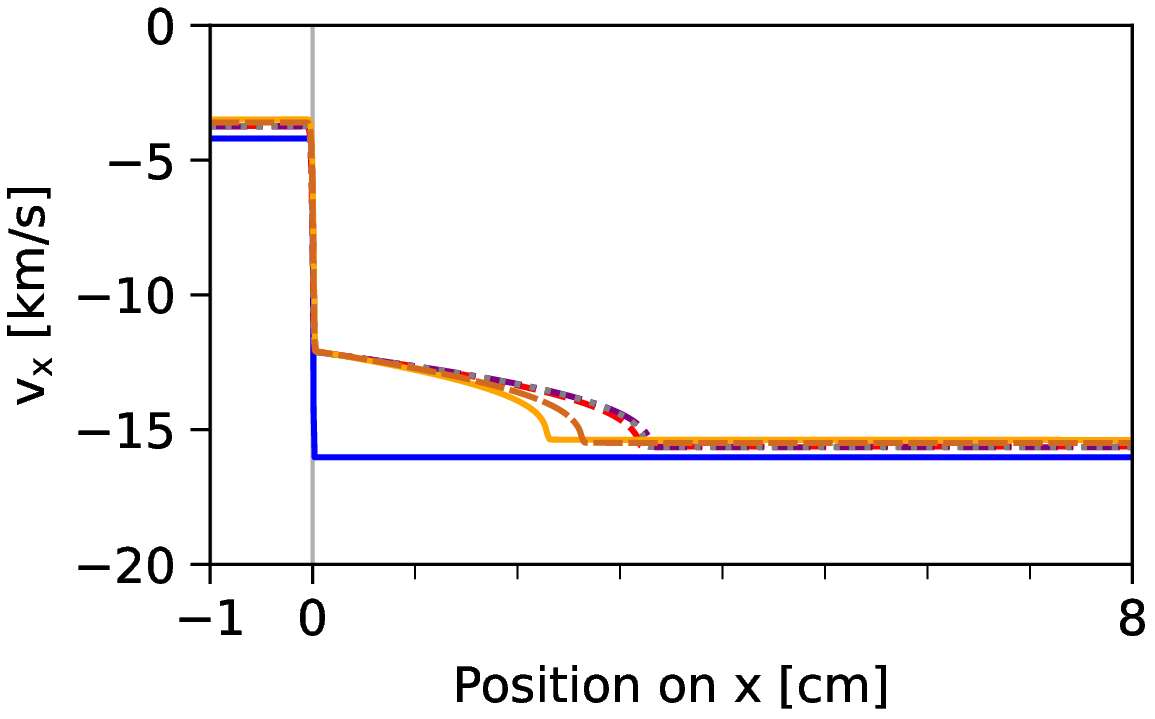}
        \caption{Velocity.}
        \label{fig:vx_M8}
    \end{subfigure}
    \hfill
    \begin{subfigure}[t]{0.45\textwidth}
        \centering
        \includegraphics[width=\textwidth]{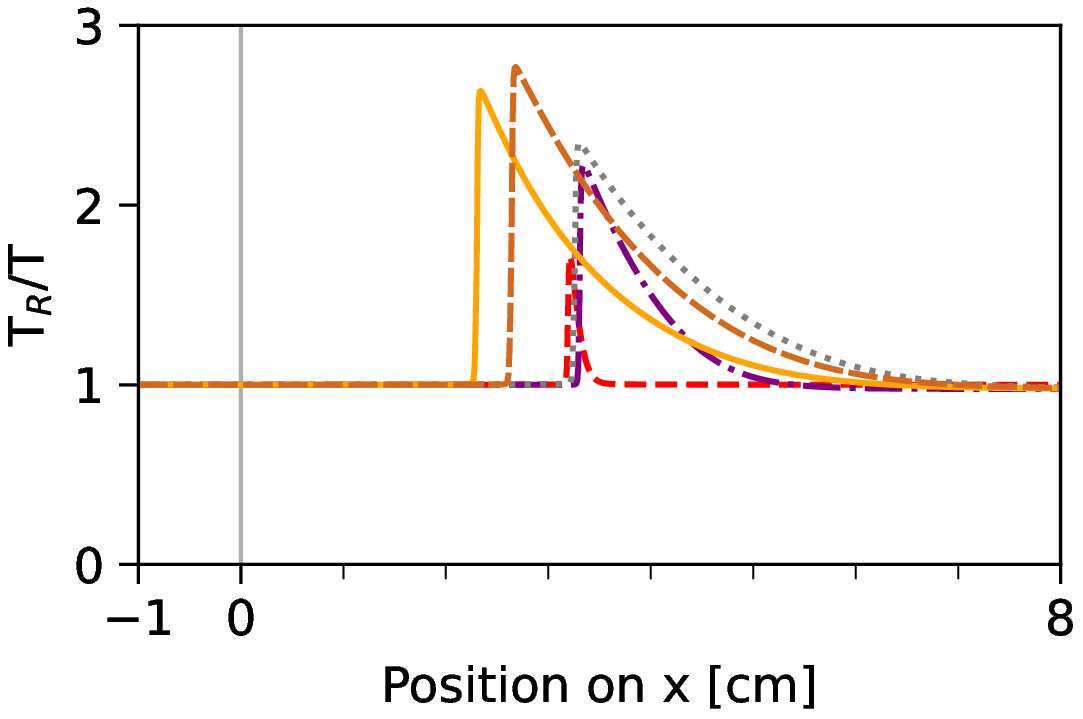}
        \caption{Ratio of radiative temperature to gas temperature.}
        \label{fig:TronT_M8}
    \end{subfigure}
    \hfill
    \begin{subfigure}[t]{0.45\textwidth}
        \centering
        \includegraphics[width=\textwidth]{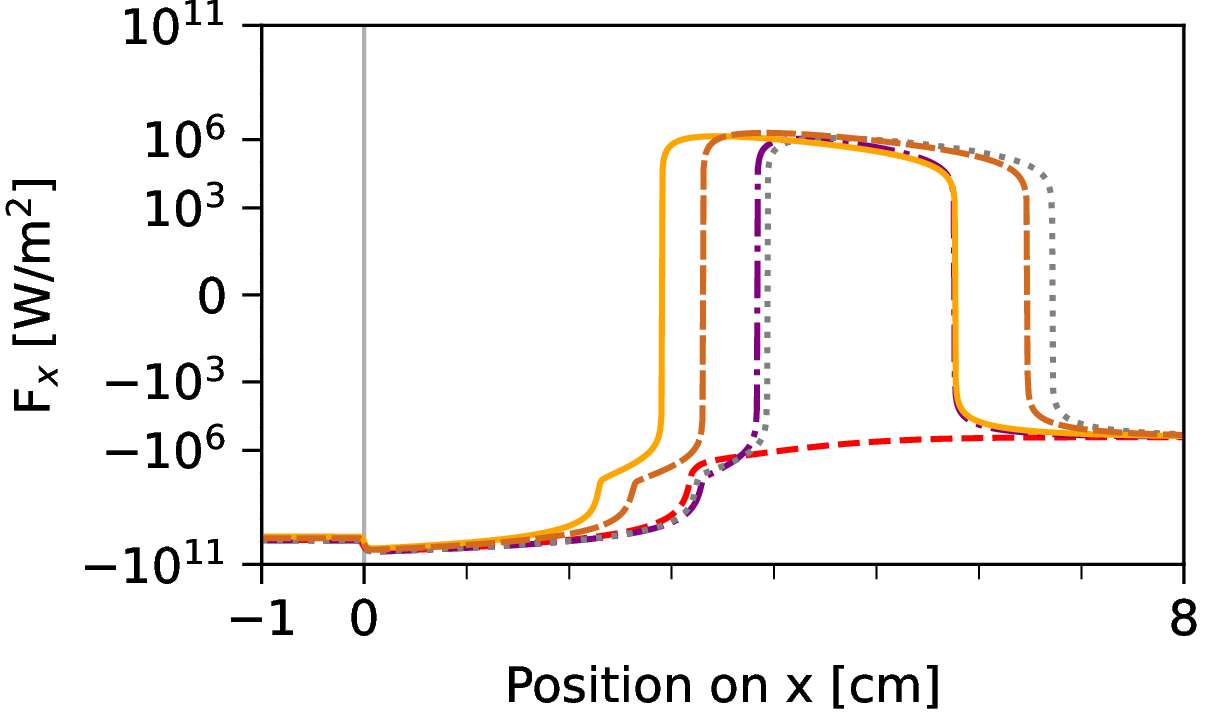}
        \caption{Radiative flux.}
        \label{fig:Fx_M8}
    \end{subfigure}
    \begin{center}
        \begin{subfigure}[t]{0.55\textwidth}
            \includegraphics[width=\textwidth]{Images/chapitre_4/label_curves_M8.jpg}
        \end{subfigure}
    \end{center}
    \caption{Profiles of the physical quantities for the \quotes{Mach~8} simulation. The position \mbox{$x=0$ cm} corresponds to the location of the shock. All these data are expressed in the shock frame.}
    \label{fig:physics_choc_M8}
\end{figure}

In these simulations, the upstream and downstream media are optically thick (with an optical depth of \mbox{$\tau \sim 10^4$} downstream and \mbox{$\tau \sim 1$ to $10^2$} upstream, for both cases considered). From the simulated data, one can compute the three dimensionless parameters characterizing radiative shocks — the cooling parameter $\chi$, the Boltzmann number $\mathrm{Bo}$, and the Mihalas parameter $\mathrm{R}$ — using equations~\eqref{eq:cooling_param_epais}, \eqref{eq:bolzmann_param}, and \eqref{eq:mihalas_param}, in order to gain insight into the nature of the shocks under study. For the \quotes{Mach~4} shock, the values of these parameters are \mbox{$\chi \approx 2~000$}, \mbox{$\mathrm{Bo} \approx 0.2$}, and \mbox{$\mathrm{R} \approx 3~000$}, whereas for the \quotes{Mach~8} shock, the values are \mbox{$\chi \approx 100$}, \mbox{$\mathrm{Bo} \approx 0.01$}, and \mbox{$\mathrm{R} \approx 200$}. These results show that both configurations correspond to optically thick radiative shocks dominated by radiative flux, with the \quotes{Mach~8} case representing a particularly strong example. The simulations, whose final profiles of hydrodynamic and radiative quantities are shown in figures~\ref{fig:physics_choc_M4} and \ref{fig:physics_choc_M8}, demonstrate that a detailed description of the spectral nature of radiation, made possible by using narrow groups in the M1-multigroup model, significantly modifies three essential aspects of the shock: its propagation speed, the hydrodynamic properties of the downstream medium, and the structure of the radiative precursor.

\starsect{Influence on the shock speed}

\begin{figure}[hbt!]
    \begin{subfigure}[t]{0.45\textwidth}
        \centering
        \includegraphics[width=\textwidth]{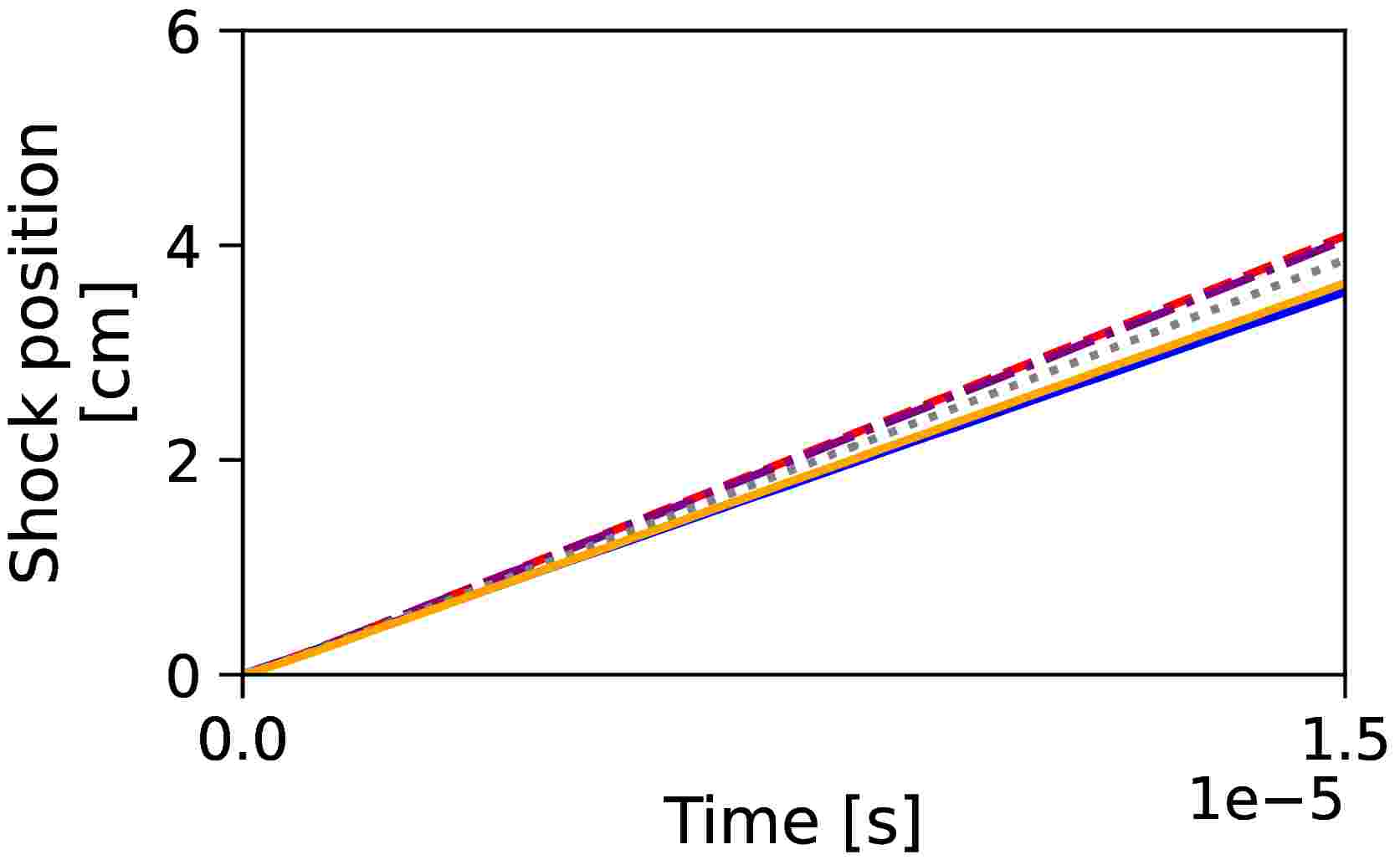}
        \caption{Simulation \quotes{Mach~4}.}
        \label{fig:shockPos_M4}
    \end{subfigure}
    \hfill
    \begin{subfigure}[t]{0.45\textwidth}
        \centering
        \includegraphics[width=\textwidth]{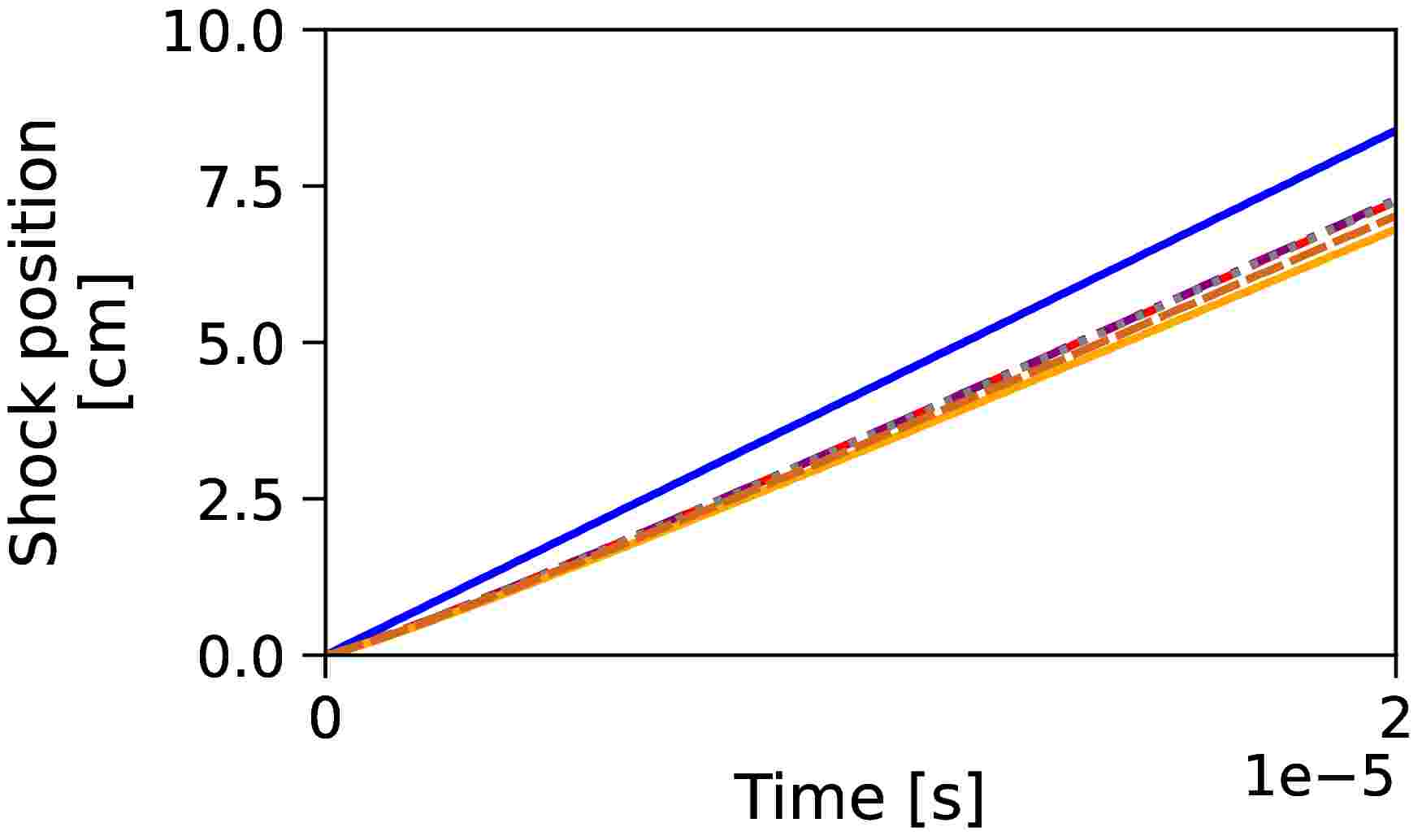}
        \caption{Simulation \quotes{Mach~8}.}
        \label{fig:shockPos_M8}
    \end{subfigure}
    \begin{center}
        \begin{subfigure}[t]{0.55\textwidth}
            \includegraphics[width=\textwidth]{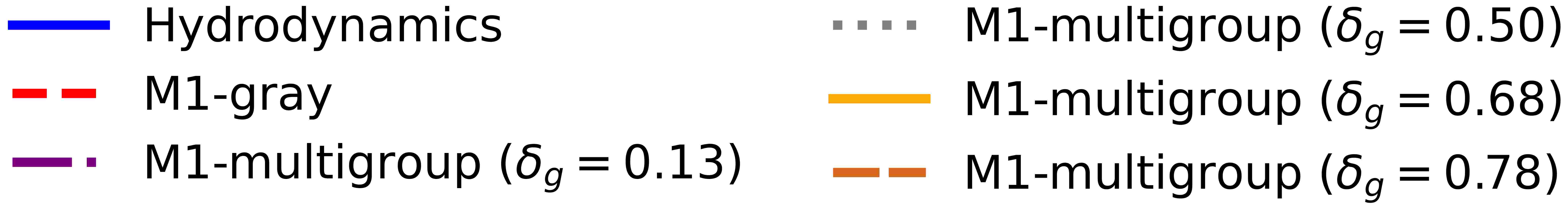}
        \end{subfigure}
    \end{center}
    \caption{Evolution of the shock position as a function of simulated time.}
    \label{fig:shockPos}
\end{figure}

First, we determined the position of the shock by identifying it as the point where the density gradient reaches its maximum. From this estimate, we deduced the shock velocity by plotting, in figure~\ref{fig:shockPos}, the evolution of its position as a function of physical time for the \quotes{Mach~4} and \quotes{Mach~8} configurations. One observes that, at the beginning of the simulations, the shock decelerates (smaller slope on the curves), and then, beyond a certain time, the position of the shock front evolves linearly with respect to time. To avoid bias associated with this initial deceleration, we fitted a linear function to the last 50\% of the time data. This regression makes it possible to estimate the final shock velocity, denoted $v_{shock}$.

\begin{figure}[hbt!]
    \begin{subfigure}[t]{0.48\textwidth}
        \centering
        \includegraphics[width=\textwidth]{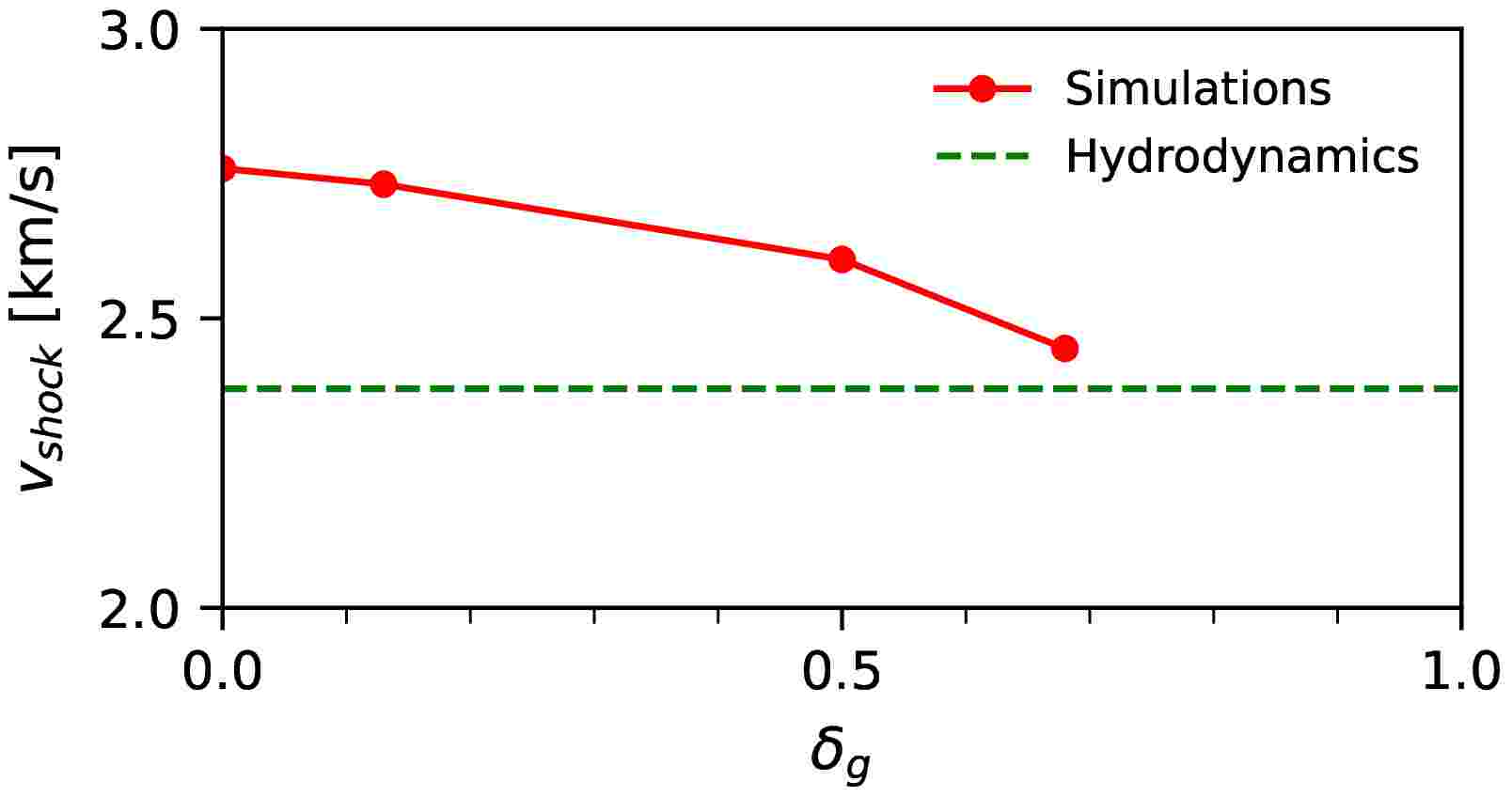}
        \caption{\quotes{Mach~4}.}
    \end{subfigure}
    \hfill
    \begin{subfigure}[t]{0.48\textwidth}
        \centering
        \includegraphics[width=\textwidth]{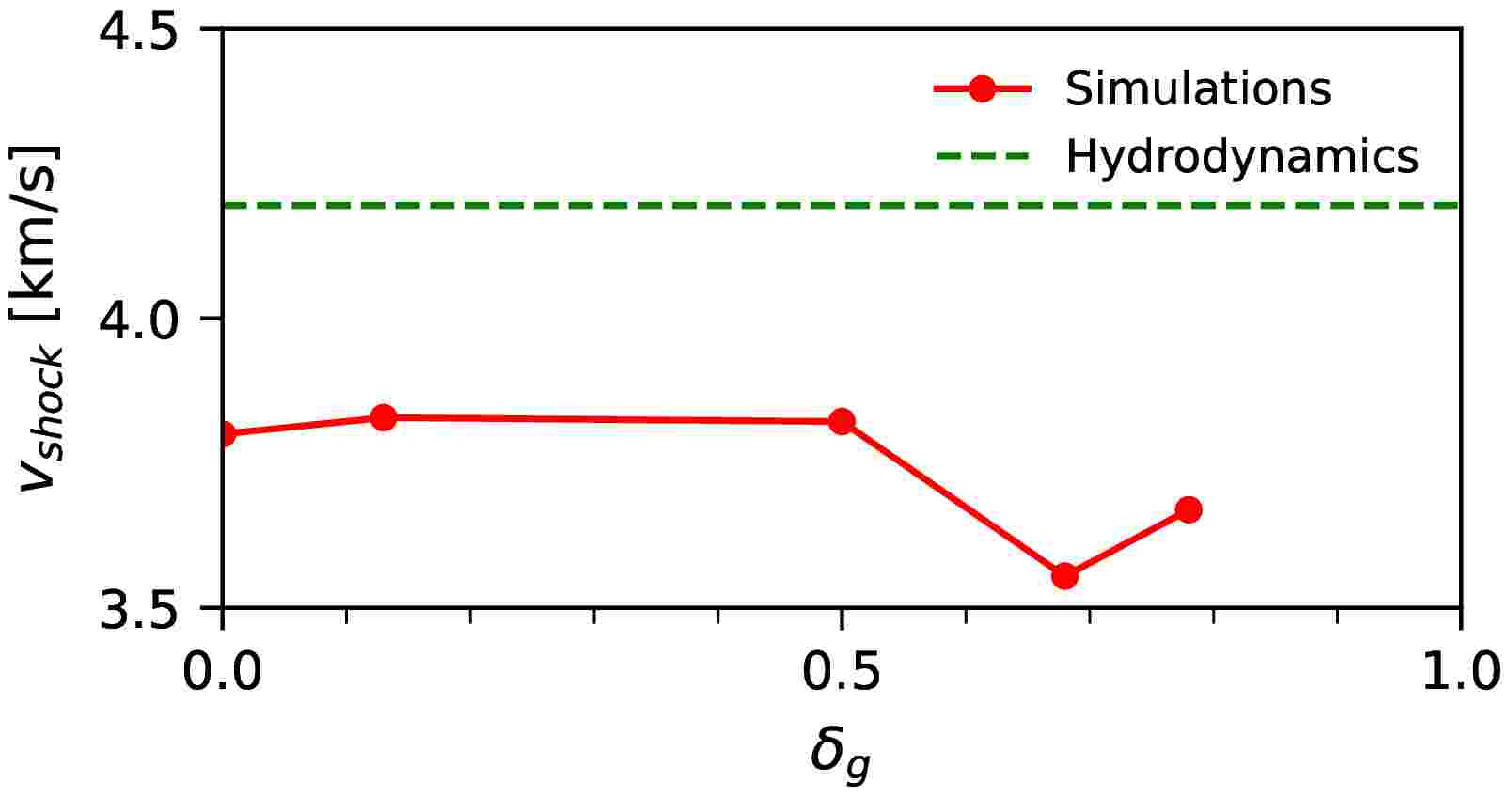}
        \caption{\quotes{Mach~8}.}
    \end{subfigure}
    \caption{Evolution of the shock velocity as a function of the group narrowness $\delta_g$. The case with \mbox{$\delta_g=0$} corresponds to the simulation using the M1-gray model.}
    \label{fig:shock_vel}
\end{figure}

We then plotted in figure~\ref{fig:shock_vel} the evolution of the shock velocity $v_{shock}$ obtained as a function of the group narrowness $\delta_g$ in the different simulations. This analysis reveals that radiation affects the shocks differently depending on the configuration: in the \quotes{Mach~4} simulation, the shock is accelerated relative to the hydrodynamic case, whereas in the \quotes{Mach~8} simulation, it is decelerated. However, reducing the width of the groups in the M1-multigroup model (increasing the group narrowness $\delta_g$) systematically leads to a reduction in shock velocity in both cases, which results in a decrease in the Mach number, since the sound speed of the upstream medium does not change. The only exception occurs between the simulations with \mbox{$\delta_g = 0.68$} and \mbox{$\delta_g = 0.78$}, where an acceleration of the shock is observed. However, the shock velocity in the M1-multigroup simulation with \mbox{$\delta_g = 0.78$} remains lower than in all the other simulations.

\starsect{Influence on downstream hydrodynamic quantities}

\begin{figure}
    \begin{subfigure}[t]{0.48\textwidth}
        \centering
        \includegraphics[width=\textwidth]{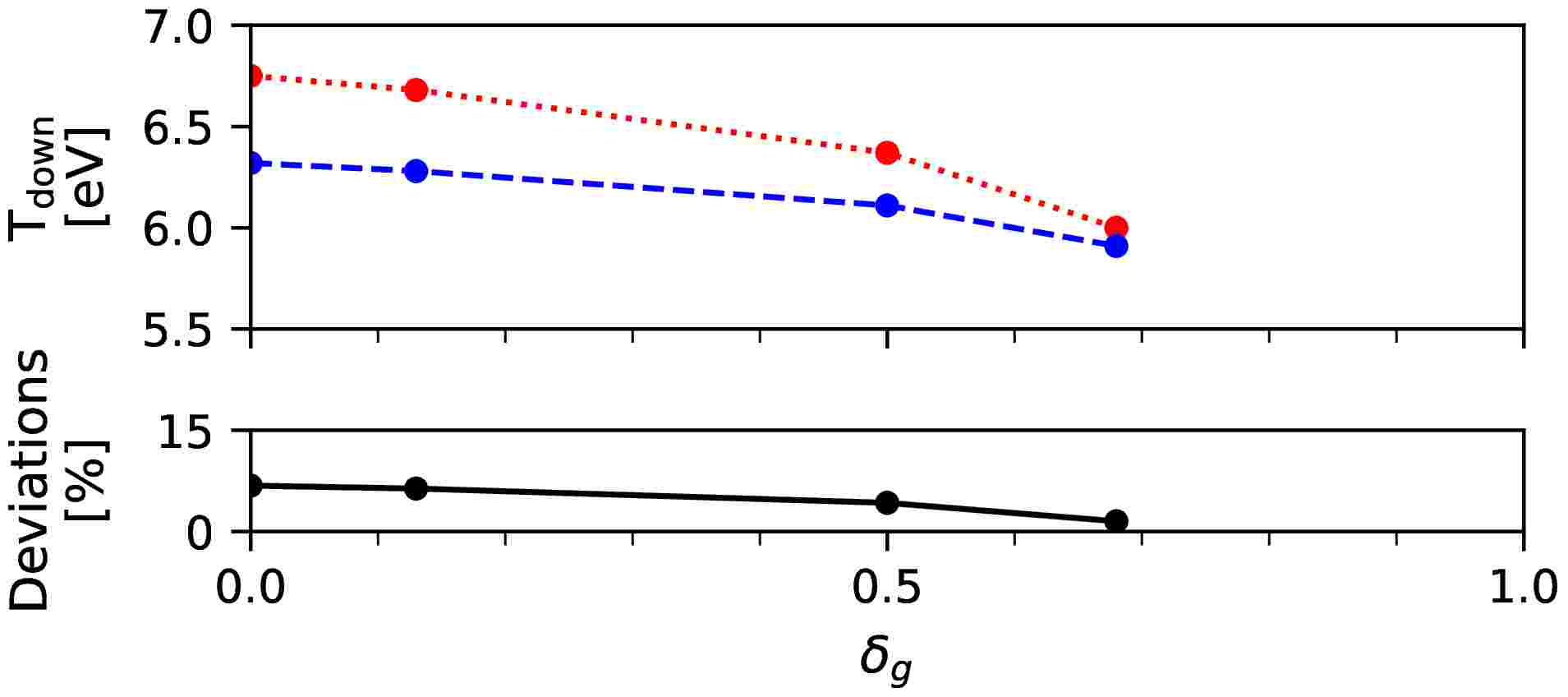}
        \caption{Downstream temperature in the \quotes{Mach~4} simulation.}
    \end{subfigure}
    \hfill
    \begin{subfigure}[t]{0.48\textwidth}
        \centering
        \includegraphics[width=\textwidth]{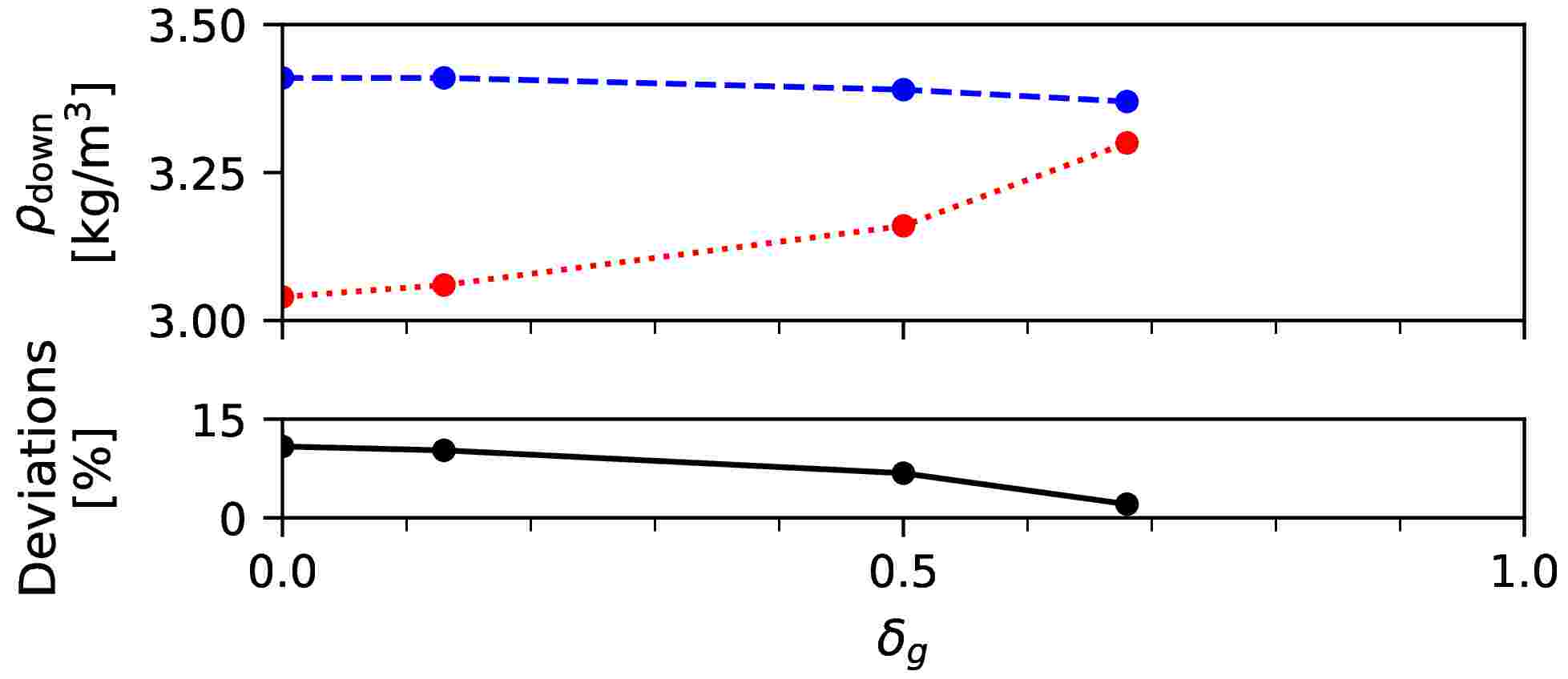}
        \caption{Downstream density in the \quotes{Mach~4} simulation.}
    \end{subfigure}
    \hfill
    \begin{subfigure}[t]{0.48\textwidth}
        \centering
        \includegraphics[width=\textwidth]{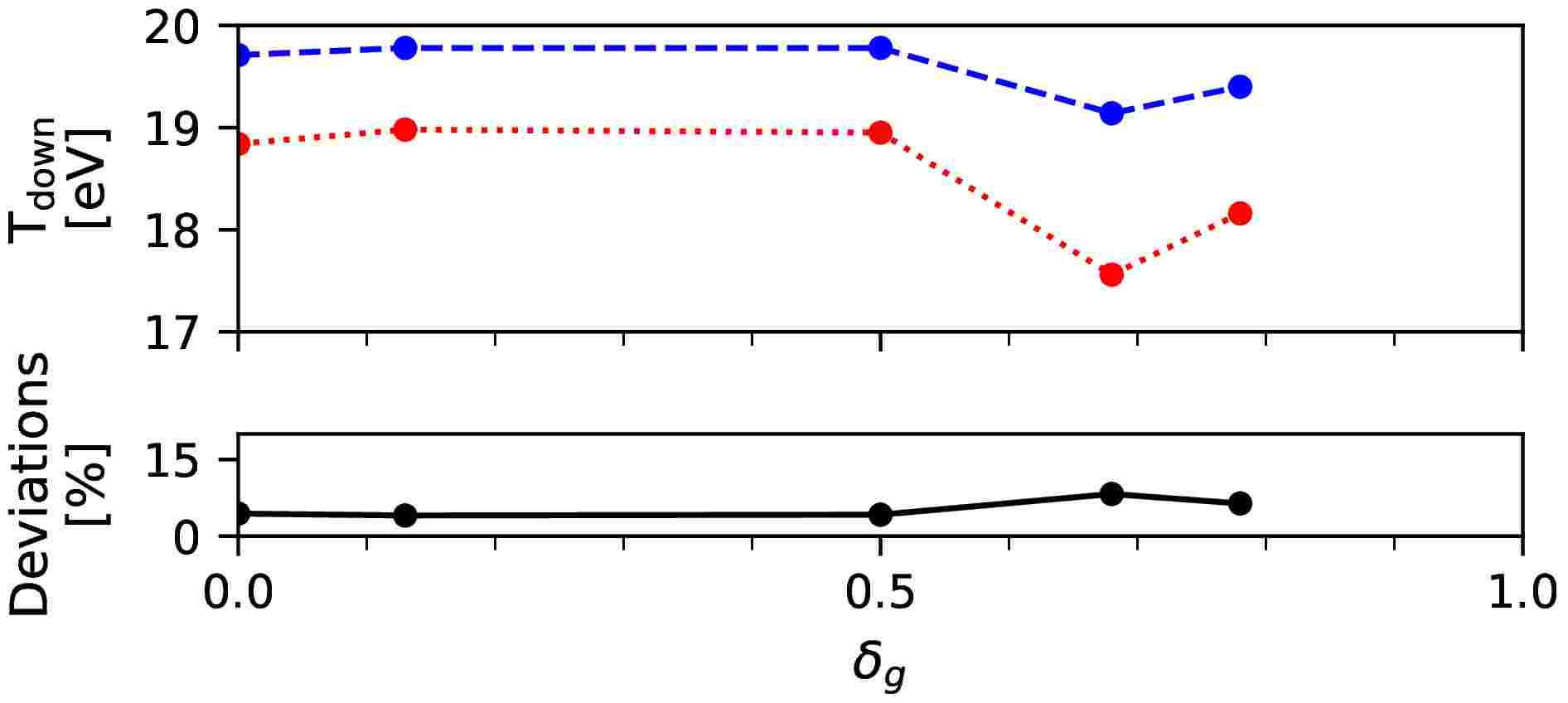}
        \caption{Downstream temperature in the \quotes{Mach~8} simulation.}
    \end{subfigure}
    \hfill
    \begin{subfigure}[t]{0.48\textwidth}
        \centering
        \includegraphics[width=\textwidth]{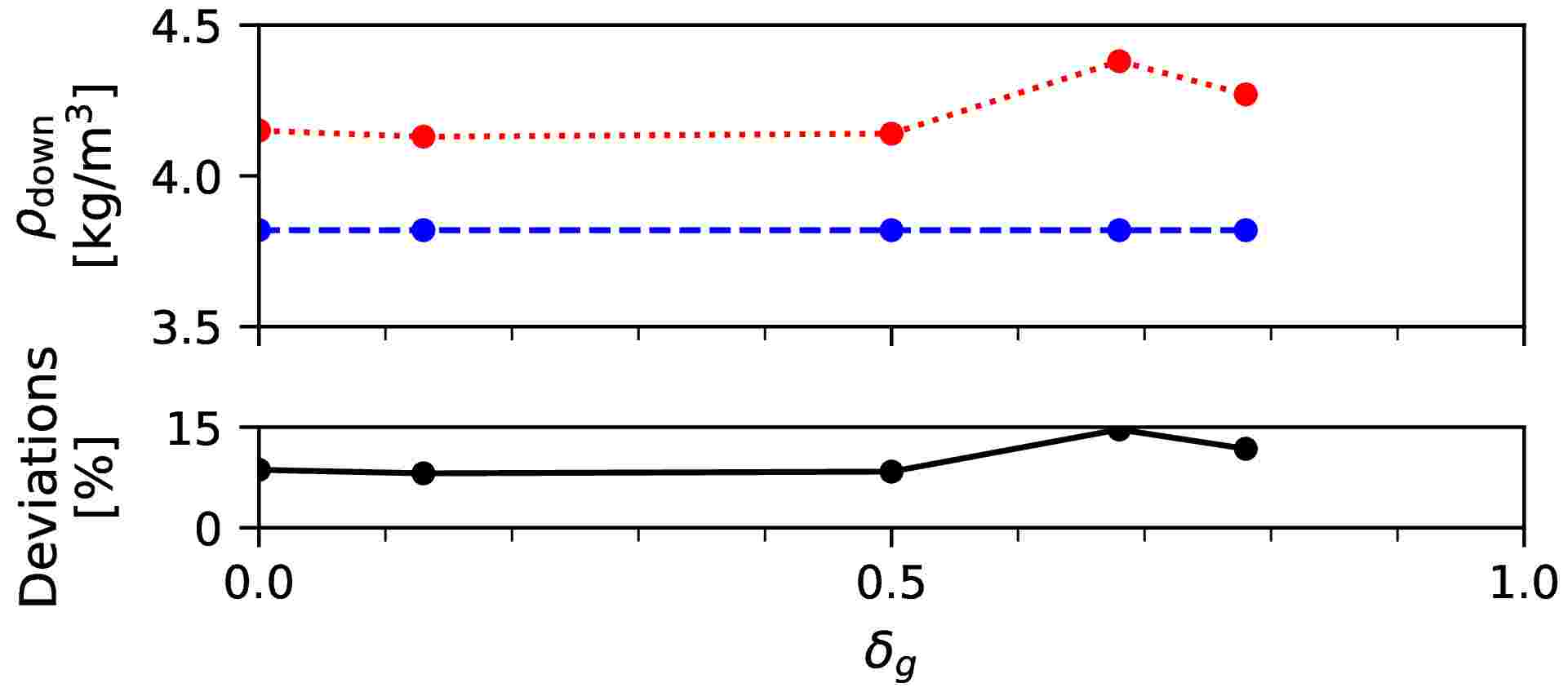}
        \caption{Downstream density in the \quotes{Mach~8} simulation.}
    \end{subfigure}
    \begin{center}
        \begin{subfigure}[t]{0.3\textwidth}
            \includegraphics[width=\textwidth]{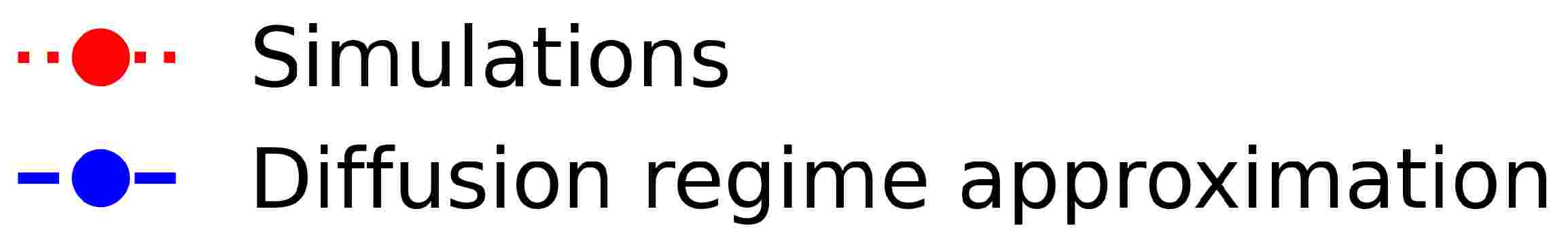}
        \end{subfigure}
    \end{center}
    \caption{Comparison of the downstream hydrodynamic quantities obtained from the simulations performed here with those predicted by the analytical model based on the diffusion regime approximation. The simulation with \mbox{$\delta_g=0$} corresponds to the simulation using the M1-gray model. The predicted value from the diffusion regime approximation changes because the shock velocity, and therefore the Mach number, evolves as a function of the group narrowness $\delta_g$.}
    \label{fig:hydro_qtt}
\end{figure}

Since this framework provides the downstream hydrodynamic quantities without bias from the initial conditions, I compared the temperature and density obtained in the simulations with the predictions of an approximate analytical model based on the diffusion regime, which is applicable to optically thick radiative shocks~\cite{bouquet_2000}. This model is widely used to provide a first estimate of the downstream hydrodynamic quantities of a radiative shock, or to initialize stationary-shock simulations (see figure~\ref{fig:hydro_qtt}).

Because the shock velocity (and therefore the associated Mach number) decreases as the group width increases (that is, as $\delta_g$ decreases), this has a direct impact on the hydrodynamic quantities predicted by the analytical diffusion regime model, which explains the variations observed in the blue curves in figure~\ref{fig:hydro_qtt}.

However, despite these variations, one finds that this approximate model introduces deviations ranging from 1 to 8\% in the gas temperature and from 3 to 15\% in the density, compared with the values obtained from the simulations. For the \quotes{Mach~4} configuration, the downstream medium is systematically hotter and less dense than predicted by the diffusion model, and the discrepancies decrease as the group narrowness $\delta_g$ increases. Conversely, in the \quotes{Mach~8} configuration, the downstream medium is systematically colder and denser, and the deviation with the approximate model increases as the group narrowness increases.

This highlights the challenges involved in accurately modeling the jump conditions of a radiative shock, and emphasizes the need for caution when relying on analytical expressions derived from the diffusion regime, particularly for simulation initialization or boundary-condition tuning.

\starsect{Influence on the radiative precursor}

\begin{figure}
    \begin{subfigure}[t]{0.49\textwidth}
        \centering
        \includegraphics[width=\textwidth]{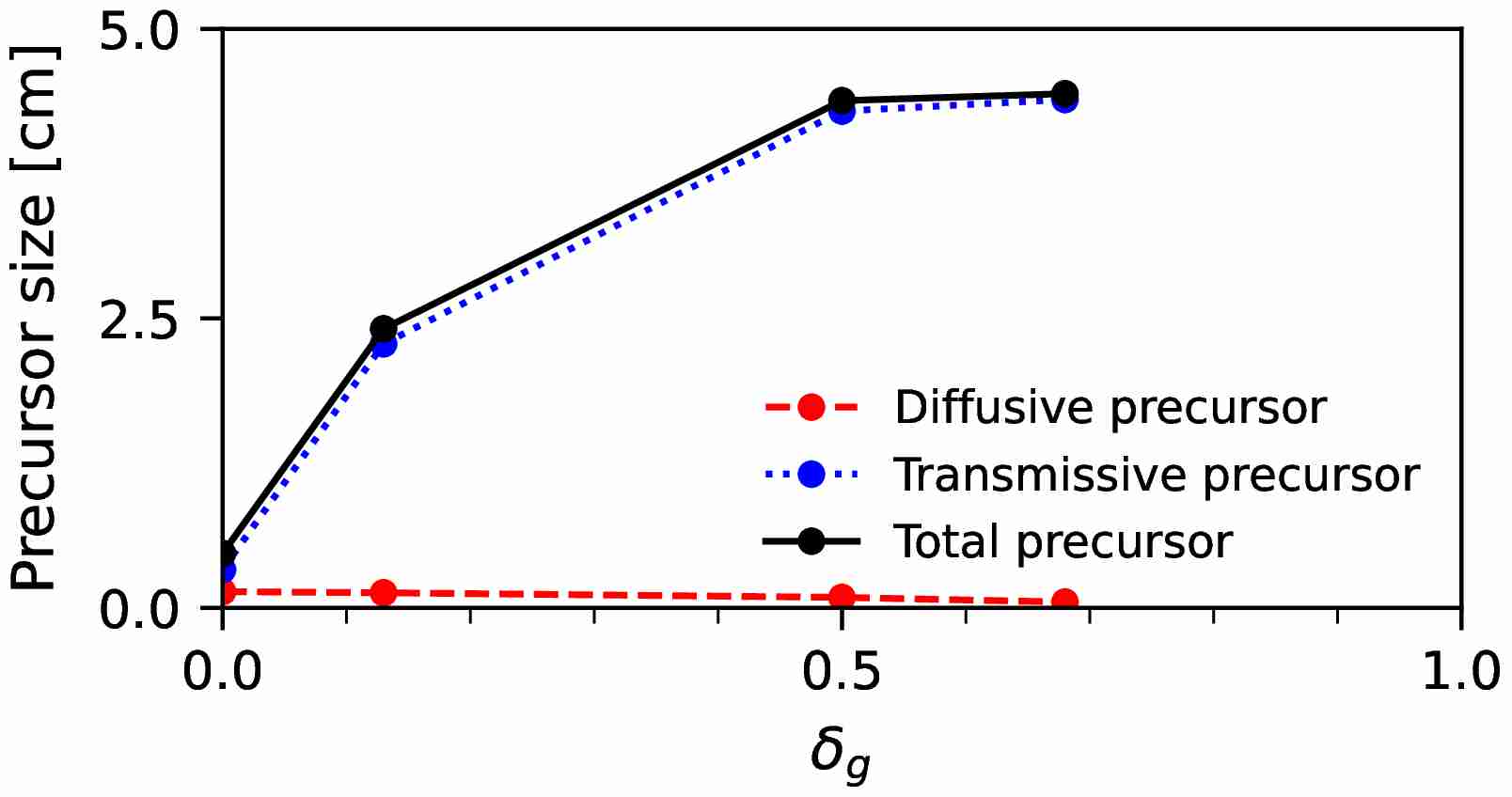}
        \caption{\quotes{Mach~4}.}
    \end{subfigure}
    \hfill
    \begin{subfigure}[t]{0.49\textwidth}
        \centering
        \includegraphics[width=\textwidth]{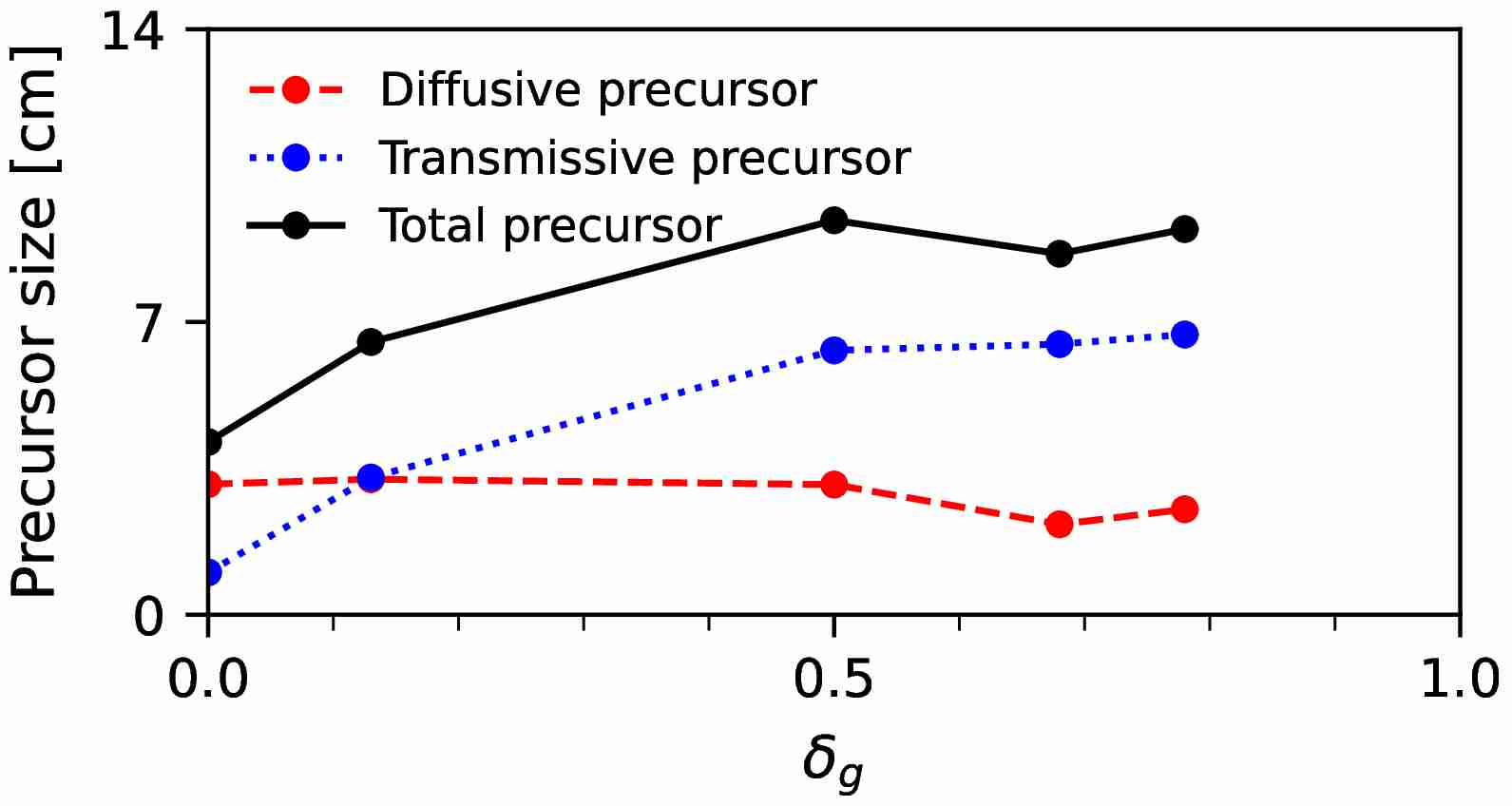}
        \caption{\quotes{Mach~8}.}
    \end{subfigure}
    \caption{Size of the radiative precursor as a function of the group narrowness $\delta_g$. The simulation with \mbox{$\delta_g=0$} corresponds to the simulation using the M1-gray model.}
    \label{fig:precursor}
\end{figure}

We define the diffusive precursor as the region upstream of the shock where the radiation and gas temperatures are approximately equal, that is, where \mbox{$\left |\mathrm{T}_R/\mathrm{T} - 1 \right | < 10^{-3}$}. The transmissive precursor, located upstream of the diffusive precursor, is characterized by a significant difference between the radiation and gas temperatures, that is, \mbox{$\left |\mathrm{T}_R/\mathrm{T} - 1 \right | \geq 10^{-3}$} (see figures~\ref{fig:TronT_M4} and \ref{fig:TronT_M8}). Figure~\ref{fig:precursor} illustrates the evolution of the size of the different radiative precursors as a function of the spectral narrowness of the groups, characterized by the parameter $\delta_g$.

In both simulated cases, one observes that the transmissive precursor expands when radiative effects are modeled with greater accuracy, at the expense of the diffusive precursor and the upstream medium. This suggests that the M1-gray model overestimates the equilibrium between gas and radiation in the radiative precursor. Moreover, it is observed that the radiative flux in the transmissive precursor becomes positive, whereas it remains negative in other regions (see figures~\ref{fig:Fx_M4} and \ref{fig:Fx_M8}). This behavior appears to be a general trend that is not well captured by the M1-gray model, particularly in the \quotes{Mach~8} simulation.

\starsect{Computational time}

\begin{table}[ht]
    \centering
    \begin{tabular}{L{5.5cm} C{2cm} l C{0.5cm} C{2cm} l}
        \hline
        \hline
        \TBstrut Simulation & \multicolumn{2}{c}{\quotes{Mach~4}} & & \multicolumn{2}{c}{\quotes{Mach~8}} \\
        \hline
        \TBstrut \textbf{Hydrodynamics}     & $7.6$          & s         & & $17$           & s \\
        \hdashline
        \Tstrut \textbf{M1-gray}             & $749.5$        & h         & & $1~821.8$   & h \\
        \Bstrut                              & $31.2$         & days      & & $75.9$       & days \\
        \hdashline
        \Tstrut \textbf{M1-multigroup}      & $5~298.9$     & h         & & $5~102.5$   & h \\
        \Bstrut $\delta_g=0.14$ - 4 groups  & $220.8$     & days      & & $212.6$      & days\\
        \hdashline
        \Tstrut \textbf{M1-multigroup}      & $52~847.5$    & h         & & $44~414.0$  & h \\
        \Bstrut $\delta_g=0.48$ - 9 groups  & $2~202.0$   & days      & & $1~850.6$   & days \\
        \hdashline
        \Tstrut \textbf{M1-multigroup}      & $121~099.6$   & h         & & $113~203.4$ & h \\
        \Bstrut $\delta_g=0.68$ - 16 groups & $5~045.8$   & days      & & $4~716.8$   & days \\
        \hdashline
        \Tstrut \textbf{M1-multigroup}        & \multicolumn{2}{c}{--} & & $191~132.5$ & h\\
        \Bstrut $\delta_g=0.78$ - 23 groups & \multicolumn{2}{c}{--} & & $7~963.9$    & days \\
        \hline
        \hline
    \end{tabular}
    \caption{Computational cost of the simulations in CPU seconds, hours, and days.}
    \label{tab:cout_simulations}
\end{table}

In order to study in more detail the influence of the spectral nature of radiation on radiative shocks, it is necessary to perform a larger number of simulations, varying in particular the shock strength (that is, the Mach number), the opacity, and the different classes of radiative shocks presented in Section~\ref{sec:intro_choc_rad}. Although optimizing the computation of the Eddington factor in the M1-multigroup model has made such simulations feasible, they remain extremely expensive in terms of computational time, as shown in table~\ref{tab:cout_simulations}. In this work, we parallelized the calculations, using up to 400 processors for the most demanding cases. Despite this, a more systematic and extensive exploration of the spectral effects of radiation on the structure of radiative shocks remains out of reach for a code such as \gls{hades}. This limitation becomes even more pronounced when one is interested in multidimensional (2D/3D) radiative effects. This constraint stems from a fundamental limitation of the explicit schemes used in \gls{hades}, namely the \gls{cfl} stability condition, which imposes a time step defined by:
\begin{equation}
    \label{eq:cfl}
    \Delta t \le \Delta x / v
\end{equation}

In purely hydrodynamic simulations, the velocity $v$ corresponds to the fluid velocity $||\vectorr{v}||$, whereas in radiative hydrodynamics it is replaced by the speed of light $c$, which is much larger (\mbox{$c \gg ||\vectorr{v}||$}). As a consequence, the time step must be drastically reduced, which significantly increases the duration of the simulations, as illustrated in table~\ref{tab:cout_simulations}. Thus, the simulation that describes radiative transfer with the greatest accuracy required 7~963.9 CPU days. After parallelization on 400 processors, it still lasted about 20 days, which remains particularly long for this type of simulation.

It therefore becomes essential to explore other approaches that allow these radiative shocks to be simulated more efficiently. In this perspective, we undertook to investigate the use of \glspl{pinn} to perform radiative-hydrodynamics simulations. The following chapter presents my first attempts at shock simulation using this neural-network architecture.

\section{Synthesis}

In this chapter, I presented my simulation results on the structure of a radiative shock, modeling radiative transfer using the M1-multigroup model, which makes it possible to account finely for the spectral nature of light. This approach led to three main results:
\begin{enumerate}
    \item The propagation speed of the shock decreases as the radiation spectrum is modeled with greater accuracy;
    \item The predictions obtained from the diffusion regime approximation for the jump conditions show notable discrepancies: errors of 1–8\% in the gas temperature and 3–15\% in the density;
    \item The more refined the spectral modeling, the larger the radiative precursor becomes, in particular the nonequilibrium region, which benefits the most from this extension.
\end{enumerate}

These results highlight the crucial importance of accurate radiation–matter interaction modeling in simulations of radiative shocks. They also shed light on the limitations of the diffusion-regime approximation often used to initialize such simulations — an approximation that can introduce significant errors in the upstream conditions and lead to nonphysical propagation of the shock inside the computational domain.

However, simulations carried out with the \gls{hades} code remain particularly expensive in computational time, notably because of the \glslink{cfl}{\glsxtrlong{cfl}} condition, which must be satisfied to ensure the stability of the explicit schemes in the code. This condition forces the use of a very small time step to maintain numerical stability when solving the radiative-hydrodynamics equations.
\clearemptydoublepage

\chapter{Extrapolation of radiative shock simulations with Physics-Informed Neural Networks} \label{ch:chapitre5}

\initialletter{A}mong recent deep-learning approaches developed to reproduce or extrapolate physical simulations, some stand out for their ability either to explicitly incorporate physical constraints or to directly learn the operators governing the system dynamics. \glsreset{pinn}\gls{pinn}s~\cite{raissi_2019} rely on the explicit incorporation of the governing equations into the loss function: the neural network is therefore trained to provide an approximate solution to these equations. Neural Operators (such as DeepONet or the Fourier Neural Operator)~\cite{li_2021, lu_2021_deeponet, kovachki_2023} adopt a different strategy by learning the integral operator that maps initial conditions and physical parameters to the solution at a later time. These methods offer remarkable generalization capability to new regimes or geometries, though at the cost of requiring accurate reference data from simulations for training. Finally, hybrid or reduced-order approaches (Hybrid Surrogates / ROM-AI)~\cite{hesthaven_2018, fresca_2021, san_2018} consist in projecting simulated fields into a lower-dimensional space obtained through a decomposition method (such as Proper Orthogonal Decomposition (POD) or an autoencoder), and then training a neural network to reproduce the temporal evolution of these latent variables. This strategy makes it possible to reconstruct full physical fields at reduced cost while preserving the essential physical coherence and structure of the system. In this work, we chose to explore the \gls{pinn} technique to extrapolate simulations of radiative shocks. Its formulation, directly based on the governing equations, has the advantage of not relying on any prior database of numerical simulations, making it a particularly suitable approach for our objective.

\gls{pinn}s constitute an innovative \gls{ai}-based approach for solving problems governed by \gls{pde}s. Introduced by Raissi et al. (2019)~\cite{raissi_2019}, it extends the pioneering work of Lagaris et al. (1998)~\cite{lagaris_1998}. Although this method can be used to solve inverse problems, the focus here is restricted to forward simulation, which is our primary objective. This approach has already been successfully applied to the modeling of shocks in pure hydrodynamics~\cite{jagtap_2020b, mao_2020, papados_2021} and, more recently, to radiative transfer~\cite{mishra_2021}, but never yet to radiative hydrodynamics. In this chapter, we therefore investigate the potential of this new approach to facilitate and accelerate the modeling of radiative shocks.

\section{Introduction to PINNs} \label{sec:intro_PINNs}

\begin{figure}
    \begin{center}
        \begin{minipage}[t]{\linewidth}
            \centering
            \includegraphics[width=\textwidth]{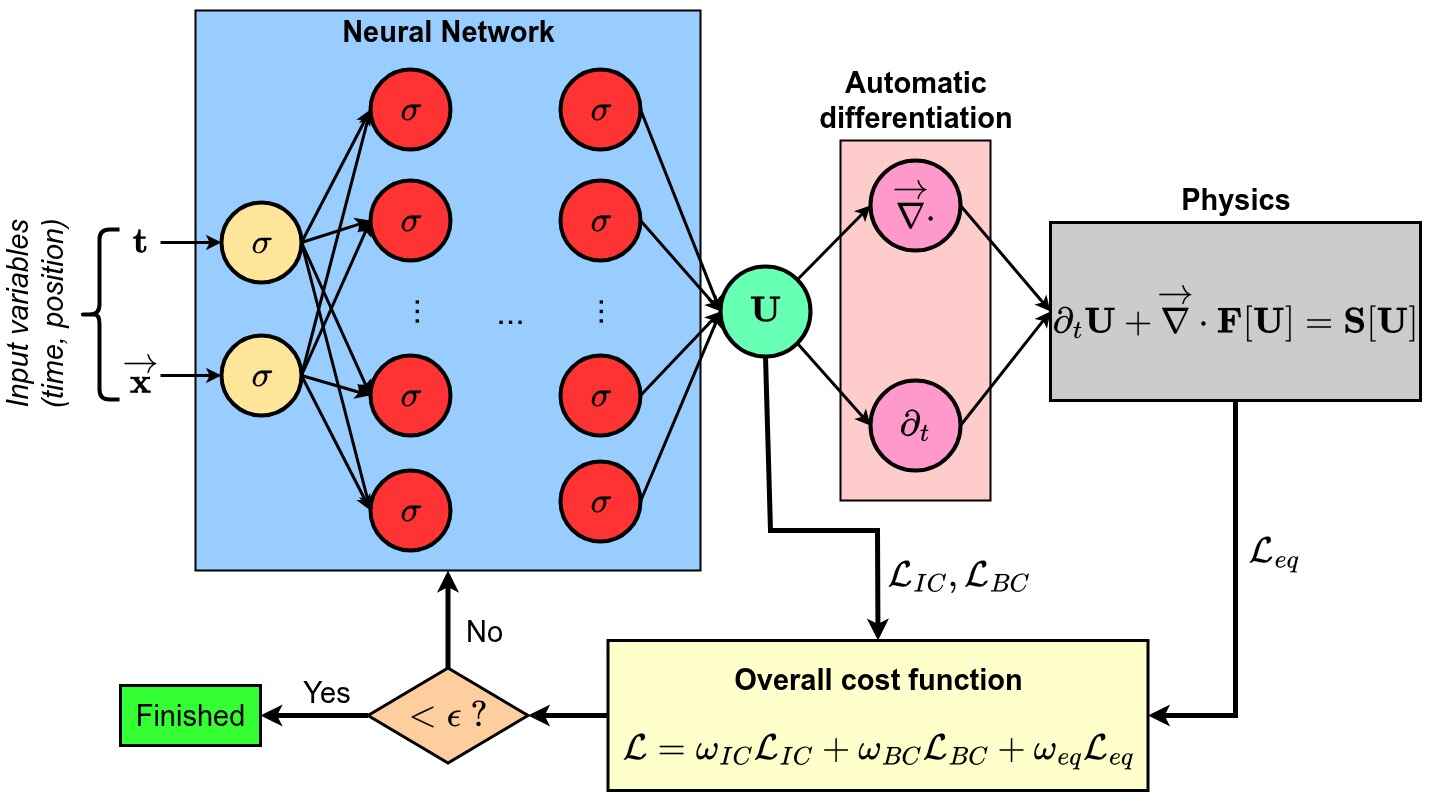}
        \end{minipage}
         \caption{Illustrative diagram of \glslink{pinn}{Physics-Informed Neural Networks}.}
        \label{fig:PINN_diagram}
    \end{center}
\end{figure}

The principle of \gls{pinn}s is to approximate the solution of a \gls{pde} using a \textbf{neural network} (typically an \gls{mlp}), whose training relies on minimizing a loss function composed of several terms: the \textbf{\gls{pde} residual}, the initial conditions, and the boundary conditions when they are specified. The spatial and temporal derivatives required to evaluate the residuals of the \gls{pde} are computed using \textbf{automatic differentiation}~\cite{baydin_2017}, a procedure based on the fact that any computer program, no matter how complex, can be decomposed into a sequence of elementary arithmetic operations and standard functions. By recursively applying the chain rule to this sequence, it is possible to obtain derivatives of arbitrary order with machine precision. Training of the neural network proceeds as long as the \textbf{total loss function} remains above a predefined threshold, or as long as its decrease remains significant. The operation of \gls{pinn}s is summarized in figure~\ref{fig:PINN_diagram}.

\gls{pinn}s are characterized by their great flexibility: they can be implemented with various types of neural network architectures, ranging from \gls{mlp}s (see Section~\secref{sec:MLP}) to more advanced structures such as \gls{cnn}, \gls{rnn}, \gls{lstm}, \gls{gan}, \gls{bnn}, or even transformers~\cite{zhao_2024}. For a detailed overview of their applications, the reader may refer to the review by Cuomo et al. (2022)~\cite{cuomo_2022}. 

\noindent In the context of this work, the focus is placed on the solution of hyperbolic \gls{pde} systems, which can be expressed in the following general form~:
\begin{equation}
    \label{eq:PDE_formulation}
    \left\{
    \begin{array}{lll}
        &\forall (t, \vectorr{x}) \in \closeinterv{t_0}{t_f} \times \Omega,~&\partial_t \vectorIA{U}(t,\vectorr{x}) + \nablav \cdot \vectorIA{F}[\vectorIA{U}(t,\vectorr{x})] = \vectorIA{S}[\vectorIA{U}(t,\vectorr{x})] \;\;\mathpunct{,}\\
        &\forall \vectorr{x} \in \Omega,~&\vectorIA{U}(t_0,\vectorr{x}) = \vectorIA{U}_0(\vectorr{x}) \;\;\mathpunct{,}\\
       & \forall (t, \vectorr{x}) \in \closeinterv{t_0}{t_f} \times \partial\Omega,~&\mathcal{B}[\vectorIA{U}(t,\vectorr{x})] = \vectorIA{B}(t,\vectorr{x}) \;\;\mathpunct{.}\\
    \end{array}
    \right.
\end{equation}

Here, $\vectorIA{U}$ denotes the function representing the state of the physical system, $\Omega$ is the spatial domain, and $\closeinterv{t_0}{t_f}$ is the time interval. The terms $\vectorIA{F}[\vectorIA{U}]$ and $\vectorIA{S}[\vectorIA{U}]$ correspond respectively to the fluxes and to the source term (which may represent a gain or a loss) associated with the quantity $\vectorIA{U}$. The functions $\vectorIA{U}_0(\vectorr{x})$ and $\vectorIA{B}(t,\vectorr{x})$ denote respectively the initial conditions and the boundary conditions prescribed on the boundary of the domain $\partial \Omega$. Moreover, $\mathcal{B}$ represents an operator applied on the boundary function, which is typically a derivative for Neumann boundary conditions or the identity operator for Dirichlet conditions. In the case of the Euler equations~\eqref{eq:euler}, these terms take the following form:
\begin{align*}
    \vectorIA{U} = 
    \begin{pmatrix}
        \rho \\
        \rho \vectorr{v} \\
        \mathrm{E}
    \end{pmatrix} &&;&&
    \vectorIA{F}[\vectorIA{U}] = 
    \begin{pmatrix}
        \rho \vectorr{v}\\
        \rho \vectorr{v} \otimes \vectorr{v} + p \identity   \\
        (\mathrm{E}+p) \vectorr{v}
        \end{pmatrix}
        &&;&& 
        \vectorIA{S}[\vectorIA{U}] = 0  \;\;\mathpunct{.}
\end{align*}

In the case of radiative hydrodynamics, using the M1-gray model for the description of radiation, these terms become~\eqref{eq:hydro_rad_gray}:
\begin{align*}
    \vectorIA{U} = 
    \begin{pmatrix}
        \rho \\
        \rho \vectorr{v} \\
        \mathrm{E} \\
        \mathrm{E}_R \\
        c^{-2} \vectorr{F_R}
    \end{pmatrix} &&;&&
    \vectorIA{F}[\vectorIA{U}] = 
    \begin{pmatrix}
        \rho \vectorr{v}\\
        \rho \vectorr{v} \otimes \vectorr{v} + p \identity \\
        (\mathrm{E}+p) \vectorr{v} \\
        \vectorr{F_R} \\
        \tensorr{P}_R \\
    \end{pmatrix}
    &&;&& 
        \vectorIA{S}[\vectorIA{U}] = 
        \begin{pmatrix}
        0\\
        \vectorr{S} \\
        c S^0 \\
        -c S^0 \\
        -\vectorr{S} \\
    \end{pmatrix} \;\;\mathpunct{.}
\end{align*}

Unlike traditional numerical methods, which rely on an explicit discretization of space and time, \gls{pinn}s adopt a so-called mesh-free approach: the solution is approximated continuously by a neural network, without an underlying mesh. This makes it possible to avoid certain classical limitations such as discretization errors, numerical stiffness, or stability constraints on the time step.

The originality of \gls{pinn}s lies mainly in two key elements: the definition of their loss function and the strategy used to select training points in the space–time domain. These aspects will be discussed in greater detail in the following sections. For a detailed overview of methodological variants, limitations, and possible improvements, the reader may refer to the Master's thesis of Wagenaar (2023)~\cite{wagenaar_2023}.

\subsection{Cost function} \label{sec:loss}

In \gls{pinn}s, the loss function plays a central role. It guides the learning of the neural network by penalizing deviations between the solution approximated by the network and the constraints imposed by the physical problem — differential equations, initial conditions, or even boundary conditions when these are used.

For a system of \gls{pde}s of the form~\eqref{eq:PDE_formulation}, we denote by $\vectorIA{U}_\theta(t,\vectorr{x})$ the approximation of the solution by a neural network parameterized by a vector $\theta$ (the weights and biases). The total loss function $\mathcal{L}$ can then be expressed as a weighted sum of several contributions:
\begin{equation}
    \label{eq:loss_general}
    \mathcal{L} = \omega_{IC} \mathcal{L}_{IC} +  \omega_{BC} \mathcal{L}_{BC} +  \omega_{eq} \mathcal{L}_{eq} \;\;\mathpunct{,}
\end{equation}

\noindent where:
\begin{itemize}
    \item $\mathcal{L}_{IC}$ is the loss on the initial conditions,
    \item $\mathcal{L}_{BC}$ is the loss on the boundary conditions,
    \item $\mathcal{L}_{eq}$ measures the residual of the differential equations (\gls{pde}),
    \item $\omega_{IC}$, $\omega_{BC}$, and $\omega_{eq}$ are weights adjusting the relative importance of each term.
\end{itemize}

In the literature, it is common to impose $\omega_{IC} \gg \omega_{eq}$ in order to force the network to properly satisfy the initial conditions before minimizing the residual of the physical equations~\cite{papados_2021}. A typical value is $\omega_{IC} = 100~\omega_{eq}$. As for $\omega_{BC}$, few authors specify its optimal value, but it is generally chosen such that $\omega_{IC} \ge \omega_{BC} > \omega_{eq}$.

\starsect{Initial conditions}

The loss function $\mathcal{L}_{IC}$ quantifies the error between the prescribed initial values and those predicted by the network. Evaluated over a set of $N_{IC}$ points $(t_0, \vectorr{x}_i)$, it is written as:
\begin{equation}
    \label{eq:loss_IC}
    \mathcal{L}_{IC} = \frac{1}{N_{IC}} \sum_{i=1}^{N_{IC}} \left \{ \vectorIA{U}_\theta(t_0, \vectorr{x}_i) - \vectorIA{U}_0(\vectorr{x}_i) \right \}^2 \;\;\mathpunct{,}
\end{equation}

\noindent where $\vectorIA{U}_0$ denotes the set of initial values of the function being sought.

\starsect{Boundary conditions}

The formulation of the term $\mathcal{L}_{BC}$ depends on the type of boundary condition considered. Suppose that $N_{BC}$ points are used to evaluate the error associated with these conditions. For example, in the case of a Neumann condition imposing a zero normal derivative on the boundary, one may use:
\begin{equation}
    \label{eq:loss_BC1}
    \mathcal{L}_{BC} = \frac{1}{N_{BC}} \sum_{i=1}^{N_{BC}} \partial_{\vectorr{n}} \vectorIA{U}_\theta(t_i, \vectorr{x}_i) ^2 \;\;\mathpunct{,}
\end{equation}

\noindent where $\partial_{\vectorr{n}}$ is the derivative along the outward normal direction at the boundary. In the case of a Dirichlet condition imposing a specific value on the boundary, one may use the loss function:
\begin{equation}
    \label{eq:loss_BC2}
    \mathcal{L}_{BC} = \frac{1}{N_{BC}} \sum_{i=1}^{N_{BC}} \left \{ \vectorIA{U}_\theta(t_i, \vectorr{x}_i) - \vectorIA{U}_{\partial \Omega} (\vectorr{x}_i) \right \}^2 \;\;\mathpunct{,}
\end{equation}

\noindent where $\vectorIA{U}_{\partial \Omega} (\vectorr{x}_i)$ is the imposed value at the boundary point $\vectorr{x}_i$.

\starsect{\glslink{pde}{\glsentrylong{pde} (\glsentryshort{pde})}}

The term $\mathcal{L}_{eq}$ penalizes the failure of the neural network to satisfy the \gls{pde}s. Considering $N_{eq}$ points in the space–time domain, the standard formulation is:
\begin{equation}
    \label{eq:loss_PDE}
    \mathcal{L}_{eq} = \frac{1}{N_{eq}} \sum_{i=1}^{N_{eq}} \left \{ \partial_t \vectorIA{U}_{\theta,i} + \nablav \cdot \vectorIA{F}[\vectorIA{U}_{\theta,i}] - \vectorIA{S}[\vectorIA{U}_{\theta,i}] \right \}^2 \;\;\mathpunct{,}
\end{equation}

\noindent where $\vectorIA{U}{\theta,i} = \vectorIA{U}\theta(t_i,\vectorr{x}_i)$ is the neural-network prediction at the space–time coordinates $(t_i,\vectorr{x}_i)$. This formulation enables a pointwise evaluation of the differential-equation residual.

\begin{figure}
    \begin{center}
        \begin{minipage}[t]{\linewidth}
            \centering
            \includegraphics[width=\textwidth]{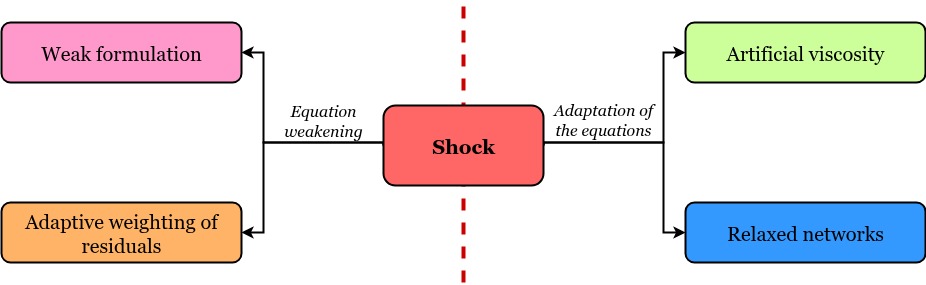}
        \end{minipage}
         \caption{Overview of different adaptations of the differential equation loss $\mathcal{L}_{eq}$ for addressing the shock problem.}
        \label{fig:loss_context}
    \end{center}
\end{figure}

However, it shows its limitations in cases where the solution contains discontinuities, such as shocks in the Euler equations. Since standard activation functions are continuous, the solution learned by the network will also be continuous, making it difficult to approximate a discontinuous profile. Moreover, at discontinuities, the local \gls{pde}s are no longer strictly valid, as mentioned in Section~\secref{sec:hydrodynamique}. This amounts to looking for an entropic solution, whereas a shock corresponds to a weak non-entropic solution. To overcome these difficulties, several strategies have been proposed, summarized in figure~\ref{fig:loss_context}.
\begin{enumerate}
    \item \textbf{Weak formulation}\\
    A first approach consists of using an integral formulation of the \gls{pde}s~\cite{kharazmi_2021, deryck_2024}, which mitigates the effects of discontinuities:
    \begin{equation}
        \label{eq:loss_PDE_VPINN}
        \mathcal{L}_{eq} = \frac{1}{N_{eq}} \sum_{i=1}^{N_{eq}} \left \{ \int \nolimits_\Omega \int \nolimits_{t_0}^{t_f} \left ( \partial_t \vectorIA{U}_\theta(t,\vectorr{x}) + \nablav \cdot \vectorIA{F}[\vectorIA{U}_\theta(t,\vectorr{x})] - \vectorIA{S}[\vectorIA{U}_\theta(t,\vectorr{x})] \right ) \phi_i(t, \vectorr{x}) \dif t \dif \vectorr{x} \right \}^2 \;\;\mathpunct{,}
    \end{equation}
    where $\phi_i$ are test functions that vanish outside the domain. By taking $\phi_i$ as Dirac distributions, one recovers the classical formulation~\ref{eq:loss_PDE}. This approach improves robustness to discontinuities but significantly increases computational cost.
    
    \item \textbf{Adaptive weighting of residuals}\\
    The impact of discontinuities can also be reduced by weighting the importance of the differential equations according to the gradient, using a factor~$\lambda_i$:
    \begin{equation}
        \label{eq:loss_PDE_WE}
        \mathcal{L}_{eq} = \frac{1}{N_{eq}} \sum_{i=1}^{N_{eq}} \lambda_i \left \{ \partial_t \vectorIA{U}_\theta(t_i,\vectorr{x})_i + \nablav \cdot \vectorIA{F}[\vectorIA{U}_{\theta,i}] - \vectorIA{S}[\vectorIA{U}_{\theta,i}] \right \}^2 \;\;\mathpunct{,}
    \end{equation}
    Two possible choices for $\lambda_i$:
    \begin{enumerate}
        \item \textbf{Liu (2023)~\cite{liu_2023}:} reduction of the importance of compressible regions
        \begin{equation}
            \label{eq:l_Liu}
            \lambda_i = \frac{1}{1 + \epsilon \left ( \left | \nablav \cdot \vectorr{v} \right | - \nablav \cdot \vectorr{v} \right )} \;\;\mathpunct{,}
        \end{equation}
        where $\epsilon$ is a tunable hyperparameter and $\vectorr{v}$ the fluid's velocity,
        \item \textbf{Ferrer-Sánchez (2024)~\cite{ferrer-sanchez_2024}:} more general expression for the one-dimensional case:
        \begin{equation}
            \label{eq:l_FS}
            \lambda_i = \frac{1}{1 + \alpha_1 \left |\partial_x \rho \right |^{\eta_1} + \alpha_2 \left |\partial_x v \right |^{\eta_2} + \alpha_3 \left |\partial_x p \right |^{\eta_3}} \;\;\mathpunct{,}
        \end{equation}
        where $\alpha_1$, $\alpha_2$, $\alpha_3$, $\eta_1$, $\eta_2$, $\eta_3$ are hyperparameters to be tuned.
    \end{enumerate}
     This approach is fast and significantly improves shock prediction, but requires careful tuning of the hyperparameters, and it may result to poor predictions of the shock dynamics.
    
    la\item \textbf{Artificial viscosity}\\
    Another option consists in adding a diffusive term to the \gls{pde}s to smooth the discontinuity:
    \begin{equation}
        \label{eq:loss_PDE_visc}
        \mathcal{L}_{eq} = \frac{1}{N_{eq}} \sum_{i=1}^{N_{eq}} \left \{ \partial_t \vectorIA{U}_{\theta,i} + \nablav \cdot \vectorIA{F}[\vectorIA{U}_{\theta,i}] - \vectorIA{S}[\vectorIA{U}_{\theta,i}] - \nu(t_i, x_i) \Delta \vectorIA{U}_{\theta,i} \right \}^2 \;\;\mathpunct{,}
    \end{equation}
    with $\nu$ a viscosity coefficient, which may be constant, parameterized, or learned~\cite{coutinho_2023}. This method requires computing second derivatives, which increases the training cost.
    
    \item \textbf{Relaxed networks}\\
    Finally, some approaches propose relaxing the strict definition of the fluxes~\cite{zhou_2024}, by predicting the fluxes separately using a second neural network:
    \begin{equation}
        \label{eq:loss_PDE_relax_tot}
        \mathcal{L}_{eq} = \mathcal{L}_{eq}^* + \epsilon_{dissip} \mathcal{L}_{dissip} \;\;\mathpunct{,}
    \end{equation}
    where $\epsilon_{dissip}$ is a tunable hyperparameter and:
    \begin{align}
        \mathcal{L}_{eq}^*  &= \frac{1}{N_{eq}} \sum_{i=1}^{N_{eq}} \left \{  \partial_t \vectorIA{U}_{\theta,i} + \nablav \cdot \vectorIA{F}_{\theta,i} - \vectorIA{S}[\vectorIA{U}_{\theta,i}] \right \}^2 \;\;\mathpunct{,} \label{eq:loss_PDE_relax_eq}\\
        \mathcal{L}_{dissip} &=  \frac{1}{N_{eq}} \sum_{i=1}^{N_{eq}} \left \{ \vectorIA{F}_{\theta,i} - \vectorIA{F}[\vectorIA{U}_{\theta,i}] \right \}^2 \;\;\mathpunct{,} \label{eq:loss_PDE_relax_dissip}
    \end{align}
    where $\vectorIA{F}_{\theta,i}$ is predicted directly by a network. In some cases, this formulation improves the quality of the solution, at the price of increased computational cost and additional hyperparameters to tune.
\end{enumerate}

\subsection{Selecting data}

Although \gls{pinn}s provide a continuous solution in the space–time domain, it is nevertheless necessary to discretize this domain in order to evaluate the different components of the loss function. Three distinct datasets must be generated to train the neural network:
\begin{itemize}
    \item one set for evaluating the initial conditions,
    \item one set for evaluating the boundary conditions,
    \item one set for evaluating the \gls{pde}s inside the domain.
\end{itemize}

The first two sets are generally created by random sampling, which is often sufficient to capture the essential information. By contrast, the selection of the points used for evaluating the \gls{pde}s requires particular attention.

As in classical numerical methods, where mesh quality strongly influences the accuracy of the solution, the choice and distribution of training points (in both space and time) have a crucial impact on the convergence and performance of \gls{pinn}s. However, unlike traditional methods, this constraint is not as explicit: simply increasing the number of points does not necessarily improve the solution. Their distribution must be carefully designed according to the expected physical dynamics.

This stems from the formulation of the \gls{pinn} loss function, which relies primarily on minimizing the residuals of the differential equations rather than explicitly comparing with a reference solution. This approach may lead to difficulties in correctly satisfying initial or boundary conditions and in ensuring proper propagation of information throughout the domain. In some cases, it may even violate the principle of causality, thereby disturbing the temporal propagation of the initial conditions~\cite{wang_2024}.

\begin{figure}
    \begin{subfigure}[t]{0.32\textwidth}
        \centering
        \includegraphics[width=\textwidth]{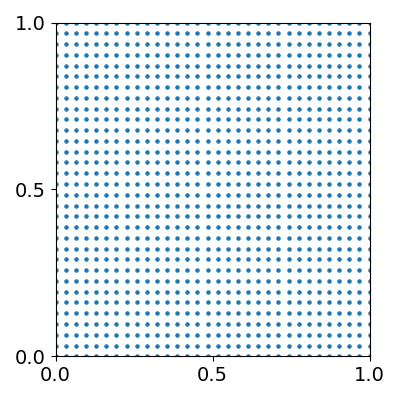}
        \caption{Grid sampling}
        \label{fig:grid}
    \end{subfigure}
    \hfill
    \begin{subfigure}[t]{0.32\textwidth}
        \centering
        \includegraphics[width=\textwidth]{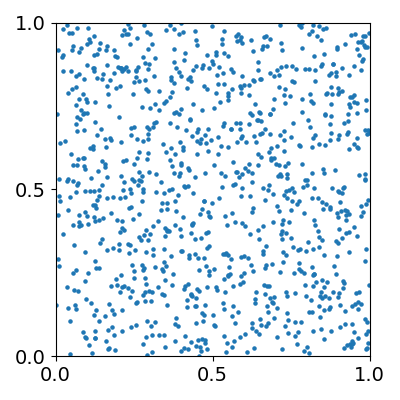}
        \caption{Uniform random sampling}
        \label{fig:uniform}
    \end{subfigure}
    \hfill
    \begin{subfigure}[t]{0.32\textwidth}
        \centering
        \includegraphics[width=\textwidth]{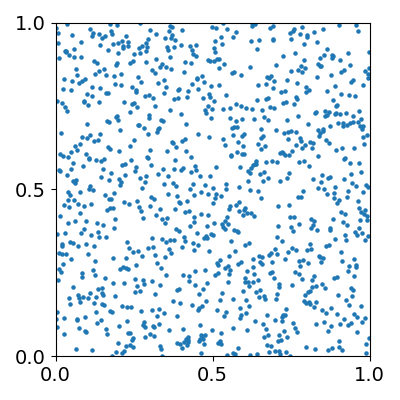}
        \caption{Latin hypercube sampling}
        \label{fig:lhs}
    \end{subfigure}
    \hfill
    \begin{subfigure}[t]{0.32\textwidth}
        \centering
        \includegraphics[width=\textwidth]{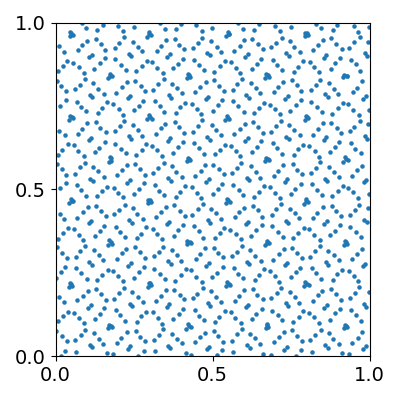}
        \caption{Sobol sampling}
        \label{fig:sobol}
    \end{subfigure}
    \hfill
    \begin{subfigure}[t]{0.32\textwidth}
        \centering
        \includegraphics[width=\textwidth]{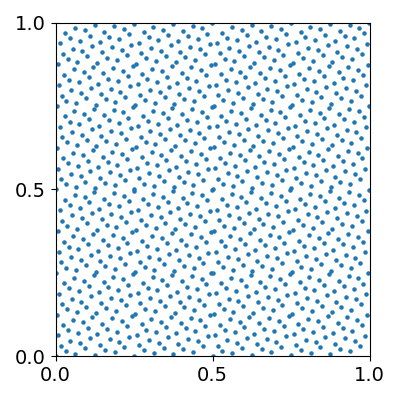}
        \caption{Hammersley sampling}
        \label{fig:hammersley}
    \end{subfigure}
    \caption{Examples of sampling strategies used in \glsxtrshort{pinn}s}
    \label{fig:echantillonnages}
\end{figure}

To address these limitations, several sampling strategies have been developed in the literature. They can be grouped into three broad categories:
\begin{enumerate}
    \item \textbf{Static non-adaptive strategies}

    These methods use a fixed set of training points, chosen before the start of training.
    \begin{itemize}
        \item \textbf{Grid sampling (figure~\ref{fig:grid}):} consists in regularly selecting spatio-temporal points $(t,x)$, similarly to a classical Cartesian mesh. This method is simple but generally yields the poorest results.
        \item \textbf{Uniform random sampling (figure~\ref{fig:uniform}):} points are selected independently and randomly according to a uniform distribution. It slightly improves results compared to grid sampling, but carries a significant risk of undersampling in certain regions.
        \item \textbf{Latin hypercube sampling (figure~\ref{fig:lhs}):} ensures better statistical coverage of the domain than uniform sampling, while retaining randomness. Although widely used in the literature~\cite{raissi_2019, rao_2020, wu_2023}, its performance remains comparable to that of uniform sampling.
        \item \textbf{Sobol sampling (figure~\ref{fig:sobol}):} a quasi-random, low-discrepancy method that aims to minimize the variance of the integral estimator in the loss function~\eqref{eq:loss_general}. It provides a more homogeneous coverage of the space, accelerates convergence, improves accuracy, and ensures better representation of the global domain.
        \item \textbf{Hammersley sampling (figure~\ref{fig:hammersley}):} another quasi-random, low-discrepancy method, constructed from a Hammersley sequence. It offers a particularly uniform distribution of points in the training space, while remaining simple to generate. Its coverage properties are comparable to those of the Sobol sequence and it is often used as an efficient alternative for reducing variance and improving the convergence of \glspl{pinn}.
    \end{itemize}
    
    \item \textbf{Non-adaptive resampling strategies}

    In this approach, the training points are periodically renewed (for example, every $n$ epochs) using one of the previous schemes. This update helps avoid overfitting to a fixed configuration and significantly improves solution quality~\cite{wu_2023};
    
    \item \textbf{Adaptive strategies:}

    These methods dynamically modify the sampling according to the evolution of training or detected errors.
    \begin{itemize}
        \item \textbf{Residual-based sampling:} points are selected according to the local magnitude of the \gls{pde} residuals, emphasizing poorly resolved regions~\cite{lu_2021}. This method is effective but may under-represent certain regions of the domain. 
        \item \textbf{Residual- and gradient-based sampling:} extends the previous method by also accounting for the local amplitude of solution gradients~\cite{mao_2023}. It is particularly useful for capturing strong-gradient phenomena such as shocks.
        \item \textbf{Probabilistic sampling:} instead of selecting only high-residual points, this approach samples points according to a probability distribution proportional to the local error~\cite{nabian_2021}. It introduces beneficial diversity during training, maintaining better domain coverage while still favoring regions with large errors.
        \item \textbf{Neural-network-based sampling:} uses a generative model to learn an optimal sampling distribution, thereby improving the resolution of complex regions of the domain~\cite{tang_2023}.
    \end{itemize}
\end{enumerate}

To assess the relevance of this approach and identify the methodological choices best suited to radiative-hydrodynamics simulations, I will, in the following section, present in detail the strategies adopted and the tests I performed.

\section{Extrapolation of radiative shock simulations with \texorpdfstring{\glslink{pinn}{Physics-Informed Neural Networks}}{PINNs}} \label{sec:PINN}

In this section, I present the preliminary experiments carried out to evaluate the relevance of \gls{pinn}-based neural networks for simulating radiative-hydrodynamics phenomena. \glspl{pinn} remain an actively developed approach, particularly regarding the treatment of discontinuities such as those that arise in shock waves. This difficulty makes their application to radiative hydrodynamics especially challenging.

The work presented here is therefore exploratory and will require further investigation before it can be fully integrated into radiative-hydrodynamics simulation codes. I would like to thank Corentin Pecontal, an intern at the Observatoire de la Côte d'Azur, for his invaluable assistance in carrying out these experiments while I was in the process of writing this manuscript.

\begin{figure}
    \begin{subfigure}[t]{0.48\textwidth}
        \centering
        \includegraphics[width=\textwidth]{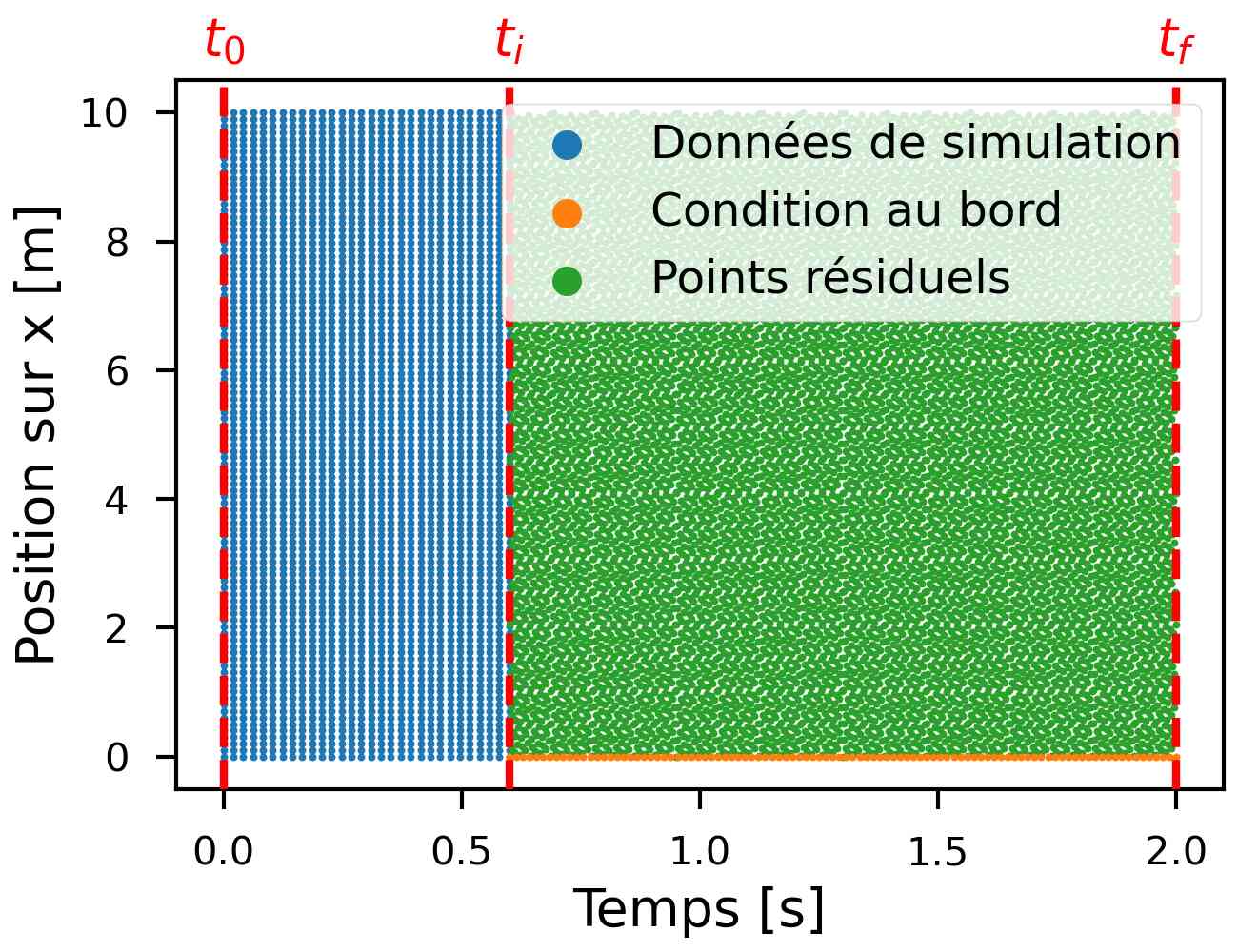}
        \caption{Purely hydrodynamic case.}
        \label{fig:data_hydro}
    \end{subfigure}
    \hfill
    \begin{subfigure}[t]{0.48\textwidth}
        \centering
        \includegraphics[width=\textwidth]{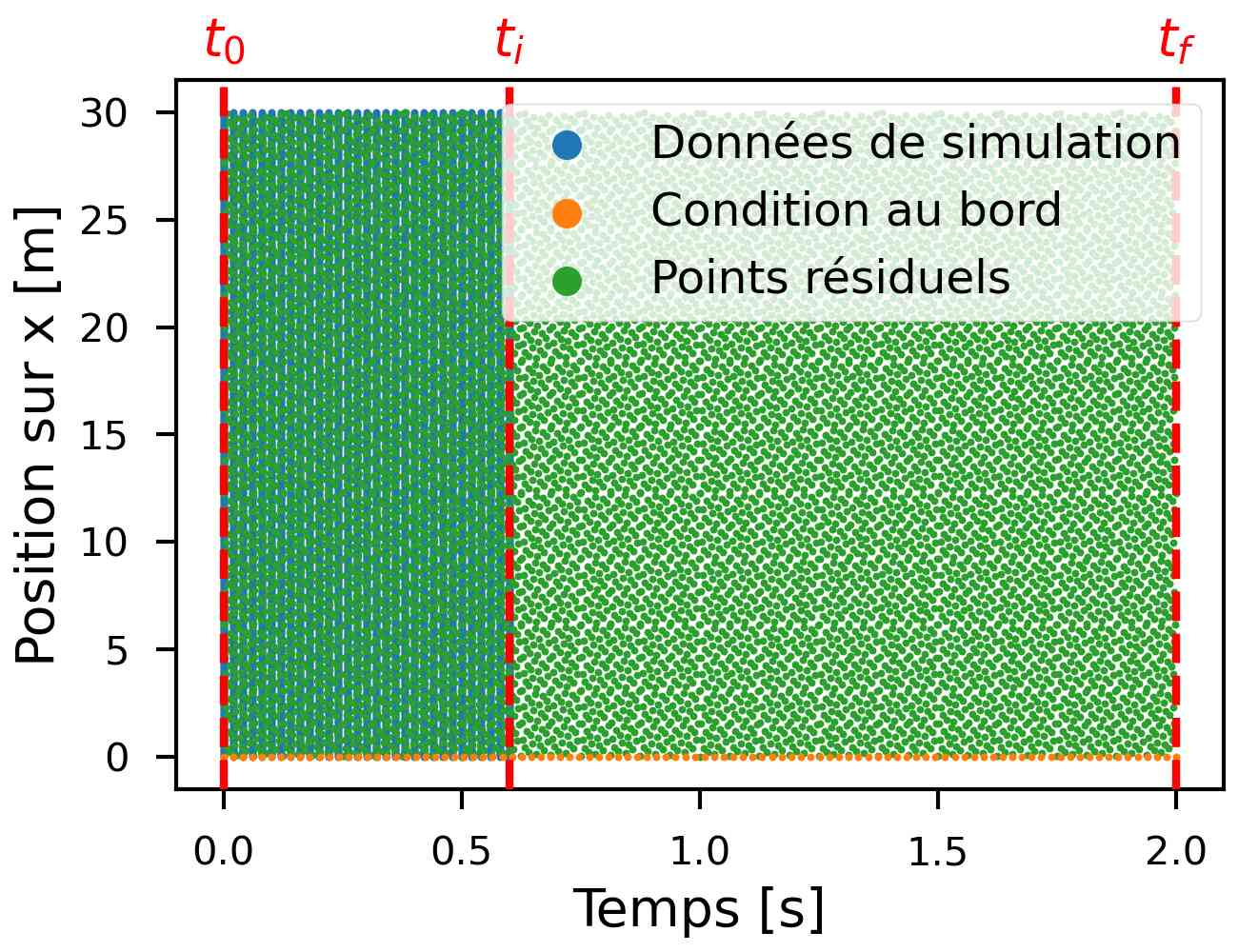}
        \caption{Radiative hydrodynamics case.}
        \label{fig:data_hydrorad}
    \end{subfigure}
    \caption{Distribution of the data used to train the \glsxtrshort{pinn} neural network. The blue points correspond to data originating from the reference simulation, the orange points to the locations where boundary conditions are imposed, and the green points to the positions where the \gls{pde} residuals are evaluated. Time and space are rescaled according to the transformations defined in the sections devoted to the equations in Sections~\secref{sec:PINN_hydro} and \secref{sec:PINN_hydrorad}.}
    \label{fig:PINN_data}
\end{figure}

The strategy adopted here differs slightly from that presented in Section~\secref{sec:PINN}. Rather than being limited to a set of initial conditions specified at time $t_0$, we assume that a dataset from a reference simulation is available, covering a time interval from $t_0$ to $t_i$, sampled at $N_t$ regularly spaced output times. The objective is then to extrapolate the evolution of the system beyond $t_i$ using a \gls{pinn}. To train the network, three datasets are used, corresponding to the sets shown in figure~\ref{fig:PINN_data}:
\begin{enumerate}
    \item a dataset extracted from a reference simulation, acting as the starting point (replacing the classical initial conditions mentioned in Section~\secref{sec:intro_PINNs}). These data cover the time interval $\closeinterv{t_0}{t_i}$;
    \item a dataset imposing the boundary conditions, selected over the time interval $\closeinterv{t_i}{t_f}$ for the purely hydrodynamic test (figure~\ref{fig:data_hydro}), and over $\closeinterv{t_0}{t_f}$ for the radiative-hydrodynamics test (figure~\ref{fig:data_hydrorad});
    \item a set of points on which the \gls{pde} residuals are evaluated, distributed over the time interval $\closeinterv{t_i}{t_f}$ for the purely hydrodynamic test (figure~\ref{fig:data_hydro}), and over $\closeinterv{t_0}{t_f}$ for the radiative-hydrodynamics test (figure~\ref{fig:data_hydrorad}).
\end{enumerate}

We consider here a physical configuration similar to that described in Section~\secref{sec:description_simu}, with the only difference that we study a shock characterized by a higher Mach number. This choice is motivated by our ultimate objective, which is to simulate stationary radiative shocks at very high Mach number, a study that could not be performed in Section~\secref{sec:description_simu} due to computational cost constraints.

The system considered corresponds to a fluid initially characterized by a density $\rho$, a temperature $\mathrm{T}$, and a velocity $v$ directed toward a wall, leading to the formation of a shock that propagates in the opposite direction over time. The fluid, composed of argon, is modeled as a monatomic ideal gas, with adiabatic index \mbox{$\gamma = 5/3$} and mean molecular mass \mbox{$\mu m_H = 39.948~\text{u}$}. The initial conditions of the fluid are summarized in Table~\ref{tab:init_pinn}. In this purely hydrodynamic framework, the resulting shock corresponds to a Mach number of about \mbox{$M \simeq 16.1$}. The only explicitly imposed boundary condition concerns the wall located at \mbox{$x = 0~\text{m}$}, where the fluid velocity is kept equal to zero.

\begin{table}[ht]
    \centering
    \begin{tabular}{C{2.9cm} C{2.9cm} C{2.9cm} C{2.9cm}}
        \hline
        \hline
        Density [kg/$\mathrm{m^3}$] & Velocity [m/s] & Temperature [K] & Temperature [eV] \\
        \hline
        \Tstrut 1 &  -50~000 & 50~000 & 4.3 \\
        \hline
        \hline
    \end{tabular}
    \caption{Initial conditions of the test case used to evaluate the \glsxtrshort{pinn}s.}
    \label{tab:init_pinn}
\end{table}

\subsection{Hydrodynamic tests} \label{sec:PINN_hydro}

First, we performed a series of tests in a purely hydrodynamic regime, with the aim of determining the learning strategy of the neural network and identifying the factors that influence the quality of the extrapolation. This section successively presents the formulation of the equations used, the selected loss function, the data distribution, the network architecture, and finally an analysis of the influence of the learning parameters on the results obtained. The study domain is defined over the spatial interval \mbox{$x \in \closeinterv{x_{\min}}{x_{\max}} = \closeinterv{0}{10}~\text{m}$} and the temporal window \mbox{$t \in \closeinterv{t_0}{t_f} = \closeinterv{0}{2 \times 10^{-4}}~\text{s}$}.

\starsect{The equations}

First of all, we reformulate the Euler equations~\eqref{eq:euler} in a way that is easier to handle with a \gls{pinn}, by explicitly expanding the derivatives of density, temperature, and velocity:
\begin{equation}
    \label{eq:euler_pinn}
    \begin{cases}
        \partial_t \rho + v \partial_x \rho + \rho \partial_x v &=0 \\
        \partial_t v  + v \partial_x v + \mathcal{R} \left ( \mathrm{T} \frac{\partial_x \rho}{\rho} + \partial_x \mathrm{T} \right ) &= 0\\
        \partial_t \mathrm{T} + v \partial_x \mathrm{T} + (\gamma-1) \mathrm{T} \partial_x v &= 0
    \end{cases} \;\;\mathpunct{,}
\end{equation}

\noindent where $\gamma$ is the adiabatic index and $\mathcal{R} = k_B / (\mu m_H)$ is the specific gas constant.

As is often the case in hydrodynamics simulations, a rescaling of the physical quantities is applied in order to reduce differences in magnitude. This normalization facilitates training of the neural network by allowing it to predict values close to unity. The residuals of equations~\eqref{eq:euler_pinn} then read:
\begin{equation}
    \label{eq:euler_pinn_redim}
    \begin{cases}
        \text{res}_\rho(\tilde{t},\tilde{x}) = \partial_{\tilde{t}} \tilde{\rho} + \tilde{v} \partial_{\tilde{x}} \tilde{\rho} + \tilde{\rho} \partial_{\tilde{x}} \tilde{v} \\
        \text{res}_v(\tilde{t},\tilde{x}) = \partial_{\tilde{t}} \tilde{v}  + \tilde{v} \partial_{\tilde{x}} \tilde{v} + \widetilde{\mathcal{R}} \left ( \tilde{\mathrm{T}} \frac{\partial_{\tilde{x}} \tilde{\rho}}{\tilde{\rho}} + \partial_{\tilde{x}} \tilde{\mathrm{T}} \right )\\
        \text{res}_\mathrm{T}(\tilde{t},\tilde{x}) = \partial_{\tilde{t}} \tilde{\mathrm{T}} + \tilde{v} \partial_{\tilde{x}} \tilde{\mathrm{T}} + (\gamma-1) \tilde{\mathrm{T}} \partial_{\tilde{x}} \tilde{v}
    \end{cases} \;\;\mathpunct{,}
\end{equation}

\noindent where the quantities with a tilde are redimensionalized according to:
\begin{align*}
    &\tilde{t} = t/\mathring{t} \;\;\mathpunct{,} && \tilde{x} = x/(\mathring{v} \mathring{t}) \;\;\mathpunct{,} && \tilde{\rho} = \rho / \mathring{\rho} \;\;\mathpunct{,} && \tilde{v} = v / \mathring{v} \;\;\mathpunct{,}  && \tilde{\mathrm{T}} = \mathrm{T} / \mathring{\mathrm{T}} \;\;\mathpunct{,} && \widetilde{\mathcal{R}} = \mathcal{R} \mathring{\mathrm{T}} / \mathring{v}^2 \;\;\mathpunct{,}
\end{align*}

\noindent The constants $\mathring{t}$, $\mathring{\rho}$, $\mathring{v}$, and $\mathring{\mathrm{T}}$ denote the rescaling factors associated with time, density, velocity, and temperature, respectively. In the experiments presented here, we used the following values: $\mathring{t} = 10^{-4}$, $\mathring{\rho} = 1$, $\mathring{v} = 10^{4}$, and $\mathring{\mathrm{T}} = 10^{5}$. For simplicity of notation, we will henceforth omit the tildes, and all quantities will be considered rescaled by default, unless stated otherwise.

\starsect{The cost function}

The cost function used to train the neural network comprises three distinct components: (i) the \gls{pde} residuals, denoted $\mathcal{L}_{PDE}$, (ii) the error between the network predictions and the simulation data, denoted $\mathcal{L}_{sim}$, and (iii) the error on the imposed boundary condition, denoted $\mathcal{L}_{BC}$.

The first component, $\mathcal{L}_{PDE}$, assesses the consistency of the predictions with the underlying physical equations. It is computed using the previously defined residuals, and incorporates the adaptive weighting strategy proposed by Liu (2023)~\cite{liu_2023}, designed to mitigate the impact of the shock discontinuity during training. It is written as:
\begin{align*}
    \mathcal{L}_{PDE} = \frac{1}{N_{PDE}}\sum_{i=1}^{N_{PDE}} \left \{ \frac{\text{res}_\rho(t_i,x_i)^2 + \text{res}_v(t_i,x_i)^2 + \text{res}_\mathrm{T}(t_i,x_i)^2 }{1 + \lambda (|\partial_{\tilde{x}} v_i| - \partial_{\tilde{x}} v_i)} \right \} \;\;\mathpunct{,}
\end{align*}

\noindent where $N_{PDE}$ is the number of residual evaluation points located at $(t_i,x_i)$, $\text{res}_\rho$, $\text{res}_v$, and $\text{res}_{\mathrm{T}}$ denote the \gls{pde} residuals evaluated from the network outputs, and $\lambda$ is a hyperparameter that will be discussed later.

The second component, $\mathcal{L}_{sim}$, measures the discrepancy between the network predictions and the simulation data, using logarithmic or hyperbolic transformations adapted to each variable. It is written as:
\begin{align*}
    \mathcal{L}_{sim} = \frac{1}{N_{sim}}\sum_{i=1}^{N_{sim}} &\left \{  \left [ \log(\rho_{pred}(t_i, x_i)) - \log(\rho_{sim}(t_i, x_i)) \right ]^2 + \right.\\
    &\left. \left [ \log(\mathrm{T}_{pred}(t_i, x_i)) - \log(\mathrm{T}_{sim}(t_i, x_i)) \right ]^2 + \right.\\
    &\left. \left [ \asinh(v_{pred}(t_i, x_i)) - \asinh(v_{sim}(t_i, x_i)) \right ]^2 \right \} \;\;\mathpunct{,}
\end{align*}

\noindent where $N_{sim}$ denotes the number of simulation points $(t_i,x_i)$, and where the functions $\log$ and $\asinh$ correspond respectively to the base-10 logarithm and the inverse hyperbolic sine. These transformations homogenize the relative influence of physical quantities, which often differ by several orders of magnitude, in the error computation.

The third component, $\mathcal{L}_{BC}$, enforces the boundary condition that the velocity must be zero at \mbox{$x = 0~\mathrm{m}$}. It is defined as:
\begin{align*}
    \mathcal{L}_{BC} = \frac{1}{N_{BC}}\sum_{i=1}^{N_{BC}} \asinh(v_{pred}(t_i, 0))^2  \;\;\mathpunct{,}
\end{align*}

\noindent where $N_{BC}$ is the number of time instants $t_i$ at which the condition is evaluated. The use of the $\asinh$ function here ensures consistency with the formulation of $\mathcal{L}_{sim}$.

The total loss function combining these three contributions is:
\begin{equation}
    \label{eq:loss_total}
    \mathcal{L} = \omega_{PDE} \mathcal{L}_{PDE} + \omega_{sim} \mathcal{L}_{sim} + \omega_{BC} \mathcal{L}_{BC} \;\;\mathpunct{,}
\end{equation}

\noindent where the coefficients $\omega_{PDE}$, $\omega_{sim}$, and $\omega_{BC}$ control the relative importance of each term. In this study, we chose the values $\omega_{PDE} = 1$, $\omega_{sim} = 100$, and $\omega_{BC} = 10$, so as to initially orient training toward learning from the simulation data, then impose the boundary conditions, and finally progressively encourage the physical extrapolation of the model through the \gls{pde}s.

\starsect{The data}

\noindent For each component of the loss function, a specific dataset is used (see figure~\ref{fig:data_hydro}):
\begin{enumerate}
    \item \textbf{$\boldsymbol{\mathcal{L}_{sim}}$, simulation data (set of blue points in figure~\ref{fig:data_hydro}):} it is evaluated over the entire spatial domain, but only over the time interval \mbox{$\closeinterv{t_0}{t_i}$}, corresponding to the supervised learning phase. Sampling is regular in space, with $100$ points uniformly distributed in space and $\mathrm{N_T}$ snapshots uniformly distributed in time over \mbox{$\closeinterv{t_0}{t_i}$}. The corresponding data come from the analytical solution of the problem, presented in Appendix~\secref{appendice:choc};

    \item \textbf{$\boldsymbol{\mathcal{L}_{PDE}}$, residual points (set of green points in figure~\ref{fig:data_hydro}):} this component measures the \gls{pde} residuals over the extrapolation region, i.e., for \mbox{$t \in \closeinterv{t_i}{t_f}$} and for all \mbox{$x \in \closeinterv{x_{\min}}{x_{\max}}$}. A total of $10\,000$ \mbox{$(t,x)$} points are randomly sampled in this domain using a Hammersley quasi-random sequence;

    \item \textbf{$\boldsymbol{\mathcal{L}_{BC}}$, boundary conditions (set of orange points in figure~\ref{fig:data_hydro}):} this component enforces the boundary condition at \mbox{$x = 0$} during the extrapolation phase. We used $100$ time points, equally spaced over the interval \mbox{$t \in \closeinterv{t_i}{t_f}$}, all located at \mbox{$x = 0$}.
\end{enumerate}

\noindent This procedure relies on two free parameters defining the training sets: the number of time snapshots $\mathrm{N_T}$ used for supervision on $\closeinterv{t_0}{t_i}$, and the ratio \mbox{$\alpha = (t_i - t_0)/(t_f - t_0)$}, which determines the fraction of the temporal interval allocated to the supervised phase relative to the extrapolation phase. An example configuration of these datasets is illustrated in figure~\ref{fig:data_hydro}, with the choices \mbox{$\mathrm{N_T} = 30$} and \mbox{$\alpha = 0.3$}. The influence of these two parameters on the performance of the \gls{pinn} will be analyzed in the following.

\starsect{The neural network architecture}

\begin{figure}
    \begin{center}
        \begin{minipage}[t]{0.8\linewidth}
            \centering
            \includegraphics[width=\textwidth]{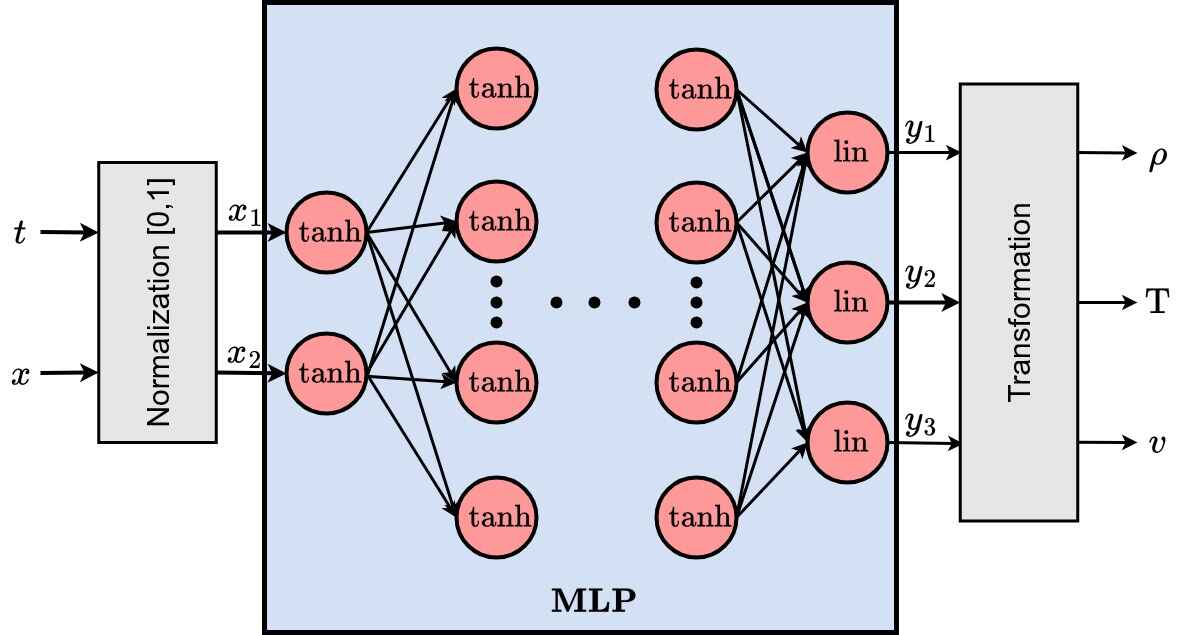}
        \end{minipage}
         \caption{Architecture of the neural network used to represent the hydrodynamic shock with the \glsxtrshort{pinn} strategy. It is composed of three parts: an input-normalization layer, an \glsxtrshort{mlp} neural network, and an output-transformation layer. $\tanh$ and $\mathrm{lin}$ correspond respectively to the hyperbolic-tangent activation function and the identity function.}
        \label{fig:PINN_architecture}
    \end{center}
\end{figure}

The neural network architecture used in these experiments consists of three distinct parts (see figure~\ref{fig:PINN_architecture}): an input-normalization layer, an \gls{mlp} network, and an output-transformation layer.

The purpose of input normalization is to map the temporal and spatial variables into the interval $\closeinterv{0}{1}$, so as to avoid biases in the training of the \gls{mlp} arising from scale differences. The transformations applied to the raw inputs $(t,x)$ are:
\begin{align}
    x_1 &= \frac{t - t_0}{t_f - t_0}\\
    x_2 &= \frac{x - x_{\min}}{x_{\max} - x_{\min}}
\end{align}

The core of the network, based on an \gls{mlp} architecture, consists of an input layer with two neurons, each using the hyperbolic-tangent activation function. This layer is followed by a variable number of intermediate hidden layers, all with the same number of neurons and also using the hyperbolic-tangent activation. Finally, the output layer contains three neurons and uses the identity activation function.

The transformation applied to the outputs of the \gls{mlp} serves two purposes: (i) to guarantee the positivity of physical quantities such as density $\rho$ and temperature $\mathrm{T}$; (ii) to introduce adjustable multiplicative and additive factors for each output, enabling adaptation to the specific orders of magnitude of the predicted variables. The raw outputs of the \gls{mlp}, denoted $(y_1, y_2, y_3)$, are therefore transformed as:
\begin{align}
    \rho &= |w_\rho|~\mathrm{SP}(y_1) + |b_\rho| \;\;\mathpunct{,}\\
    \mathrm{T} &= |w_T|~\mathrm{SP}(y_2) + |b_T| \;\;\mathpunct{,}\\
    v &= |w_v|~y_3 + |b_v| \;\;\mathpunct{,}
\end{align}

\noindent where $\mathrm{SP}$ denotes the SoftPlus function, which guarantees the positivity of its output, and where the coefficients $w_\rho$, $w_T$, $w_v$, $b_\rho$, $b_T$, and $b_v$ are trainable parameters optimized during the training of the neural network.

\begin{figure}
    \begin{center}
        \begin{minipage}[t]{0.7\linewidth}
            \centering
            \includegraphics[width=\textwidth]{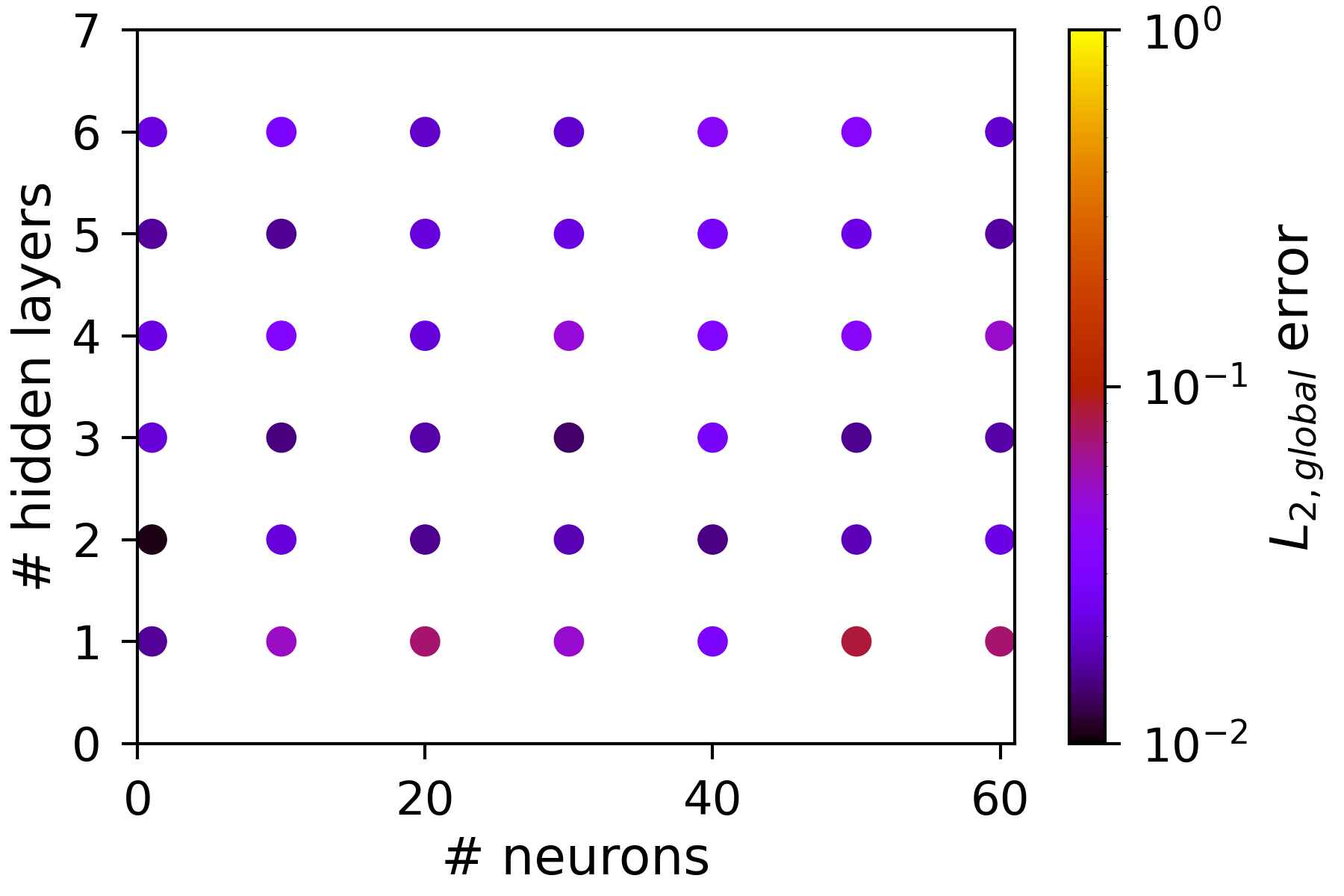}
        \end{minipage}
         \caption{Evaluation of the $L_{2,global}$ error for different numbers of hidden layers and different numbers of neurons per hidden layer.}
        \label{fig:choix_architecture}
    \end{center}
\end{figure}

In order to definitively fix the architecture of the \gls{mlp}, we evaluated its ability to represent a solution exhibiting a sharp discontinuity, characteristic of the shock under consideration, by varying both the number of hidden layers and the number of neurons per hidden layer. To do so, we restricted the training to a purely supervised phase (parameter $\alpha = 1$, cost function $\mathcal{L}=\mathcal{L}_{IC}$), using only data taken from an independent test set composed of regularly spaced points over the space–time domain \mbox{$(t, x) \in \closeinterv{t_0}{t_f} \times \closeinterv{x_{\min}}{x_{\max}}$}. More precisely, we considered a regular Cartesian grid of $256 \times 256$ points. The density, velocity, and temperature profiles used as references come from the analytical solution of the problem, presented in Appendix~\secref{appendice:choc}. To assess the performance of the neural networks, we evaluated the following relative quadratic errors, computed over this test set:
\begin{align}
    L_{2,\rho} &= \frac{\sum_{i=1}^{N_{test}} \left \{ \rho_{pred}(t_i, x_i) - \rho_{sim}(t_i, x_i)\right \}^2}{\sum_{i=1}^{N_{test}} \rho_{sim}(t_i, x_i)^2} \;\;\mathpunct{,} \label{eq:L2r}\\
    L_{2,\mathrm{T}} &= \frac{\sum_{i=1}^{N_{test}} \left \{ \mathrm{T}_{pred}(t_i, x_i) - \mathrm{T}_{sim}(t_i, x_i)\right \}^2}{\sum_{i=1}^{N_{test}} \mathrm{T}_{sim}(t_i, x_i)^2} \;\;\mathpunct{,} \label{eq:L2T}\\
    L_{2,v} &= \frac{\sum_{i=1}^{N_{test}} \left \{ v_{pred}(t_i, x_i) - v_{sim}(t_i, x_i)\right \}^2}{\sum_{i=1}^{N_{test}} v_{sim}(t_i, x_i)^2} \;\;\mathpunct{,} \label{eq:L2v}
\end{align}

\noindent where $X_{pred}$ denotes the quantities predicted by the neural network, $X_{sim}$ the reference quantities obtained from the analytical solution, and $N_{test} = 65~536$ the total number of test points. A global error, denoted $L_{2,global}$, can also be defined as the arithmetic mean of the three previous errors:
\begin{equation}
    L_{2,global} = \frac{L_{2,\rho} + L_{2,\mathrm{T}} + L_{2,v}}{3} \;\;\mathpunct{.} \label{eq:L2glob_hyd}
\end{equation}

The neural networks were trained successively using the \gls{adam} optimizer for $10~000$ epochs (with a learning rate \mbox{$\eta_{\mathrm{Adam}} = 5 \times 10^{-3}$}), followed by the \gls{lbfgs} optimizer for an additional $2~000$ iterations (with a learning rate \mbox{$\eta_{\mathrm{LBFGS}} = 0.1$}). The global errors $L_{2,\mathrm{global}}$ obtained for the different architectures are shown in figure~\ref{fig:choix_architecture}. The results show that all tested configurations are able to represent the discontinuity with relatively low error, except for certain configurations with only a single hidden layer. Consequently, and as a compromise between expressiveness and training cost, we retained an architecture with three hidden layers containing one neuron each for the remainder of the experiments.

\starsect{The impact of the different training parameters}

\begin{figure}
    \begin{subfigure}[t]{0.49\textwidth}
        \centering
        \includegraphics[width=\textwidth]{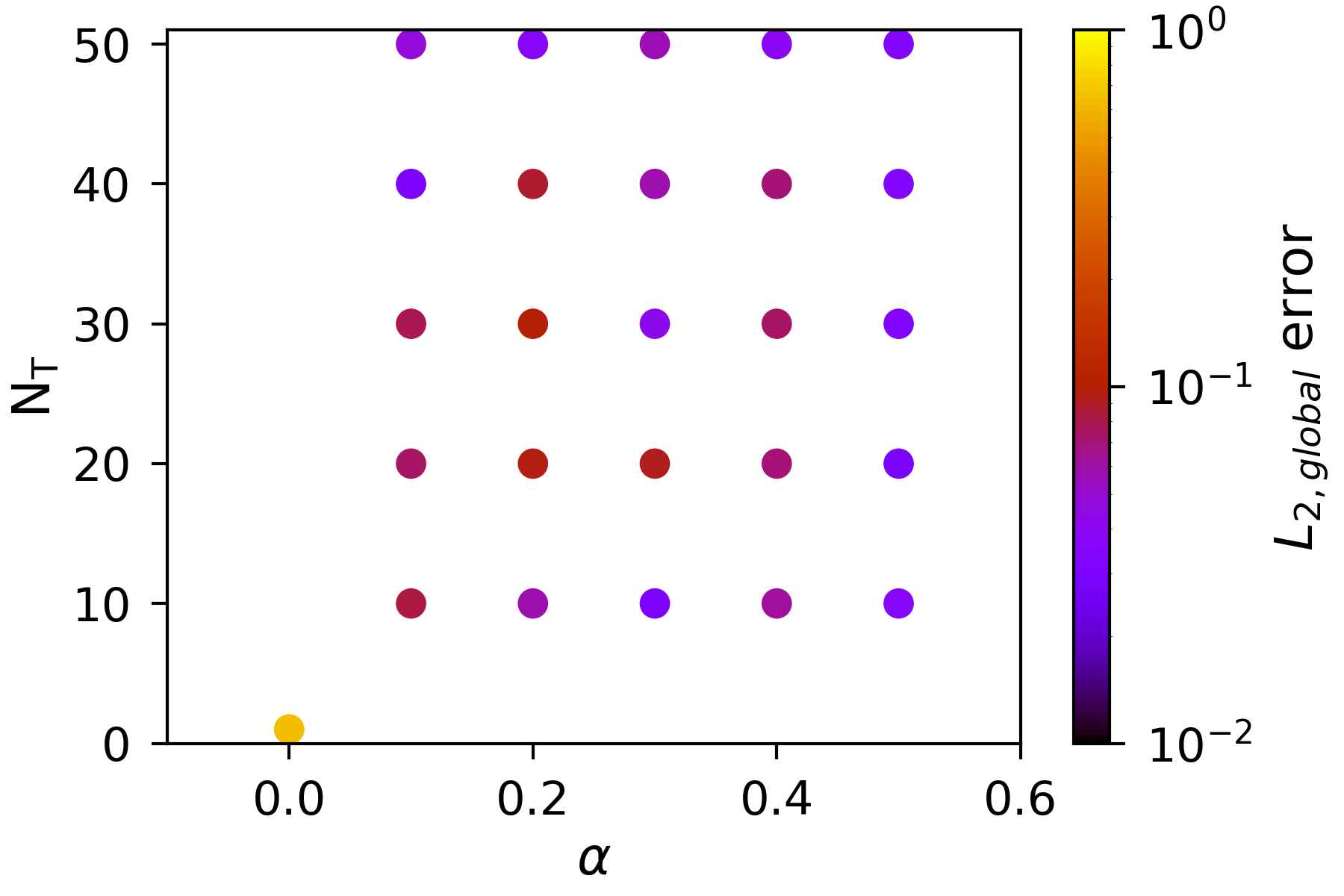}
        \caption{$\lambda = 10^{2}$}
    \end{subfigure}
    \hfill
    \begin{subfigure}[t]{0.49\textwidth}
        \centering
        \includegraphics[width=\textwidth]{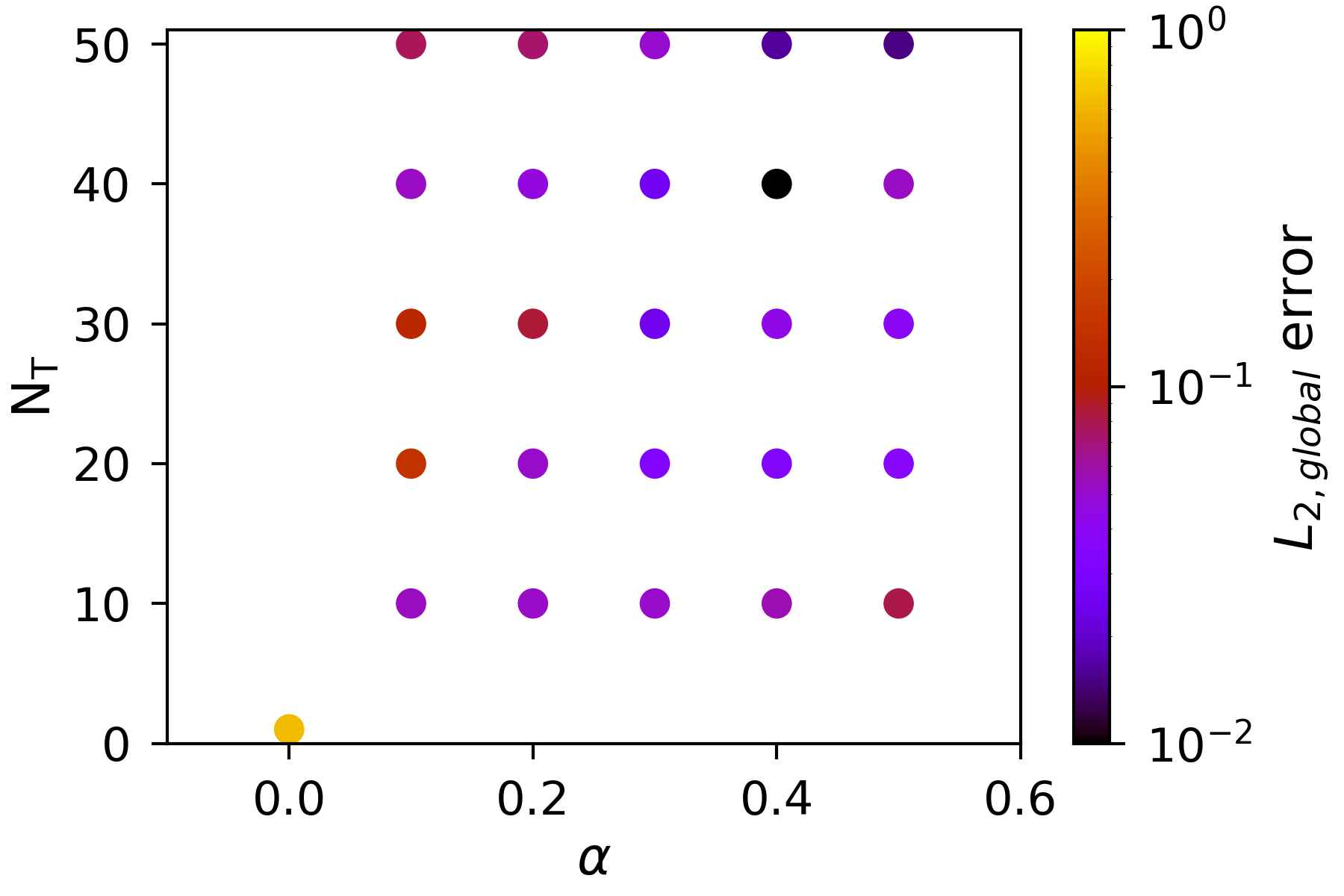}
        \caption{$\lambda = 10^{3}$}
    \end{subfigure}
    \hfill
    \begin{subfigure}[t]{0.49\textwidth}
        \centering
        \includegraphics[width=\textwidth]{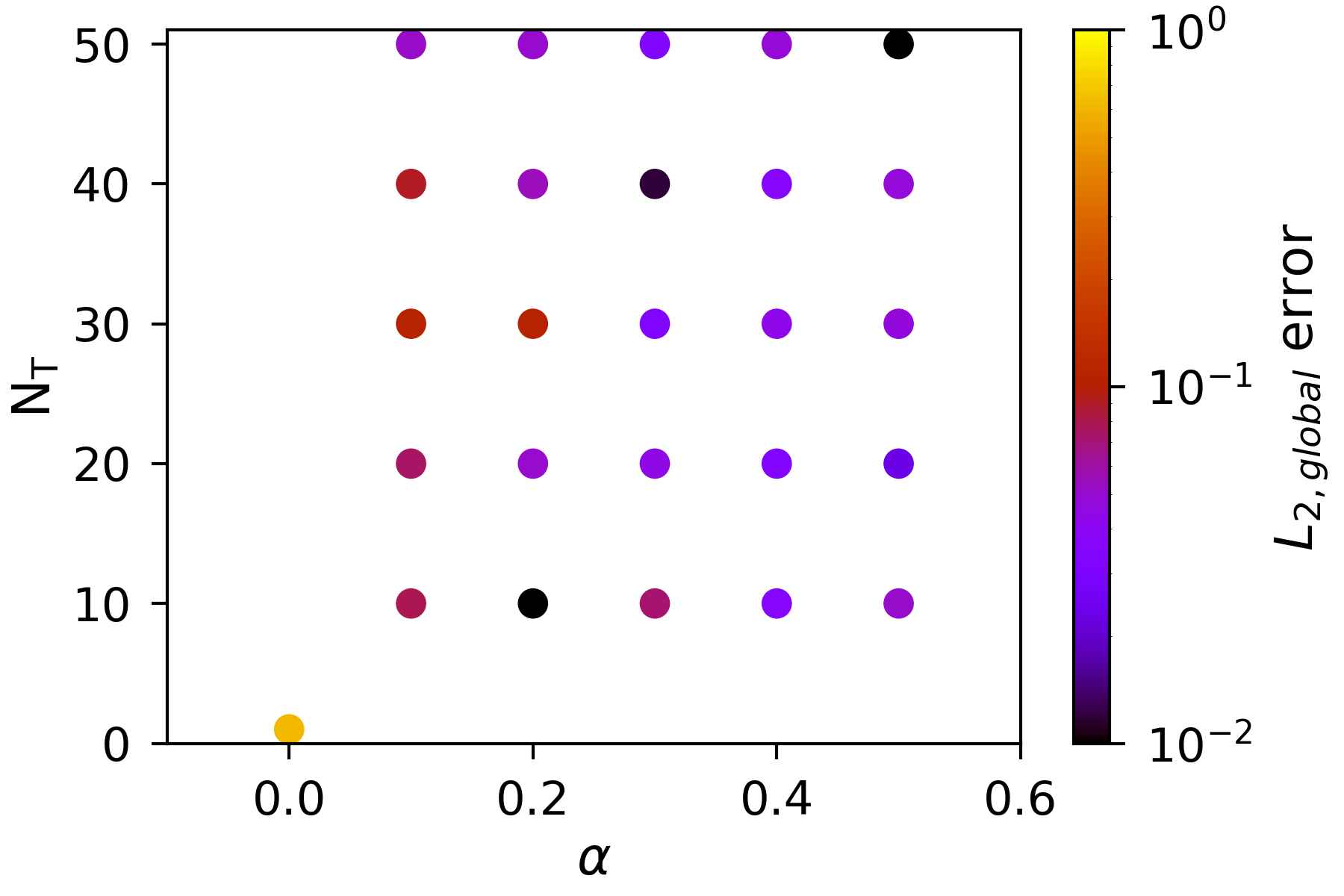}
        \caption{$\lambda = 10^{4}$}
    \end{subfigure}
    \hfill
    \begin{subfigure}[t]{0.49\textwidth}
        \centering
        \includegraphics[width=\textwidth]{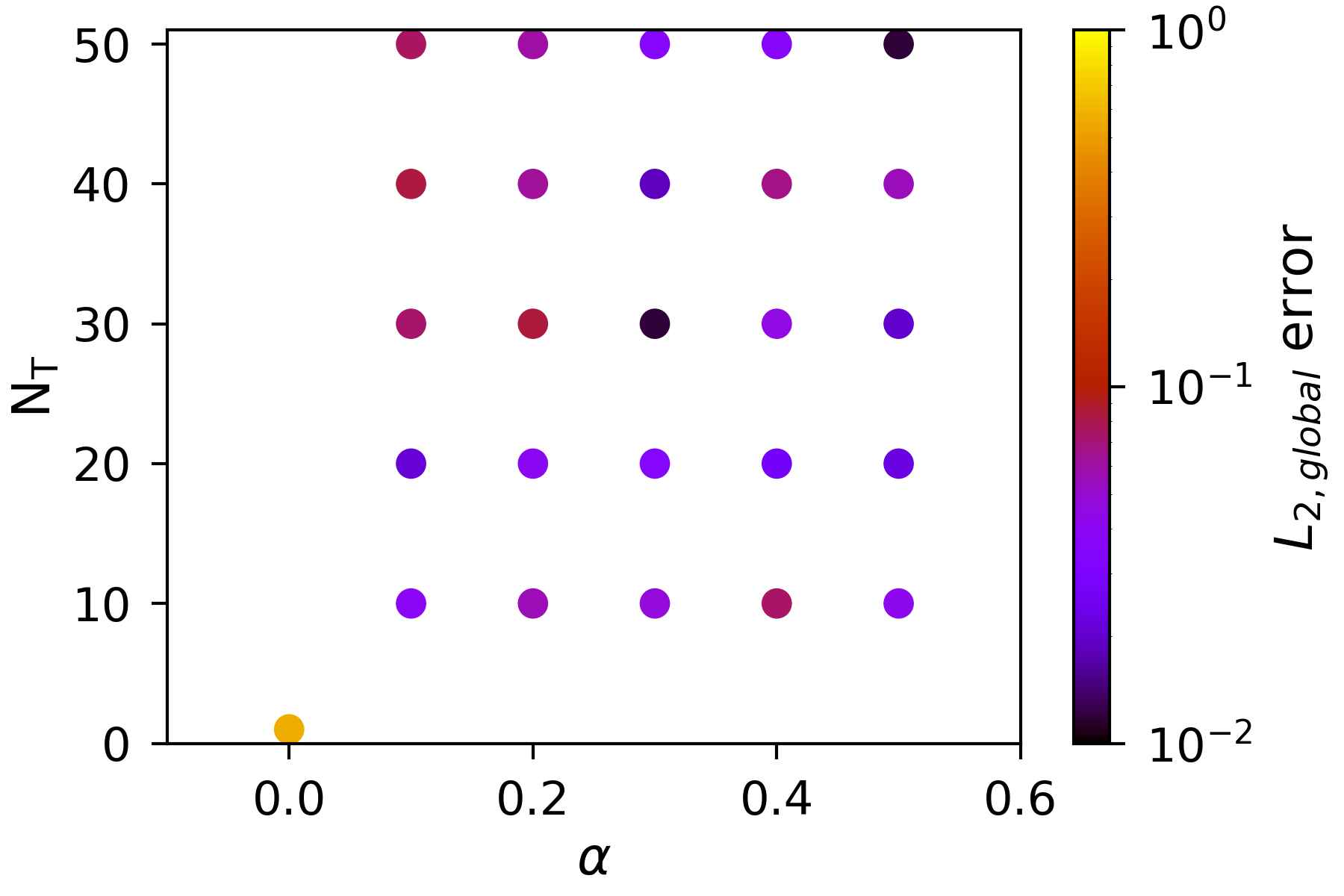}
        \caption{$\lambda = 10^{5}$}
    \end{subfigure}
    \caption{Global $L_{2}$ error for different values of the hyper-parameter $\lambda$, while varying the number of temporal sampling points $\mathrm{N_T}$ and the ratio $\alpha = (t_i - t_0)/(t_f - t_0)$. The simulation with $\mathrm{N_T}=1$ and $\alpha = 0$ corresponds to the classical \glsxtrshort{pinn} strategy, with the adaptive \glsxtrshort{pde} weighting proposed by Liu (2023).}
    \label{fig:params_impact}
\end{figure}

Let us summarize here the three study parameters introduced in the previous sections, whose influence on the performance of the neural networks is analyzed: (i) the hyper-parameter $\lambda$, which appears in the adaptive weighting of the \gls{pde} residuals, (ii) the number of temporal sampling points $\mathrm{N_T}$ used in the simulation data, and (iii) the ratio \mbox{$\alpha$}, which quantifies the relative duration of the supervised phase (training on simulation data) within the total temporal training window.

To study the impact of these parameters, we carried out a systematic exploration by training several neural networks for different combinations of values. Each training run followed the same protocol: an initial optimization phase using the \gls{adam} algorithm for $10~000$ epochs (learning rate \mbox{$\eta_{\mathrm{Adam}} = 10^{-3}$}), followed by a refinement stage using the \gls{lbfgs} algorithm for an additional $2~000$ epochs (learning rate \mbox{$\eta_{\mathrm{LBFGS}} = 10^{-2}$}). For each configuration tested, the retained model is the one that reached the lowest value of the global loss function $\mathcal{L}$.

\begin{figure}
    \begin{subfigure}[t]{0.49\textwidth}
        \centering
        \includegraphics[width=\textwidth]{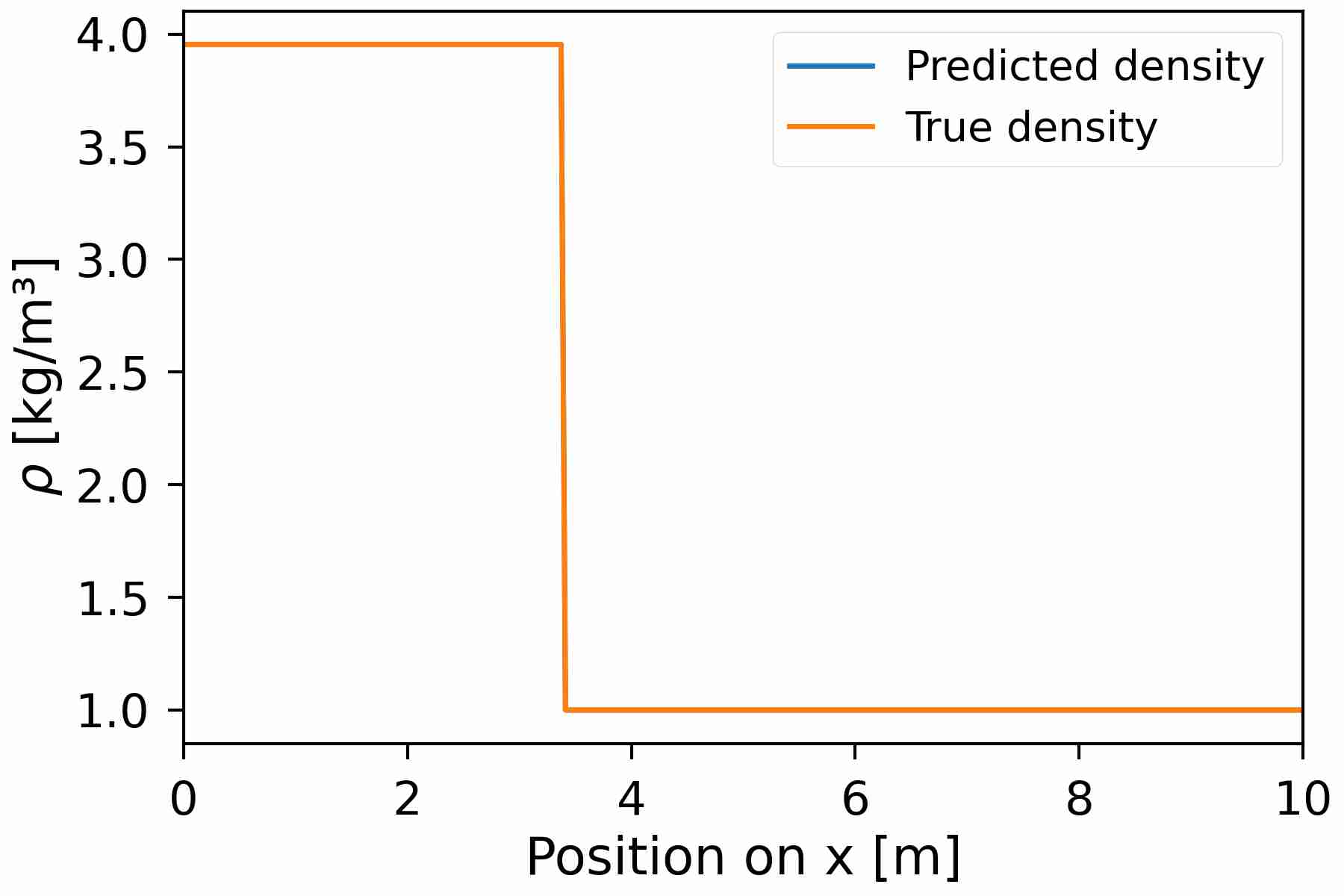}
        \caption{Density}
    \end{subfigure}
    \hfill
    \begin{subfigure}[t]{0.49\textwidth}
        \centering
        \includegraphics[width=\textwidth]{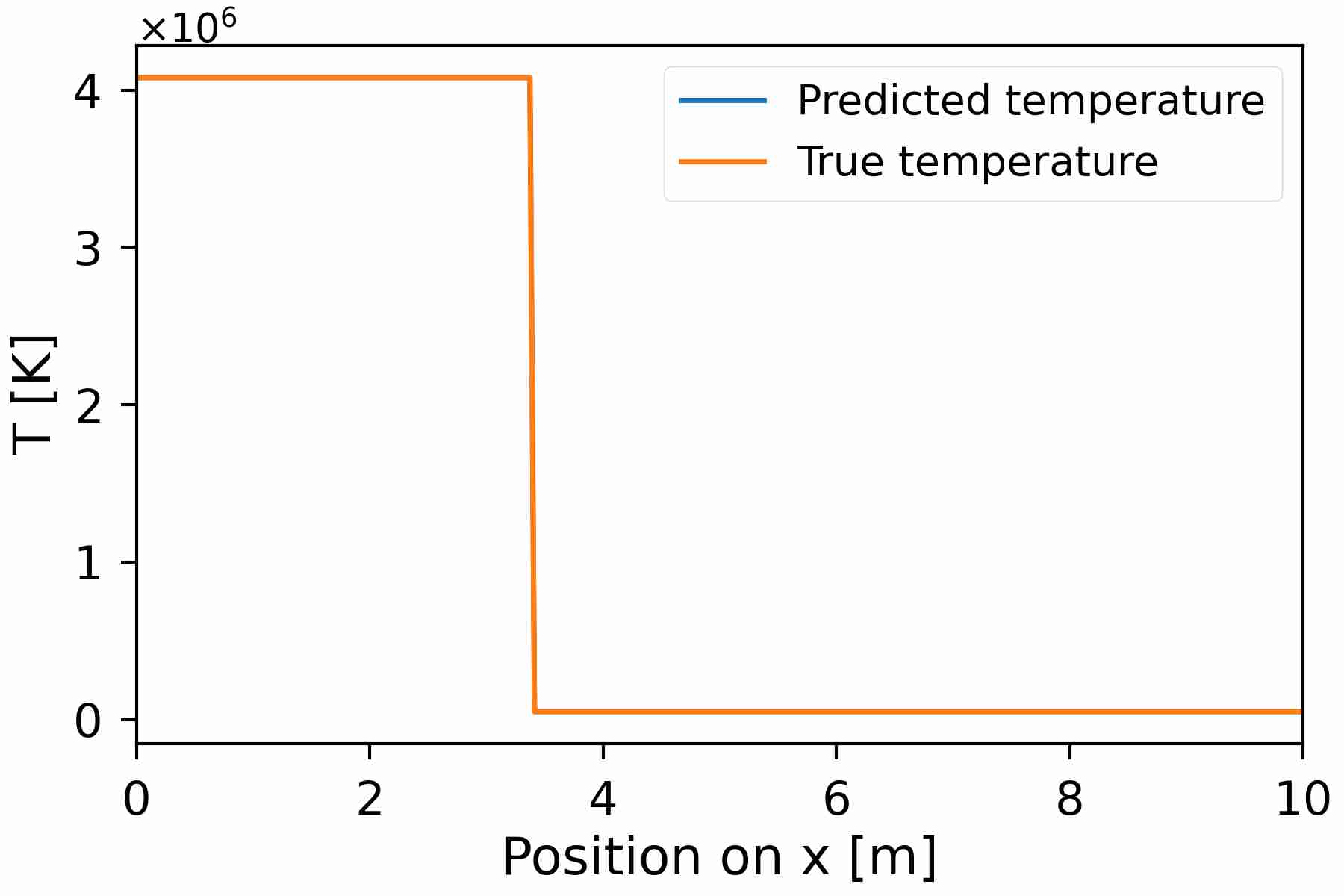}
        \caption{Temperature}
    \end{subfigure}
    \hfill
    \begin{center}
        \begin{subfigure}[t]{0.49\textwidth}
            \centering
            \includegraphics[width=\textwidth]{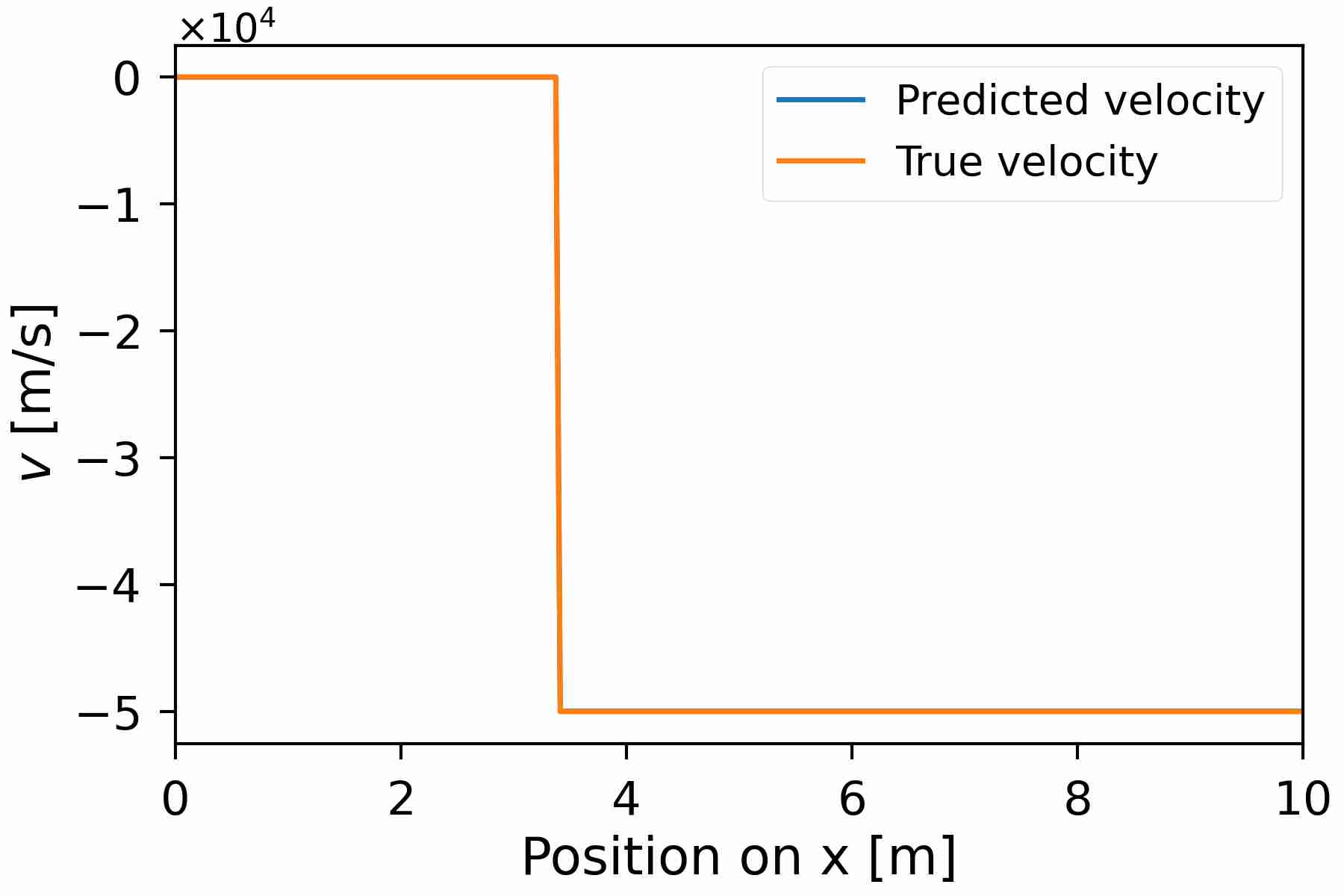}
            \caption{Velocity}
        \end{subfigure}
    \end{center}
    \caption{Comparison of the density, temperature, velocity, radiation temperature, and reduced flux profiles predicted by the neural network and obtained from the simulation, at the final time \mbox{$t_f = 2 \times 10^{-4}~\text{s}$}. The network was trained with the parameters \mbox{$\alpha = 0.1$}, \mbox{$\mathrm{N_T} = 40$}, and \mbox{$\lambda = 10^{3}$}. The variables are shown in their rescaled form.}
    \label{fig:pred_hydro}
\end{figure}

In accordance with the methodology described above, we quantified the accuracy of the predictions using the global error $L_{2,\mathrm{global}}$, computed on the same test set. The results, shown in figure~\ref{fig:params_impact}, indicate that increasing either $\mathrm{N_T}$ or the ratio $\alpha$ tends to significantly improve the quality of the predictions. This highlights a key point: a small value of $\alpha$, corresponding to a short supervised phase, can be compensated for by a higher temporal density of simulation data. In other words, it is not necessary to extend the duration of the supervised phase to ensure reliable extrapolation, provided that it contains a sufficient number of temporal samples. Moreover, the analysis reveals the crucial role of the hyper-parameter $\lambda$: the best performances are consistently obtained for large values, typically \mbox{$\lambda = 10^3$}, suggesting that a stronger weighting of the differential residuals is essential to properly capture the extrapolated system dynamics. As illustrated in figure~\ref{fig:pred_hydro}, the neural network accurately reproduces the shock structure when using $\lambda = 10^{3}$, $\alpha = 0.1$, and $\mathrm{N_T} = 40$. The two curves are indistinguishable, demonstrating the quality of the extrapolation.

Finally, it is worth noting that the so-called \quotes{classical} \gls{pinn} configurations, namely \mbox{$\alpha = 0$} and \mbox{$\mathrm{N_T} = 1$}, systematically yield the poorest results, regardless of the value of $\lambda$. This behavior can be explained by the adaptive weighting strategy proposed by Liu (2023)~\cite{liu_2023}: although it effectively mitigates the impact of discontinuities during training, it tends, in turn, to weaken the representation of shock dynamics when the \gls{pde}-based constraints are not sufficiently informed by supervised data.

\subsection{Radiative hydrodynamic tests}  \label{sec:PINN_hydrorad}

We then applied the same approach in order to extrapolate a radiative shock simulation, while keeping the same initial conditions as before. However, this part of the work remains incomplete due to technical difficulties, that I describe below together with the methodological choices adopted.

The computational domain considered here extends spatially over \mbox{$x \in \closeinterv{x_{\min}}{x_{\max}} = \closeinterv{0}{30}~\text{m}$} and temporally over \mbox{$t \in \closeinterv{t_0}{t_f} = \closeinterv{0}{2 \times 10^{-4}}~\text{s}$}. The enlargement of the spatial domain was motivated by the need to correctly capture the entire radiative precursor. In this configuration, photon diffusion in the material is neglected, and we assume that their mean free path is independent of both radiation frequency and the hydrodynamic conditions of the fluid. This assumption makes it possible to assign the same value to the Planck and Rosseland mean opacities, namely \mbox{$\kappa_R = \kappa_P = 1/\ell$}, where $\ell$ denotes the photon mean free path. For the case studied, we set $\ell = 1~\text{m}$.

In what follows, I successively present the formulation of the radiative hydrodynamics equations employed, the details of the loss function used, the training data adopted, the neural network architecture, and finally the difficulties encountered with the solution currently obtained.

\starsect{The equations}

First, let us rewrite the radiative hydrodynamics equations~\eqref{eq:hydro_rad_mg}, adopting the M1-gray model for the description of radiation, in order to make explicit the time derivatives of the following physical variables: the density $\rho$, the fluid velocity $v$, the gas temperature $\mathrm{T}$, the radiation temperature $\mathrm{T}_R$, and the reduced flux $\mathrm{f}_R$:
\begin{equation}
    \label{eq:hydrorad_pinn}
    \begin{cases}
        \partial_t \rho + v \partial_x \rho + \rho \partial_x v &=0 \\
        \partial_t v + v \partial_x v + \mathcal{R} \left ( \mathrm{T} \frac{\partial_x \rho}{\rho} + \partial_x \mathrm{T} \right ) &= S_v\\
        \partial_t \mathrm{T}   + v \partial_x \mathrm{T} + (\gamma-1) \mathrm{T} \partial_x v &= c S_\mathrm{T}\\
        \partial_t \mathrm{T}_R + c \left \{ \mathrm{f}_R \partial_x \mathrm{T}_R + \frac{\mathrm{T}_R}{4} \partial_x \mathrm{f}_R \right \} &= c S_{\mathrm{T}_R}\\
        \partial_t \mathrm{f}_R + c \left \{ \left ( \chi_R' - \mathrm{f}_R \right ) \partial_x \mathrm{f}_R + \frac{4 \left ( \chi_R - \mathrm{f}_R^2 \right )}{\mathrm{T}_R} \partial_x \mathrm{T}_R \right \} &= c S_{\mathrm{f}_R}
    \end{cases} \;\;\mathpunct{,}
\end{equation}

\noindent where $c$ is the speed of light, $\chi_R$ is the Eddington factor of the M1-gray model, and $\chi_R'$ its first derivative. These two quantities are given by:
\begin{align}
    \chi_R &= \frac{3 + 4 \mathrm{f}_R^2}{5 + 2 \sqrt{4 - 3 \mathrm{f}_R^2}  \;\;\mathpunct{,}}\\
    \chi_R' &=  \frac{2 \mathrm{f}_R}{\sqrt{4-3\mathrm{f}_R^2}}  \;\;\mathpunct{,}
\end{align}

\noindent Moreover, $S_v$, $S_\mathrm{T}$, $S_{\mathrm{T}_R}$ and $S_{\mathrm{f}_R}$ denote the source terms that describe the interaction between radiation and matter. Under the assumptions made concerning the opacities, these source terms can be written, to order~$1$ in $v/c$, as:
\begin{align}
    S_v &= \frac{a_R \mathrm{T}_R^4}{\rho \ell} \left [ \mathrm{f}_R - \frac{v}{c} \left \{ \chi_R + \left (\frac{\mathrm{T}}{\mathrm{T}_R} \right )^4 \right \} \right ] \;\;\mathpunct{,} \label{eq:src_v}\\
    S_\mathrm{T} &=  \frac{(\gamma - 1) a_R \mathrm{T}_R^4}{\rho \mathcal{R} \ell} \left [ 1 - \left (\frac{\mathrm{T}}{\mathrm{T}_R} \right )^4 - \frac{2 v \mathrm{f}_R}{c} \right ] \;\;\mathpunct{,} \label{eq:src_T}\\
    S_{\mathrm{T}_R} &= -\frac{\mathrm{T}_R}{4 \ell} \left [ 1 - \left (\frac{\mathrm{T}}{\mathrm{T}_R} \right )^4 - \frac{v \mathrm{f}_R}{c} \right ] \;\;\mathpunct{,}  \label{eq:src_Tr}\\
    S_{\mathrm{f}_R} &= -\frac{1}{\ell} \left [ \mathrm{f}_R \left (\frac{\mathrm{T}}{\mathrm{T}_R} \right )^4 - \frac{v}{c} \left \{ \chi_R - \mathrm{f}_R^2 + \left (\frac{\mathrm{T}}{\mathrm{T}_R} \right )^4 \right \} \right ] \;\;\mathpunct{,}  \label{eq:src_fr}
\end{align}

\noindent where $a_R$ is the radiation constant. As before, the neural networks are trained using the rescaled version of these radiative hydrodynamics equations. The residuals of the differential equations then take the form:
\begin{equation}
    \label{eq:hydrorad_pinn_res}
    \begin{cases}
        \text{res}_\rho(\tilde{t},\tilde{x})& = \partial_{\tilde{t}} \tilde{\rho} + \tilde{v} \partial_{\tilde{x}} \tilde{\rho} + \tilde{\rho} \partial_{\tilde{x}} \tilde{v} \\
        \text{res}_v(\tilde{t},\tilde{x})& = \partial_{\tilde{t}} \tilde{v} + \tilde{v} \partial_{\tilde{x}} \tilde{v} + \tilde{\mathcal{R}} \left ( \tilde{\mathrm{T}} \frac{\partial_{\tilde{x}} \tilde{\rho}}{\tilde{\rho}} + \partial_{\tilde{x}} \tilde{\mathrm{T}} \right ) - \tilde{S}_v\\
        \text{res}_{\mathrm{T}}(\tilde{t},\tilde{x})& = \tilde{c}^{-1} \partial_{\tilde{t}} \tilde{\mathrm{T}} + \frac{\tilde{v}}{\tilde{c}} \partial_{\tilde{x}} \tilde{\mathrm{T}} + \frac{(\gamma-1) \tilde{\mathrm{T}}}{\tilde{c}} \partial_{\tilde{x}} \tilde{v} - \tilde{S}_\mathrm{T}\\
        \text{res}_{\mathrm{T}_R}(\tilde{t},\tilde{x})& = \tilde{c}^{-1} \partial_{\tilde{t}} \tilde{\mathrm{T}}_R + \mathrm{f}_R \partial_{\tilde{x}} \tilde{\mathrm{T}}_R + \frac{\tilde{\mathrm{T}}_R}{4} \partial_{\tilde{x}} \mathrm{f}_R - \tilde{S}_{\mathrm{T}_R}\\
        \text{res}_{\mathrm{f}_R}(\tilde{t},\tilde{x}) &= \tilde{c}^{-1} \partial_{\tilde{t}} \mathrm{f}_R +  \left ( \partial_{\mathrm{f}_R} \chi_R - \mathrm{f}_R \right ) \partial_{\tilde{x}} \mathrm{f}_R + \frac{4 \left ( \chi_R - \mathrm{f}_R^2 \right )}{\tilde{\mathrm{T}}_R} \partial_{\tilde{x}} \tilde{\mathrm{T}}_R - \tilde{S}_{\mathrm{f}_R}
    \end{cases} \;\;\mathpunct{,}
\end{equation}

\noindent where the rescaled source terms are:
\begin{align*}
    \tilde{S}_v &= \frac{\tilde{a}_R~\tilde{\mathrm{T}}_R^4}{\tilde{\rho}~\tilde{\ell}} \left [ \mathrm{f}_R - \frac{\tilde{v}}{\tilde{c}} \left \{ \chi_R + \left (\frac{\tilde{\mathrm{T}}}{\tilde{\mathrm{T}}_R} \right )^4 \right \} \right ] \;\;\mathpunct{,}\\
    \tilde{S}_\mathrm{T} &=  \frac{(\gamma - 1)~\tilde{a}_R~\tilde{c}~\tilde{\mathrm{T}}_R^4}{\tilde{\rho}~\tilde{\mathcal{R}}~\tilde{\ell}} \left [ 1 - \left (\frac{\tilde{\mathrm{T}}}{\tilde{\mathrm{T}}_R} \right )^4 - \frac{2 \tilde{v} \mathrm{f}_R}{\tilde{c}} \right ] \;\;\mathpunct{,}\\
    \tilde{S}_{\mathrm{T}_R} &= \frac{\tilde{\mathrm{T}}_R}{4~\tilde{\ell}} \left [ 1 - \left (\frac{\tilde{\mathrm{T}}}{\tilde{\mathrm{T}}_R} \right )^4 - \frac{\tilde{v} \mathrm{f}_R}{\tilde{c}} \right ] \;\;\mathpunct{,}\\
    \tilde{S}_{\mathrm{f}_R} &= \frac{1}{\tilde{\ell}} \left [ \mathrm{f}_R \left (\frac{\tilde{\mathrm{T}}}{\tilde{\mathrm{T}}_R} \right )^4 - \frac{\tilde{v}}{\tilde{c}} \left \{ \chi_R - \mathrm{f}_R^2 + \left (\frac{\tilde{\mathrm{T}}}{\tilde{\mathrm{T}}_R} \right )^4 \right \} \right ] \;\;\mathpunct{.}
\end{align*}

\noindent The quantities with a tilde are rescaled variables defined as:
\begin{align*}
    &\tilde{t} = t/\mathring{t} \;\;\mathpunct{,} && \tilde{x} = x/(\mathring{v} \mathring{t}) \;\;\mathpunct{,} && \tilde{\rho} = \rho / \mathring{\rho} \;\;\mathpunct{,} && \tilde{v} = v / \mathring{v} \;\;\mathpunct{,}  && \tilde{\mathrm{T}} = \mathrm{T} / \mathring{\mathrm{T}} \;\;\mathpunct{,} && \widetilde{\mathcal{R}} = \mathcal{R} \mathring{\mathrm{T}} / \mathring{v}^2 \;\;\mathpunct{,}\\
    &\tilde{\mathrm{T}}_R = \mathrm{T}_R / \mathring{\mathrm{T}} \;\;\mathpunct{,} && \tilde{c} = c / \mathring{v} \;\;\mathpunct{,} && \tilde{a}_R = a_R \mathring{\mathrm{T}}^4 / (\mathring{\rho} \mathring{v}^2) \;\;\mathpunct{,}
\end{align*}

\noindent and $\mathring{t}$, $\mathring{\rho}$, $\mathring{v}$ and $\mathring{\mathrm{T}}$ are scaling constants. In what follows, we adopt the values $\mathring{t} = 10^{-4}$, $\mathring{\rho} = 10$, $\mathring{v} = 10^{4}$ and $\mathring{\mathrm{T}} = 10^{5}$. To simplify the notation, we will again omit the tildes hereafter; all quantities will be implicitly considered rescaled unless stated otherwise.

\starsect{The cost function}

As in the hydrodynamic case, the loss function includes the \gls{pde} residual component $\mathcal{L}_{PDE}$, the prediction error relative to the simulation data $\mathcal{L}_{sim}$, and the boundary-condition error $\mathcal{L}_{BC}$. These three components can be written as:
\begin{align*}
    \mathcal{L}_{PDE} = \frac{1}{N_{PDE}}\sum_{i=1}^{N_{PDE}} &\left \{ \frac{\text{res}_\rho(t_i,x_i)^2 + \text{res}_v(t_i,x_i)^2 + \text{res}_\mathrm{T}(t_i,x_i)^2 + \text{res}_{\mathrm{T}_R}(t_i,x_i)^2 + \text{res}_{\mathrm{f}_R}(t_i,x_i)^2}{1 + \lambda (|\partial_{\tilde{x}} v_i| - \partial_{\tilde{x}} v_i)}\right \} \;\;\mathpunct{,}\\
    \mathcal{L}_{sim} = \frac{1}{N_{sim}}\sum_{i=1}^{N_{sim}} &\left \{  \left [ \log(\rho_{pred}(t_i, x_i)) - \log(\rho_{sim}(t_i, x_i)) \right ]^2 + \right.\\
    &\left. ~~\left [ \log(\mathrm{T}_{pred}(t_i, x_i)) - \log(\mathrm{T}_{sim}(t_i, x_i)) \right ]^2 + \right.\\
    &\left. ~~\left [ \asinh(v_{pred}(t_i, x_i)) - \asinh(v_{sim}(t_i, x_i)) \right ]^2  + \right.\\
    &\left. ~~\left [ \log(\mathrm{T}_{R,pred}(t_i, x_i)) - \log(\mathrm{T}_{R,sim}(t_i, x_i)) \right ]^2 + \right.\\
    &\left. ~~\left [ \asinh(\alpha_{\mathrm{f}_R} \mathrm{f}_{R,pred}(t_i, x_i)) - \asinh(\alpha_{\mathrm{f}_R} \mathrm{f}_{R,sim}(t_i, x_i)) \right ]^2 \right \} \;\;\mathpunct{,}\\
    \mathcal{L}_{BC} = \frac{1}{N_{BC}}\sum_{i=1}^{N_{BC}} & \left \{ \asinh(v_{pred}(t_i, 0))^2 + \asinh(\alpha_{\mathrm{f}_R} \mathrm{f}_{R,pred}(t_i, 0))^2 \right \} \;\;\mathpunct{,}
\end{align*}

\noindent where the quantities $X_{pred}$ correspond to the values predicted by the neural network, and the quantities $X_{sim}$ denote the values taken from the simulation data. The parameter $\lambda$ is a training hyper-parameter to be tuned, whereas $\alpha_{\mathrm{f}_R}$ is a factor introduced to increase the weight of the reduced flux in the loss function and, as we will see in the section devoted to the architecture, it also helps the neural network predict values with the correct order of magnitude. Indeed, the values of $\mathrm{f}_R$ are very small, which tends to make their learning more difficult; introducing $\alpha{\mathrm{f}_R}$ therefore compensates for this imbalance. Empirically, we observed that choosing $\alpha{\mathrm{f}_R} = 10^{5}$ yields the best results.

The total loss function retains the same form as that used in the purely hydrodynamic test (see equation~\eqref{eq:loss_total}), and we adopted the same weights \mbox{$\omega_{sim}=10$}, \mbox{$\omega_{PDE}=1$}, and \mbox{$\omega_{BC}=1$}, for the same reasons as in the purely hydrodynamic case.

\starsect{The data}

We performed this simulation using the \gls{hades} code, in a plane-parallel configuration. The spatial domain is discretized into \mbox{$3\,000$ cells} along the $x$–axis and \mbox{$6$ cells} along the $y$–axis. One hundred output times were stored, uniformly distributed over the full duration of the simulation. The box dimensions are \mbox{$\closeinterv{x_{min}=0}{x_{max}=30}$ m} in the $x$ direction and \mbox{$\closeinterv{y_{min}=0}{y_{max}=0.06}$ m} in the $y$ direction, and the evolution is computed from \mbox{$t_0 = 0$ s} up to \mbox{$t_f = 2 \times 10^{-4}$ s}.

As in the purely hydrodynamic case, three distinct datasets are used to train the neural network (see figure~\ref{fig:data_hydrorad}):

\begin{enumerate}
    \item \textbf{\gls{hades} simulation data} (blue points).\\
    These are used to evaluate the cost term $\mathcal{L}_{sim}$. This dataset covers the entire spatial domain and all times within \mbox{$\closeinterv{t_0}{t_i}$}. The network is trained on $\mathrm{N_T}$ timesteps selected in this interval, and all spatial grid points are retained (\mbox{$N_x = 3\,000$}).
    
    \item \textbf{Points for the radiation hydrodynamics equations} (green points).\\
    They are used to evaluate the PDE residuals in the cost term $\mathcal{L}_{PDE}$. These points are distributed over the entire spatial domain and over all times in $\closeinterv{t_0}{t_f}$. We sampled \mbox{$10\,000$ points} randomly in this domain using a quasi-random Hammersley sequence.
    
    \item \textbf{Points for boundary conditions} (orange points).\\
    These are used in the cost term $\mathcal{L}_{BC}$. The points are located at \mbox{$x = 0$ m} and span all times in \mbox{$\closeinterv{t_0}{t_f}$}. We use \mbox{$100$ time points} in this dataset.
\end{enumerate}

In this case, we chose to extend the last two datasets to the entire temporal domain (\mbox{$t \in \closeinterv{t_0}{t_f}$}), rather than restricting them only to the extrapolation interval (\mbox{$t \in \closeinterv{t_i}{t_f}$}), as was done in the purely hydrodynamic test. This modification proved essential: it allows the neural network to correctly reproduce the simulation data and to satisfy the physical equations for times \mbox{$t \in \closeinterv{t_0}{t_i}$}. Without this constraint, the extrapolation obtained beyond $t_i$ becomes completely unphysical. Two parameters therefore need to be explored, as in the purely hydrodynamic case:
\begin{itemize}
    \item \mbox{$\alpha = (t_i - t_0)/(t_f - t_0)$}, which sets the fraction of the time window devoted to supervised training;
    \item $\mathrm{N_T}$, the number of simulation timesteps used in this supervised phase.
\end{itemize}

\starsect{The neural network architecture}

\begin{figure}
    \begin{center}
        \begin{minipage}[t]{0.8\linewidth}
            \centering
            \includegraphics[width=\textwidth]{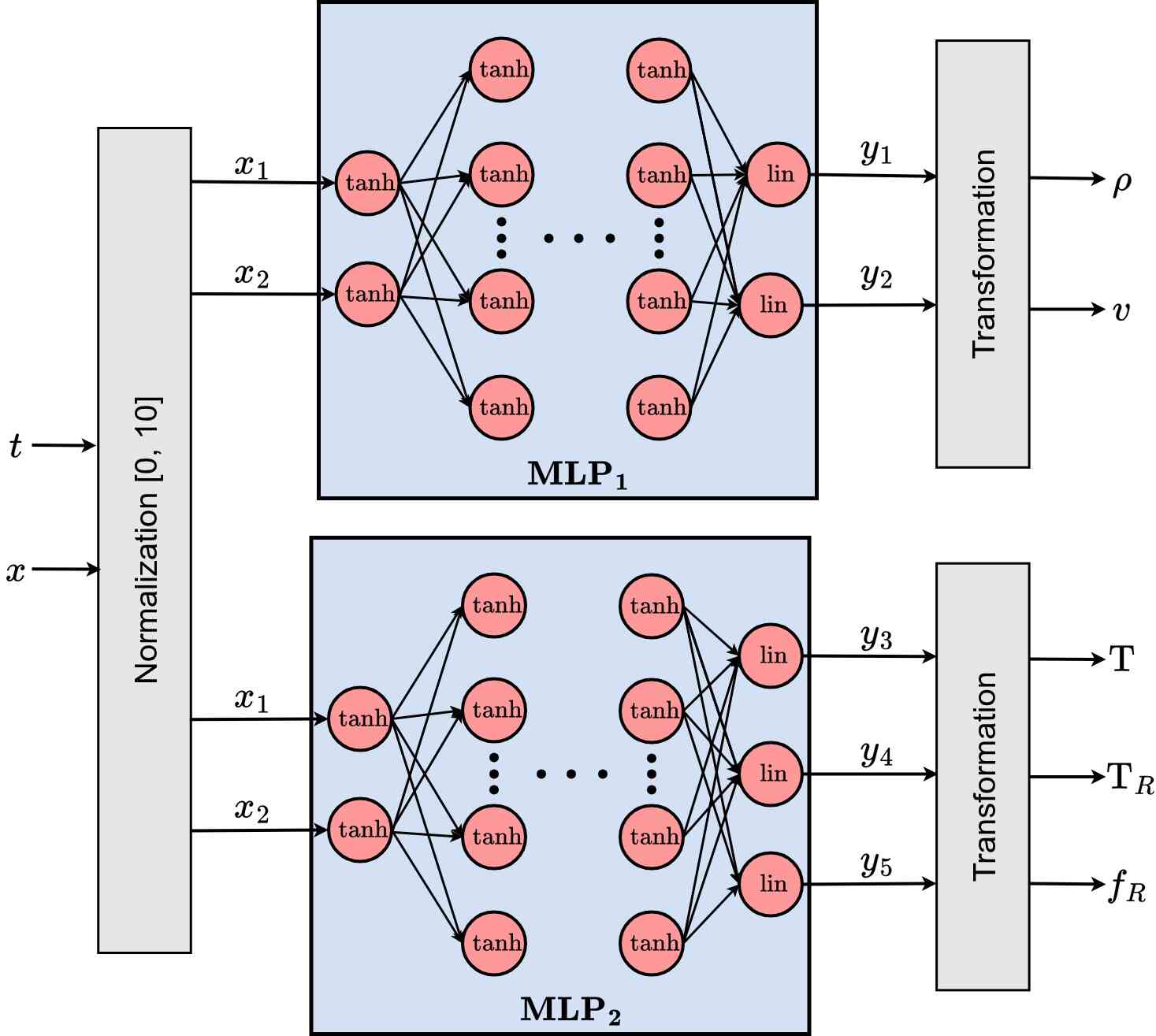}
        \end{minipage}
         \caption{Neural-network architecture used to represent the radiative shock with the \glsxtrshort{pinn} strategy. It is composed of two \glsxtrshort{mlp} neural networks: the first predicts the density and the velocity, while the second predicts the gas temperature, the radiation temperature, and the reduced flux. Each of these networks consists of three parts: an input-normalization layer, an \glsxtrshort{mlp} core, and an output-transformation layer. The symbols $\tanh$ and $\mathrm{lin}$ denote the hyperbolic-tangent activation function and the identity function, respectively.}
        \label{fig:PINN_architecture_hydrorad}
    \end{center}
\end{figure}

For these experiments, we used an architecture composed of two independent \gls{mlp} neural networks (see figure~\ref{fig:PINN_architecture_hydrorad}):
\begin{itemize}
    \item the first predicts the fluid density $\rho$ and velocity $v$,
    \item the second predicts the gas temperature $\mathrm{T}$, the radiation temperature $\mathrm{T}_R$, and the reduced flux $\mathrm{f}_R$.
\end{itemize}

We opted for this separation of the neural networks because of the similarity observed between the profiles of $\rho$ and $v$ on the one hand, and of $\mathrm{T}$ and $\mathrm{T}_R$ on the other. Moreover, we chose to assign the prediction of the reduced flux $\mathrm{f}_R$ to the network responsible for predicting $\mathrm{T}$ and $\mathrm{T}_R$. Indeed, our tests showed that, for a sufficiently expressive architecture, it is possible to obtain an accurate prediction of this quantity despite its different profile.

Each network consists of three modules: an input-normalization layer, the \gls{mlp} core, and finally an output-transformation layer. The input normalization layer rescales the spatio-temporal variables into a common interval $\closeinterv{0}{10}$, which facilitates training by limiting scale effects. This strategy proved more effective than the one used in the purely hydrodynamic case, and is written:
\begin{align}
    x_1 &= 10 \frac{t - t_0}{t_f - t_0}\\
    x_2 &= 10 \frac{x - x_{\min}}{x_{\max} - x_{\min}}
\end{align}

The output transformation layer of the \gls{mlp}s serves the same purposes as in the purely hydrodynamic case: ensuring the positivity of physical quantities and adapting their scales. The transformations are defined as:
\begin{align}
    \rho  &= |w_\rho|~\mathrm{SP}(y_1) + |b_\rho|            \;\;\mathpunct{,}\\
    v &= |w_v|~y_2 + |b_v|                               \;\;\mathpunct{,}\\
    \mathrm{T} &= |w_T|~\mathrm{SP}(y_3) + |b_T|                  \;\;\mathpunct{,}\\
    \mathrm{T}_R &= |w_T|~\mathrm{SP}(y_4) + |b_T|                  \;\;\mathpunct{,}\\
    \mathrm{f}_R &= \frac{|w_{\mathrm{f}_R}|~y_5}{\alpha_{\mathrm{f}_R}} \;\;\mathpunct{.}
\end{align}

\noindent Here, $\mathrm{SP}$ denotes the SoftPlus activation, chosen to guarantee positivity of the relevant output variables. The coefficients $w_\rho$, $w_T$, $w_v$, $w_{\mathrm{f}R}$ and their associated biases $b\rho$, $b_T$, $b_v$ are free parameters optimized during training. In particular, the scaling factor $\alpha_{\mathrm{f}_R}$, already introduced in the loss term $\mathcal{L}_{sim}$, is intended to help the network predict the very small values often taken by the reduced flux $\mathrm{f}_R$.

To evaluate prediction quality, we introduce the following relative squared errors, computed on a test set containing all \gls{hades} simulation data:
\begin{align}
    L_{2,\mathrm{T}_R} &= \frac{\sum_{i=1}^{N_{test}} \left \{ \mathrm{T}_{R,pred}(t_i, x_i) - \mathrm{T}_{R,sim}(t_i, x_i)\right \}^2}{\sum_{i=1}^{N_{test}} \mathrm{T}_{R,sim}(t_i, x_i)^2} \;\;\mathpunct{,}\\
     L_{2,\mathrm{f}_R} &= \frac{\sum_{i=1}^{N_{test}} \left \{ \mathrm{f}_{R,pred}(t_i, x_i) - \mathrm{f}_{R,sim}(t_i, x_i)\right \}^2}{\sum_{i=1}^{N_{test}} \mathrm{f}_{R,sim}(t_i, x_i)^2} \;\;\mathpunct{,}
\end{align}

\noindent where the quantities indexed \quotes{pred} come from the neural network predictions, and those indexed \quotes{sim} come from the test set. The total number of samples is $N_{test}=303\,000$. From these individual errors, we define aggregated errors:
\begin{align}
     L_{2,MLP_1} &= \frac{L_{2,\rho} + L_{2,v}}{2} \;\;\mathpunct{,}\\
     L_{2,MLP_2} &= \frac{L_{2,\mathrm{T}} + L_{2,\mathrm{T}_R} + L_{2,\mathrm{f}_R}}{3} \;\;\mathpunct{,}\\
     L_{2,global} &= \frac{2 L_{2,MLP_1} + 3 L_{2,MLP_2}}{5} \;\;\mathpunct{,}
\end{align}

\noindent where $L_{2,\rho}$, $L_{2,v}$, and $L_{2,\mathrm{T}}$ were defined in equations~\eqref{eq:L2r}, \eqref{eq:L2v}, and \eqref{eq:L2T}. $L_{2,MLP_1}$ evaluates the performance of $\mathrm{MLP}_1$, $L_{2,MLP_2}$ that of $\mathrm{MLP}_2$, and $L_{2,global}$ measures the overall performance of the strategy.

In this architecture, two hyper-parameters remain to be chosen for each network: the number of hidden layers and the number of neurons per hidden layer. As in the previous study, we explored several configurations, training each architecture on the full simulation dataset (setting $\alpha = 1$ and $\mathrm{N_T}=101$). Training proceeds in two stages:
\begin{itemize}
    \item a first phase using the \gls{adam} optimizer for $10\,000$ epochs (learning rate $\eta_{\mathrm{Adam}} = 5 \times 10^{-3}$),
    \item followed by a second phase using the \gls{lbfgs} optimizer for $2\,000$ additional iterations (learning rate $\eta_{\mathrm{LBFGS}} = 10^{-2}$).
\end{itemize}

\begin{figure}
    \begin{subfigure}[t]{0.49\textwidth}
        \centering
        \includegraphics[width=\textwidth]{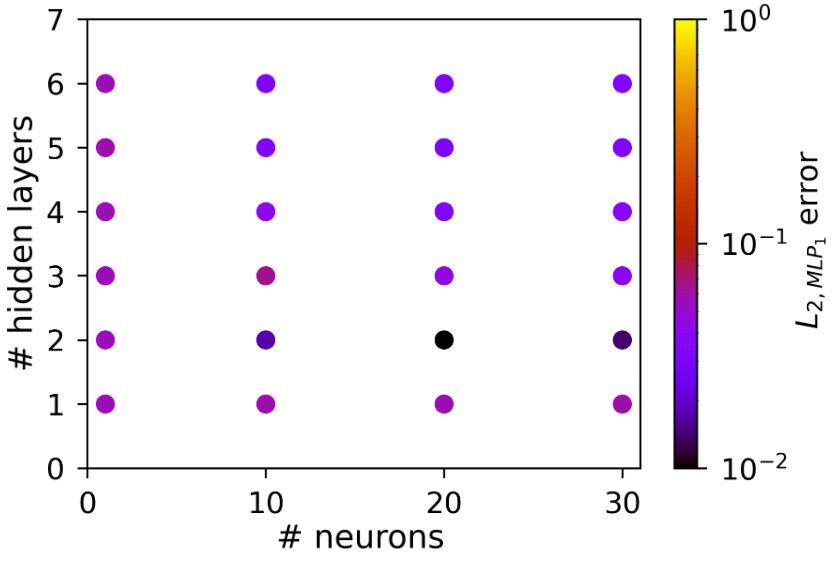}
        \caption{Evaluation of the $L_{2,MLP_1}$ error for different numbers of hidden layers and different numbers of neurons per hidden layer.}
        \label{fig:choix_MLP1}
    \end{subfigure}
    \hfill
    \begin{subfigure}[t]{0.49\textwidth}
        \centering
        \includegraphics[width=\textwidth]{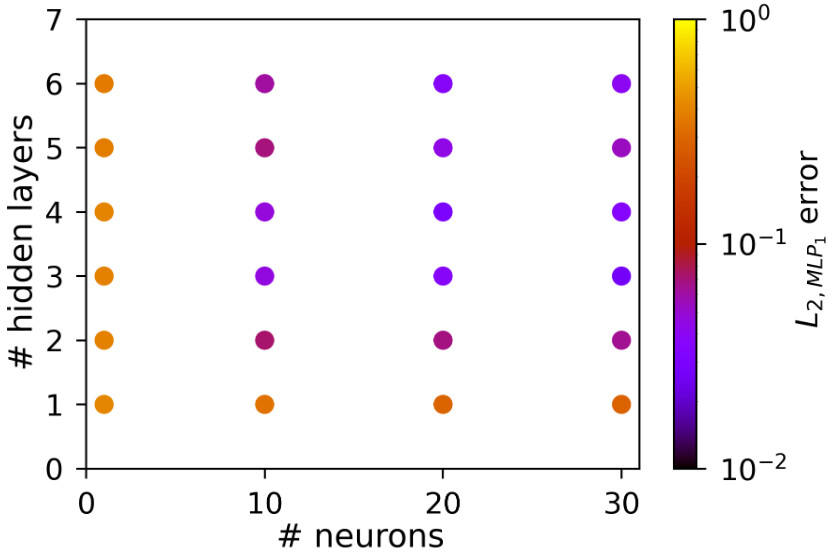}
        \caption{Evaluation of the $L_{2,MLP_2}$ error for various numbers of hidden layers and neurons per hidden layer.}
        \label{fig:choix_MLP2}
    \end{subfigure}
    \caption{Evaluation of the $L_2$ error of the $\text{MLP}_1$ and $\text{MLP}_2$ neural networks as a function of the number of hidden layers and the number of neurons per hidden layer.}
    \label{fig:choix_MLP_hydrorad}
\end{figure}

\begin{figure}
    \begin{subfigure}[t]{0.46\textwidth}
        \centering
        \includegraphics[width=\textwidth]{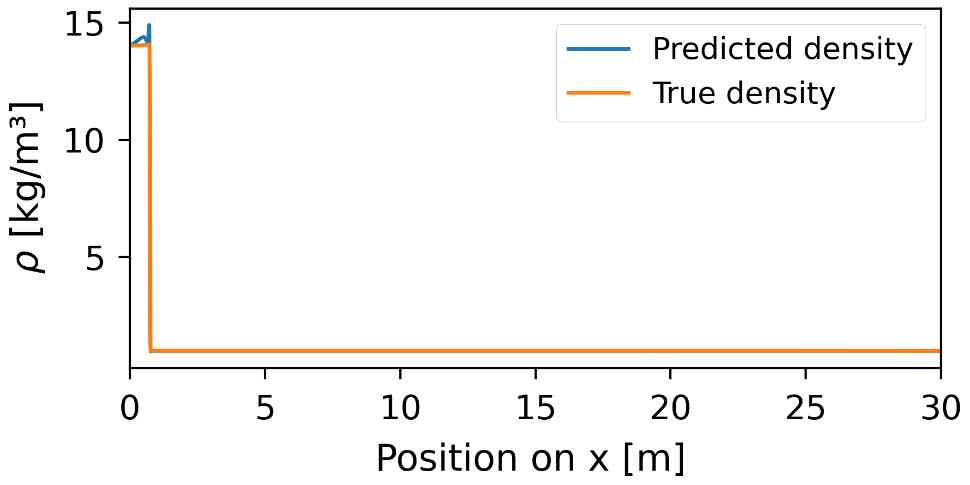}
        \caption{Density}
    \end{subfigure}
    \hfill
    \begin{subfigure}[t]{0.46\textwidth}
        \centering
        \includegraphics[width=\textwidth]{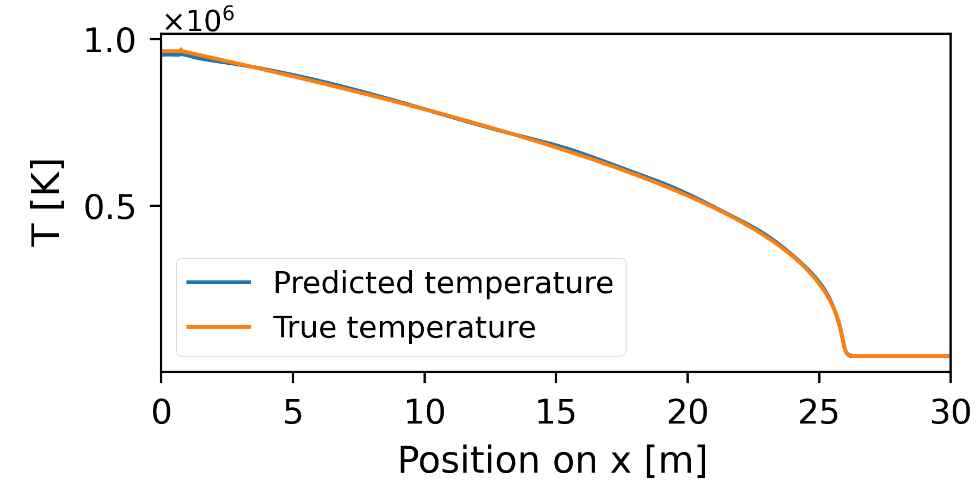}
        \caption{Gas temperature}
    \end{subfigure}
    \hfill
    \begin{subfigure}[t]{0.46\textwidth}
        \centering
        \includegraphics[width=\textwidth]{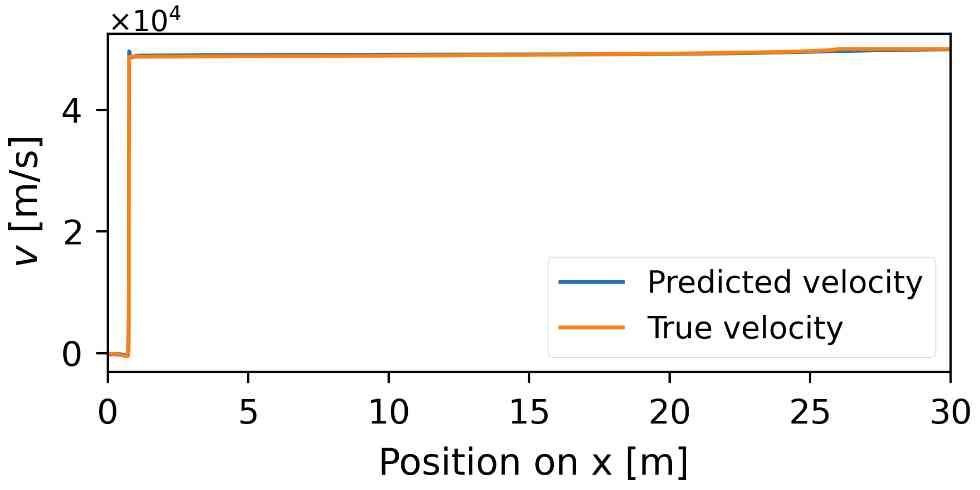}
        \caption{Velocity}
    \end{subfigure}
    \hfill
    \begin{subfigure}[t]{0.46\textwidth}
        \centering
        \includegraphics[width=\textwidth]{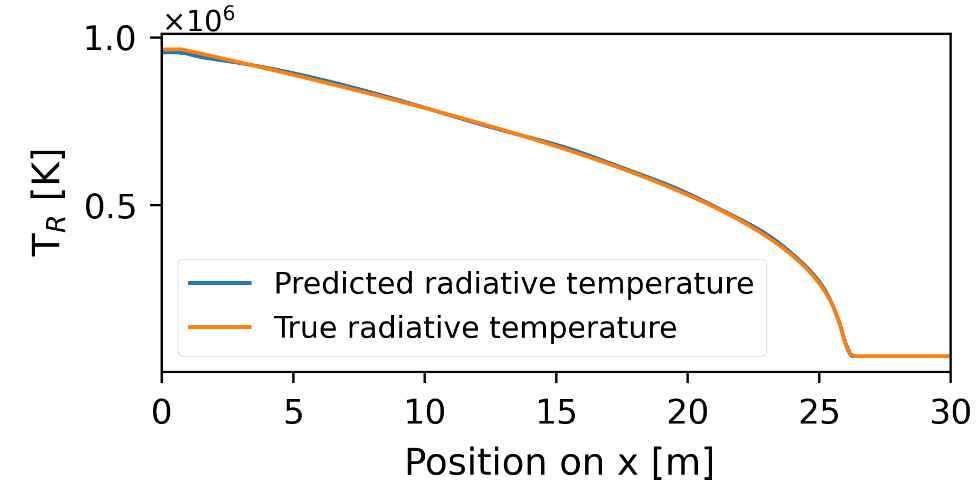}
        \caption{Radiative temperature}
    \end{subfigure}
    \begin{center}
        \begin{subfigure}[t]{0.46\textwidth}
            \centering
            \includegraphics[width=\textwidth]{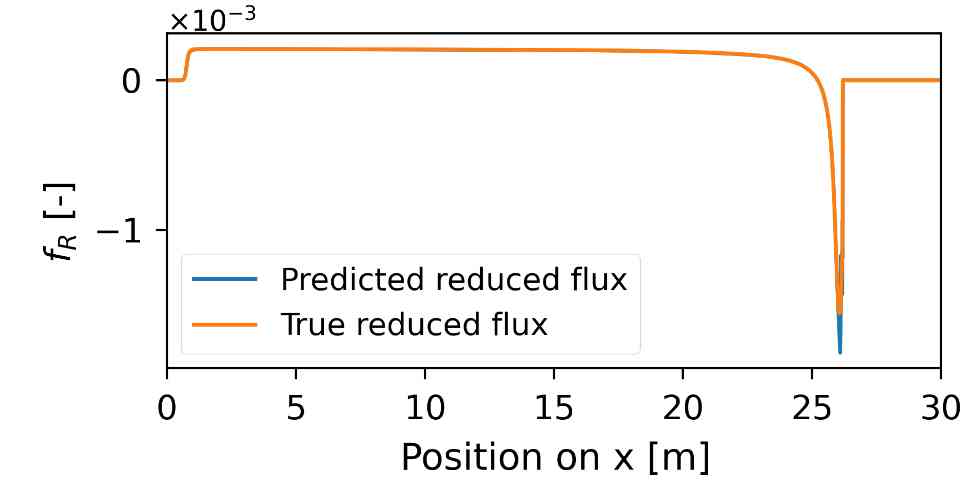}
            \caption{Reduced flux}
        \end{subfigure}
    \end{center}
    \caption{Comparison of the density, temperature, velocity, radiative temperature, and reduced flux profiles predicted by the neural network and obtained from the simulation, at the final time \mbox{$t_f = 2 \times 10^{-4}~\text{s}$}. The network was trained exclusively on simulation data.}
    \label{fig:trained_data}
\end{figure}

Once the training phases were completed, we evaluated the prediction errors in order to analyze the performance of the different neural-network architectures. To do so, we examined separately the errors $L_{2,MLP_1}$ and $L_{2,MLP_2}$, associated respectively with each network. For $\text{MLP}1$, the results shown in figure~\ref{fig:choix_MLP1} indicate that the error $L_{2,MLP_1}$ decreases as both the number of hidden layers and the number of neurons per layer increase. Based on these observations, we selected an architecture composed of two hidden layers, each containing 20 neurons, since this configuration yielded a low error on the test set. Similar conclusions were reached for the $\text{MLP}_2$ network, as illustrated by the results in figure~\ref{fig:choix_MLP2}. We therefore also adopted, for this network, an architecture with two hidden layers of 20 neurons each.

\noindent Figure~\ref{fig:trained_data} shows that the neural networks trained from the simulation data are, overall, able to correctly reproduce the profiles of the physical quantities considered. The predictions are satisfactory for the fluid velocity, the temperatures $\mathrm{T}$ and $\mathrm{T}_R$, as well as for the reduced flux $\mathrm{f}_R$. However, the density is less accurately captured, particularly in the shocked region of the fluid.

\begin{figure}
    \begin{subfigure}[t]{0.32\textwidth}
        \centering
        \includegraphics[width=0.96\textwidth]{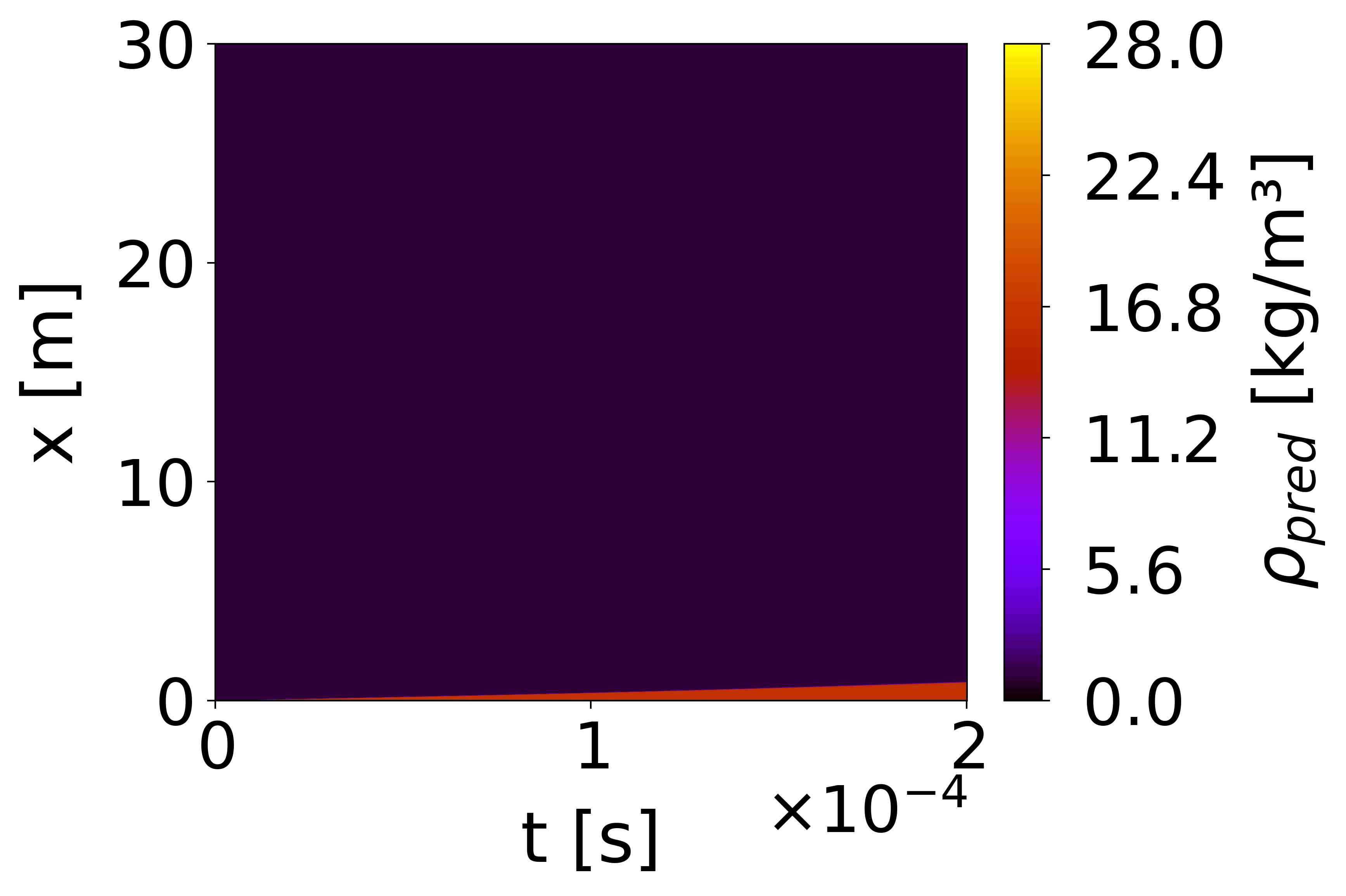}
        \caption{Predicted density.}
    \end{subfigure}
    \hfill
    \begin{subfigure}[t]{0.32\textwidth}
        \centering
        \includegraphics[width=0.96\textwidth]{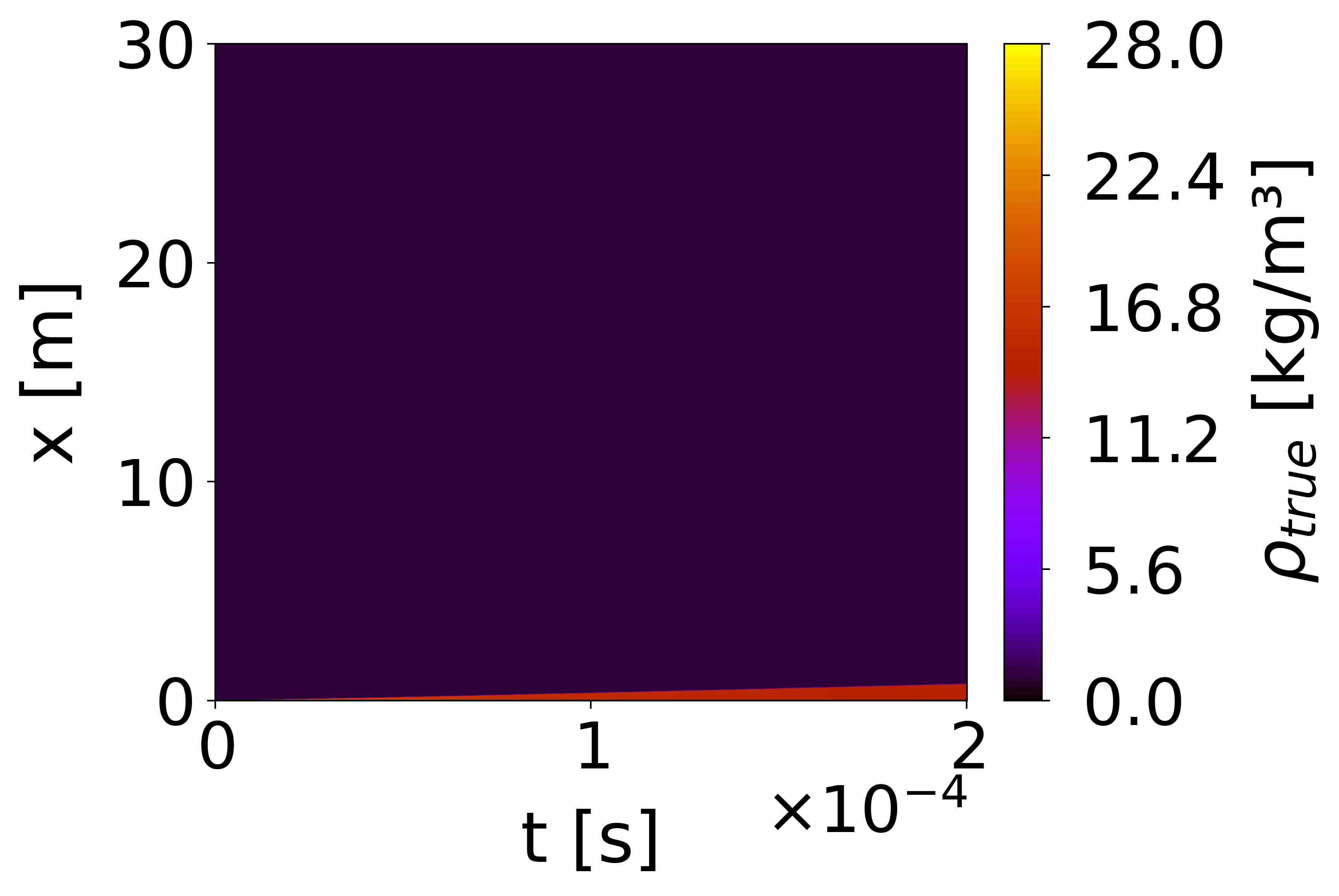}
        \caption{Simulated density.}
    \end{subfigure}
    \hfill
    \begin{subfigure}[t]{0.32\textwidth}
        \centering
        \includegraphics[width=0.96\textwidth]{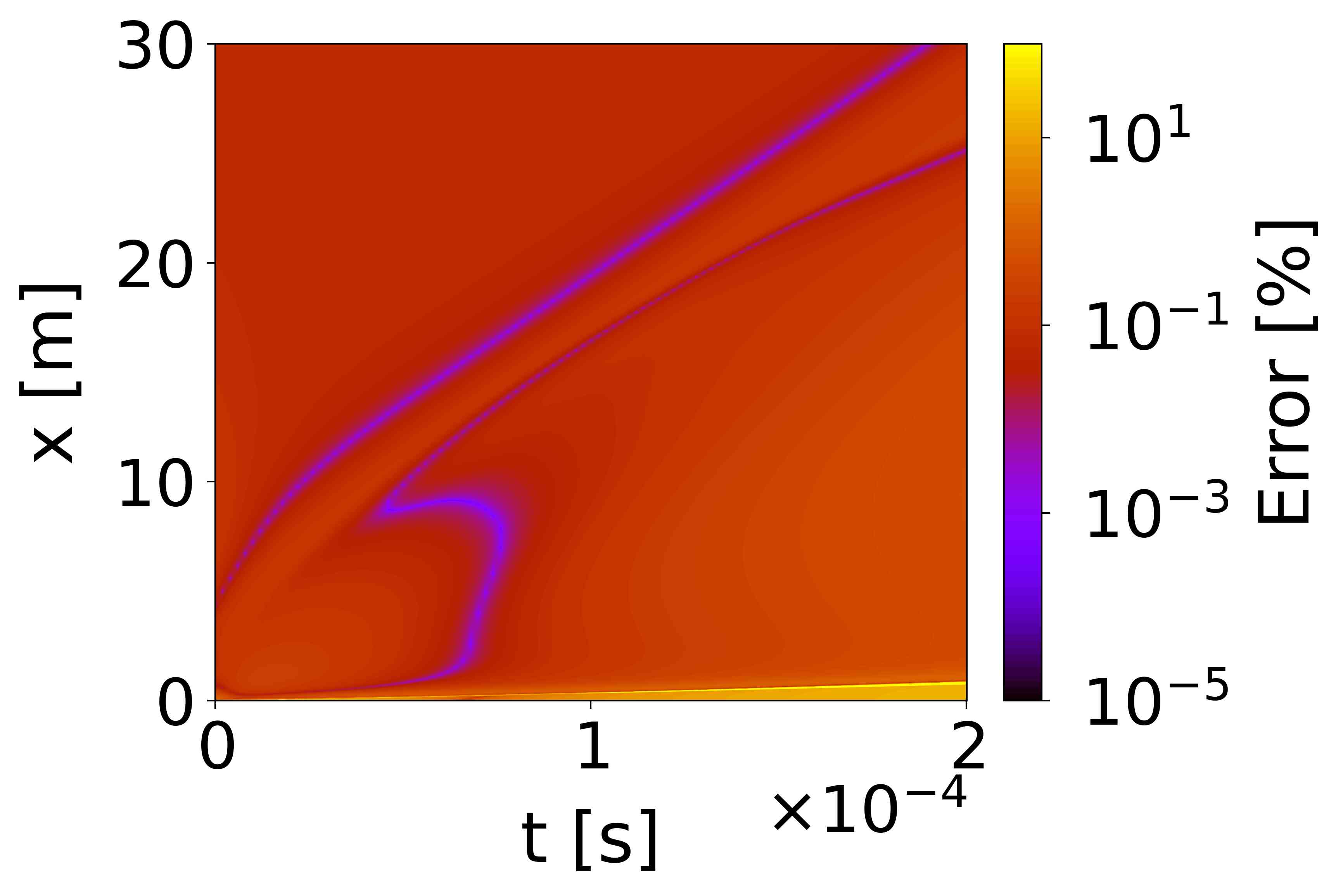}
        \caption{Error on the density.}
    \end{subfigure}
    \hfill
    \begin{subfigure}[t]{0.32\textwidth}
        \centering
        \includegraphics[width=0.96\textwidth]{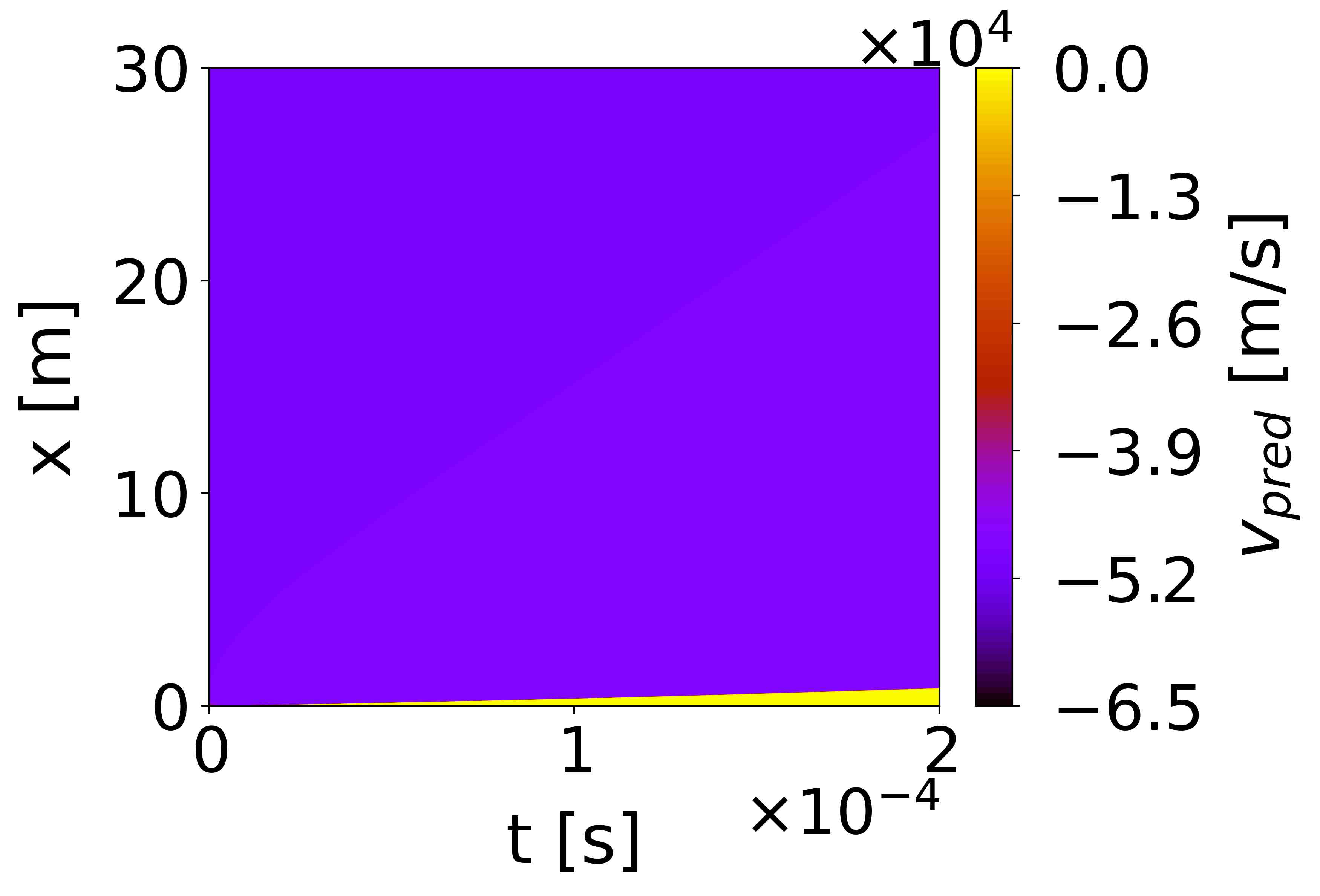}
        \caption{Predicted velocity.}
    \end{subfigure}
    \hfill
    \begin{subfigure}[t]{0.32\textwidth}
        \centering
        \includegraphics[width=0.96\textwidth]{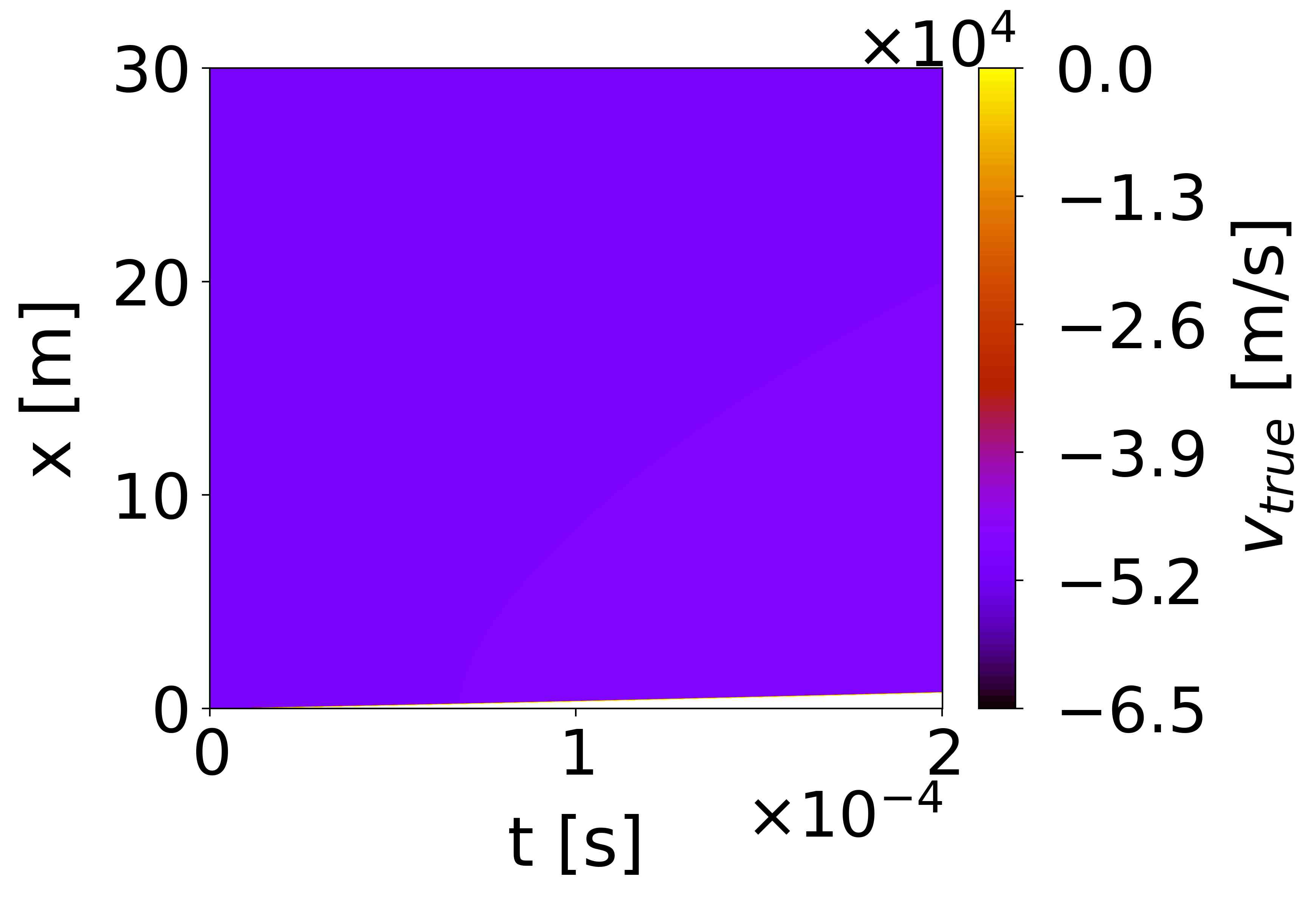}
        \caption{Simulated velocity.}
    \end{subfigure}
    \hfill
    \begin{subfigure}[t]{0.32\textwidth}
        \centering
        \includegraphics[width=0.96\textwidth]{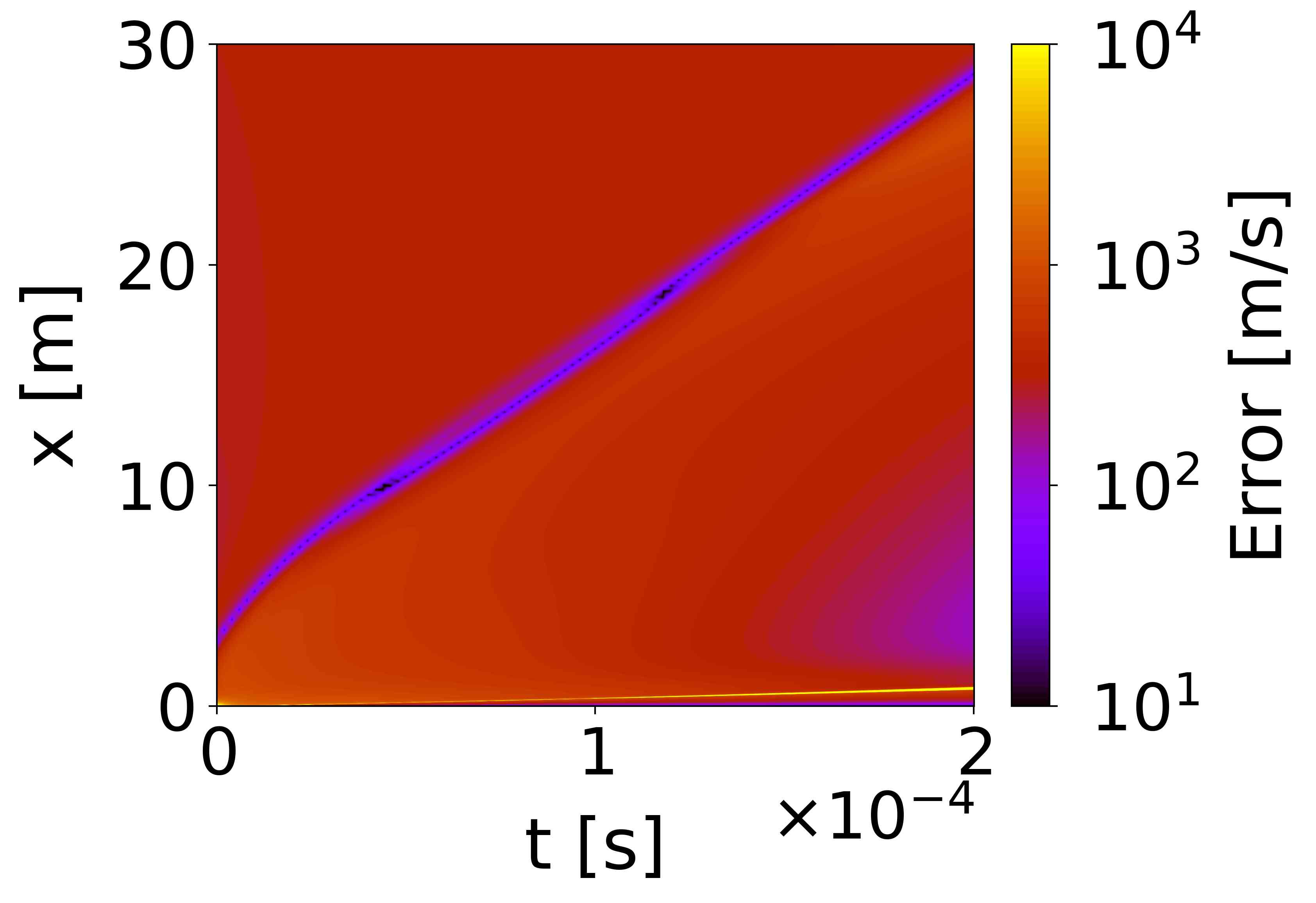}
        \caption{Error on the velocity.}
    \end{subfigure}
    \hfill
    \begin{subfigure}[t]{0.32\textwidth}
        \centering
        \includegraphics[width=0.96\textwidth]{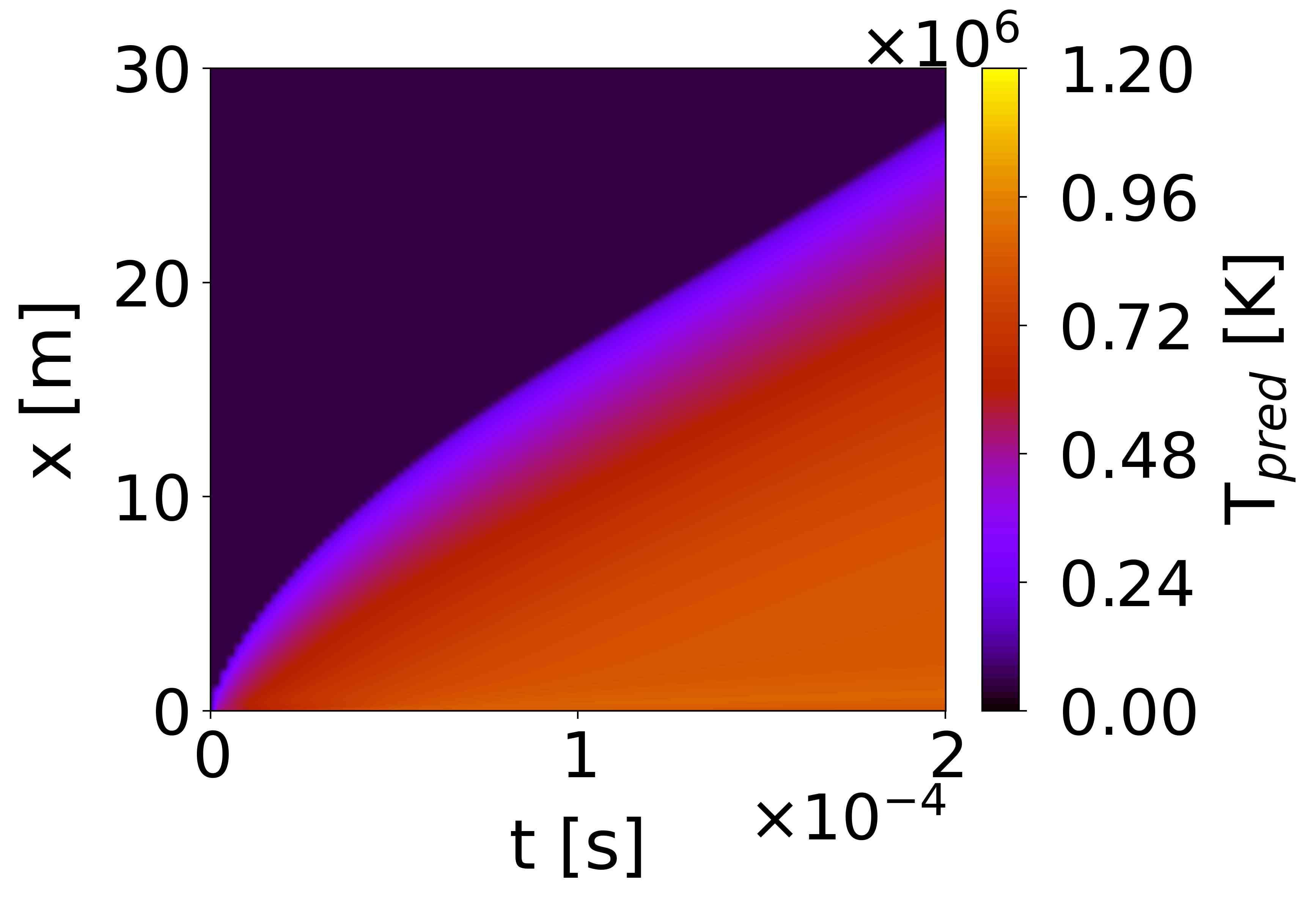}
        \caption{Predicted temperature.}
    \end{subfigure}
    \hfill
    \begin{subfigure}[t]{0.32\textwidth}
        \centering
        \includegraphics[width=0.96\textwidth]{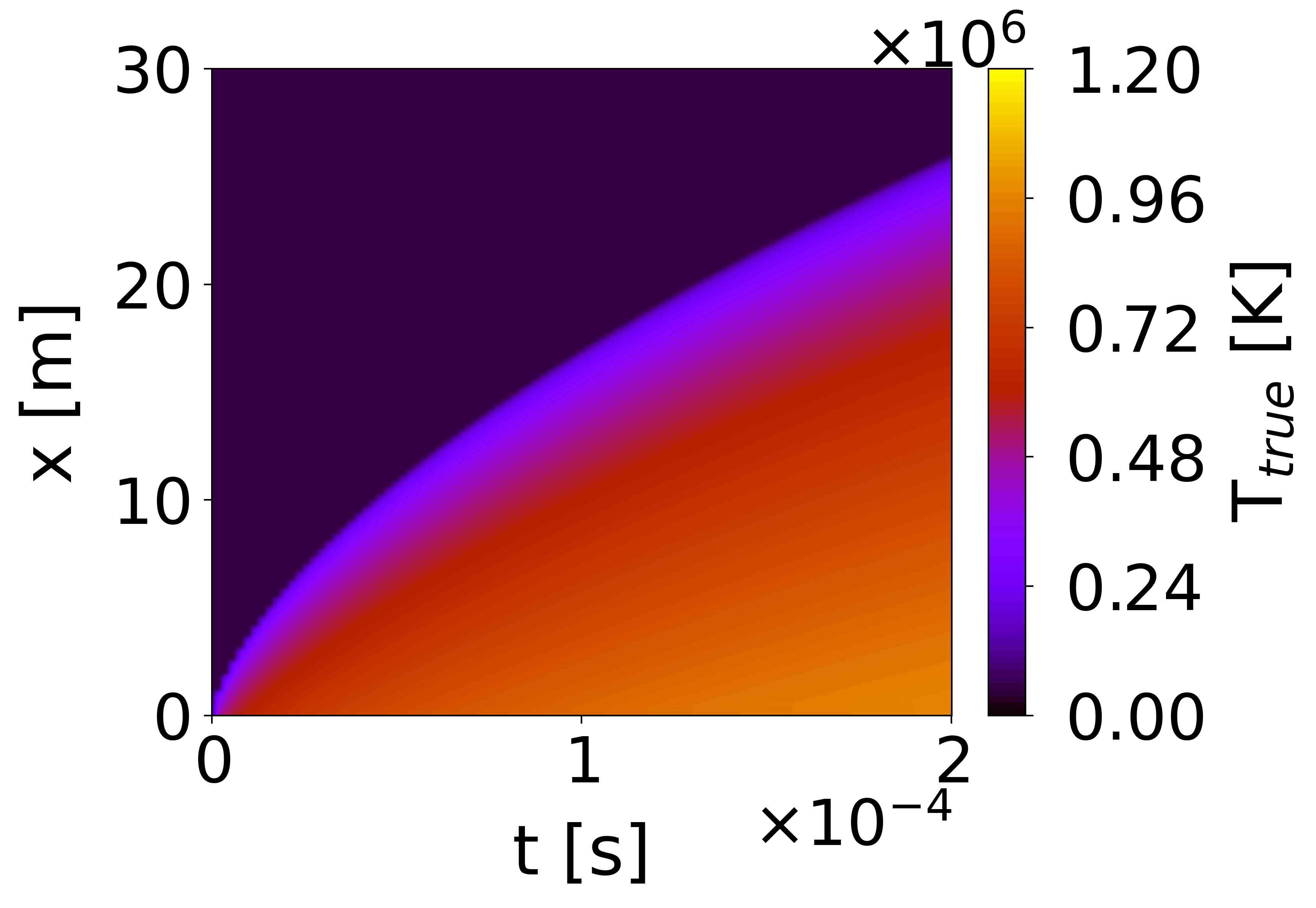}
        \caption{Simulated temperature.}
    \end{subfigure}
    \hfill
    \begin{subfigure}[t]{0.32\textwidth}
        \centering
        \includegraphics[width=0.96\textwidth]{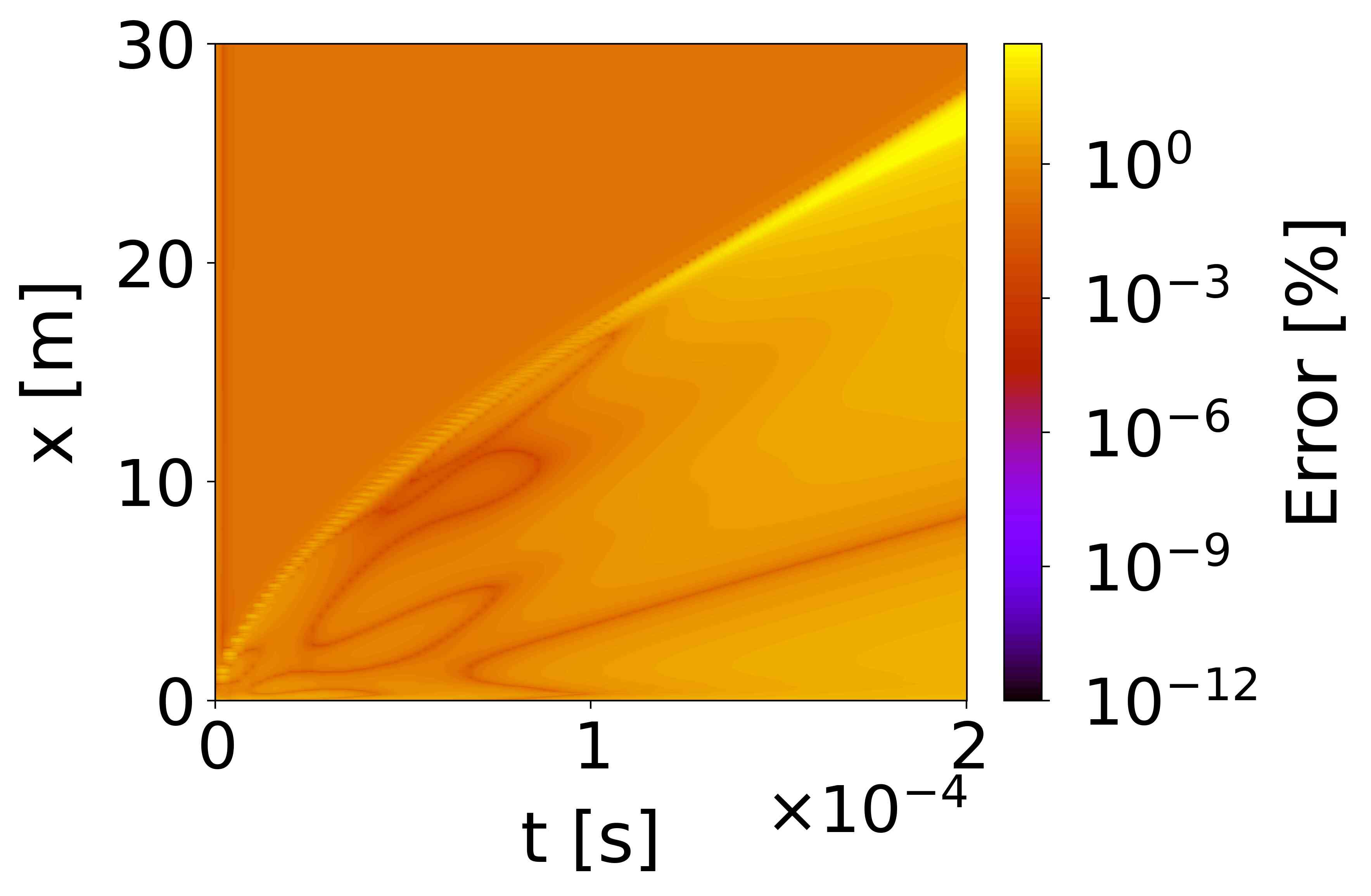}
        \caption{Error on the temperature.}
    \end{subfigure}
    \hfill
    \begin{subfigure}[t]{0.32\textwidth}
        \centering
        \includegraphics[width=0.96\textwidth]{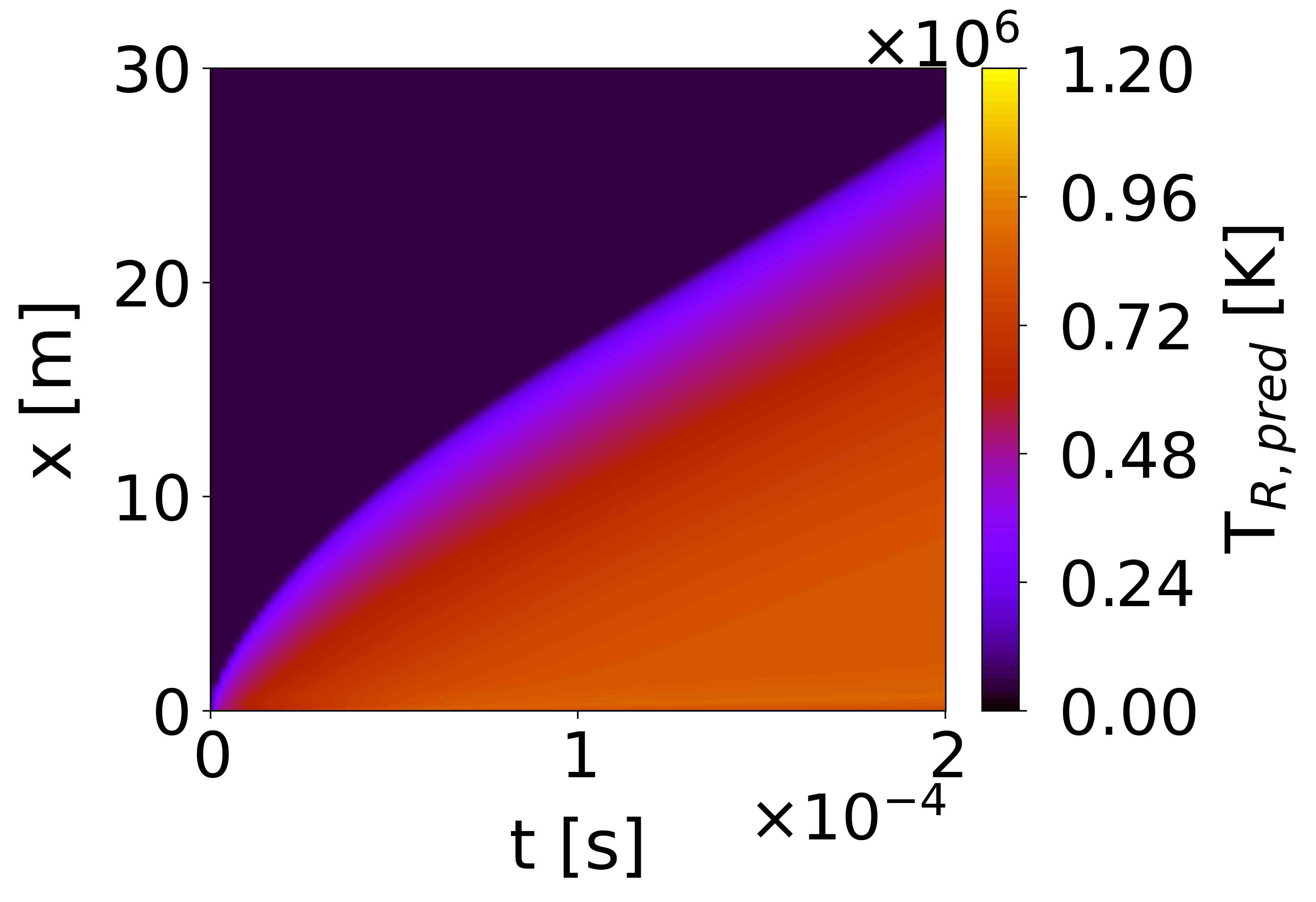}
        \caption{Predicted radiative temperature.}
    \end{subfigure}
    \hfill
    \begin{subfigure}[t]{0.32\textwidth}
        \centering
        \includegraphics[width=0.96\textwidth]{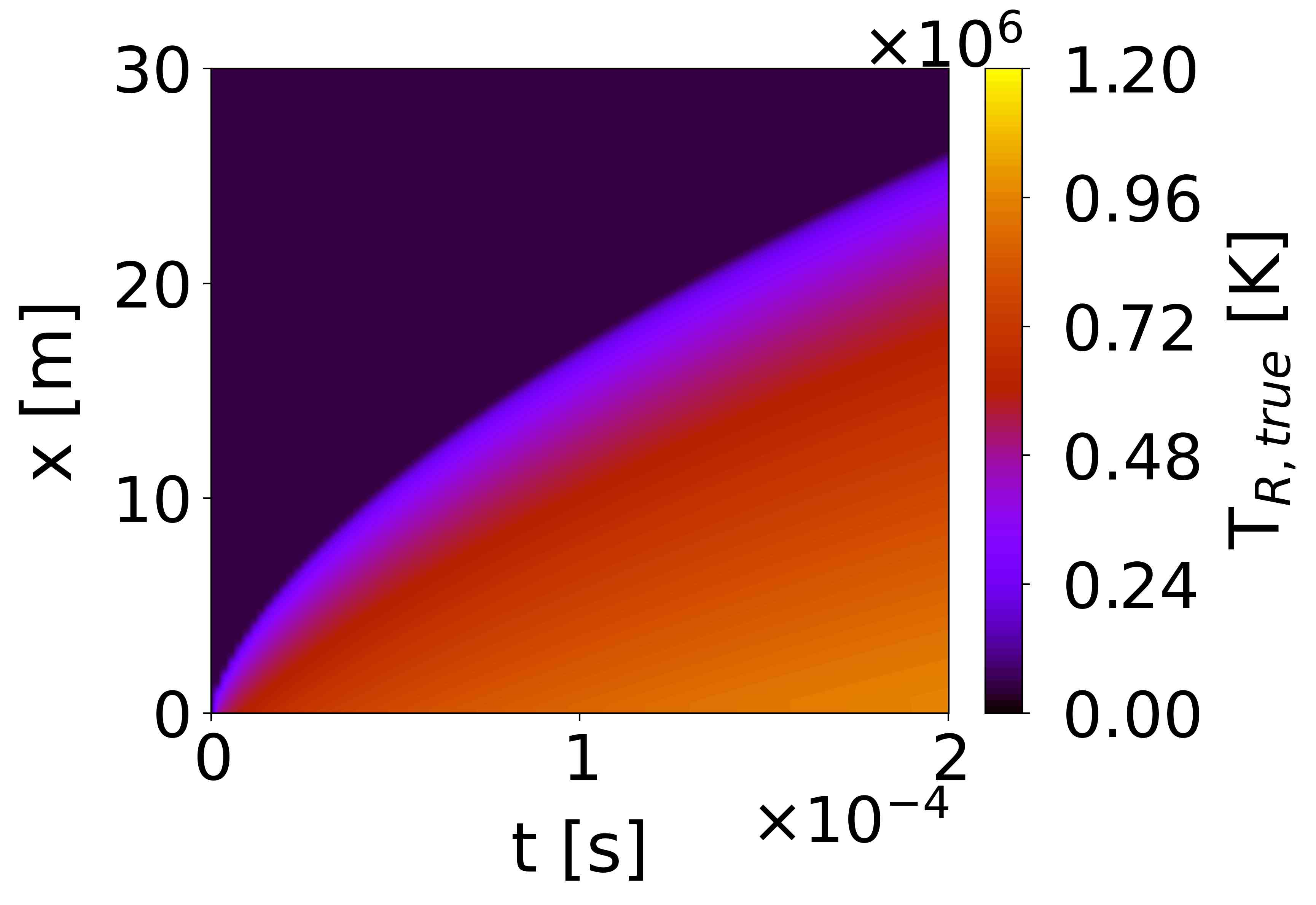}
        \caption{Simulated radiative temperature.}
        \label{fig:Tr_vraie}
    \end{subfigure}
    \hfill
    \begin{subfigure}[t]{0.32\textwidth}
        \centering
        \includegraphics[width=0.96\textwidth]{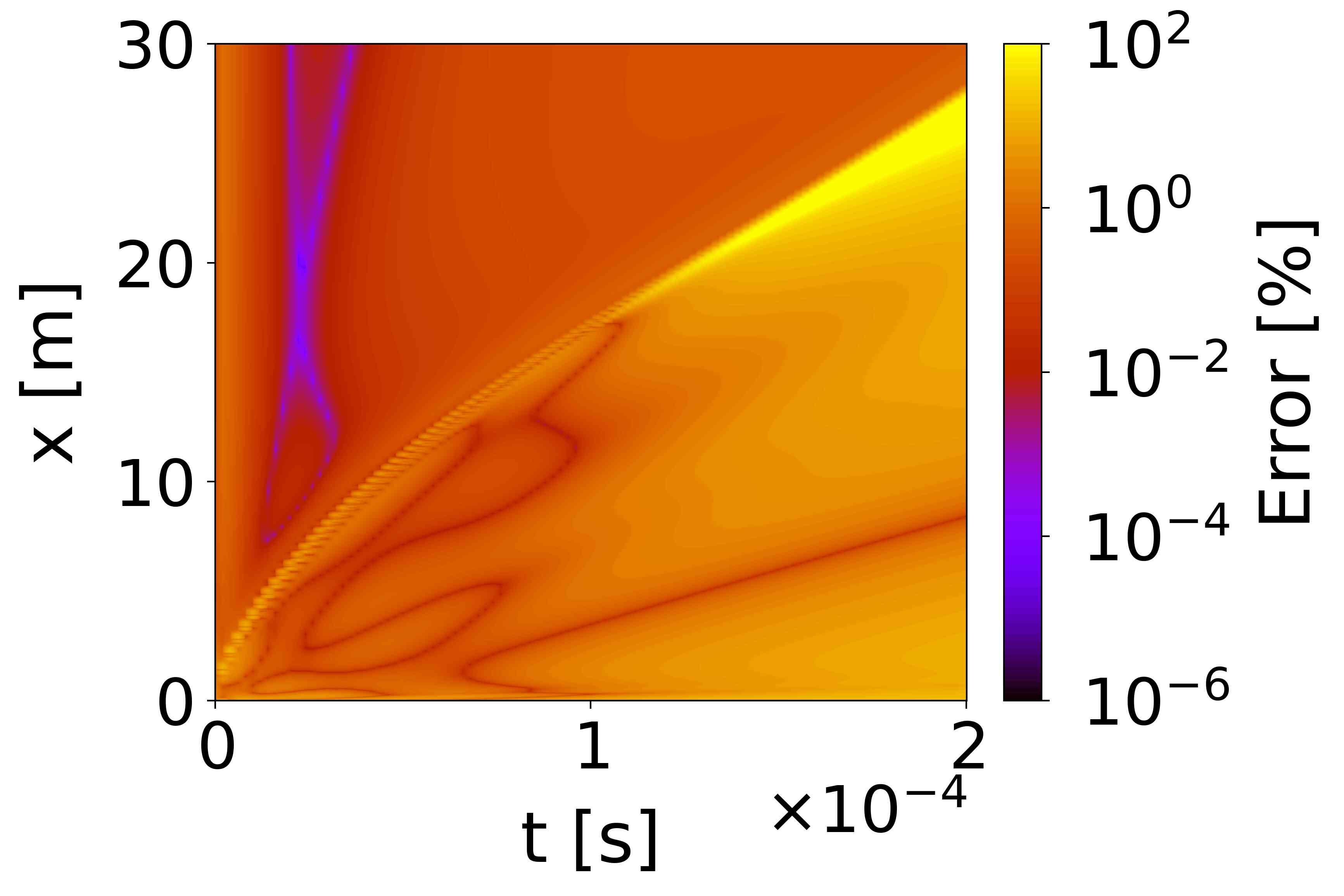}
        \caption{Error on the radiative temperature.}
    \end{subfigure}
    \hfill
    \begin{subfigure}[t]{0.32\textwidth}
        \centering
        \includegraphics[width=0.96\textwidth]{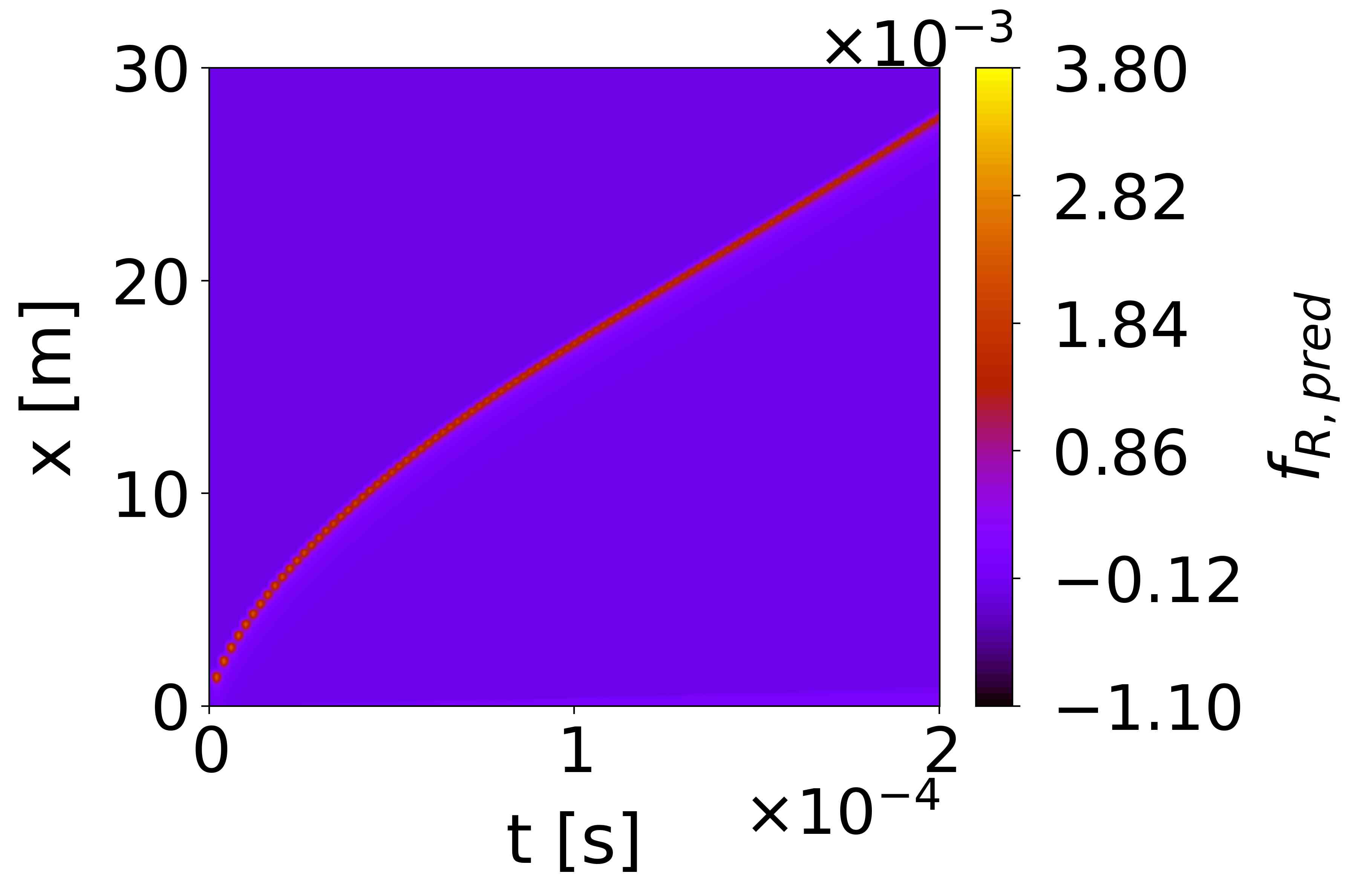}
        \caption{Predicted reduced flux.}
    \end{subfigure}
    \hfill
    \begin{subfigure}[t]{0.32\textwidth}
        \centering
        \includegraphics[width=0.96\textwidth]{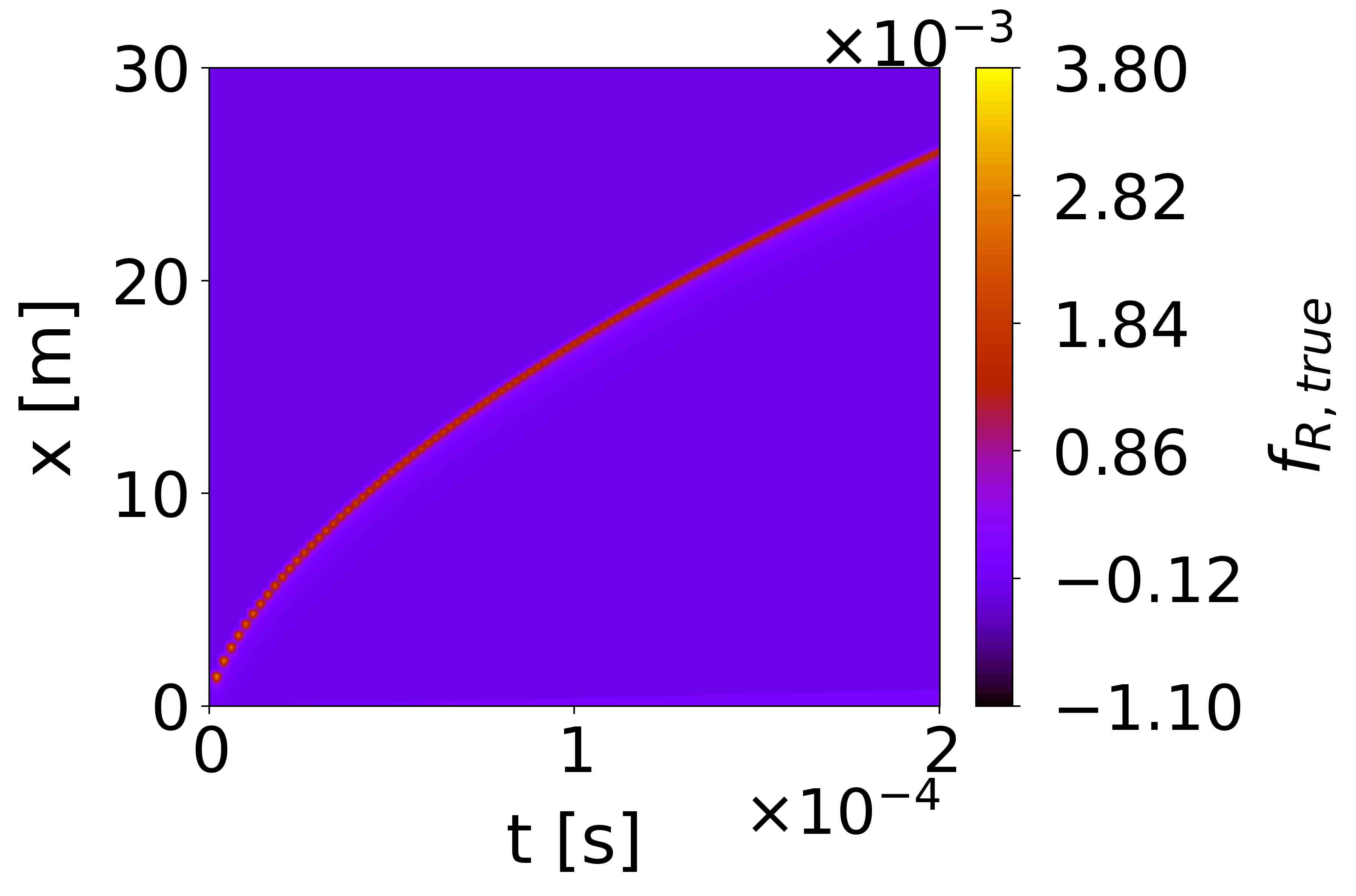}
        \caption{Simulated reduced flux.}
    \end{subfigure}
    \hfill
    \begin{subfigure}[t]{0.32\textwidth}
        \centering
        \includegraphics[width=0.96\textwidth]{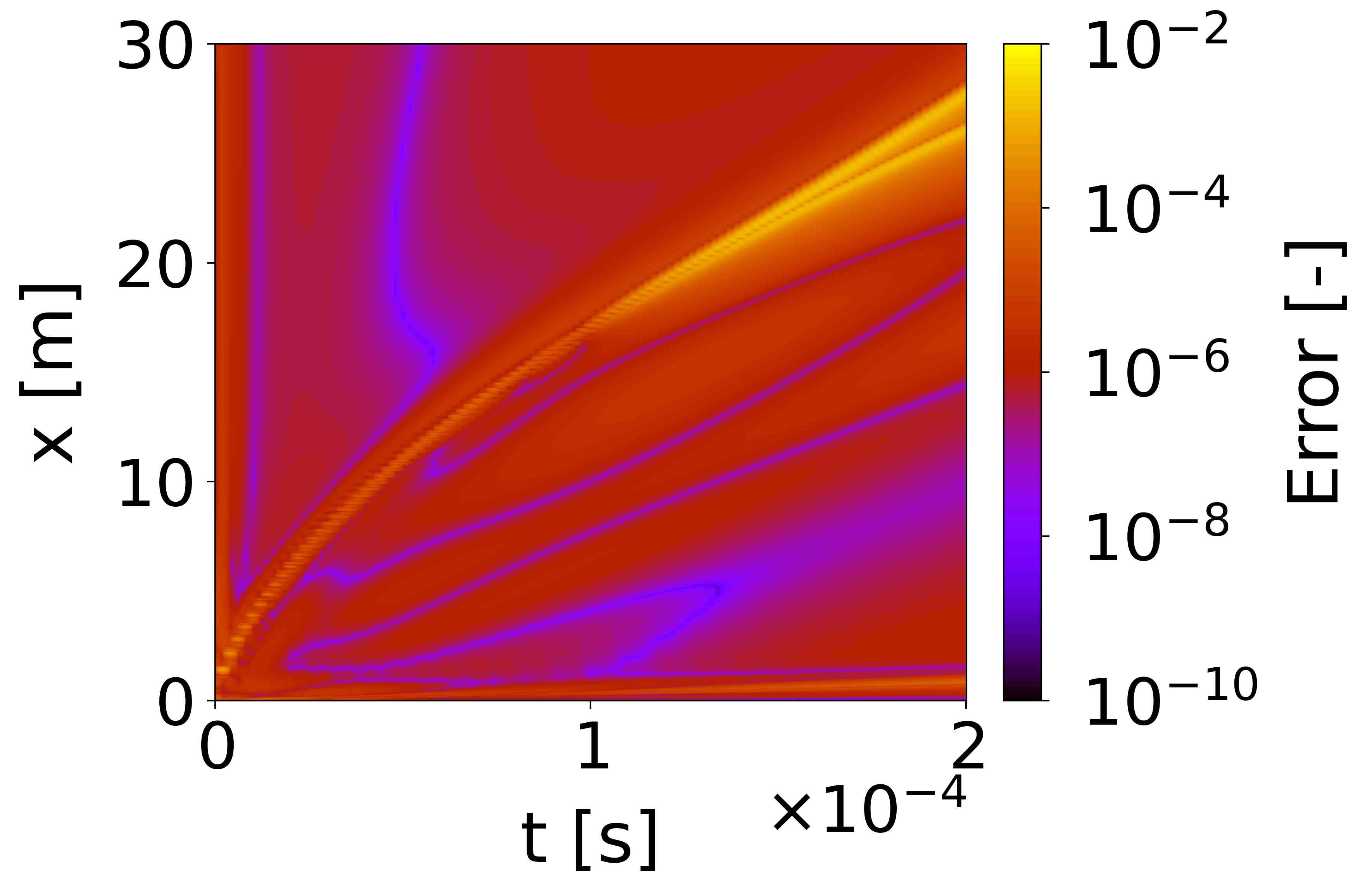}
        \caption{Error on the reduced flux.}
    \end{subfigure}
    \caption{Comparison between the prediction of the \glsxtrshort{pinn} neural network and the data from the \glsxtrshort{hades} simulation. The neural network was trained on the \glsxtrshort{hades} simulation data for times \mbox{$t \in \closeinterv{0}{1 \times 10^{-4}}$ s}.}
    \label{fig:comp_pinn}
\end{figure}

\starsect{Training difficulties observed with the \glsxtrshort{pinn} strategy}

We performed a test with the parameters \mbox{$\lambda = 10^{3}$}, \mbox{$\alpha = 0.5$} and \mbox{$\mathrm{N_T} = 50$}, representative of the best results I obtained so far. The analysis of the \gls{pinn} neural-network predictions (see figure~\ref{fig:comp_pinn}) shows that it faithfully reproduces the dynamics over the time interval \mbox{$t \in \closeinterv{0}{1 \times 10^{-4}}$ s} (in physical units), i.e. the region on which it was trained using data from the \gls{hades} code.

However, beyond this interval, at times such that \mbox{$t \in \closeinterv{1 \times 10^{-4}}{2 \times 10^{-4}}$ s}, although the network is still able to correctly capture the propagation of the shock, two limitations clearly appear: (i) it does not correctly reproduce the dynamics of the radiative precursor front, which it extrapolates almost linearly whereas it should slow down; (ii) it provides inaccurate predictions of the hydrodynamic and radiative values downstream of the shock.

These difficulties can be explained by two major characteristics of the physical situation studied: (i) the presence of a shock, which introduces a discontinuity in the hydrodynamic fields; (ii) the very stiff nature of the equations, particularly at the edge of the radiative precursor, where the source terms reach high values, typically \mbox{$S_v \approx 10^{3}$ m/$\mathrm{s^2}$}, \mbox{$c S_\mathrm{T} \approx 10^{11}$ K/s}, \mbox{$c S_{\mathrm{T}_R} \approx 10^{12}$ K/s} and \mbox{$c S_{\mathrm{f}_R} \approx 10^{5}$ $\mathrm{s^{-1}}$}, to be compared with the values of the physical quantities in this region, which are \mbox{$v \approx 10^{4}$ m/s}, \mbox{$\mathrm{T} \approx \mathrm{T}_R \approx 10^{4}$ K}, and \mbox{$\mathrm{f}_R \approx 10^{-3} \in \closeinterv{-1}{1}$}.

In the experiment presented here, the evolution equations for the temperature $\mathrm{T}$, the radiative temperature $\mathrm{T}_R$, and the reduced flux $\mathrm{f}_R$ were divided by the speed of light $c$ (see equation~\eqref{eq:hydrorad_pinn_res}). This normalization artificially reduces the stiffness and makes the network training considerably easier. However, we believe that it decreases the relative importance of the corresponding equations in the loss function and may be one of the causes of the poor representation of the radiative precursor dynamics, since these three quantities are precisely those most strongly influenced by the precursor. Without this weighting by $c$, training becomes extremely difficult, to the point that the neural network is no longer even able to reproduce even the simulation data.

Moreover, in this work, we applied the adaptive-residual weighting method proposed by Liu (2023)~\cite{liu_2023}, which improves the representation of discontinuities in the hydrodynamic quantities. Nevertheless, it remains necessary to develop more suitable strategies capable of simultaneously handling equation stiffness and discontinuities, in order to better capture the structure of the radiative precursor. Designing such an approach represents an essential direction for future work in this field.

Finally, in future developments, it will be crucial to ensure that the reduced flux $\mathrm{f}_R$ and the ratio $\mathrm{T}/\mathrm{T}_R$ are predicted with very high accuracy by the network, since these quantities directly enter the source terms $S_v$, $c S_\mathrm{T}$, $c S_{\mathrm{T}_R}$ and $c S_{\mathrm{f}_R}$. As can be seen from their expressions (see equations~\eqref{eq:src_v}–\eqref{eq:src_fr}), even a small error in the estimation of $\mathrm{f}_R$ or $\mathrm{T}/\mathrm{T}_R$ can lead to significant deviations in the estimation of these source terms.

\section{Synthesis}

In order to overcome the limitation imposed by the \gls{cfl} condition of the \gls{hades} code, I explored the use of \gls{pinn} neural networks with the objective of extrapolating, to later times, data produced by pre-existing numerical simulations, in particular those obtained with \gls{hades}.

In the case of pure hydrodynamics, this approach proved satisfactory: it makes it possible to efficiently extrapolate the simulated profiles and to faithfully recover the structure of the hydrodynamic shock. In contrast, the results obtained for radiative shocks remain, at this stage, less satisfactory. This difficulty appears to be linked to the combination of two factors: on the one hand, the presence of a pronounced discontinuity at the shock front, and on the other hand, the strong stiffness of the source terms describing the interactions between radiation and fluid within the radiative precursor. These terms are extremely sensitive to the ratio between the gas temperature and the radiation temperature, $\mathrm{T}/\mathrm{T}_R$, as well as to the reduced flux $\mathrm{f}_R$, which complicates the training of the neural network.

To improve the quality of extrapolation in the context of radiative shocks, it therefore appears necessary to develop a method capable of simultaneously handling both the shock discontinuity and the stiffness of these source terms, in order to ensure a coherent and stable representation of the fluid–radiation coupling.
\clearemptydoublepage

\chapter{Conclusions and perspectives} \label{ch:conclusions}
\addcontentsline{toc}{chapter}{Conclusions and perspectives}

\section*{Conclusions}

The simulation code \gls{hades} is a powerful and accurate tool for studying the interactions between a fluid and radiation, particularly thanks to its ability to accurately model radiative transfer using two different models: the M1-gray model and the M1-multigroup model. However, this accuracy comes at a high computational cost, especially in the case of the M1-multigroup model. One of the main bottlenecks lies in the computation of the Eddington factor in the M1-multigroup model, an essential ingredient of the model's closure relation but one for which no analytical expression exists. Until now, three approaches were possible: search algorithms, which are very accurate but extremely costly in computational time; interpolation on precomputed databases, which is fast and accurate but requires prior knowledge of the radiative quantities and their orders of magnitude, a condition rarely met, especially in a multigroup context; and finally, the use of the analytical Eddington factor from the M1-gray model in multigroup simulations, a fast but very inaccurate approach.

After a detailed study of the dependence of the Eddington factor on the radiative quantities and on the frequency bounds of a considered group, I developed a method based on the use of \glslink{ai}{artificial intelligence}, in which neural networks were trained using accurate data produced by search algorithms. This approach turns out to be about $3~000$ times faster than these same search algorithms, while being significantly more accurate than using the factor from the M1-gray model. Moreover, it is fully general, since it requires no prior knowledge of the characteristics of the radiation, unlike interpolation techniques.

This improvement of the \gls{hades} code enabled me to conduct a detailed study of the influence of the spectral character of radiation on the dynamics and structure of radiative shocks. I highlighted three major effects induced by fine spectral modelling with the M1-multigroup model, compared with the M1-gray model: a slower shock, a lower temperature and a higher density in the shocked region, as well as an enlargement of the radiative precursor, especially of its non-equilibrium zone. Furthermore, the comparison between the jump relations obtained from the M1-multigroup simulations and those predicted by a model based on the diffusion approximation reveals non-negligible discrepancies, on the order of 3–15\% for the density and 1–8\% for the temperature. These results emphasise the need for caution when using simplified analytical models based on the diffusion approximation.

Beyond numerical considerations, these results have major astrophysical implications, since radiative shocks are ubiquitous throughout the stellar life cycle. During star formation, for instance, the accretion shocks associated with the appearance of the first hydrostatic core play a crucial role in determining the thermal structure of protostars. Our simulations indicate that when spectral effects are treated more accurately, the energy dissipation is greater than predicted by the M1-gray model, implying cooler and more compact protostars.

In supernova remnants, hydrodynamic instabilities such as the Vishniac instability~\cite{miniere_2014_these} or the detonation-wave instability observed experimentally by Grun~\cite{grun_1991} are strongly influenced by radiative losses, which modify the dynamics of the shocked layers. A faithful modelling of radiative transport, made possible by the M1-multigroup formalism, could help reproduce the filamentary structures that are observed but not yet obtained in current simulations.

Finally, in binary systems containing a magnetised accreting white dwarf, simulations based on parametrised cooling functions predict quasi-periodic oscillations in the light curve, resulting from a radiative-shock cooling instability~\cite{busschaert_2013}. However, these models struggle to reproduce observations across the full electromagnetic spectrum~\cite{bonnet-bidaud_2015}, which may stem from an overly simplified treatment of radiative processes. The M1-multigroup model developed in this work represents a methodological advance that could improve the realism of such simulations.Nevertheless, although such results are now accessible with the \gls{hades} code, a systematic study of radiative shocks across a wide variety of configurations, particularly for varying Mach numbers, different opacity laws, or in 2D geometry, remains difficult, if not impossible, due to the still very high computational cost. This limitation arises mainly from the Courant–Friedrichs–Lewy condition imposed by the explicit schemes used in \gls{hades}, which requires extremely small time steps to ensure the stability of the radiative transfer resolution.

To overcome this obstacle, I explored the potential of using \gls{pinn} neural networks, which differ from classical architectures by explicitly embedding the physical equations within their loss function. The goal is to enable temporal extrapolation of simulations at a cost far lower than that required to continue the computation directly with \gls{hades}. To assess the potential of this approach, I first tested it on the case of a strong hydrodynamic shock. The results showed that it yielded a faithful extrapolation of the simulation data, including at times significantly beyond those initially provided.

However, its application to radiative shocks proved more challenging: the extrapolation performance is, at this stage, noticeably less satisfactory. This difficulty arises from the intrinsic complexity of the problem, which combines the presence of a discontinuity with the pronounced stiffness of the differential equations. This situation highlights the need for further methodological development to adapt the \gls{pinn} approach effectively to the specificities of radiative hydrodynamics.

\section*{Perspectives}

The application of artificial intelligence to radiative hydrodynamics simulation is a field rich with promising perspectives. Much work remains to be done to fully exploit these approaches, both from a methodological standpoint and to improve the physical understanding of the modelled phenomena. As an illustration, several research directions deserve to be explored in the coming years:

\begin{itemize}
    \item Improving \gls{pinn}-based strategies to allow for more reliable extrapolation of radiative hydrodynamics simulations, particularly in complex contexts such as radiative shocks. A complementary line of research would be to develop uncertainty or prediction-error estimators, using probabilistic architectures such as \emph{Bayesian Neural Networks}, for instance, in order to quantify the confidence in extrapolated predictions;
    
    \item Exploring the use of neural operators (\emph{Neural Operators}), which enable learning solution mappings of parametrised problems in operator form, for example by directly predicting the next time step from the current state as a replacement for classical numerical schemes. Such an approach can significantly reduce the computational cost of simulation codes and is already being used in other simulation frameworks, as suggested in recent work by Azizzadenesheli et al. (2024)~\cite{azizzadenesheli_2024};

    \item Investigating the possibility of using machine learning methods to improve the computation of the mean opacities $\kappa_E$ and $\kappa_F^{(i)}$ appearing in the M1-gray and M1-multigroup models. The goal would be to avoid the common approximation of replacing them with Planck and Rosseland means, and instead compute more realistic weighted opacities from opacity tables (e.g. TOPS), depending on the local shape of the specific intensity in the M1 model;

    \item Considering the use of neural networks trained on experimental data or on data from reference codes that solve the specific intensity (for example via characteristics or Monte Carlo methods) in order to provide a more accurate and realistic estimate of the closure relation (Eddington tensor). Such an approach could potentially overcome some intrinsic limitations of the M1 model, such as the poor representation of the interaction of two light beam, by enriching the formalism with learned closure relations better suited to complex regimes in 2D or 3D configurations.
\end{itemize}
\clearemptydoublepage


\appendix
\setcounter{chapter}{0}
\renewcommand{\thechapter}{\Alph{chapter}}  

\chapter{Basic concepts on the Planck function}
\label{appendice:fct_Planck}

\initialletter{I}n this appendix, we detail several calculation elements related to the Planck function, which will be particularly useful in Appendix~\secref{appendice:Radiation_numerique}, devoted to the numerical aspects of radiative transfer. Before delving into these calculations, let us recall that the Planck function describes the spectral distribution of the energy intensity emitted by a blackbody in thermal equilibrium at a given temperature. In dimensionless form, it can be written as follows:
\begin{equation}
    \label{eq:b_func}
    b(x) = \frac{15}{\pi^4} \frac{x^3}{e^{x} - 1} \;\;\mathpunct{,}
\end{equation}

We will first present the analytical expressions of the successive derivatives of the function $b$, then detail the computation of its antiderivative, as well as the way to approximate certain of its integrated forms, which are required for the numerical treatments of radiative transfer.

\section{Derivatives of the Planck function} \label{subappendice:derivees_planck}

Let us begin by presenting the first three derivatives of the Planck function, which appear in particular in asymptotic expansions and certain numerical approximations. The three first derivatives of $b$ are written as follows:
\begin{align*}
    b' (x) =& \frac{15}{\pi^4} x^2 \left ( \frac{3 - x}{(e^{x} - 1)} - \frac{x}{(e^{x} - 1)^2} \right ) \;\;\mathpunct{,}\\
    b'' (x) =& \frac{15}{\pi^4} x \left ( \frac{2 x^2}{(e^x - 1)^3} + 3x \frac{x-2}{(e^x - 1)^2} + \frac{x^2 - 6x + 6}{e^x - 1} \right ) \;\;\mathpunct{,}\\
    b^{(3)} (x) =& - \frac{15}{\pi^4} \left ( \frac{6 x^3}{(e^x - 1)^4} + 6 x^2 \frac{2 x - 3}{(e^x - 1)^3} + x \frac{7 x^2 - 27 x + 18}{(e^x - 1)^2} + \frac{x^3 - 9x^2 + 18 x - 6}{e^x - 1} \right ) \;\;\mathpunct{.}
\end{align*}

\section{Antiderivatives of the Planck function} \label{subappendice:primitive_planck}

Let us now introduce the antiderivative $\mathcal{B}$ of the Planck function $b$, which is expressed in the following manner:
\begin{equation}
    \label{eq:b_int_func}
    \mathcal{B}(x) = \int_0^x  b(z) \dif z \;\;\mathpunct{.}
\end{equation}

\noindent This antiderivative admits an exact expression involving \textit{polylogarithm functions}:
\begin{equation}
    \label{eq:b_int_sol}
    \mathcal{B}(x) = \frac{15 e^{-x}}{\pi} \left [ -6 \Li{4}(e^{-x}) - 6 x \Li{3}(e^{-x}) - 3 x^2 \Li{2}(e^{-x}) - x^3 \Li{1}(e^{-x})\right ] + \mathcal{C}^P \;\;\mathpunct{,}
\end{equation}

\noindent where $\mathcal{C}^P$ is an integration constant. By convention, it is set to \mbox{$\mathcal{C}^P = 1$}, so that \mbox{$\mathcal{B}(+\infty)=1$}. The polylogarithm function of order $n$, denoted $\Li{n}$, is defined for \mbox{$|x| < 1$} by the series:
\begin{equation}
    \label{eq:polylog}
    \Li{n}(x) = \sum_{k=1}^{\infty} \frac{x^k}{k^n} \;\;\mathpunct{.}
\end{equation}

However, this exact expression cannot be used directly in numerical contexts due to the infinite nature of the series. It is therefore necessary to truncate it to a finite number of terms, denoted $N_P$, which leads to using the following approximation:
\begin{equation}
    \label{eq:polylog_approx}
    \Li{n}^{N_P}(x) = \sum_{k=1}^{N_P} \frac{x^k}{k^n} \;\;\mathpunct{.}
\end{equation}

\begin{figure}
    \begin{subfigure}[t]{0.49\textwidth}
        \centering
        \includegraphics[width=\textwidth]{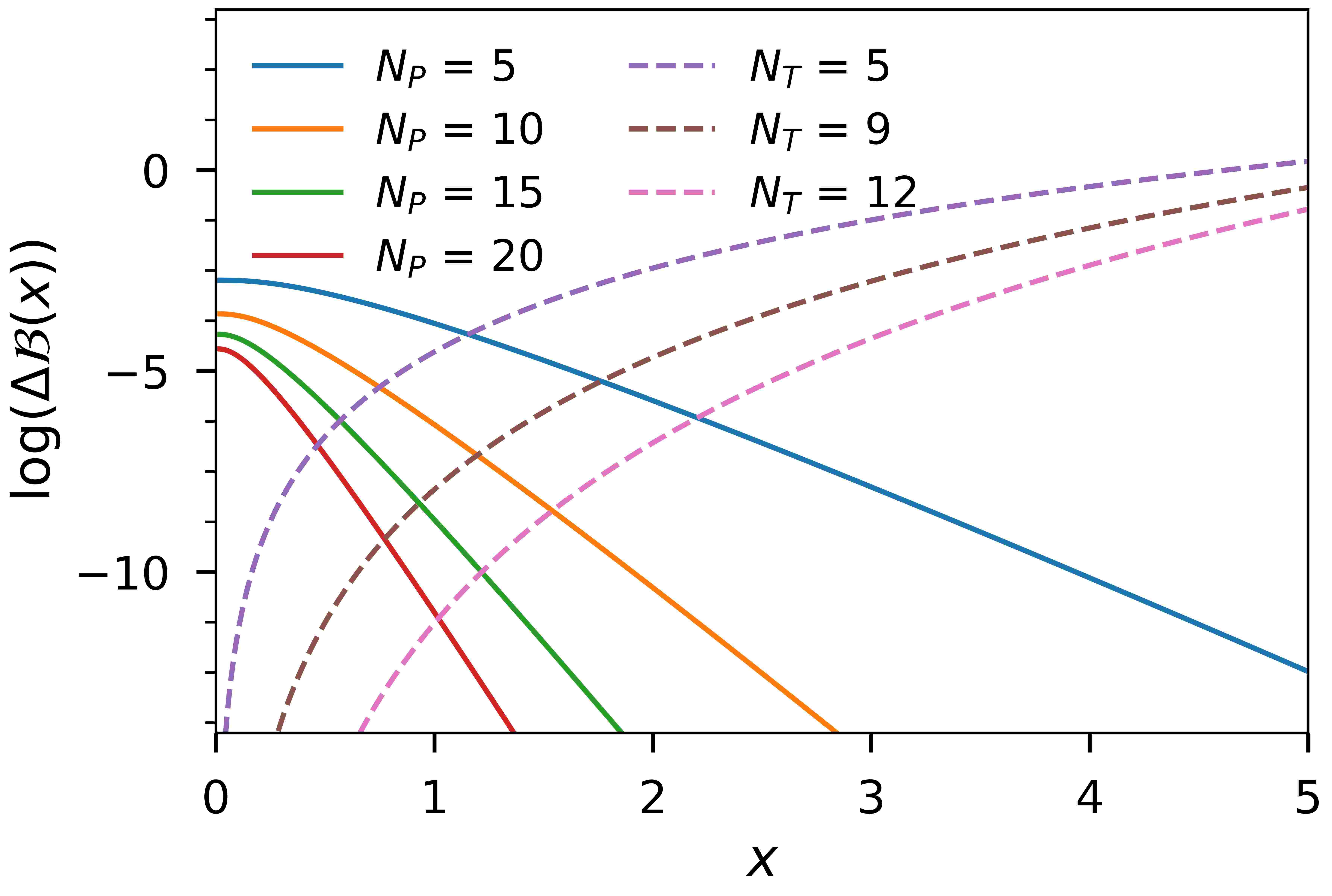}
        \caption{Error on the function $\mathcal{B}$.}
        \label{fig:B_error}
    \end{subfigure}
    \hfill
    \begin{subfigure}[t]{0.49\textwidth}
        \centering
        \includegraphics[width=\textwidth]{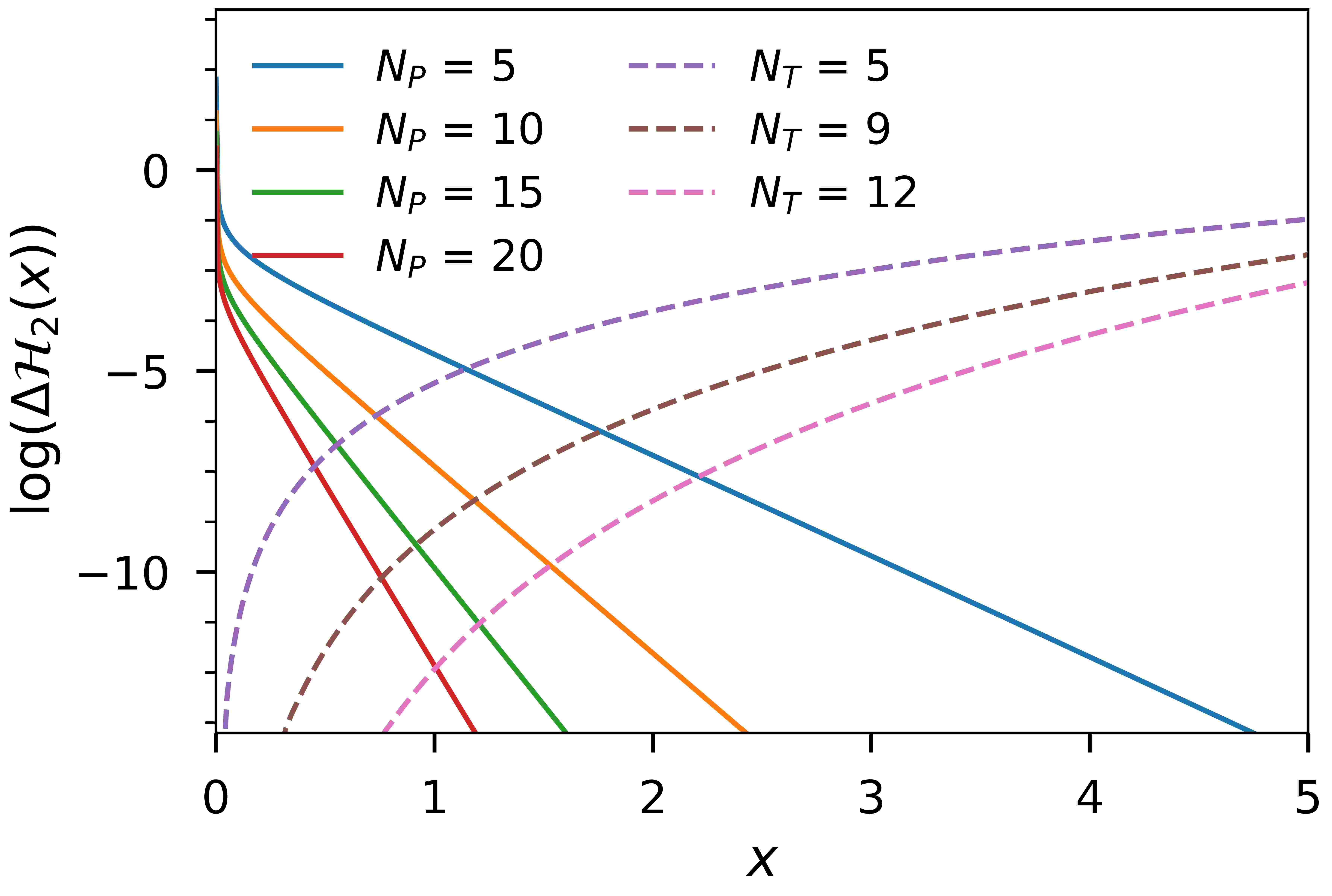}
        \caption{Error on the function $\mathcal{H}_2$.}
        \label{fig:H2_error}
    \end{subfigure}
    \hfill
    \begin{subfigure}[t]{0.49\textwidth}
        \centering
        \includegraphics[width=\textwidth]{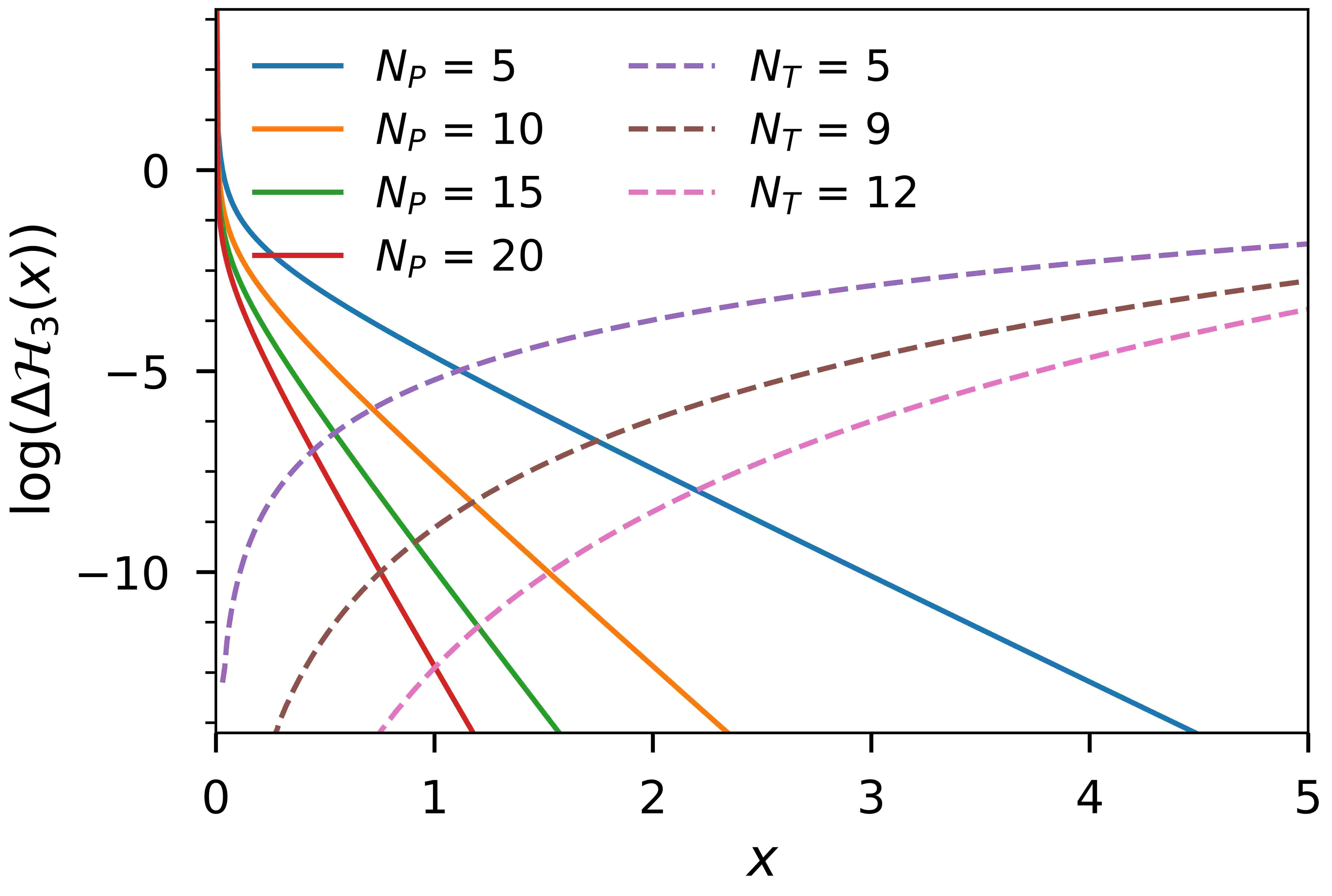}
        \caption{Error on the function $\mathcal{H}_3$.}
        \label{fig:H3_error}
    \end{subfigure}
    \hfill
    \begin{subfigure}[t]{0.49\textwidth}
        \centering
        \includegraphics[width=\textwidth]{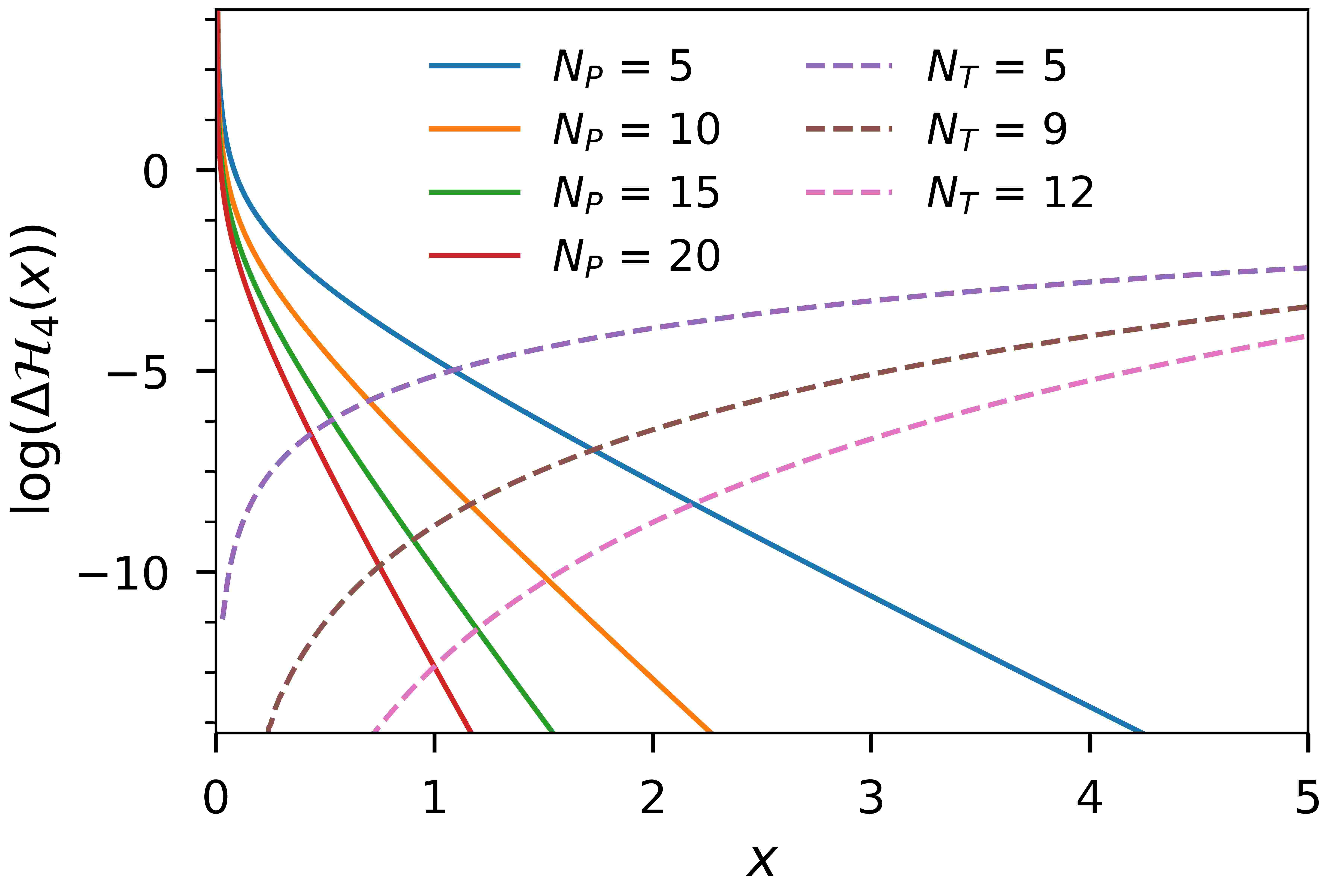}
        \caption{Error on the function $\mathcal{H}_4$.}
        \label{fig:H4_error}
    \end{subfigure}
    \caption{Logarithm of the relative error on the estimation of the functions $\mathcal{B}$, $\mathcal{H}_2$, $\mathcal{H}_3$ and $\mathcal{H}_4$, obtained using two methods: the approximation by polylogarithms truncated at $N_P$ terms (solid lines), and the approximation by a Taylor expansion of order $N_T$ for $\mathcal{B}$ (dashed lines). The error is computed with respect to the truncated polylogarithm approximation, using $N_P=100~000$.}
    \label{fig:errors_integrals}
\end{figure}

This approximation is, however, problematic in certain cases, particularly for derived quantities such as the integral of \mbox{$z \mapsto \mathcal{B}(z)/z^4$}, used in the computation of the radiative energy. Indeed, the truncation may introduce a non-physical divergence, as can be seen in figure~\ref{fig:errors_integrals}. To remedy this difficulty, Hung Chinh Nguyen (2011)~\cite{nguyen_2011_these} proposes an alternative method based on the Taylor expansion of the Planck function close to \mbox{$x=0$}. This approach makes it possible to obtain an analytical approximation of $\mathcal{B}$ by term-by-term integration of the Taylor expansion of $b$:
\begin{equation}
    \label{eq:b_taylor_approx}
    b_T^{N_T}(x) = \frac{15}{\pi^4} \sum_{k=2}^{N_T} T_k x^k \;\;\mathpunct{,}
\end{equation}

\noindent where $N_T$ is the order of the retained series, and the coefficients $T_k$ are those of the Taylor expansion of the function $b$ and are such that:
\begin{align*}
    & T_2 = 1 \;\;\mathpunct{,} && T_3 = -\frac{1}{2} \;\;\mathpunct{,} && T_4 = \frac{1}{12} \;\;\mathpunct{,} && T_5 = 0   \;\;\mathpunct{,}\\
    & T_6 = -\frac{1}{720} \;\;\mathpunct{,} && T_7 = 0 \;\;\mathpunct{,} && T_8 = \frac{1}{30240} \;\;\mathpunct{,} &&T_9 = 0 \;\;\mathpunct{,}\\
    & T_{10} = -\frac{1}{1209600} \;\;\mathpunct{,} && T_{11} = 0 \;\;\mathpunct{,} && T_{12} = \frac{1}{47900160} \;\;\mathpunct{.} &&
\end{align*}

\noindent The integration of this Taylor expansion then provides an approximation of $\mathcal{B}$, valid for small values of $x$:
\begin{equation}
    \label{eq:b_int_taylor_approx}
    \mathcal{B}_T^{N_T}(x) = \frac{15}{\pi^4} \sum_{k=3}^{N_T+1} \frac{T_{k-1} x^k}{k} + \mathcal{C}^T \;\;\mathpunct{.}
\end{equation}

\noindent where $\mathcal{C}^T$ is an integration constant that we set to zero in order to ensure that \mbox{$\mathcal{B}_T^{N_T}(0) = 0$}. By combining the two approaches, Taylor expansion of order $N_T$ for $x < x^*$ and polylogarithm functions truncated at $N_P$ for \mbox{$x > x^*$}, it is possible to construct an accurate global approximation of the function $\mathcal{B}$. For example, by taking \mbox{$N_P = 20$}, \mbox{$N_T = 12$}, and a transition point \mbox{$x^* = 1$}, the maximum approximation error remains below $10^{-11}$, as illustrated in figure~\ref{fig:B_error}.

\section{Functions derived from the Planck function} \label{subappendice:H234_planck}

Let us introduce a family of antiderivatives associated to the Planck function $\mathcal{B}$, defined by:
\begin{equation}
    \label{eq:fct_Hi}
    \mathcal{H}_i (x) = \int \frac{\mathcal{B}(z)}{z^i} \dif z \;\;\mathpunct{.}
\end{equation}

\begin{figure}
    \begin{subfigure}[t]{0.49\textwidth}
        \centering
        \includegraphics[width=\textwidth]{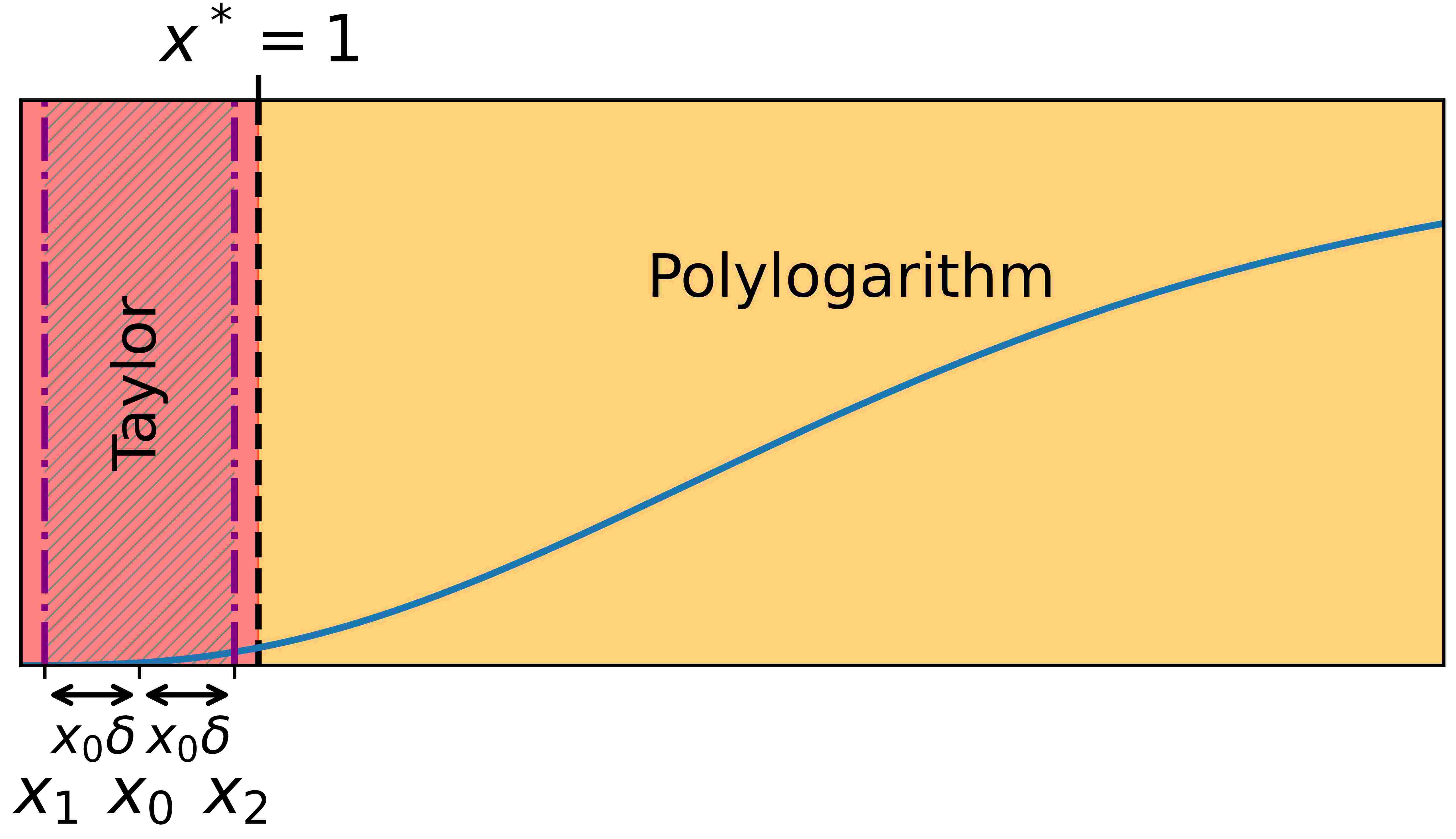}
        \caption{Computation of the integrals $\mathcal{H}_i(x_1, x_2)$ in the case where $x_1 < x^*$ and $x_2 < x^*$.}
    \end{subfigure}
    \hfill
    \begin{subfigure}[t]{0.49\textwidth}
        \centering
        \includegraphics[width=\textwidth]{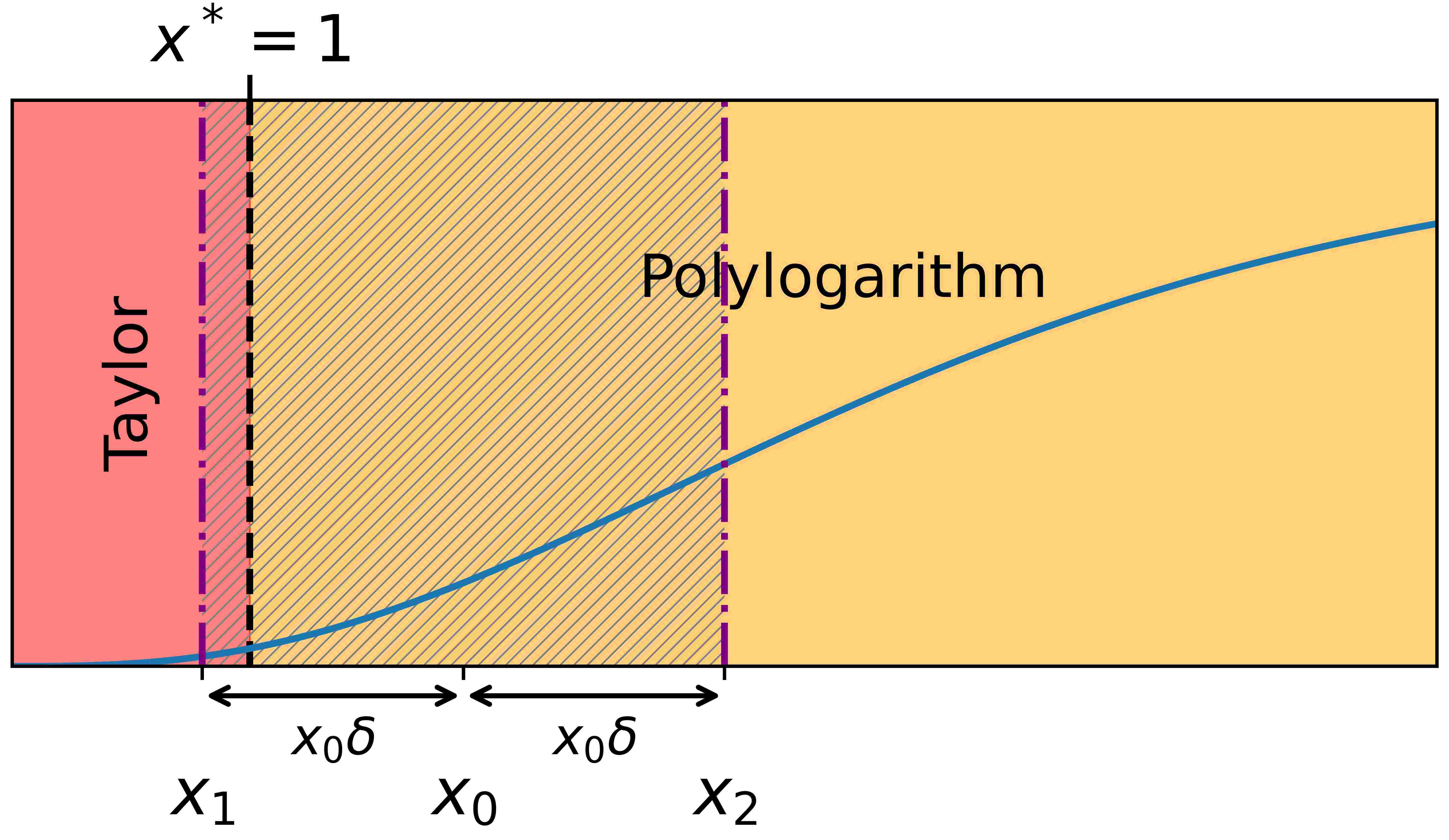}
        \caption{Computation of the integrals $\mathcal{H}_i(x_1, x_2)$, in the case where $x_1<x^*$ et  $x_2>x^*$.}
    \end{subfigure}
    \hfill
    \begin{center}
        \begin{subfigure}[t]{0.49\textwidth}
            \centering
            \includegraphics[width=\textwidth]{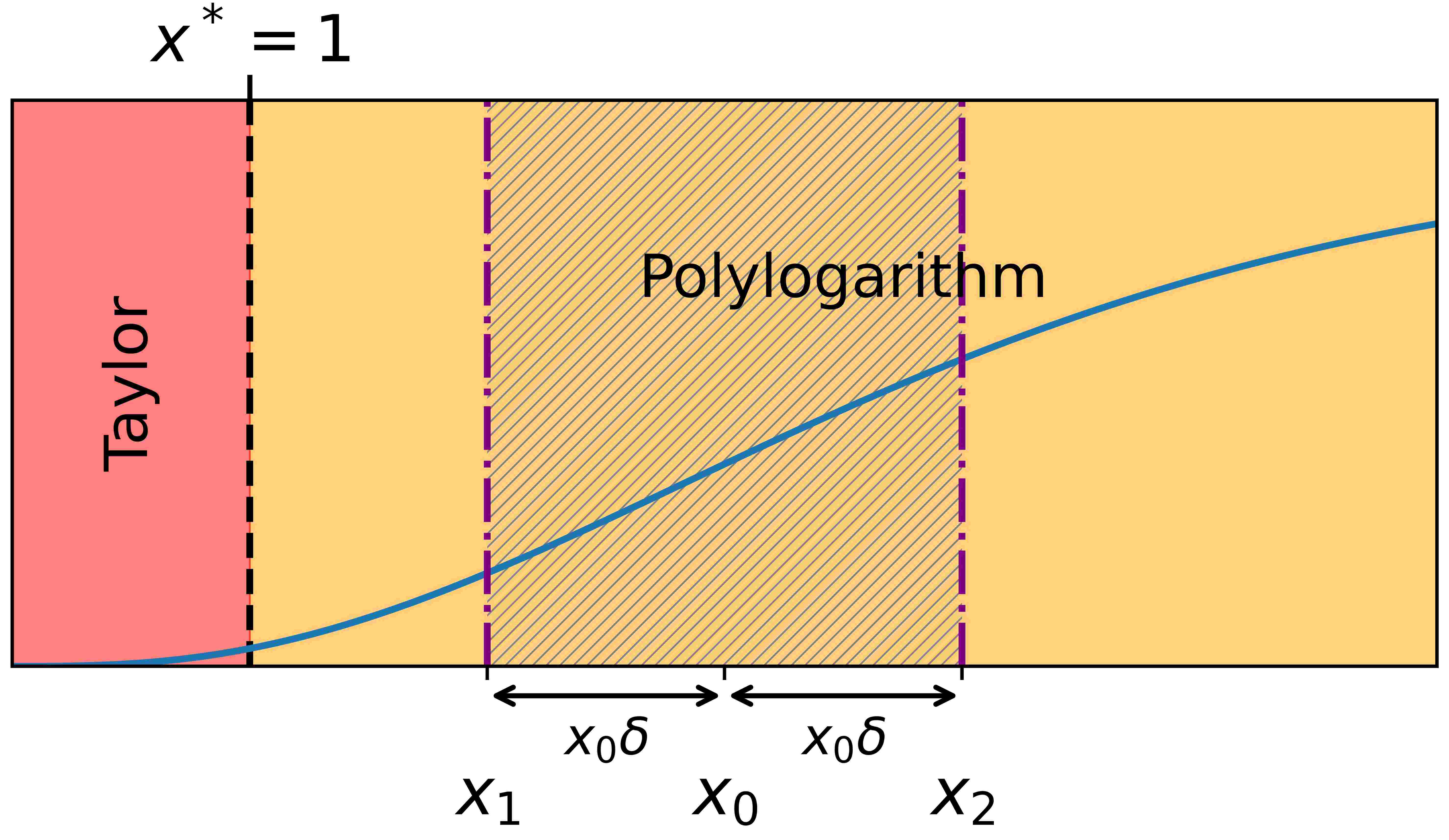}
            \caption{Computation of the integrals $\mathcal{H}_i(x_1, x_2)$, in the case where $x_1>x^*$ et  $x_2>x^*$.}
        \end{subfigure}
    \end{center}
    \caption{Explanatory diagram of the computation of the integrals $\mathcal{H}_i$, using Taylor expansions, truncated polylogarithm functions, or a combination of both.}
    \label{fig:integral_calc}
\end{figure}

\indent In this section, I revisit and expand upon the results presented by Hung Chinh Nguyen (2011)~\cite{nguyen_2011_these}, which I have further developed in order to enable a more efficient computation of these integrals in the case where the integration bounds $x_1$ and $x_2$ are very close. I will also present the expressions of the successive derivatives of these integrals.\\

\noindent \textbf{\ding{118} Numerical computation of the functions $\bm{\mathcal{H}_4}$, $\bm{\mathcal{H}_3}$ and $\bm{\mathcal{H}_2}$}\

\noindent Using the approximations of the Planck function $\mathcal{B}$ through truncated polylogarithm functions (equation~\eqref{eq:polylog_approx}) or through a Taylor series expansion (equation~\eqref{eq:b_taylor_approx}), the functions $\mathcal{H}_i$ can be estimated using the following expressions:
\begin{equation}
    \label{eq:Hi_approx}
    \mathcal{H}_i (x) = 
    \begin{cases}
        \mathcal{H}_{i,P}^{N_P}(x) & \text{if $x \le x^*$} \\
        \mathcal{H}_{i,T}^{N_T}(x) & \text{otherwise}
    \end{cases} \;\;\mathpunct{.}
\end{equation}

\noindent where the subscripts $P$ and $T$ respectively denote the truncated polylogarithm functions and the Taylor expansion, and $x^*$ is the matching point between their two domains of validity. This computation is illustrated in figure~\ref{fig:integral_calc}. For the specific cases of the functions $\mathcal{H}_2$, $\mathcal{H}_3$, and $\mathcal{H}_4$, one obtains the following explicit expressions for the truncated polylogarithm functions:
\begin{align*}
    \mathcal{H}_{2,P}^{N_P}(x) &= \frac{15}{\pi^4} \left [ 6 x^{-1}\Li{4}^{N_P}(e^{-x}) + 4 \Li{3}^{N_P}(e^{-x}) + x \Li{2}^{N_P}(e^{-x}) \right ] - \frac{1}{x} + \mathcal{C}_2^P \;\;\mathpunct{,} \\
    \mathcal{H}_{3,P}^{N_P}(x) &= \frac{15}{\pi^4} \left [ 3 x^{-2} \Li{4}^{N_P}(e^{-x}) + 3 x^{-1} \Li{3}^{N_P}(e^{-x}) + \Li{2}^{N_P}(e^{-x}) \right ] - \frac{1}{2 x^2} + \mathcal{C}_3^P \;\;\mathpunct{,}\\
    \mathcal{H}_{4,P}^{N_P}(x) &= \frac{15}{\pi^4} \left [ 2 x^{-3} \Li{4}^{N_P}(e^{-x}) + 2 x^{-2} \Li{3}^{N_P}(e^{-x}) + x^{-1} \Li{2}^{N_P}(e^{-x}) \right ] - \frac{1}{3 x^3} + \mathcal{C}_4^P \;\;\mathpunct{,} \\
\end{align*}

\noindent and, for the Taylor expansions:
\begin{align*}
    \mathcal{H}_{2,T}^{N_T}(x) &= \frac{15}{\pi^4} \sum_{k=2}^{N_T} \frac{T_k x^k}{k (k+1)} + \mathcal{C}_2^T \;\;\mathpunct{,} \\
    \mathcal{H}_{3,T}^{N_T}(x) &= \frac{15}{\pi^4} \sum_{k=1}^{N_T-1} \frac{T_{k+1} x^k}{k (k+2)} + \mathcal{C}_3^T  \;\;\mathpunct{,}\\
    \mathcal{H}_{4,T}^{N_T}(x) &= \frac{15}{\pi^4} \left [ \frac{T_2 \ln(x)}{3} + \sum_{k=1}^{N_T-2} \frac{T_{k+2} x^k}{k (k+3)} \right ] + \mathcal{C}_4^T \;\;\mathpunct{.} \\
\end{align*}

\noindent The terms $\mathcal{C}_i^P$ and $\mathcal{C}_i^T$ are integration constants. As for the function $\mathcal{B}$, one can verify that choosing \mbox{$N_P = 20$}, \mbox{$N_T = 12$}, and \mbox{$x^* = 1$} ensures excellent accuracy (maximum error below $10^{-12}$; see figure~\ref{fig:errors_integrals}). For the computation of an integral of the form \mbox{$\displaystyle \int_{x_1}^{x_2} \frac{\mathcal{B}(z)}{z^i} \dif z$} with \mbox{$x_2 \ge x_1$}, we introduce:
\begin{align*}
    &x_0 = \frac{x_2 + x_1}{2} \;\;\mathpunct{,} && x_T = \frac{x^* + x_1}{2} \;\;\mathpunct{,} && x_P = \frac{x_2 + x^*}{2} \;\;\mathpunct{,} \\
    &\delta = \frac{x_2 - x_1}{2 x_0} \;\;\mathpunct{,} && \delta_T = \frac{x^* - x_1}{2x_T} \;\;\mathpunct{,} && \delta_P = \frac{x_2 - x^*}{2x_P} \;\;\mathpunct{,} \\
\end{align*}

\noindent and we write:
\begin{equation*}
    \mathcal{H}_i (x_1 ; x_2) = \int \nolimits_{x_1}^{x_2} \frac{\mathcal{B}(z)}{z^i} \dif z = \delta e^{-x_1} x_0^{1-i} \mathcal{H}_i^* (x_1 ; x_2)  \;\;\mathpunct{,}
\end{equation*}

\noindent where the function $\mathcal{H}_i^*$ depends on the domain in which the bounds $x_1$ and $x_2$ lie:
\begin{equation*}
    \mathcal{H}_i^* (x_1 ; x_2) = 
    \begin{cases}
        \mathcal{H}_{i,P}^*(x_1 ; x_2) & \text{if $x_1 > x^*$ and $x_2 > x^*$} \\
        \mathcal{H}_{i,T}^*(x_1 ; x_2) & \text{if $x_1 \le x^*$ and $x_2 \le x^*$} \\
        \frac{\delta_T}{\delta} \left ( \frac{x_0}{x_T} \right )^{i-1} \mathcal{H}_{i,T}^*(x_1 ; x^*) + e^{x_1 - x^*}\frac{\delta_T}{\delta} \left ( \frac{x_0}{x_P} \right )^{i-1} \mathcal{H}_{i,P}^*(x^* ; x_2) & \text{otherwise} 
    \end{cases}  \;\;\mathpunct{.}
\end{equation*}

\noindent Letting $x_l$ and $x_h$ denote the lower and upper bounds (equal respectively to $x_1$ or $x^*$, and to $x_2$ or $x^*$), and letting $x_m$ and $\delta_m$ be variables such that \mbox{$x_m = (x_h + x_l)/2$} and \mbox{$\delta_m = (x_h - x_l)/(2 x_m)$}, the functions $\mathcal{H}_{i,P}^{*}$ and $\mathcal{H}_{i,T}^{*}$ are given by:
\begin{align*}
    \mathcal{H}_{i,P}^*(x_l ; x_h) = e^{x_l} \delta_m^{-1} x_m^{i-1} \left ( \mathcal{H}_{i,P}^{N_P}(x_h) -  \mathcal{H}_{i,P}^{N_P}(x_l) \right ) \;\;\mathpunct{,}\\
    \mathcal{H}_{i,T}^*(x_l ; x_h) = e^{x_l} \delta_m^{-1} x_m^{i-1} \left ( \mathcal{H}_{i,T}^{N_T}(x_h) - \mathcal{H}_{i,T}^{N_T}(x_l) \right ) \;\;\mathpunct{.}
\end{align*}

For the functions $\mathcal{H}_2$, $\mathcal{H}_3$, and $\mathcal{H}_4$, the differences $\mathcal{H}_{i,P}^*$ can be efficiently evaluated by expanding the truncated polylogarithm functions:
\begin{align*}
    \mathcal{H}_{2,P}^*(x_l; x_h) =& - \frac{15}{\pi^4} \sum_{k=1}^{N_P} e^{-(k-1) x_l} \left [ 6 \frac{h^1_k(x_m, \delta_m)}{k^4} + 4 x_m \frac{\Delta_k(x_m, \delta_m) }{k^3} + x_m^2 \frac{h^2_k(x_m, \delta_m)}{k^2}\right ] + \frac{2 e^{x_l}}{1-\delta_m^2}  \;\;\mathpunct{,} \\
    \mathcal{H}_{3,P}^*(x_l; x_h) =& - \frac{15}{\pi^4} \sum_{k=1}^{N_P} e^{-(k-1) x_l} \left [ 3 \frac{h^3_k(x_m, \delta_m) }{k^4} + 3 x_m \frac{h^1_k(x_m, \delta_m)}{k^3} + x_m^2 \frac{\Delta_k(x_m, \delta_m) }{k^2}\right ] + \frac{2 e^{x_l}}{(1-\delta_m^2)^2} \;\;\mathpunct{,} \\
    \mathcal{H}_{4,P}^*(x_l; x_h) =& - \frac{15}{\pi^4} \sum_{k=1}^{N_P} e^{-(k-1) x_l} \left [ 2 \frac{h^4_k(x_m, \delta_m) }{k^4} + 2 x_m \frac{h^3_k(x_m, \delta_m) }{k^3} + x_m^2 \frac{h^1_k(x_m, \delta_m)}{k^2}\right ] + \frac{2 e^{x_l} (3+\delta_m^2)}{3 (1-\delta_m^2)^3} \;\;\mathpunct{,}
\end{align*}

\noindent where the auxiliary functions $h^j_k$ are defined by:
\begin{align*}
    h^1_k(x_m, \delta_m) &= \frac{\Delta_k(x_m, \delta_m)  + \Sigma_k(x_m, \delta_m) }{1-\delta_m^2} \;\;\mathpunct{,} \\
    h^2_k(x_m, \delta_m) &= \Delta_k(x_m, \delta_m) - \Sigma_k(x_m, \delta_m) \;\;\mathpunct{,} \\
    h^3_k(x_m, \delta_m) &= \frac{(1+\delta_m^2) \Delta_k(x_m, \delta_m) + 2 \Sigma_k(x_m, \delta_m) }{(1-\delta_m^2)^2} \;\;\mathpunct{,} \\
    h^4_k(x_m, \delta_m) &= \frac{(1+3\delta_m^2) \Delta_k(x_m, \delta_m) + (3+\delta_m^2)\Sigma_k(x_m, \delta_m) }{(1-\delta_m^2)^3} \;\;\mathpunct{.} 
\end{align*}

\noindent The functions $\Delta_k$ and $\Sigma_k$ are given by:
\begin{align}
    \Delta_k(x_m, \delta_m)  &= \frac{1 - e^{-2 k x_m \delta_m}}{\delta_m} \;\;\mathpunct{,}  \label{eq:D_fct}\\
    \Sigma_k(x_m, \delta_m)  &= 1 + e^{-2k x_m \delta_m} \;\;\mathpunct{.} \label{eq:S_fct}
\end{align}

\noindent For \mbox{$k x_m \delta_m \ll 1$}, one may use the approximation:
\begin{equation*}
    \Delta_k(x_m, \delta_m) \approx 2 k x_m \left ( 1 - k x_m \delta_m  +\frac{2(k x_m \delta_m)^2}{3} \right ) \;\;\mathpunct{.}
\end{equation*}

\noindent Finally, the differences $\mathcal{H}_{i,T}^*$ can also be efficiently evaluated using the following expressions:
\begin{align*}
    \mathcal{H}_{2,T}^*(x_l; x_h) =&  \frac{15 e^{x_l}}{\pi^4} \sum_{k=3}^{N_T+1} \sum_{l=0}^{\entiere{\frac{k-2}{2}}} \binom{k-1}{2l+1}  \frac{2 T_{k-1} x_m^k \delta_m^{2l}}{(k-1) k} \;\;\mathpunct{,} \\
    \mathcal{H}_{3,T}^*(x_l; x_h) =&  \frac{15 e^{x_l}}{\pi^4} \sum_{k=3}^{N_T+1} \sum_{l=0}^{\entiere{\frac{k-3}{2}}} \binom{k-2}{2l+1} \frac{2 T_{k-1}x_m^k \delta_m^{2l}}{(k-2) k} \;\;\mathpunct{,} \\
    \mathcal{H}_{4,T}^*(x_l; x_h) =&  \frac{15 e^{x_l}}{\pi^4} \left \{ \sum_{k=4}^{N_T+1} \sum_{l=0}^{\entiere{\frac{k-4}{2}}} \binom{k-3}{2l+1} \frac{2 T_{k-1} x_m^k \delta_m^{2l}}{(k-3) k } + \frac{T_2 x_m^3}{3 \delta_m} \ln\left (\frac{1+\delta_m}{1-\delta_m}\right ) \right \} \;\;\mathpunct{,}
\end{align*}

\noindent where $\binom{N}{k}$ denotes the binomial coefficient “$N$ choose $k$”, and the function $\entiere{\bcdot}$ is the floor function. For small values of $\delta_m$, the logarithm in $\mathcal{H}_{4,T}$ can be approximated by:
\begin{equation*}
    \frac{1}{\delta_m} \ln\left (\frac{1+\delta_m}{1-\delta_m}\right ) \approx 2 \left ( 1 + \frac{\delta_m^2}{3} \right ) \;\;\mathpunct{.}
\end{equation*}\\

\noindent \textbf{\ding{118} Computation of combined expressions of the functions $\bm{\mathcal{H}_4}$, $\bm{\mathcal{H}_3}$, and $\bm{\mathcal{H}_2}$}\\

Let $x_1$ and $x_2$ be two real numbers such that \mbox{$x_2 \ge x_1$}, and let \mbox{$x_0 = (x_1 + x_2)/2$} and \\ \mbox{$\delta = (x_2 - x_1)/(2x_0)$}. In the following appendix, we will need to evaluate the following combinations:
\begin{align}
    \mathcal{E}(x_1 ; x_2) &= x_0^3 \mathcal{H}_4(x_1 ; x_2) \;\;\mathpunct{,} \label{eq:Eg_int}\\
    \mathcal{F}(x_1 ; x_2) &= x_0^2 \mathcal{H}_3(x_1 ; x_2) - x_0^3 \mathcal{H}_4(x_1 ; x_2) \;\;\mathpunct{,} \label{eq:fg_int}\\
    \mathcal{P}(x_1 ; x_2) &= x_0 \mathcal{H}_2(x_1 ; x_2) - 2 x_0^2 \mathcal{H}_3(x_1 ; x_2) + x_0^3 \mathcal{H}_4(x_1 ; x_2) \;\;\mathpunct{.} \label{eq:chig_int}
\end{align}

\noindent In practice, these quantities are evaluated using the following expressions, valid for any configuration of the bounds $x_1$ and $x_2$ relative to a threshold value $x^*$:
\begin{align*}
    \mathcal{E} (x_1 ; x_2) &= \delta e^{-x_1} \mathcal{E}^* (x_1 ; x_2)  \;\;\mathpunct{,}\\
    \mathcal{F} (x_1 ; x_2) &= \delta^3 e^{-x_1} \mathcal{F}^* (x_1 ; x_2) \;\;\mathpunct{,}\\
    \mathcal{P} (x_1 ; x_2) &= \delta^3 e^{-x_1} \mathcal{P}^* (x_1 ; x_2) \;\;\mathpunct{.}
\end{align*}

\noindent with:
\begin{align*}
    \mathcal{E}^* (x_1 ; x_2) &= 
    \begin{cases}
        \mathcal{H}_{4,P}^*(x_1 ; x_2) & \text{if $x_1 > x^*$ and $x_2 > x^*$} \\
        \mathcal{H}_{4,T}^*(x_1 ; x_2) & \text{if $x_1 \le x^*$ and $x_2 \le x^*$} \\
        \mathcal{H}_4^*(x_1 ; x_2) & \text{otherwise} 
    \end{cases}\;\;\mathpunct{,}\\[5pt]
    \mathcal{F}^* (x_1 ; x_2) &= 
    \begin{cases}
        \mathcal{F}_P^*(x_1 ; x_2) & \text{if $x_1 > x^*$ and $x_2 > x^*$} \\
        \mathcal{F}_T^*(x_1 ; x_2) & \text{if $x_1 \le x^*$ and $x_2 \le x^*$} \\
        \delta^{-2} \left ( \mathcal{H}_3^*(x_1 ; x_2) - \mathcal{H}_4^*(x_1 ; x_2) \right ) & \text{otherwise} 
    \end{cases}\;\;\mathpunct{,}\\[5pt]
    \mathcal{P}^* (x_1 ; x_2) &= 
    \begin{cases}
        \mathcal{P}_P^*(x_1 ; x_2) & \text{if $x_1 > x^*$ and $x_2 > x^*$} \\
        \mathcal{P}_T^*(x_1 ; x_2) & \text{if $x_1 \le x^*$ and $x_2 \le x^*$} \\
        \delta^{-2} \left ( \mathcal{H}_2^*(x_1 ; x_2) - 2 \mathcal{H}_3^*(x_1 ; x_2) + \mathcal{H}_4^*(x_1 ; x_2) \right ) & \text{otherwise} 
    \end{cases} \;\;\mathpunct{.}
\end{align*}

\noindent Thus, the integral $\mathcal{E}^*$ is already known, since it is computed in the same way as $\mathcal{H}_4^*$. We still need to derive explicit the expressions of $\mathcal{F}^*$ and $\mathcal{P}^*$. The quantities $\mathcal{F}_P^*$, $\mathcal{F}_T^*$, $\mathcal{P}_P^*$, and $\mathcal{P}_T^*$ are defined from the functions $\mathcal{H}_{i,P}^*$ and $\mathcal{H}_{i,T}^*$ as follows:
\begin{align*}
    \mathcal{F}_P^*(x_1 ; x_2) &= \mathcal{H}_{2,P}^*(x_1 ; x_2) - \mathcal{H}_{4,P}^*(x_1 ; x_2) \;\;\mathpunct{,}\\
    \mathcal{F}_T^*(x_1 ; x_2) &= \mathcal{H}_{3,T}^*(x_1 ; x_2) - \mathcal{H}_{4,T}^*(x_1 ; x_2) \;\;\mathpunct{,}\\
    \mathcal{P}_P^*(x_1 ; x_2) &= \mathcal{H}_{2,P}^*(x_1 ; x_2) - 2 \mathcal{H}_{2,P}^*(x_1 ; x_2) + \mathcal{H}_{4,P}^*(x_1 ; x_2) \;\;\mathpunct{,}\\
    \mathcal{P}_T^*(x_1 ; x_2) &= \mathcal{H}_{2,T}^*(x_1 ; x_2) - 2 \mathcal{H}_{3,T}^*(x_1 ; x_2) + \mathcal{H}_{4,T}^*(x_1 ; x_2) \;\;\mathpunct{.}
\end{align*}

From the explicit expressions of the functions $\mathcal{H}_{i,P}^*$ given earlier, one obtains the following efficient formulas for the computation of $\mathcal{F}_P^*$ and $\mathcal{P}_P^*$:
\begin{align*}
    \mathcal{F}_P^*(x_1 ; x_2) =& \frac{15}{\pi^4} \sum_{k=1}^{N_P} e^{-(k-1) x_1} \left [ \frac{z^1_k(x_0, \delta)}{k^4} + x_0 \frac{z^2_k(x_0, \delta)}{k^3} + x_0^2 \frac{z^3_k(x_0, \delta)}{k^2} \right ] - \frac{8 e^{x_1}}{3 (1-\delta^2)^3} \;\;\mathpunct{,}\\
    \mathcal{P}_P^*(x_1 ; x_2) =& - \frac{15}{\pi^4} \sum_{k=1}^{N_P} e^{-(k-1) x_1} \left [ \frac{q^1_k(x_0, \delta)}{k^4} + x_0 \frac{q^2_k(x_0, \delta)}{k^3} + x_0^2 \frac{h^1_k(x_0, \delta)}{k^2} \right ] + \frac{2 e^{x_1} (1+3\delta^2)}{3 (1-\delta^2)^3} \;\;\mathpunct{,}
\end{align*}

\noindent where the auxiliary function $h_k^1$ was introduced in the previous section, and where the functions $z_k^i$ and $q_k^i$ are defined as follows:
\begin{align*}
    z^1_k(x_0, \delta) &= \frac{8 (\Sigma_k(x_0, \delta) + \Delta_k(x_0, \delta)) + (1-\delta^2)(1+3\delta^2)\mathcal{T}_k(x_0, \delta)}{(1-\delta^2)^3} \;\;\mathpunct{,}\\
    z^2_k(x_0, \delta) &= \frac{(1-\delta^2)\mathcal{T}_k(x_0, \delta) + 4\Delta_k(x_0, \delta)}{(1-\delta^2)^2} \;\;\mathpunct{,}\\
    z^3_k(x_0, \delta) &= \frac{\Delta_k(x_0, \delta)}{1-\delta^2} \;\;\mathpunct{,}\\
    q^1_k(x_0, \delta) &= 2 \frac{(1+3\delta^2)\left ( \Delta_k(x_0, \delta) +  \Sigma_k(x_0, \delta) \right ) - (1-\delta^2)(1-3\delta^2) \mathcal{T}_k(x_0, \delta)}{(1-\delta^2)^3} \;\;\mathpunct{,}\\
    q^2_k(x_0, \delta) &= \frac{4 \delta^2 \Delta_k(x_0, \delta)}{(1-\delta^2)^2} \;\;\mathpunct{.}
\end{align*}

\noindent The functions $\Sigma_k$ and $\Delta_k$ are given respectively by equations~\eqref{eq:S_fct} and~\eqref{eq:D_fct}, and the function $\mathcal{T}_k$ is defined by:
\begin{equation}
    \label{eq:T_fct}
    \mathcal{T}_k(x_0, \delta) = \frac{k x_0 \Sigma_k(x_0, \delta)-\Delta_k(x_0, \delta)}{\delta^2} \;\;\mathpunct{.} 
\end{equation}

\noindent For small values of \mbox{$k x_0 \delta$}, this function admits the following approximation:
\begin{equation*}
     \mathcal{T}_k(x_0, \delta) \approx \frac{2 (k x_0)^3}{3}\left( 1 - k x_0 \delta  + \frac{3 (k x_0 \delta )^2}{5}\right) \;\;\mathpunct{.} 
\end{equation*}

\noindent Similarly, the explicit expressions of $\mathcal{H}_{i,T}^*$ make it possible to efficiently evaluate $\mathcal{F}_T^*$ and $\mathcal{P}_T^*$:
\begin{align*}  
    \mathcal{F}_T^*(x_1 ; x_2) =& \frac{15 e^{x_1}}{\pi^4} \left \{ \sum_{k=2}^{\entiere{\frac{N_T}{2}}} \frac{2 x_0^{2k+1} T_{2k} \delta^{2(k-2)}}{(2k-1) (2k+1)} + \sum_{k=6}^{N_T+1} \sum_{l=0}^{\entiere{\frac{k-6}{2}}} \binom{k-2}{2l+3} \frac{4 x_0^k T_{k-1} (l+1) \delta^{2l}}{(k-3) (k-2) k} - \right.\\
    & \left. \frac{T_2 x_0^3}{3 \delta^3} \left [\ln\left (\frac{1+\delta}{1-\delta}\right )  - 2 \delta\right ]\right \} \;\;\mathpunct{,} \\    
    \mathcal{P}_T^*(x_1 ; x_2) =& \frac{15 e^{x_1}}{\pi^4} \left \{ \sum_{k=2}^{\entiere{\frac{N_T+1}{2}}}\frac{x_0^{2k} T_{2k-1} \delta^{2(k-2)}}{(2k-1) k} + \sum_{k=2}^{\entiere{\frac{N_T}{2}}}\frac{2 (2k-3) x_0^{2k+1} T_{2k} \delta^{2(k-2)}}{(2k-1) (2k+1)} + \right.\\
    &\left. \sum_{k=6}^{N_T+1} \sum_{l=0}^{\entiere{\frac{k-6}{2}}} \binom{k-1}{2l+3}  \frac{4 (l+1) (2l+1) x_0^k T_{k-1} \delta^{2l}}{(k-3) (k-2) (k-1) k } + \frac{T_2 x_0^3 }{3 \delta^3} \left [ \ln\left (\frac{1+\delta}{1-\delta}\right ) - 2\delta \right ]\right \} \;\;\mathpunct{.}
\end{align*}

\noindent For small values of $\delta$, the logarithmic term is approximated by:
\begin{equation*}
    \frac{1}{\delta^3} \left \{\ln\left ( \frac{1+\delta}{1-\delta} \right ) - 2\delta \right \} \approx \frac{2}{3} \left ( 1 +\frac{3 \delta^2}{5}\right ) \;\;\mathpunct{.}
\end{equation*}

The main advantage of the formulations developed in this section for the quantities $\mathcal{E}$, $\mathcal{F}$, and $\mathcal{P}$ lies in the fact that they explicitly highlight the proportionality factors in $\delta$ and $e^{-x_1}$. This explicit factorization facilitates the computation of these quantities, and in particular that of the ratios \mbox{$\mathcal{F}/\mathcal{E}$} and \mbox{$\mathcal{P}/\mathcal{E}$}, which will be used in Appendix~\secref{appendice:Radiation_numerique}.

However, in the case where \mbox{$x_1 < x^*$} and \mbox{$x_2 > x^*$}, no such explicit dependence on $\delta$ can be extracted for $\mathcal{F}$, $\mathcal{P}$, and $\mathcal{E}$. This limitation then degrades the accuracy of the computations of the ratios \mbox{$\mathcal{F}/\mathcal{E}$} and \mbox{$\mathcal{P}/\mathcal{E}$} when $\delta$ becomes very small.\\

\noindent \textbf{\ding{118} Derivatives of the functions $\bm{\mathcal{H}_4}$, $\bm{\mathcal{H}_3}$, and $\bm{\mathcal{H}_2}$}\\

Let us now present the odd-order derivatives of the functions $\mathcal{H}_2$, $\mathcal{H}_3$, and $\mathcal{H}_4$. The first, third, and fifth derivatives of the function $\mathcal{H}_2$ are written as follows:
\begin{align*}
    &\mathcal{H}_2'(x)      &&= \frac{\mathcal{B}(x)}{x^2}  \;\;\mathpunct{,}&&\\
    &\mathcal{H}_2^{(3)}(x) &&= \frac{h_2^{(3)}(x)}{x^4} &&= \frac{x^2 b'(x) - 4 x b(x) + 6 \mathcal{B}(x))}{x^4}  \;\;\mathpunct{,}\\
    &\mathcal{H}_2^{(5)}(x) &&= \frac{h_2^{(5)}}{x^6}    &&= \frac{x^4 b^{(3)}(x) - 8 x^3 b''(x) + 36 x^2 b'(x) - 96 x b(x) + 120 \mathcal{B}(x)}{x^6}  \;\;\mathpunct{,}
\end{align*}

\noindent those of the function $\mathcal{H}_3$ are given by:
\begin{align*}
    &\mathcal{H}_3'(x) &&= \frac{\mathcal{B}(x)}{x^3}  \;\;\mathpunct{,}&&\\
    &\mathcal{H}_3^{(3)}(x) &&= \frac{h_3^{(3)}}{x^5} &&= \frac{x^2 b'(x) - 6 x b(x) + 12 \mathcal{B}(x))}{x^5}  \;\;\mathpunct{,}\\
    &\mathcal{H}_3^{(5)}(x) &&= \frac{h_3^{(5)}}{x^7} &&= \frac{x^4 b^{(3)}(x) - 12 x^3 b''(x) + 72 x^2 b'(x) - 240 x b(x) + 360 \mathcal{B}(x)}{x^7}  \;\;\mathpunct{,}
\end{align*}

\noindent while those of the function $\mathcal{H}_4$ are written as follows:
\begin{align*}
    &\mathcal{H}_4'(x) &&= \frac{\mathcal{B}(x)}{x^4}  \;\;\mathpunct{,}&&\\
    &\mathcal{H}_4^{(3)}(x) &&= \frac{h_4^{(3)}(x)}{x^6} &&= \frac{x^2 b'(x) - 8 x b(x) + 20 \mathcal{B}(x))}{x^6} 
 \;\;\mathpunct{,} \\
    &\mathcal{H}_4^{(5)}(x) &&= \frac{h_4^{(5)}}{x^8} &&= \frac{x^4 b^{(3)}(x) - 16 x^3 b''(x) + 120 x^2 b'(x) - 480 x b(x) + 840 \mathcal{B} (x)}{x^8} \;\;\mathpunct{.}
\end{align*}

\noindent In the expressions above, the functions $h_i^{(j)}$ denote the numerators of the corresponding derivatives, defined by the general relation $h_i^{(j)}(x) = x^{i+j-1} \mathcal{H}_i^{(j)}(x)$.
\clearemptydoublepage

\chapter{Numerical calculation of the radiative quantities}
\label{appendice:Radiation_numerique}

\initialletter{A}s presented in Chapter~\secref{ch:chapitre2}, the computation of the integrals associated with the radiation energy, radiation flux, and radiation pressure is essential in the framework of the M1-multigroup model, as it enables a precise evaluation of its closure relation. These integrals make it possible to compute the reduced flux $\mathrm{f}_g$ and, more importantly, the Eddington factor $\chi_g$, from the Lagrange multipliers \mbox{$(\alpha_{0,g}, \beta_g)$}. An accurate determination of $\chi_g$ relies on evaluating these integrals in search algorithms, described in Appendix~\secref{appendice:algorithmes_de_recherche}. This approach is largely based on the work of Hung Chinh Nguyen (2011)~\cite{nguyen_2011_these}, which I have adapted and, in some cases, improved in order to increase accuracy when the initial estimates proved insufficient.

\begin{figure}
    \begin{center}
        \begin{minipage}[t]{0.75\linewidth}
            \centering
            \includegraphics[width=\textwidth]{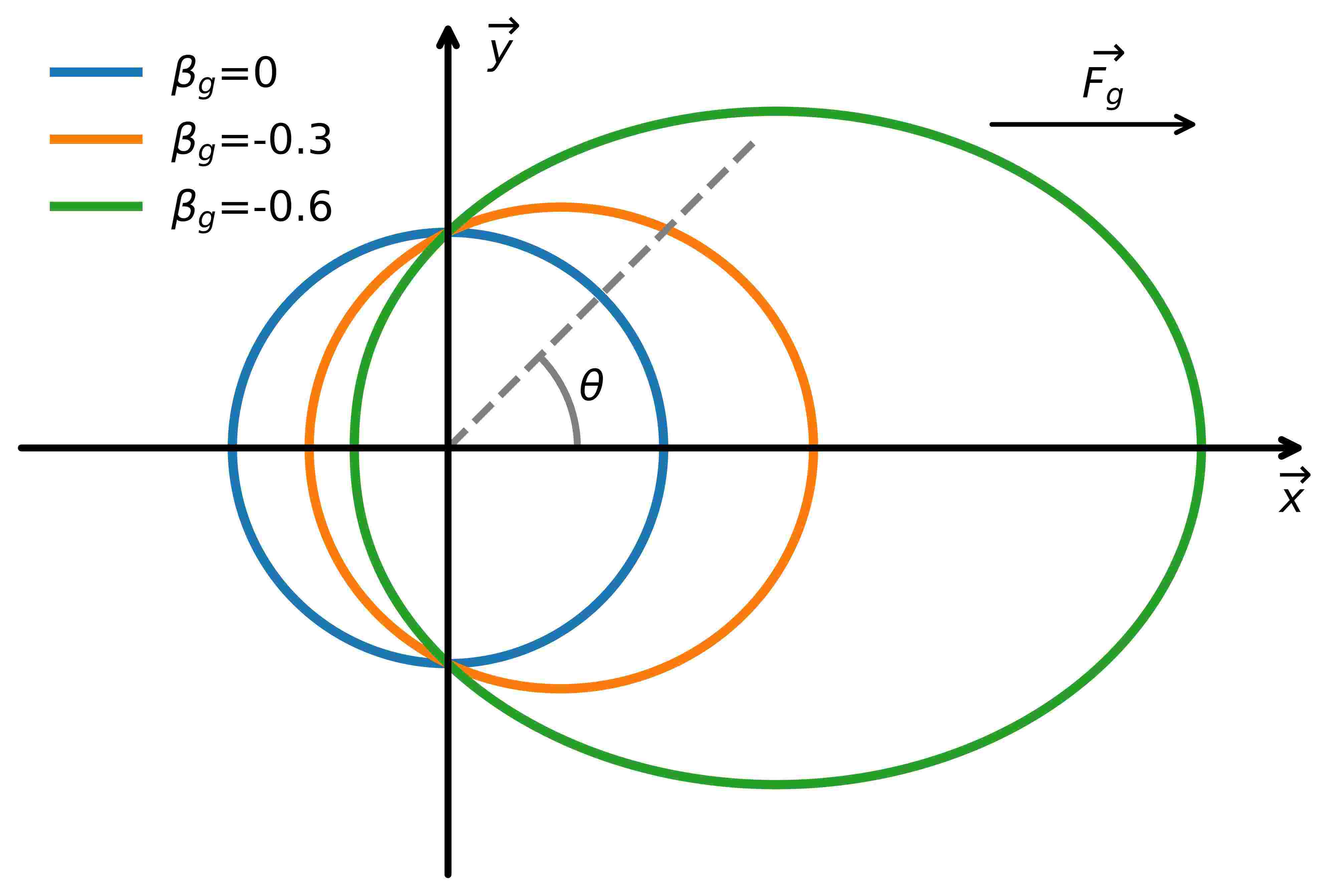}
            \caption{Form of the specific intensity $\mathcal{I}_{\nu,g}$ for different values of $\beta_g$, assuming \mbox{$h \nu \alpha_{0,g} / k_B = 1$}. The reference frame is chosen such that the $x$-axis corresponds to the direction of the radiative flux $\vectorr{F}_g$.}
            \label{fig:I_nug}
        \end{minipage}
    \end{center}
\end{figure}

The objective of this section is to present in detail the numerical computation of the integrals of the radiative quantities associated to group $g$: the energy $\mathrm{E}_g$, the flux $\vectorr{F}_g$, and the pressure $\tensorr{P}_g$. To simplify notation, we denote by $\nu_1$ and $\nu_2$ the frequency bounds of the considered group. By convention, the $x$-axis is chosen as the propagation direction of the radiative flux (see figure~\ref{fig:I_nug}). Let us first recall the expression of the specific intensity~\eqref{eq:I_specifique_mg}, which, in the framework of the M1-multigroup model, can be written as:
\begin{equation*}
    \mathcal{I}_\nu(\theta ; \alpha_{0,g}, \beta_g) = \sum_{g=1}^{\mathcal{G}} \indicatrix_g(\nu) ~\mathcal{I}_{\nu,g}(\theta ; \alpha_{0,g}, \beta_g) \;\;\mathpunct{,}
\end{equation*}

\noindent where $\mathcal{I}_{\nu,g}$ denotes the specific intensity within frequency group $g$, given by:
\begin{equation*}
     \mathcal{I}_{\nu,g}(\theta ; \alpha_{0,g}, \beta_g) = \frac{2 h \nu^3}{c^2} \left [ \exp\left ( \frac{h \nu}{k_B} \alpha_{0,g} (1 + \beta_g \cos(\theta)) \right ) -1 \right ]^{-1} \;\;\mathpunct{.}
\end{equation*}

The parameters $\alpha_{0,g}$ and $\beta_g$ are the Lagrange multipliers associated with group $g$, with $\alpha_{0,g} \in \mathbb{R}^+$ and $\beta_g \in \openinterv{-1}{1}$. The angle $\theta$ denotes the angle between a considered direction and the propagation axis of the radiative flux (see figure~\ref{fig:I_nug}). In the adopted reference frame, the radiative flux and radiative pressure are expressed as follows:
\begin{equation*}
    \begin{cases}
        \vectorr{F_g} &= F_g~\vectorr{x} = \sgn{F_g} \mathrm{f}_g~c~\mathrm{E}_g~\vectorr{x} \;\;\mathpunct{,}\\
        \tensorr{P}_g &= 
        \begin{bmatrix}
            \mathrm{P}_g & 0 \\
            0             & \mathrm{P}_g^*
        \end{bmatrix} = 
        \begin{bmatrix}
            \chi_g & 0 \\
            0             & \frac{1-\chi_g}{2}
        \end{bmatrix}~\mathrm{E}_g \;\;\mathpunct{,}
    \end{cases}
\end{equation*}

\noindent where $\mathrm{f}_g$ and $\chi_g$ denote the reduced flux and the Eddington factor, respectively. These two quantities vary within the following intervals: $\mathrm{f}_g \in [0;1[$ and $\chi_g \in [1/3;1[$. As indicated in equations~\eqref{eq:Er_g_calc}, \eqref{eq:Fr_g_calc}, and~\eqref{eq:Pr_g_calc}, the computation of the radiative quantities can then be written as:
\begin{equation*}
    \begin{cases}
        \mathrm{E}_g &= \frac{4 h \pi}{c^3} \int \nolimits_{\nu_1}^{\nu_2} \int \nolimits_{-1}^{1} \nu^3 \left [ \exp\left (\frac{h \nu}{k_B} \alpha_{0,g} ( 1+ \beta_g \mu )\right) - 1 \right ]^{-1}~\mathrm{d}\mu~\mathrm{d}\nu \;\;\mathpunct{,}\\
        F_g &= \frac{4 h \pi}{c^2} \int \nolimits_{\nu_1}^{\nu_2} \int \nolimits_{-1}^{1} \nu^3 \mu \left [ \exp\left (\frac{h \nu}{k_B}  \alpha_{0,g} ( 1+ \beta_g \mu )\right) - 1 \right ]^{-1}~\mathrm{d}\mu~\mathrm{d}\nu  \;\;\mathpunct{,}\\
        \mathrm{P}_g &= \frac{4 h \pi}{c^3} \int \nolimits_{\nu_1}^{\nu_2} \int \nolimits_{-1}^{1} \nu^3 \mu^2 \left [ \exp\left (\frac{h \nu}{k_B}  \alpha_{0,g} ( 1+ \beta_g \mu )\right) - 1 \right ]^{-1}~\mathrm{d}\mu~\mathrm{d}\nu  \;\;\mathpunct{.}
    \end{cases}
\end{equation*}

These expressions can also be rewritten in a dimensionless form by introducing the following redimensionalized quantities:
\begin{equation*}
    \begin{cases}
        \widehat{\mathrm{E}}_g &= \frac{1}{2} \int \nolimits_{-1}^{1} \frac{\mathcal{B}(x_1 ; x_2)}{\hat{\alpha}_0^4 (1 + \beta_g \mu)^4}~\mathrm{d}\mu \;\;\mathpunct{,} \\
        \widehat{F_g} &= \frac{1}{2} \int \nolimits_{-1}^{1} \mu \frac{\mathcal{B}(x_1 ; x_2)}{\hat{\alpha}_0^4 (1 + \beta_g \mu)^4}~\mathrm{d}\mu  \;\;\mathpunct{,} \\
        \widehat{\mathrm{P}}_g &= \frac{1}{2} \int \nolimits_{-1}^{1} \mu^2 \frac{\mathcal{B}(x_1 ; x_2)}{\hat{\alpha}_0^4 (1 + \beta_g \mu)^4}~\mathrm{d}\mu  \;\;\mathpunct{.}
    \end{cases}
\end{equation*}

\noindent Here, the redimensionalized quantities are defined as follows:
\begin{align*}
    \widehat{\mathrm{E}}_g = \mathrm{E}_g/a_R \refmark{T}^4 \;\;\mathpunct{,} && \widehat{F_g} = F_g/a_R c \refmark{T}^4 \;\;\mathpunct{,} && \widehat{\mathrm{P}}_g = \mathrm{P}_g/a_R \refmark{T}^4 \;\;\mathpunct{,} && \hat{\alpha}_0 = \alpha_{0,g} \refmark{T} \;\;\mathpunct{,}
\end{align*}

where $\refmark{T}$ denotes a reference temperature and $a_R$ the radiation constant. $\mathcal{B}$ is the integral of the Planck function $b$ evaluated between $x_1$ and $x_2$ (see Appendix~\secref{appendice:fct_Planck}). The bounds $x_1$ and $x_2$ of the integral, are given by \mbox{$x_1 = \hat{\nu_1} \alpha_0 \left( 1 + \beta_g \mu \right)$} and \mbox{$x_2 = \hat{\nu_2} \alpha_0 \left( 1 + \beta_g \mu \right)$}, where \mbox{$\hat{\nu} = h\nu / (k_B \refmark{T})$}. To simplify notations, we will hereafter omit the \quotes{$\hat{}$} when referring to dimensionless quantities.

Finally, introducing $z = \nu \alpha_0 (1 + \beta_g \mu)$, the dimensionless radiative quantities can be expressed as follows:
\begin{equation*}
    \begin{cases}
        \mathrm{E}_g &= E_{\nu_2} - E_{\nu_1} \;\;\mathpunct{,} \\
        F_g &= F_{\nu_2} - F_{\nu_1} \;\;\mathpunct{,} \\
        \mathrm{P}_g &= \mathrm{P}_{\nu_2} - \mathrm{P}_{\nu_1}  \;\;\mathpunct{,}
    \end{cases}
\end{equation*}

\noindent where~:
\begin{align*}
    E_{\nu_j} &= \frac{\nu_j^3}{2 \alpha_0 \beta_g} \int \nolimits_{z_1(\nu_j)}^{z_2(\nu_j)} \frac{\mathcal{B}(z)}{z^4}~\mathrm{d}z \;\;\mathpunct{,}\\
    F_{\nu_j} &= \frac{\nu_j^2}{2 \alpha_0^2 \beta_g^2} \int \nolimits_{z_1(\nu_j)}^{z_2(\nu_j)} \frac{\mathcal{B}(z)}{z^3}~\mathrm{d}z - \frac{\nu_j^3}{2 \alpha_0 \beta_g^2} \int \nolimits_{z_1(\nu_j)}^{z_2(\nu_j)} \frac{\mathcal{B}(z)}{z^4}~\mathrm{d}z \;\;\mathpunct{,}\\
    \mathrm{P}_{\nu_j} &= \frac{\nu_j}{2 \alpha_0^3 \beta_g^3} \int \nolimits_{z_1(\nu_j)}^{z_2(\nu_j)} \frac{\mathcal{B}(z)}{z^2}~\mathrm{d}z - \frac{2 \nu_j^2}{2 \alpha_0^2 \beta_g^3} \int \nolimits_{z_1(\nu_j)}^{z_2(\nu_j)} \frac{\mathcal{B}(z)}{z^3}~\mathrm{d}z + \frac{\nu_j^3}{2 \alpha_0 \beta_g^3} \int \nolimits_{z_1(\nu_j)}^{z_2(\nu_j)} \frac{\mathcal{B}(z)}{z^4}~\mathrm{d}z  \;\;\mathpunct{.}
\end{align*}

\noindent and where $z_1(\nu) = \alpha_0 \nu (1 - \beta_g)$ and $z_2(\nu) = \alpha_0 \nu (1 + \beta_g)$. Here, one recognizes the functions $\mathcal{H}_2$, $\mathcal{H}_3$, and $\mathcal{H}_4$, introduced in Appendix~\secref{subappendice:H234_planck}. These terms can also be rewritten in the following form:
\begin{align*}
    E_{\nu_j} &= \frac{\mathcal{E}\left (z_1(\nu_j) ; z_2(\nu_j) \right ) }{2\alpha_0^4 \beta_g}  \;\;\mathpunct{,}\\
    F_{\nu_j} &= \frac{\mathcal{F}(z_1(\nu_j) ; z_2(\nu_j))}{2 \alpha_0^4 \beta_g^2}  \;\;\mathpunct{,}\\
    \mathrm{P}_{\nu_j} &= \frac{\mathcal{P}(z_1(\nu_j) ; z_2(\nu_j))}{2 \alpha_0^4 \beta_g^3} \;\;\mathpunct{.}
\end{align*}

\noindent Thus, the reduced flux and the Eddington factor can be expressed as:
\begin{align}
    \mathrm{E}_g &= \frac{\mathcal{E}\left (z_1(\nu_2) ; z_2(\nu_2) \right ) - \mathcal{E}\left (z_1(\nu_1) ; z_2(\nu_1) \right )}{2\alpha_0^4 \beta_g}   \;\;\mathpunct{,} \label{eq:Eg_num} \\
    \mathrm{f}_g &= \left | \frac{1}{\beta_g} \frac{\mathcal{F}(z_1(\nu_2) ; z_2(\nu_2)) - \mathcal{F}(z_1(\nu_1) ; z_2(\nu_1))}{\mathcal{E}\left (z_1(\nu_2) ; z_2(\nu_2) \right ) - \mathcal{E}\left (z_1(\nu_1) ; z_2(\nu_1) \right )} \right | \;\;\mathpunct{,} \label{eq:fg_num} \\
    \chi_g &= \frac{1}{\beta_g^2} \frac{\mathcal{P}(z_1(\nu_2) ; z_2(\nu_2)) - \mathcal{P}(z_1(\nu_1) ; z_2(\nu_1))}{\mathcal{E}\left (z_1(\nu_2) ; z_2(\nu_2) \right ) - \mathcal{E}\left (z_1(\nu_1) ; z_2(\nu_1) \right )}  \;\;\mathpunct{,} \label{eq:chig_num}
\end{align}

\begin{figure}
    \begin{center}
        \begin{minipage}[t]{0.75\linewidth}
            \centering
            \includegraphics[width=\textwidth]{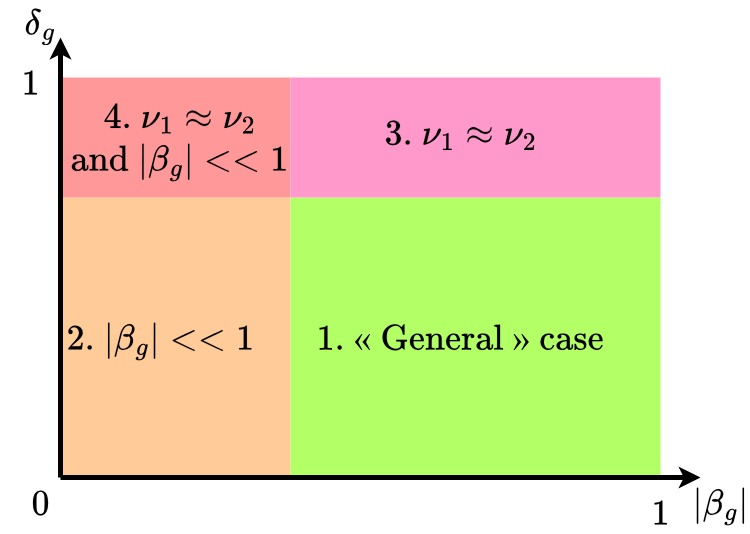}
            \caption{Different cases for computing the quantities $\mathrm{E}_g$, $\mathrm{f}_g$, and $\chi_g$. $\delta_g$ denotes the group narrowness and depends on the frequency bounds of the group according to the expression $\delta_g = \nu_1 / \nu_2$.}
            \label{fig:domains_calc}
        \end{minipage}
    \end{center}
\end{figure}

where the functions $\mathcal{E}$, $\mathcal{F}$, and $\mathcal{P}$ are defined by equations~\eqref{eq:Eg_int}, \eqref{eq:fg_int}, and \eqref{eq:chig_int}. Four cases can be distinguished for the computation of these quantities (see figure~\ref{fig:domains_calc}):
\begin{enumerate}
    \item \textbf{\quotes{General} case:} This corresponds to the non-isotropic case (i.e., $|\beta_g| > 0$) in which the frequency bounds of the group, $\nu_1$ and $\nu_2$, are well separated;

    \item \textbf{Case close to isotropy ($\boldsymbol{\beta_g \ll 1}$):} In this case, since $z_1(\nu) \approx z_2(\nu)$, the terms are generally well computed thanks to the expansions presented in Appendix~\secref{appendice:fct_Planck}, except when $z_1(\nu_1) < 1$ and $z_2(\nu_1) > 1$ and/or $z_1(\nu_2) < 1$ and $z_2(\nu_2) > 1$\footnote{The threshold value $x^*$ used here is 1.}.  
    We however assume that the frequency bounds $\nu_1$ and $\nu_2$ remain well separated;

    \item \textbf{Case close to the spectral limit ($\boldsymbol{\nu_1 \approx \nu_2}$):} In this situation, differences according to the frequencies bounds, such as $\mathcal{E}\!\left(z_1(\nu_2); z_2(\nu_2)\right) - \mathcal{E}\!\left(z_1(\nu_1); z_2(\nu_1)\right)$, become very small and are likely to be numerically inaccurate;

    \item \textbf{Case close to both isotropy and the spectral limit:} In this case, the difficulties of the two previous cases occur simultaneously.
\end{enumerate}

In this work, I addressed only Cases~1 and~2, while Cases~3 and~4, which correspond to regimes close to the spectral limit, were not considered. This limitation required restricting the group narrowness $\delta_g$ to values smaller than 0.9999. I now present the method used to compute the quantities $\mathrm{E}_g$, $\mathrm{f}_g$, and $\chi_g$, first in a general framework and then under an approximation close to isotropy.

\section{\quotes{General} case} \label{subappendice:general}

To accurately compute the quantities $\mathrm{E}_g$, $\mathrm{f}_g$, and $\chi_g$, we replace the functions $\mathcal{E}$, $\mathcal{F}$, and $\mathcal{P}$ by the functions $\mathcal{E}^*$, $\mathcal{F}^*$, and $\mathcal{P}^*$, defined by the equations (see Appendix~\secref{subappendice:H234_planck}):
\begin{align*}
    \mathcal{E}^* (x_1 ; x_2) &= \delta^{-1} e^{x_1} \mathcal{E} (x_1 ; x_2)  \;\;\mathpunct{,}\\
    \mathcal{F}^* (x_1 ; x_2) &= \delta^{-3} e^{x_1} \mathcal{F} (x_1 ; x_2) \;\;\mathpunct{,}\\
    \mathcal{P}^* (x_1 ; x_2) &= \delta^{-3} e^{x_1} \mathcal{P} (x_1 ; x_2) \;\;\mathpunct{.}
\end{align*}

\noindent The equations~\eqref{eq:Eg_num}, \eqref{eq:fg_num}, and \eqref{eq:chig_num} can then be rewritten as:
\begin{align}
    \mathrm{E}_g &= \frac{e^{-z_1(\nu_1)}}{2\alpha_0^4} \left [e^{-X_{1{,}2}} \mathcal{E}^*\left (z_1(\nu_2) ; z_2(\nu_2) \right ) - \mathcal{E}^*\left (z_1(\nu_1) ; z_2(\nu_1) \right ) \right ] \;\;\mathpunct{,} \label{eq:Eg_num2} \\
    \mathrm{f}_g &= \left | \beta_g \frac{e^{-X_{1{,}2}} \mathcal{F}^*(z_1(\nu_2) ; z_2(\nu_2)) - \mathcal{F}^*(z_1(\nu_1) ; z_2(\nu_1))}{e^{-X_{1{,}2}} \mathcal{E}^*\left (z_1(\nu_2) ; z_2(\nu_2) \right ) -\mathcal{E}^*\left (z_1(\nu_1) ; z_2(\nu_1) \right )} \right | \;\;\mathpunct{,} \label{eq:fg_num2} \\
    \chi_g &= \frac{e^{-X_{1{,}2}} \mathcal{P}^*(z_1(\nu_2) ; z_2(\nu_2)) - \mathcal{P}^*(z_1(\nu_1) ; z_2(\nu_1))}{e^{-X_{1{,}2}} \mathcal{E}^*\left (z_1(\nu_2) ; z_2(\nu_2) \right ) - \mathcal{E}^*\left (z_1(\nu_1) ; z_2(\nu_1) \right )} \;\;\mathpunct{,} \label{eq:chig_num2}
\end{align}

\noindent where $X_{1,2} = \alpha_0 (\nu_2 - \nu_1)(1 - \beta_g)$. For the computation of the functions $\mathcal{E}^*$, $\mathcal{F}^*$, and $\mathcal{P}^*$, we rely on the expressions developed in Appendix~\secref{appendice:fct_Planck}. This formulation offers several numerical advantages: it avoids any division by $\beta_g$, thereby eliminating the risk of divergence or error amplification when $\beta_g$ becomes small; it also prevents the multiplication of terms by decaying exponentials, which are numerically rounded to zero when $\alpha_0 \nu$ is large, making the calculations infeasible. However, in cases where $z_1(\nu_1) < 1 < z_2(\nu_1)$ or $z_1(\nu_2) < 1 < z_2(\nu_2)$, the expressions used inevitably involve divisions by $\beta_g$. These configurations lead to amplified numerical errors and a significant loss of accuracy as $\beta_g$ tends to zero, as illustrated in figures~\ref{fig:fg_noDL} and \ref{fig:chig_noDL}. To overcome this limitation, it is necessary to resort to an alternative method that is better suited to this regime.

\section{Case close to isotropy} \label{subappendice:isotrope}

In the near-isotropic case, $\beta_g$ takes very small values, and thus one can compute the Taylor expansion of the radiation energy, the reduced flux, and the Eddington factor with respect to $\beta_g$. Introducing $\xi_j = \alpha_0 \nu_j$, the Taylor expansions of the integral functions $\mathcal{H}_2$, $\mathcal{H}_3$, and $\mathcal{H}_4$ are:
\begin{align*}
    \mathcal{H}_2(z_1(\nu_j) ; z_2(\nu_j)) &= \frac{2~\beta_g}{\xi_j} \mathcal{B}(\xi_j) + \frac{\beta_g^3}{3 \xi_j} h_2^{(3)}(\xi_j) + \frac{\beta_g^5}{60 \xi_j} h_2^{(5)}(\xi_j) + \mathcal{O}(\beta_g^6) \;\;\mathpunct{,} \\
    \mathcal{H}_3(z_1(\nu_j) ; z_2(\nu_j)) &= \frac{2~\beta_g}{\xi_j^2} \mathcal{B}(\xi_j) + \frac{\beta_g^3}{3 \xi_j^2} h_3^{(3)}(\xi_j) + \frac{\beta_g^5}{60 \xi_j^2} h_3^{(5)}(\xi_j) + \mathcal{O}(\beta_g^6) \;\;\mathpunct{,} \\
    \mathcal{H}_4(z_1(\nu_j) ; z_2(\nu_j)) &= \frac{2~\beta_g}{\xi_j^3}~\mathcal{B}(\xi_j) + \frac{\beta_g^3}{3 \xi_j^3} h_4^{(3)}(\xi_j) + \frac{\beta_g^5}{60 \xi_j^3} h_4^{(5)}(\xi_j) + \mathcal{O}(\beta_g^6) \;\;\mathpunct{,}
\end{align*}

\noindent where the functions $h_i^{(j)}$ are defined in Appendix~\secref{subappendice:H234_planck}.  
Thus, the Taylor expansions of the functions $\mathcal{E}$, $\mathcal{F}$, and $\mathcal{P}$ can be written as:
\begin{align*}
    \mathcal{E}(z_1(\nu_j) ; z_2(\nu_j)) &= 2~\beta_g~\mathcal{B}(\xi_j) + \frac{\beta_g^3}{3} h_4^{(3)}(\xi_j) + \frac{\beta_g^5}{60} h_4^{(5)}(\xi_j) + \mathcal{O}(\beta_g^6) \;\;\mathpunct{,} \\
    \mathcal{F}(z_1(\nu_j) ; z_2(\nu_j)) &=\frac{\beta_g^3}{3} f_3(\xi_j) + \frac{\beta_g^5}{60} f_5(\xi_j) + \mathcal{O}(\beta_g^6) \;\;\mathpunct{,} \\
    \mathcal{P}(z_1(\nu_j) ; z_2(\nu_j)) &= \frac{2 \beta_g^3}{3} \mathcal{B}(\xi_j) + \frac{\beta_g^5}{60} p_5(\xi_j) + \mathcal{O}(\beta_g^6) \;\;\mathpunct{,}
\end{align*}

\noindent where the auxiliary functions $f_i$ and $p_i$ are given by:
\begin{align*}
    &f_3(x) &&= h_4^{(3)}(x) - h_3^{(3)} (x) &&= 2x b(x) - 8 \mathcal{B}(x) \;\;\mathpunct{,} \\
    &f_5(x) &&= h_4^{(5)}(x) - h_3^{(5)} (x) &&=  4 \left [ x^3 b''(x) - 12 x^2 b'(x) + 60 x b(x) - 120 \mathcal{B} (x) \right ] \;\;\mathpunct{,} \\
    &p_5(x) &&=  h_2^{(5)}(x) - 2 h_3^{(5)}(x) + h_4^{(5)}(x) &&= 12 \left [ x^2 b'(x) - 8 x b(x) + 20 \mathcal{B}(x) \right ] \;\;\mathpunct{,} 
\end{align*}

\begin{figure}
    \begin{subfigure}[t]{0.49\textwidth}
        \centering
        \includegraphics[width=\textwidth]{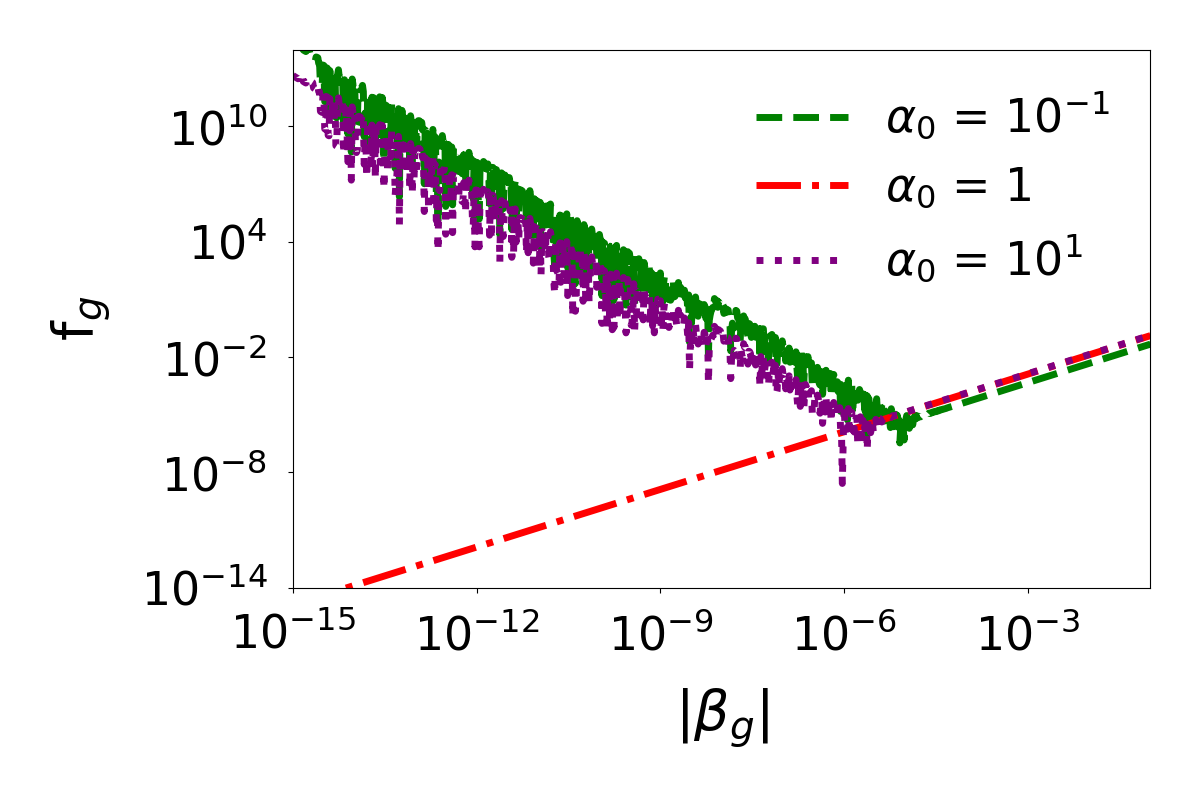}
        \caption{Evolution of the reduced flux without using the Taylor expansions for small values of $|\beta_g|$.}
        \label{fig:fg_noDL}
    \end{subfigure}
    \hfill
    \begin{subfigure}[t]{0.49\textwidth}
        \centering
        \includegraphics[width=\textwidth]{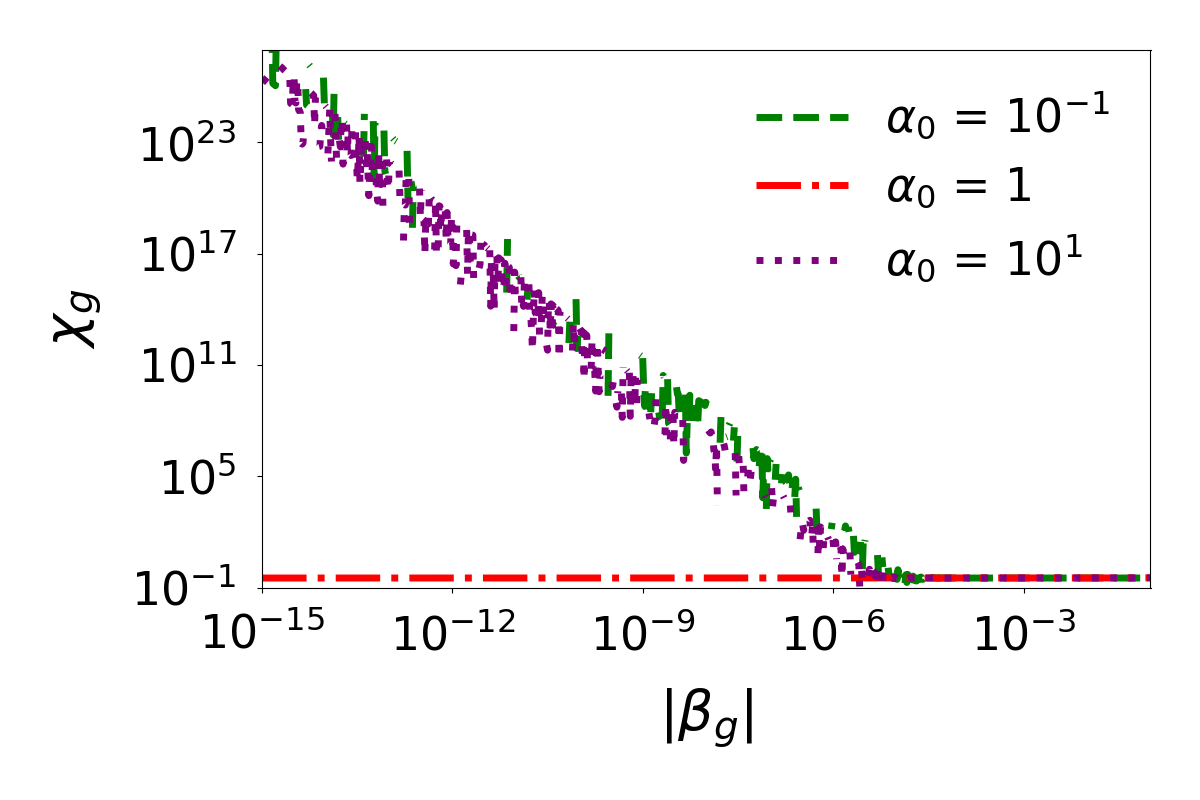}
        \caption{Evolution of the Eddington factor without using the Taylor expansions for small values of $|\beta_g|$.}
        \label{fig:chig_noDL}
    \end{subfigure}
    \hfill
    \begin{subfigure}[t]{0.49\textwidth}
        \centering
        \includegraphics[width=\textwidth]{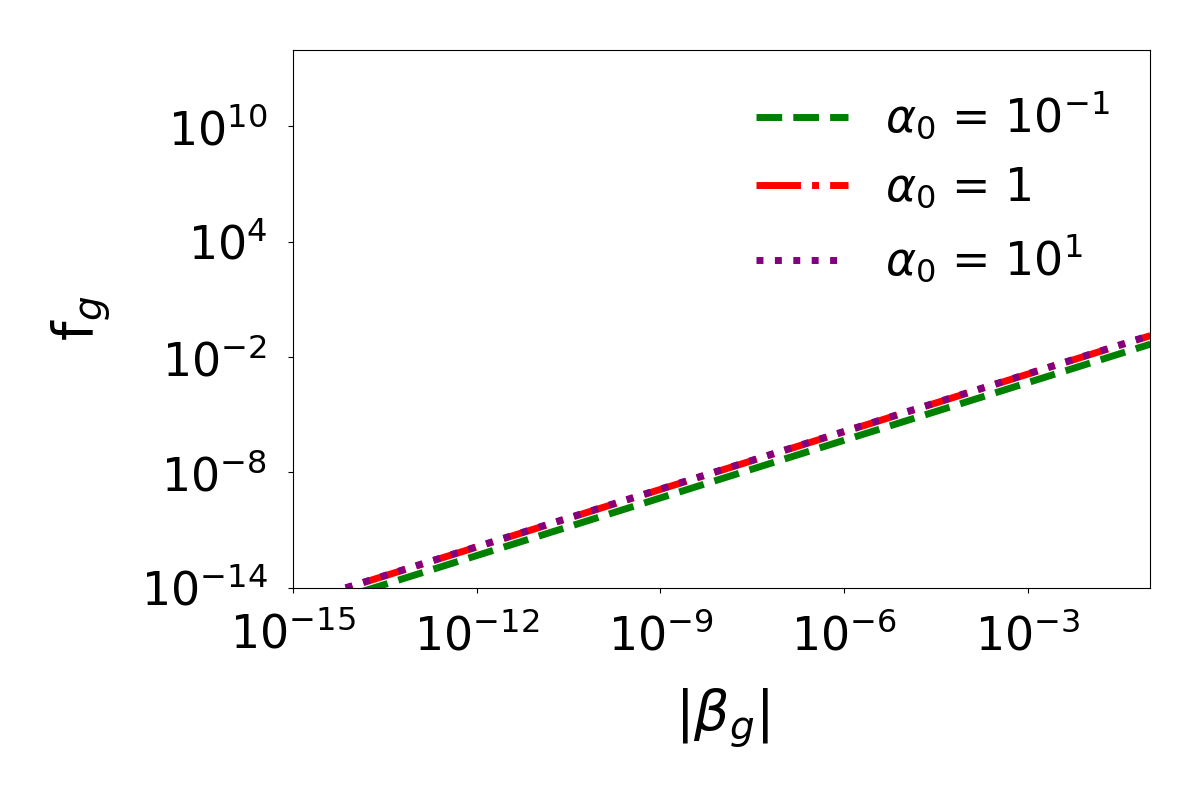}
        \caption{Evolution of the reduced flux using the Taylor expansions for small values of $|\beta_g|$.}
        \label{fig:fg_DL}
    \end{subfigure}
    \hfill
    \begin{subfigure}[t]{0.49\textwidth}
        \centering
        \includegraphics[width=\textwidth]{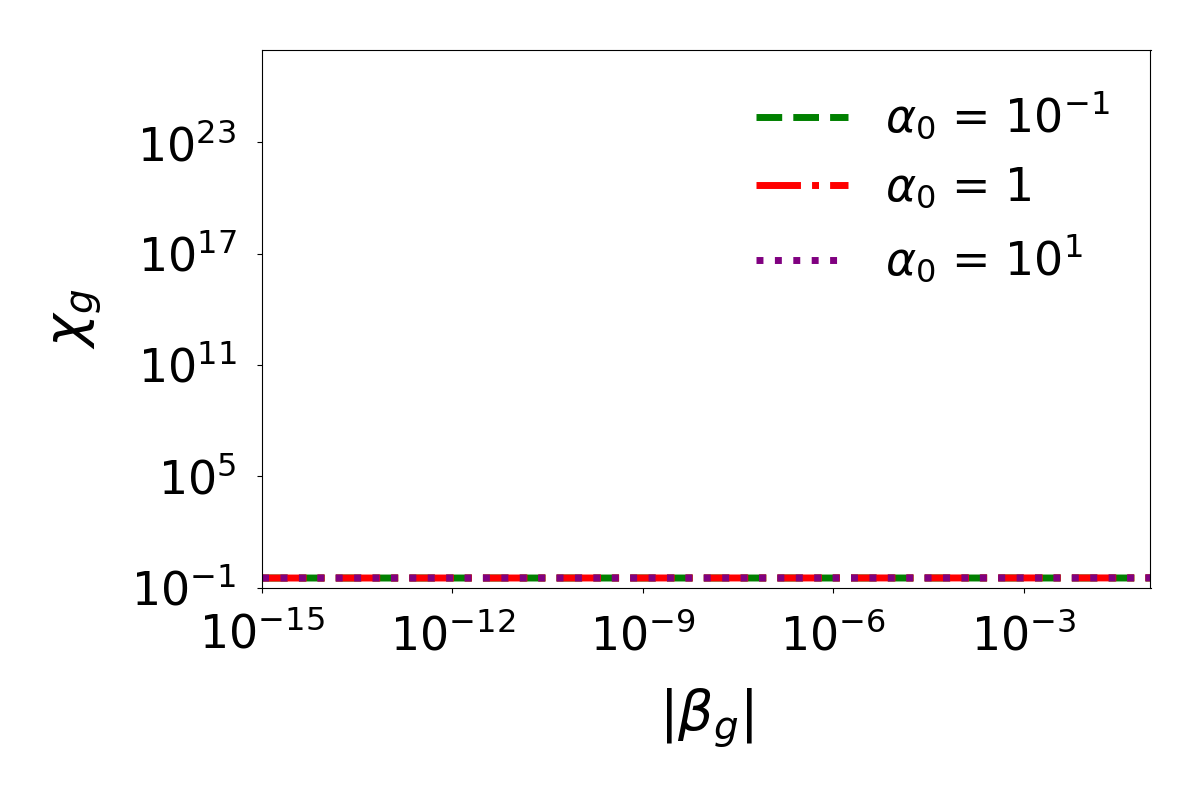}
        \caption{Evolution of the Eddington factor using the Taylor expansions for small values of $|\beta_g|$.}
        \label{fig:chig_DL}
    \end{subfigure}
    \caption{Evolution of the reduced flux and the Eddington factor as functions of $|\beta_g|$, with and without the use of the Taylor expansions presented in Section~\secref{subappendice:isotrope}, for different values of $\alpha_0$ and using the frequency bounds \mbox{$\nu_1 = 0.1$} and \mbox{$\nu_2 = 10$}.}
    \label{fig:DL_corr}
\end{figure}

\noindent and the successive derivatives of the Planck function $b$ are given in Appendix~\secref{subappendice:derivees_planck}. Thus, a second-order Taylor expansion of the quantities of interest $\mathrm{E}_g$, $\mathrm{f}_g$, and $\chi_g$ can be derived:
\begin{align}
    \mathrm{E}_g &= \frac{1}{\alpha_0^4} \left \{ \mathcal{B}(\xi_2) - \mathcal{B}(\xi_1)  + \frac{\beta_g^2}{6} \left ( h_4^{(3)}(\xi_2) - h_4^{(3)}(\xi_1) \right ) \right \}\\
    \mathrm{f}_g &= \left |\frac{\beta_g}{6} \frac{f_3 (\xi_2)-f_3 (\xi_1)}{\mathcal{B}(\xi_2)-\mathcal{B}(\xi_1)} \right |\\
    \chi_g &= \frac{1}{3} \left \{ 1 + \frac{\beta_g^2}{120} \frac{3 \left [ p_5(\xi_2)-p_5(\xi_1) \right ] - 20 \left [ h_4^{(3)}(\xi_2) - h_4^{(3)}(\xi_1) \right ]}{\mathcal{B}(\xi_2)-\mathcal{B}(\xi_1)} \right \}
\end{align}

As illustrated in figure~\ref{fig:DL_corr}, using the Taylor expansions prevents the amplification of numerical errors when $|\beta_g|$ is small. In practice, these expressions are therefore used to evaluate the radiative quantities $\mathrm{E}_g$, $\mathrm{f}_g$, and $\chi_g$ in this regime.
\clearemptydoublepage

\chapter{Computing the Eddington factor using search algorithms}
\label{appendice:algorithmes_de_recherche}

\initialletter{I}n a simulation performed with the \gls{hades} code, based on the M1-multigroup model, each spectral group $g$ is characterized at a given time by two fundamental quantities: the reduced flux $\mathrm{f}_g$ and the radiative energy $\mathrm{E}_g$. One of the main objectives of the model is to determine, for each group, the Eddington factor $\chi_g$, which closes the radiative system by relating the radiative pressure tensor to the radiative energy. The three quantities, $\mathrm{E}_g$, $\mathrm{f}_g$, and $\chi_g$, are not independent. Indeed, they can all be expressed in terms of two fundamental parameters, denoted $\alpha_{0,g}$ and $\beta_g$, known as Lagrange multipliers. These multipliers naturally arise within a variational framework that consists in minimizing the entropy of the radiation field under constraints (energy and flux conservation), in accordance with the minimum entropy principle. The associated numerical procedure, which allows one to evaluate the integral expressions involving these multipliers, is detailed in Appendices~\secref{appendice:fct_Planck} and~\secref{appendice:Radiation_numerique}, where the spectral integration methods and the approximations employed are presented.

According to the work of Turpault (2003)~\cite{turpault_2003_these}, it has been shown that for any physically admissible pair \mbox{($\mathrm{E}_g$, $\mathrm{f}_g$)}, there exists a unique pair of Lagrange multipliers \mbox{($\alpha_{0,g}$, $\beta_g$)} that allows one to reconstruct the radiation field compatible with these constraints. It is therefore natural to devise a search algorithm aimed at recovering this unique pair of multipliers, and subsequently deducing the corresponding Eddington factor $\chi_g$. In this section, we assume that the frequency bounds of group $g$ are fixed at $\nu_1$ and $\nu_2$. The integral expressions depend explicitly on these bounds as well as on the multipliers. For clarity, we introduce the following functional notations:
\begin{itemize}
    \item \mbox{$\mathrm{E}_g \left( \alpha_{0,g}, \beta_g \right)$}: radiation energy associated to group $g$,
    \item \mbox{$\mathrm{f}_g \left( \alpha_{0,g}, \beta_g \right)$}: reduced flux associated to group $g$,
    \item \mbox{$\chi_g \left( \alpha_{0,g}, \beta_g \right)$}: corresponding Eddington factor.
\end{itemize}

These functions are defined from the spectral integrals of the radiation field and are evaluated numerically using the methods described in Appendix~\secref{appendice:Radiation_numerique}. Finally, we denote by $\mathrm{E}_g^*$ and $\mathrm{f}_g^*$ the target values of the radiation energy and the reduced flux, obtained from the simulation, for which we seek to determine the corresponding Eddington factor $\chi_g$. The search for the Lagrange multipliers therefore constitutes an essential intermediate step in the consistent calculation of this factor.

\section{Search algorithms} \label{subappendice:algorithmes_de_recherche}

Historically, a first algorithm was proposed by Hung Chinh Nguyen in his PhD thesis (2011)~\cite{nguyen_2011_these}, in the context of the development of the \gls{hades} code. This algorithm relies on a hybrid approach that combines the robustness of the bisection method with the fast convergence of Newton's method. More precisely, the bisection method is used to determine a suitable value of the Lagrange multiplier $\beta_g$, ensuring that the radiative flux constraint is satisfied, while Newton's method is then employed to efficiently compute the multiplier $\alpha_{0,g}$, associated with energy conservation. The entire procedure is summarized in the algorithm presented below.

\begin{tcolorbox}[colback=gray!5!white,colframe=blue!75!black,title=Algorithm C.1~: Bissection-Newton algorithm]
    We seek to determine the pair \mbox{($x$, $\beta$)}, with \mbox{$x=\ln(\alpha_0)$}, such that \mbox{$\ln\!\left(\mathrm{E}_g(e^x, \beta)\right) = \ln(\mathrm{E}_g^*)$} and \mbox{$\mathrm{f}_g(e^x, \beta) = \mathrm{f}_g^*$}. We denote by $i$ the index of the main iteration, by $j$ the index of the dissection or Newton method iterations, and by $x^{(i)}$ and $\beta^{(i)}$ the multipliers estimated at iteration $i$. The procedure is as follows:
    \begin{itemize}
        \item \textbf{Special case:} If $\mathrm{f}_g^* = 0$, then the Eddington factor is directly $\chi_g = 1/3$,
        \item \textbf{Otherwise}, we proceed through successive iterations:
        \begin{enumerate}
            \item \textbf{Initialization:} the multipliers are initialized from the values obtained with the M1-gray model (see Dubroca and Feugeas, 1999~\cite{dubroca_1999}).
            \item \textbf{Iterative loop:}
            \begin{itemize}
                \item \textit{Step 1 — Bisection method for $\beta^{(i+1)}$:}\\
                Fixing $x^{(i)}$, we search for $\beta^{(i+1)}$ such that \mbox{$\mathrm{f}_g\!\left(e^{x^{(i)}},\, \beta^{(i+1)}\right) = \mathrm{f}_g^*$}. The search is carried out using the bisection method over the initial interval $[0; 1[$. The search stops when the condition
                \[
                \left| \left ( \beta_{j+1}^{(i+1)} - \beta_{j}^{(i+1)} \right ) \Big/ \beta^{(i+1)}_j \right | \le \epsilon_D
                \]
                
                \item \textit{Step 2 — Newton search for $x^{(i+1)}$:}\\
                Fixing $\beta^{(i+1)}$, we search for $x^{(i+1)}$ such that \mbox{$\ln\!\left(\mathrm{E}_g\!\left(\alpha_0^{(i+1)}, \beta^{(i+1)}\right)\right) = \ln\!\left(\mathrm{E}_g^*\right)$},
                where \mbox{$\alpha_0^{(i+1)} = e^{x^{(i+1)}}$}. A Newton algorithm in $\mathbb{R}$ is used. The search stops when is met the condition 
                \[
                \left | \left ( \alpha_{0,j+1}^{(i+1)} - \alpha_{0,j}^{(i+1)} \right ) \Big/ \alpha_{0,j}^{(i+1)} \right | \le \epsilon_N
                \]
                
                \item \textit{Step 3 — Stopping criterion:}\\
                Steps~1 and~2 are repeated as long as the sum of the relative differences
                \[
                \left | \left (\alpha_0^{(i+1)} - \alpha_0^{(i)} \right ) \Big/ \alpha_0^{(i)} \right| + \left | \left ( \beta^{(i+1)} - \beta^{(i)} \right ) \Big/ \beta^{(i)} \right |
                \]
                remains greater than $\epsilon_T$.
            \end{itemize}
        \end{enumerate}
        \item Once convergence is reached, the corresponding Eddington factor $\chi_g$ is computed.
    \end{itemize}

    The parameters $\epsilon_D$, $\epsilon_N$, and $\epsilon_T$ are small constants setting the tolerances for the steps of the bisection method, Newton's method, and of the global algorithm, respectively. In this work, they have been set to $10^{-8}$, $10^{-8}$, and $10^{-5}$.
\end{tcolorbox}

The initial search algorithm, although functional, exhibits a notable limitation: it determines the Lagrange multipliers sequentially, estimating them one at a time. This approach slows down the computation and reduces its efficiency. To address this issue, I developed a new algorithm based on a \textit{line search} strategy, capable of estimating both multipliers, $\alpha_{0,g}$ and $\beta_g$, simultaneously. By exploiting the structure of the problem and evaluating the constraints jointly, this method significantly improves the convergence speed while maintaining good accuracy on the associated radiative quantities.
\begin{tcolorbox}[colback=gray!5!white,colframe=blue!75!black,title=Algorithm C.2~: Line search algorithm ]
    We seek to determine the pair \mbox{($x$, $y$)}, with \mbox{$x = \ln(\alpha_0)$} and \mbox{$y = y_0\, \sigma^{-1}(\beta)$}, such that the relations \mbox{$\ln\!\left(\mathrm{E}_g(e^x, \sigma(y/y_0))\right) = \ln(\mathrm{E}_g^*)$} and \mbox{$\ln\!\left(\mathrm{f}_g(e^x, \sigma(y/y_0))\right) = \ln(\mathrm{f}_g^*)$} are satisfied. We denote by $i$ the index of the main iteration, and by $x^{(i)}$ and $y^{(i)}$ the values of $x$ and $y$ estimated at iteration~$i$. The procedure is as follows:
    \begin{itemize}
        \item \textbf{Special case:} if $\mathrm{f}_g^* = 0$, then the Eddington factor is directly $\chi_g = 1/3$,
        \item \textbf{Otherwise}, we proceed through successive iterations:
        \begin{enumerate}
            \item \textbf{Initialization:} the multipliers are initialized from the values obtained with the M1-gray model (see Dubroca and Feugeas, 1999~\cite{dubroca_1999}).
            \item \textbf{Iterative loop:} we search for $x^{(i+1)}$ and $y^{(i+1)}$ using a \textit{line search} algorithm, such that the sum of squared logarithmic residuals
            \[
            \left[\ln\!\left(\mathrm{f}_g\!\left(\alpha_0^{(i+1)}, \beta^{(i+1)}\right)\right) - \ln(\mathrm{f}_g^*)\right]^2 + \left[\ln\!\left(\mathrm{E}_g\!\left(\alpha_0^{(i+1)}, \beta^{(i+1)}\right)\right) - \ln(\mathrm{E}_g^*)\right]^2
            \]
            vanishes, where $\alpha_0^{(i+1)} = e^{x^{(i+1)}}$ and $\beta^{(i+1)} = \sigma(y^{(i+1)}/y_0)$.           
            \item \textbf{Stopping criterion:} the algorithm stops when the sum of the relative variations
            \[
            \left | \left (\alpha_{0,g}^{(i+1)} - \alpha_{0,g}^{(i)} \right )/\alpha_{0,g}^{(i)} \right | + \left | \left (\beta_g^{(i+1)} - \beta_g^{(i)} \right )/\beta_g^{(i)} \right |
            \]
            becomes smaller than a small constant $\epsilon$.
        \end{enumerate}
        \item Once convergence is achieved, the corresponding Eddington factor $\chi_g$ is computed.
    \end{itemize}

    The parameter $\epsilon$ is a small constant that sets the tolerance of the global algorithm, while $y_0$ is a normalization parameter, here set respectively to $10^{-5}$ and $100$. The function $\sigma$ denotes the sigmoid function.
\end{tcolorbox}

\section{Comparison of the search algorithms} \label{subappendice:comparaison_algorithmes_de_recherche}

\begin{figure}
    \begin{subfigure}[t]{0.48\textwidth}
        \centering
        \includegraphics[width=\textwidth]{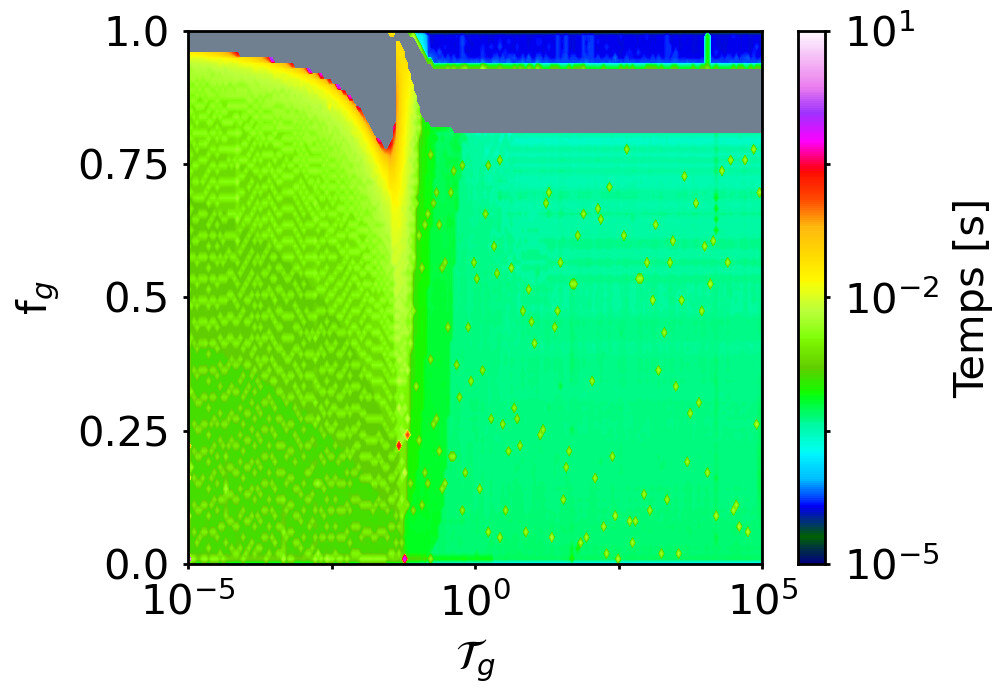}
        \caption{Computation time of the Eddington factor using Algorithm C.1: Bisection–Newton algorithm.}
        \label{fig:time_DN}
    \end{subfigure}
    \hfill
    \begin{subfigure}[t]{0.48\textwidth}
        \centering
        \includegraphics[width=\textwidth]{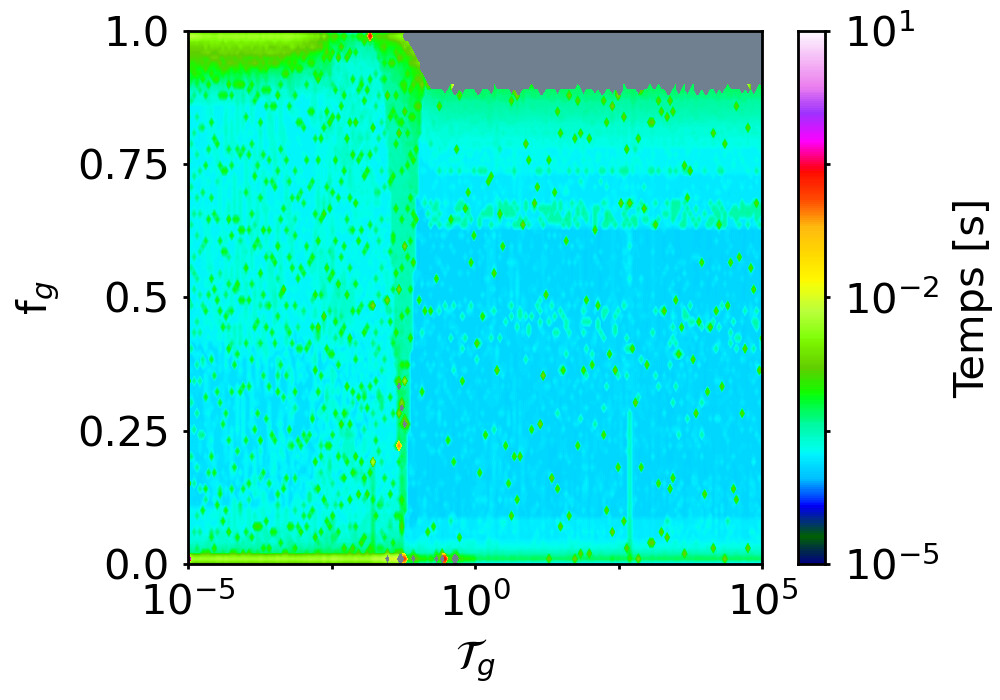}
        \caption{Computation time of the Eddington factor using Algorithm C.2: Line-search algorithm.}
        \label{fig:time_LS}
    \end{subfigure}
    \caption{Computation time of the Eddington factor as a function of the reduced flux $\mathrm{f}_g$ and the dimensionless radiative temperature $\mathcal{T}_g$, for the rescaled frequency bounds \mbox{$\nu_1 = 0.1$} and \mbox{$\nu_2 = 10$}. The results are obtained using the search algorithms C.1 and C.2. The gray areas indicate cases in which the algorithm did not converge.}
    \label{fig:time_search}
\end{figure}

The figure~\ref{fig:time_search} illustrates the computation time required to evaluate the Eddington factor as a function of the reduced flux $\mathrm{f}_g$ and the dimensionless radiative temperature $\mathcal{T}_g$\footnote{The dimensionless radiative temperature $\mathcal{T}_g$ is related to the radiation energy $\mathrm{E}_g$; its precise definition is given in Section~\secref{sec:dependence_chig}.}. A precise analysis of this figure raises two main observations:

\begin{enumerate}
    \item The \textbf{line search} algorithm yields an overall shorter computation time than the \textbf{Bisection–Newton} algorithm, while providing comparable estimates of the Eddington factor;
    \item The \textbf{line search} algorithm also proves more robust: it converges across a wide range of conditions, unlike the Bisection–Newton method, whose convergence may fail in certain regions of the parameter space. The convergence failures of the line search are mainly located in the region where $\mathrm{f}_g \approx 1$ and $\mathcal{T}_g \gtrsim 1$. However, in this specific region, the radiative quantities can be evaluated using the analytical approximations given by equations~\eqref{eq:Tg_H}, \eqref{eq:fg_H}, and \eqref{eq:chig_H}, detailed in Section~\secref{sec:dependence_chig}, which makes it possible to circumvent these numerical difficulties.
\end{enumerate}

Owing to its superior performance, both in numerical efficiency and reliability, the line search algorithm was therefore adopted for all computations presented in this work.
\clearemptydoublepage

\chapter{Hydrodynamic shock on a wall}
\label{appendice:choc}

\begin{figure}
    \centering
    \includegraphics[width=0.6\textwidth]{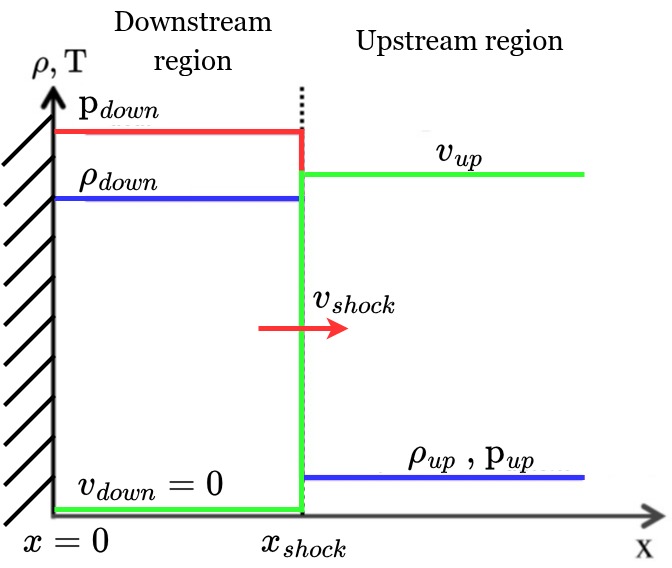}
    \caption{Diagram representing the structure of a hydrodynamic shock propagating from a wall, located in $x=0$ at a given moment.}
    \label{fig:wall_shock_explain}
\end{figure}

\initialletter{I}n this appendix, I develop the analytical solution of the hydrodynamic case used as a reference in chapter~\secref{ch:chapitre4}. The physical configuration of the problem is recalled in figure~\ref{fig:wall_shock_explain}.

We first determine the hydrodynamic quantities in the downstream region from the upstream conditions. To this end, we use the Rankine–Hugoniot jump relations. In the particular case where the fluid velocity in the downstream region is zero ($v_{down} = 0$), these relations are written:
\begin{align}
    (\rho_{up} - \rho_{down}) v_{shock}  &= \rho_{up} v_{up} \;\;\mathpunct{,} \\
    \rho_{up} v_{shock} v_{up} &= \rho_{up} v_{up}^2 + \mathrm{p}_{up} - \mathrm{p}_{down} \;\;\mathpunct{,} \\
    (\mathrm{E}_{up} - \mathrm{E}_{down}) v_{shock} &= (\mathrm{E}_{up} + \mathrm{p}_{up}) v_{up} \;\;\mathpunct{,}
\end{align}

\noindent where $E$ denotes the total energy of the fluid, given by the expression:
\begin{equation}
    \mathrm{E} = \frac{1}{2} \rho v^2 + \frac{p}{\gamma-1} \;\;\mathpunct{,}
\end{equation}

The resolution of this system makes it possible to express the velocity of the shock $v_{shock}$, as well as the density and pressure in the downstream medium, as a function of the upstream quantities. We obtain:

\begin{align}
    v_{shock} &= -\frac{(\gamma-3) \mathrm{M}_{lab} + \sqrt{\left \{ (\gamma+1) \mathrm{M}_{lab} \right \}^2 + 16}}{4 \mathrm{M}_{lab}} v_{up} \;\;\mathpunct{,} \\
    \rho_{down} &= \frac{(\gamma+1) \mathrm{M}_{lab} + \sqrt{\left \{ (\gamma+1) \mathrm{M}_{lab} \right \}^2 + 16}}{(\gamma-3) \mathrm{M}_{lab} + \sqrt{\left \{ (\gamma+1) \mathrm{M}_{lab} \right \}^2 + 16}} \rho_{up} \;\;\mathpunct{,} \\
    \mathrm{p}_{down} &= \frac{4 + \gamma \mathrm{M}_{lab} \left \{ (\gamma+1) \mathrm{M}_{lab} + \sqrt{\left ( (\gamma+1) \mathrm{M}_{lab} \right \}^2 + 16} \right \}}{4} \mathrm{p}_{up} \;\;\mathpunct{,}
\end{align}

\noindent where the parameter $M_{lab}$ (analogous to a Mach number, but expressed in the laboratory reference frame) is defined by:
\begin{equation}
    \mathrm{M}_{lab} = \frac{|v_{up}|}{\sqrt{\gamma p_{up}/\rho_{up}}} \;\;\mathpunct{.}
\end{equation}

\noindent This number should not, however, be confused with the Mach number of the shock itself, which characterizes the shock strength and is given by:
\begin{equation}
    \mathrm{M} = \frac{|v_{up} - v_{shock}|}{\sqrt{\gamma p_{up}/\rho_{up}}} \;\;\mathpunct{,}
\end{equation}

Assuming that the fluid behaves as an ideal gas, the downstream temperature can also be expressed as a function of the upstream temperature:
\begin{equation}
    \mathrm{T}_{down} = \frac{p_{down}}{p_{up}} \frac{\rho_{up}}{\rho_{down}} \mathrm{T}_{up}
\end{equation}

Finally, denoting by $t_0$ the initial time of the simulation, and by $\mathcal{H}$ the Heaviside function (equal to $1$ for $x \geq 0$ and $0$ otherwise), the full analytical solution of this problem is given by:
\begin{equation}
    \begin{cases}
        \rho &= \rho_{down} + (\rho_{up}-\rho_{down}) \mathcal{H} [x - v_{shock} (t-t_0)] \;\;\mathpunct{,}\\
        v &= v_{down} + (v_{up}-v_{down}) \mathcal{H} [x - v_{shock} (t-t_0)] \;\;\mathpunct{,}\\
        \mathrm{T} &= \mathrm{T}_{down} + (\mathrm{T}_{up}-\mathrm{T}_{down}) \mathcal{H} [x - v_{shock} (t-t_0)] \;\;\mathpunct{.}\\
    \end{cases}
\end{equation}
\clearemptydoublepage

\refstepcounter{chapter} \label{ch:bibliography}
\raggedright
\nocite{*}
\printbibliography 
\clearemptydoublepage
\end{document}